Universität Hamburg
DER FORSCHUNG | DER LEHRE | DER BILDUNG

# Airport Capacity and Performance in Europe

## A study of transport economics, service quality and sustainability

Cumulative dissertation
with the aim to obtain the academic degree
of a „doctor rerum oeconomicarum (Dr. rer. oec.)“
(according to the doctoral regulations of the
Faculty of Business Administration
dated July 9, 2014)

at the Faculty of Business Administration
(Hamburg Business School)
Moorweidenstr. 18 in 20148 Hamburg, Germany
of Universität Hamburg

submitted by

Branko Bubalo

born on June 30th, 1977 in Berlin, Germany
resides at Yorckstraße 48 in 10965 Berlin.

Berlin, August 13th, 2020

AIRPORT CAPACITY AND PERFORMANCE IN EUROPE – B. BUBALO

| | |
|---|---|
| Vorsitzender: | Prof. Dr. Knut Haase |
| Erstgutachter: | Prof. Dr. Stefan Voß |
| Zweitgutachter: | Prof. Dr. Joachim Daduna |
| | |
| Datum der Disputation: | 18. Dezember 2020 |

To Claudia, Richard, and Henri

## Acknowledgements

*Quel voyage*! I would like to thank the following people for joining my journey and for their support in completing this piece of work. I had the pleasure to meet and talk to countless smart people in the field of economics and air transportation. First and foremost, I want to thank my long-time mentors Prof. Jürgen Müller and Prof. Joachim Daduna from Berlin School of Economics and Law for their continuous moral and scientific support. Prof. Daduna introduced me to my doctoral supervisor Prof. Stefan Voß from University of Hamburg, who I want to thank for his trust in my talent and for supporting my scientific endeavor. Prof. Voß helped me publish some of my initial papers and invited me to his conferences. He and his secretary Mrs. Julia Bachale always provided literature and immediate response whenever I needed their help in finding a specific source of information or for solving bureaucratic hurdles. In this context I also want to thank Mrs. Elke Thoma from graduate school for pushing us doctoral students, such as my highly regarded colleague Dr. Frederik Schulte, to the finish line by perpetual motivation and constructive help in solving even bigger bureaucratic hurdles. Mrs. Thoma made me aware of several interesting doctoral workshops, where the one at MIT in Boston, USA, held by Prof. Moshe Ben-Akiva had the greatest impact on me. From University of Hamburg I would also like to thank Prof. Knut Haase for providing last-minute support to finishing my PhD studies.

Over the years I had the pleasure to meet and talk to the most important academics in my field. As an undergraduate student I worked in a student research project together with my respected colleagues such as Dr. Tolga Ülkü and Dr. Vanessa Liebert. This project was partly managed by Prof. Hans-Martin Niemeier from University of Applied Sciences in Bremen, who I want to thank for constantly pushing us students to the scientific boundaries. At various occasions he introduced me to his professional network, such as at workshops, seminars, and especially at the ATRS and EAC conferences. At these occasions I met Prof. Hansjochen Ehmer of IUBH University of Applied Sciences Bad Honnef and Prof. Frank Fichert from Worms University of Applied Sciences who provided immediate feedback whenever I presented initial results at the early stages of my research.

I still follow the advice from great Prof. David Gillen from University of British Columbia, who was a frequent visitor of our workshops in Berlin, that we young scientists need to "be aggressive on the data" to gain insights from our datasets. At a conference I met a former colleague of Prof. Gillen, Prof. Victor Hugo Valdés Cervantes from Universidad Anáhuac México, who is the brightest, funniest, and politest person, a great motivator, and now a friend of my family.

I appreciate the fruitful and inspiring discussions with brilliant Prof. Nicole Adler from Hebrew University of Jerusalem and Prof. Gianmaria Martini, and his colleague Dr. Davide Scotti from University of Bergamo, who gave me great confidence that I was on the right track with my research. A group of scientists from University of Westminster is very noteworthy. I met Prof. Anne Graham as well as Prof. Nigel Dennis and Dr. Andrew Cook from her team on several occasions. They introduced me to the field of air transportation capacity and delay, its costs and propagation. I was delighted that Prof. Graham agreed to be my second supervisor for my diploma thesis. I want to deeply thank her for her time, efforts, and motivation.

Furthermore, I want to thank Prof. Vojin Tosić and his colleagues Dr. Bojana Mircović and Dr. Radosav Jovanović from University of Belgrade for their valuable recommendations. This group of academics showed interest in my earliest research on airport capacity and demand and saw its potential.

I thank my colleagues from my airport simulation research network. In the SIMMOD User Group Meetings, organized by an extraordinary team consisting of Dr. Christoph Schneider from Munich airport, Mr. Karl-Heinz Keller from SITA and Stefan Wendeberg from Frankfurt airport, I gained valuable insights about the application of simulations to airport capacity related problems. The first persons that helped me a great deal with my work with airport simulations were Mr. John Zinna from the FAA and Mr. Gregory Bredford from AirportTools, who provided the modelling tools I needed for my simulation studies. Later my dear colleague Mr. Eric Boyajian from ATAC gave me valuable advice when time-and-time-again I ran into problems in my simulations or had troubles with software bugs.

When I worked for some Norwegian institutions, I had the pleasure to work together with a group of intelligent people from the local airport operator Avinor: Mr. Bjarne Aurstad, Mr. Svein Bratlie, Mr. Kjell-Arne Sakshaug and Mr. Erik Gløersen. I want to thank Mr. Sakshaug for giving me permission to publish my report on airport capacity enhancements at Oslo-Gardermoen airport.

My journey sometimes took unexpected turns. I was asked by Prof. Gunnar Prause and Dr. Anatoli Beifert from University of Applied Sciences, Technology, Business and Design in Wismar to participate in a research project on the potential of an air cargo market in the Baltic Sea Region. Eventually I wrote the final report of a three-year research project, that was funded by the European Regional Development Fund.

My latest research was commissioned by different South Korean institutions. Several projects were ongoing for which my colleagues and me were asked to give our advice and opinions, such as about the master planning of Incheon, Gimhae and Jeju-Island airports. In this context I want to thank Prof. Byung Jong Kim from Korea Aerospace University and his former doctoral student Dr. Jaesung Park for their kind hospitality and their trust in our analysis.

Finally, I want to sincerely thank my dear colleague, collaborator, and "partner-in-crime" Prof. Alberto Gaggero for his vision and drive to conduct research on a high level. When I mentioned my skills regarding the programming of web crawlers for collecting data automatically from the Internet, he saw immediate potential where to apply such bots. His dedication in finishing our papers on competition, delay, and delay propagation in networks cannot be underestimated.

Over the course of three years I gave lectures on two colleges. I want to thank the “next generation,” my more than 200 undergraduate students, in my courses on Operations Management, Distribution Management, Supply-Chain Management and Business Informatics for participating and for giving me great positive feedback despite my rather unconventional approach to teaching.

Branko Bubalo

University of Hamburg, Institute of Information Systems, August 2020

## Zusammenfassung

Das Ziel dieser Dissertation ist es, einen Überblick über die operative und finanzielle Leistungsfähigkeit von Flughäfen in Europa zu bieten. In *Benchmarking*-Studien werden Flughäfen anhand von Kennzahlen nach technischen und wirtschaftlichen Gesichtspunkten beurteilt und mit anderen Flughäfen verglichen. Das besondere Interesse liegt hierbei vor allem in der Frage, mit welchen Kennzahlen die Wahrnehmung des *Quality of Service* aus Sicht der Passagiere für die Dienstleistungen an einem Flughafen am besten gemessen werden kann.

Bei der Verwendung rein *quantitativer* Variablen gibt es Einschränkungen, da die Leistungsfähigkeit eines Flughafens nicht allein nach seiner Produktivität, zum Beispiel gemessen an der Anzahl der abgefertigten Flugzeuge oder Passagiere pro Jahr, bewertet werden sollte. An großen Verkehrsflughäfen werden kritische Entscheidungen und wichtige Prozesse von Menschen getroffen bzw. gesteuert. Viele Prozesse werden auch maschinell und durch Informationstechnik unterstützt. Unzählige kleine Interaktionen in diesen Prozessen summieren sich über die Zeit zu mess- und verwertbaren Größen.

Es können auch *qualitative* Aspekte des Flughafenbetriebs gemessen werden, wenn geeignete Variablen zu diesem Zweck ausgewählt werden und die notwendigen Daten verfügbar sind. Die Leistungskennzahl *Verspätung* wird im Rahmen der Bewertung von Flughäfen eingeführt, da sie die vom Passagier wahrgenommene Servicequalität am deutlichsten ausdrückt. Sie misst die durchschnittliche Zeit, die ein Passagier an einem Flughafen durchschnittlich durch *Warten* verliert.

Ein Flughafen ist ein komplexes System, das aus vielen verschiedenen vernetzten *Warteschlangensystemen* besteht. Passagiere werden immer mehr Platz in den Terminals benötigen und erwarten eine schnellere Bearbeitung an den verschiedenen Abfertigungs- und Kontrollpunkten eines Flughafens, was bedeutet, dass Wartezeiten und Warteschlangenlängen weiter reduziert werden müssen.

In den Flughafenterminals können sich schnell lange Warteschlangen bilden, wenn die Servicekapazität nicht mit der Nachfrage Schritt halten kann. An verschiedenen Stellen innerhalb der Prozesskette benötigen viele Passagiere unterschiedliche Dienstleistungen in schneller Folge. Innerhalb des Terminals wird pro Station normalerweise nur ein Passagier oder eine kleinere Gruppe von Passagieren gleichzeitig bedient, wodurch der Durchsatz begrenzt ist und Engpässe entstehen können. Üblicherweise sind parallele Abfertigungsplätze für die verschieden Services vorhanden, damit Passagierströme schnell abgefertigt werden können.

Verspätungen können sich in kurzer Zeit zu erheblichen Beträgen aufsummieren, insbesondere dann, wenn Kapazitäten begrenzt sind und die Nachfrage nach einer Dienstleistung hoch ist. Bei der Ermittlung von *Verspätungskosten*, können die Gesamtkosten der Wartezeiten für alle Prozesse im System bestimmt werden. Diese Verspätungskosten, die bei Erreichen der Kapazität progressiv steigen, sollten dann in Kosten-Nutzen-Analysen einbezogen werden, um geeignete Maßnahmen zu ergreifen. Gegebenenfalls können diese Kosten auch weitere Diskussionen mit den Interessengruppen, wie zum Beispiel Kreditgebern, Eigentümern, der Politik oder der Öffentlichkeit, hinsichtlich langfristig zu planenden Infrastrukturmaßnahmen auslösen.

Immer dann, wenn zeitliche Entwicklungen vorhergesagt oder verschiedene Zukunftsszenarien miteinander verglichen werden sollen, können mit entsprechenden Tools gezielte Flughafensimulationen durchgeführt werden. *Schnellzeitsimulationen* helfen dabei, Ineffizienzen bei der Abfertigung von Flugzeugen aufzudecken. Auf der Basis von Simulationsmodellen ist es in der Planungsphase möglich, die Auswirkungen von Veränderungen in der vorhandenen Infrastruktur auf die Abläufe zu modellieren und vor Baubeginn zu bewerten.

Für eine genaue Analyse der Abläufe ist es notwendig, regelmäßig in kurz aufeinander folgenden Zeitabständen jeweils die Anzahl der ankommenden und abfliegenden Passagiere und die Starts und Landungen zu beobachten. Diese Aufkommen können erheblich von Minute zu Minute und von Stunde zu Stunde variieren. Entscheidungen bezüglich des Zeitrahmens und des Aggregationsgrades müssen entsprechend dem Ziel der Analyse getroffen werden.

Die vorliegende Arbeit bietet einen kompakten Überblick über die mittel- und langfristige Kapazitäts- und Masterplanung an Flughäfen. Da ein Flughafen die nötigen Kapazitäten für die aktuelle und prognostizierte Anzahl von Flügen und Passagieren bereitstellen muss, wird die Entwicklung der Luftverkehrsnachfrage im Rahmen von Prognosen untersucht und es werden die Möglichkeiten der rechtzeitig zu erfolgenden Kapazitätserweiterungen evaluiert. Die präsentierten Methoden können verwendet werden, um zukünftige Szenarien mit unterschiedlichen Entwicklungen von Nachfrage und Kapazität abzuleiten.

Darüber hinaus wird ein Schwerpunkt auf die Messung der *Servicequalität* gelegt, die in Zukunft eine wichtige Rolle im Luftverkehr spielen wird. Flughäfen werden Engpässe durch Bereitstellung ausreichender Kapazitäten bei der Abfertigung von Passagieren und Flugzeugen verringern müssen. Hierbei gibt es verschiedene Möglichkeiten, die Kapazität eines Flughafens zu erweitern. Kapazitätsreserven können in manchen Fällen kurzfristig nur durch Änderungen der Prozesse freigesetzt werden, in anderen Fällen können Kapazitätsengpässe nur durch technische Maßnahmen reduziert werden. Sollte der Bau von zusätzlicher Infrastruktur erforderlich werden, so muss dies langfristig geplant werden.

Die Lade- und Entladevorgänge von Großraumflugzeugen können die betreffende Infrastruktur erheblich belasten. Je mehr Passagiere in kurzer Zeit in das Terminalgebäude strömen, desto eher können Engpässe und lange Warteschlangen auftreten. Passagiere müssen die Sicherheitskontrolle passieren, ihre Pässe überprüfen lassen, ihr Gepäck einchecken oder abholen und auf ihren Abflug am Gate warten.

Auch die Flugzeuge benötigen wichtige Dienste, wie zum Beispiel die Genehmigung der Flugsicherung, bestimmte Routen oder Landebahnen zu nutzen. Die Bodenkontrolle bestimmt die Rollwege, die Parkposition oder das Gate für jeden Flug. Das bedeutet, dass es wichtig ist, den Flugbetrieb kontinuierlich zu beobachten, um herauszufinden, zu welchen Zeiten und in welchem Maße jeweils die Terminal-, Luftseiten- und Luftraumkapazität besonders ausgelastet ist.

Angesichts der aktuellen Coronavirus-Pandemie wird ein höheres Bewusstsein der Passagiere für den persönlichen Schutz prognostiziert. Mit diesem steigenden Anspruch steigt für die Flughäfen auch die Herausforderung, ausreichende Abstandsregeln zwischen Passagieren, die in einer Warteschlange stehen oder auf eine andere

Dienstleistung warten, zu gewährleisten. Der Platzbedarf pro Passagier im Flughafengebäude wird durch diese Abstandsregeln erheblich steigen. Diese Entwicklung wird sich vor allem dann verschärfen, wenn der Luftverkehr wieder wächst.

Da sich die Pandemie derzeit weiterhin in vielen Teilen der Welt ausbreitet, sind behördlich geforderte Schutz- und Hygienemaßnahmen im Flughafengebäude oder während des Fluges ein geeignetes Mittel, Infektionsketten zu unterbrechen und die Verbreitung des Virus zu verlangsamen. Es wird mehr Kontrollen innerhalb der Prozesskette geben, als es in der Vergangenheit der Fall war. So können zukünftig Gesundheits- und Körpertemperaturprüfungen zum Zeitpunkt der Abreise sowie bei der Ankunft zur „neuen Normalität“ werden.

Die Nachfrage wird wieder auf das Niveau vor der Corona-Pandemie ansteigen, wenn im täglichen Luftverkehr die erforderlichen Vorsichtsmaßnahmen eingehalten werden. Augenblicklich ist es unmöglich vorherzusagen, wann dies der Fall sein wird, aber erste Anzeichen, dass die Fluggesellschaften ihren Betrieb wieder aufnehmen und ihre Streckenpläne ab Herbst 2020 voraussichtlich erweitern werden, sind bereits zu erkennen.

Bei luftseitig ausgelasteten oder bald überlasteten Flughäfen orientieren sich die Bemühungen darauf, die Belegungszeit der Start- und Landebahnen durch die Flugzeuge zu verringern und Prozesse effizienter zu organisieren. Die verfügbare Kapazität der Bahnen wird erhöht, wenn sie von einer größeren Anzahl von Flugzeugen im Zeitverlauf genutzt werden können. Eine Möglichkeit die jeweilige Belegungszeit der Start- und Landebahnen zu reduzieren kann beispielsweise darin bestehen, Schnellabrollwege zu bauen, auf denen Flugzeuge die Bahn mit größerer Geschwindigkeit verlassen können als bei regulären Rollbahnen.

Kennzahlen werden für teils sehr komplexe und äußerst dynamische Prozesse gebildet, um die zu untersuchenden Probleme auf möglichst wenige signifikante Variablen zu vereinfachen. Es muss *a priori* definiert werden, welche Informationen ein Kennwert enthält und auf welcher Datengrundlage dieser gebildet wurde. Ein *Benchmarking* von Flughäfen unterschiedlichster Größe, verschiedenster Eigentümerverhältnisse und in Ländern mit den unterschiedlichsten Rahmenbedingungen, ist eine große Herausforderung. Nach erfolgreicher Durchführung und genauer Analyse kann ein Benchmarking zu sehr wertvollen Schlussfolgerungen und Handlungsempfehlungen führen, die das Flughafenmanagement bei der Entscheidungsfindung unterstützen können.

Grundlage für die rechnerische Ermittlung der Kennzahlen sind systematisch erhobene Daten verschiedener Flughäfen. Einige Daten sind frei zugänglich, andere sind nur intern in den Unternehmen vorhanden und werden vertraulich behandelt. Für manche Studien war es erforderlich, Daten von Grund auf neu zu erheben, beispielsweise durch den Versand von Fragebögen an die Flughäfen. Für die Luftverkehrsbranche werden umfangreiche Mikro- und Makrodaten veröffentlicht. Diese Daten können mit Wirtschafts- und demografischen Daten verknüpft werden. Bezüglich der Quellen, ob Daten oder Literatur, ist es ratsam die Qualität im Rahmen der Möglichkeiten zu prüfen und *Plausibilitätsprüfungen* zu machen.

Zur Informationsgewinnung werden Daten der betrieblichen Abläufe über Zeiträume wie Minuten, Stunden, Tage, Wochen oder Monate zusammengefasst. Finanzdaten werden basierend auf den analysierten Gewinn- und Verlustrechnungen pro Betriebsjahr gruppiert und gegebenenfalls bezüglich Kaufkraft, Inflation und Wechselkurs bereinigt.

Wenn ein Flughafen finanziell erfolgreicher ist als ein anderer und im Vergleich zu seinen Mitbewerbern eine hervorragende Servicequalität bietet, muss es das Ziel sein die Gründe dafür herauszufinden und über ausreichende quantitative Informationen für diese außergewöhnliche Leistung zu verfügen. Dadurch sind andere Flughäfen in der Lage, sich an einem solchen *Best-Practice* Flughafen ein Beispiel zu nehmen, um ihre Performance zu verbessern.

Neben Veränderungen im operativen Bereich spielen in der vorliegenden Arbeit auch die tiefgreifenden wirtschaftlichen Veränderungen im Luftverkehr eine Rolle. In diesem Zusammenhang erfolgt ein Rückblick auf die wirtschaftlichen Entwicklungen der letzten dreißig Jahre. Ab Mitte der neunziger Jahre stieg die Nachfrage nach Flugreisen explosionsartig, angefacht durch die Deregulierung des Luftverkehrs und den dadurch zunehmenden *Wettbewerb*. Insbesondere der Markteintritt der *Low-Cost-Carrier* brachte den Verbrauchern mehr Reisemöglichkeiten zu günstigen Ticketpreisen.

Damit das Low-Cost-Geschäftsmodell funktioniert, müssen die Durchlaufzeiten auf den Flughäfen so kurz wie möglich sein. Störungen, die insbesondere auch von anderen Fluggesellschaften verursacht werden, müssen, soweit möglich, vermieden werden. Das Geschäftsmodell ist besonders erfolgreich bei Direktverbindungen auf dem Kurzstreckenmarkt, auf dem die Passagiere Flugverbindungen zu einem geringen Preis erhalten sowie allerdings auch weniger Borddienste geboten bekommen.

Die marktführenden Low-Cost-Carriers zeichnen sich durch eine sehr gute Pünktlichkeit ihrer Flüge aus. In Bezug auf das Flugangebot profitieren die Kunden von einer verbesserten Konnektivität, nicht nur auf dem inländischen Markt, sondern auch bei internationalen und innereuropäischen Verbindungen, da durch mehr Konkurrenten am Luftverkehrsmarkt das Angebot insbesondere direkt erreichbarer Reiseziele steigt.

Weitere betriebliche und finanzielle *Dichtevorteile* ergeben sich, wenn Flugverbindungen auf einige wichtige Hub- oder Basisflughäfen konzentriert werden. Passagiere haben, insbesondere wenn sie über diese Drehkreuze reisen, mehr Auswahlmöglichkeiten hinsichtlich der direkten Nonstop-Ziele. Gleichzeitig haben *Full-Service-Carrier* an den Hubflughäfen oder Low-Cost-Carrier an ihren Stützpunkten mehr Möglichkeiten, Verzögerungen oder Flugausfälle zu verringern, die durch verspätet ankommende Flüge, schlechtes Wetter oder durch lokale Betriebsstörungen, zum Beispiel aufgrund technischer Ausfälle, verursacht werden. Im Allgemeinen erleben Passagiere dort auch eine bessere Servicequalität.

In der aktuellen Diskussion fehlte bisher eine Maßzahl für die Netzwerkstruktur, mit der die Leistung von Flughäfen und Fluggesellschaften, die in verschiedenen Netzwerkstrukturen operieren, verglichen werden kann. Ziel war es deshalb, ein Maß zu finden, das die Netzwerkstruktur auf einer Skala zwischen 0 und 1 beschreibt, wobei „0“ ein *sternförmiges Netzwerk* beschreibt und „1“ ein *vollständig vermaschtes Netzwerk* darstellt.

Viele der Low-Cost-Carrier betreiben eine andere Streckennetzstruktur als die etablierten Full-Service-Carrier. Um die Betriebskosten auf einen notwendigen Umfang zu begrenzen und Personal und Maschinen im Tagesverlauf möglichst hoch auszulasten, nutzen viele erfolgreiche Low-Cost-Carrier günstig gelegene Sekundärflughäfen als Stützpunkt für ihren Betrieb. Diese Flughäfen bieten niedrigere Landegebühren, bedienen einen ähnlichen Markt und sind in der Regel nicht so überlastet, wie primäre Hub-Flughäfen. Hieraus ergibt sich, dass die Flugnetze der Low-Cost-Carrier in der Regel eine *Punkt-zu-Punkt-Struktur* aufweisen.

Im Vergleich erzielen Full-Service-Carrier Leistungsvorteile durch die Bündelung von Flugverbindungen an Drehkreuzen durch Zubringerflüge, wodurch sich in diesen Fällen *Hub-and-Spoke-Strukturen* in den Flugnetzen bilden. Im Rahmen von Allianzen unterstützen sich auch konkurrierende Fluggesellschaften insbesondere an den Drehkreuzen bei den betrieblichen Abläufen und bei dem Serviceangebot für Passagiere.

Untersuchungen zeigen, dass die durchschnittlichen Verspätungen, in von den Fluggesellschaften betriebenen Netzen sehr unterschiedlich ist. Vorteile resultieren auch aus einer höheren Verfügbarkeit von Flugzeugen, Reservebesatzungen sowie Wartungseinrichtungen und Ersatzteilen an Hub-Flughäfen oder an Stützpunkten, was letztlich zu weniger Verspätungen oder Flugausfällen führt.

Bezüglich der betrieblichen und finanziellen Leistungsfähigkeit spielt die Flughafengröße eine wesentliche Rolle. Skaleneffekte sorgen für eine bessere Rentabilität der großen Flughäfen im Vergleich zu kleineren. Es wurde deshalb der Frage nachgegangen, ab welcher Nachfrage ein Flughafen profitabel ist. Beispielsweise erreichen untersuchte Flughäfen in Frankreich bereits die Gewinnschwelle bei einer Nachfrage von nur circa 300.000 Passagieren pro Jahr. Im Durchschnitt erzielt ein Flughafen allerdings erst ab einer Nachfrage von einer Million Passagieren pro Jahr Gewinne, gemessen am *Betriebsergebnis vor Zinsen und Steuern pro Passagier*.

So ist es für Fluggesellschaften und Flughäfen schwierig, ein tragfähiges Geschäft aufzubauen, wenn die lokale Nachfrage nach Flugleistungen gering ist. Wirtschaftlich attraktive Umsteigeverkehre spielen in diesen Fällen kaum eine Rolle, bei größeren Drehkreuzen dagegen sind diese ein wesentlicher Wirtschaftlichkeitsfaktor. Zusätzlicher Nachfrage könnte allerdings auch durch attraktive Flughafengebühren und eine gute Infrastruktur generiert werden. Wesentliche lokale Voraussetzungen für das Wachstum eines Flughafens sind allerdings eine starke und dynamische lokale Wirtschaft, sowie eine prosperierende Industrie.

Abgelegene Regionen sind in einem gewissen Maß von der Erreichbarkeit per Flugzeug abhängig, beispielsweise wenn der Transportzugang per Schiene, Straße oder über das Wasser nicht ausreichend gegeben ist. Ein Grund hierfür ist häufig die Topografie in verschiedenen Ländern, die sich kaum für den Bau komplexer Bodentransportnetze eignet. Zum Funktionieren eines Staates muss das Hinterland mit den Zentren verbunden sein, damit auch für die Bevölkerung in den Regionen öffentliche Dienstleistungen und auch kommerzielle Leistungsangebote erreichbar sind.

Da die Nachfrage in vielen Fällen zu niedrig ist, um wirtschaftlich tragfähige Verbindungen zu etablieren, wird Luftverkehr in manchen strukturschwachen

Regionen subventioniert. Beispiele hierfür sind Inseln, die durch entsprechende Flugangebote mit dem Festland verbunden werden. Die Betriebskosten pro Passagier sind auf diesen Routen vergleichsweise hoch und können durch den Ticketpreis kaum gedeckt werden. Daher vergibt der Staat im Rahmen der Daseinsfürsorge für solche Strecken entsprechende *Subventionen*, die zu erhöhten sozialen Kosten und zu verbundenen gesamtgesellschaftlichen Wohlfahrtsverlusten führt.

Wie das Beispiel Norwegen zeigt, stiegen die Lufttransportkosten im Vergleich zu anderen Ländern überproportional, vor allem durch falsch gesetzte Anreize und durch geringen Wettbewerb bei öffentlichen Ausschreibungen. So machen Fluggesellschaften, die die Ausschreibung gewonnen haben und die ausgeschriebenen Strecken bedienen, in einigen Fällen höhere Betriebskosten im Vergleich zum Marktniveau geltend. Das kann ein Beleg dafür sein, dass diese Fluggesellschaften ihre *quasi-monopolistische* Position im Wettbewerb ausnutzen, darüber hinaus sind die Ausschreibungsanforderungen für diese Flugrouten in Bezug auf Flugzeuggröße und Ausrüstung so streng festgelegt, dass es kaum zu einem Wettbewerb im Europäischen Markt kommt und nur eine oder sehr wenige Fluggesellschaft(en) als Anbieter in Frage kommen.

Im regionalen Luftverkehr kann sich die Zusammensetzung der Nutzlast in Flugzeugen unterscheiden, um gegebenenfalls eine geringe Passagiernachfrage durch Einbeziehung der Luftfracht zu kompensieren. Das Platzangebot auf den jeweiligen Routen sollte sich allerdings, soweit möglich, nach wirtschaftlichen Überlegungen richten. Es gibt allerdings auch reine Frachtflüge, aber in der Regel wird in Passagierflugzeugen Reisegepäck und zusätzliche Fracht gemeinsam transportiert, damit die Kapazität des Gepäckbereichs möglichst vollständig genutzt wird.

Passagiere und Fracht können zu einer Maßzahl aggregiert werden, der *Workload Unit*, bei der ein Passagier einschließlich Gepäck genau einhundert Kilogramm Fracht entspricht. Ein Nachteil dieser Kennzahl besteht darin, dass die unterschiedliche Infrastruktur, die für die Abwicklung der Prozesse im Passagier- oder Frachtbereich erforderlich ist, nicht berücksichtigt wird. Daher sollte diese Kennzahl nur mit Einschränkungen verwendet werden. In der Regel ist auch zu empfehlen, die Passagier- und Frachtnachfrage getrennt zu analysieren.

Darüber hinaus wurde untersucht, ob bestimmte regionale Verbindungen das wirtschaftliche Potenzial für Luftfrachtflüge haben und welche Nachfrage bestehen kann. Exklusive Luftfrachtrouten und Frachtflughäfen sind nur dann wirtschaftlich sinnvoll, wenn ausreichend luftfrachtaffine Güter im Einzugsbereich der Flughäfen gefertigt oder bereitgehalten werden. Darüber hinaus kann eine Luftfrachtnachfrage für den Transport von verderblichen Gütern, wie zum Beispiel Agrarprodukten und Blumen sowie von hochwertigen Gütern, wie zum Beispiel elektronischen und pharmazeutischen Produkten, bestehen.

Ein aktuell diskutiertes Problem sind die negativen Auswirkungen der Luftfahrt auf die Umwelt. So sind bereits Trends in Richtung eines *Green Travel* und auch zu Extremformen wie dem *Flight Shame* zu erkennen, wenn Touristen während ihrer Reise umweltfreundliche Transportmittel nutzen und sozial verträglichere Unterkünfte wählen sowie Flüge gänzlich vermeiden. Die jüngere Generation setzt zum Teil neue

Standards und fordert, dass die Umweltverschmutzung durch den Luftverkehr und damit auch dessen ökologischer Fußabdruck deutlich reduziert wird.

Mit Blick hierauf, wird die Machbarkeit einer Reduzierung von Triebwerksemissionen auf einem Flughafen in der vorliegenden Arbeit untersucht. So werden verschiedene Reihenfolgen des Push-Back-Services untersucht, der von Flugzeugen benötigt wird, um ihre Parkposition zu verlassen. Mit Hilfe von Simulationen kann das Flughafenmanagement bei der Umsetzung verbesserter Abläufe bei der Flugzeugabfertigung unterstützt werden, mit dem Ziel die Verspätungen und Gesamtemissionen zu verringern. Es ist mit den heute verfügbaren Systemen möglich, innerhalb von Sekunden zahlreiche Iterationen eines Simulationsmodells durchzurechnen.

Die Push-Back-Sequenzen werden gemäß ihrer berechneten Verspätungszeiten, Rollzeiten und voraussichtlichen Emissionen innerhalb eines kurzen Zeithorizonts sortiert. Durch die Wahl der geeignetsten Sequenz können die Gesamtverspätungen und Emissionen der Flugzeuge am Boden reduziert werden. Der Schwerpunkt liegt hierbei auf der Berücksichtigung einer Reihe gasförmiger Emissionen wie Kohlenwasserstoffe, Kohlenmonoxid, Kohlendioxid und Stickoxide.

Es wird erwartet, dass sich durch eine verbesserte Re-Sequenzierung der Push-Back-Dienste eine signifikante Reduzierung der Gasemissionen erreichen lässt. Eine solche Verringerung der Gesamtemissionen verbessert die Umweltleistung von Flughäfen im Rahmen eines Benchmarkings. Wenn zum Beispiel zwei Flughäfen eine ähnliche Leistung, gemessen an der Anzahl der Passagiere, erzielen, sollte der Flughafen mit der besseren Umweltleistung im Vergleich zum anderen Flughafen einen höheren Stellenwert einnehmen.

## Abstract

The purpose of this dissertation is to present an overview of the operational and financial performance of airports in Europe. In *benchmarking* studies, airports are assessed and compared with other airports based on key indicators from a technical and economic point of view. The interest lies primarily in the question which key figures can best measure the perception of *quality of service* from the point of view of the passenger for the services at an airport.

There are limitations when using purely *quantitative* variables because the performance of an airport should not be assessed solely based on its productivity, for example in terms of the number of aircraft or passengers handled per year. At large commercial airports, critical decisions and important processes are made and controlled by people. Many processes are also supported mechanically and by information technology. Countless small interactions in these processes add up over time to measurable and comparable quantities.

*Qualitative* aspects of airport operations can also be measured if suitable variables are selected for this purpose and the necessary data are available. The performance indicator *delay* is introduced as part of the assessment of airports, since it most clearly expresses the service quality perceived by the passenger. It measures the average time that a passenger at an airport loses *waiting*.

An airport is a complex system that consists of a network of many different *queuing systems*. Passengers will need more and more space in the terminals and expect faster processing at the various check points of an airport, which means that waiting times and queue lengths must be reduced further.

Queues can quickly form in the airport terminals if the service capacity cannot keep up with demand. At various points in the process chain, many passengers need different services in quick succession. Within the terminal, normally only one passenger or a smaller group of passengers is served per service desk at the same time, which limits throughput and can lead to bottlenecks. There are usually parallel counters and service stations for the various services so that passenger flows can be processed quickly.

Delays can add up to substantial amounts in a short period of time, especially when capacities are limited and the demand for a service is high. When applying *costs of delay*, the total costs of waiting times can be determined for all processes in the system. These delay costs, which increase progressively when capacity is reached, should then be included in cost-benefit analyzes to take appropriate measures. If necessary, these costs can also trigger further discussions with interest groups, such as lenders, owners, politicians, or the public, regarding infrastructure measures to be planned in the long term.

Whenever developments over time are to be predicted or different future scenarios are to be compared, airport simulations can be carried out using the appropriate tools. *Fast-time simulations* help reveal inefficiencies in aircraft handling. Based on simulation models, it is possible in the planning phase to model the effects of changes in the existing infrastructure on the processes and to evaluate them before the start of construction.

For a precise analysis of the processes, it is necessary to observe the number of arriving and departing passengers and the take-offs and landings frequently in short succession. These volumes can vary considerably from minute to minute and from hour to hour. Decisions regarding the time frame and the degree of aggregation must be made according to the goal of the analysis.

The present work offers a compact overview of medium and long-term capacity and master planning at airports. Since an airport has to provide the necessary capacities for the current and *forecast* number of flights and passengers, the development of air traffic demand is examined in the context of forecasts and the possibilities of timely capacity expansions to be implemented are evaluated. The presented methods can be used to derive future scenarios with different developments in demand and capacity.

There will also be a focus on measuring service quality, which will play an important role in air traffic in the future. Airports will have to reduce bottlenecks by providing sufficient passenger and aircraft handling capacities. There are various ways of expanding the capacity of an airport. In some cases, capacity reserves can only be freed by changing the processes; in other cases, capacity bottlenecks can only be reduced by technical measures. If the construction of additional infrastructure becomes necessary, this must be planned in the long term.

The loading and unloading processes of wide-body aircraft can put considerable strain on the infrastructure in question. The more passengers flow into the terminal building in a short time, the more likely bottlenecks and long queues can occur. Passengers must go through security, have their passports checked, check in or claim their luggage and wait for their departure at the gate.

The aircraft also require important services, such as the approval of air traffic control to use certain routes or runways. The ground control determines the taxiways, the parking position, or the gate for each flight. This means that it is important to continuously monitor flight operations to find out at what times and to what extent the terminal, airside and airspace capacity is utilized.

In view of the current coronavirus pandemic, a higher level of passenger awareness of personal safety is predicted. With this increasing demand, the challenge for airports to ensure sufficient distance rules between passengers who are in a queue or waiting for another service also increases. The space requirement per passenger in the airport building will increase considerably due to these distance rules. This development will intensify especially when air traffic grows again.

Since the pandemic is currently continuing to spread in many parts of the world, demanded protective and hygienic measures in the airport building or during the flight are suitable means of interrupting the infection chains and slowing the spread of the virus. There will be more controls within the process chain than has been the case in the past. In the future, health, and body temperature tests at the time of departure and on arrival can become the “new normal.”

Demand will rise back to the level before the corona pandemic if the necessary precautionary measures are followed in daily air traffic. Currently it is impossible to predict when this will happen, but the first signs that the airlines will resume operations and are likely to expand their route plans from autumn 2020 are already evident.

When airports are congested or soon congested, efforts are being made to reduce the occupancy of the runways by the aircraft and to organize processes more efficiently. The available capacity of the runways is increased if they can be used by a larger number of aircraft over time. One way to reduce the occupancy time of the runways can be, for example, to build high-speed taxiways on which planes can leave the runway at a higher speed than with normal taxiways.

Key figures are created for sometimes complex and extremely dynamic processes to simplify the problems to be investigated to as few significant variables as possible. It must be defined *a priori* what information a characteristic value contains and on what data basis it was created. Benchmarking airports of different sizes, different ownership structures and in countries with different framework conditions is a major challenge. After successful implementation and detailed analysis, benchmarking can lead to valuable conclusions and recommendations for action, which can support airport management in the decision-making process.

The basis for the mathematical determination of the key figures is systematically collected data from various airports. Some data are freely accessible, others are only available internally in the company and are treated confidentially. For some studies it was necessary to collect data from scratch, for example by sending questionnaires to the airports. Extensive micro and macro data are published for the aviation industry. This data can be linked to economic and demographic data. Regarding the sources, whether data or literature, it is advisable to check the quality within the scope of the possibilities and to carry out *plausibility checks*.

To obtain information, data from operational processes are summarized over periods such as minutes, hours, days, weeks, or months. Financial data are grouped based on the analyzed profit and loss accounts per year and, if necessary, adjusted for purchasing power, inflation, and exchange rate.

If an airport is financially more successful than another and offers an excellent quality of service compared to its competitors, the goal must be to find out the reasons for this and to have sufficient quantitative information for this extraordinary performance. This enables other airports to take an example from such a *best practice* airport to improve their performance.

In addition to changes in the operational area, the profound economic changes in air traffic also play a role in the present work. In this context, we look back at the economic developments of the past thirty years. From the mid-1990s, the demand for air travel exploded, fueled by the deregulation of air traffic and the resulting increase in competition. In particular, the market entry of the *Low-Cost Carriers* brought consumers more travel options at low ticket prices.

For the low-cost business model to be viable, the throughput times at the airports must be as short as possible. Disruptions that are caused in particular by other airlines must be avoided as far as possible. The business model is particularly successful in the case of direct connections on the short-haul market, on which passengers receive flight connections at a low price, but also receive fewer on-board services.

The market-leading Low-Cost Carriers are characterized by a great punctuality of their flights. With regard to the offered flights, customers benefit from improved connectivity, not only on the domestic market, but also on international and intra-European connections, as more competitors in the aviation market increase the number of destinations that can be reached directly.

Further operational and financial advantages in terms of *economies of density* result if flight connections are concentrated on hub or base airports. Passengers, especially when traveling through these hubs, have more choices regarding direct non-stop destinations. At the same time, *Full-Service Carriers* at the hub airports or Low-Cost Carriers at their bases have more options to reduce delays or flight cancellations caused by delayed flights, bad weather or local operational disruptions, e.g. due to technical failures. In general, passengers at these airports experience better service quality.

The current discussion has so far lacked a measure of the network structure with which the performance of airports and airlines operating in different network structures can be compared. The aim was therefore to find a measure that describes the network structure on a scale between 0 and 1, with "0" describing a *star network* and "1" representing a *fully connected network*.

Many of the Low-Cost Carriers operate a different route network structure than the established Full-Service Carriers. To limit operating costs to a necessary extent and to utilize personnel and machines as much as possible throughout the day, many successful Low-Cost Carriers use appropriately located secondary airports as a base for their operations. These airports offer lower landing fees, serve a similar market, and are typically not as congested as primary hub airports. It follows from this that the Low-Cost Carrier flight networks generally have a point-to-point structure.

In comparison, Full-Service Carriers achieve performance advantages by bundling flight connections at hubs through feeder flights, which in these cases creates hub-and-spoke structures in the flight networks. As part of alliances, competing airlines also support each other, particularly at the hubs, in the operational processes and in the range of services for passengers.

Research shows that the average delays in airline operated networks vary widely. Advantages also result from a higher availability of aircraft, reserve crews as well as maintenance facilities and spare parts at hub airports or at bases, which consequently leads to fewer delays or flight cancellations.

The size of the airport plays an important role in terms of operational and financial performance. *Economies of scale* ensure that large airports are more profitable than smaller ones. The question was therefore asked as to what demand made an airport profitable. For example, examined airports in France are already reaching the break-even point with a demand of only around 300,000 passengers per year. On average, an airport only makes a profit, measured in terms of *earnings before interest and taxes* per passenger, when demand is above circa one million passengers per year.

It is difficult for airlines and airports to build a sustainable business if local demand for flight services is low. Economically attractive transfer traffic hardly plays a role in these cases, however, at larger hubs these are an essential economic factor. Additional demand could also be generated by attractive airport fees and a good infrastructure. However, essential local prerequisites for the growth of an airport are a strong and dynamic local economy and a prospering industry.

Remote regions are dependent to a certain extent on accessibility by plane, for example if there is insufficient access by rail, road, or water. One reason for this is often the topography in different countries, which is hardly suitable for the construction of complex ground transport networks. For a state to function, the hinterland must be connected to the centers so that public and commercial services are also reachable to the population in the regions.

Since demand is in many cases too low to establish economically viable connections, air transport is subsidized in some structurally weak regions. Examples of this are islands that are connected to the mainland by corresponding flight offers. The operating costs per passenger on these routes are comparatively high and can hardly be covered by the ticket price. For this reason, the state grants appropriate *subsidies* for services of general interest, which leads to increased social costs and associated welfare losses.

As the example of Norway shows, air transport costs rose disproportionately compared to other countries, due to incorrectly set incentives and low competition in public tenders. For example, airlines that have won the tender and operate the routes in some cases claim higher operating costs compared to the market level. This can be evidence that these airlines are exploiting their *quasi-monopolistic* competitive position, and the tender requirements for these flight routes are so strict in terms of aircraft size and equipment that there is hardly any competition in the European market and only one or very few airlines are considered as providers.

In regional air traffic, the composition of the payload in aircraft can differ to compensate for a low passenger demand by transporting additional air cargo. However, the payload on the respective routes should, as far as possible, be based on economic considerations. However, there are also pure cargo flights, but usually luggage and additional cargo are transported together in passenger aircraft so that the capacity of the baggage area is used as fully as possible.

Passengers and freight can be aggregated into a measure, the *Workload Unit*, in which a passenger including luggage corresponds to exactly one hundred kilograms of freight. A disadvantage of this key figure is that the different infrastructure that is required for the handling of processes in the passenger or freight area is not considered. This key figure should therefore only be used with caution. As a rule, it is advisable to analyze passenger and freight demand separately.

In addition, it was examined whether certain regional connections have the economic potential for air cargo flights and which demand can exist. Exclusive air freight routes and cargo airports only make economic sense if sufficient air freight-related goods are manufactured or made available in the catchment area of the airports. In addition, there may be air freight demand for the transportation of perishable goods, such as agricultural products and flowers, as well as high-quality goods, such as electronic and pharmaceutical products.

A currently discussed problem is the negative impact of aviation on the environment. Trends in the direction of *green travel* and extreme forms such as the *flight shame* can already be seen in tourism, when tourists use environmentally friendly means of transport, choose more socially acceptable accommodations, and avoid flights altogether. The younger generation is setting new standards and demands that air pollution and thus the environmental footprint of air transportation be significantly reduced.

The feasibility of reducing aircraft engine emissions at an airport is examined in the present work. Different sequences of the push-back service that planes need to leave their parking position are examined. With the help of simulations, airport management can be supported in the implementation of improved processes in aircraft handling, with the aim of reducing delays and total emissions. With the systems available today, it is possible to calculate numerous iterations of a simulation model within seconds.

The push-back sequences are sorted according to their calculated delay times, taxi times and expected emissions within a short time horizon. By choosing the most advantageous sequence, the total delays, and the aircraft ground emissions can be reduced. The focus here is on considering several gaseous emissions such as hydrocarbons, carbon monoxide, carbon dioxide and nitrogen oxides.

It is expected that improved re-sequencing of the push-back services will result in a significant reduction in gas emissions. Such a reduction in total emissions improves the environmental performance of airports through benchmarking. For example, if two airports perform similarly in terms of number of passengers, the airport with the better environmental performance should rank higher than the other airport.

## Contents

## Figures

## Tables

## List of abbreviations

| | |
|---|---|
| ACI | Airport council international |
| ACRP | Airport Cooperative Research Program |
| ADS-B | Automatic dependent surveillance - broadcast |
| AEA | Association of European airlines |
| AFIS | Aerodrome flight information service |
| AI | Artificial intelligence |
| AIP | Aeronautical information publication |
| AirTOp | Air Traffic Optimization |
| ANSP | Air navigation service provider |
| ARC | Aachen research center |
| ASK | Available seat kilometers |
| ATAC | Airborne Tactical Advantage Company |
| ATC | Air traffic control |
| ATM | Air transport movement / Air traffic management |
| ATS | Air transport system |
| ATRS | Air transport research society |
| BAA | BAA Limited |
| BER | Berlin-Brandenburg International airport (opening date 10.31.2020) |
| BRIC | Brazil, Russia, India, and China |
| BSR | Baltic sea region |
| BTO | Baltic transport outlook |
| CAST | Comprehensive Airport Simulation Technology |
| Cat. I | Category I weather conditions, here: 200 ft. decision height and 500 ft. minimum visibility. |
| CBA | Cost-benefit analysis |
| CDM | Collaborative decision making |
| CFMU | Central flow management unit at Eurocontrol |
| CN | Combined Nomenclature |
| CO | Carbon monoxide |
| $CO_2$ | Carbon dioxide |
| COVID-19 | Coronavirus Disease 2019 |
| CPLEX | IBM ILOG CPLEX Optimization Studio |
| CRPK | Costs per revenue-passenger kilometer |
| CVP | Cost-volume-profit |
| DEA | Data envelopment analysis |
| DHC6 | de Havilland Canada DHC-6 aircraft |
| DHC8 | de Havilland Canada DHC-8 / Bombardier Dash 8 aircraft |
| DLR | Deutsches Zentrum für Luft- und Raumfahrt / German Aerospace Center |
| EAC | European Aviation Conference |
| EAD | European AIS database |
| EAS | Essential air services |
| EBIT | Earnings before interest and taxes |
| EBITDA | Earnings before interest, taxes, depreciation, and amortization |
| EC | European commission |
| ERDF | European regional development fund |
| EPNdB | Effective perceived noise in decibels |
| EQC | Enhanced Quota Count |
| ELFAA | European low fares airline association |
| ETS | Emission trading system |
| EUR | Euro |
| FAA | Federal Aviation Administration |
| FCFS | First-come first-served |
| FOI | Totalförsvarets forskningsinstitut (Swedish Defense Research Agency) |
| FSC | Full-service carrier |
| GAP | German airport performance research project |
| GARS | German aviation research society |
| GDP | Gross domestic product |
| GPS | Global positioning system |
| H | Heavy aircraft class (more than 136 tons MTOW) |
| HC | Hydrocarbon |
| IATA | International air transport association |
| ICAO | International civil aviation organization |
| ICCL | International conference on computational logistics |
| IFR | Instrument flight rules |
| ILS | Instrument landing system |
| IMF | International Monetary Fund |
| L | Light aircraft class (less than seven tons MTOW) |
| LNCS | Lecture notes in computer science |
| LNS | Large neighborhood search |
| LOS/LoS | Level-of-service |
| LCC | Low-cost carrier |
| M | Medium aircraft class (seven to 136 tons MTOW) |
| MCT | Minimum connecting time |
| METAR | Meteorological aviation routine weather report |
| MI | Mix index |
| MIME | Market-based impact mitigation for the environment |
| MPO | Mixed parallel operations |
| MSWV | Ministerium für Stadtentwicklung, Wohnen und Verkehr des Landes Brandenburg |
| MTOW | Maximum take-off weight |
| NASA | National Aeronautics and Space Administration |
| NATS | National Air Traffic Services United Kingdom/NATS Holding |
| NEXTGEN | The next generation air transportation system |
| NM | Nautical mile (1 NM = 1.852 km) |
| NOK | Norwegian kroners |
| $NO_x$ | Nitrogen oxides |
| OAG | Official airline guide |
| OECD | Organization for economic co-operation and development |
| Ops | Operations, i.e. flights |
| PANS-RAC | Procedures for air navigation services - rules of the air & air traffic |
| PAX | Passenger(s) |
| PFP | Partial factor productivity |
| PPP | Purchasing power parity |
| PPR | Prior permission required |
| PRM | Precision runway monitor |
| PSO | Public service obligation |
| QC | Quota Count |
| RET | Rapid exit taxiway |
| ROI | Return on investment |
| ROT | Runway occupancy time |
| RPK | Revenue-passenger kilometer |
| RRPK | Revenue per revenue-passenger kilometer |
| SAGE | System for assessing aviation’s global emissions |
| SARS-CoV-2 | Severe acute respiratory syndrome caused by novel coronavirus |
| SC | Schedules conference |
| SES | Single European sky |
| SESAR | Single European sky ATM research programme |
| SF340 | Saab 340 aircraft |
| SID | Standard instrument departure route |
| SIMMOD | Airport and airspace simulation model |
| SITA | Société Internationale de Télécommunication Aéronautique |
| STAR | Standard terminal arrival route |
| STOL | Short take-off and landing runway |
| TAAM | Total Airspace and Airport Modeler |
| TEAM | Tactical enhanced arrival mode |
| TØI | Transportøkonomisk institutt (Institute of transport economics) |
| TMA | Terminal control (or Maneuvering) area |
| TRB | Transportation research board |
| VBA | Visual basic for applications |
| VFR | Visual flight rules |
| VIF | Variance inflation factor |
| VRP | Vehicle routing problem |
| WATS | World air transport statistics |
| WHO | World Health Organization |
| WLU | Workload unit |
| WTC | Wake turbulence category |

## Synopsis

At the time of writing we are amidst a global pandemic which affects the life of many people. The outcome of this crisis is yet unclear. The number of infected people with the novel coronavirus, labeled SARS-CoV-2 by the World Health Organization (WHO) in late January 2020, and the related death toll related to the severe paths of its illness varies from country to country.

Most countries have stay-at-home orders in place to limit the social interaction between people and, therefore, the spread of the virus. Latest figures suggest a fatality rate of between 1% to 4% of all infected, which is a more than ten-times greater figure than the fatality rate of, e.g., the seasonal flu. As of today August 13th, 2020, according to the numbers collected by the John Hopkins University[1] globally more than 20.6 million people have been tested positive to SARS-CoV-2 since the beginning of the pandemic, more than 750,000 people have died. The virus infection causes the respiratory disease Covid-19 (Corona Virus Disease 2019) that may lead to severe pneumonia and may cause similar symptoms as SARS (Severe Acute Respiratory Syndrome), which spread in Asia in the years 2002 and 2003.

As far as we currently know, the disease emerged from the Chinese city Wuhan and quickly spread to its surrounding Hubei province in late December 2019. Soon the virus spread globally through trade and travel to many other countries. The first flight connections to China were halted in late January 2020. By mid-March international air travel came to an almost complete standstill. Many fleets were grounded and airlines struggle against bankruptcy.

The International Air Transport Association (IATA) estimated in mid-April that in 2020 global air travel will drop considerably by about 55% compared to 2019, which means a decline in airline passenger revenues by $314 billion.[2] The global economy, its production and supply-chains are interrupted, many people face reduced work or unemployment and in general a socially, psychologically and economically difficult time. We observe closed schools, kindergartens, restaurants, bars, small businesses, cinemas, theaters, and many more places were people used to interact closely and socially. In several countries it is assumed that the peak of new daily infections is surpassed, because the social distancing rules were followed by the people, therefore, the human-to-human transmission of the virus is limited, and the exponential growth of newly infected patients is either reduced or stopped.

We assume that the economy will suffer for a longer period, but not longer than the third quarter of 2020. However, we expect air travel demand to increase sharply, as soon as travel restrictions and mandatory quarantine times of up to two weeks after arrival in a foreign country, are lifted in more and more countries. We think, summer 2021 is a reasonable timeframe, when air travel demand and passenger numbers will be back at the 2019 level. The SARS pandemic and the financial crisis in 2009 have shown, that air transport demand will return quickly, when a global crisis affecting air transport demand is beyond its peak.

---

1 https://coronavirus.jhu.edu/map.html [last accessed August 6th, 2020]

2 https://www.iata.org/en/pressroom/pr/2020-04-14-01/ [last accessed April 14th, 2020]

Air transportation had been a growing global industry and will continue to flourish when the current pandemic is defeated. This industry requires intelligent management efforts to make its operations more efficient and its environmental footprint more sustainable. This dissertation highlights some recent research topics and long-lasting problems in European air transportation.

We rely on quantitative approaches to our research to which we apply different mathematical tools. We gather data that we deem necessary to confront our problems and that aid our reasonings and recommendations. If we fail in collecting all required data, we need to carefully choose inputs that we can get a hold of to answer at least partially our questions. In forecasting, i.e. predicting some future actions and reactions, it is even more important to make sound assumptions about future changes and developments. Frequently forecasts depend on a handful of critical factors, that might influence future states.

We base our judgements on mostly publicly available or, say, freely accessible data. We are aware that errors in measurements can occur, that is why we do our own cross- and plausibility checks, to filter unreliable data, outliers, or transmission errors. Sometimes we had different sources for similar data, so we could compare the quality of data from one source to the other. One such example is flight schedule data, that can be obtained from various sources. We use flight schedule data as a main input to our simulations to model dynamics at an airport and to predict future developments. As an output of the simulation we get performance measures that change over time. Therefore, we can make quantitative judgements on the costs and benefits of future infrastructural changes at an airport, especially the interplay between airport capacity, demand, and delay.

The construction and expansion of airport infrastructure typically requires enormous amounts of capital expenditure. Furthermore, on the one hand the air transport system needs to be supplied with adequate machinery, such as aircraft, pushback trucks, aircraft tracking, navigation and communication tools, and other supporting equipment. On the other hand, to run such a complex system causes significant investment, operating, maintenance and environmental costs. This is the main reason why the public, business and science show keen interest to operate the air transport system in the most economical, which often means in the most (cost and ecological) efficient manner, to give customers and the society the maximum benefit and to provide the environment the largest respite. To increase productivity and efficiency in air transportation often means making airports more profitable, more fuel and energy efficient, reducing noise and gaseous emissions from operations, is our main motivation for this work.

Air transport carriers provide the service to *fly from origin A to destination B*. We exchange a ticket price or service fee with the offered transport service. We as passengers trust in some basic level of safety, security, and service quality. Air transport users rely on international agreements, laws, and standards and upon compliance to them.

In Chapter 4 we show that a low ticket price, as e.g. offered by Low-Cost Carriers (LCCs), does not mean to sacrifice *service quality*, expressed in terms of time efficiency, at all. On average, we measure lower delays and therefore a better service quality, for services offered by LCCs. Additional services above the basic, such as the transportation of luggage or the serving of food and beverages on board during the flight, could indeed result in higher ticket prices.

Globally air transportation is a growing industry which grew 4% to 5% on average during the past four decades. According to the *Airbus Global Market Forecast 2018-2037* air transportation will continue to grow by 4.4% annually during the next two decades. That means a doubling of demand every 16 to 17 years. In the past any shocks to the demand development by political instability, terror or economic crises were overcome within a short time. We observe that growth in air transportation demand is tightly connected to growth of gross domestic product (GDP), however, over time we see larger volatility, i.e. larger up and down swings, and higher average growth rates in demand development, than we see in GDP.

In parallel to global economic growth, especially in the Asian markets, the standard of living improved internationally as well. Today modern communication devices, such as smartphones are available and affordable to the broad public. These devices which are usually connected to the Internet allow global communication, even in the remotest regions.

So, what are the motives why people travel from *A* to *B*? *Mobility* follows hard on the heels of communication and information. We are increasingly virtually and socially interconnected. In general, urban mobility and individual transport is increasing. People demand to move freely and seamlessly from one place to another. We find this development within cities or between cities and metropolitan areas. Bus connections, train connections, number of private cars or number of trips per bike, car, bus, train, or aircraft are growing. The recent emergence of bike and car sharing businesses can only dampen the increase in privately owned cars and urban road traffic. Many cities have great problems with road congestion ("traffic jams") during peak times ("rush hours").

At the same time this development underlines the demand of the current generation for instantly available transport alternatives, such as state-of-the-art public transportation, private Taxi services, such as Uber, car, bike, e-bike or e-scooter sharing services to keep waiting times for an available mode and transport duration to the next destination to a minimum.

Time is valued higher than ever before in our fast-paced societies, so the sooner and quicker we ride, the better. It will be more and more common to instantly pay for such transport services by an app installed on our smartphones.

But what drives mobility globally? Whole generations nowadays prefer a different country than their home country to develop professionally, academically, or creatively. Through the Internet people gather enough information to get an "abstract" impression of places and cultures in other countries. This leads to people wanting to travel to and to explore places away from home. Many people move to other countries for a certain amount of time or even permanently. We observe an increasing number of *expatriates*, who work, live, or study abroad, away from their native country.

Expatriates from developing countries in many cases support their relatives or friends in their home countries financially. When visiting their native countries expats give their relatives and friends an idea of life abroad from their own first-hand experience. This could lead to more and more emigrating people when the potential future and quality of live is better abroad.

All people want to live in a peaceful, stable, and constitutional society in which every individual can develop and work freely. Such desire is the main motive for people to leave, or even flee, their hometown and country to take the chance to move and live in another, perhaps better, place. Especially the "brain drain", i.e. the loss of qualified, creative, educated, young, strong men and women, is bitter for some developing countries, because the migration of these forces will have detrimental long-term economic effects. Nevertheless, most people feel connected to their native home country and region which leads to regular mutual visits and to global mobility in general.

In many developed countries for example in Europe, in North America or in Asia it is essential to provide the best education for young people, so enterprises have access to a skilled and innovative work force. This work force can be enhanced by people emigrating to other countries to advance their skills or by people immigrating from other countries that bring skills which are needed for the national economy.

Even in our European society life is not all that well, and we, like any other society, struggle with social problems, such as inequality and low birth rate. Costs of living is remarkably high in the bigger cities. Therefore, financial pressure and poverty is increasing in urban societies. The regions outside of metropolitan areas struggle with a high unemployment rate, with economic decline and a decreasing number of local businesses. Many member states of the European Union and other countries in Europe need to provide financial support in form of subsidies for local industries and small and medium-sized enterprises. These subsidies are required to stabilize the local economy and to keep the rural areas internationally competitive and attractive.

In Chapter 6 and Chapter 7 we highlight the need to subsidize remote regions in Europe in order to provide basic accessibility by air transport with the local or national capitals or hub airports and to keep fares affordable for the general public.

In addition there exist a large number of historic and recent international trade agreements and relationships, which lead to bilateral visits and therefore an increase in global mobility, which produces a basic level of demand for international air transport, for example from Europe to the capitals of Latin-America, Africa or the Middle East and *vice versa*.

In Asia we see other reasons for increasing mobility, especially regarding air transportation. With the technological and economic boom in South-East and East Asia and in the Middle East during the last two decades an affluent class emerged, who can now afford to travel for leisure by air transport. Therefore, these groups are in the position to reach destinations domestically or internationally, e.g. to build business partnerships, to make family visits or to travel to far away countries.

*Tourism* is a large and increasing industry. Only one such example is the growing Chinese tourism, which poses new challenges for neighboring countries. Popular places in the region, for example holiday islands Jeju-do in South Korea or Bali in Indonesia, are increasingly being visited by Chinese tourists among other destinations, such as Thailand, Vietnam, or Cambodia.

At the same time fares for flight tickets dropped considerably over the last decade, particularly due to an increasing number of LCCs on the market. This development accelerated demand for air transport even further and made flight tickets affordable to even more people.

The problem that surfaces is that airport capacity and air traffic infrastructure is not growing at the same pace as demand. We observe a growing number of airports that have difficulties coping with serving a growing passenger demand with the given infrastructure. This situation eventually leads to bottlenecks in the global air transport system. In Europe it takes many years, often at least one to two decades, to build additional runway infrastructure to expand airports. It is somewhat easier to expand terminal infrastructure, but even such projects may require additional road or rail infrastructure that takes long approval processes prior construction. Growing public opposition makes the planning and approval processes even more cumbersome.

For this dissertation we conducted several studies to highlight the current situation regarding air transport performance in Europe. We put special emphasis on small and large airport operational and financial performance (Chapters 5, 6 and 7) and we highlight methods to support airport master planning by use of computer simulations and quantitative modelling (Chapters 8, 9 and 10). For the most part we built our own datasets, because, in contrast to the U.S., most data required for an analysis of European airports does not lie in the public domain.

With our use of simulation tools, it becomes obvious that capacity is related to operational delay, i.e. waiting time, in a system. In Chapter 1 succeeding this introduction we will present a framework for parts of my work, which sets the agenda for finding the right measures to solve airport capacity related problems. Delay is inversely related to *level of service (LoS),* which means when delay increases, LoS decreases. Since delay could accumulate quickly and increases exponentially during periods of traffic congestion, LoS deteriorates non-linearly as well.

In Chapter 1.2 we will introduce the policies that lead to unleashed air transport demand, first in the U.S., later in Europe and in Asia after opening, privatizing and deregulating the market. This development started in the mid-seventies and continued in the late last century and in the early new Millennium.

It becomes more and more difficult, especially for European hubs, to expand proportional to demand, so capacity of airport infrastructure is reached at many places during peak periods. We present methods to define “design peak periods” during which representative observations can be made. For example, schedules for modeling purposes are typically extracted during peak periods, when airport infrastructure is under pressure from simultaneously maneuvering aircraft and flows of passengers.

We find that different airports operate similar runway configurations during their historical evolution. Therefore, in Chapter 1 we investigate if benchmarks from one airport can be used for other airports planning or operating a similar layout. We compare international parallel runway airports, which may present a template for decision making of airport management, when choices need to be made regarding different expansion scenarios.

The first chapter is titled “Airport evolution and capacity forecasting” and was submitted by editor Professor Anne Graham’s request to *Research in Transportation Business & Management* Volume 1, Issue 1, in 2011. During the review process the manuscript was rejected by the reviewers. Time was too close to the publishing deadline for any major revisions, so the manuscript remained unpublished. Chapter 1 is a completely revised version of the original working paper. We updated some figures and statistics. Furthermore, we included recommendations from the reviewers.

Chapter 2 is a precursor to our operational performance and simulation studies. We introduce the main measures and indicators and the elementary basics of queuing theory, such as *Little's Law*.

We give further recommendations on how to gather required data, such as flight schedules, how to finetune the simulation and how to improve the experimental design. For example, it is necessary to define a system boundary, when we study a system, because processes and systems are so highly connected that a complete system analysis would be close to impossible. In the case of our simulations, we generally use the Terminal Maneuvering Area (TMA) around an airport as system boundary. The TMA is roughly defined by a 50 km radius airspace around an airport. On the boundary of the TMA aircraft enter or terminate the simulation. Models including several airports, which are interconnected by routes, are currently uncommon. We have built a nationwide draft simulation model for Norway, but without some sort of automation to build the model and to enter the input data, such models are currently too cumbersome to develop and are limited in their practical use.

In Chapter 2 we furthermore put an emphasis on the main performance indicator for airlines and airports – *punctuality*. We present punctuality not as a single number, but as a full distribution. Another important distribution, which also could be used to make our simulation studies more "realistic", is the distribution of interarrival time, i.e. headway, of aircraft approaching the airport.

Chapter 2 was first published as a working paper within the German Airport Performance (GAP) student research project, later a slightly revised version was published in June 2011 in the *Aerlines* magazine, volume 50, under the title "Airport punctuality, congestion and delay – The scope for benchmarking".

Chapter 3 focuses on airport capacity and how this could be a good measure of maximum operational productivity. We benchmark European airports with similar characteristics, e.g. based on financial, operational, or environmental indicators, so we can compare them better against each other. We establish a methodology to classify airports with a similar runway configuration, and we calculate capacity based on documents from the Federal Aviation Authority (FAA). Furthermore, we develop a simple procedure of finding representative design peak traffic samples, since we want to compare day-to-day operations at different airports. The simulation models in Part III: "Airport Capacity and Operations" of this dissertation require such *design peak day* traffic samples for the baseline scenarios.

The third chapter was published under the title "Airside productivity of selected European airports" in the *conference proceedings of the Air Transport Research Society* (ATRS) World Conference 2010 in Porto.

In the fourth chapter we will discuss a development, which propelled the growth of air transportation. The market of the "traditional" air carriers is under pressure due to competition from the so called "low fare airlines" or LCCs. These airlines offer cheap ticket prices for mainly intracontinental connections. The biggest airlines in Europe in this segment are for example, Ryanair, EasyJet, or Air Berlin and in the U.S for example

Southwest Airlines or jetBlue. My colleague Alberto Gaggero from the University of Pavia and me shed light on the question, if increasing competition by LCCs has a positive impact on the quality of service at an airport, measured in average delay, for all users. We have tried to answer this question for 100 European airports, by taking meteorological data into account and excluding weather related delay. For our study we collected information on daily flights and meteorological conditions for these airports over a period of 20 months. We programmed a web crawler for this purpose to collect and structure the data automatically. The data set includes weather characteristics, scheduled and actual departure or arrival times, aircraft type, origin and destination and flight number. The code for the web crawler is presented in the Addendum.

Chapter 4 was published in October 2015 in *Transport Policy*, Volume 43, published by Elsevier with the title "Low-cost carrier competition and airline service quality in Europe."

Chapter 5 emerged from an airport benchmarking study in which the author participated. The study was commissioned and financed by the Norwegian Ministry of Transport and Communications (*Samferdselsdepartementet*) and was supervised and coordinated by Jürgen Müller from Berlin School of Economics and Law and Hans-Martin Niemeier from University of Applied Sciences Bremen. In this project we compared the developments of different costs and revenues at European airports over several years against Norwegian airports, which to a large extend are operated by *Avinor*, a publicly owned and privately managed company.

We were particularly interested in the question, how large the "critical mass", i.e. the demand measured in passengers per year, needs to be, when operations can be expected to become profitable. Our main measure of profitability is *Earnings before Interests and Taxes* (EBIT). We call the crossing point of (zero) profits and demand level the *break-even point*. We assume that it is theoretically possible that an airport with a demand greater than the critical mass can operate profitably.

We analyzed financial and operational data of the years 2002 to 2010 from 139 European airports in ten countries to identify the most profitable organizations for each passenger level.

We found that airport exist which are operating profitable with less than 100,000 passengers per year, for example the airports Aurillac (AUR) and Bergerac-Roumaniere (EGC) in France. In Italy we found examples of profitable airport with less than 350,000 passengers per annum. These are *best practice* airports being profitable with demands below the widely assumed (e.g. according to analyses published by Deutsche Bank Research) average demand levels of between 500,000 and 2 million passengers per year. It is frequently the case that such best practice airports have discovered a favorable cost and revenue structure for themselves. Eventually it is our goal to reveal the "recipe" which makes such airports profitable to apply it on other airports with a similar level of demand. For example, we may reveal creative business ideas which generate additional revenue flows.

By applying an algorithm, we could graph and follow the development of the *profitability frontier* or *envelope* over a couple of years. This curve is defined by benchmark airports with the maximum profits (or losses) per each passenger level. We observed in this kind of presentation how the "benchmark airports" (i.e. such airports with a similar demand level, processes, cost and revenue structure etc.) which achieve the best profitability compared to their peers, change over time. We assume that other airports with approximately the same number of annual passengers would be able to achieve the same profits (or losses) like the best practice benchmarks.

The fifth chapter with the title "Benchmarking European airports based on a profitability envelope" was published by Springer in 2012 as part of the series *Lecture Notes in Computer Science*, Volume 7555, Computational Logistics.

Chapter 6 is a reprint of the final report for the EU research project "Baltic.AirCargo.Net" which was written by the author. The report was based on latest flight schedules, data from my previous projects and reports from within the project. The research project was managed by the University of Applied Sciences: Technology, Business and Design in Wismar, Germany.

We tried to identify scheduled air cargo routes and networks in the Baltic Sea Region (BSR) which have growth potential. This region was out of political interest more broadly defined than what are commonly considered to be the Baltic States, since we included, for example, Belarus, Western airports of Russia (e.g. exclave Kaliningrad or Saint Petersburg) and airports in Norway.

We used statistical data on commodity trade among BSR countries and with Non-BSR countries. Our aim was to identify industries that rely on air cargo and justify higher transportation costs. Our main measure was *value per weight*. We discovered that especially high-tech equipment and electronics with a high value per ton is predestined for air transportation.

We make recommendations for new routes and future industries that develop the need for air transport services. Currently trade between Russia and Germany dominates the region, especially regarding the air cargo routes between Moscow and Frankfurt. We benchmarked different charges and its structure and schemes among airports in the region.

Chapter 6 was published in 2014 by Berliner Wissenschafts-Verlag (BWV) under the title "Economic Outlook for the Air Cargo Market in the Baltic Sea Region" as a book chapter in *Air Cargo Role for Regional Development and Accessibility in the Baltic Sea Region – Handbook of the EU projects Baltic Bird and Baltic.AirCargo.Net in the framework of the Baltic Sea Region Programme 2007–2013*.

In Chapter 7, like in Chapter 5, the basic considerations developed out of the study for the Norwegian Ministry of Transport and Communications. Internal discussions revealed a dilemma that subsidies for certain flight connections, so called Public Service Obligation (PSO) routes, increase every year and that there exist too few competitors for serving these routes. The biggest restrictions to the Norwegian low demand markets pose the Short Take-off and Landing (STOL) runways with lengths less than 1,200 meters.

From the ministry we received information about the total amount of subsidies in different administrative regions and for different route networks among airports in Norway. Further information is stated in the "invitation to tender" documents which are published on the ministry website for applicants to serve the offered routes. In these publications we find revenue and passenger data on each offered route in form of *origin-destination matrices*. We use these tables for our further calculations, in which we tried to solve the questions: how large the unit costs are, and which subsidies are required for certain PSO routes.

By using the “*goal seeking*” methodology and with our own problem structure we were able to trace back the operating costs for any PSO route between low demand airports in Norway. We could then compare these costs with “market prices”, i.e. unit costs, of other carriers in Europe.

We found that the reclaimed costs of the single operator of a remote PSO network in Norway were far above comparable market prices. One could conjecture that there is evidence for market abuse by a monopoly carrier.

The seventh chapter was published by Springer in March 2013 in the journal *NETNOMICS*, Volume 13, Issue 2, in under the title *“*Social costs of public service obligation routes - calculating subsidies of regional flights in Norway.”

Chapter 8 resulted from collaboration with Joachim Daduna from Berlin School of Economics and Law. As part of my diploma thesis, among other topics, I calculated the capacity of new Berlin-Brandenburg International airport (BER). By using a simulation modelling software (SIMMOD), which is developed by the U.S. Federal Aviation Authority (FAA) and is applied for air transport planning, I was able to simulate airside processes at future BER airport, for example on the two parallel runways. Together with Professor Daduna we extended the initial model and analyzed the impact of different *aircraft mixes* on runway capacity. Furthermore, we wanted to forecast the consequences of *traffic growth* on the airside processes at BER airport until capacity is reached and when this will approximately occur.

For solving these questions, we stepwise increased todays combined demand at existing airports Schönefeld (SXF) and Tegel (TXL) in the simulation and measured the resulting aircraft *throughput* und *average delay*, i.e. waiting time, at critical locations on the airfield. In this way we create a function of average delay depending on average hourly and daily demand for the currently planned infrastructure at BER airport.

Delay represents an important indicator for service quality of an airport. For the two independent parallel runways modelled in segregated mode we calculated a capacity of 76 flights per hour at a predefined level of service of 6 minutes per flight. This means that the installed runway capacity at BER is scaled too low and will be reached during peak periods shortly after opening the airport in October 2020. We compared and validated our results against data from London-Heathrow and Munich airports. In real daily operations additional delay occurs because of manifold reasons, for example because of inclement weather, understaffed air traffic control, late passengers, technical problems, misdirected luggage, and cargo or due to other processes.

The eighth chapter was published in January 2012 by Springer under the title “Airport capacity and demand calculations by simulation—the case of Berlin-Brandenburg International Airport” in the journal *NETNOMICS*, Vol. 12, Issue 3.

Chapter 9 is the result of a consultant project, where I was asked by Norwegian airport operator *Avinor* to conduct a simulation study of Oslo-Gardermoen (OSL) airport. Shortly before I was commissioned with the simulation study a masterplan until the year 2050 was published for OSL, which includes several expansion scenarios. In a team meeting we decided to reduce the simulation study to the most plausible expansion scenarios. This planned infrastructure, additional terminals, piers, taxiways and runways were included by the author into an airport simulation model. In this way we were able to activate certain infrastructure in the model when certain capacities (parking stands, taxiway, or runway) are reached or certain expansion scenarios shall be tested.

Since this study required a detailed model, especially regarding pushback and taxiing processes, we detected many hotspots and bottlenecks in the airport layout. Therefore, recommendations could be made regarding the quality of the plans, scenarios, and possible improvements.

At the time when we build the model it became evident that a study was needed by the airport management to evaluate the capacity gain, when a pair of rapid exit taxiways (RETs) at the two runways is added. The investment and construction costs require objective justification. We were able to quantify the capacity and service quality increase when a pair of RETs is added to the initial airport layout of the base year 2017.

In this study we used average delay as our main performance indicator. However, we chose a "conservative" level of service of *5 minutes of average delay per flight* as our threshold value which defines the practical capacity at OSL airport. After installing the RETs on the two independent parallel runways operated in *mixed mode,* we calculated a practical runway capacity of up to 100 movements per (peak) hour, which means a 10% capacity increase compared to the scenario without RETs.

The ninth chapter with the title *"*Simulating airside capacity enhancements at Oslo-Gardermoen airport post-2017 - Measuring the effect on level of service by adding a pair of high-speed runway exits*"* was published in December 2016 as an *internal scientific report* within *Avinor*.

Chapter 10 is the result from collaboration between my colleagues Frederik Schulte and Professor Stefan Voß of University of Hamburg. We examine the question of how much taxiing fuel emissions are emitted by the (departing) aircraft per (design) day, and if and to what extent we can reduce these emissions through improved operations and alternative push-back order sequences.

We use a combination of optimization methods (skill VRP - Vehicle Routing Problem) and simulation techniques and present our calculations using the example of Oslo Airport. We can calculate used amounts of fuel based on the accumulated taxiing times we get from the simulation. Thus, it is possible to investigate different sequences of departing aircraft in terms of externalities, delays, and emissions.

The results so far reveal that emissions can be reduced by up to 10% through an altered push-back sequence, depending on the initial sequence (which may already be optimal). Further studies on this topic will follow. Our tables on gaseous emissions factors depending on the amount of burned fuel by aircraft engines during the taxiing process can be used and extended by other airports for environmental benchmarking studies.

The tenth chapter entitled "Reducing Airport Emissions with Coordinated Pushback Processes: A Case Study" was submitted to the 8th International Conference on Computational Logistics (ICCL) in Southampton, UK, October 18-20, 2017 and was published by Springer Publishing in *Lecture Notes in Computer Science* (LNCS), vol. 10572.

The eleventh and final chapter of this dissertation examines an important aspect in air transportation that yields a huge field of research. We will investigate network effects and performance advantages of different type of *network structure*. We will see that the topology of networks, such as *hub-and-spoke* (H&S) or *point-to-point* (P2P) networks, has an impact on the service level, measured in average delay. We introduce a simple measure of graph density from graph theory to quantify the extend of interconnectivity between airports in the air transport network. Then we apply this measure to the networks served by carriers in Europe. We further distinct between LCCs and FSCs and analyze the differences in performance and network structure. We measure how delay that is caused at an origin propagates through the network.

In our view *delay* represents a negative externality from production in the service industry. Therefore, our goal is to reduce delay through managerial changes in operations. Aircraft starting their operating day punctual with no delay tend to build up delay from earlier flights of that aircraft or inherit delay from other flights over the course of the day, especially during busy periods. If we find ways to minimize delay, which means we find the causes of delay and try to minimize their impact on scheduled operations, we usually reduce the operating costs and environmental footprint of commercial aviation.

Our main finding is that aircraft tend to propagate less delay, if the flight originates from a *hub* airport that is very frequented in comparison to other airports in the same network. We found measurable differences between the main European carriers in network structure and performance.

The original manuscript of Chapter 11 was presented at the ATRS world conference in July 2018 in Seoul, South Korea. We included the revised version of our paper from June 11th, 2019 that was originally submitted to the journal *Transportation Research Part A: Policy and Practice* and is currently under revision.

Over the course of preparing this work we planted the seed for several follow-up studies written by our international colleagues. Our work was cited and referenced by several academics and practitioners in the field. In the case of Berlin-Brandenburg airport (Chapter 8) our work proved to be an early valid recount of the real situation. A couple of years after our study the problems with the construction and dimensioning of the airport became obvious to the broader public. The new airport is still not yet opened and associated setbacks and cost escalations make daily headlines in the news. The opening of the BER is now scheduled for October 31st, 2020.

We are satisfied that we could provide answers to international transport ministries, airport operators and other practitioners and decision makers with our research. Recent work includes consultation and research on airports in France, Sweden, and South Korea and on delay propagation in air transport networks in Asia, the U.S. and especially in Europe.

# Part I: An Introduction to Airport Benchmarking

Part I of this dissertation gives an overview about the developments and problems in air transportation in the past thirty years and delivers global concepts how to tackle the main managerial problems. One of the approaches is the technique of airport benchmarking which has some major drawbacks in how it was hitherto executed. The following chapters highlight how airport benchmarking could be refined by making use of capacity related measures, such as runway capacity and delay. For the first time airport benchmarking not only relies on historic data, but by conducting airport simulations future airport layouts can be compared to existing configurations. Therefore, it is possible to connect past performance observations at airports with future forecasts, financial outlooks and master plans.

# 1 Airport evolution and capacity forecasting

Branko Bubalo

**Abstract.** The growth of airports is typically limited by landside or airside capacity. From a commercial perspective, large stable volumes of passengers are desired to pass through the airport facilities, but demand fluctuates daily and hourly. Management at congested airports must work under certain trade-off conditions, where runway throughput could be affected by a growing number of average passengers per flight, because Heavy aircraft (offering many seats) require further separations minima between succeeding flights. Consequently, the sequencing of batches from the same aircraft category must take place. Decisions and timing regarding airport expansion must be based on reviewing scenarios which consider the mix of present and future aircraft types and growth of traffic, long before certain level-of-service thresholds are exceeded. An emphasis is placed on methods of assessing baseline and future peak time traffic volumes and level-of-service through observations and simulation including the concept of simultaneous occupation of space.

## 1.1 Introduction

In recent years it could be observed that large infrastructure investments were taking place in many countries worldwide. In developing countries and markets, infrastructure is built on a large scale. With increasing income and wealth, people and their economies develop the need for energy, goods, information and mobility. This leads to the progressive installation of domestic or continental telecommunication, energy transmission and transportation networks. Because economies and markets are strongly interlinked, in times of globalization they cannot be viewed in isolation.

Therefore, as our own (e.g. European or North American) need for new infrastructure grows and approaches some level of maturity, this is not the case for many countries, e.g. in Asia. On the other hand, the substantial growth of the economies in countries like Brazil, Russia, India, and China (BRIC) will have a huge impact on national air transportation and its infrastructure (FAA 2007). Prior to the global financial crisis, the Challenges of Growth report (EUROCONTROL 2008) estimated for a most likely scenario a near doubling of 2007 traffic levels until 2030. In the future it is expected that an increasing flow of traveling passengers originating from e.g. BRIC countries will transfer or terminate at European airports (de Neufville 1995, p. 6).

Besides instant communication people in (transformed) industrialized countries demand rapid transportation for business or leisure needs. Airline profits not only in the domestic European air transport market are largely driven by frequent business travel, although tourism or personal (leisure) travel still accounts for most of the demand on many routes. In general, profitability of airlines, even at times of high load factors, looks bleak, mainly as a result of low fares and even more so at times when fuel prices are increasing and volatile.

This situation will lead to further consolidation or failures of airlines (ACRP 2010, p. 8). It is not clear, if new market entrants, especially Low-Cost Carriers (LCCs), will outweigh the reduction of competitors that go bankrupt. The airline industry will face further economic challenges. We find a lot of competition in an industry that gives consumers a great utility, but at the same time earns its operating profits at the margin.

## 1.2 Air transportation deregulation and Low-Cost Carrier competition

Since the mid-seventies in the U.S. and during the 1990s in Europe, regulations concerning prices, routes and the scheduling process for air travel were gradually abandoned, allowing for more freedom for different business strategies and leading to increased competition among airlines. LCCs entered the European short- and medium-haul market in the late 1990s with airlines like Ryanair and EasyJet, which were able to offer much lower fares than their established counterparts, and consequently captured a large share of the market. This development brought more competition and opportunities, but also at the same time huge challenges to various portions of the European air transport system. As de Neufville (2008) points out, there will be a war over prices and capacities (**Table 1**) not only among airlines and airline alliances, but also between major hub airports.

**Table 1.** Airlines before and after deregulation (Source: de Neufville 2008).

| Choice | Before Deregulation | After Deregulation | Implications of Deregulation |
|---|---|---|---|
| Routes | Strictly controlled | Freedom to change | Loss of secure tenure |
| Prices | Set by formula | Freedom to change | Price wars |
| Frequency of flights | Controlled | Freedom to set schedules | Capacity wars |
| Aircraft type | Often controlled | Freedom to choose | Capacity wars |

With success of the LCCs a network of *secondary airports* (de Neufville 2005) is evolving across Europe. Airports, which used to be operated in a safely regulated climate under federal authority and supported by subsidies, must now learn to compete against privatized and rapidly growing secondary airports. Therefore, most international European airports are going through a somewhat painful transformation process towards becoming modern profit-oriented businesses (Graham 2005, p. 99; IATA 2004, p. 109).

Secondary airports not only offer specialized services for LCCs, but also for other customers, e.g. business aviation, general aviation, cargo or the military (de Neufville 2000). Typically, especially in the U.S. and in Japan, in metropolitan or general catchment areas with large populations, multi airport systems are in place (de Neufville and Odoni 2003, pp. 129; de Neufville 2005; Bonnefoy, de Neufville and Hansman 2010), which are able to serve this wide variety of airport clientele. Consequently, it may be anticipated that competition will intensify on comparable routes from different airports in those same regions.

Already today, when traveling from Rome to London, the difference in time, distance and convenience is negligible, whether the route Rome-Ciampino to London-Stansted airport (served mainly by LCCs) or Rome-Fiumicino to London-Heathrow airport (served mainly by flag carriers) is chosen, but the difference in ticket fare is significant. In the greater London area, there are five international airports Heathrow, Stansted, Gatwick, London-City and Luton. Furthermore, the greater London region has the highest density of airports and airstrips in Europe, which may serve as additional reliever airports in the future. Paris is served by three airports, Charles-de-Gaulle, Orly und Le Bourget. In contrast, the multi airport system of Berlin with Tegel, Schönefeld and Tempelhof airports will be fully replaced by one single airport Berlin-Brandenburg International (BER) (Bubalo and Daduna 2011).

Although reliever airports will experience strong growth, the main European hubs will dominate the air transport system and will need adequate airport capacity. A hub is a main international airport which links the hinterland and national routes, the spokes, with international connections. We therefore speak of a hub-and-spoke network in Europe. Over time, with new routes and airports, this will transform into different layers of hub-and-spoke, point-to-point or hybrid type networks (de Neufville 2005).

## 1.3 Air transportation economics

It is a great relief for the air transportation industry that the global economy is back on track and that demand for air transportation in 2010 reached previous levels of 2008. Only the price for jet fuel remains almost 3-fold higher when compared to 2000, at currently 107 US-Dollars per barrel (IATA 2011).

Fuel prices are not the only dark cloud on the horizon, however. Additional taxes and regulations, e.g. regarding environmental issues, mainly noise, carbon dioxide ($CO_2$) and nitrogen oxides ($NO_x$) emissions, place further financial pressure on airlines, which in turn look for opportunities to reduce costs in their balance sheets. Thus, airlines try to lower the aircraft landing charges at airports, in some cases threatening to divert their traffic to alternative locations (Walters 1978, p. 132) in an effort to force airports into cooperating.

**Fig. 1** shows how the main stakeholders in commercial air transportation – the passengers, airlines, and airports - are connected financially. Ticket fares paid by passengers for scheduled services are the dominating source of revenue for European airlines, accounting for over 90% of the revenues. Additional revenue increasingly comes from cargo services. On the cost side, airlines must work with very high (variable) direct operating costs, mainly due to the necessary fuel, flight crew and maintenance for their flights (Wensveen 2007, p. 304). The split between direct and indirect costs for members of the Association of European Airlines (AEA) is approximately 60% to 40%. Airport charges account for about 8% of direct operating costs, but together with air navigation charges these amounts represent 16% of the direct costs or about 10% of the total airline costs.

The airport charges present a major source of income for the airports and feed directly into the aircraft-related and passenger-related revenues as part of general aeronautical revenues. The aircraft-related charges or revenues are paid directly by the airlines to the airports on an aircraft maximum take-off weight (MTOW) basis and are usually subject to negotiation. Other charges are collected as published in the airport charges manuals for services such as aircraft gate stand or parking space, provision of fuel; towing, aircraft maintenance and sanitation (Walters 1978, p. 133). In the case of passenger-related charges or revenues, these are collected by the airlines for the airports on a per passenger basis for passenger services mainly inside the terminal facilities.

Data from the Airport Council International (ACI) Europe (2010) suggests a split of 47% to 53% between aeronautical and non-aeronautical revenues among its member airports in 2008. Non-aeronautical revenues result from offering additional services to airport customers. These additional services include e.g. shopping, car parking, food and beverages, and car rental facilities. The various types of customers, domestic, leisure, international and business passengers, request many additional services which are provided by airports. For many airports the terminals represent strong revenue generators.

As noted earlier, around 50% of total airport revenue is generated from non-aeronautical (commercial) activities, mainly from providing space for shops, restaurants, offices, conference rooms and even hotels. Airports have become vital socio-economic centers where passengers enjoy spending time and money. Then too the close proximity to air transport is increasingly beneficial to many local and regional businesses. These factors combine as high-demand airports evolve into larger entities, which are increasingly closely linked to the immediately surrounding region. Such airports could be described as airport city, *Aerotropolis*[1] (a term coined by Kasarda) or airport region (**Fig. 2**).

[1] http://www.aerotropolis.com/ [last accessed on 31/07/2018]

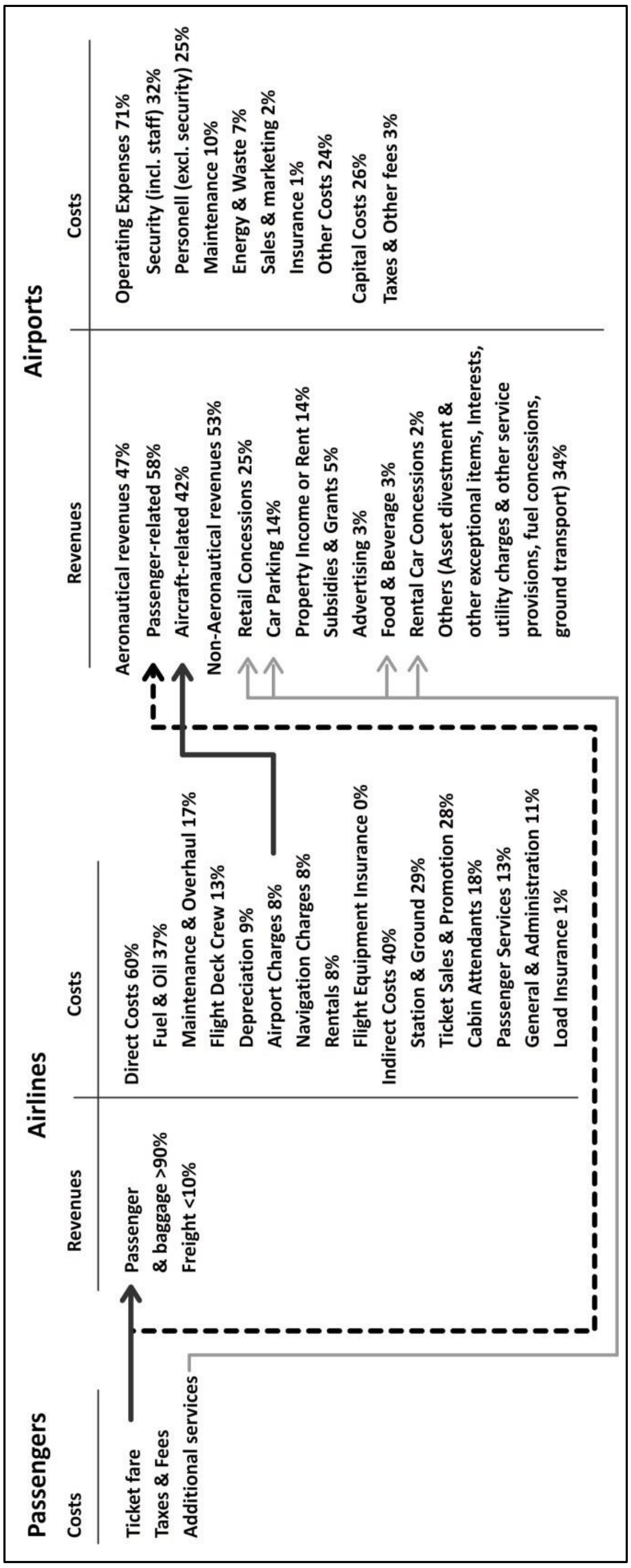

**Fig. 1.** Sources of Revenue and Costs in the Air Transportation Industry (compiled from AEA 2008 and ACI Europe 2010)

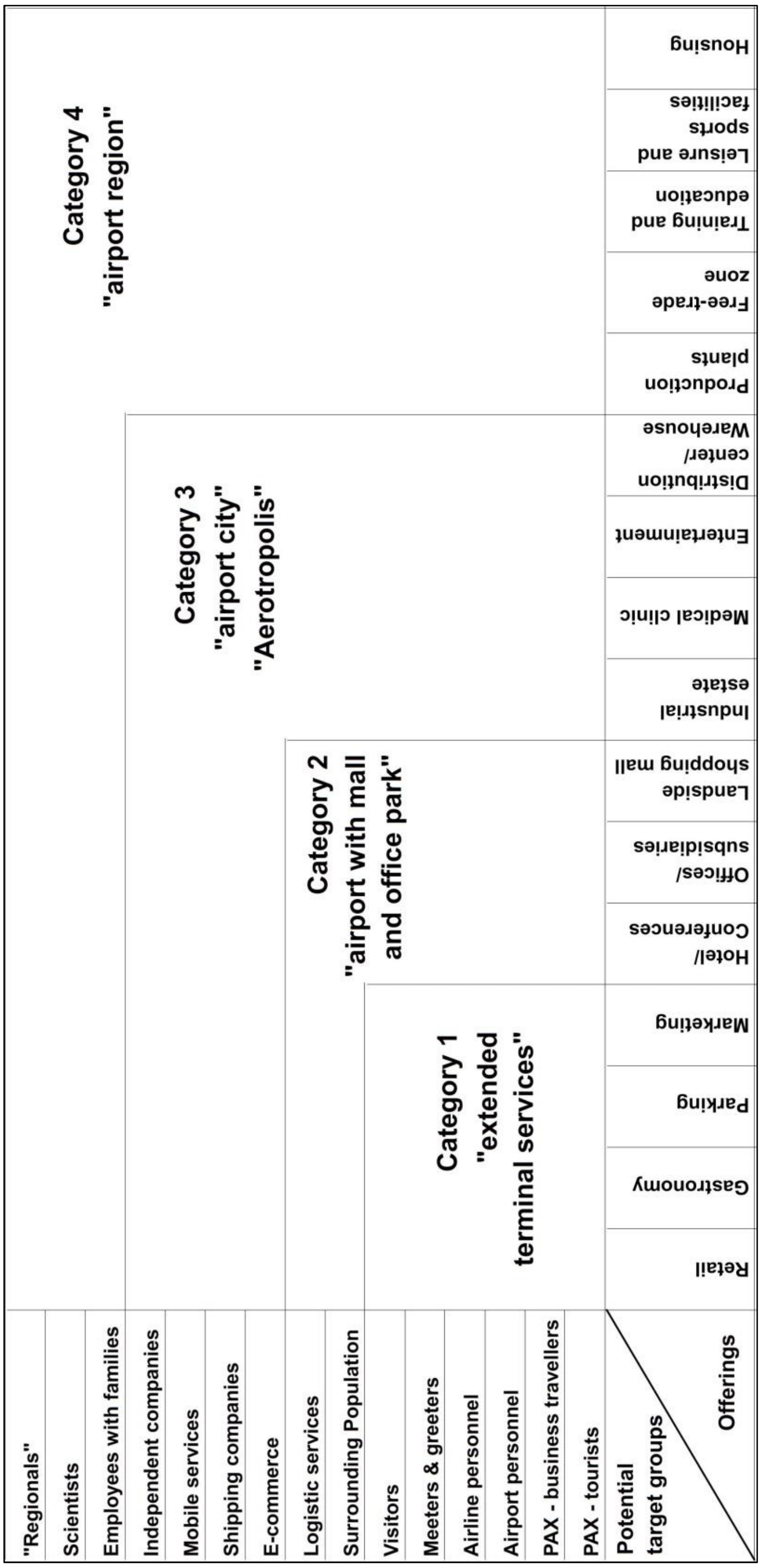

**Fig. 2.** Potential target groups and offerings at airports (based on Leutenecker 2008)

Airports handling a large share of freight not only need to guarantee quick transit times, but also need to provide state-of-the-art facilities for sorting, packing, storing, and distributing shipments of logistic companies (Leutenecker 2008). However, globally we still find many examples of commercially underdeveloped airports with only basic services, such as duty-free and souvenir shops or restaurants.

The operating expenditures of European airports include costs for labour (25%), airport maintenance (10%), energy and waste (7%), and adequate safety and security (32%). Operating costs account for 71% of total costs, whereas capital costs for investments in airport infrastructure accounts only for 26% of the total costs (ACI Europe 2010).

### 1.4 Airport demand

Passengers rarely have a great deal of choice regarding accessible airports, and certainly most metropolitan hub airports have a virtual monopoly for serving a large densely populated area with routes to international destinations. Furthermore, connecting or transfer passenger traffic is a strongly competitive market among hub airports in Europe, especially regarding intercontinental long-haul high-yield routes. It is indeed difficult to find accurate data about the amount of transfer passengers (de Neufville and Odoni 2003, p. 134) shared among European airports, which would allow estimates to be made about the additional income generated by this group, e.g. by retail or food and beverage sales. By conducting surveys more information about the preferences of transfer passengers is gathered.

The data from ACI provides insight into the number of international passengers at the top global airports (**Table 2**), which are arguably equally interesting as a target group in their own right from a purely commercial point of view (IATA 2004). Surprisingly **Table 2** shows many European airports at top of the list among the largest global hubs ranked by the number of international passengers in 2008, i.e. London-Heathrow (1.), Paris-Charles-de-Gaulle (2.), Amsterdam (3.) and Frankfurt (5.). In the U.S. we find airports serving even more total passengers and total flights than what is shown in **Table 2**, but this market is largely driven by domestic demand, such as at Atlanta-Hartsfield, Dallas-Fort Worth or Chicago-O'Hare airports.

Even at a major internationally recognized airport like New York-John F. Kennedy international passengers make up only 47% of the total passengers at that airport. Of course when dividing the market at European airports into domestic intra-EU flights and international extra-EU flights, only three airports (Paris-Charles-de-Gaulle, London-Heathrow and Frankfurt-am-Main) can be identified as having number of flights with passengers originating or terminating outside of Europe larger than 50% (Eurostat 2008).

**Table 2.** Main global hubs by the number of international passengers (ACI Europe 2008, Eurostat 2008)

| City | IATA Code | ICAO Code | International Passengers in millions | Total Flights in thousands | Passengers per Flight | International Passengers as Percentage of Total Passengers | Intra-EU Passengers as Percentage of Total Passengers | Extra-EU Passengers as Percentage of Total Passengers |
|---|---|---|---|---|---|---|---|---|
| London Heathrow | LHR | EGLL | 61.3 | 478.5 | 140 | 91% | 33% | 58% |
| Paris Charles de Gaule | CDG | LFPG | 55.8 | 559.8 | 109 | 92% | 42% | 50% |
| Amsterdam Schiphol | AMS | EHAM | 47.3 | 446.6 | 106 | 100% | 56% | 44% |
| Hong Kong Chek Lap Kok | HKG | VHHH | 47.1 | 309.7 | 155 | 99% | - | - |
| Frankfurt am Main | FRA | EDDF | 46.7 | 485.8 | 110 | 87% | 37% | 50% |
| Dubai International | DXB | OMDB | 36.6 | 270.4 | 138 | 98% | - | - |
| Singapore Changi | SIN | WSSS | 36.3 | 234.8 | 161 | 96% | - | - |
| Tokyo Narita | NRT | RJAA | 32.3 | 194.4 | 172 | 97% | - | - |
| London Gatwick | LGW | EGKK | 30.4 | 263.7 | 130 | 89% | 57% | 32% |
| Bangkok Suvarnabhumi | BKK | VTBS | 30.1 | 249.4 | 155 | 78% | - | - |
| Madrid Barajas | MAD | LEMD | 29.8 | 469.7 | 108 | 59% | 36% | 23% |
| Seoul Incheon | ICN | RKSI | 29.6 | 212.6 | 142 | 98% | - | - |
| Munich Franz Josef Strauss | MUC | EDDM | 24.5 | 432.3 | 80 | 71% | 44% | 27% |
| Dublin | DUB | EIDW | 22.6 | 211.9 | 111 | 96% | 85% | 11% |
| New York John F. Kennedy | JFK | KJFK | 22.4 | 441.4 | 108 | 47% | - | - |
| Zurich | ZRH | LSZH | 21.4 | 275.0 | 80 | 97% | 66% | 31% |
| Rome Leonardo da Vinci-Fiumicino | FCO | LIRF | 21.4 | 346.7 | 101 | 61% | 37% | 24% |
| London Stansted | STN | EGSS | 20.0 | 193.3 | 116 | 89% | 83% | 6% |
| Copenhagen Kastrup | CPH | EKCH | 19.4 | 264.1 | 81 | 90% | 62% | 28% |
| Vienna Schwechat | VIE | LOWW | 19.0 | 292.7 | 67 | 96% | 63% | 33% |
| Toronto Pearson | YYZ | CYYZ | 18.4 | 430.6 | 75 | 57% | - | - |
| Brussels | BRU | EBBR | 18.3 | 258.8 | 71 | 99% | 65% | 34% |
| Manchester | MAN | EGCC | 18.1 | 204.8 | 105 | 85% | 55% | 30% |
| Kuala Lumpur | KUL | WMKK | 17.8 | 211.2 | 130 | 65% | - | - |
| Barcelona | BCN | LEBL | 17.4 | 321.5 | 94 | 58% | 46% | 12% |
| Istanbul Atatürk | IST | LTBA | 17.1 | 276.1 | 104 | 60% | - | - |
| Los Angeles | LAX | KLAX | 16.7 | 622.5 | 96 | 28% | - | - |

## 1.5 Economies of scale

Financially airports rely heavily on the amount of passenger traffic passing through their facilities and using their services. In the past airports were mainly concerned about convenience of service for the primary carriers stationed there, which were normally their largest clients. For this reason, they were less interested in airline load factors or available seats per flight. Today, it is generally recognized that airports benefit directly from strong demand and the economies of scale which are needed by airlines to achieve profits through large volume, highly utilized and highly frequented scheduled flights (Walters 1978, p. 131; de Neufville 2000, p. 5; IATA 2004, p. 109).

However, airports share a similar financial risk, when passenger demand on individual routes and flights is declining. A superior *level of service* (LOS) at airports is a precondition for airlines to meet their turnaround times and to maximize aircraft utilization. Both airlines and airports increasingly have a keen interest for making travelling seamless and enhancing the travel experience for the passengers, which ideally leads to repeated visits or connections, by, for example, business travelers (customer retention). Regarding selected UK airports a trend can be observed when spare capacity at airports is shrinking, but demand is rising (Mott MacDonald 2010). Since additional demand cannot be satisfied by airlines by increasing the frequency of flights at selected airports with capacity limitations, the growth of such markets is only possible by means of higher load factors or larger aircraft, resulting in more average passengers per flight (IATA 2004, p. 91).

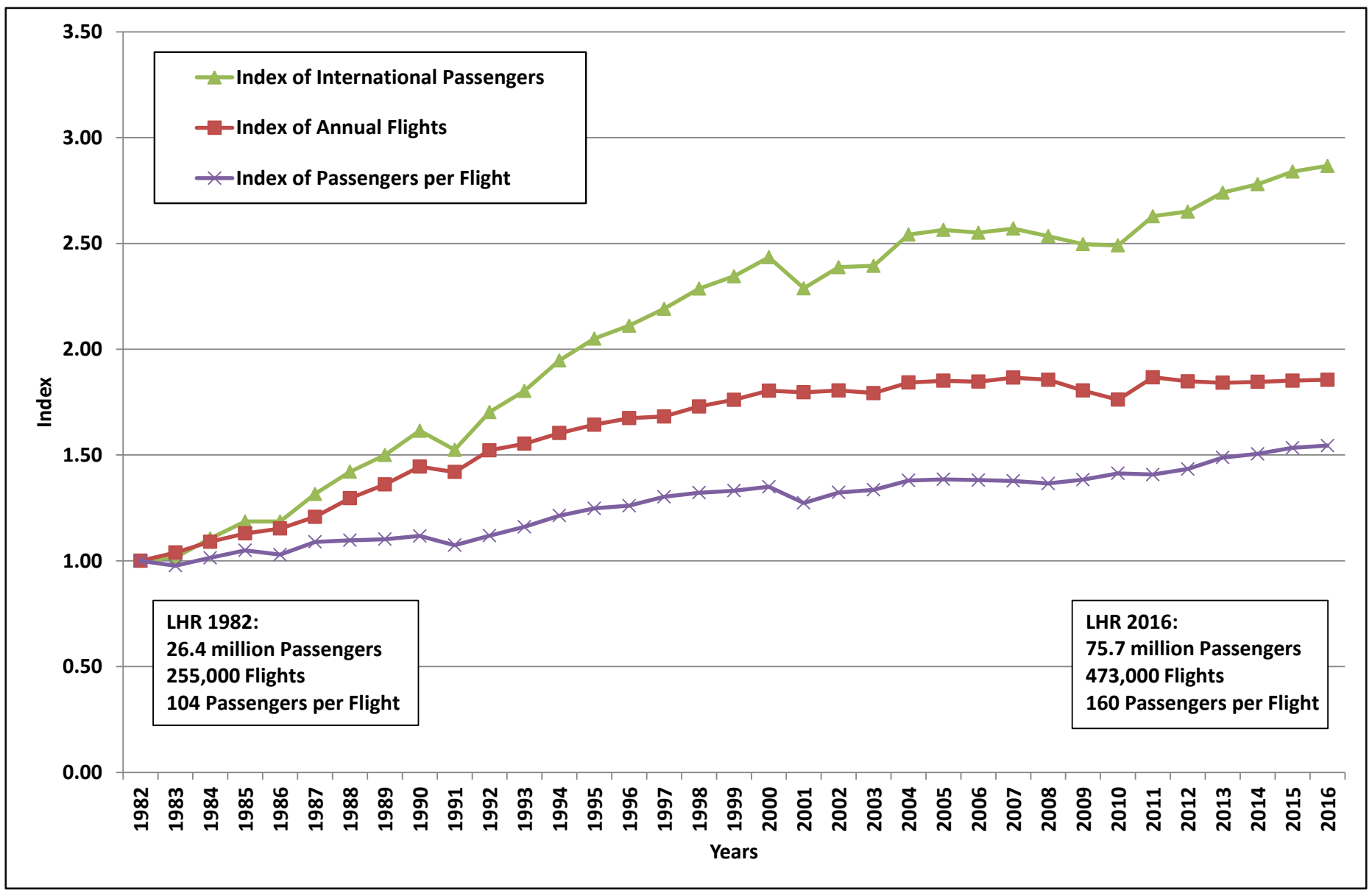

**Fig. 3.** London Heathrow Trend in Annual Passengers and Flights from 1982 to 2016 (Source: UK Civil Aviation Authority)

When studying the long-term trend in passengers, movements and passengers per flight at London-Heathrow airport over the last three decades (**Fig. 3**), the direct effect of the number of average passengers per flight on the resulting number of annual passengers can be observed. Especially between 1991 and 2000 small continuous increases in the number of average passengers per flight, parallel to small increases in annual aircraft movements, led to a substantial boost in passenger numbers at London-Heathrow from 40 to 64 million. From 1982 to 2016, passenger numbers at Heathrow almost tripled (from 26.4 to 75.6 million), and number of flights increased by 85% (from 255,000 to 473,000). Passengers per flight grew by a moderate 54% (from 104 to 160).

In absence of available data before 2003, the relation between available seats and passengers on board, the (seat) load factor, at London Heathrow airport is assumed to have remained constant over the last decades, at around 75%.

## 1.6 Estimating capacity

Like other modes of transport and especially regarding scheduled services, demand at airports fluctuates by the hour, the day, the week, the month, and the year. Therefore, demand and capacity are typically expressed by these time bases, such as for example flights or passengers per hour. The capacity of an airport on the airside (runways, apron, aircraft parking space) is mainly determined by the aircraft mix and associated separation minima between aircraft types of different weight and wake turbulence categories, and runway configuration (FAA 1983). On the landside (passenger or cargo facilities and airport access) capacity is limited by the available space and processing speeds of various stations in - and outside of - the terminal. This includes not only security checks and passport control but also extends especially to baggage handling (IATA 2004).

If capacity or service rate is not sufficient and cannot meet the fluctuating demand, then excess demand results in the build-up of waiting queues and delays, and consequently lower LOS (**Fig. 4**). Normally queues dissolve in quieter periods, especially during midday hours ("fire breaks"), and therefore do not continue to increase indefinitely. When the demand rate is higher than the service rate over consecutive hours, the whole airport system is operating in an unsustainable manner and enters a region of poor LOS. Under this condition queues tend to increase exponentially.

An assessment of capacity for an airport is commonly based on a design day schedule and involves a (design) peak hour analysis, often separately for arriving and departing traffic or flows of terminating, originating, and transferring passengers. For European airports, a separate analysis of international intra- and extra-EU, Schengen, and domestic passenger flows should be made.

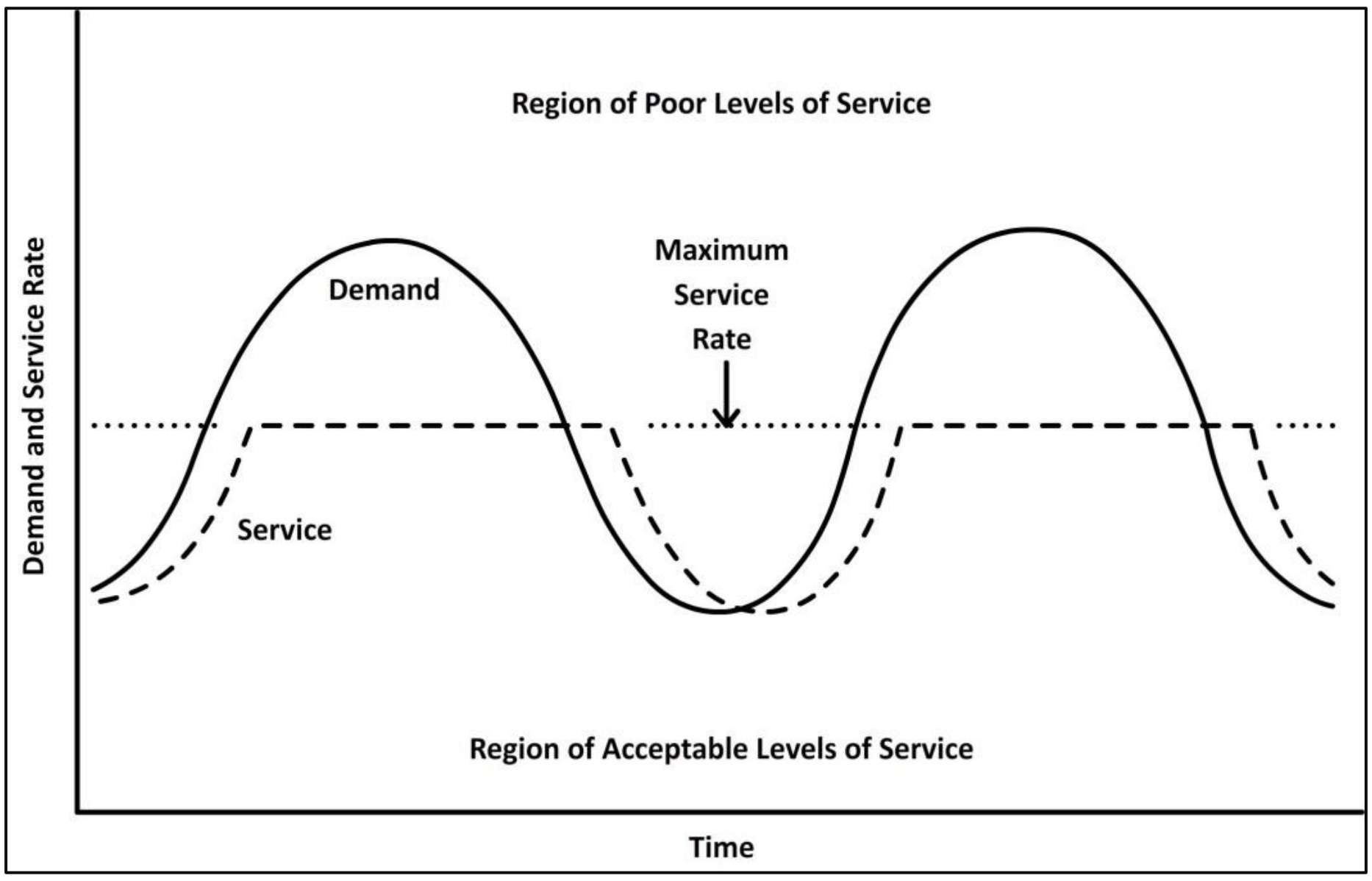

**Fig. 4.** Rate of demand and service (Source: TRB 1975)

Finding the number of flights and the passenger volume in the design peak hour at a particular airport can certainly be a data intensive task, especially depending on the chosen definition for the design peak hour, which has been variously defined by various international transport-related institutions and ministries as the standard busy rate, the typical peak hour, the busy hour, etc. (de Neufville and Odoni 2003, pp. 851). However, the design peak hour should be understood to satisfy only one precondition: that it should not represent an absolute peak, but rather a busy period which recurs during 10 to 30 days (de Neufville and Odoni 2003, p. 853) throughout the year. Therefore, the design day and the design peak hour can be estimated in the following straight-forward way.

## 1.7 Design peak period assumptions

The Central Flow Management Unit (CFMU) of EUROCONTROL publishes weekly reports with the demand pattern of average daily flights and minutes of en-route delay (2009 dark red line, 2010 light green line) each week in European airspace for the running year (**Fig. 5**), with data from 2009 (dark blue columns) and 2010 (light yellow columns). It shows the seasonal variation of demand and related delay between winter and summer.

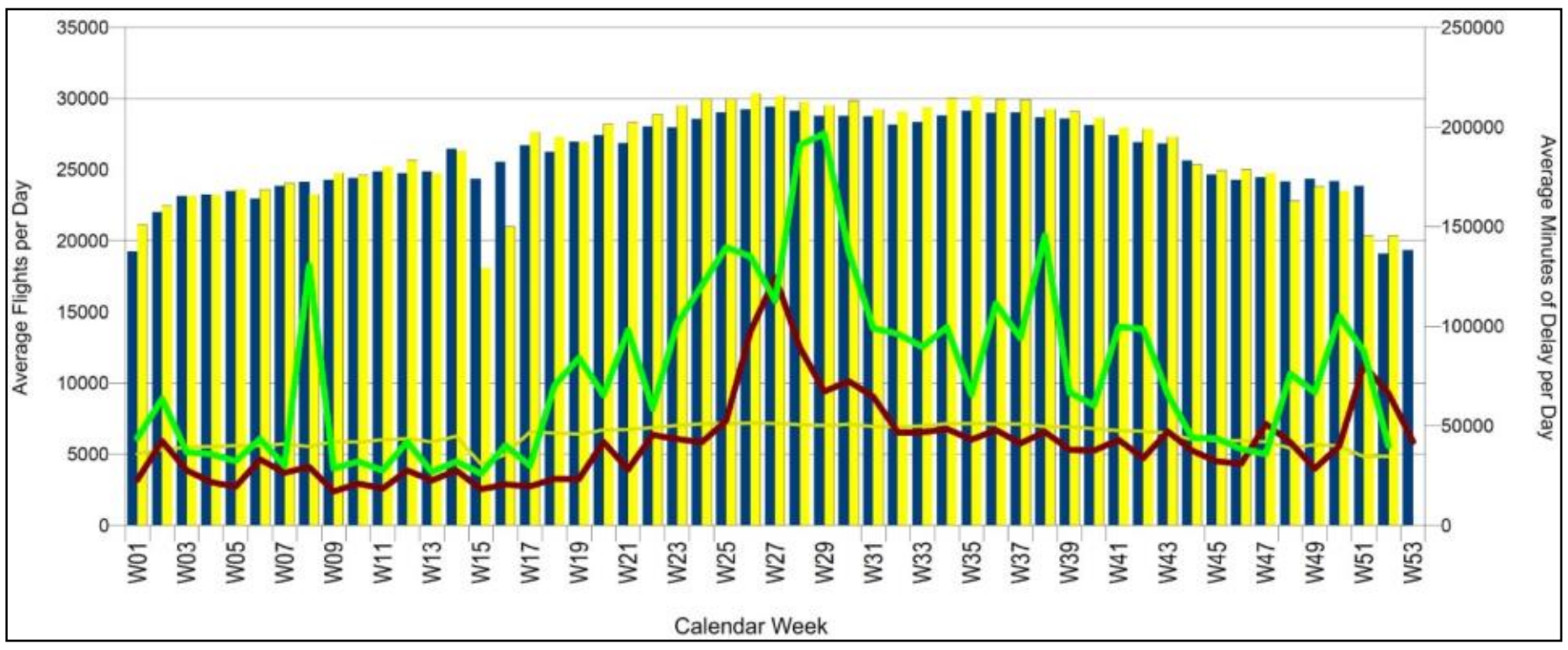

**Fig. 5.** Pattern of average daily European flights and delay (EUROCONTROL 2010)

With the demand pattern displayed in **Fig. 5**, it is possible to select a given week having average traffic above or below a given threshold. Observations have shown that this pattern does not significantly change over the years. By narrowing the candidate weeks needed for eventually isolating a design (peak) day or hour for peak period analysis, the time and data effort needed to find the proper schedule for an airport analysis is significantly reduced.

Therefore it is plausible to assume, at least for the larger European hub airports, that the design peak day or hour can be isolated from selected traffic data samples especially for weeks 24 to 27 (usually mid-June to the beginning of July) and for weeks 34 to 38 (usually in September) (**Fig. 5**), these being the two most important busy periods.

As **Table 3** also shows, the busiest day of the week is frequently a Friday. However, as general peak period characteristics are desired, absolute peaks should be avoided. Accordingly, a design day other than a Friday is chosen. This could be a Monday or Thursday (which tend to be the second busiest days), but it also depends on the day-to-day variation of traffic throughout the week. Hence it would make little sense to choose a Thursday as design day for an airport under investigation, when it is known that peak demand and congestion can be found on Saturdays (ACRP 2010, p. 91).

A collection of operated airport schedules for different representative days of the week, the month, or the year, is certainly an important prerequisite for capacity analyses and especially for detailed demand pattern analyses, e.g. regarding seasonal variations. Design day flight schedules also serve as main input for airport simulations.

**Table 3.** Peak Days and Peak Daily Flights in European Airspace (modified from EUROCONTROL CFMU 2011)

| Year | Date | Calendar Week | Flights in Europe |
|---|---|---|---|
| 2010 | Fri 02/07/2010 | 26 | 32,575 |
| | Fri 10/09/2010 | 36 | 32,341 |
| | Fri 09/07/2010 | 27 | 32,334 |
| | Fri 18/06/2010 | 24 | 32.247 |
| | Thu 01/07/2010 | 26 | 32,198 |
| 2009 | Fri 26/06/2009 | 26 | 34,476 |
| | Fri 03/07/2009 | 27 | 33.895 |
| | Fri 10/07/2009 | 28 | 33,833 |
| | Fri 28/08/2009 | 35 | 33,383 |
| | Fri 11/09/2009 | 37 | 33.342 |
| 2008 | Fri 27/06/2008 | 26 | 34,476 |
| | Thu 26/06/2008 | 26 | 33,895 |
| | Fri 13/06/2008 | 24 | 33,833 |
| | Thu 19/06/2008 | 25 | 33,383 |
| | Fri 04/07/2008 | 27 | 33,342 |
| 2007 | Fri 31/08/2007 | 35 | 33,506 |
| | Fri 29/06/2007 | 26 | 33.480 |
| | Fri 14/09/2007 | 37 | 33,371 |
| | Fri 07/09/2007 | 36 | 33,279 |
| | Fri 21/09/2007 | 38 | 32.971 |
| 2006 | Fri 15/09/2006 | 37 | 31,914 |
| | Fri 01/09/2006 | 35 | 31,841 |
| | Fri 30/06/2006 | 26 | 31,686 |
| | Fri 08/09/2006 | 36 | 31,553 |
| | Fri 22/09/2006 | 38 | 31,550 |
| 2005 | Fri 17/06/2005 | 24 | 30,663 |
| | Fri 01/07/2005 | 26 | 30,569 |
| | Fri 02/09/2005 | 35 | 30,469 |
| | Fri 16/09/2005 | 37 | 30.338 |
| | Fri 09/09/2005 | 36 | 30,169 |

### 1.8 Airport congestion and level of service

When looking at operated design day schedules from airports, we can gain insight into the characteristic peaking of the demand pattern at an airport (de Neufville and Odoni 2003, p. 856). These transportation system demand patterns are plotted over time, by the hour of the day, where the information about individual peaks is used to dimension and design the related server or part of the infrastructure. In this article only the most critical parts of the airport system will be examined: The airport terminal, the runway system, and the immediate airspace.

### Terminal demand, capacity and LOS

The pattern of distributed hourly seats by flown aircraft at Amsterdam-Schiphol airport is exemplarily presented in **Fig. 6**. Over the course of the day, the pattern of demand at Schiphol airport fluctuating between arriving seats and departing seats is clearly recognizable. It is equally clear that resources need to be shifted within the same terminal space. Indeed, strong arrival peaks with up to 9,500 seats per hour, between say 08:00 and 09:00, and departure peaks with up to 11,000 seats per hour, between say 10:00 and 11:00 can be observed. At around 14:00 both patterns form the highest simultaneous peak in seat volume, approximately 16,000 seats per hour.

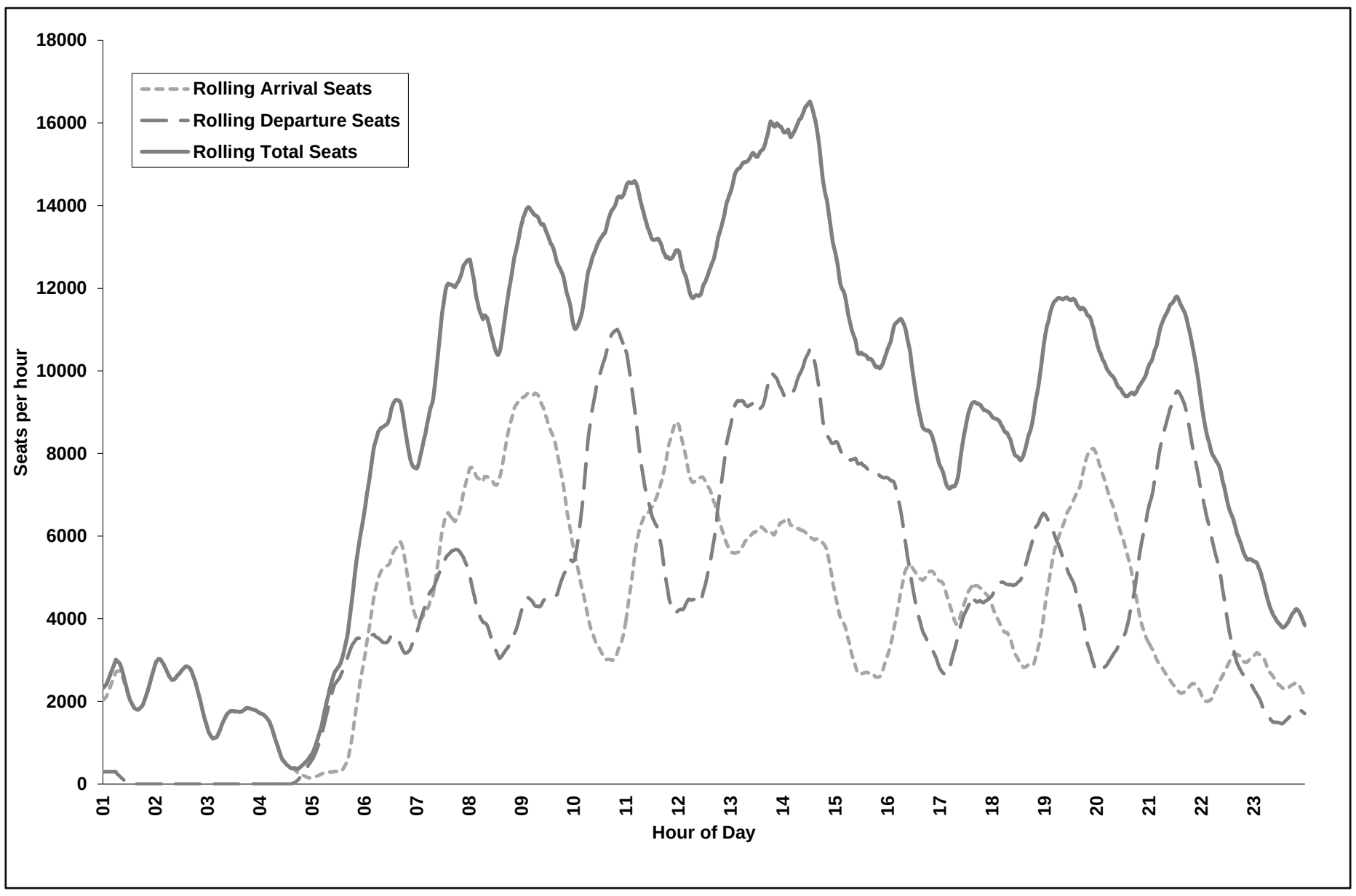

**Fig. 6.** Hourly seat distribution at Amsterdam-Schiphol airport on design peak day 2008

What had hitherto not been recognized, are the actual load factors on these flights (ACRP 2010, p. 88), but first order approximations for the design (peak) day can be made by applying seat load factors of between 75% and 85%, resulting in total peak passenger volumes of between 12,000 and 13,500 passengers per hour. It is important to note, that the actual dwell time (de Neufville and Odoni 2003, pp. 639) of passengers using the airport facilities at the same time should be factored into this result.

Thus the terminal space should be dimensioned and designed according to the number of passengers simultaneously occupying the volume of space by applying

*Little's Law*: *The number of simultaneously present passengers in the terminal equals design passenger volume* (e.g. per hour) *multiplied by dwell time* (e.g. in hours) (Little 2011).

The dwell time of passengers is, however, obviously not easy to assess without large-scale observations of passenger flows. Given the difficulties in obtaining data on arriving and departing passenger dwell time, the minimum connecting time (MCT) for transfer passenger could be used as an approximation (minimum) dwell time in the airport terminal buildings, but keeping in mind that not all processes required for originating and terminating passengers are included in this figure.

For example, Schiphol airport has a high rate of international passengers, of which many might be connecting, and it serves as a connecting hub for KLM airline. Schiphol achieves an MCT of between 40 minutes (international-domestic connections) and 1 hour and 20 minutes (international-international connections) or say 1 hour on average (assuming an equal share of connections, however a larger share for international connections is much more likely in this case). These MCT include all the processing times needed for passengers and their luggage to transfer from an arrival gate to the departure gate. Since in our example at Schiphol airport the average MCT and approximated average dwell time per passenger is about 1 hour, the passenger volume in the terminal facilities during the design hour is equal to the volume of passengers simultaneously present.

Furthermore, space is dimensioned and provided according to a predefined LOS for each terminal facility, which includes check-in, security, passport control, departure gates, etc. (IATA 2004, pp. 179). Thus, space standards vary between 1.0 and 2.5 square meters per passenger, depending on the desired LOS for the specific facility (de Neufville and Odoni 1992; IATA 2004, pp. 179).

An excellent (A) or high (B) LOS should be targeted for stable flows, few delays and high levels of comfort. What is even more interesting for the individual passenger is not only the available space in a particular queue or departure/arrival hall, but rather how long he or she will have to *wait* (**Table 4**).

In **Table 4** the desired maximum waiting times are suggested for different airport terminal facilities (IATA 2004, p. 189). Nowadays the development is such that electronic advance ticketing over the internet and carry-on luggage are making facilities like the check-in counters or baggage conveyor systems gradually obsolete. If neither sub-optimal space requirements nor waiting times can be achieved, IATA defines such facilities as “under-provided” (IATA 2014).

**Table 4.** Level-of-Service maximum waiting time guidelines (modified from IATA 2004 and IATA 2014)

| Level of Service and Maximum Waiting Time Guidelines (in minutes) | | Check-In Economy | Check-In Business class | Passport Control Inbound | Passport Control Outbound | Baggage Claim | Security |
|---|---|---|---|---|---|---|---|
| A | An excellent level-of-service. Conditions of free flow, no delays and excellent levels of comfort | 0-12 | 0-3 | 0-7 | 0-5 | 0-12 | 0-3 |
| B | High level-of-service. Conditions of stable flow, very few delays and high levels of comfort | | | | | | |
| C | Good level-of-service. Conditions of stable flow, acceptable delays and good levels of comfort | | | | | | |
| D | Adequate level-of-service. Conditions of unstable flow, acceptable delays for short periods of time and adequate levels of comfort | 12-30 | 3-5 | 7-15 | 5-10 | 12-18 | 3-7 |
| E | Inadequate level-of-service. Conditions of unstable flow, unacceptable delays and inadequate levels of comfort | | | | | | |
| F | Unacceptable level-of-service. Conditions of cross-flows, system breakdowns and unacceptable delays; an unacceptable level of comfort. | | | | | | |

When airport management has a desire to translate design hourly passenger volume into anticipated annual figures at airports certain design hour factors come into use (Kanafani 1981; de Neufville and Odoni 2003, pp. 851; Janić 2007, p. 61). The design hour factor is the percentage share of design hour volume to annual volume. It is important to realize that with the increasing annual volume of traffic or passengers the trend of design hour factor to annual volume is strongly decreasing (**Fig. 7**). This effect is largely due to the strong peaking of low volume and underutilized airports, in contrast to airports with high volume and constant traffic flow. **Fig. 7** highlights the range of design hour factors for 60 airports in Europe.

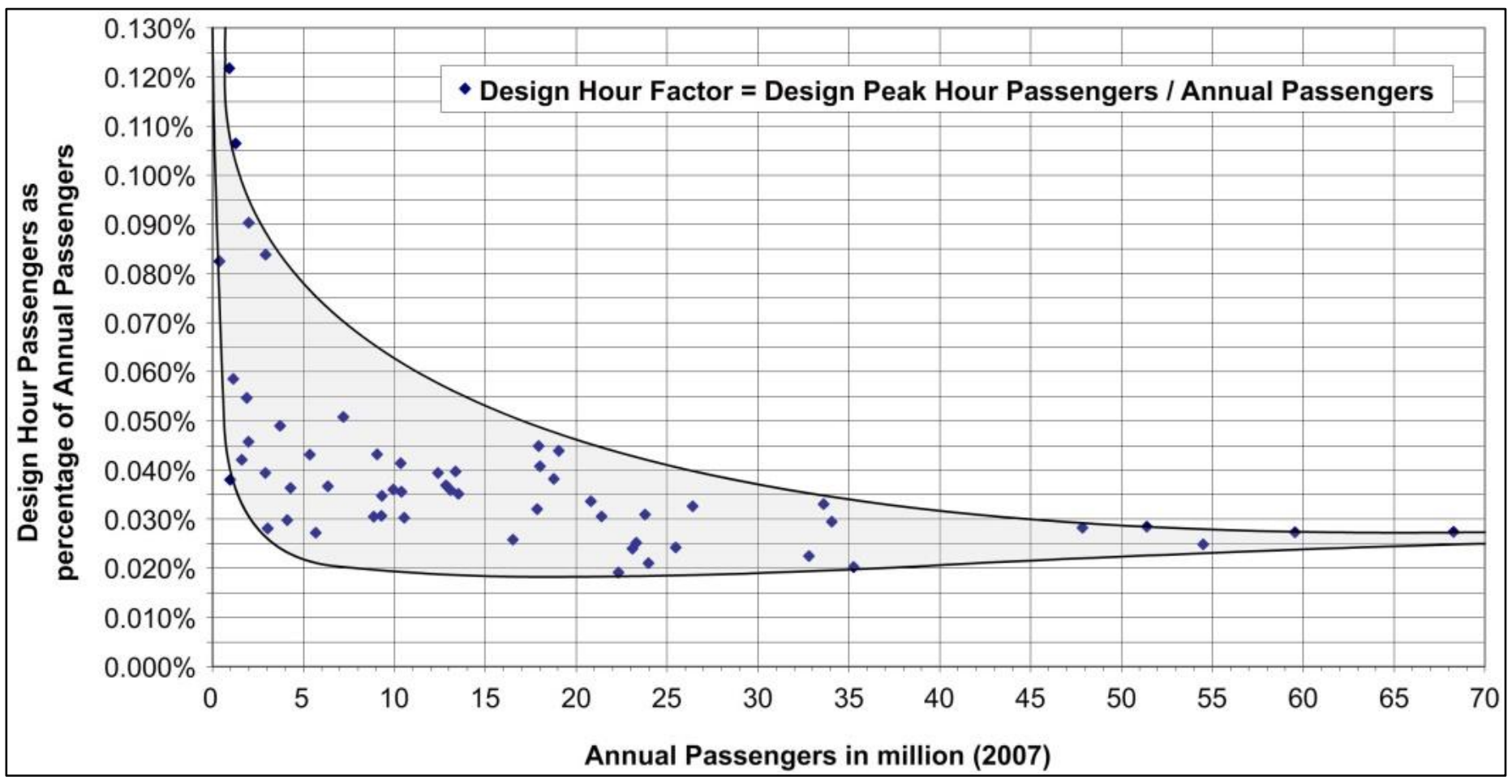

**Fig. 7.** Trend and range of design hour factor to annual passengers at European airports in 2007 (Data from Eurostat, Official Airline Guide, Flightstats.com)

Checking the consistency of these conversion factors is critical when assuming volumes of traffic or passengers in airport expansion planning in the future (Kanafani 1981; de Neufville and Odoni 2003, p. 859). For example, an airport currently serving 10 million passengers might have a volume of 4,000 passengers on the design peak hour, resulting in a conversion factor of 0.04%. Now the forecast ten years into the future predicts an annual volume of 15 million passengers, thus resulting in a decreasing design hour factor of 0.035% (**Fig. 7**) and in an hourly volume of 5,250 passengers per hour.

A good overall representation is the "assumptions cube" shown in **Fig. 8**. It is an extension of the "assumptions rectangle" (Kanafani 1981), by the dimension "utilization", which means we added the relationships between demand vs. capacity to the picture. Kanafani (1981) includes in his original model the dimension "time" (annually vs. hourly) and "functional area" (airside vs. landside). We learned from his paper that it is crucial in airport master planning to check the figures and assumptions carefully.

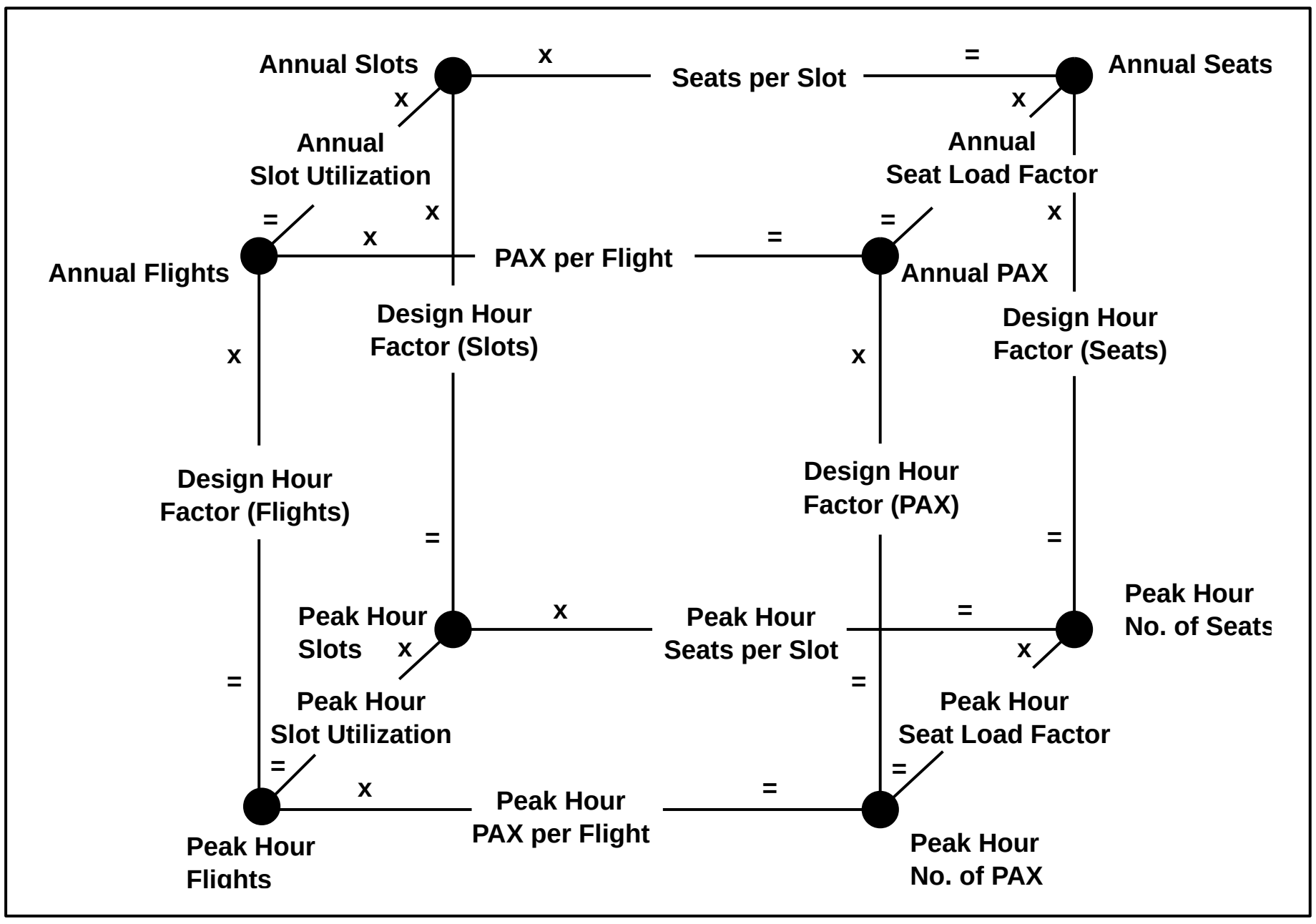

**Fig. 8.** Assumptions cube (own illustration, based on Kanafani 1981)

In our case, capacity on the airside at *coordinated*, i.e. (partly) congested airports is expressed by the *number of available slots*, i.e. maximum number of flights, per time period. In the case of unconstrained airports, the number of slots can be replaced by the airport capacity, for example by runway or parking stand capacity.

On the landside inside the terminal capacity is expressed by the *maximum number of passengers per time period*, which can be served by various servers and stations, such as passport control and check-in counters.

When aircraft arrive or depart, passengers request to be served almost simultaneously within a short period of time. During times of day when traffic is high it is difficult for an airport to serve each request immediately. Therefore, queues will form, and passengers will have to wait a certain amount of time before being served. It is common that the number of passengers per time period is determined by average aircraft size, i.e. by the number of flights, available seats on arriving and departing aircraft and related (expected or actual) load factor on each flight.

**Runway demand, capacity and LOS**

The most critical issue with runways and the expansion of runway capacity is the long lead time for approval, planning and constructing such a fundamental piece of infrastructure. Additionally, expansion projects at existing airports are strongly opposed by environmental groups and residents in the vicinity of an airport. Narita international airport presents a good example of a failed airport expansion project, due to successful local opposition. Located 60 kilometers east of Tokyo this airport was initially planned to serve the city's international flights, but local opposition forced the government to reduce its plans from three runways to eventually one runway. It took over three decades of negotiation for a second short runway to go into operation in 2002.

Since runways are depreciated over up to 40 years (see annual reports of Fraport and Schiphol Group), the construction costs for building a new runway should certainly not pose an insurmountable obstacle for airports with a critical mass of demand. Compared to some observed construction costs for airport terminals, which can easily reach billions of Euros (e.g. London-Heathrow Terminals 1, 3 and 5) (IATA 2003, p. 359), costs for building new runways seem reasonably low at up to approximately 300 million Euro (e.g. Schiphol airport's fifth runway Polderbaan) (IATA 2003, p. 185).

Runways may be viewed as "lumpy investments" (Walters 1978, p. 136), since capacity (i.e. supply) cannot be adjusted to demand very easily, however there are still certain (smaller) steps of progression. Runways can evolve from simple airstrips for gliders and small propeller planes, to medium-sized 3-kilometer long 45-meter wide runways for most commercial aircraft, and on to full-length 4-kilometer long 60-meter wide asphalt runways for very large aircraft such as the Airbus A380.

Furthermore the navigational and surveillance equipment installed at an airport and at a particular runway end, ranges from none installed, to the typical instrument landing system (ILS), to the high-end, state-of-the-art, precision runway monitor (PRM) equipment installed to allow independent landings on close-spaced parallel runways (< 210 meters lateral separation). Therefore, construction costs to expand current capacity can vary quite substantially for different types of runways and their configuration (Butler 2008, p. 5).

Runway demand and capacity is, analogous to terminal demand and capacity, measured in arrivals, departures, or total flights per unit of time (usually per [rolling] 15 minutes or one hour) (Janić 2007, p. 268).

When regarding a runway system as a queueing system, it is well known that the reciprocal of the demand rate is the interarrival time (de Neufville and Odoni 2003, pp. 822), which is the weighted average interval between all arriving and departing flights demanding service at the runway(s). Consequently the inverse of the *runway occupancy time* (ROT) (for a single runway) (Horonjeff 2010, pp. 497) or minimum interarrival time (for a runway system) is the throughput capacity or service rate, which is the weighted average minimum physically possible interval between peak demands in a near saturated queueing system. One of the main differences between demand and service rate is the distribution of flights, where the former is characterized by an incoming fluctuating Poisson distribution above and below capacity and the latter is characterized by an outgoing organized flow (see **Fig. 4**). Graphically these flows can be shown with cumulative diagrams (de Neufville and Odoni 2003, pp. 819). As explained in Section 1.6, ideally demand is sufficiently lower than capacity and can be served at all times, but if demand cannot be served immediately, for example due to other aircraft blocking the runway, waiting queues and delays develop.

For example, an airport with a single runway has a peak demand of 40 flights per hour, resulting in an average interarrival time of one aircraft every 90 seconds. Since field observations reveal that mixed takeoffs and landings (of mostly medium sized aircraft) occupy a given runway only for a minimum of 70 seconds, resulting in a capacity of about 51 flights per hour, a demand rate of 40 flights per hour could be served. If future peak hour demand at that airport is estimated at 65 flights per hour and the average interarrival time is 55 seconds and thus less than the capacity threshold of 70 seconds, this demand can only be accommodated on a two- runway system (such as two independent parallel runways). Small aircraft require as little as 40 seconds or less ROT, so capacity and thus requirements for additional infrastructure can vary substantially with the aircraft type mix.

Obviously, runway capacity is determined by either ROT, the number of landings compared to departures, separation minima due to wake turbulences, aircraft type mix, and prevailing weather conditions at an airport (Horonjeff 2010, p. 489). Means to minimize ROT and to increase runway capacity at an airport with regard to the predominant aircraft mix include optimized arrival and departure aircraft sequencing, adequate locating of runway exits, such as 30-degree angle rapid exits, and optimized distances from runway thresholds to the runway exits.

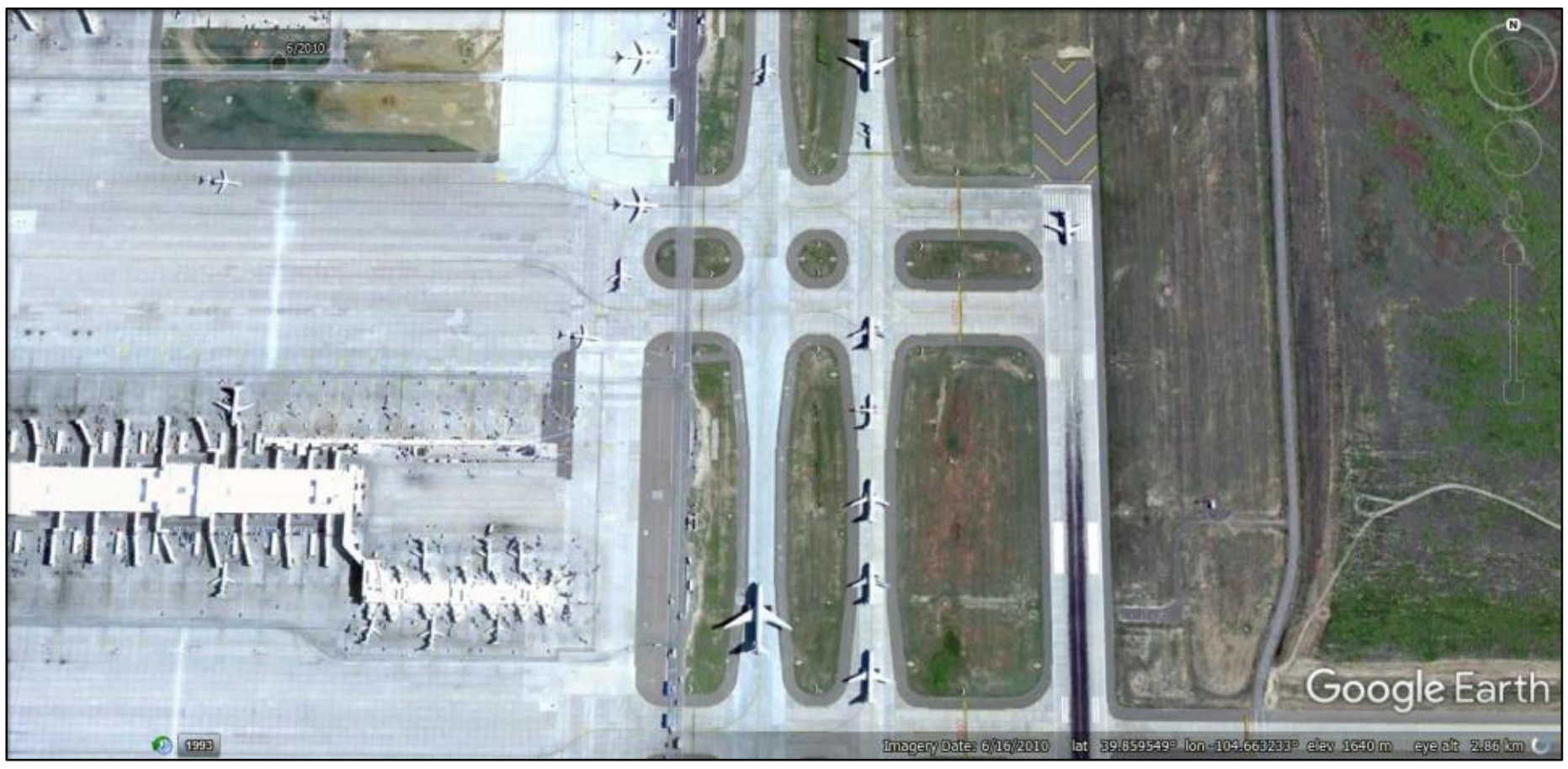

**Fig. 9.** Bottleneck situation at departure queue for runway 17R at Denver airport (Source: Google Earth 2010)

As one can see in the aerial photograph of the runway departure queue at Denver airport on June 16, 2010 (**Fig. 9**), at least eight aircraft are visible waiting for departure at runway 17R. Calculating roughly up to two minutes between following departures, the last aircraft will have to wait at least 16 minutes for takeoff clearance. Because this waiting time in the queue is effectively wasted time and costs airlines enormous amounts of money (for schedule deterioration, passenger compensation, additional fuel and crew costs), this kind of operational bottleneck must be avoided or at the very least minimized by airport and air traffic flow management.

When the demand for runway service is greater than the throughput capacity, departing aircraft must wait in the departure queue and arriving aircraft have to wait in the airspace holding stack. Delays can build up very rapidly in periods of congestion (de Neufville and Odoni 2003, pp. 444). In Fig. 10 the evolution of daily demand to LOS, measured in average delay per flight, is presented for far spaced independent parallel runways (greater than 1,310 meters lateral separation). This theoretical relationship (Horonjeff 2010; p. 488) is reproduced using data from SIMMOD simulations, based on the original design schedules of London Heathrow and BER airports, of various traffic mixes, types of operation (segregated mode and segregated mode plus mixed mode during the peaks), conservative and less conservative separation minima (e.g. 2.5 nautical miles on final approach) and stepwise increasing levels of demand (Bubalo and Daduna 2011). Such simulations are used to predict the ultimate capacity of a runway system and to forecast the impacts of future demands (Horonjeff 2010, p. 152).

As it can be seen in **Fig. 10**, when daily demand increases on the parallel runways, the average delay per flight increases exponentially. For example, London-Heathrow airport operates under a LOS of 10 minutes of average delay per flight (NATS 2007, p. 5), whereas Munich airport operates under a level of service of about 5 minutes of average delay per flight. The practical implication is, significantly more flights are available at Heathrow on each day (**Fig. 10**).

While Heathrow airport could offer between 1,150 and 1,400 slots per day, or 1,250 slots on average, on its independent parallel runways, Munich could only offer between 950 and 1,200 slots, or 1,050 slots on average. It is therefore the airport management's judgment call to decide on how many slots the airport wants to offer under the trade-off which LOS it wants to maintain for its customers (de Neufville and Odoni 2003, pp. 847).

In general, a working airport system is driven mainly by sheer growing demand. Changes in the current aircraft mix towards a more desired split of shares resulting in capacity (or environmental) benefits can mainly be influenced by airport management using incentives such as aircraft landing charges. These incentives should ultimately result in airline schedule changes. Alternatively airport management may advance operational procedures by flow optimization methods in cooperation with local air traffic control (ATC) (by sequencing similar aircraft types to minimize average separation minima between aircraft) or may expand infrastructure to minimize ROT, for example by building sufficient rapid runway exits or by installing different departure queue locations for different aircraft types.

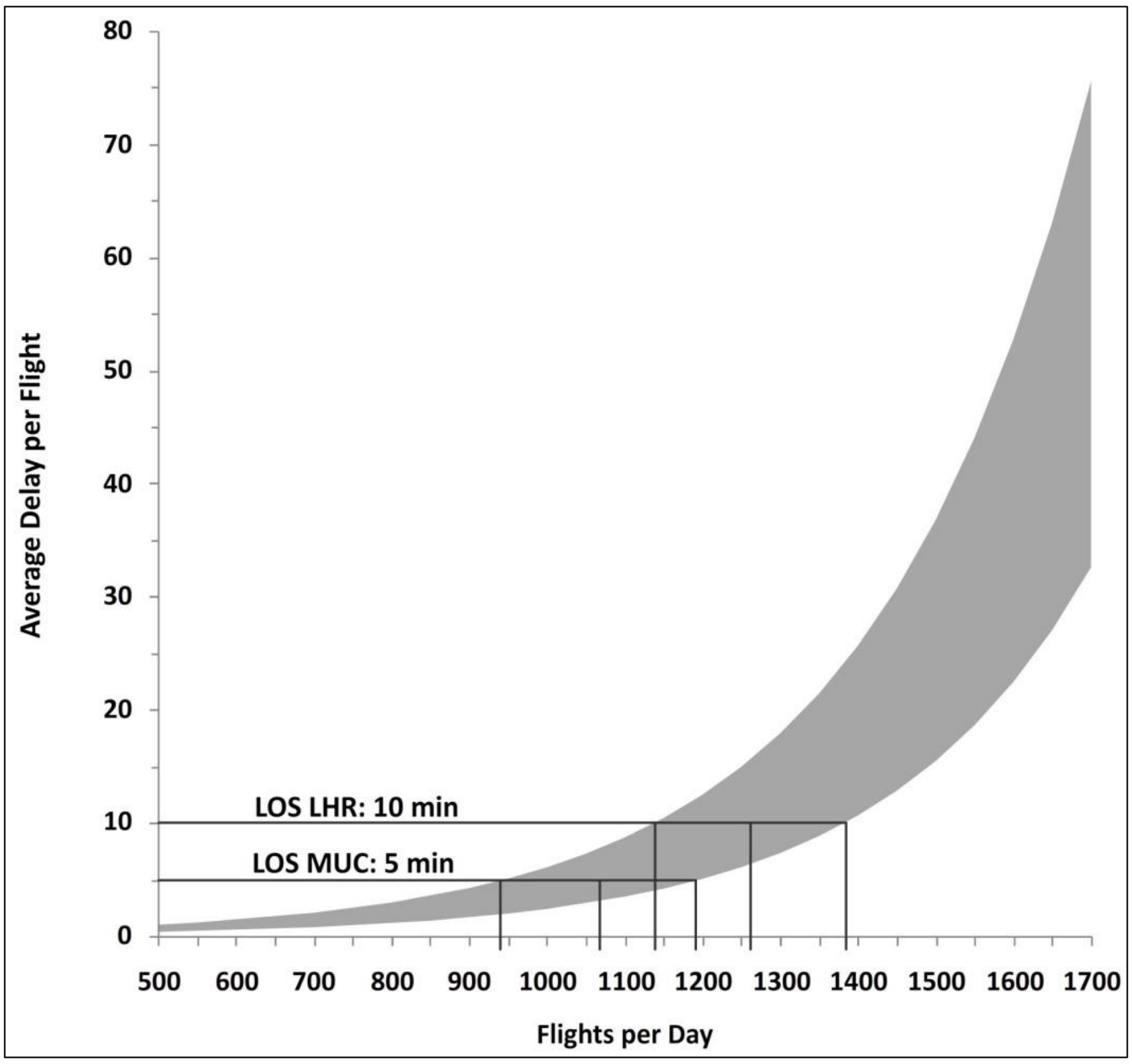

**Fig. 10.** Relationship between Daily Demand and LOS for Far Parallel Runways

### Airspace demand, capacity and LOS

As noted earlier, if runway capacity is not sufficient, arriving aircraft must queue in airspace holding stacks or lock into the holding pattern, until they are granted a landing slot by ATC. Therefore, coordination and collaborative decision making (CDM) between airport and local ATC is vital to expand airport capacity without resorting to costly measures like the outright construction of additional infrastructure (Butler 2008).

The role of ATC is mainly to direct air traffic safely to and from an airport through airspace. Here the rules are mainly determined by international standards from the International Civil Aviation Organization (ICAO) published in Doc 4444, the Procedures for Air Navigation Service – Rules of the Air and Air Traffic Services (PANS-RAC) (ICAO 2001). The main task should be to organize the fluctuating flows of arriving aircraft at the arrival fixes and to convert these flights into a continuous concentrated flow of landing aircraft towards the runway. This should be manageable in all weather conditions (snow, wind, rain, fog, etc.), but certainly under instrument flight rules (IFR) that allow instrument landings under inclement weather conditions with a minimum visibility of one nautical mile (1.85 kilometers) and a cloud ceiling of 245 meters.

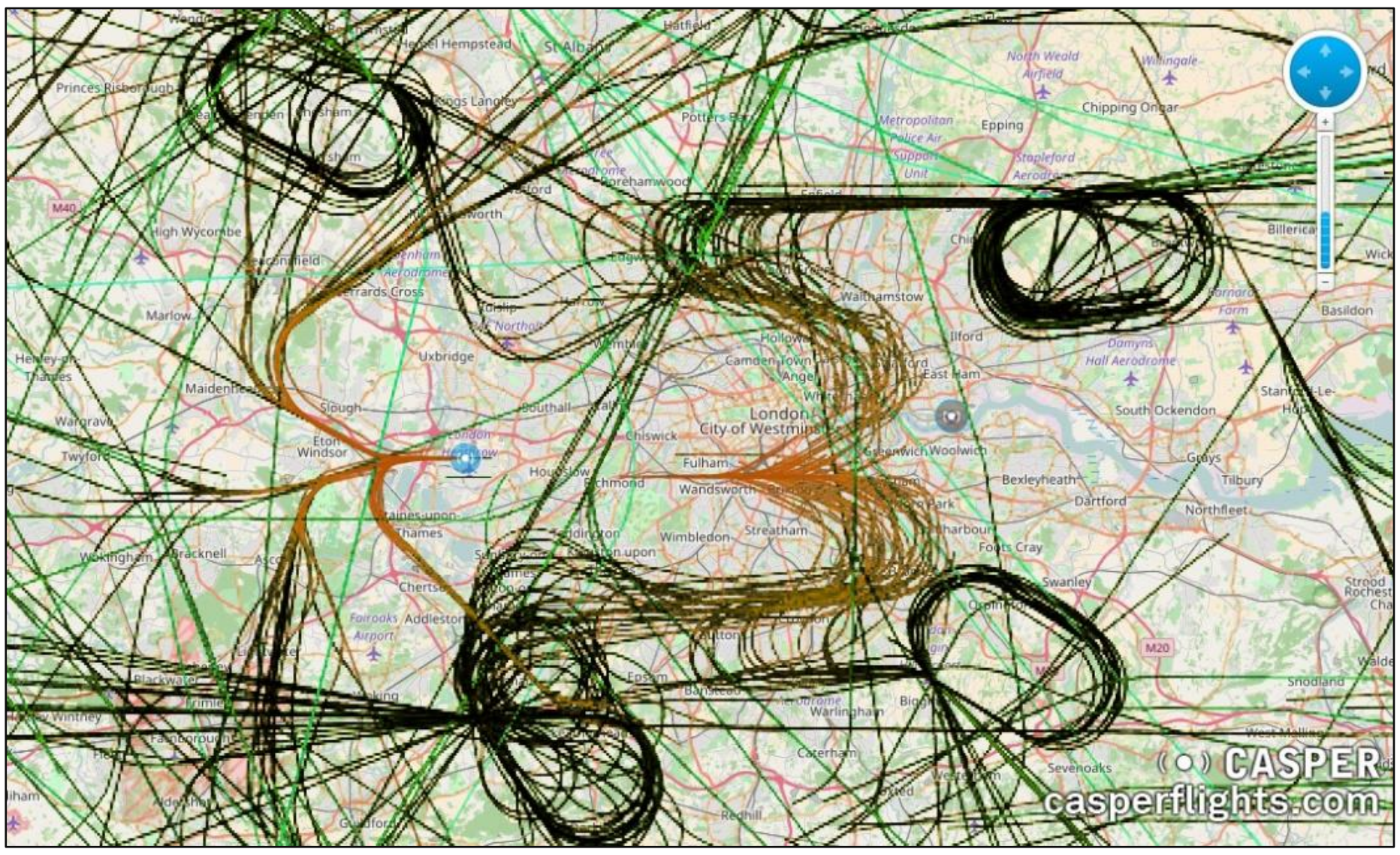

**Fig. 11.** Airspace and flight tracks around London Heathrow airport on June 27th, 2014 between 8:00 and 9:30 (Source: Casperflights.com 2016)

In the immediate airspace of London Heathrow airport, four main holding stacks are in place, where the aircraft circle (and descend) until they get clearance to land (**Fig. 11**). The flight tracks in **Fig. 11**, constructed by flight position signals from Automatic Dependent Surveillance Broadcast (ADS-B) transmitters on board modern aircraft, show clearly how the airspace around London-Heathrow is operated.

Although tracks from an afternoon winter schedule are displayed in the figure, large queueing can be observed even in off-peak periods at Heathrow airport, suggesting that this airport is operating well above a stable long-term sustainable capacity and is highly congested throughout the year.

### 1.9 Airport evolution and future challenges

It will be interesting to see in which direction air transportation evolves in the future. All imaginable scenarios are possible, but the timeframes for investments are highly uncertain (Button 2004, pp. 23). Current research projects within the Single European Sky (SES) Air Traffic Management (ATM) Research Programme (SESAR), which is initiated and coordinated by the European Commission and EUROCONTROL, try to identify performance determinants and bottlenecks in the flow of European air traffic (EUROCONTROL 2009). State-of-the-art technology is expected to be reviewed for implementation in air transportation. Surveillance and positioning of air traffic through satellites, like the Global Positioning System (GPS), or through advanced radar technology has only just begun to reveal its full potential.

Opposition and legal action by environmental groups against building of such large infrastructures as airports and particularly runways will increase, further extending the planning horizons. The discussion about extending the largest hub in Europe – London-Heathrow – has been dragging on for years. Heathrow airport has experienced congestion over the last two to three decades. That private company Heathrow belongs to the Spanish infrastructure managing company *Ferrovial* has not helped to gain support for an airport extension. Just recently further plans to expand airport capacity at Heathrow by a third runway were approved by the government (Plimmer and Ford 2018).

**Fig. 12** depicts how the development of Heathrow was anticipated just a few years ago. Between 2000 and 2005, extended operating hours were to have been implemented, then in 2007 mixed-mode operations were to be introduced and finally in 2015 a new third runway would go into operation (Janić 2007, pp. 14).

Unfortunately, not a single measure has materialized so far, leaving Heathrow just as saturated as it was before with a capacity of about 90 to 95 flights per hour. To ease some congestion during peak periods at Heathrow parallel approaches on both runways are operated under the Tactically Enhanced Arrival Measures (TEAM) (BAA 2009, p. 17).

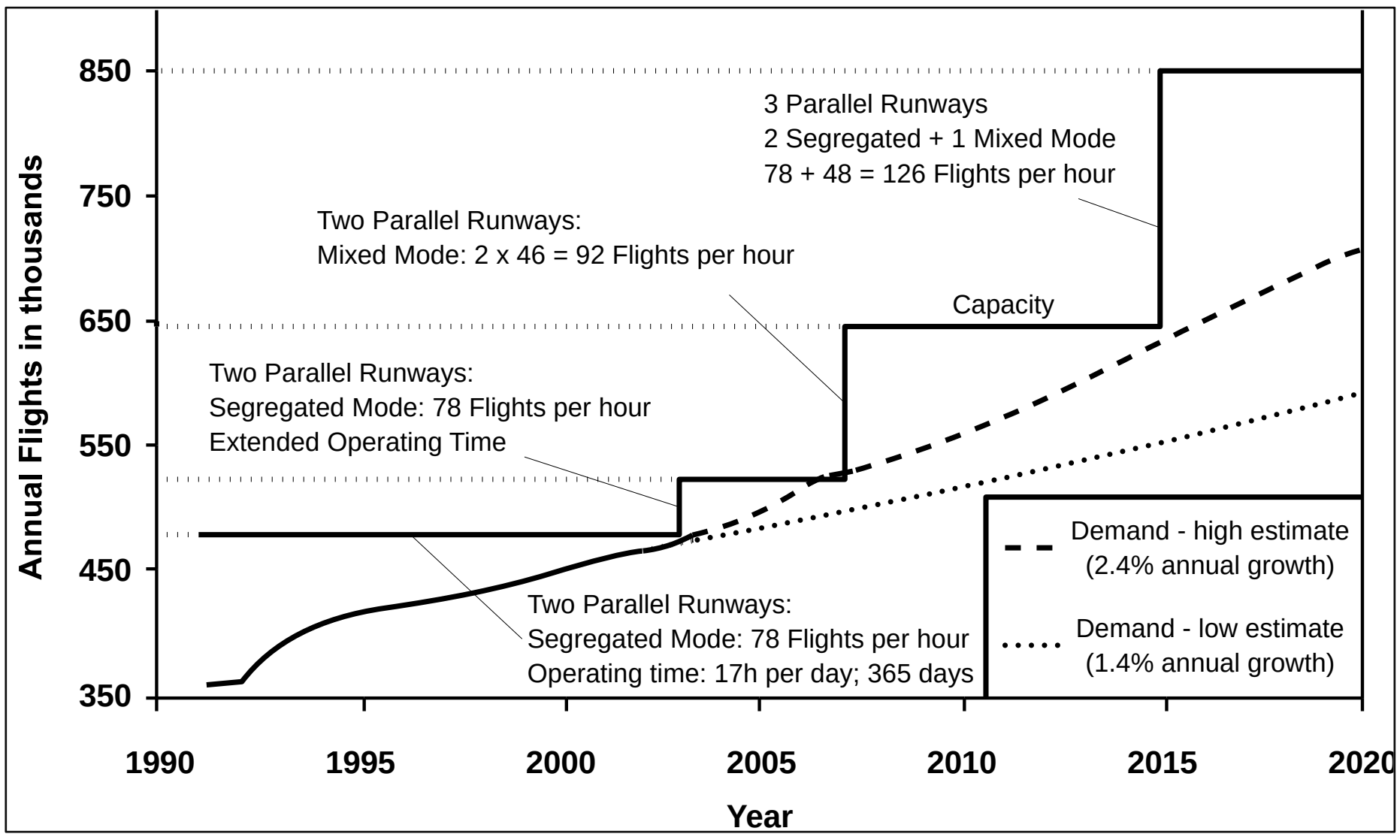

**Fig. 12.** Possible long-term scenarios for London Heathrow airport by adjusting capacity to demand (Janić 2007)

Heathrow is an extreme example of an airport lacking adequate expansion of capacity. But in Europe there are other hubs that are taking the opportunity to develop much faster, commonly many years ahead of demand. Paris' Charles-de-Gaulle airport has undergone major investments in the last two decades, pushing the capacity from 72 flights per hour on a far parallel runway system in 1990 (SRI International 1990, pp. 72) to currently 120 flights per hour on a four parallel runway system.

Madrid's Barajas airport has expanded even further since the mid-1980s. Before 1990 Madrid-Barajas had a peak hour capacity of 30 flights per hour, serving a demand of 13.2 million passengers with 139,000 flights on a crossing runway system (SRI International 1990, pp. 46) (**Fig. 13**).

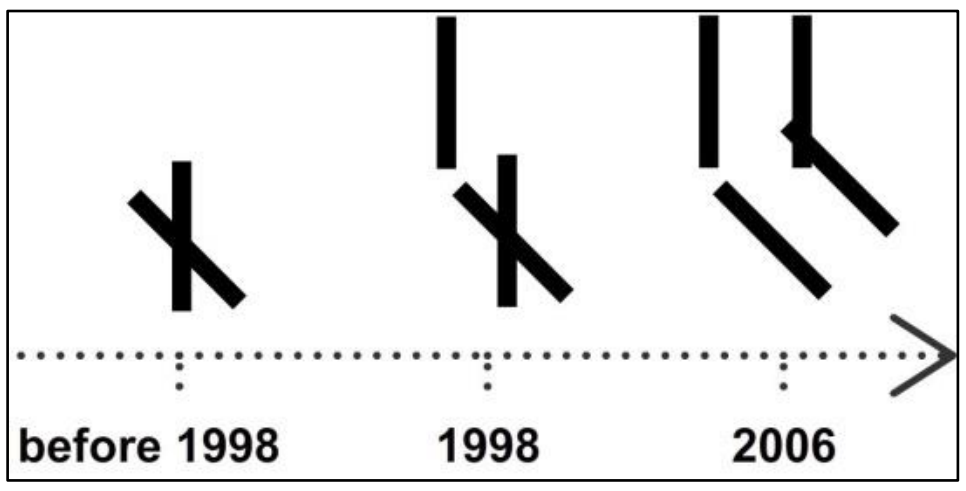

**Fig. 13.** Evolution path of Madrid Barajas airport

By 1998 when another runway was added, Barajas airport had already improved peak hour capacity to 50 flights per hour (IATA 1998). With this third runway in place from 1998, parallel operations started at Barajas, pushing peak hour capacity to 74 operations per hour, for a demand of 28 million passengers and 306,000 flights (IATA 2003). Finally, in 2006 the master plan was completed with the addition of two further runways and the closing of an older one Madrid-Barajas has currently two pairs of far spaced parallel runways, which achieve well beyond 100 movements per hour for a demand surpassing 45 million passengers and 450,000 flights.

Comparing developments in the U.S. to the development of airports in Europe, much more could be on the horizon. For example, in Europe there is only one airport which could manage triple independent parallel IFR landings - that is Amsterdam Schiphol airport (AMS). Although triple and even quadruple simultaneous IFR landings are still being under study by the FAA (McNerney and Hargrove 2007), U.S. airports especially are pushing development in that direction to be able to attain further capacity benefits. Currently the maximum hourly capacity can be observed at Atlanta's Hartsfield airport with 200 peak hour flight and over 20,000 distributed seats per hour.

**Fig. 14** shows how the evolution path of airport systems and runway configurations may look like over time. When runways are added to an existing system, the degrees of freedom for the airport management are limited on the question where a new runway can be placed. The trend goes towards placing new runways parallel to the main landing and departure runways, but this is not always the case. A classification of such planned runway configurations makes a benchmarking against existing airports with a similar configuration possible.

For example, an airport with two independent parallel runways may make the decision to add another runway. This new runway may be added to the existing system in several different ways, parallel, crossing or diverting. Then counterparts for each possible planned configuration may be found internationally as a comparison. Capacity, productivity, and performance may be benchmarked against expected parameters in the airport expansion masterplan. We think most information that is needed to compare one airport to a suitable counterpart is available in publicly accessible places, especially on the Internet. We find financial statements of airports and airlines, satellite imagery, airport diagrams and charts, schedules, aircraft position data and ticket prices readily available.

It will be easier to find international counterparts and best practice airports with, say, three or less runways. Airports with four and more runways are still rare and show increasingly complex configurations. Here in Europe for example Schiphol airport (AMS) in the Netherlands has six runways, of which three are the main runways and the other three are crosswind runways and alternative runways for smaller aircraft. In the U.S. we see airport masterplans like the complete overhaul of Chicago O'Hare airport (ORD), which results in a system with eight active runways. Such cases are rather unique and call for careful assumptions and calculations.

Sophisticated runway operations depend largely on the type of equipment in place and on the skills of the air traffic controllers. Currently radar screens are updated about every 5 seconds, which leads to delayed reaction times and imprecise aircraft localization. In the future aircraft will be equipped with Global Positioning System (GPS) signal receivers allowing them to be located by transmitted ADS-B signals in 3-dimensional space over time. Radar screen update frequency may improve to one second or less between images. Furthermore, wake turbulence may be accurately predicted, which would allow a dynamic separation between succeeding landing or departing aircraft on the same (or close parallel) runway under different wind conditions. Separations of 2.5 nautical miles or less at any time between aircraft may seem reasonable to achieve (certainly in a harmonic sequence of aircraft of the same wake turbulence categories [WTC]).

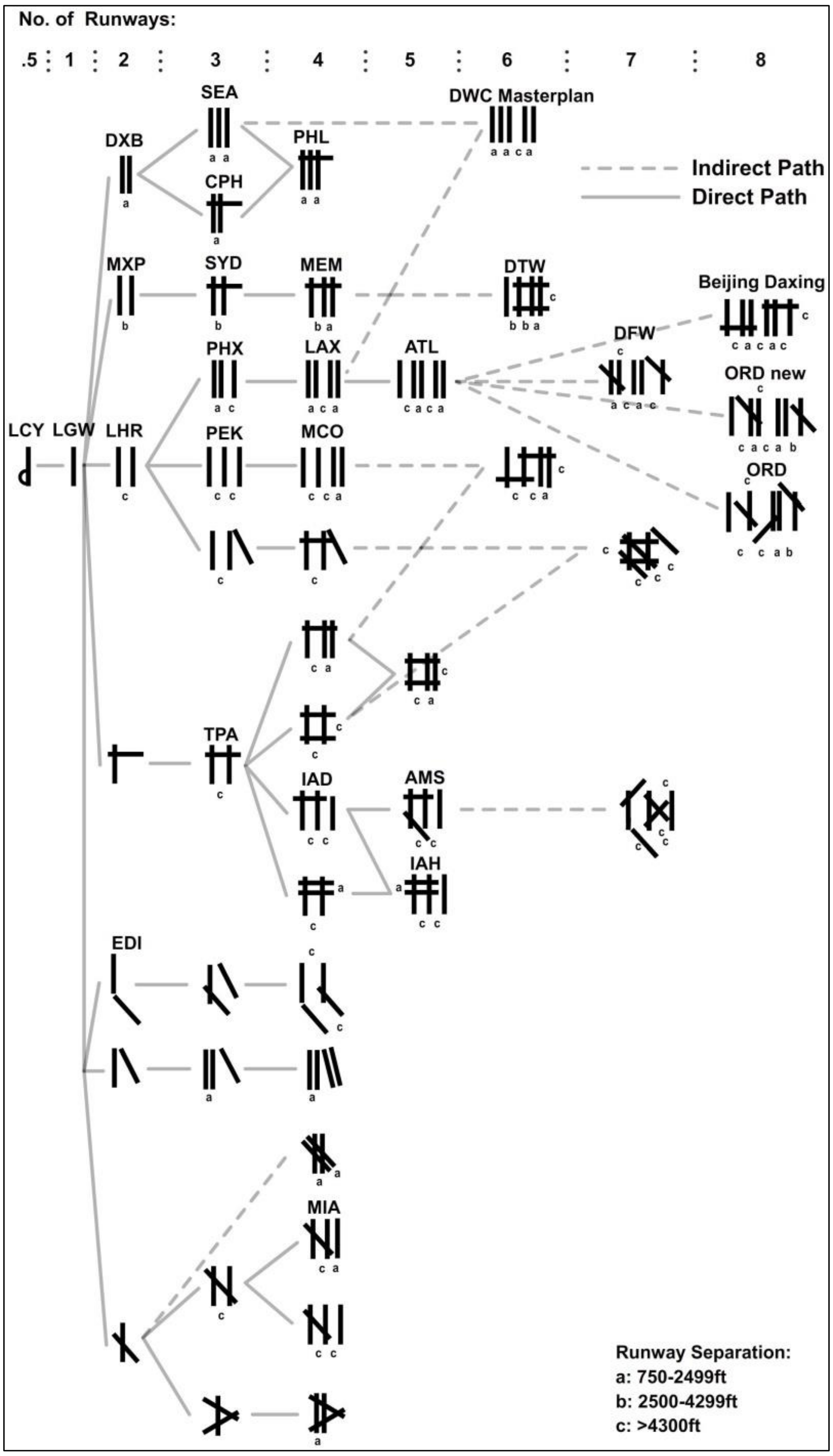

**Fig. 14.** Airport classification system and evolution paths (selected airports)

## 1.10 Conclusion

It is a long way to reach the goals for seamless travelling. Large scale initiatives like NextGen in the U.S. or SES in Europe aim to accelerate the provision of capacity ahead of the increase of demand (EUROCONTROL 2003). Operational performance and bottlenecks on the ground and in the airspace need to be determined and resolved (EUROCONTROL 2009). In the near term many international airports and especially the main hubs will experience peak congestion, leading to aircraft delays which propagate through the air transport network. In Europe, the most prominent examples of congested and near saturated airports like London Heathrow, Frankfurt am Main, and London Gatwick put a lot of effort into technical measures to raise capacity of its current airport system. Therefore, traffic levels and capacity enhancement procedures at those airports can serve as a benchmark for airports with similar configurations. Frankfurt received a new fourth runway in winter 2011. On the contrary London Heathrow can only grow through increasing load factors and seat capacities offered by airlines, because a new third runway is highly unlikely. Internationally Tokyo Haneda airport is comparable to the configuration of London Heathrow[1]. However, Haneda airport achieves the same level of passengers of almost 70 million, but with significantly fewer daily flights, but with a higher payload of up to 200 passengers per flight on average.

The trend towards increasing load factors, aircraft size and passengers per flight can only be influenced by airport management through incentives in the airport charges schemes aiming at discriminating certain aircraft types. Ideally a multi-airport system could be established around metropolitan areas, which would allow the provision of services for specific airlines and aircraft types and would create a win-win situation for stronger and weaker players regarding market split and traffic relief. Nevertheless, airports continue to grow commercially mainly though income from retail concessions and other non-aeronautical revenue, such as parking fees. This development is driven by the average spending per passenger at the airport.

Moreover, this article presents critical requirements in planning future capacity and demand levels. In the terminals the concept of dwell time and the underlying trends when converting annual figures to peak hourly volumes should be integrated for consistent airport capacity expansion planning. Order-of-magnitude assumptions on peak volumes can be conducted with the presented guideline material.

Further research should head towards finding the best practice airport, regarding sustainable capacity and LOS, through airport benchmarking of similar airport configurations and/or airport simulations. Here the emphasis should be put on simultaneous IFR operations on independent parallel runways in the short-term and on close-spaced parallel runways in the long-term. In the U.S. studies are under way researching the feasibility of quadruple parallel IFR operations (McNerney and

---

[1] At the time of writing Haneda had three runways of which two independent parallel runways were mainly operated and the third was used as an alternative runway under strong side winds or due to noise abatement procedures during the night; a fourth runway was added in 2012 to support the alternative configuration and to spread noise more on the sea instead of the city.

Hargrove 2006). However, such procedures require large scale implementation of sophisticated aircraft and ATC technology.

# 2 Airport punctuality, congestion, and delay: The scope for benchmarking

Branko Bubalo

**Abstract.** The benchmarking of airport performance increasingly requires level-of-service (LOS) indicators for a fair comparison among members of the same peer group. For a true performance analysis such inclusion of quality measures for the provided service is necessary to differentiate airports with similar pure output quantities, for example number of passengers. Variation of scheduled versus actual departure and arrival times could substantially cause accumulating operating costs for carriers and could furthermore pose the risk and inconvenience of missed connections for the passengers. This article examines determinants of flight delays at airports. Thereby we present performance indicators such as slot capacity utilization, queueing time and punctuality. The essence of underlying phenomena in queueing theory such as *Little's Law,* arrival and departure distributions, and cumulative throughput and demand diagrams are briefly explained. This work's aim is the exploration of ways of measuring and observing performance quality from actual flight schedules with a focus on usability for subsequent airport benchmarking and traffic modeling.

## 2.1 Introduction

It is largely recognized that transportation - and this of course includes air transportation - is vital for the economic development of countries and markets globally. Developing societies not only require telecommunication to connect with the rest of the world, but they additionally develop a need for free mobility of people and goods, which serves as a main driver for trade and economic prosperity. In transportation ultimately what drives the choice of one mode over the other is finding the fastest, cheapest, safest, most reliable and (nowadays) environmentally friendliest way to move something from point *A* to point *B.*

In contrast to most information networks, which can be expanded quite rapidly, and which constantly deliver significant advances in their stages of technological evolution, transportation networks cannot be expanded so well ahead of demand. Procedural changes are constantly being anticipated by stakeholders in a variety of network industries to maximize throughput and efficiency. Because of the magnitude in dimension and long-term impact, infrastructure such as roads, rail tracks, bridges, ports and certainly airports conflicts significantly with our personal sphere, neighborhoods, and natural environments.

Particularly airports in Europe are faced with stagnating investments in fundamental infrastructure, as environmental concerns are becoming more important for the public and for the political agenda. In our industrialized economies, politicians as well as our managers refrain from taking unpopular decisions such as building new airports or runways, because many people could be impacted, and a significant amount of (tax) money would be required to provide direct or support infrastructure.

It can be assumed that potential demand at, say, Frankfurt or London Heathrow airport is substantial and cannot be met during peak times. During the last two decades both airports have tweaked out the maximum flights given the currently installed capacity. In the case of London-Heathrow a third runway to expand airside capacity significantly is planned. However, it is currently doubtful if it will be built. The consequence is an everyday exercise in traffic flow optimization by continuously minimizing server (i.e. runway, apron or terminal) occupancy times and maximizing punctuality (Eurocontrol, 2005). An aircraft must load, unload and turnaround at an ever-increasing faster pace.

In the first part of this article, the declared capacity/available slots of an airport will be related to actual levels of demand to derive the capacity utilization as a *measure of congestion*. This will be exercised on a set of single runway airports.

In the second part, the focus is on developing performance measures related to level-of-service (LOS), such as schedule adherence (i.e. punctuality, variability, and delay), which should be considered in any airport performance or benchmarking study.

Furthermore this article discusses an airport's most critical limit, the runway capacity, which is defined as the *maximum service rate* per time unit ($\mu$) and its inverse, the *minimum service time* ($1/\mu$) (this minimum time interval between following arriving or departing aircraft is also called the *headway*) (Gosling et al., 1981).

The airports London-Heathrow (LHR) and Tokyo-Haneda (HND) airports have been chosen for benchmarking, as they are comparable in terms of amount of annual passengers and mainly operating far-spaced parallel runways (a fourth take-off runway has been added recently at Haneda airport, which most likely will change procedures and capacity considerably). These airports also represent perfect counterparts in terms of punctuality.[1]

Based on airport timetables, consisting of information on actual and scheduled gate times and the basics of queueing theory, recommendations for the critical relationships and determinants for benchmarking and performance are derived.

### 2.2 Declared airport capacity, demand and utilization

Although there are still many airports scattered across Europe which could be transformed into civil international airports in the future, this circumstance is of little benefit to the big hub airports which are ultimately needed to accommodate large-scale passenger flows to and from international connections and regional airports. During day-to-day operation airport capacity is either sufficiently available to accommodate any current demand level and "organic", i.e. anticipated, increase in demand, or it may be limited and must therefore be coordinated by an airport slot coordinator. This latter is the case for all "congested" Level 2 and 3 coordinated airports in Europe, which according to the International Air Transport Association (IATA) world scheduling

[1] https://www.travelweekly.com/Travel-News/Airline-News/OAG-Tokyo-Haneda-the-worlds-most-punctual-large-airport [accessed on September 13th, 2019]

guidelines (WSG) must declare their capacity to the slot coordinator. Each of the airport stakeholders (airport operator, coordinator, airlines, and air traffic control) must work towards maximizing the capacity of the processes under his jurisdiction, thereby bringing more available capacity to the table as a result of streamlined processes. During the bi-annual scheduling conference initiated by the IATA, the declared capacity is used as a reference for the scheduling process for the seasonal airline schedules. Each scheduled flight at the constrained airports is assigned a landing and take-off right - a *slot*.

At the currently most capacity constrained airports in Europe, Frankfurt am Main (FRA), London-Heathrow and London-Gatwick (LGW), there is hardly any idle slots available for growth and/or unscheduled flights (such as general aviation, charter, military or government flights). Benchmarking has shown that there are many examples of European airports where we can find capping of capacity at much lower levels than what would be operationally feasible under instrument flight rules (IFR) (see Bubalo, 2009) (**Table 5**). In general, there is little, if any, available evidence why capacity is declared at exactly the chosen levels.

For example, we could benchmark the airports London-Gatwick, London-Stansted (STN) and Stuttgart (STR) simply based on the *single runway* airport configuration. When looking at the numbers in **Table 5**, we realize that the declared hourly and daily capacity, i.e. available slots, in the main operating hours between 06:00 and 23:00 differs among this peer group, with 797, 733 and 714 daily slots respectively, and 50, 50 and 42 peak hourly slots respectively. However, one could argue that the maximum capacity should be equal, since these European airports must operate arrivals and departures on the runway under *mixed mode*. Based on the best-practice, as at the single-runway airport London Gatwick in the example with a capacity of 797 daily and 50 maximum hourly slots, the implication would be that these hourly and daily slots are potentially achievable, given the same technology and controller experience, for all airports within the same peer group.

So why is runway capacity not declared at the same levels for all "mature" single runway airports, such as the ones described? If demand is huge and growing, it should be in the public interest to expand the facilities. It will be explained below that setting capacity at certain levels is only reasonable in connection with LOS standards, i.e. measures of service quality and congestion, which differ among airports.

The relation between actual demand and slot capacity is the capacity utilization ($\lambda/\mu$), which is a strong first indicator of congestion (particularly for utilization rates above 75%). For the airports Gatwick, Stansted and Stuttgart this means a peak daily demand of 678, 408 and 370 flights in 2009 respectively and consequently a peak daily slot utilization of 85%, 56% and 52% respectively, and a peak hour demand of 49, 38 and 35 flights and peak hourly slot utilization of 98%, 76% and 83%.

**Table 5.** Peak Daily and Hourly Slot Capacity and Utilization for Selected Single Runway Airports (Data from Eurostat, Slot Coordination and Flightstats.com)

| Airport | IATA | Passengers in millions | Flights in thousands | Load Factor | Passengers per flight | Daily Capacity | Daily Capacity Utilization | Slots per hour | Peak Hourly Capacity Utilization | Runway Service Time (Inverse of hourly slots) in seconds |
|---|---|---|---|---|---|---|---|---|---|---|
| | | 2010 | 2010 | 2010 | 2010 | 2009 | 2009 | 2009 | 2009 | 2009 |
| London-Gatwick | LGW | 31.4 | 233.5 | 78.7% | 134 | 797 | 85% | 50 | 98% | 72 |
| London-Stansted | STN | 18.6 | 143.0 | 76.9% | 130 | 733 | 56% | 50 | 76% | 72 |
| Stuttgart | STR | 9.3 | 111.7 | 71.0% | 83 | 714 | 52% | 42 | 83% | 86 |
| Birmingham | BHX | 8.6 | 84.8 | 74.1% | 101 | 680 | 45% | 40 | 70% | 90 |
| Berlin-Schoenefeld | SXF | 7.3 | 65.5 | 75.3% | 111 | 448 | 41% | 26 | 65% | 138 |
| London-City | LCY | 2.8 | 60.0 | 63.6% | 47 | 646 | 37% | 38 | 95% | 95 |
| Average | | 13.0 | 116.4 | 73.3% | 101 | 670 | 53% | 41 | 81% | 92 |
| Std. Dev. | | 10.4 | 65.1 | 5.4% | 33 | 120 | 17% | 9 | 13% | 25 |

These numbers decreased to rather low levels in 2009 due to the global financial crisis, but demand stabilized slightly in 2010. However, demand levels at these airports in April 2011 still show evident signs of recession, with current levels of 647, 372 and 301 daily flights (81%, 51% and 42% capacity utilization), respectively, and 45, 35 and 27 peak hourly flights (87%, 70% and 64% capacity utilization), respectively.[1]

## 2.3 Airports as systems of queueing systems

For a better understanding of airport capacity and congestion problems, queueing theory gives great insights. As de Neufville and Odoni (2003: pp. 819-863) point out, all airport facilities can be described a system of queueing systems, where arrivals at a service facility are randomly distributed, waiting lines form and users are therefore delayed and must wait before being served.

### Free flow for free mobility

In the simplest case, vehicles, carrying "information" (such as passengers), flow in a *first-come-first-served* (FCFS) sequence through a hub and spoke (node and link) network from origin to destination. If an aircraft traversing a link reduces speed to maintain a minimum separation to a preceding object, then this decrease in velocity could feed back to the last object in a queue, when there are insufficient buffers to compensate for speed differences induced by leading flights. These *reactionary delays* propagate to all following flights in sequence without enough buffer times and therefore could result in additional costs to aircraft arriving hours later. An example of this process might be cars on a single lane road approaching a slow-moving tractor in front. This will, in a short time, create a long queue behind the tractor; but in road traffic in general there are admittedly opportunities to overtake a slower vehicle that do not exist at or in the vicinity of airports.

To avoid queues altogether, each individual flight needs to be able to move freely and seamlessly, *as if it were the only aircraft operating in the system*. In general airport capacity must be seen as a continuous flow of aircraft passing through the airport and airspace system, where each flight requests service at the airport and is in most cases served immediately or added to the end of a waiting queue in a FCFS discipline (de Neufville and Odoni, 2003: pp. 367-407). ATC manages the landing and take-off sequence and handles communication with pilots. The critical common approach path is a defined space leading to the landing-end of the runway, which is shared by all aircraft approaching a runway during IFR conditions (**Fig. 15**).

Collectively **Fig. 15**, **Fig. 16** and **Fig. 17** show the different stages of a flight approaching an airport. **Fig. 15** illustrates the flight approaching an entry gate of the terminal airspace area, where flights arriving from different directions are concentrated and separated for the final approach on the common approach path (Swart, 2003). Because of the bundling of flights, these are significant potential points of congestion and aircraft may be held at the entry in holding airspace. **Fig. 16** is the typical graphic representation of the time and space separations applied by ATC and the varying aircraft speeds during the final approach beyond the entry gate in a so-called time-space diagram (Trani, 2005). The prime concern at airports is the maintaining of safety separations, because of the risk of an encounter with *wake turbulences* caused by the wings of a leading airplane. The airport management in collaboration with ATC seeks to *minimize* these separations to increase airport capacity to just about the legal safety minima (**Table 6**). The last stage of a flight is the touchdown, where various deceleration maneuvers are conducted, before the aircraft exits the runway onto the taxiway system (**Fig. 17**). Here the airport management could influence the runway occupancy time (ROT) by building adequate runway exits for the prevalent aircraft categories.

[1] https://www.flightglobal.com/news/articles/recession-obscures-european-airport-capacity-crunch-354731/ [accessed on September 13th, 2019]

Since the capacity is ultimately limited by the safety requirement of not allowing more than one aircraft on the runway at any one time, the capacity of a runway under mixed mode operation is in practice therefore mainly determined by the mean acceleration and deceleration speeds of departing and arriving aircraft and hence the headways between the aircraft (**Table 6**). London-Gatwick airport with a runway service time of 72 seconds per flight or a service rate, i.e. capacity, of 50 flights per hour is a prime example of an airport which utilizes its single runway to the maximum and allows very little idle runway time between subsequent flights (see **Table 5**).

As we can see in **Table 6** not only the mandatory separation minima have an influence on overall capacity, but also the different speeds and the mix of different aircraft categories based on maximum take-off weight (MTOW). Today, the lowest separation between aircraft is 2.5 nautical miles (NM) under IFR on final approach (about 10 NM distance from the runway) given the required controller training and experience, runway configuration and radar equipment. In this case headways as low as 55 to 65 seconds and consequently capacities as high as 55 to 65 flights per hour could be achieved for a single runway.

Combinations of Heavy preceding Light aircraft in the sequence of arrivals or departures imply the most significant loss of precious server time and hence capacity, therefore a homogeneous mix of aircraft, which require only 3 NM on average, is most efficient in terms of capacity utilization (**Table 6**).

**Table 6.** Variation in headway and capacity due to different aircraft weights, separation minima and speeds

| Aircraft Category (1) | | | Radar Separation (1) | | Radar Separation Headway at Take-off Speed | | Inverse Radar Separation (Capacity) | |
|---|---|---|---|---|---|---|---|---|
| Preceding aircraft | Succeeding aircraft | MTOW | Nautical Miles (NM) | Kilometers (km) | 250 km/h | 300 km/h | 250 km/h | 300 km/h |
| Heavy | Heavy | >136 tons | 4 NM | 7.4 km | 106 sec | 89 sec | 34 Ops/hr | 40 Ops/hr |
| | Medium | 7-136 tons | 5 NM | 9.3 km | 134 sec | 112 sec | 27 Ops/hr | 32 Ops/hr |
| | Light | < 7 tons | 6 NM | 11.1 km | 160 sec | 133 sec | 23 Ops/hr | 27 Ops/hr |
| Medium | Light | < 7 tons | 5 NM | 9.3 km | 134 sec | 112 sec | 27 Ops/hr | 32 Ops/hr |
| Other combinations and same categories | | | 3 NM | 5.6 km | 81 sec | 67 sec | 44 Ops/hr | 54 Ops/hr |
| Minimum on final approach | | | 2.5 NM | 4.6 km | 66 sec | 55 sec | 55 Ops/hr | 65 Ops/hr |

Notes:

(1) ICAO Doc 4444-RAC/501 (PANS-RAC) Part V Section 16 & Part VI Section 7.4

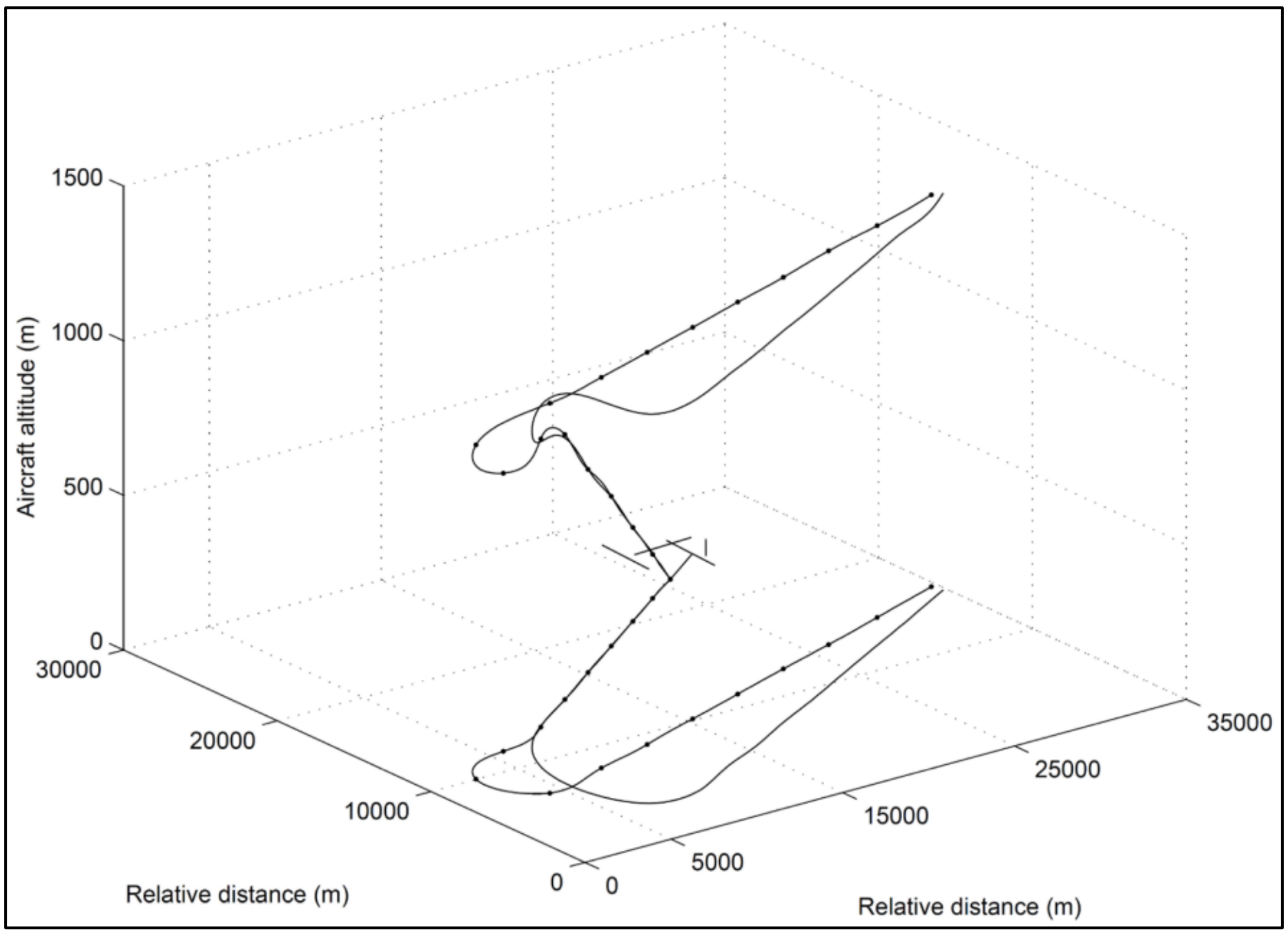

**Fig. 15.** Two subsequent flights sharing the same common approach path at Amsterdam Schiphol in 3-D space and as 2-D ground projection (Swart, 2003).

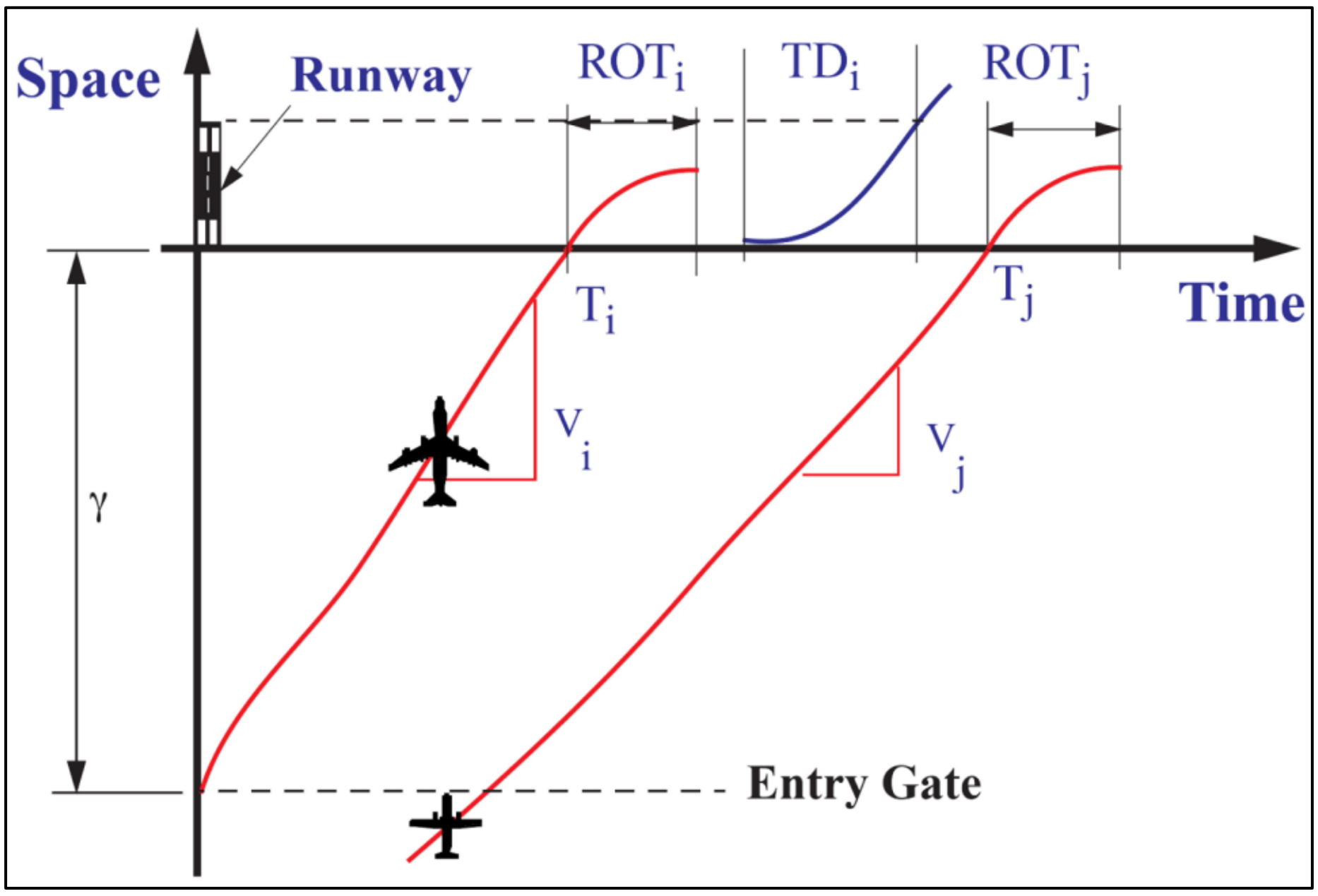

**Fig. 16.** Occupancy of time-space in runway occupancy time for landings and in take-off distance for departures on a mixed mode single runway (Trani, 2019).

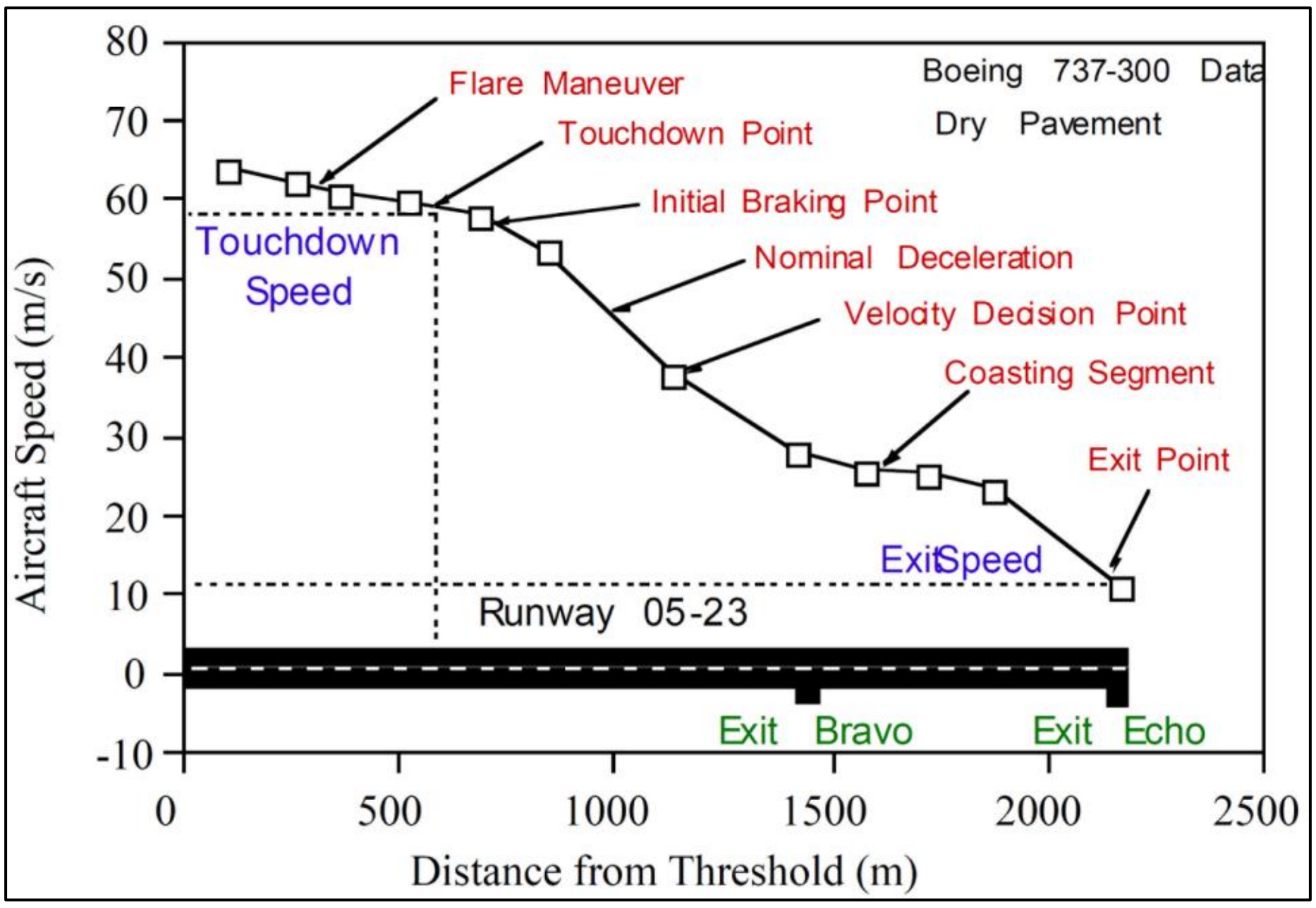

**Fig. 17.** Deceleration speeds during phases of a landing on a runway (Trani et al., 1995)[1]

### The significance of delays

Delays occurring during airport operations frequently become the center of attention (Vaze, 2009). In any "costs" of time, such as lost productivity from business travelers, these escaped opportunities increase non-linearly and accumulate quickly among the passengers concerned. Delays foremost induce considerable operating costs for airlines, such as fuel and crew costs. Since random variation of traffic results in delays, these can accumulate even under normal conditions. Any overload conditions over prolonged periods during the day will result in significant delays. A rule-of-thumb is, therefore, that on average a server/queueing system, e.g. a runway, should never be utilized more than 75% of its capacity. Providing buffer- and idle server time between flights remains necessary for queues to dissolve after periods of dense traffic.

The buildup of aircraft delay for landing aircraft at airports was first studied theoretically by Bowen and Pearcey (1948). However, delays not only occur at runways and their respective holding areas, they can occur at any bottleneck (i.e. point of congestion) in the process chain of an airport as well. This may be the runway, the apron, the terminal facilities, or any other "server", serving passenger, freight, or aircraft.

[1] I am indepted to Prof. Trani for his kind permission to use his figures in my dissertation.

## Punctuality and traffic variability

Since air transportation is primarily a scheduled service, airlines significantly depend on punctuality of arrivals and do not appreciate a lot of variation in their operations, mainly to ensure the “turn-around” (unloading, refueling, loading, etc.) is accomplished on schedule to clear the aircraft for subsequent departure (EUROCONTROL, 2005). Moreover, from a commercial perspective airport management is interested in the timely freeing of space for subsequent arrivals with new passengers.

Variability of traffic is driven by probability distributions, whether from human, technical or natural variation. Technical variation is understood as disturbances resulting from different aircraft types, regarding weight, flown distance and speed. Although weather and human factors are said to be unpredictable, even here certain regular patterns are widely recognized or currently under study, e.g. the study of seasonal effects.

Whereas in **Fig. 18** we have Tokyo Haneda (HND) as a prime example of a punctual airport which exhibits very little variation in schedule, mainly due to serving short-haul domestic routes and a homogeneous mix of mostly Heavy type (see **Table 6**) aircraft, we see to the contrary London Heathrow (LHR) (**Fig. 19**) an airport with a poor punctuality record, serving long-distance markets and a broader spectrum of aircraft types. At Haneda, 63% of the flights are on-time to the minute, whereas at Heathrow only about 5% of the flights were on-time during the days sampled. Other days of observation have been chosen to underline the differences in each case.

For airline scheduling, airport flow management but also for benchmarking purposes punctuality is expressed in *percentage of flights delayed less than 15 minutes*. Again, both airports presented certainly show extreme differences, with 98% of the flights being “on-time” at Haneda airport, i.e. delayed less than 15 minutes, and consequently only 2% of the flights severely delayed to only about 70% of the daily flights being on-time at Heathrow and up to 30% of flights severely delayed (**Fig. 18** and **Fig. 19**). On average (50% of the flights) we observe no delays at Haneda and about 6 to 7 minutes per flight at Heathrow. Here the early arrivals are balanced against the delayed flights, which otherwise would result in much higher delays of around 12 minutes per flight at Heathrow. In practice, various literature suggests to never exceed an average LOS of 4 to 5 minutes.

In these examples the deviation from the scheduled times ranges from about 20 minutes before scheduled time to 20 minutes after scheduled time at Haneda airport, and from about 50 minutes before scheduled time to 70 minutes after scheduled time (thereby omitting some extreme outliers) at London-Heathrow. Recent studies take v*ariance* from the average as a measure of traffic variability from expected travel time.

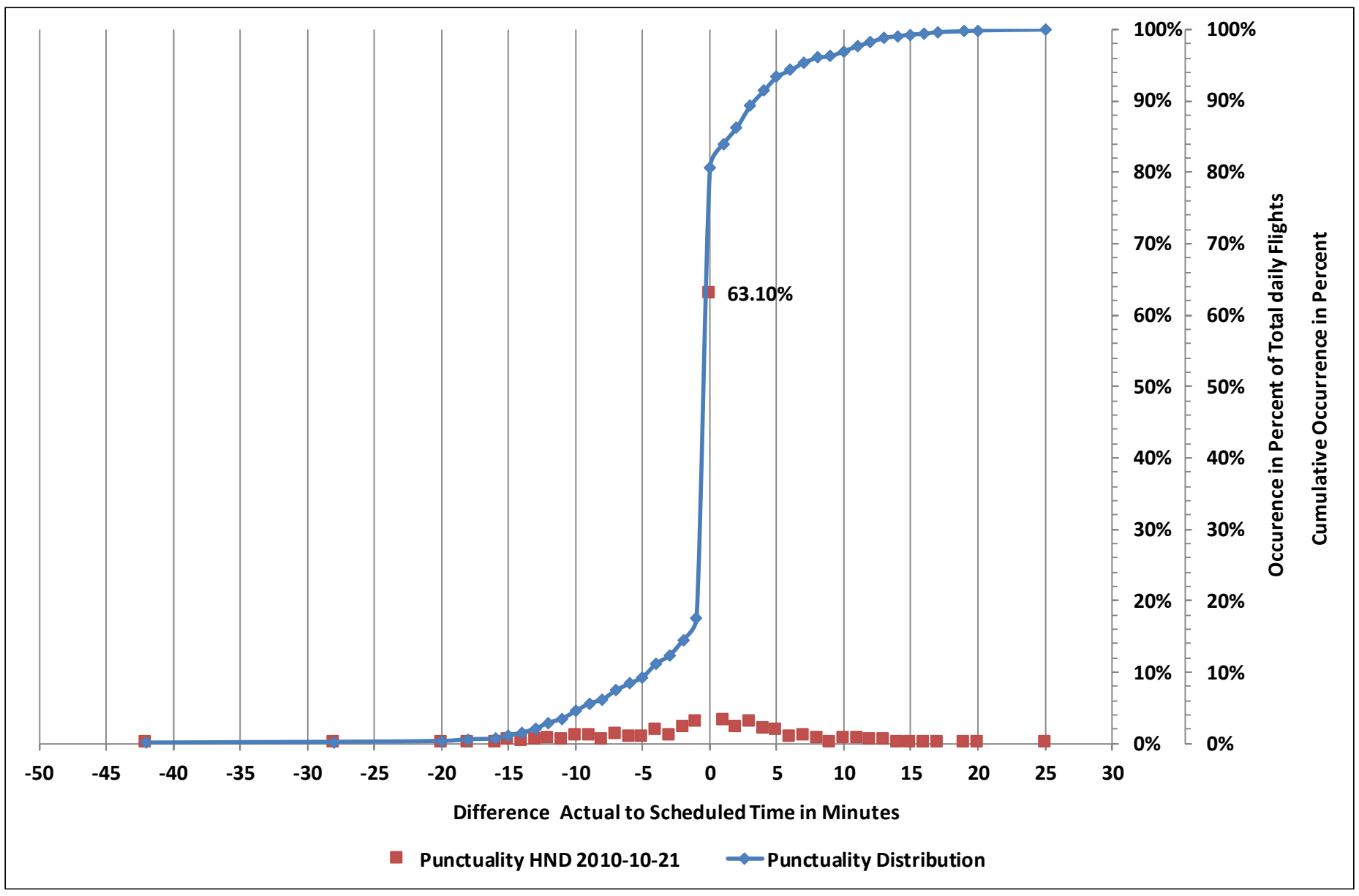

**Fig. 18.** Punctuality at Tokyo Haneda airport in October 2010

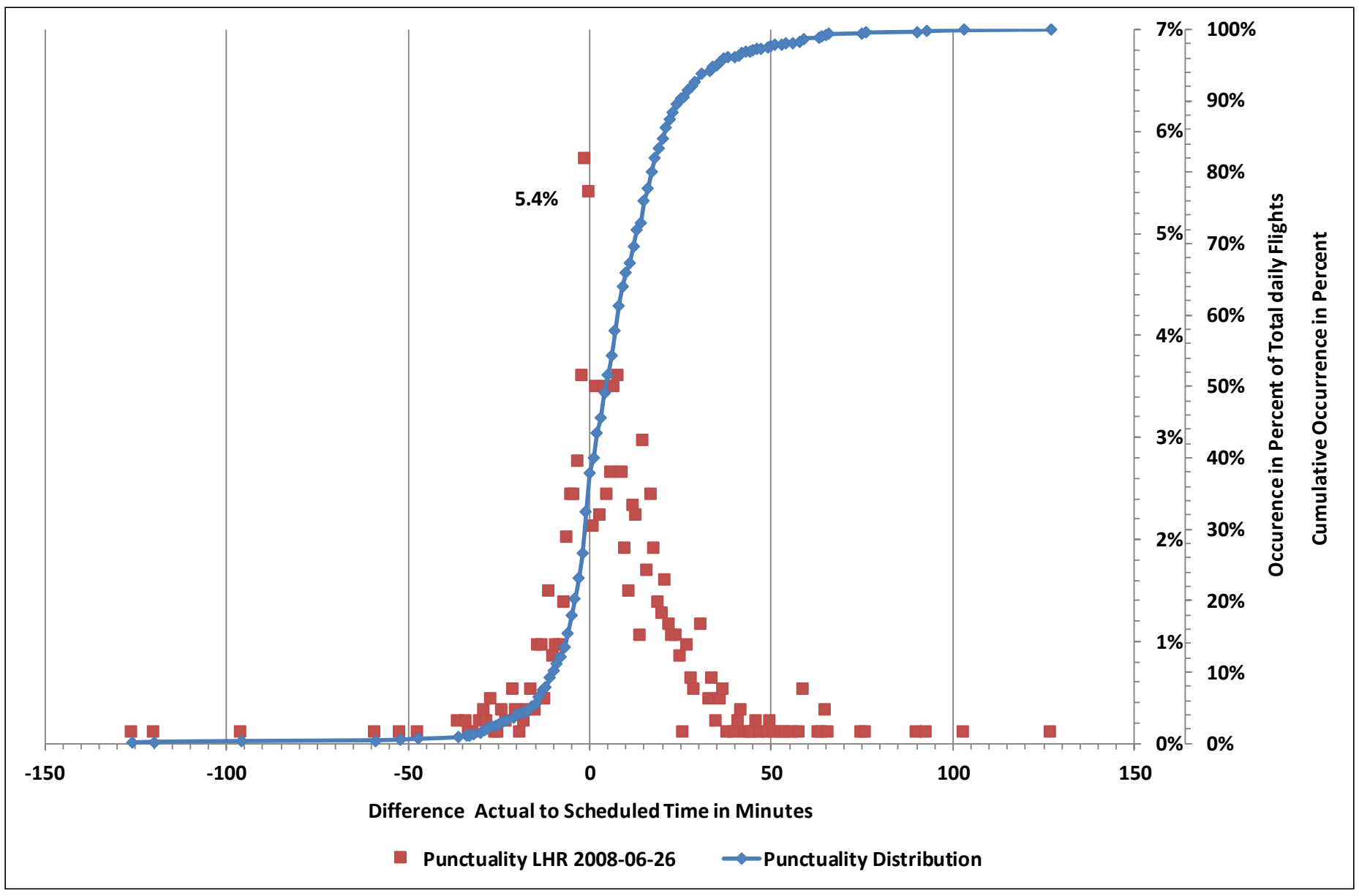

**Fig. 19.** Punctuality at London Heathrow airport on a Peak Day in June 2008

### Little's Law and cumulative diagrams

An outstanding contribution by J.D.C. Little provides the proof and broad applicability for approximating the average waiting time in a queue from the number of people in a queue and people arriving at the queue in a particular time by "Little's Law" (Little and Graves, 2008):

$$\textit{(Mean) Waiting time (W) (in minutes)} = \frac{\textit{(Mean) Number of Objects in the Queue (L)}}{\textit{(Mean) Number of Objects Arriving } (\lambda) \textit{ (per minute)}}.$$

Little's Law is particularly important for calculating an unknown in a queueing system, when the other two variables are known. In practice it could be easier (or cheaper) to control or to observe some variables than others, for example regarding a management decision to invest in surveillance equipment to monitor the service quality by observing number of people or objects in a queue. One astonishing fact that derives from Little's Law is that ultimately an airport system can be broken down into smaller units of interconnected queuing systems, consisting of waiting lines and persons in service, and resulting in an average waiting time for the passenger (Little and Graves 2008), where the output (passengers or aircraft) of one queue serves as an input for the next queue. Especially in the terminal facilities, passengers have to stay in the system for further processing, so no "information" is lost and passenger flows can be aggregated to be served from a few servers, i.e. runways, or disaggregated to be served by many servers, i.e. security lanes and check-in counters.

Little's formula is explained by a vivid example: For reasons of business travelling one needs to make a flight on short notice and has already arrived at the airport late. As one passes the check-in area, you become annoyed to find a queue some 75 meters long waiting in front of the security check. As the flight is scheduled in 30 minutes, it is vital to know how long you will have to wait to get through security. (For simplicity we will assume that everybody is getting inside, that people have already been entering the security check for some time and that the length of the queue remains stable for the time you are waiting).

One starts with some simple observations and order-of-magnitude assumptions: Observing the queue for, say, 4 minutes, one discovers that 10 people arrive in the queue per minute on average and that there is 1 meter between rows of people and one row consisting of two and a half people on average (Single individuals arrive as well as small groups of people). One quickly calculates that there are 188 people in the queue (2.5 people x 75 meters / 1 meter). While doing this you have moved 15 meters forward. Now there are some 60 meters of queue in front of you with about 150 people. Quickly applying the knowledge of Little's Law, one learns that it takes further 15 minutes waiting time (150 people in the queue divided by 10 people arriving per minute) before entering the security check.

Passing through security check would take a few, say 5, additional minutes. From experience one knows that the walk to the gate will not take longer than another 5 minutes. So now you can relax knowing you will make it to the gate in 25 minutes and hence to your flight on time.

What does this have to do with airport capacity you might ask? Well, this intuitive example can as well be applied to many operational problems, dealing with flows of passengers, cargo and/or aircraft at the airport. If a LOS is defined by airport management (e.g. a *maximum average delay of around 4 minutes per flight* or maximum length of a queue), it is possible to control and manage the airport queuing systems, and to translate this value to a certain maximum sustainable level. When this level is surpassed, an additional server would need to be opened to maintain the LOS or service time would need to decrease. Such control and management systems would collect the necessary information from sensors and collectors in the airport process chain such as light barriers, cameras or wireless signal detectors for subsequent identification of objects and current state of condition of a queue or system.

The distribution of arrivals in a queuing system has been observed by Erlang (1909) for telephone connections to a call center. Similarly, for airports as queueing systems, **Fig. 20** and **Fig. 21** show the distribution of intervals between actual succeeding arrivals and departures at London-Heathrow and Tokyo-Haneda airports, with scheduled flights shown separately. It is obvious that airlines prefer to schedule flights in bunches within the same minute at airports, or at 5- or 10-minute intervals. The distribution of actual flights does not fit the scheduled flights distribution, which means the inter-arrival times exhibit quite significant variations as compared to the schedule. At London-Heathrow (**Fig. 20**) this is far more obvious than at Tokyo-Haneda airport (**Fig. 21**).

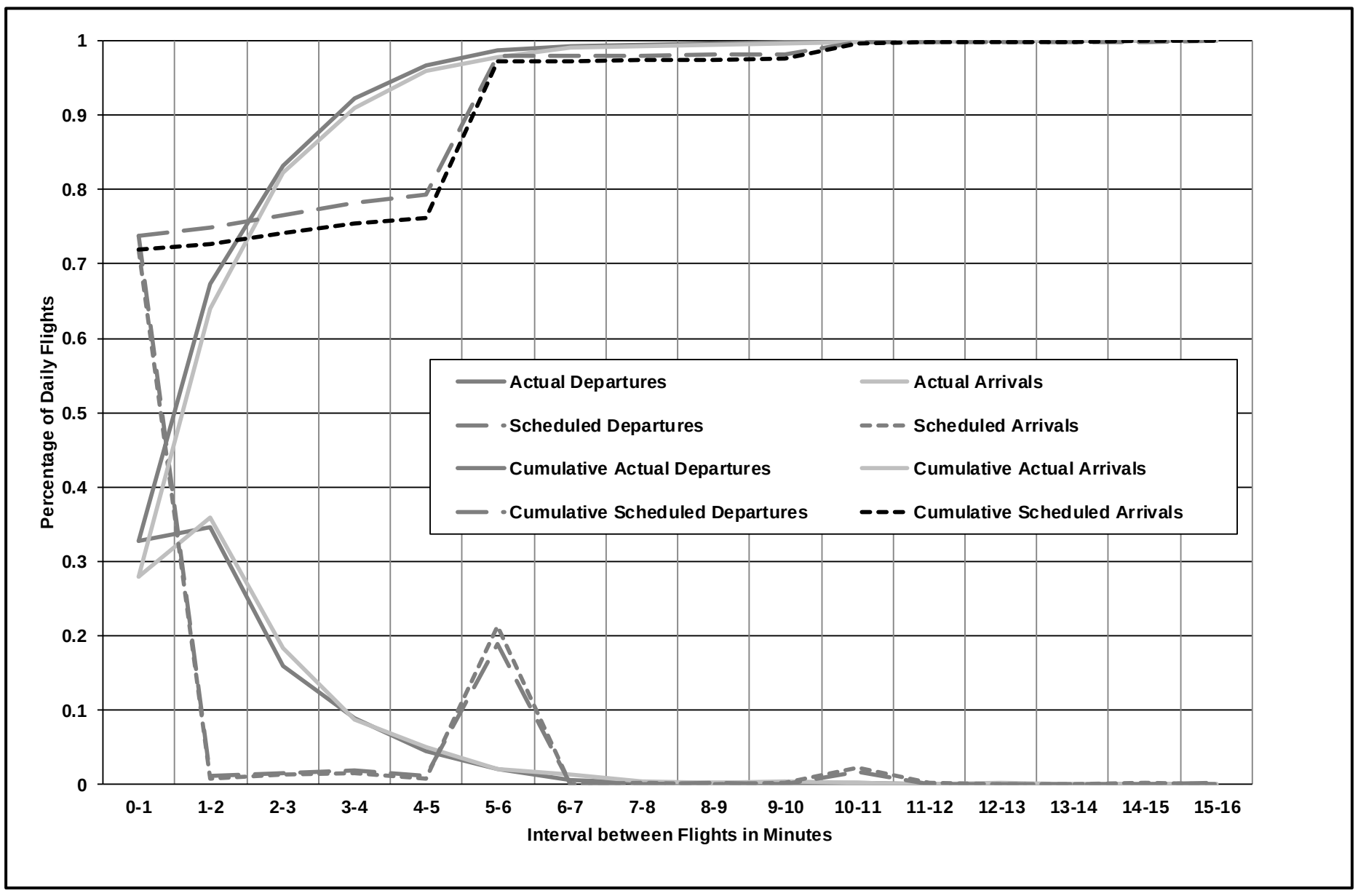

**Fig. 20.** Scheduled and actual interval times of flight at London-Heathrow airport

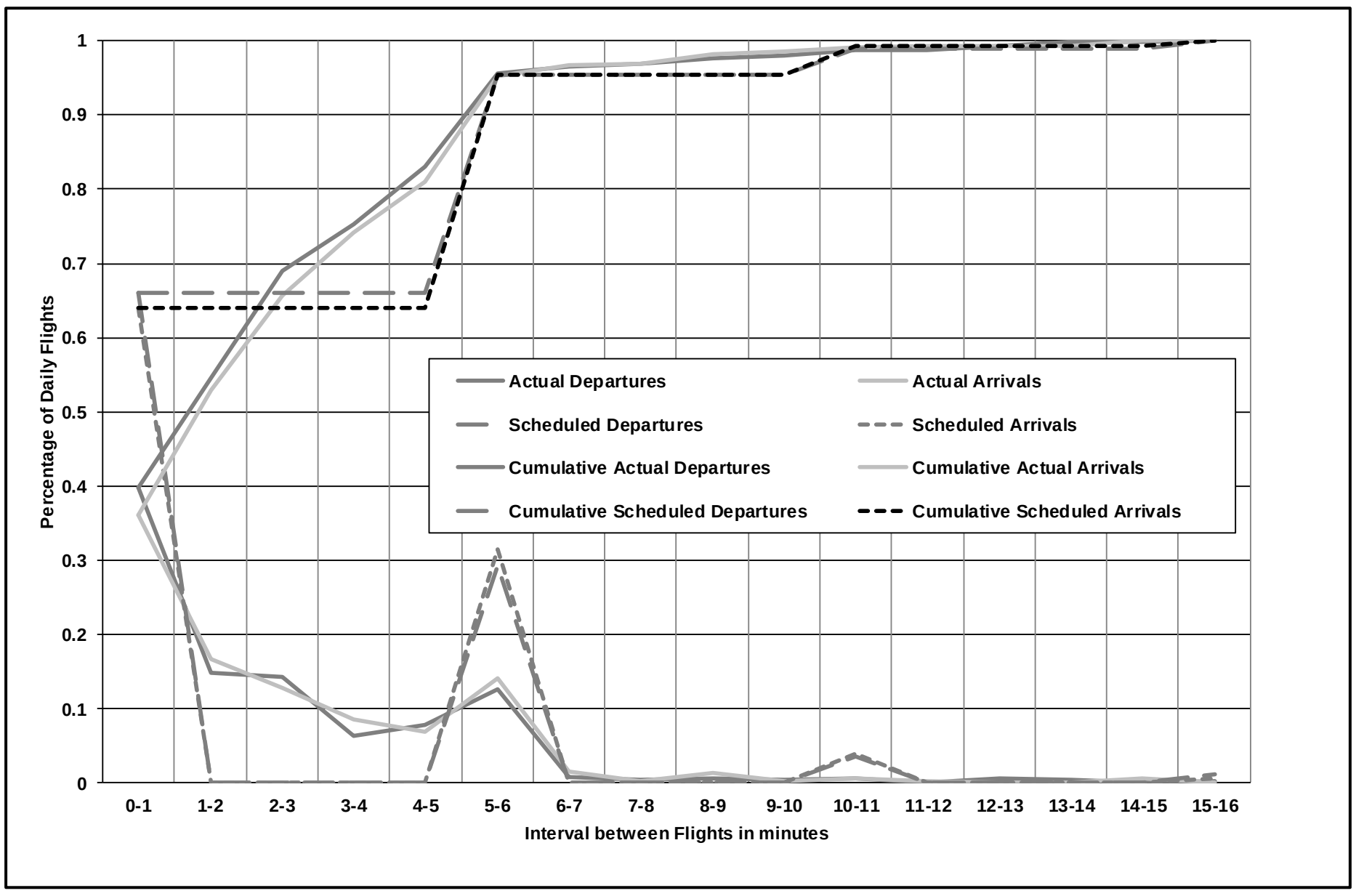

**Fig. 21.** Scheduled and actual interval times of flights at Tokyo-Haneda airport

At Heathrow about 75% of the daily arrivals and departures are scheduled within a minute of each other; 20% of the flights are scheduled at 5-minute intervals; and about 3% are scheduled at 10-minute intervals. Of these flights only about one in three are actually operated within the same minute (corresponding to a demand rate of at least 60 flights per hour per runway), 35% of all flights are operated at intervals of more than one minute, but less than 2 minutes (corresponding to a demand rate of 30 to 60 flights per hour per runway) and about 15% of the flights are operated under 3 minutes (corresponding to a demand rate of 20 to 30 flights per hour per runway).

On average subsequent arrivals or departures are served in intervals of 60- and, say, 90-seconds (corresponding to a demand rate of about 40 flights per hour. This corresponds in order-of-magnitude to the inverse of the slot capacity of arrivals (44 per hour) and departures (44 per hour) at Heathrow of 82 and 82 seconds, respectively (**Fig. 20**).

In contrast, at Haneda airport about 65% of the daily flights are scheduled in bunches of flights within a minute of each other, 30% are scheduled in intervals of between 5 and 6 minutes and about 5% are scheduled in intervals of less than 11 minutes. Of these scheduled flights, up to 40% are actually operated within a minute of each other, 15% are operated at an interval of between 1 and 2 minutes and surprisingly around 13% (compared to Heathrow with only about 3%) are operated with a 5 to 6 minutes headway between flights.

Haneda manages to reduce its delays by allowing larger breaks between arrivals and departures to relax the traffic flow. On average flights seem to be operated within 90 seconds and 2-minute intervals, which would correspond to a demand rate of 30 to 40 flights per hour per departure or arrival (**Fig. 21**).

Since the flight schedule data only allows the calculation of the headway or interval between flights by the minute, it is not possible to derive more accurate distributions. The data already suggests that arriving and departing flights follow approximately the same distribution. Consequently, this means that airport management is not flexible enough to reduce the variation of the incoming flights in favor of the scheduled departing flights, by, for example, adjusting the turn-around times according to the incoming delays.

However, it should be mentioned that the traffic mix at Haneda is different compared to that of London-Heathrow, with a higher percentage of Heavy aircraft. Hence Haneda manages to operate the airport with fewer flights but more average passengers per flight to achieve the same number of annual passengers as London-Heathrow (**Table 7**).

During the main operating hours, the separation minima between the sequence of arrivals and departures at Tokyo-Haneda and London-Heathrow airport are applied to each flight according to its MTOW and turbulence category (given in **Table 6**; distances of 2.5 NM have not been assigned in this example). This gives us a weighted average minimum distance of 3.84 NM for the flows in and out of Haneda airport and 3.65 NM for the flows at Heathrow airport. Depending on the most likely average speed of the aircraft, these translate for example for average approach and departure speeds of 250 kilometers per hour into hypothetical total airport capacities of about 70 and 74 flights per hour, respectively over the main operating hours, and for 300 kilometers per hour into capacities of about 84 and 89 flights per hour, respectively (**Table 7**).

This illustrates the calculation of maximum *capacity* in *average aircraft speed divided by average spacing* for arrivals and departures just by summing the minimum distances over a certain period of time. This means capacity is directly proportional to aircraft speed and inverse proportional to the minimum separation (Gosling et al., 1981: p. 51).

Some argue that Little’s Law is of small value when dealing with airports and fluctuations of daily and hourly demand, because processes in the airport are never stable over time and relevant queues rarely disappear completely (Vaze, 2009). Furthermore, the servers may operate above capacity at overload levels and therefore show rapidly increasing delays. At airports, the end of each operating day marks a natural break from further arriving demand, which gives room for any queues and accumulations of delayed flights to dissolve. To visualize the determinants of Little’s formula and to further understand the fluctuation of demand, service rate and length of queue, we will now look at cumulative diagrams.

Generally applicable to traffic congestion problems are the *Newell*- or cumulative-diagrams. The data required to plot such diagrams are usually provided by the output of simulation programs (**Fig. 22** and **Fig. 23**) but can also be observed given the financial resources, technology and manpower. Various sources, such as de Neufville and Odoni (2003) and Little and Graves (2008), point out the importance of such cumulative diagrams in revealing periods of heavy congestion, during which customers are queued and therefore delayed.

**Table 7.** Descriptive data and capacity estimates for Tokyo-Haneda and London-Heathrow airports from aircraft sequences in actual flight schedules (Source: ACI, Flightstats.com)

| Airport | PAX (2008) | Flights (2008) | PAX per Flight | % Heavy | % Medium | Daily Flights (08 - 20h) | | Avg. Separation Minima in Nautical Miles | | Capacity in Flights per hour at AC Speed: 250 km/h | | 300 km/h | | 250 km/h | 300 km/h |
|---|---|---|---|---|---|---|---|---|---|---|---|---|---|---|---|
| | in millions | in thousands | | | | ARR | DEP | ARR | DEP | ARR | DEP | ARR | DEP | Total | Total |
| Tokyo Haneda | 66.8 | 339.6 | 197 | 60% | 40% | 363 | 373 | 3.83 | 3.85 | 35 | 35 | 42 | 42 | 70 | 84 |
| London Heathrow | 67.0 | 478.5 | 140 | 40% | 60% | 532 | 556 | 3.63 | 3.66 | 37 | 37 | 45 | 44 | 74 | 89 |

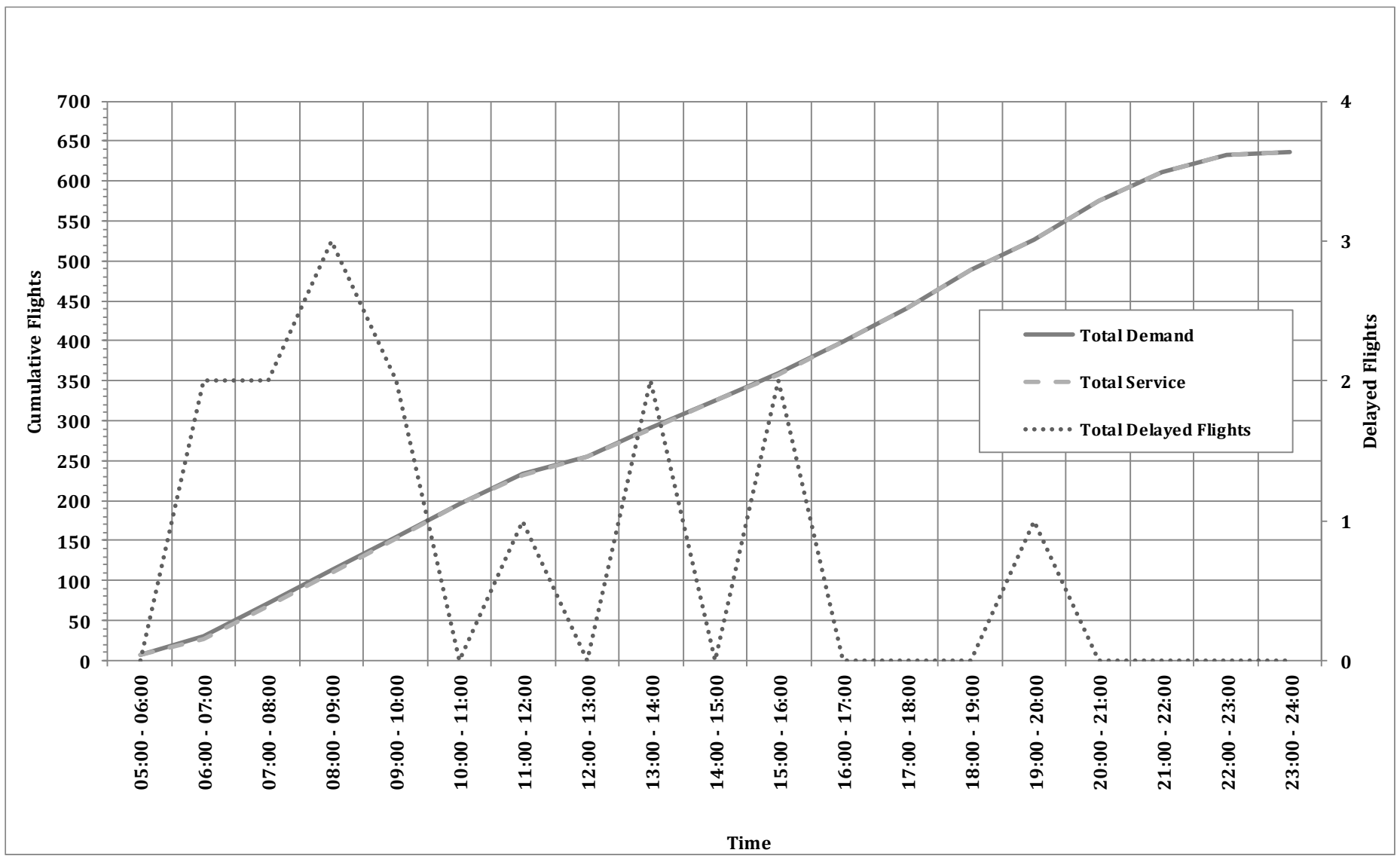

**Fig. 22.** Cumulative diagrams of Berlin-Brandenburg International (BER) airport baseline simulation output

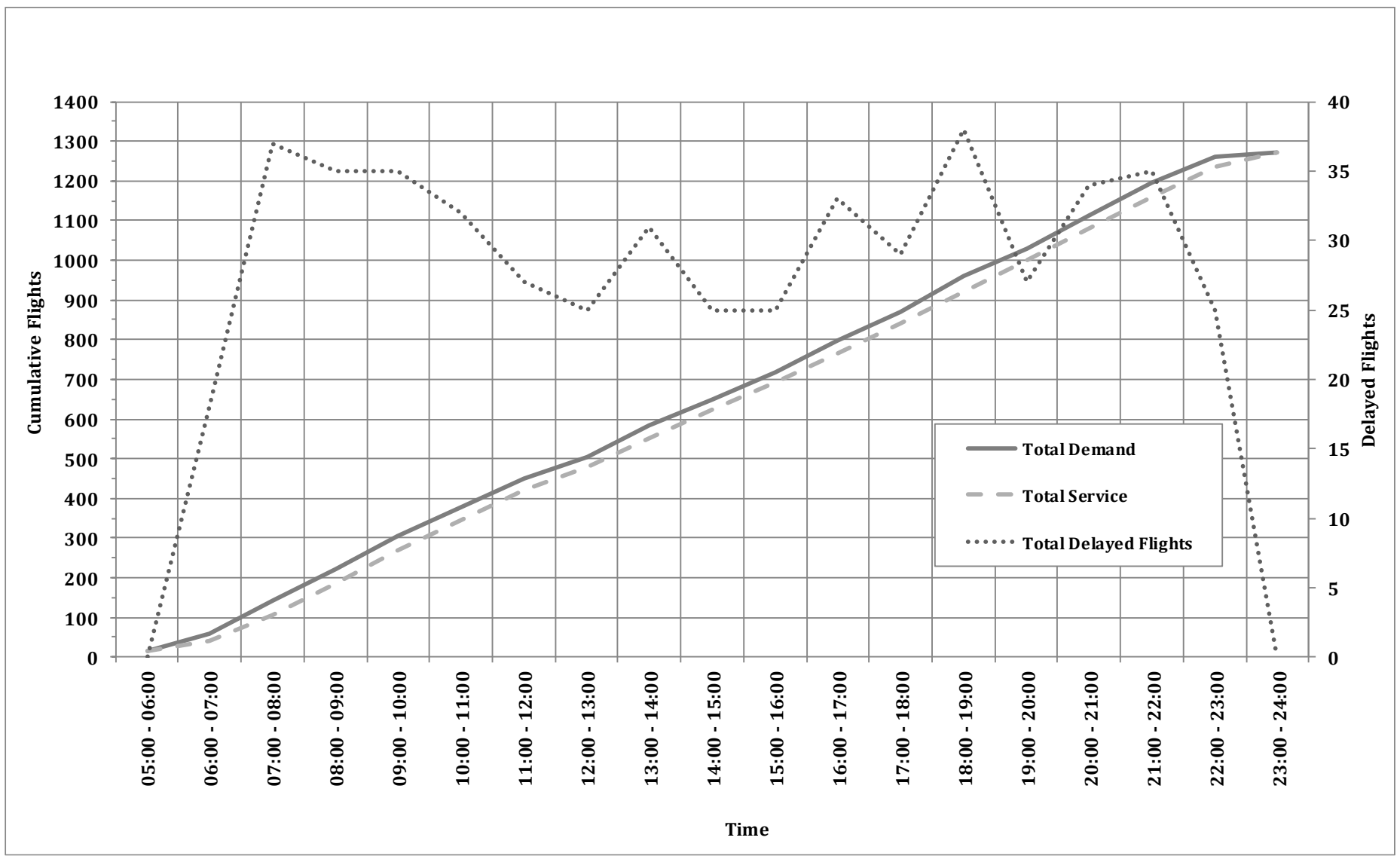

**Fig. 23.** Cumulative diagrams of Berlin-Brandenburg International (BER) airport 100% growth scenario simulation output

The three plotted functions in **Fig. 22** and **Fig. 23** that are primarily required for interpretation stem from a simulation study of the Berlin-Brandenburg International (BER) airport (currently under construction and scheduled for a October 31st, 2020 opening), which was conducted with the airport/airspace simulation environment SIMMOD (developed by the Federal Aviation Administration [FAA] in the U.S.). The graphs illustrate the cumulative flights over time of day (demand rate), the flights requesting service over time of day (service rate, i.e. capacity) and the difference of both (vertical distance), the accumulating flights in queue. In the simulation study of BER, first the baseline demand of 640 daily and 48 peak hourly flights of the combined schedule of the two airports to be phased out, Berlin-Tegel and Berlin-Schönefeld, was simulated; this revealed negligible delays (**Fig. 22**). Then in a second step the baseline flights in the simulated schedule were doubled on the independent parallel runway configuration under construction at BER in segregated mode (**Fig. 23**).

The interpretation of the cumulative diagrams of the simulation output reveals that the 100% increase over the baseline demand and up to 90 peak hourly flights will be too much for BER to handle. Subsequent studies have shown that the sustainable/practical capacity would be reached at a level 80% above the baseline demand, at about 1100 daily and around 80 peak hourly flights, because from this point the LOS of 5 minutes of average delay per flight is clearly exceeded.

Compared to the baseline, **Fig. 23** shows how the demand and service flows diverge and the number of delayed flights increases. Almost all the daily flights, but mainly the departures, from 07:00 onwards are queued and therefore delayed. The hypothetical waiting queue could reach a length of up to around 40 aircraft, and the average waiting time remains high at about 20 to 30 minutes per aircraft (measured in horizontal distance between demand and service function) (**Fig. 23**). The slope of each cumulative graph defines the rate in aircraft per unit of time.

## 2.4 Conclusion

In this article the main determinants of air traffic punctuality and congestion at airports have been presented, regarding maintaining a sustainable LOS. We, as air transportation customers, expect seamless service and high schedule adherence, so capacity and service facilities must be expanded and planned in line with and even slightly ahead of demand.

To make some order-of-magnitude calculations of congestion and delay, it is a good start for an analysis to begin with some basic capacity utilization figures from actual demand and declared capacity (**Table 5**). Furthermore, airport management should closely monitor the LOS of arriving and departing flights compared to schedule (**Fig. 18** and **Fig. 19**). Here the most popular determinant of punctuality is the percentage of flights delayed less than 15 minutes. From the same figures another determinant for LOS in average delay per flight can be isolated. This is certainly an inexpensive way for airport management to calculate adequate capacity for its airport and subsystems regarding LOS, i.e. service quality, because if schedule adherence is high during normal operation it could well be assumed that capacity equals demand at any given time.

When capacity is planned and slots are distributed, random distributions of the demand should be considered in the airport schedules to reduce queuing. As was shown for Haneda airport (**Fig. 21**), buffers should be implemented or slots should be restricted to relax the traffic flow from time to time during the day, to allow for queues to dissolve and for punctuality to return. It would therefore be reasonable for airport management to influence the scheduling of flights, smooth the traffic flow and minimize congestion delays by assigning flights to a particular minute by introducing intermediate intervals of 2 to 5 minutes between flights, according to the actual arrival and departure distributions (**Fig. 20** and **Fig. 21**). Here further research and a range of case-studies should deliver fruitful insights.

The application of Little's Law and the cumulative diagrams of arriving and departing aircraft from a queueing system would deliver answers about its state of congestion (**Fig. 22** and **Fig. 23**). Here the objective for airport management should be in balancing the delay to an average of less than 5 minutes per flight. The cumulative diagrams show that when overload situations occur, immediately waiting queues and therefore delays accumulate disproportionally. However, cumulative diagrams require the most information (especially regarding the customers arriving at and leaving a queuing system over time) from field observations and a large amount of preparation and calibration time to create a simulation scenario. Based on flight schedules and observed random distributions, we are now able to simulate both airport operations and future scenarios with increased "realism".

Ideally queues rarely form, and the service rate is equal to the demand rate, as was shown for the BER airport baseline scenario.

Results of such an airport capacity analysis including LOS could not only be reasonably represented by cumulative diagrams but should nowadays (to reach a broader audience) be presented in the form of (4-D [3-D plus time]) animations of the queues as well. Adding data from Geographic Information Systems (GIS) (which could combine various sources of information such as satellite imagery, floor plans, population densities, noise or gaseous emission footprints etc.) to the mix of planning tools already discussed give those responsible for airport management a greater capacity for planning by looking at the environmental impact and externalities involved (**Fig. 24**). In this manner the determinants of airport flight delays can be utilized to develop practical performance indicators to aid not only in planning capacity for the future but also to aid in increasing passenger and airport community satisfaction.

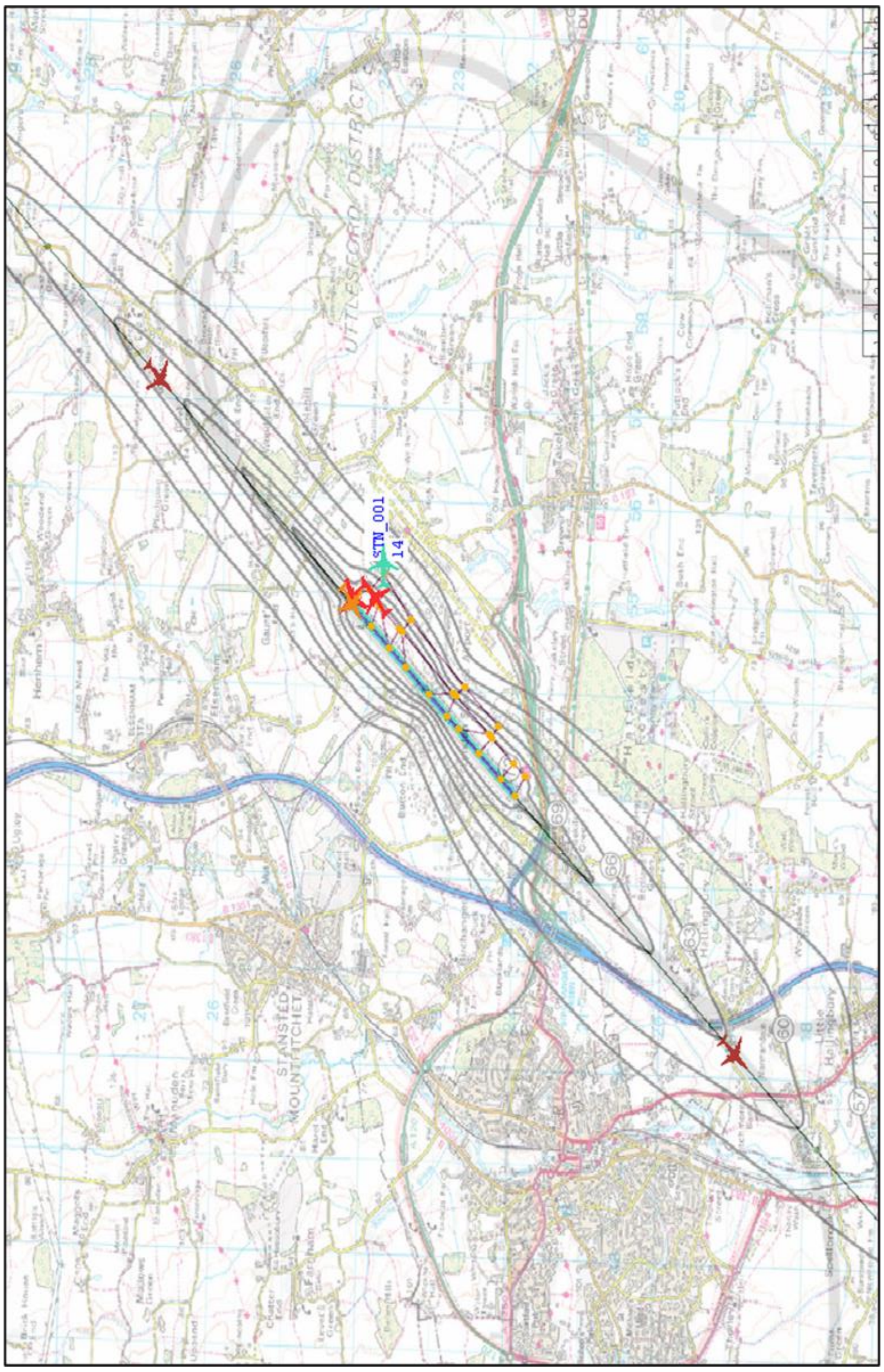

**Fig. 24.** Overlay of Noise and Population maps on airside traffic animation (and associated departure queue) at Stansted airport (Sources: AirportTools, BAA)

# 3 Airside productivity of selected European airports[1]

Branko Bubalo

**Abstract.** Productivity and benchmarking studies on airports vary in terms of methodologies used, variables chosen and airport sample size. The usage of certain operational airport parameters on the airside, especially the sole number of runways, as indicators of capacity used in previous studies is questioned. It can be shown that such indicators tend to give contradictory and unreasonable results. Instead one must take other important determinants of airport infrastructure and capacity into account, such as preferential runway(s) in use and runway configuration, apron capacity, aircraft parking positions or the daily pattern of demand.

This paper focuses on results of previous airport benchmarking studies on productivity, capacity, and delay. Order-of-magnitude approximations for airport airside capacity are delivered together with data samples of European airports, which could be used for forecasts or serve as benchmarks for other airports with similar characteristics. It has been found that apart from the fundamental role of runway capacity for airports, the environmental capacity might play a larger role in the future to guarantee long-term sustainable aviation. Air transportation in Europe made a commitment to reduce fuel-burn, hence $CO_2$-Emissions, aircraft and engine noise and delays under its Single European Sky (SES) initiative and the inclusion of aviation in the Emission Trading System (ETS) from 2010 on.

Firstly, a distinction of the main quantitative outputs of production of airports will be made and three distinctive airport peer groups will be established for comparison. Secondly, the relationship between demand and airside capacity will be explained, with an emphasis on runway capacity and available slots, and its utilization on an hourly and annual basis. Thirdly, externalities of production, especially delay, are analyzed in detail by simulating airside traffic in SIMMOD. The paper finishes with concluding remarks. Further research about noise and $CO_2$-Emissions will be carried out in the future to derive benchmarks for environmental capacity.

[1] This article has been published in the proceedings of the Air Transport Research Society (ATRS) World Conference 2010, Porto, Portugal.

## 3.1 Introduction

Today's airports are vast and expensive infrastructures, which have considerable positive and negative impact on population and environment. An airport and its (growing) traffic could seriously affect the local neighbors in the vicinity by decreasing air quality and increasing aircraft noise. On the other hand - regionally and globally - airports represent the most important interfaces for various transport modes and cargo transport. Especially the demand for business aviation, tourism and perishable goods make air transportation the mode of choice for the transportation and logistics industry. Consumer behavior towards air travel has also changed in the last ten years. With the rapid development of telecommunication and information technology, specifically the Internet, and the simultaneous emergence of low fare airlines, the demand for air travel has considerably increased and makes flying sought after and affordable for the broader public. Even though the September 11 attacks in the U.S. in 2001, the SARS epidemic in 2003, and the recent global banking crisis in 2009 caused temporary adverse effects to the increasing demand for air travel, the trend towards further future growth is clear.

In the past we have seen almost unconstrained growth in the western world, which started in the late 1940's and continued until the end of the century. This has been largely fueled by the deregulation of air transportation in the 70s in the U.S. and in the 90s in Europe. The North American and European markets and main routes have now matured quite considerably. Therefore, most of the future growth of demand for air transportation will happen in Asia and the Middle East with the increasing wealth and education in those regions. China, India and the oil-rich countries on the Arabian Peninsula invest billions of Euros in airport infrastructure to boost and support their economic development. This does not imply that European air transportation demand will stagnate at the current level; it will rather grow at a lower rate of around 3-4% annually in terms of total flights and at around 4-5% annually in terms of total passengers. This would still result in a doubling of traffic and passengers in the next 16 to 20 years.

The question which arises is therefore: *Has Europe sufficient and flexible airport capacity to serve future demand?*

The following sections will present the results of recent benchmarking studies and approaches in estimating current demand and airside capacity, the utilization of infrastructure, and the level of congestion. To support a noise cap and trade system with noise quantities, and since environmental capacity can play a vital role in airport development, a method of turning noise into a tradable unit will be presented on the basis of current Quota Count (*QC*) Systems and Noise Certification Data.

## 3.2 Airport productivity and demand

The main output of "production" of an airport can be divided into the following streams of "products", which must be analyzed separately:

- Airside productivity output: Number of aircraft landings and take-offs or movements.
- Landside productivity output: Number of passengers and cargo.

We divide between airside and landside productivity. At the airside aircraft are served and handled between the runways, apron and parking positions. At the landside passenger and cargo is handled among parked aircraft, the gates, terminals and cargo facilities. So eventually the flow of airside movements on the runway(s) and apron splits into streams of passengers and cargo inside the landside facilities. The airside, the runway system particularly, is a critical requirement for the operation of an airport. A new runway at an airport needs not only vast amounts of investment, but also a timely forward-planning frame of approximately ten years, e.g. for planning, legal approval, acquisitions, and construction.

The division between airside and landside can also be made with airport revenues, where a large portion of aviation (airside) revenues comes from the aircraft landing, parking, handling, and central infrastructure charges, but also from noise and emission charges. Non-aviation (landside) revenues on the contrary come from passenger services and consumption, through lease and rents from shops, retail, food and beverages, and passenger parking. Many privatized airports in Europe nowadays generate approximately half of their revenues from non-aeronautical sources, which make them less dependent on airline landing charges, but more dependent on a constant flow of consuming passengers. The financial term for covering airport infrastructure costs with not only aeronautical revenues, but also with non-aeronautical revenues, is the "dual till" approach (Czerny 2006). To enrich the passenger experience of spending time and money at an airport, huge investments in attractive terminal design are being made.

Perhaps the most important prerequisite for airside analysis of airports is the projected or actual flight schedule of each airport. With available flight schedule data over longer periods of time, but including at least one representative day, many important observations on productivity, runway efficiency, traffic mix, and traffic variability or seasonality can be made. The International Air Transport Association (IATA) suggests the first step for estimating capacity and demand of an airport should be simple busy-period traffic observations (IATA 1981). Therefore, the first diagram to plot would be the traffic load plot by hour of day, which gives insights about the daily periods of peak demand. Peak period traffic information is required for detailed capacity planning of the airside, for example runways, and the landside, for example passenger facilities.

To illustrate the importance of the daily traffic load plots and their ease in providing information, **Fig. 25** shows the interruptions of airport traffic at London-Heathrow due to the European airspace closure between April 15th and April 22nd 2010 resulting from the Eyjafjallajoekull volcano ash cloud.

At the 66th IATA Annual General Meeting 2010 in Berlin, Giovanni Bisignani, CEO and Director General of IATA, commented in his speech entitled "State of the Air Transport Industry" on the impact of this week-long airspace closure for the European and global economy: "April gave us a vivid picture of life without aviation: 10 million people stranded. Hotels and convention centers empty. Seafood and flowers rotting, and just-in-time production delayed. The volcano cost the economy $5 billion, far more than the $1.8 billion of lost airline revenue. The eruption was a wake-up call. The message was clear: without air connectivity, modern life is not possible. Aviation is vital." (IATA 2010)

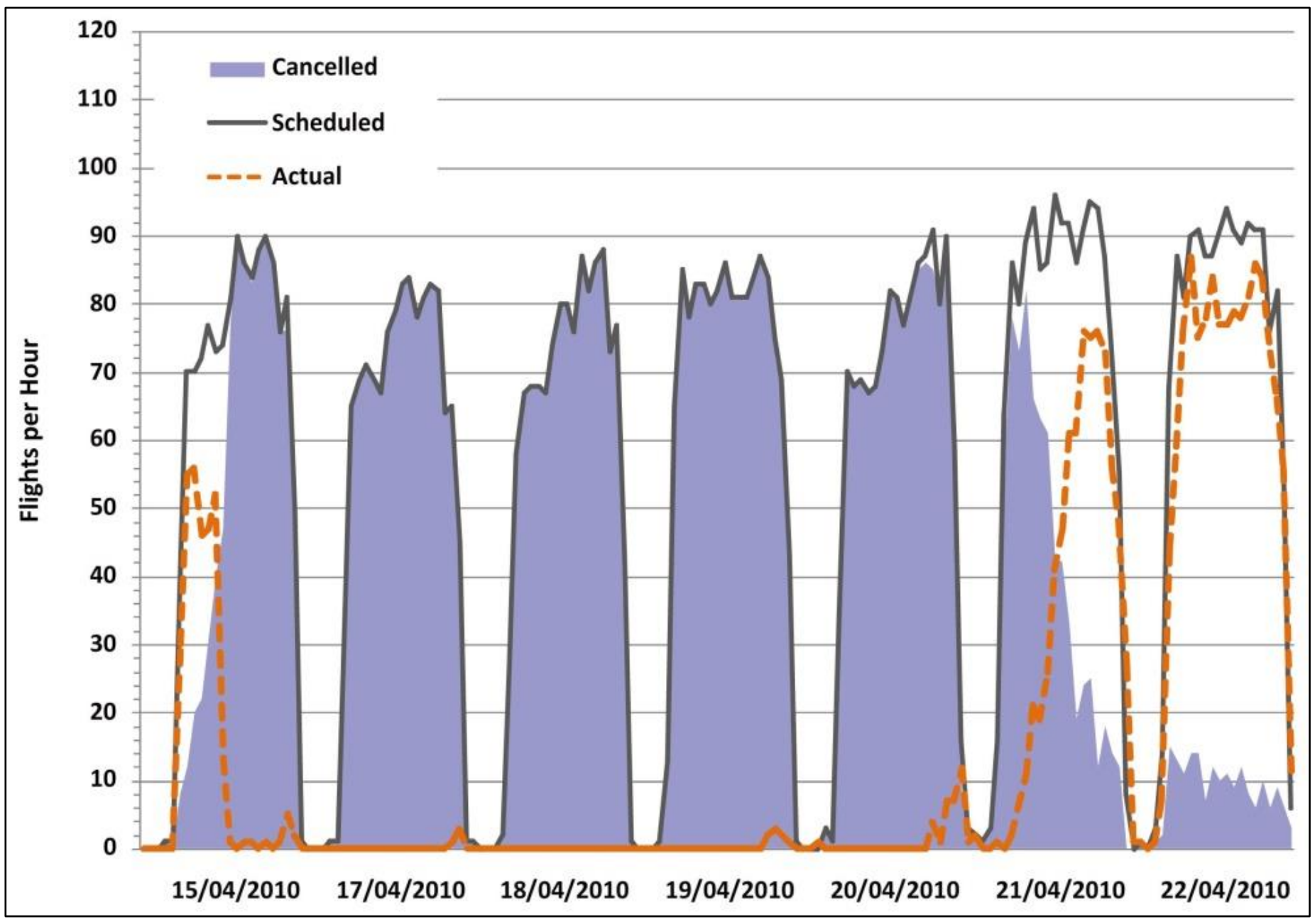

**Fig. 25.** Volcanic ash-cloud disruption at London-Heathrow airport in April 2010 (Source: Bubalo 2010)

It must be noted that for a complete picture of overall productivity of an airport, the airside and the landside must be assessed together; otherwise the picture will lack consistency and might result in ambivalent conclusions.

Assessment of landside productivity would include service quality measures of the processes inside the terminal, which are not easy to calculate without detailed information on the processing speeds of the various servers, e.g. check-in counters or baggage claim units. A slight hint on the efficient use of terminal capacity comes from the Minimum Connecting Times (MCT), which are used in the flight booking process to adequately connect transit flights. **Table 8** shows the MCT for connections among and inside airports in the greater London area. It is striking that international connection flights from one terminal to the other at London Heathrow (LHR) airport can span between 45 minutes (Terminal 1 to Terminal 1) and 2 hours (Terminal 1 and Terminal 5). Therefore, a large city like London with an airport system of four airports might offer more comfortable connections over the alternative airports London-Stansted (STN), London-Gatwick (LGW) or London-City (LCY).

**Table 8.** Minimum Connecting Times at Greater London Area Airports (Data from Amadeus Selling Platform 2009)

| Connection | From Terminal | To Terminal | Dom/Dom in minutes | Dom/Int in minutes | Int/Dom in minutes | Int/Int in minutes |
|---|---|---|---|---|---|---|
| LCY-LCY | **-** | | **30** | **30** | **30** | **30** |
| LGW-LGW | **N-** | **N** | **45** | **45** | **45** | **45** |
| LHR-LHR | **1-** | **1** | **45** | **45** | **45** | **45** |
| STN-STN | **-** | | **45** | **45** | **45** | **45** |
| LHR-LHR | **4-** | **4** | **90** | **45** | **45** | **45** |
| LGW-LGW | **S-** | **S** | **40** | **45** | **60** | **55** |
| LHR-LHR | **2-** | **2** | **---** | **---** | **---** | **60** |
| LHR-LHR | **4-** | **1** | **60** | **60** | **75** | **60** |
| LHR-LHR | **3-** | **3** | **60** | **60** | **60** | **60** |
| LHR-LHR | **5-** | **5** | **60** | **60** | **60** | **60** |
| LHR-LHR | **1-** | **4** | **90** | **60** | **60** | **60** |
| LHR-LHR | **3-** | **2** | **---** | **75** | **75** | **75** |
| LGW-LGW | **S-** | **N** | **75** | **75** | **75** | **75** |
| LHR-LHR | **2-/3-** | **1** | **75** | **75** | **75** | **75** |
| LHR-LHR | **1-/2-/3-/4-** | **TN** | **90** | **90** | **90** | **90** |
| LHR-LHR | **2-/3-** | **4** | **90** | **90** | **90** | **90** |
| LHR-LHR | **1-/2-/3-/4-** | **5** | **120** | **120** | **120** | **120** |
| LGW-LHR | **-** | | **150** | **150** | **150** | **150** |
| LCY-LHR | **-** | | **180** | **180** | **180** | **180** |
| STN-LGW | **-** | | **180** | **180** | **180** | **180** |
| LHR-STN | **-** | | **200** | **200** | **200** | **200** |
| STN-LHR | **-** | | **200** | **200** | **200** | **200** |
| LHR-LTN | **-** | | **205** | **205** | **205** | **205** |
| LGW-LCY | **-** | | **210** | **210** | **210** | **210** |
| LHR-LCY | **-** | | **210** | **210** | **210** | **210** |
| LCY-LTN | **-** | | **240** | **240** | **240** | **240** |
| LCY-STN | **-** | | **240** | **240** | **240** | **240** |
| STN-LTN | **-** | | **240** | **240** | **240** | **240** |

### 3.3 Airport peer groups for benchmarking

When dealing with different airports in size, location and stage of maturity, it becomes obvious that comparisons among airports – *benchmarking* – is a difficult undertaking. This is even truer for financial or economical comparisons, where different landing fees, accounting standards, national laws and regulations, levels of outsourcing, and level of privatization frequently distort the results. Various papers point out these complexities and offer promising solutions (Graham 2008).

From an engineering perspective on airport benchmarking, one can already conclude that these difficulties exist in other forms, but that for a large portion of European airports comparisons of the airside operation are indeed possible.

The main limitation on runway operations at airports results from the safety separation between successive landing and departing aircraft on the same runway and the lateral separation between those using parallel runways, due to wake turbulence created by the wingtips of the aircraft. Encountering wake turbulence from preceding aircraft during the critical landing and take-off phase can seriously impact on the stability of an aircraft in the air. Therefore, air traffic control applies specific separation requirements to aircraft of different sizes and weights. A “Small” aircraft (< 7 tons) following a “Large” aircraft (7-136 tons) requires a safety distance of approximately 5 nautical miles. A “Large” aircraft following a “Small” aircraft requires only a separation of approximately 3 nautical miles (NATS 2010; Horonjeff 2010). So, the mix and sequencing of aircraft types obviously has a direct effect on runway capacity.

**Table 9.** Traffic Mix and Mix Index of Selected European Airports (Bubalo 2009, Official Airline Guide 2009)

| **Airport** | Airport Code | **Small** | **Large** | **Heavy** | **Mix Index** |
|---|---|---|---|---|---|
| **Amsterdam** | AMS | 0% | 83% | 17% | 135% |
| **Athens** | ATH | 0% | 95% | 5% | 110% |
| **Birmingham** | BHX | 0% | 98% | 2% | 103% |
| **Brussels** | BRU | 0% | 90% | 10% | 121% |
| **Cologne** | CGN | 0% | 98% | 2% | 103% |
| **Copenhagen** | CPH | 0% | 96% | 4% | 107% |
| **Dusseldorf** | DUS | 0% | 97% | 3% | 106% |
| **Frankfurt** | FRA | 0% | 76% | 24% | 147% |
| **Hannover** | HAJ | 0% | 100% | 0% | 100% |
| **London-City** | LCY | 0% | 100% | 0% | 100% |
| **London-Gatwick** | LGW | 0% | 91% | 9% | 118% |
| **London-Heathrow** | LHR | 0% | 66% | 34% | 168% |
| **London-Luton** | LTN | 0% | 99% | 1% | 102% |
| **Munich** | MUC | 0% | 94% | 6% | 111% |
| **Nice** | NCE | 46% | 54% | 1% | 56% |
| **Oslo** | OSL | 0% | 100% | 0% | 100% |
| **Palma de Mallorca** | PMO | 0% | 100% | 0% | 100% |
| **London-Stansted** | STN | 0% | 99% | 1% | 102% |
| **Stuttgart** | STR | 0% | 99% | 1% | 101% |
| **Vienna** | VIE | 0% | 96% | 4% | 108% |
| **Zurich** | ZRH | 0% | 90% | 10% | 120% |

**Table 9** shows the mix of different aircraft classes at some European airports. Since “Heavy” aircraft (>136 tons) in the mix of an airport strongly influence its overall throughput, a mathematical expression, called the Mix Index, has been adopted from the U.S. Federal Aviation Administration, which underlines the significance of the ”Heavy” class. The Mix Index (MI) adds to the (usually predominant) percentage share of "Large" aircraft, an additional three-fold percentage share of specifically "Heavy" aircraft, therefore the formula reads:

$$Mix\ Index = 3x\ (\%\ class\ \text{“}Heavy\text{”}\ aircraft) + (\%\ class\ \text{“}Large\text{”}\ aircraft).$$

Tretheway (2006) and Forsyth (2004) suggest isolating the potential peer groups of airports, where among peers a benchmarking analysis can be made. Earlier work (Bubalo 2008) isolates such peer groups according to the primary runway system used and their configuration and the corresponding airport runway capacity. The underlying analysis of the capacity of runway systems at US airports has been conducted by the FAA and is documented in the Advisory Circular "Airport Capacity and Delay" (FAA 1995). The results presented in that document are used to isolate peer groups based on similar traffic mix and maximum productivity of runway operations, namely the annual and hourly capacity of an airport.

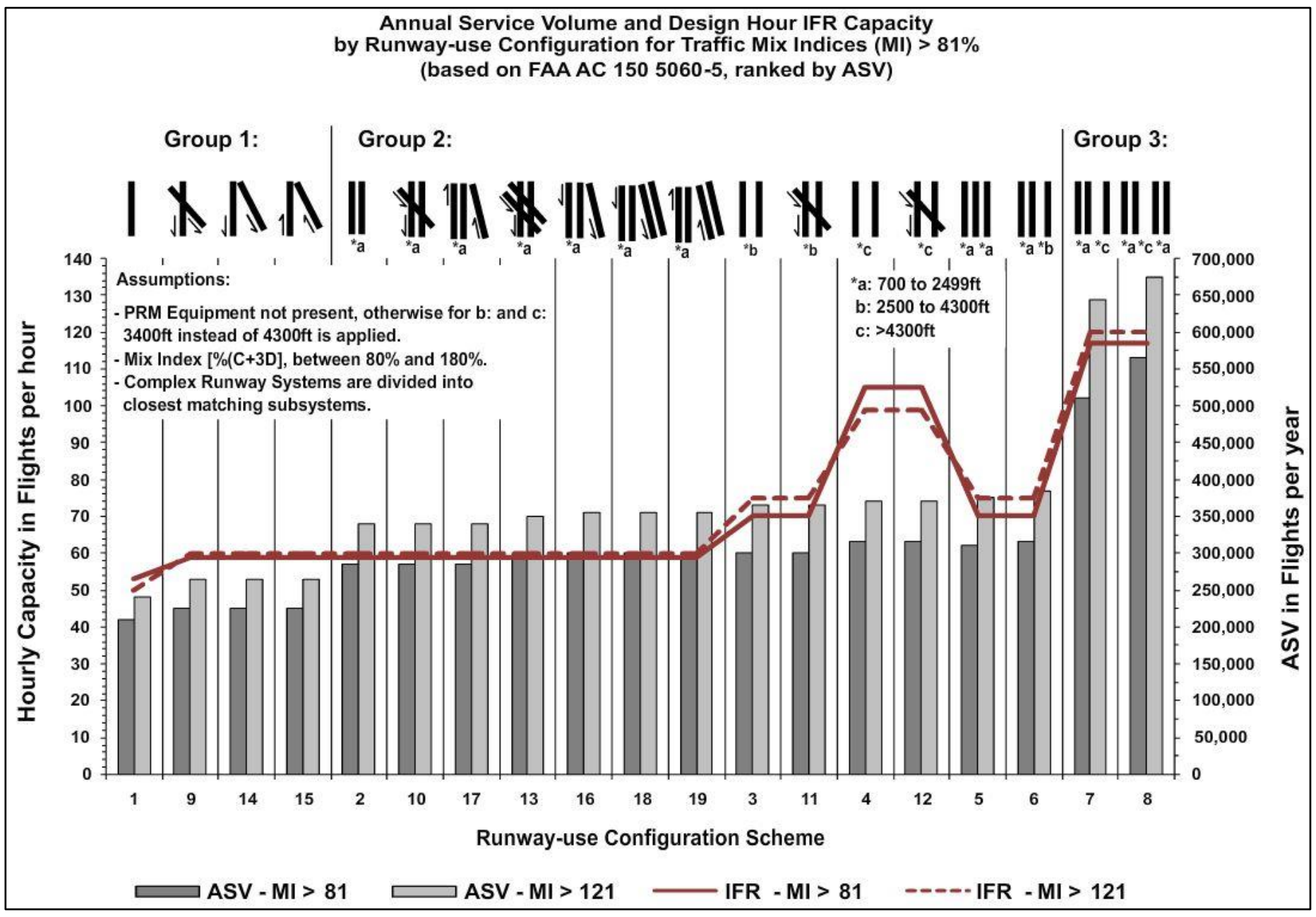

**Fig. 26.** Peer Groups of Airports Based on Annual and Hourly Capacity (Source: FAA 1995; Bubalo 2008)

The Federal Aviation Administration (FAA) in the U.S. developed a simple technique to estimate order-of-magnitude annual and hourly capacity (FAA 1995; Horonjeff 2010, p. 532). As **Fig. 26** shows, three main groups with approximately similar annual capacity have been isolated. Group 1 represents airports with a single runway, which might have an additional crosswind runway to accommodate changing wind direction. The extra crosswind runway will therefore not increase the overall runway capacity.

Group 2 represents airports with parallel runways, which are less than 4300 feet/1.3 kilometers apart. Most of the airports in **Fig. 26** with separation indication a) 700 to 2500 ft. separation and b) 2500 to 4300 ft. separation can only be operated coordinately due to safety regulations. Wake turbulence vortices caused by aircraft on one of the dependent runways can be blown laterally by winds and possibly impact on aircraft approaching the parallel runway. Exceptions are configuration 4 and 12 with indication c) more than 4300 ft. separation, which allow independent operation of the runways and therefore have higher hourly capacities than their peers.

Group 3 includes all runway systems with complex configurations. This group's configurations have a minimum of two independent parallel runways and one additional close spaced parallel runway on one side (configuration 7) (**Fig. 26**).

When looking at the annual capacities from the FAA methodology perspective and that of the annual operations of different European airports plotted by number of runways as shown in **Fig. 27**, we find proof that discussing exclusively the number of runways in use as an input parameter for productivity analyses with linear equation-solving (e.g. Data Envelopment Analysis) leaves too much variation in the upper and lower capacity limits to deliver adequate results. For example, a doubling of number of runways would <u>not</u> result in a doubling of capacity!

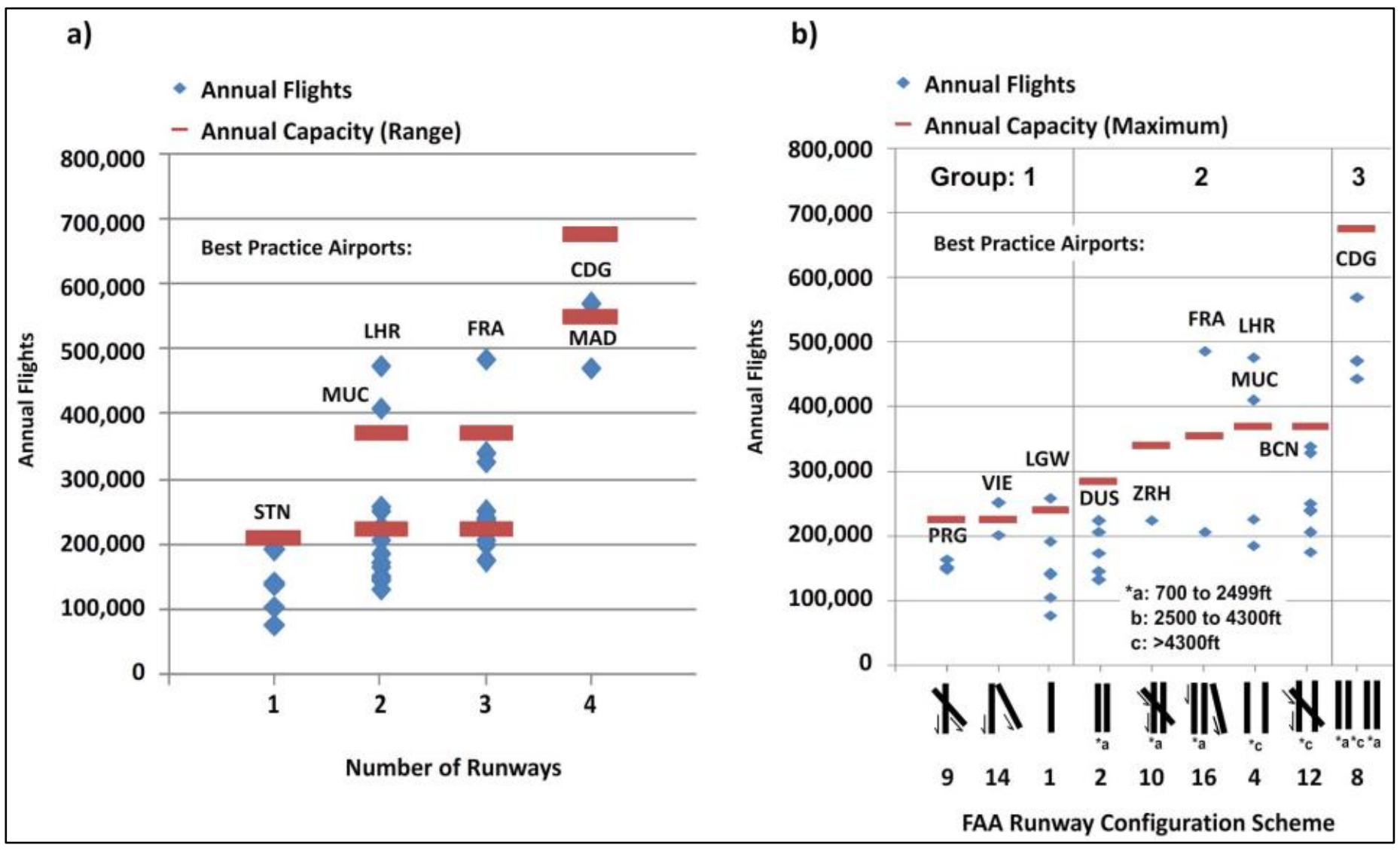

**Fig. 27.** Comparison of annual capacity and demand by a) number of runways and b) by FAA runway configuration number (Bubalo 2010).

London Gatwick airport (LGW) is an example in which the drawback of earlier analyses can be nicely shown. In a strict sense, London Gatwick has two runways, but the airport's Aeronautical Information Publication (AIP) states that the airport only operates one runway under its preferential runway-use system and uses the other runway for aircraft taxiing and emergencies.

By applying the configuration 1 (single runway-in-use) from the FAA methodology, instead of simply counting the number of runways, returns the runway capacity for London Gatwick airport (**Fig. 27a**) of about 240,000 annual flights. The FAA methodology requires the study of more detailed technical information of the operational characteristics of the airports chosen for study. AIP information can be downloaded from the European AIS Database (EAD) from EUROCONTROL for any European airport.

**Fig. 27b** gives a more diverse picture of different groups and subgroups of airports and their configuration and capacity. The direct effect on capacity of additional runways added to an airport could be estimated either by capacity approximations using the FAA methodology or by the isolation of best practices. **Fig. 27** highlights the best practice airports in Europe for each runway configuration, which alternatively can be used as a benchmark. Frankfurt and London Heathrow are extreme examples of airports which operationally exceed their respective approximated capacity, but these are the same airports that experience most of the annual delays (**Table 10**), a fact which suggests that operations at these locations reach capacity fairly frequently. London Gatwick (LGW) airport is a prominent example of a highly productive, but severely congested, single-runway airport. Even globally the 260,000 flights per year in 2007 are without comparison.

Especially on busy days this extraordinary performance of London-Gatwick can be easily observed. By comparison the biggest single-runway airport in the U.S., San-Diego airport (SAN), reaches a far lower number of hourly operations than its European counterpart. As can be seen in the hourly arrival and departure plot of busy-day traffic (in so-called “Gilbo diagrams”) of San-Diego and London-Gatwick airports in **Fig. 28**, San-Diego airport can only serve 41 hourly operations (20 arrivals and 21 departures simultaneously per hour, which are obtained by constructing or imagining a 45 degree line from the origin to the identical maxima of the identical scale of x and y), whereas London-Gatwick reaches an impressive maximum of 61 actual served flights (29 arrivals and 32 departures) during the peak day traffic (Gilbo 1993).

Of course **Table 10** reveals that this high productivity comes at the expense of delay, which in the case of London-Gatwick resulted in ca. 80,000 minutes (or over 1,330 hours) of delay in 2006. Using the “cost per minute of delay” estimations from the Eurocontrol document “Standard-Inputs for Cost-Benefit Analysis” (Eurocontrol 2009) it is possible to derive annual delay costs for the top 21 congested airports in Europe (**Table 10**). The value of time, estimated at 42 Euro per minute of delay, results in an approximate total of some 3.3 million Euros at London-Gatwick in 2006 (network delay propagation, i.e. “knock-on effects”, are not included).

It is not surprising that London-Heathrow airport tops **Table 10**, causing an enormous delay for the international air transport network of 715,761 minutes, which is 9-fold the delay compared to London-Gatwick (Rank 16$^{th}$), resulting in an approximate annual delay cost of 30 million Euros.

**Table 10.** Annual delays recorded by CFMU and calculated delay costs of European airports in 2006 (Eurocontrol 2009)

| | Airport Name | IATA | ICAO | Experienced Delay in Minutes (2006) | Annual Delay Costs at 42€ per Minute |
|---|---|---|---|---|---|
| **1.** | LONDON HEATHROW | LHR | EGLL | 715761 | 30,061,962 |
| **2.** | FRANKFURT MAIN | FRA | EDDF | 671693 | 28,211,106 |
| **3.** | MILANO MALPENSA | MXP | LIMC | 626853 | 26,327,826 |
| **4.** | WIEN | VIE | LOWW | 534717 | 22,458,114 |
| **5.** | ROMA FIUMICINO | FCO | LIRF | 464088 | 19,491,696 |
| **6.** | MADRID BARAJAS | MAD | LEMD | 388094 | 16,299,948 |
| **7.** | MUENCHEN | MUC | EDDM | 343938 | 14,445,396 |
| **8.** | ZURICH | ZRH | LSZH | 248709 | 10,445,778 |
| **9.** | PARIS ORLY | ORY | LFPO | 242897 | 10,201,674 |
| **10.** | ISTANBUL - ATATUERK | IST | LTBA | 216167 | 9,079,014 |
| **11.** | SCHIPHOL | AMS | EHAM | 151918 | 6,380,556 |
| **12.** | COPENHAGEN/KASTRUP | CPH | EKCH | 124148 | 5,214,216 |
| **13.** | LONDON CITY | LCY | EGLC | 111567 | 4,685,814 |
| **14.** | PRAHA RUZYNE | PRG | LKPR | 105861 | 4,446,162 |
| **15.** | PARIS CH DE GAULLE | CDG | LFPG | 81062 | 3,404,604 |
| **16.** | LONDON GATWICK | LGW | EGKK | 79190 | 3,325,980 |
| **17.** | ROMA CIAMPINO | CIA | LIRA | 60362 | 2,535,204 |
| **18.** | MANCHESTER | MAN | EGCC | 59495 | 2,498,790 |
| **19.** | TEGEL-BERLIN | TXL | EDDT | 55816 | 2,344,272 |
| **20.** | LONDON STANSTED | STN | EGSS | 53408 | 2,243,136 |
| **21.** | PALMA DE MALLORCA | PMI | LEPA | 44508 | 1,869,336 |
| | Total | | | **5,380,252** | **225,970,584** |

## 3.4 Airport capacity and delay

But what exactly is the importance of airport capacity? Airport capacity represents the limit of productivity under current conditions in a specific time, usually per hour, per day, per month or per year. An airport operator should make clear that the airport operates and serves demand below a *practical capacity*, where an acceptable level-of-service of e.g. four minutes average delay per daily flight, is guaranteed for the airport users. The practical or sustainable capacity should never be exceeded for longer periods.

As it can be seen in **Fig. 30**, the closer an airport operates towards its ultimate or "physical" throughput capacity, the stronger delays increase beyond an acceptable level of service, and eventually theoretically delays reach infinity, which means flights never leave the gate or wait an infinite time in the holding pattern in the airspace. Therefore, the arriving and landing aircraft have priority over departing aircraft, due to limited fuel reserves which allow waiting in the holding stack in the airspace only for a certain period of maybe 20 to 30 minutes maximum.

At congested airports, which are slot coordinated, the amount of hourly capacity must be declared by the airport operator (IATA 2010b). The declared capacity is the common denominator of all processes at an airport involved in serving passengers, aircraft, or cargo. Ideally the declared capacity is close to the practical capacity.

The temporary collapse of the airport system can be seen during weather events like snow, fog, heavy rain, and winds, when an airport's airside (runway) capacity could be greatly reduced due to poor visibility and lateral winds which exceed the safety limit. This can happen quite frequently in some regions of Europe.

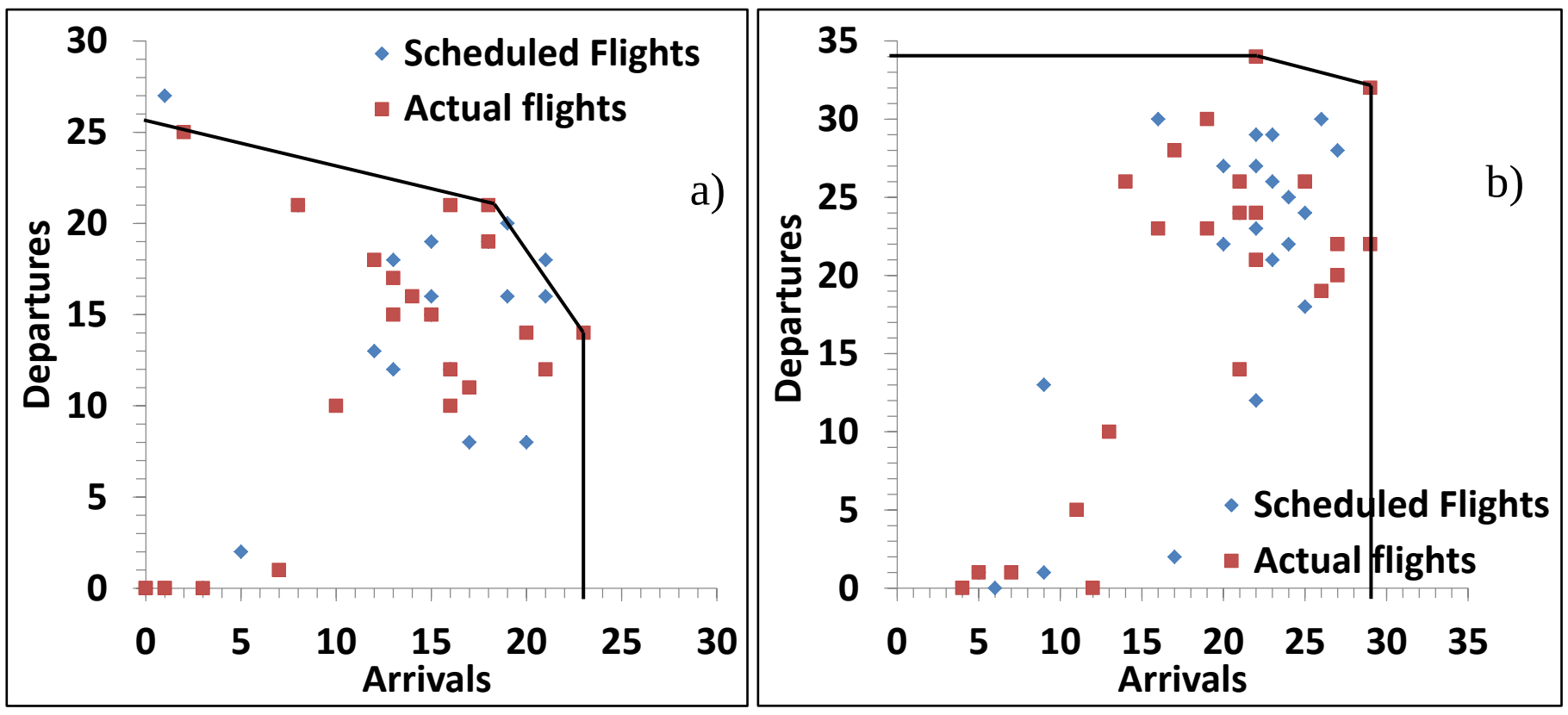

**Fig. 28.** Gilbo diagrams of single runway airports a) San Diego, USA, and b) London Gatwick on design peak day 2008 (Data from Flightstats.com)

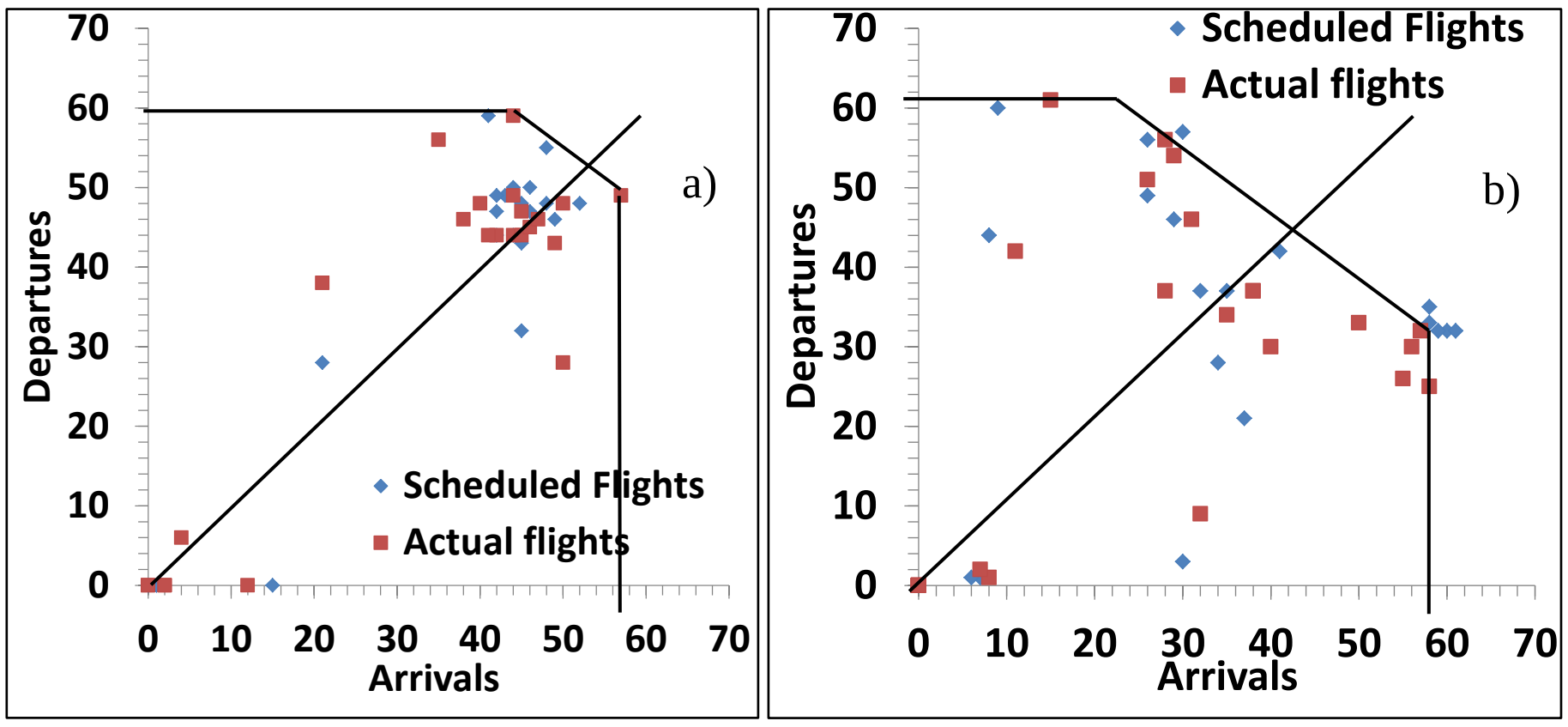

**Fig. 29.** Gilbo diagrams for independent parallel runway airports a) London Heathrow and b) Munich on design peak day 2008 (Data from Flightstats.com).

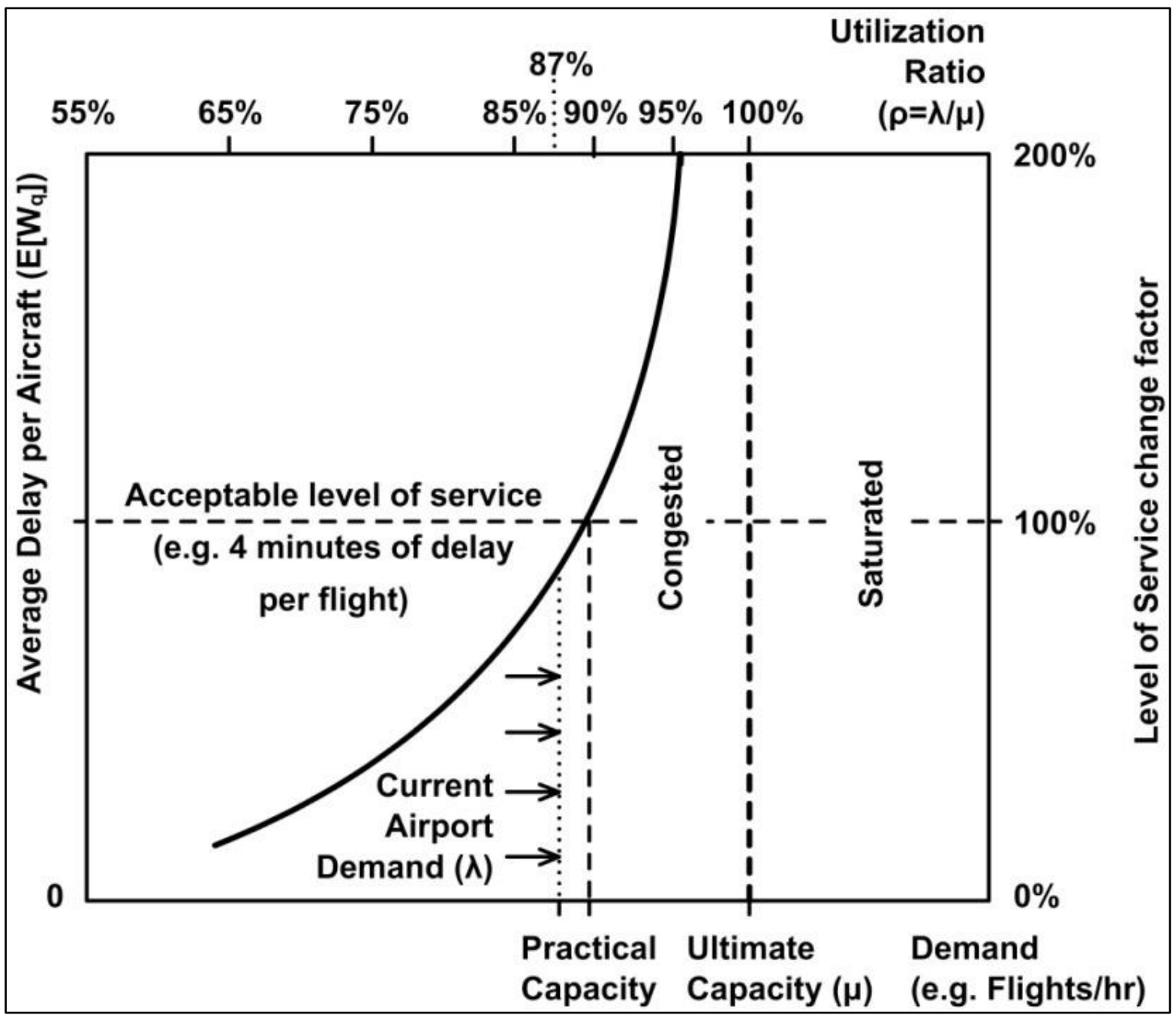

**Fig. 30.** Fundamental relationship between demand, capacity, and delay (Modified from: Horonjeff 2010, p. 488).

It is indeed always possible for demand to exceed capacity for short periods of time, due to fluctuations of demand at the airport. The situation becomes more critical when capacity is utilized more than 100% over a minimum of one hour and measurable waiting queues and delay develop (Horonjeff 2010).

As we can see from **Fig. 30**, it can make a huge difference in service quality, e.g. average delay per flight, when an airport operates at a capacity utilization of 65%, 75%, 85%, or even higher. Practical capacity usually serves as declared capacity for the slot coordinator and should never exceed 85-90% of the ultimate capacity during consecutive busy hours, otherwise the airport system becomes unstable and sensitive to changes in demand or available capacity, e.g. due to unscheduled flights, runway incursions, or weather.

The practical capacity for the maximum sustainable landing and departures at a given airport can be estimated by constructing the Capacity Envelope in the Gilbo Diagram of that airport (**Fig. 29**). With data of many operating hours this kind of diagram provides a precise picture of how many arrivals and departures are maximally possible under current conditions. In the case of the Gilbo diagrams of London-Heathrow and Munich airports (**Fig. 29**), the practical capacity is 100 flights per hour (50 arrivals and 50 departures per hour) for London and 82 flights per hour (41 arrivals and 41 departures simultaneously per hour) for Munich. The Gilbo diagram for Munich reveals that that airport achieves its best operational performance with a 64% (57 arrivals per hour) and 36% (32 departures per hour) share of arrivals to departures, resulting in a total of 89 hourly flights. The Gilbo diagrams can be modified to include the frequency of occurrence of each plotted point. This has been done by Kellner (2009) to derive so-called "density plots". The density plots can then be used to isolate outliers and to establish frequency thresholds, e.g. occurrences 95% of the time.

## 3.5 Traffic load diagrams

For a general overview of demand and capacity, the traffic load diagrams over time-of-day are interesting. The hourly demand can be monitored and quantified with these plots. Certainly, the diagrams offer even more information by including hourly capacity data, e.g. hourly capacity under Instrument Flight Rules (IFR) from the FAA methodology, declared capacity or available slots. **Fig. 31** gives an example of such a diagram for Paris' Charles-de-Gaule (CDG) airport on a design peak day 2009, where it can be observed that this airport has a high productivity of about 100 flights per hour and uses its slot capacity of about 110 Slots per hour very efficiently. In contrast the traffic at Charles-de-Gaule airport in 2008 was much higher, exceeding slot capacity and even exceeding the IFR capacity of 120 flights per hour.

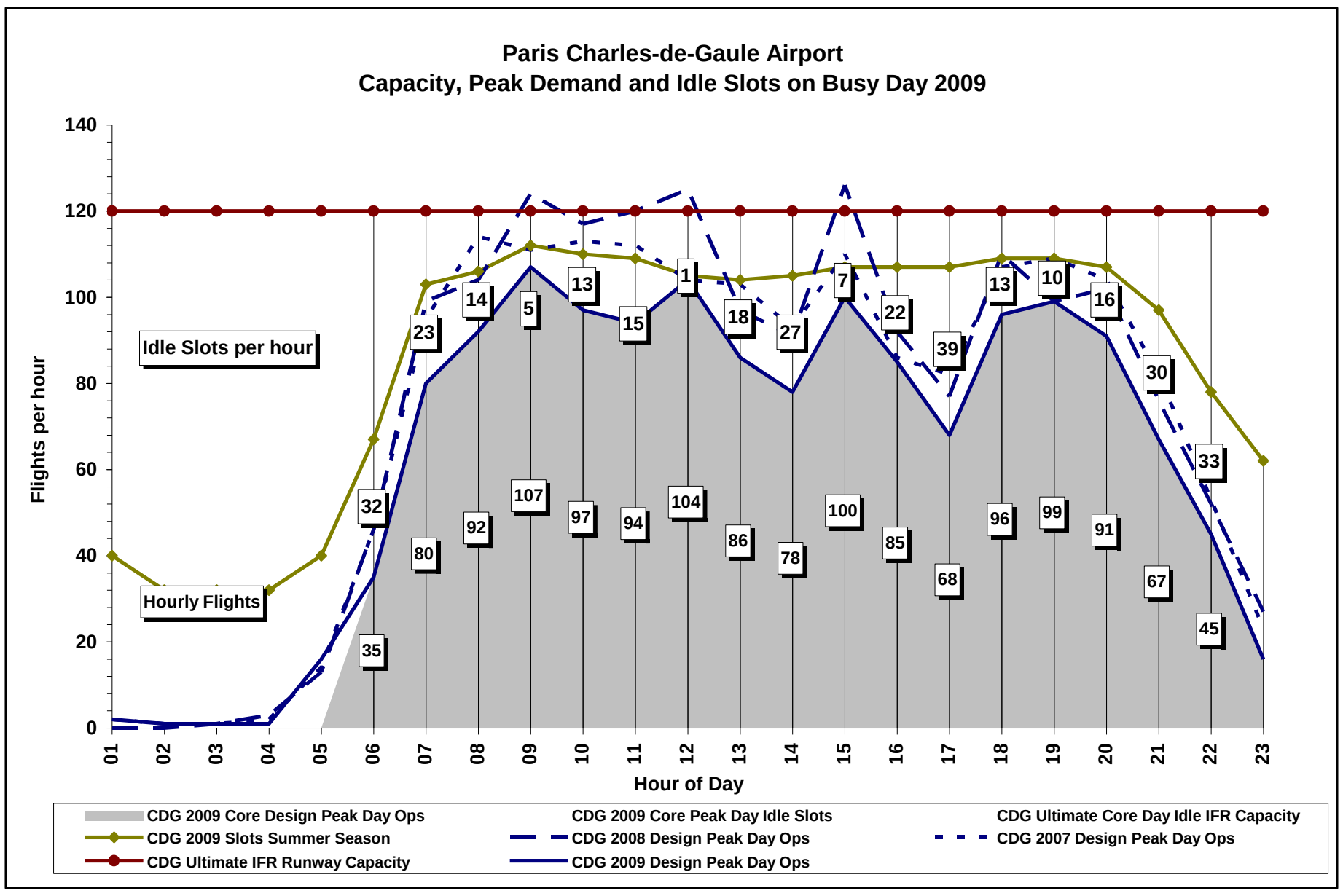

**Fig. 31.** Traffic load and capacity diagram for Paris CDG airport (Source: Bubalo 2010 with data from Flightstats.com)

The expression for the number of used slots and/or IFR capacity is *Capacity Utilization*, which is the quotient of demand divided by capacity. The annual capacity utilization is the annual operations divided by the annual service volume, and the hourly capacity utilization is the design peak hour demand divided by the available slots per hour or estimated IFR capacity. For a selection of capacity utilization figures please refer to the basic airport data **Table 11**.

### 3.6 Conclusion

As it can be shown for London-Heathrow and Gatwick the high productivity of both airports results in huge amounts of experienced delay and delay costs for the airport users. It is ongoing research that demonstrates how this externality of highly productive airports can be included in productivity and efficiency analyses to be able to make fair assumptions and comparisons of airports about service quality and externalities.

Another crucial externality of air transportation is *annoyance* caused by aircraft noise on the community in the vicinity of airports. Cumulative long-term aircraft noise is made responsible for all kinds of stress symptoms, which could even lead to decreased life expectancy due to cardiovascular diseases (Greiser 2007). Aircraft noise and community annoyance and the resulting health effects are currently being studied at the EU level. Another important research project (MIME-Market-based Impact Mitigation for the Environment) is currently looking into the feasibility of transforming noise/annoyance into tradable noise permits for a market-based approach to reducing aircraft noise.

The difficulty of including these externalities into recent econometric analyses for calculating airport productivity has been the non-linear relationship between number of operations and delay or noise. A second issue is the maximization of output (e.g. annual operations) and the simultaneous minimizing of “unwanted” output (e.g. noise and delays) in input/output analysis.

The approach to airport productivity analysis taken in this paper concentrates more on the fundamental understanding of externalities of air transportation than on conducting a holistic analysis.

An earlier capacity study of 20 European single-runway airports, which simulated airside airport operations and increasing growth of demand in SIMMOD (A popular airport *SIMulation and MODeling* software developed by the FAA), gave great insight into queuing problems at airports in general. In the SIMMOD study the relationship between capacity, demand and delay could be clearly observed (**Fig. 30**).

**Fig. 32** shows the trend of delay-per-flight from the results of all 20 simulated airports. To get an impression of the spread of simulated delays in each of the modeled 2850 operating flight-hours, an upper and lower boundary, in addition to the moving average and exponential trend for delays, has been plotted. Although heteroscedasticity can be observed in the plotted data, the trend lines provide enough evidence for an over-proportional increase of delays with an increasing number of operations per hour at single runway airports as has been presumed earlier from the theoretical relationship between demand, capacity, and delay depicted in **Fig. 30**.

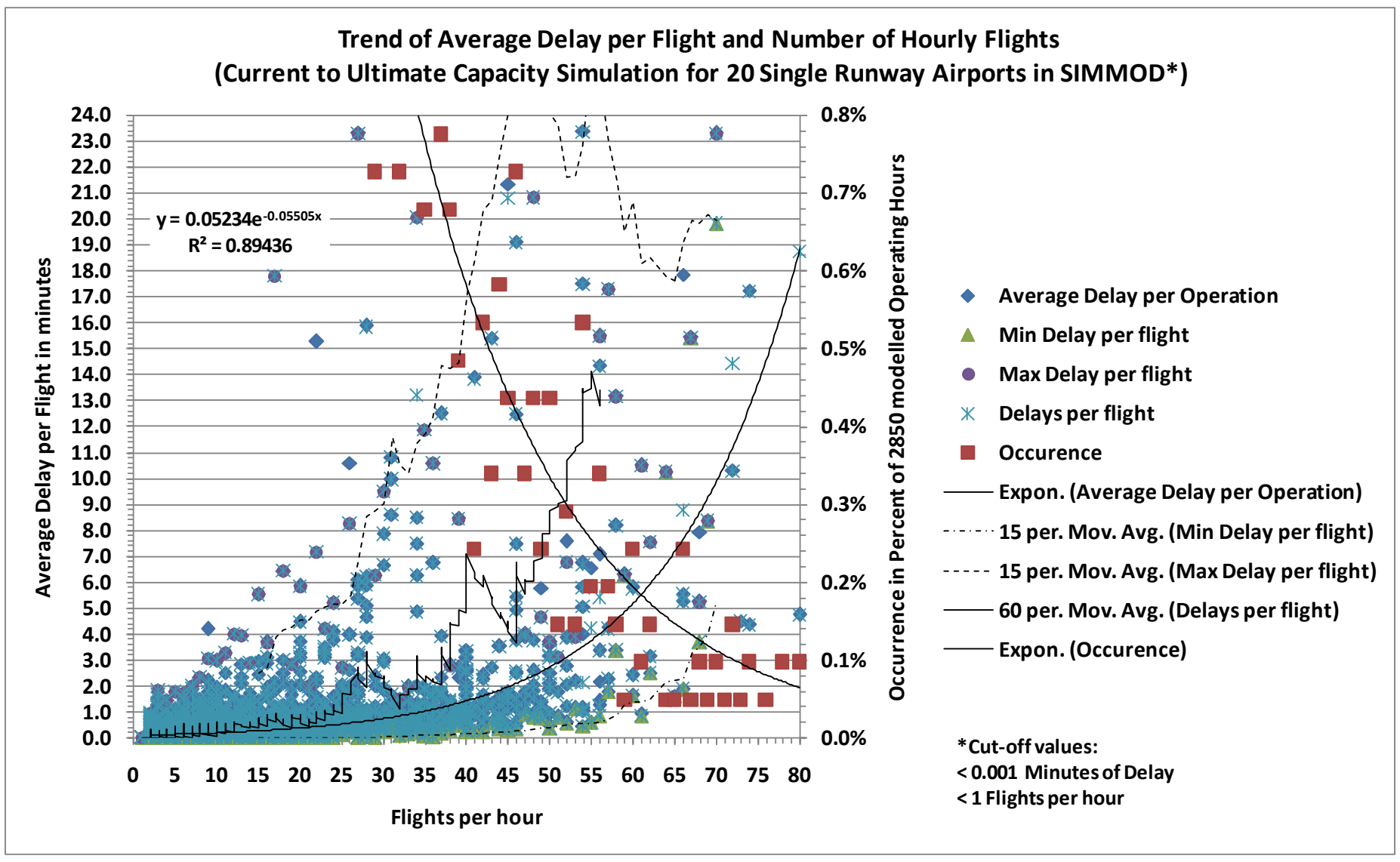

**Fig. 32.** Trend of Average Delay per Flight for European Single-runway airports (Source: Bubalo 2009)

The observed simulation data included traffic data of congested and soon-to-be-congested single-runway airports, like London-Gatwick, London-City, London-Stansted, Birmingham or Stuttgart airports (Bubalo 2009). By steadily increasing traffic in the simulations from a baseline scenario to an ultimate growth scenario, I wanted to calculate the ultimate capacity of each modeled airport. The results shown in **Fig. 32** give an indication of the maximum throughput of single-runway airports in Europe. It is interesting to note that from a relatively low level of demand of between 35 to 40 flights per hour, delays and random effects start increasing strongly. At 40 Flights per hour the lower boundary reveals 0.3 minute of average delay per flight, the moving average over a constant period of 60 data points reveals 6 minutes per flight and the upper boundary reveals 14 minutes per flight. At 50 flights per hour the situation becomes more critical, when we observe a minimum of 0.5 minutes, on average 8 minutes, and a maximum of over 24 minutes of delay.

It should be noted that the discontinuity of the trend curves could be due to airports being excluded from the simulation at high growth scenarios due to grid locks occurring from lack of apron capacity.

Ultimately it can safely be concluded that 60 flights per hour represent the ultimate capacity of single-runway airports in Europe, above which unpredictability is predominant. It is quite remarkable that an airport like London-Gatwick had a peak throughput of 61 flights per hour in 2008; under those circumstances there could be no margin-of-safety or latent capacity for further growth. One can only imagine the exceptionally high workload for air traffic control (ATC) staff and the extremely tight sequencing of aircraft for arrivals and departures at London's Gatwick airport.

Since the (hourly) practical capacity is about 85 to 90% of the ultimate capacity of 60 flights per hour, we derive a practical capacity of between 51 and 54 flights per hour on average for European single-runway airports. This figure is almost identical with the order-of-magnitude figures from the FAA methodology.

Further research will achieve greater insight into simulating and modeling of even more complex airport systems with more than one runway. A study on parallel-runway airports is currently in progress.

Noise and $CO_2$-emissions will be included in future capacity studies, as soon as new results and legislation in these fields is published. A promising approach for quantifying noise can be found using the "Enhanced Quota Count (*EQC*) System" (Figlar and Gjestland 2009), which enables the transformation of noise measures into tradable units and permits, by applying a continuous function adopted from the Quota Count (*QC*) system. Such *QC* system is already in place at one of the most congested European hub airports – London-Heathrow.

The advantage of calculating the *EQC* is that it is based on aircraft certification noise levels, which must be provided to authorities in the aircraft and aircraft engine certification process. This noise is measured during the take-off and landing phase and for overflights. The measure is called *effective perceived noise* and is expressed in decibels (*EPNdB*).

According to Figlar and Gjestland (2009) the *EQC* is calculated by *linearizing* the *EPNdB*, because noise is measured on a logarithmic scale, i.e. it increases exponentially, and cannot therefore simply be added or averaged. Like the current technique how to calculate the *QC* Figlar and Gjestland (2009) use a penalty factor *p* for time-of-day[1] and a normalization constant of 0.25 (at EBNdB of 84dB). The formulas to calculate *EQC*s are the following:

$$EQC_{take-off} = 0.25 * 10^{0.1(EPNdB_{take-off} - 84 + p)} \quad (1)$$

$$EQC_{landing} = 0.25 * 10^{0.1(EPNdB_{landing} - 84 + p)} \quad (2)$$

Evaluation of the mitigation of carbon emissions from aircraft fuel burn with the ETS must be analyzed further as additional information becomes available after the first trading round.

[1] According to the European Environment Agency (EEA) we find the following definitions for different daytimes: *Day* is the time between 7 and 19 and receives no penalty, *evening* is the time from 19 to 23 and receives a penalty of 5dB, and *night* is the time between 23 and 7 and receives a penalty of 10dB. [https://www.eea.europa.eu/help/glossary/eea-glossary/lden]

**Table 11.** Basic data of selected European airports

| | | | | | Annual Demand | | Annual Capacity | Capacity Utilization | | Peak Hourly Demand | | Hourly Capacity Utilization | |
|---|---|---|---|---|---|---|---|---|---|---|---|---|---|
| | | | | | 2007 | 2007 | 2007 | 2007 | 2009 | 2008 | 2009 | 2008 | 2009 |
| **Group** | **Airport** | **IATA** | **No of Runways** | **FAA Runway config. No.** | **Annual Passengers (in millions)** | **Annual Flights** | **Annual Service Volume** | **Annual Capacity Utilization** | **Slots per hour Summer Season** | **Flights per hour** | **Flights per hour** | **Slot Utilization** | **Slot Utilization** |
| 3 | Paris Charles-de-Gaule | CDG | 4 | 8 | 59.55 | 569,281 | 675,000 | 84% | 105 | 126 | 107 | 120% | 102% |
| 3 | Madrid-Barajas | MAD | 4 | 8 | 51.40 | 470,315 | 565,000 | 83% | 100 | 112 | 112 | 112% | 112% |
| 3 | Amsterdam | AMS | 6 | 8 | 47.85 | 443,677 | 635,000 | 70% | 108 | 111 | 106 | 103% | 98% |
| 2 | Frankfurt/Main | FRA | 3 | 16 | 54.50 | 486,195 | 355,000 | 137% | 83 | 89 | 87 | 107% | 105% |
| 2 | London-Heathrow | LHR | 2 | 4 | 68.28 | 475,786 | 370,000 | 129% | 86 | 103 | 90 | 120% | 105% |
| 2 | Munich | MUC | 2 | 4 | 34.07 | 409,654 | 315,000 | 130% | 90 | 93 | 92 | 103% | 102% |
| 2 | Barcelona | BCN | 3 | 12 | 32.81 | 339,020 | 315,000 | 108% | 60 | 80 | 74 | 133% | 123% |
| 2 | Rome-Fiumiciano | FCO | 3 | 12 | 33.62 | 328,213 | 315,000 | 104% | 90 | 103 | 100 | 114% | 111% |
| 2 | Copenhagen | CPH | 3 | 12 | 21.4 | 250,170 | 315,000 | 79% | 83 | 70 | 62 | 84% | 75% |
| 2 | Brussels | BRU | 3 | 12 | 17.93 | 240,341 | 370,000 | 65% | 74 | 71 | 67 | 96% | 91% |
| 2 | Paris-Orly | ORY | 3 | 12 | 26.42 | 238,384 | 315,000 | 76% | 70 | 63 | 60 | 90% | 86% |
| 2 | Oslo | OSL | 2 | 4 | 19.04 | 226,221 | 315,000 | 72% | 60 | 60 | 49 | 100% | 82% |
| 2 | Zurich | ZRH | 3 | 10 | 20.81 | 223,707 | 340,000 | 66% | 66 | 57 | 57 | 86% | 86% |
| 2 | Dusseldorf | DUS | 2 | 2 | 17.85 | 223,410 | 285,000 | 78% | 47 | 51 | 58 | 109% | 123% |
| 2 | Manchester | MAN | 2 | 2 | 22.33 | 206,498 | 285,000 | 72% | 46 | 69 | 51 | 150% | 111% |
| 2 | Istanbul | IST | 3 | 16 | 25.49 | 206,188 | 300,000 | 69% | 40 | 44 | 47 | 110% | 118% |
| 2 | Stockholm-Arlanda | ARN | 3 | 12 | 18.01 | 205,251 | 315,000 | 65% | 80 | 61 | 50 | 76% | 63% |

**Table 11.** Basic data of selected European airports (cont.)

| | | | | | Annual Demand | | Annual Capacity | Capacity Utilization | | Peak Hourly Demand | | Hourly Capacity Utilization | |
|---|---|---|---|---|---|---|---|---|---|---|---|---|---|
| | | | | | 2007 | 2007 | 2007 | 2007 | 2009 | 2008 | 2009 | 2008 | 2009 |
| **Group** | **Airport** | **IATA** | **No of Runways** | **FAA Runway config. No.** | **Annual Passengers (in millions)** | **Annual Flights** | **Annual Service Volume** | **Annual Capacity Utilization** | **Slots per hour Summer Season** | **Flights per hour** | **Flights per hour** | **Slot Utilization** | **Slot Utilization** |
| 2 | Palma de Mallorca | PMI | 2 | 4 | 23.10 | 184,605 | 315,000 | 59% | 60 | 44 | 45 | 73% | 75% |
| 2 | Helsinki | HEL | 3 | 12 | 13.10 | 174,751 | 315,000 | 55% | 80 | 41 | 44 | 51% | 55% |
| 2 | Nice | NCE | 2 | 2 | 10.38 | 173,584 | 260,000 | 67% | 50 | 52 | 48 | 104% | 96% |
| 2 | Berlin-Tegel | TXL | 2 | 2 | 13.37 | 145,451 | 285,000 | 51% | 52 | 42 | 42 | 81% | 81% |
| 2 | Lyon | LYS | 2 | 2 | 7.19 | 132,076 | 285,000 | 46% | 51 | 44 | 43 | 86% | 84% |
| 1 | London-Gatwick | LGW | 2 | 1 | 35.27 | 258,917 | 240,000 | 108% | 46 | 56 | 49 | 122% | 107% |
| 1 | Vienna | VIE | 2 | 14 | 18.77 | 251,216 | 225,000 | 112% | 66 | 67 | 59 | 102% | 89% |
| 1 | Dublin | DUB | 3 | 14 | 23.31 | 200,891 | 225,000 | 89% | 46 | 44 | 43 | 96% | 93% |
| 1 | London-Stansted | STN | 1 | 1 | 23.80 | 191,520 | 210,000 | 91% | 38 | 47 | 38 | 124% | 100% |
| 1 | Prague | PRG | 2 | 9 | 12.40 | 164,055 | 225,000 | 73% | 46 | 57 | 39 | 124% | 85% |
| 1 | Hamburg | HAM | 2 | 9 | 12.85 | 151,752 | 225,000 | 67% | 53 | 44 | 38 | 83% | 72% |
| 1 | Warsaw | WAW | 2 | 9 | 9.29 | 147,985 | 225,000 | 66% | 34 | 32 | 26 | 94% | 76% |
| 1 | Lisbon | LIS | 1 | 1 | 13.52 | 141,905 | 210,000 | 68% | 36 | 37 | 34 | 103% | 94% |
| 1 | Stuttgart | STR | 1 | 1 | 10.35 | 139,757 | 210,000 | 67% | 42 | 41 | 35 | 98% | 83% |
| 1 | Birmingham | BHX | 1 | 1 | 9.32 | 104,480 | 210,000 | 50% | 40 | 29 | 28 | 73% | 70% |
| 1 | London-City | LCY | 1 | 1 | 2.91 | 77,274 | 210,000 | 37% | 24 | 36 | 36 | 150% | 150% |
| | **Mean** | | 2 | | 24.55 | 247,955 | 310,909 | 79% | 62 | 63 | 58 | 102% | 94% |

# Part II: Air Transport Economics

Part II of this dissertation highlights some economic problems in air transportation that have not yet been discussed in the prior chapters. There exists a connection between the management of airports, of airlines and of air space, which are sometimes difficult to analyze separately. One of the main drivers of air transport demand development is competition between airlines. From a passengers' perspective competition brings benefits regarding service quality and ticket price. Increasing competition can not only be found between airlines, but furthermore between airports, especially between global hubs with a large share of transfer and transit traffic, and recently even between air navigation service providers (ANSPs). The privatization of air transport organizations plays a large role in increasing competition, because organizations are now forced to operate profitably, are expected to provide excellence service quality, and are competing to attract new customers. In some situations, it is necessary that the state subsidizes air transport, when demand in some regions is too low for any company to operate routes self-sufficiently. The airports in such regions have little degrees-of-freedom to make profits from scheduled passenger air traffic alone, which leads some airports to investigate alternative sources of income through air cargo or general aviation traffic.

# 4 Low-cost carrier competition and airline service quality in Europe*

Branko Bubalo†
University of Hamburg

Alberto A. Gaggero‡
University of Pavia

**Abstract.** The authors are investigating whether a higher presence and more efficient operations of low-cost carriers (LCCs) can increase the service quality in terms of on-time performance of all the flights landing at an airport. We sample 100 European airports located in 76 metropolitan areas of diverse sizes in 19 countries on both a daily and flight-by-flight basis during the period from April 2011 to December 2012. We construct a panel dataset at the flight code level comprising about 3.5 million observations. We find that LCCs contribute to a reduction of delays for airlines and flights landing at the observed airport. From the customers' point of view and taking into consideration the level of service, we conclude that the presence of LCCs represents a positive externality for an airport. Airport management may therefore consider the proactive increase of LCCs market share in their long-term business strategies.

## 4.1 Introduction

Passengers, airlines, airport management and industry experts all consider flight delay as one of the most important measures of service quality. For instance, several reports in the aviation industry document the most punctual airlines, ranking them in descending order by an on-time performance indicator, such as the percentage of flights delayed less than 15 minutes. Airlines acknowledge the importance of being punctual and are keen to announce any improvement in their performance score.[1] They regularly implement employee bonus programs to reward achieved on-time performance within the organization (Forbes et al., 2011). In a previous paper, Forbes (2008) finds that

* We thank two anonymous referees, Marco Alderighi, Volodymyr Bilotkach, Jürgen Müller and participants of the Air Transport Research Society (ATRS) world conference 2013 in Bergamo, Italy, for helpful comments. We are grateful to Charles "Chuck" H. Eypper for careful proof-reading. This paper was initiated while Gaggero was visiting the Berlin School of Economics and Law in spring 2011. Gaggero is grateful to Professor Müller for his kind hospitality. This study was independently and privately funded by the authors. All errors are ours.

† Institute for Information Systems, University of Hamburg, Von-Melle-Park 5, 20146 Hamburg, Germany. Email address: branko.bubalo@googlemail.com. Corresponding author.

‡ Department of Economics and Management, University of Pavia, Via S. Felice 5, 27100 Pavia, Italy. Email address: alberto.gaggero@unipv.it.

[1] Punctual airlines advertise their on-time performance in their promotional campaigns as a marketing tool to retain or attract new customers. Ryanair even plays a punctuality jingle inside the aircraft as soon as the airplane lands on-time.

consumers complain more often when they fail to receive the higher quality they expect. Suzuki (2000) observes decreasing customer retention and shows that passengers after they have personally experienced delays are more likely to switch airlines for subsequent flights. Moreover, since customers consider delay as a form of product quality, decreasing on-time flight performance has a negative influence on airline fares (see Prince and Simon, 2014).

The impact of airline delay can have detrimental effects for both passengers and airlines. Due to delay passengers may miss a connecting flight, a business meeting, or a family celebration, bearing the consequences in terms of wasted time, foregoing earnings or, more generally, opportunity costs, negative utility, annoyance and dissatisfaction. In addition to worsening their on-time performance record, airlines may incur additional operating and compensation costs, since they must reroute or refund passengers, offer refreshments at best and accommodation, transportation, or cash payments at worst.

Indeed, who should bear the costs of delay is still being debated. Under various circumstances, the responsibility or cause for the delay may remain undefined, as the reasons for poor on-time performance are manifold.[1] In Europe, regulation (EC) No 261/2004 provides the legal framework for the compensation of and assistance to passengers in the event of being denied boarding, cancellation or long flight delays. However, some airlines (especially low-cost carriers, LCCs) lobby for less regulation in order to transfer part of the risk and the implied costs of delay to the final consumers in exchange for lower fares.[2] For these reasons, any policy or action apt to reduce flight delay should be promoted.

In this paper we investigate whether a different composition of competing carrier types - full-service carriers (FSCs) versus LCCs - serving an airport, might impact the on-time performance of all the flights landing at such an airport. More specifically, we argue that a higher presence of LCCs at the airport of origin may reduce the average flight delay. This is because the “no-frills” and faster aircraft turn-around policy pursued by LCCs should mitigate airport congestion and the “knock-on” effect of flight delays, which could otherwise propagate throughout the network. Much faster and more precise operations conducted by LCCs at an airport quickly free up aircraft parking positions and thereby reduce the waiting period for other flights.[3]

We investigate whether the presence of LCCs at the airport of origin can constitute a positive externality by reducing airline delay using a sample of flights operated within

[1] Airport flight protocols authorized by the International Air Transport Association (IATA) identify at least 100 different codes for causes of delay.

[2] For instance, in April 2011 Ryanair introduced an extra charge per passenger to cover the costs of flight cancellations and delays not under the direct responsibility of the airline, such as weather conditions or national strikes. Zhang and Zhang (2006) show that when carriers have market power, they can internalize congestion costs by setting a higher ticket price, so that passengers eventually bear the costs.

[3] Around 47% of delays are due to airline-related operations at airports, e.g. aircraft turnaround operations, while the remaining delays mainly stem from air traffic control, weather, and airport capacity constraints (Eurocontrol, 2001).

Europe, where the phenomenon of LCCs has emerged only in recent years. Our dataset covers 23,402 flight codes, 3,270 routes and 100 airports for a total of 3,486,376 observations. Our sample period includes daily observations from April 16th, 2011 to December 23rd, 2012. Applying panel data fixed effect techniques and controlling for the most cited factors affecting airline on-time performance, we find that a stronger presence of LCCs at the airport of origin has a positive and statistically significant effect on the on-time performance of the flights.

This effect is evident in two examples of city-pair connections serving a comparable catchment area: London-Madrid and Paris-Barcelona. More specifically, the route originating at London-Heathrow (100% served by FSCs) to Madrid-Barajas has an average delay of 17.3 min per flight, while on the route from London-Stansted (88% served by LCCs) to Madrid-Barajas the average delay is 12.4 min. On the connection Paris-Charles de Gaulle (89% served by FSCs) to Barcelona-El Prat we observed an average delay of 4.8 minutes per flight, compared to an average *early arrival* of 5.5 minutes on the route Paris-Beauvais (100% served by LCCs) to Barcelona-El Prat.

Our main objective is to present a detailed study on the topic of airline delay based on a comprehensive and unique sample of European airports; previous empirical examinations of airline delay focus almost exclusively on U.S. airports[1]. Indeed, the richness of our dataset allows us to cover both the peak and the off-peak seasons, thereby offering a complete year-round picture of airline on-time performance in Europe[2].

The choice of working with European data is further motivated by the fact that European airspace is characterized by congestion problems and insufficient airport capacity (Raffarin, 2004; Santos and Robin, 2010). As air traffic within Europe grows because of the enlargement to the east and because of a deeper integration among countries, flight delay (or congestion-related delay troubling all modes of transport) is becoming an important economic and policy issue. One of the main priorities under the European Union's research program "Horizon 2020" is, in fact, to make transport systems seamless through better mobility and less congestion.

The next section reviews the literature. This is followed in Section 4.3 by a description of the data and a brief analysis. The econometric model is presented in Section 4.4, while the results of the study are discussed in Section 4.5. Section 4.6 summarizes our findings.

## 4.2 Literature review

A large number of applied works studying on-time performance is based on data from the United States, typically from the U.S. Bureau of Transportation Statistics. Mazzeo (2003), for example, sampled 50 U.S. airports in January, April and July 2000 to study the effect of airline market concentration on flight delays. He finds that on monopoly routes the prevalence and duration of flight delays is significantly greater. Although meteorological conditions, congestion and scheduling are the main causes of delay, he is able to show by controlling for such factors that increasing competition on the route level is correlated with better on-time performance. A more recent study by Greenfield (2014) comes to similar conclusions by analyzing the top 100 airports in the U.S. serving most arrivals and departures. Based on a limited number of observations he finds that an increase in market concentration is correlated with an increase in delay. Rupp et al. (2006), on the other hand, arrives at the opposite conclusion, namely that more competition on U.S. routes worsens on-time performance. The authors reach this conclusion by using a greater degree of schedule differentiation occurring on the less competitive routes. Similar results are obtained by Prince and Simon (2014), who find a worsening on-time performance of incumbents at U.S. airports post entry or even in cases of an entry threat of an LCC, such as Southwest Airlines. They explain this trend by the incumbents' efforts to reduce costs, such as the utilization of stand-by crews or aircraft, to compete on price prior to entry by an LCC, which leads to a reduction of service quality.

[1] See the works by Mazzeo (2003), Forbes (2008), Rupp and Sayanak (2008), Rupp (2009), Forbes and Lederman (2010), Forbes et al. (2011) Ater (2012), Prince and Simon (2015).

[2] The lack of a comprehensive dataset on flight delays and the available capacity at European airports has hindered any empirical study on the European aviation market (Bel & Fageda, 2010; Santos & Robin, 2010).

We challenge the belief that increased competition reduces service quality and demonstrate that more competition by LCCs on certain city-pair markets would not only lead to a reduction in average fares, but moreover increase service quality. We believe that, as fares of FSCs and LCCs on certain (competing) routes converge, the main competitive advantage among carriers for attracting passengers is the quality dimension, measured in on-time performance (Rupp and Sayanak, 2008). This belief is supported by solid theory (Tirole, 1988) which predicts that in oligopolistic markets, where prices converge at marginal cost levels, the substitutability for similar products occurs through vertical differentiation such as product quality (see, for example, Forbes and Lederman 2010). If one competitor can offer high quality products in a particular market, the long-term equilibrium is, all else equal, established at a point of low price and maximum (technically and economically feasible) quality.

Mayer and Sinai (2003) suggest two potential causes of flight delays: hub and spoke policy and congestion externality. The former spurs hub carriers to schedule a large variety of potential connections and destinations within a relatively short time span adding convenience for, say, business travelers. Congestion externality is due to airports allowing unrestricted landings and take-offs by airlines and ignoring the fact that their marginal scheduled traffic during peak periods increases queueing and travel time for other airlines.

However, in Europe large hub airports are capacity-controlled in terms of the maximum number of available slots per time period (such as per hour) by a coordinator, and such airports participate in the bi-annual scheduling conferences organized by the International Air Transport Association (IATA). Therefore, similar to the findings of Mayer and Sinai (2003), hubbing strategies could be a primary driver of air traffic congestion in Europe as well. LCCs, to the contrary, have a point-to-point network strategy, where passengers' itineraries through one or more connecting airports do not play a vital role in their business model; thus, customer complaints about missed connecting flights are presumably negligible. LCCs face passenger compensation (and additional operating costs) in cases of long delay and, for, say, the consequent diversion of a flight to an alternate destination. FSCs have higher compensation efforts, due to the larger percentage of connecting (transit or transfer) passengers.

Rupp (2009) conducted an empirical study to determine to what degree if at all U.S. carriers internalize congestion costs using ten years of on-time performance data.[1] He reaches different conclusions based on the perspective adopted to examine flight delays. From the carrier's perspective, shorter excess travel time occurs at highly concentrated airports, suggesting that carriers internalize airport congestion. From the passenger's perspective, however, he finds that departure and arrival delays are more likely at highly concentrated airports, where airlines do not internalize passenger delay costs and behave as atomistic competitors. In another analysis of congestion patterns at U.S. hub airports, Ater (2012) investigated the relationship between airport concentration and the waves of departing and arriving flights, known as flight banks. His findings suggest that any attempt to reduce congestion externality may have a limited impact on congestion at highly concentrated airports, because congestion is already internalized by hub-airlines.

To the best of our knowledge, the number of studies on flight delays within Europe is very limited. Raffarin (2004) studied airport congestion in Europe from a theoretical perspective. She suggests an innovative approach to Air Traffic Control pricing to reduce European airspace congestion problems and the corresponding flight delays.

Santos and Robin (2010) used a sample of the main European airports during the period between 2000 and 2004 to identify the causes of flight delays. Among other things, they find that airport concentration is a significant variable in explaining flight delay. It is noteworthy that the authors, however, are not able to measure delay on a flight-by-flight basis, but are compelled to use an average delay given by the ratio between the aggregated minutes of delay by an airline in a given airport and the number of flights of the observed airlines. In the present study, we are more accurate in measuring airline on-time performance since we use scheduled and actual non-stop flight times (reported accurate to the minute) to determine the delay. Indeed, in Europe the lack of accurate flight schedule data including flight delay information is probably one of the reasons for the limited number of empirical works in this field (see Bel and Fageda, 2010; Santos and Robin, 2010).

Considering the impact of LCCs on flight delays, empirical studies are not numerous. Dresner et al. (1996) analyzed the impact of low-cost carrier Southwest Airlines on the fares on certain U.S. city-pair markets. However, they did not measure differences in service quality. Rupp and Sayanak (2008) and Prince and Simon (2014) are two other noteworthy studies in this regard. However, these works focus on the competitive situation regarding LCCs in the U.S. air travel market only.

With this paper we contribute to the literature on flight delays in a number of ways. First, as mentioned above, our empirical results are based on a data sample covering a set of European airports not examined by previous literature. For this reason, it is interesting to see how the findings of our paper compare with similar studies done using

1 For a theoretical discussion on airport congestion costs see the works by Brueckner (2005), Brueckner (2009) and Brueckner and Van-Dender (2008).

U.S. data (Mazzeo, 2003; Forbes, 2008; Rupp, 2009; Ater, 2012; Prince and Simon, 2014). Second, we estimate the effect of airport competition on flight delay and distinguish between FSCs and LCCs: to the best of our knowledge, this is the first study which makes this distinction for European airports[1]. Third, we are able to pin down with a high degree of precision meteorological conditions and their influence on weather-related delay at the time of take-off or landing. Contrary to previous empirical works, which use daily averages of the weather status (Mazzeo, 2003; Forbes and Lederman, 2010), we use weather reports published every 30 minutes. Finally, the long-time period of our sample, spanning from mid-April 2011 to the end of December 2012, covers the full extent of the peak and off-peak seasons. This distinction is not always made in the empirical literature, which often tends to sample random months in a year (Mazzeo, 2003).

## 4.3 Data

The sample used in this analysis is a combination of flight schedule and meteorological data. The data were collected from the Internet[2] using a “web crawler”, which automatically seeks out web pages and records the information requested.

We instructed our algorithm to collect flight records for the airports in our sample over a period of twenty months[3]. The data are downloaded daily with a one-day time lag to record the final status of each flight. These circa 12 million observations consist of the origin, destination, aircraft type, flight number, scheduled and actual arrival (and departure) times of all airlines landing (and departing) at those airports. We then built a reduced dataset of around 3.5 million flights scheduled on routes among the 100 airports in our sample. Based on this information we infer the delay of each flight by the difference between the scheduled and actual times of departure or arrival.

Additionally, we constructed variables measuring the market share of each airline both at airport and at route level. Additional information on flight status ("landed", "cancelled", "diverted" or "re-directed") is also gathered, however, we only include landed flights on the scheduled routes. We also disregard cargo flights. In order to avoid the inclusion of extraordinary delay or early arrivals, which may represent outlying observations, we exclude from the sample the first and the last 0.2 percentiles of the distribution of delay.

Information on the low-cost status of an airline is obtained from European Low Fares Airline Association (ELFAA). Thus, we define a carrier as an LCC if it is a member of ELFAA during the sampling period.[4]

[1] Rupp and Sayanak (2008) are the first who make this distinction regarding U.S. airports.

[2] To a large degree the data stems from publicly available information from the websites flightstats.com and wunderground.com.

[3] Our sample comprises the main European airports and extends the airport list in Bel and Fageda (2010). See the Appendix for the full list of the airports considered in our analysis.

[4] The LCCs of our sample are Blue Air, EasyJet, Flybe, Jet2, Norwegian Air Shuttle, Ryanair, Sverigeflyg, Transavia.com, Vueling and Wizz Air. We do not observe any airline entering or exiting ELFAA during the sample period.

Data on the meteorological conditions are obtained from the Meteorological Terminal Aviation Routine Weather Reports (METARs), which are commonly used by pilots in their flight preparation procedures. Weather reports are typically published on a half-hourly basis at airports or nearby weather stations and consist of general weather condition, temperature, wind direction and speed, humidity, and pressure. Similarly, to collecting airport arrival and departure times, we use a second web crawler to build the weather database. The very frequent weather sampling by METAR allows observing the meteorological situation at the point of origin or destination when the airplane is scheduled to take-off or to land.

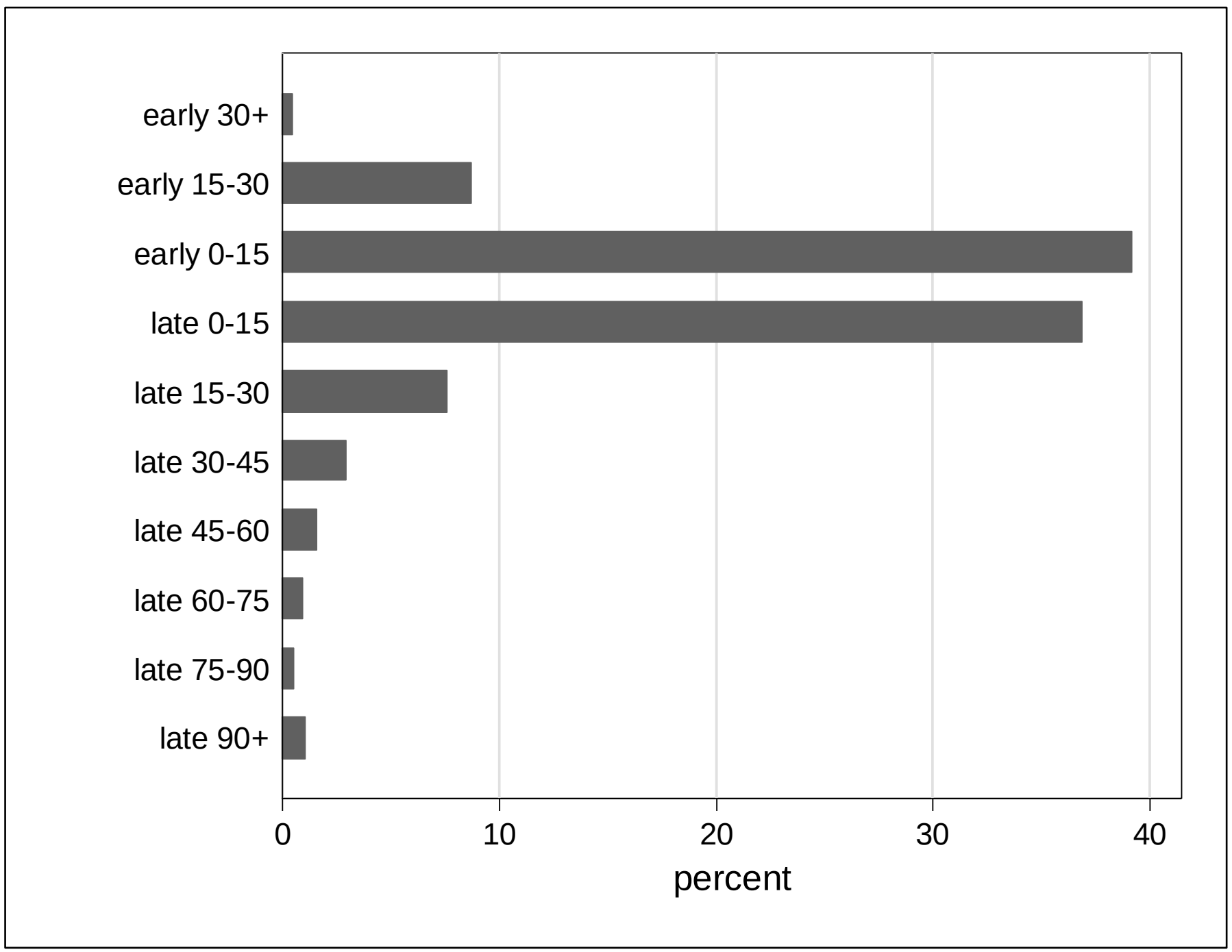

**Fig. 33.** Minutes of late/early arrivals in the sample.

With our data, it is possible to know the scheduled flying time for the flights that originated or landed at one of the 100 sampled airports. In other words, we can identify with a high degree of accuracy the weather conditions at the origin or destination airport of a flight at the time when such conditions are more likely to influence the delay. Previous empirical literature, e.g. Mazzeo (2003), uses instead an average weather status for the whole day. We deem the accuracy of meteorological information relevant because within the same day the weather conditions, such as precipitation, wind speed and direction, often change so rapidly that at certain times it can hinder the flight departure or landing, while at other times of day the weather becomes totally innocuous.[1]

[1] Think for instance of fog in the morning which clears up during the day: morning flights could

A deeper investigation into the composition of delay is given in **Fig. 33**, which displays the frequency of observations at 15-minute intervals for late or early arrival flights in the sample. Interestingly a substantial number of flights are recorded as arriving early, about 10% of the sample for early arrivals within the range of 15 to 30 minutes, and almost 40% for flights landing up to 15 minutes early. This picture confirms the suspicion that airlines allow some time buffer within each flight to be able to catch up on any delays and thus reduce the number of declared delays.

Finally, **Table 12** categorizes the 3,270 routes of our sample according to the number and type of carriers serving each route. Thus, for instance, the first row means that our sample comprises 2,098 monopoly routes, 1,527 of them are operated by an LCC and 571 by a FSC.

Overall, the table shows that there is a considerable number of routes which are not served by LCCs, about half of the sample. This feature implies a considerable amount of data variability in terms of LCCs presence, which represents a key variable of our econometric model presented in the next section.

**Table 12.** Routes by different type of carriers.

| Number of carriers serving the route | Low-cost carriers | Full-service carriers | All type of carriers |
|---|---|---|---|
| 1 | 1,527 | 571 | 2,098 |
| 2 | 106 | 702 | 808 |
| 3 | 7 | 241 | 248 |
| 4 | 0 | 89 | 89 |
| 5 | 0 | 22 | 22 |
| 6 | 0 | 5 | 5 |

## 4.4 Econometric model

From equation (1) we model flight delay as a function of LCCs presence at the airport, route and airport competition, airport congestion, weather condition, seasonality, airport-airline-route-flight fixed effects,

$$Delay_{fdt} = \alpha LCC_{ot} + \beta' Z_{fodt} + \rho_{fod} + \tau_t + \varepsilon_{fodt} \qquad (1)$$

where *f* is the flight code, *o* the airport of origin, *d* the airport of destination and *t* the day of observation.[1] The dependent variable $Delay_{fd}$ is the minutes of arrival delay of flight *f* leaving airport *o* and landing at *d*. The variable $LCC_o$ represents our variable of be severely affected (possibly cancelled or diverted), while afternoon flights could be completely unaffected.

[1] Since the pair *od* already defines the route, to lighten the notation we drop the route subscript *r* from the equation. Also, as a flight code is carrier specific, we drop the carrier subscript *c*.

interest, which is given by ratio of the flights operated by LCCs over the total number of flights at the airport of origin *(LCC* $share_o$). As a robustness check on our procedure, we substitute in place of LCC $share_o$, the ratio of the number of LCCs over the total number of airlines serving the airport of origin *(LCCs/Tot* $Carriers_o$). We expect a negative sign on α to sustain our theory that a higher presence of LCCs at the airport of origin can represent a positive externality by reducing the delay.

The vector $Z_{fodt}$ collects various controls. We include the LCCs market share at the airport of destination (*LCC* $share_d$), calculated in the same way as *LCC* $share_o$ and, in the robustness check *LCCs/Tot* $Carriers_d$, constructed in the same fashion as *LCCs/Tot* $Carriers_o$. Generally speaking, a higher presence of LCCs at the arrival airport might also be beneficial to the other airlines by reducing the queuing time for landing. However, since the knock-on effect on delay is usually stronger at the departure airport, the effect of LCCs on delay should be more evident at the airport of origin rather than at the airport of destination.

We consider the minutes of departure delay ($Delay_{fo}$) to take into account that a plane may land behind schedule because of a late departure from its airport of origin. Another relevant determinant of on-time performance is represented by the weather, both at the airport of origin and at the airport of destination, because bad weather conditions may require more operations for take-off and landing or may restrict the use of the runway(s). To control for influence of the weather on arrival delay we fix the clear-sky condition as reference category and include a set of dummy variables, one for each of the remaining weather conditions (cloudy, foggy, hazy, rainy, snowy and thundery) both at the airport of origin and destination. Thus, the estimated coefficients on the reported dummies measure the minutes of early (if negative) or late (if positive) arrival due to the observed weather condition with respect to the reference case of clear-sky. Further, to consider the delay induced by de-icing the aircraft, we include a dummy variable equal to one in case the temperature during the time of departure at the airport of origin falls below the freezing point of water, i.e. zero degrees Celsius.

Then, as in Mazzeo (2003), we include the number of flights scheduled to land at the same airport during the same hour ($Congestion_d$). This variable is meant to capture any peak effect and, more generally, airport congestion problems that are cited as a main driver of poor on-time performance. As a further measure of airport traffic, we add to our controls the number of all the flights landing in the airport within the same day (*Number landing* $flights_d$).

Finally, competition may play a role both at the route and at the airport level (Bilotkach and Pai, forthcoming; Brueckner, 2002; Mazzeo, 2003; Rupp 2009). We include the route market share of the airline (*Route market* $share_{od}$), the aggregated market shares of all the LCCs operating in route (*LCC route market* $share_{od}$), as well as the airline's market shares at the airport of origin (*Airport market* $share_o$) and at the airport of destination (*Airport market* $share_d$).

Some authors also include demand variables, such as population, income, and unemployment in the region of the airport of origin and destination (see Rupp, 2009 among others). These data, however, are usually collected on an annual basis, thereby producing repeated values for each flight code during the same year. As our sample comprises only two years, such demographical variables would probably not fit with our fixed effect estimation, which, on the contrary, depends fully on the within-group variation.

The time fixed effects are represented by the parameter $\tau_t$ and comprise a day-of the-week indicator, as well as a month indicator to address daily and seasonal demand fluctuations, respectively. The flight-airport-airline-route fixed effects are captured by the parameters $\rho_{fod}$, while the error term of the regression, assumed random with zero mean, is represented by $\varepsilon_{fodt}$.

It is worth noticing that besides comprising the entire measurable, time-invariant factors specific to the airport (e.g. hub status), to the route (e.g. distance), to the airline (e.g. national company), and to the flight (e.g. aircraft capacity), $\rho_{fod}$ controls for unobservable and thus immeasurable heterogeneity which, if ignored, could bias the estimates. For instance, LCCs may deliberately choose to fly to secondary airports.[1] Because of lower traffic volumes, these airports are more likely to experience fewer delays, *ceteris paribus*. By including route, airport, and flight fixed effects in the regression; we are able to fully control for such cases.

Standard errors are clustered by date and airport of destination to allow the residuals of different flight codes landing at the same airport during the same day to be correlated. This procedure takes into consideration possible shocks that are airport-date specific: for example, because of an accident a runway may be temporary closed, causing delay to all flights scheduled to land at the airport during the day of the accident.

The main descriptive statistics of the variables included in the analysis are reported in the Appendix. As far as the Variance Inflation Factor (VIF) test is concerned, there is no serious problem of multicollinearity associated with our econometric model; the VIF table is also shown in the Appendix.

### 4.5 Results and robustness

**Table 13** reports the results. We examine different specifications, adding some controls each time. The estimates on the variables of interest are robust to these various specifications, as well as to the different order of adding the controls (not reported here).

*LCC share$_o$* is negative and statistically significant, indicating that a higher share of LCCs at the airport of origin improves the on-time performance of all flights (and therefore airlines) departing from that airport. This result seems to stem from the faster turn-around operations by LCCs, which promptly free up parking positions and gates and thus leave more time to the other airlines to conduct their own operations. As far as the other regressors are concerned, we observe that the effect of a higher presence of LCCs at the arrival airport is not as evident as in the case of *LCC share$_o$*.

This finding is in line with the presumption that the knock-on effect on delay is usually stronger at the airport of origin rather than at the airport of destination. Departing delay is found positive and statistically significant, as expected. The estimated coefficient equal to about 0.8 indicates that ten-minute departure delay turns into an eight-minute late arrival. This result means that airlines can mitigate on average about 20% of their initial departure delay.

[1] Suppose, for example, that a LCC decides to fly to a specific airport due to a time-invariant characteristic of the airport such as the fact that the airport is located far away from the city center and for this reason the airport fees are low.

The weather variables, both at the airport of origin and at the airport of destination, behave as expected. Setting *clear sky* as the reference category, the positive sign on those variables indicates that inclement weather conditions increase the arrival delay. Interestingly, from a general glance at **Table 13**, the highest minutes of delay are reached in those cases that are expected *a priori* to have a greater influence on flying (i.e. fog, as it reduces visibility; snow as it slows down aircraft mobility and may require additional operations to clean the runway; thunder as a harbinger of severe weather conditions). Reading the estimates, fog at both origin and destination causes about 9 minutes of delay, snow about 12 minutes and thunder 15 minutes. On average, de-icing operations at the airport of origin are responsible for about two additional minutes of arrival delay.

*Airport congestion* is found positive and statistically significant, confirming that a larger number of flights landing within the same hour at an airport constitute a negative externality. Further, the positive sign on *Number landing flights* indicates that flights landing at airports with large daily traffic tend to have worse on-time performance. This result suggests that big airports, contrary to small airports, might experience more difficulties in easing the knock-on effect of built-up delay because they lack spare capacity in terms of free alternative landing slots, which may serve as a delay relief or buffer.

The positive signs on *Airport market share*$_o$ and on *Airport market share*$_d$ indicate that the higher the market share of the airline at the airport of origin and/or destination, the larger the arrival delay. This finding is in line with the presumption that passengers tend to experience significantly longer arrival delays at more concentrated airports (Rupp, 2009).

**Table 13.** Effect of LCCs on flight delay - dependent variable “ arrival delay in minutes” .

| | Model 1 - full sample | | Model 2 - full sample | | Model 3 - full sample | | Model 4 - FSCs | |
|---|---|---|---|---|---|---|---|---|
| | Coeff. | Std. err. | Coeff. | Std. err. | Coeff. | Std. err. | Coeff. | Std. err. |
| LCC share – origin | -1.057*** | (0.271) | -0.918*** | (0.268) | -0.785*** | (0.268) | -0.761** | (0.330) |
| LCC share – destination | -0.355 | (0.442) | -0.226 | (0.427) | -0.847** | (0.426) | 1.275*** | (0.494) |
| Departing delay | 0.813*** | (0.005) | 0.806*** | (0.005) | 0.805*** | (0.005) | 0.852*** | (0.006) |
| Cloudy – origin | | | 0.319*** | (0.025) | 0.318*** | (0.025) | 0.267*** | (0.027) |
| Cloudy – destination | | | 0.847*** | (0.044) | 0.843*** | (0.044) | 0.890*** | (0.047) |
| Foggy – origin | | | 3.480*** | (0.084) | 3.485*** | (0.084) | 3.080*** | (0.085) |
| Foggy – destination | | | 5.842*** | (0.302) | 5.802*** | (0.301) | 5.499*** | (0.303) |
| Hazy – origin | | | 0.918*** | (0.103) | 0.913*** | (0.103) | 0.730*** | (0.120) |
| Hazy – destination | | | 0.065 | (0.198) | 0.059 | (0.196) | -0.133 | (0.220) |
| Rainy – origin | | | 1.272*** | (0.043) | 1.269*** | (0.043) | 1.197*** | (0.047) |
| Rainy – destination | | | 2.221*** | (0.075) | 2.206*** | (0.074) | 2.347*** | (0.080) |
| Snowy – origin | | | 9.332*** | (0.203) | 9.328*** | (0.203) | 8.556*** | (0.201) |
| Snowy – destination | | | 3.165*** | (0.177) | 3.122*** | (0.176) | 3.223*** | (0.188) |
| Thundery – origin | | | 6.228*** | (0.254) | 6.250*** | (0.254) | 5.426*** | (0.246) |
| Thundery – destination | | | 9.076*** | (0.533) | 9.083*** | (0.532) | 9.452*** | (0.580) |
| De-icing operations | | | 2.005*** | (0.086) | 2.004*** | (0.085) | 1.929*** | (0.091) |
| Route market share | | | | | -0.094 | (0.091) | -0.455*** | (0.105) |
| LCC route market share | | | | | -0.631*** | (0.149) | -1.485*** | (0.216) |
| Airport market share - orig. | | | | | 0.745*** | (0.267) | -0.01 | (0.349) |
| Airport market share - dest. | | | | | 1.229*** | (0.400) | 2.963*** | (0.544) |
| Airport congestion | | | | | 0.090*** | (0.003) | 0.083*** | (0.003) |
| Number of landing flights | | | | | 0.011*** | (0.001) | 0.009*** | (0.001) |
| $R^2$ | 0.568 | | 0.574 | | 0.574 | | 0.641 | |
| Observations | 3,486,376 | | 3,486,376 | | 3,486,376 | | 2,672,722 | |

a) Robust standard errors to heteroscedasticity and serial correlation in parenthesis, clustered by date and airport of destination. One, two and three asterisks indicate significance at the 10%, 5% and 1% level, respectively.

b) All models include flight code, airline, airport and route fixed effects. Dummy variables for the day of the week and the month are included, but not reported

**Table 14.** Robustness - dependent variable " arrival delay in minutes" .

| | Model 1 - full sample | | Model 2 - full sample | | Model 3 - full sample | | Model 4 - FSCs | |
|---|---|---|---|---|---|---|---|---|
| | Coeff. | Std. err. | Coeff. | Std. err. | Coeff. | Std. err. | Coeff. | Std. err. |
| LCCs Tot Carriers – origin | -1.965*** | (0.264) | -1.861*** | (0.262) | -1.968*** | (0.262) | -0.704* | (0.413) |
| LCCs Tot Carriers – destination | -0.866** | (0.404) | -0.810** | (0.396) | -0.426 | (0.395) | 1.038* | (0.606) |
| Departing delay | 0.813*** | (0.005) | 0.806*** | (0.005) | 0.805*** | (0.005) | 0.852*** | (0.006) |
| Cloudy – origin | | | 0.320*** | (0.025) | 0.319*** | (0.025) | 0.267*** | (0.027) |
| Cloudy – destination | | | 0.848*** | (0.044) | 0.843*** | (0.044) | 0.890*** | (0.047) |
| Foggy – origin | | | 3.479*** | (0.084) | 3.485*** | (0.084) | 3.079*** | (0.085) |
| Foggy – destination | | | 5.842*** | (0.302) | 5.801*** | (0.301) | 5.501*** | (0.303) |
| Hazy – origin | | | 0.915*** | (0.103) | 0.911*** | (0.103) | 0.727*** | (0.121) |
| Hazy – destination | | | 0.065 | (0.198) | 0.057 | (0.196) | -0.128 | (0.220) |
| Rainy – origin | | | 1.273*** | (0.043) | 1.269*** | (0.043) | 1.197*** | (0.047) |
| Rainy – destination | | | 2.222*** | (0.075) | 2.206*** | (0.074) | 2.347*** | (0.080) |
| Snowy – origin | | | 9.335*** | (0.203) | 9.332*** | (0.203) | 8.556*** | (0.201) |
| Snowy – destination | | | 3.165*** | (0.177) | 3.122*** | (0.176) | 3.222*** | (0.188) |
| Thundery – origin | | | 6.232*** | (0.254) | 6.253*** | (0.254) | 5.426*** | (0.246) |
| Thundery – destination | | | 9.079*** | (0.533) | 9.085*** | (0.532) | 9.449*** | (0.580) |
| De-icing operations | | | 2.002*** | (0.086) | 2.004*** | (0.085) | 1.928*** | (0.091) |
| Route market share | | | | | -0.079 | (0.091) | -0.442*** | (0.105) |
| LCC route market share | | | | | -0.761*** | (0.150) | -1.419*** | (0.216) |
| Airport market share - dest. | | | | | 1.284*** | (0.404) | 2.546*** | (0.531) |
| Airport market share - orig. | | | | | 0.979*** | (0.270) | 0.219 | (0.335) |
| Airport congestion | | | | | 0.090*** | (0.003) | 0.083*** | (0.003) |
| Number of landing flights | | | | | 0.011*** | (0.001) | 0.009*** | (0.001) |
| $R^2$ | | 0.568 | | 0.574 | | 0.574 | | 0.641 |
| Observations | | 3,486,376 | | 3,486,376 | | 3,486,376 | | 2,672,722 |

a) Robust standard errors to heteroscedasticity and serial correlation in parenthesis, clustered by date and airport of destination. One, two and three asterisks indicate significance at the 10%, 5% and 1% level, respectively.

b) All models include flight code, airline, airport and route fixed effects. Dummy variables for the day of the week and the month are included, but not reported.

The variable *Route market share* does not appear as a relevant variable in our sample. Two interesting points, instead, can be claimed from the estimated coefficient on *LCC route market share*, which is found negatively signed and statistically significant. First, if the route is predominantly served by an LCC, the average delay is lower: this means that on average LCCs are more often on-time than FSCs. Second, if a FSC is facing a LCC on the route, it will strive to be more punctual than if it were not in competition with a LCC: This is because FSCs are more apt to compete with LCCs on quality rather than on price.

The above finding that LCCs are more punctual on average than FSCs questions whether our results might be affected by the good on-time performance of LCCs. The last column of the table reports the estimates of the subsample of FSCs flights only. The coefficient on the variable of interest, $LCC\ share_o$, is still statistically significant, maintaining the same sign and magnitude.

As a further robustness check, we replace $LCC\ share_o$ and $LCC\ share_d$ in **Table 13** with the ratio of the number of LCCs over the total number of carriers operating at the airport of origin ($LCCs/Tot\ Carriers_o$) and destination ($LCCs/Tot\ Carriers_d$) respectively. These results are reported in **Table 14**. Again, the coefficient on $LCCs/Tot\ Carriers_o$ is negative and statistically significant, confirming that the stronger the presence of LCCs at the airport of origin, the smaller the delay at the airport of destination. The estimated coefficients on the other regressors do not change qualitatively.

Finally, in order to be confident that the significance of our variables is not a consequence of the large sample size, we ran 100 regressions randomly, drawing about 20% of the observations of the entire dataset each time. $LCC\ share_o$ is always found negatively signed and, in a very large number of cases, statistically significant.[1]

At this point the reader may wonder if there is a threshold concerning the percentage of LCC flights above which the positive effect of LCCs on on-time performance becomes substantial.

**Fig. 34** draws a polynomial smoothed line of arrival delay over $LCC\ share_o$. Besides confirming the negative relationship between LCC share and arrival delay (already documented by **Table 13** and **Table 14**) the diagram reveals a very interesting pattern. There is a minimal effect on delay when LCC share is below 20%, probably because the number of LCC flights is too small to be influential. Above the 20% threshold, the effect appears and remains evident until the LCC share reaches 40%, at which point diminishing returns to LCC share begin. Further reductions of delay are only possible if an airport is primarily served by LCCs (LCC share above 70%). Finally, in line with the law of diminishing returns, the effect of LCCs is reduced when their share exceeds 80%.

[1] Also, the other regressors do not change qualitatively. A table is available upon request reporting only the estimated coefficient $LCC\ share_o$ for each subsample. This table was not included in the paper due to space considerations.

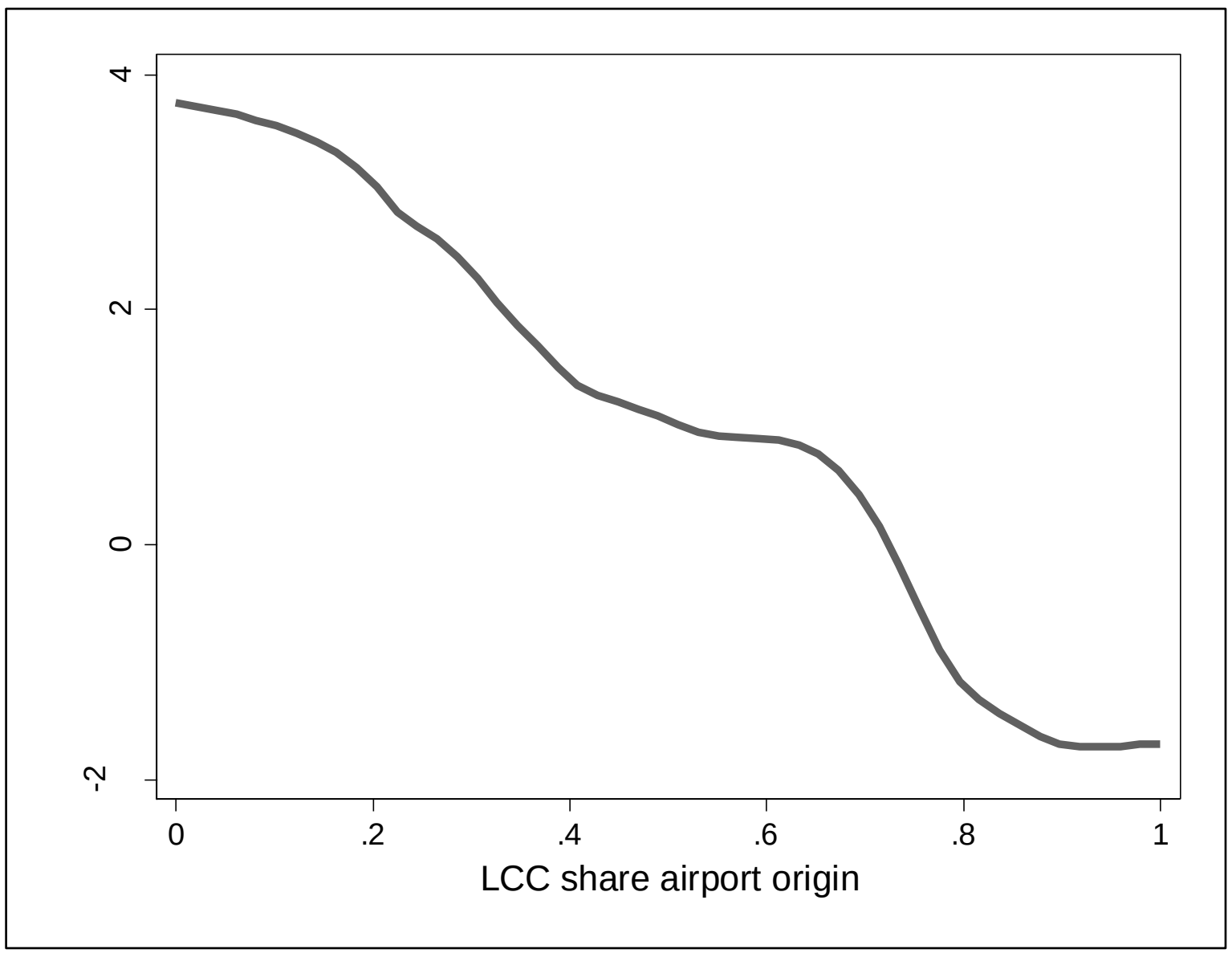

**Fig. 34.** Arrival delay and LCC share (local polynomial smoothed line).

## 4.6 Conclusions

In this paper we have analyzed the effect of LCCs on airline delays. Given their no-frills and faster turn-around policy, a priori one would expect that LCCs could manage delay more easily and eventually clear a slot more quickly. In other words, faster operations by LCCs reduce the overall waiting period for the take-off or landing slots at an airport and thus produce a positive externality to the remaining airlines, which should be more likely to land on-time.

We have examined whether a higher presence of LCCs at the departing airport may reduce the arrival delay of the flights landing at an observed airport, irrespective of the carriers' LCC or FSC characteristic. We have used a comprehensive dataset covering 23,402 flight codes, 3,270 routes and 100 European airports for a total of 3,486,376 observations. Our sample period spans from mid-April 2011 to the end of December 2012, both on a daily and on a flight-by-flight basis.

We have applied panel data fixed effect techniques and included the main drivers of flight delay among the controls. Since we have used fixed effects at the individual flight level, we are able to control for flight-specific characteristics influencing flight delays (Rupp, 2009). For instance, larger airplanes need longer embarking and disembarking operations, which may induce more delay. Since airport congestion tends to build through the day and therefore, all else being equal, early flights are more likely to be on-time, the departure time may also be a relevant explanatory variable of flight delay. These are all examples of time-invariant panel characteristics, which are absorbed into the fixed effects. Besides controlling for any observable time-invariant factors, specific to the airport, to the route, to the airline and to the flight, our fixed effects control for any unobservable and thus immeasurable heterogeneity.

We have found that on average LCCs contribute to the reduction of delay of all flights landing at a given airport and thus constitute a positive externality for all the airlines serving the airport. Our results have important policy and management implications in terms of the composition of carrier types at an airport. For the main cities in Europe the tendency has been to separate the airports between those mainly served by FSCs and those mainly served by LCCs. This is because passengers choosing one type of carrier rather than the other may have different needs. For instance, passengers flying with FSCs would probably give more importance to the airport facilities and their accessibility compared to passengers flying with LCCs, who, to the contrary, value their personal time less and are keen to pay cheaper ticket fares. The direct effect on ticket prices by LCCs' competition and entry threat on certain city-pair markets in Europe could be the next direction for research in this field. Ticket fares on routes can be recorded via a web crawler[1] from a range of service providers over the Internet.

Our findings in this study indicate that LCCs pose a threat and challenge to incumbents by competing in product quality. We show that a mixture of FSCs and LCCs within the same airport may be considered in the strategic decision-making of the airport management, since this source of intra-airport competition is beneficial in reducing airline flight delays and, therefore, in increasing the service quality and the customer satisfaction of the passengers.

1 See the program flow chart and explanation in the Addendum.

## Appendix

### Airport List

Our sample comprises the main European airports (see **Table 15**) and extends the airport list in Bel and Fageda (2010).

**Table 15.** Airports considered in the empirical analysis.

| | | | |
|---|---|---|---|
| Aberdeen (ABZ) | Dortmund (DTM) | London-City (LCY) | Palma Mallorca (PMI) |
| Alghero (AHO) | Dublin (DUB) | London-Gatwick (LGW) | Paris-Charles De Gaulle (CDG) |
| Alicante (ALC) | Durham Tees Valley (MME) | London-Heathrow (LHR) | Paris-Orly (ORY) |
| Amsterdam (AMS) | Düsseldorf (DUS) | London-Luton (LTN) | Pisa (PSA) |
| Athens (ATH) | East Midlands (EMA) | London-Stansted (STN) | Prestwick (PIK) |
| Barcelona (BCN) | Edinburgh (EDI) | Lübeck (LBC) | Reus (REU) |
| Beauvais (BVA) | Eindhoven (EIN) | Lyon (LYS) | Riga (RIX) |
| Belfast (BFS) | Florence (FLR) | Madrid (MAD) | Rome-Ciampino (CIA) |
| Bergamo (BGY) | Frankfurt (FRA) | Malaga (AGP) | Rome-Fiumicino (FCO) |
| Berlin-Schönefeld (SXF) | Geneve (GVA) | Malmo (MMX) | Saarbrücken (SCN) |
| Berlin-Tegel (TXL) | Genoa (GOA) | Malta (MLA) | Salzburg (SZG) |
| Bilbao (BIO) | Girona (GRO) | Manchester (MAN) | Southampton (SOU) |
| Birmingham (BHX) | Glasgow (GLA) | Marseille (MRS) | Stockholm-Arlanda (ARN) |
| Bologna (BLQ) | Hahn (HHN) | Milan-Linate (LIN) | Stockholm-Bromma (BMA) |
| Bournemouth (BOH) | Hamburg (HAM) | Milan-Malpensa (MXP) | Stockholm-Skavsta (NYO) |
| Bratislava (BTS) | Hanover (HAJ) | München (MUC) | Stuttgart (STR) |
| Bremen (BRE) | Ibiza (IBZ) | Naples (NAP) | Tenerife-Norte Los Rodeos (TFN) |
| Bristol (BRS) | Istanbul-Atatürk (IST) | Newcastle (NCL) | Tenerife-Sur Reina Sofia (TFS) |
| Brussels (BRU) | Istanbul-Gökcen (SAW) | Nice (NCE) | Treviso (TSF) |
| Cagliari (CAG) | Las Palmas (LPA) | Nürnberg (NUE) | Trieste (TRS) |
| Cardiff (CWL) | Leeds (LBA) | Olbia (OLB) | Turin (TRN) |
| Catania (CTA) | Leipzig (LEJ) | Oslo-Gardermoen (OSL) | Venice (VCE) |
| Charleroi (CRL) | Lille (LIL) | Oslo-Rygge (RYG) | Vienna (VIE) |
| Cologne (CGN) | Liverpool (LPL) | Oslo-Torp (TRF) | Weeze (NRN) |
| Copenhagen (CPH) | Ljubljana (LJU) | Palermo (PMO) | Zürich (ZRH) |

### Descriptive Statistics

The main descriptive statistics of the variables included in the analysis are reported in **Table 16**. The average delay is about 3 minutes, however if we exclude from the statistics those flights landing early or on-time, the average delay is above 19 minutes. We observe the maximum number of 698 daily departures at London-Heathrow airport in the U.K., whereas one daily departure was our minimum cut-off threshold, which was observed at Lübeck airport in Germany.

**Table 16.** Descriptive statistics.

| Variable | Mean | Std. Dev. | Min | Max |
|---|---|---|---|---|
| Delay | 2.72 | 21.40 | -50 | 208 |
| LCC share – origin | 0.21 | 0.24 | 0 | 1 |
| LCC share – destination | 0.22 | 0.24 | 0 | 1 |
| LCCs/Total carriers – orig. | 0.12 | 0.14 | 0 | 1 |
| LCCs/Total carriers – dest. | 0.12 | 0.14 | 0 | 1 |
| Departing delay | 8.96 | 19.66 | 0 | 1,020 |
| Cloudy – origin | 0.55 | 0.50 | 0 | 1 |
| Cloudy – destination | 0.55 | 0.50 | 0 | 1 |
| Foggy – origin | 0.02 | 0.14 | 0 | 1 |
| Foggy – destination | 0.02 | 0.13 | 0 | 1 |
| Hazy – origin | 0.01 | 0.09 | 0 | 1 |
| Hazy – destination | 0.01 | 0.09 | 0 | 1 |
| Rainy – origin | 0.09 | 0.29 | 0 | 1 |
| Rainy – destination | 0.08 | 0.27 | 0 | 1 |
| Snowy – origin | 0.01 | 0.09 | 0 | 1 |
| Snowy – destination | 0.02 | 0.13 | 0 | 1 |
| Thundery – origin | 0.00 | 0.05 | 0 | 1 |
| Thundery – destination | 0.00 | 0.05 | 0 | 1 |
| De-icing operations | 0.05 | 0.21 | 0 | 1 |
| Route market share | 0.67 | 0.29 | 0 | 1 |
| LCC route market share | 0.23 | 0.36 | 0 | 1 |
| Airport market share - orig. | 0.22 | 0.21 | 0.01 | 1 |
| Airport market share - dest. | 0.22 | 0.21 | 0.01 | 1 |
| Airport congestion | 18.72 | 13.90 | 1 | 76 |
| Number landing flights | 274.32 | 180.82 | 1 | 698 |

(a) Number of observations 3,486,376

The maximum of 17 hours was our threshold for departure delay. For example, a departing flight scheduled at 6:30, which has actually departed at 23:30, is assigned 17-hour delay; had this flight departed at 23:31, then it would be considered an early arrival by 6 hours and 59 minutes on the evening one calendar day earlier. We shortened the delay distribution by 0.2 per cent on each end to reduce such outliers. This conversion procedure was necessary to consider daylight saving time changes in different countries at remote airports and to link all data in Universal Time (UTC). In this way, we were able to produce a continuous timeline of events at the observed airports regarding actual flight operations and weather conditions (**Fig. 35**).

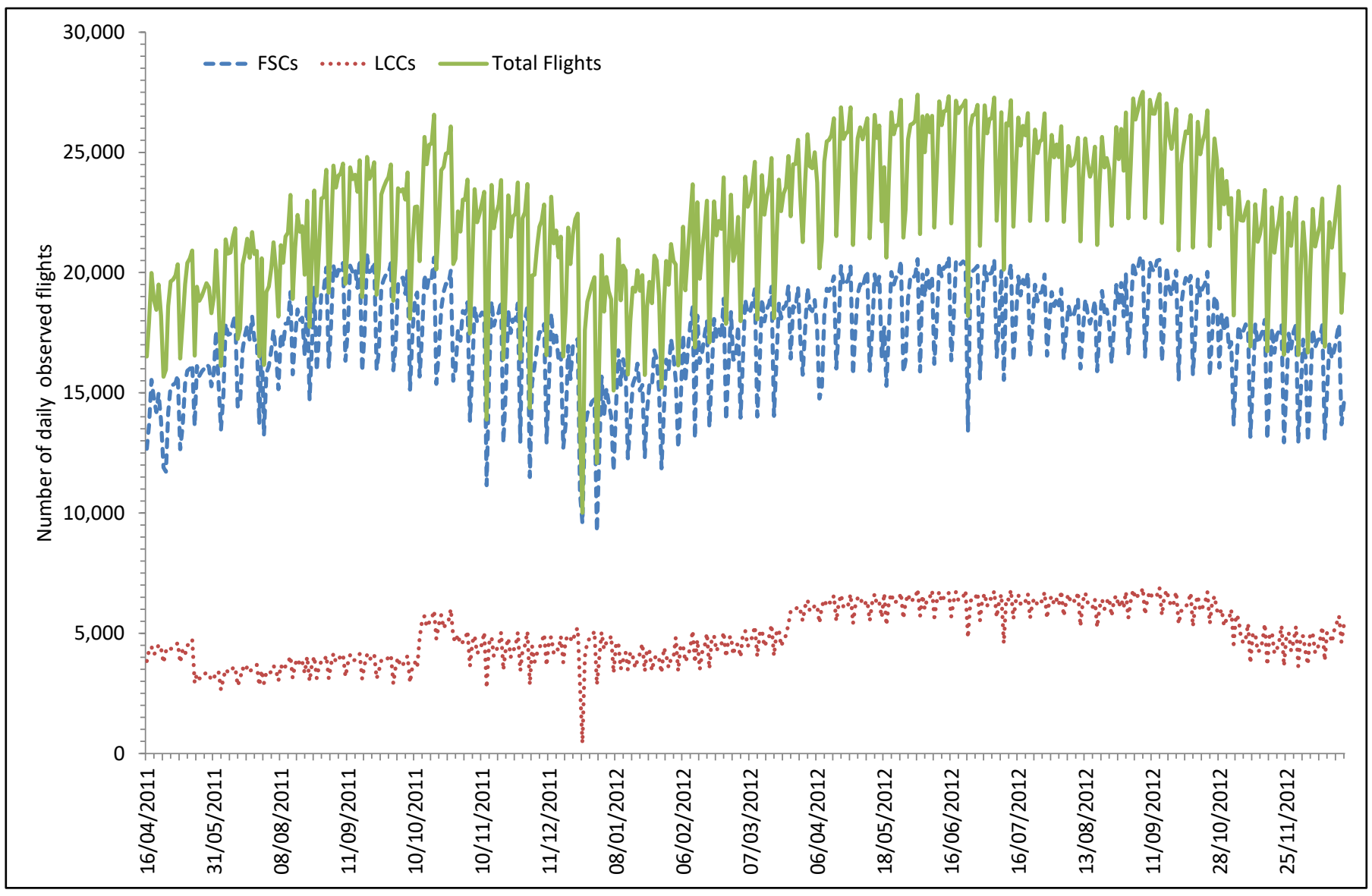

**Fig. 35.** Timeline of daily observed flights over 618 days.

## Variance Inflation Factor

As **Table 17** shows, several regressors have a Variance Inflation Factor (VIF) close to 1, meaning that there is no correlation among the kth predictor and the remaining predictor variables. Moreover, the highest value of VIF is below the rule-of-thumb threshold of 4, which warrants further investigation, and thus far from exceeding 10, which represents a sign of serious multicollinearity requiring correction.

**Table 17.** Variance inflation factor.

| Variable name | VIF | SQRT VIF | Tolerance | R-Squared |
|---|---|---|---|---|
| LCC share – origin | 2.46 | 1.57 | 0.41 | 0.59 |
| LCC share – dest. | 2.52 | 1.59 | 0.40 | 0.60 |
| LCCs/Total carriers – orig. | 2.47 | 1.57 | 0.40 | 0.60 |
| LCCs/Total carriers – dest. | 2.78 | 1.67 | 0.36 | 0.64 |
| Departing delay | 1.01 | 1.00 | 0.99 | 0.01 |
| Cloudy - origin | 1.26 | 1.12 | 0.79 | 0.21 |
| Cloudy - destination | 1.24 | 1.11 | 0.81 | 0.19 |
| Foggy - origin | 1.07 | 1.03 | 0.94 | 0.06 |
| Foggy - destination | 1.04 | 1.02 | 0.96 | 0.04 |
| Hazy - origin | 1.03 | 1.01 | 0.97 | 0.03 |
| Hazy - destination | 1.02 | 1.01 | 0.98 | 0.02 |
| Rainy - origin | 1.20 | 1.09 | 0.83 | 0.17 |
| Rainy - destination | 1.16 | 1.08 | 0.86 | 0.14 |
| Snowy - origin | 1.12 | 1.06 | 0.89 | 0.11 |
| Snowy - destination | 1.05 | 1.02 | 0.95 | 0.05 |
| Thundery - origin | 1.01 | 1.00 | 0.99 | 0.01 |
| Thundery - destination | 1.01 | 1.00 | 0.99 | 0.01 |
| De-icing operations | 1.14 | 1.07 | 0.88 | 0.12 |
| Route market share | 1.31 | 1.14 | 0.76 | 0.24 |
| LCC route market share | 1.39 | 1.18 | 0.72 | 0.28 |
| Airport market share - orig. | 1.36 | 1.17 | 0.74 | 0.26 |
| Airport market share - dest. | 1.39 | 1.18 | 0.72 | 0.28 |
| Airport congestion | 3.36 | 1.83 | 0.30 | 0.70 |
| Number landing flights | 3.69 | 1.92 | 0.27 | 0.73 |

# 5 Benchmarking European airports based on a profitability envelope - A break-even analysis

Branko Bubalo

**Abstract.** In this paper a simplified benchmarking methodology is presented. This new approach is based on the computation of a discrete envelope over distributed data points. Financial and operational data from 139 European airports in 10 countries was collected for the years 2002 to 2010. For reasons of comparability financial data is deflated to a reference price level, currency and point in time. The data requirements are reduced to the two core variables of the production process, passenger demand and profits or deficits (before interests and taxes) per year. Such data is used to isolate airport industry benchmarks based on minimum passenger levels and maximum profitability. Benchmarking can guide supranational decision-making and regulation on airport subsidy policy by stating maximum feasible profits per passenger and by estimating critical demand levels at the break-even point. In conclusion, scenario-based calculations about potential efficiency gains for underperforming airports are outlined.

## 5.1 Introduction

To this day large-scale financial or operational comparisons of airports across countries are rare and far from sufficient in guiding decision makers in a simplified manner. Given the importance of airports in a globalized economy regarding the linking of national and international destinations and the magnitude of daily travelers and shipped cargo, current research undervalues careful empirical observation and numerical description in favor of theoretical modeling. Frequently the results and implications from models for decision-making in management can only be validated with large efforts, if ever. A deeper understanding of the real processes would lead to better predictions of “what if?” scenarios compared to “what is?” baselines. However, full comprehension requires many observations until the underlying interrelationships become obvious. In our case the preliminary plotting of performance ratios from the collected data leads to the discovery of regular patterns. Access to information and full knowledge of today’s mathematical techniques allow the replication of the described model on another dataset. For the first time such vast collection of financial and operational data from 139 European airports has been holistically analyzed. This article represents only one in a line of articles (e.g. Adler, Ülkü and Yazhemsky 2012, or Bubalo 2012).

Independent of type of ownership (public, partially private, or fully privatized) airports should strive for a financial breakeven. Managerial feasibility of maximizing profitability and performance is limited by existing demand. Similar to most (private) businesses economic losses need to be minimized in order to limit subsidies or other compensations, e.g. debt from credits. It should be a common goal for airport management to maximize profits and to reach breakeven, so corporate taxes, benefitting the whole society, may be paid. There is no clear reason why the society should constantly ‘pay’ for subsidies benefitting few passengers travelling to or from loss-making airports frequently found in remote locations.

It is obvious that loss-making airports would not be able to survive in a competitive and fully private market (Gillen and Lall 1997). Airports rarely or never achieving a breakeven must therefore receive subsidies to some degree. Often this is achieved by the mechanism of cross-subsidization between profitable and non-profitable airports inside an airport portfolio of a multi-airport operator. To individual airports these subsidies come directly in form of public funds (taxpayer’s money), (low interest) grants, national capital expenditure programs or other non-operating sources of income. This article follows the definition by Adler, Liebert and Yazhemsky (2012), who define an airport as “a private [or public] production system in which society maximizes social welfare by encouraging airport management to maximize profits.” A loss of social welfare is suspected, when public funds support air transport infrastructure through subsidies, which could instead be spend on local transport, infrastructure, hospitals, or schools (Doganis and Thompson 1973)

A new partial factor productivity (PFP) approach in airport benchmarking and performance measurement (Vogel and Graham 2010) is developed driven by the collected panel data. From earlier research it was found that it is sufficient to strictly focus only on a few core variables of the production process in order to pinpoint best-practices. Reducing the data requirements has many advantages, such as better ability of data handling, extending, updating and adjusting. It is a key task in performance measurement to find the right level of aggregation of the data for any specific analysis. For this study, demand as a given Input of the production process (although passengers have Input and Output characteristics) is measured in number of total (arriving and departing) passengers. Unit profits (or losses) as an Output, on the other hand, are measured in *earnings before interests and taxes* (EBIT) per passenger (PAX). In many cases this kind of data can be directly extracted from income statements and airport operational statistics (Doganis and Thompson 1973, Doganis and Thompson 1975, Koopmans 2008, Vogel and Graham 2010). However, there exist understandable *ressentiments* against the publication and sharing of financial data in the airport industry, which requires personal appeasements, non-disclosure agreements or the usage of carefully coordinated questionnaires.

The trend and shift of the profitability frontier and function will be observed for the 139 airports over a time-frame of nine years, answering questions such as: Which benchmark or range of profits could *a priori* be expected for any airport given the local level of demand? Or how does ‘my’ airport perform over time relative to the best-practices? Where is the break-even point for the airport-industry located in a particular year, and in which direction does the break-even point and the profitability envelope shift over the years?

The paper is structured in five parts, starting with this introduction. In 5.2 the discussion continues with the theoretical background including the review of literature, overview of different methodologies and description of the dataset. 5.3 will show the application of the profitability envelope over the collected data points. 5.4 will look at some possible gains in profitability, if the identified benchmarks would be reached. 5.5 will draw some conclusions and will give an outlook on further research.

### 5.2 Theoretical background, methodology and data description

In the discussion on measuring the profitability or ‘cost efficiency’ of airports one goal with this study is to draw parallel conclusions to more sophisticated linear programming and optimization techniques, such as Data Envelopment Analysis (DEA), but with less restrictive data requirements. These techniques frequently make use of multiple inputs and outputs with strict requirements on multi-dimensional data. However, in the research community it is increasingly difficult to justify and defend straightforward and simple approaches, where the analyst preserves full control over his or her data. In recent analyses there is little progress in reducing the number of variables to only a few explanatory variables, which are convenient to communicate to practitioners and which offer intuition of the internal processes. It is common that the consequences of the results are neither appreciated nor implemented by the parties for which these are aimed for. Science has shown that even in the broad field of mechanics for practical purposes it is sufficient to work with only a few variables, such as force depending on mass and acceleration (or gravity). The suitability of some core variables is discussed in the following paragraphs.

Figure 4 exhibits a clearly recognizable correlation pattern with the elementary measures: costs, revenues, and EBIT. However, many benchmarking studies are trying to explain differences in efficiency or profitability by additional variables, such as type of ownership or capacity utilization (Bubalo and Daduna 2012), which both suit only a selected peer group of larger airports and have little relevance for smaller unprofitable airports. The previous benchmarking study involving Norwegian airports operated by the state-owned company Avinor has led to the conclusion that a public entity is indeed capable of successfully managing a system of close to 50 airports. Since the operations of Avinor are closely monitored by the auditor general office in Norway greater transparency towards the public spending is guaranteed, which is reflected by frequently published public reports about the activities of Avinor and the level of subsidies per passenger (Norwegian Ministry of Transport and Communications 2009). To the contrary in Germany, where a large degree of airports are privatized as limited companies, yet, partly owned by local authorities, it is increasingly difficult to obtain detailed financial information.

In supplement regression analyses for this recent benchmarking study including Norwegian airports it was found by Adler, Ülkü and Yazhemsky (2012) that there exists a sufficient correlation between DEA relative efficiency scores and the PFP measure of unit profits, measured in ‘EBIT per PAX.’ However, it is evident that airports need the “critical mass” (Heymann and Vollenkemper 2005) in number of passengers and consequential revenues to operate self-sufficiently. Therefore, deeming loss-making airports to be inefficient and *vice versa* profit-making airports to be efficient would not allow for a fair comparison among different sized entities.

Similar to the DEA we aim at revealing and computing the ‘relative efficiency’ compared to other similar sized peers (Adler, Ülkü and Yazhemsky 2012; Adler, Liebert and Yazhemsky 2012). Accordingly, the ratios of outputs over inputs are related to the size of airports, thus relative to comparable peers. An airport is deemed the more inefficient the further the calculated ratio is located from the related profitability benchmark.

One reason of not using EBIT or other profit measures in the DEA is the inability of the method to deal with variables which could be either negative or positive such as profits. This drawback could be circumvented in all linear programs by adding a ‘sufficiently large constant’ to the profit figures in order to make them all positive, thereby not changing the “optimum strategies” (Dorfman 1951). A DEA study using EBIT or another measure of profitability has yet to be conducted. Most existing studies focus on the input-output balance of costs and revenues and make use of additional traffic or capacity data (Adler, Ülkü and Yazhemsky 2012; Adler, Liebert and Yazhemsky 2012).

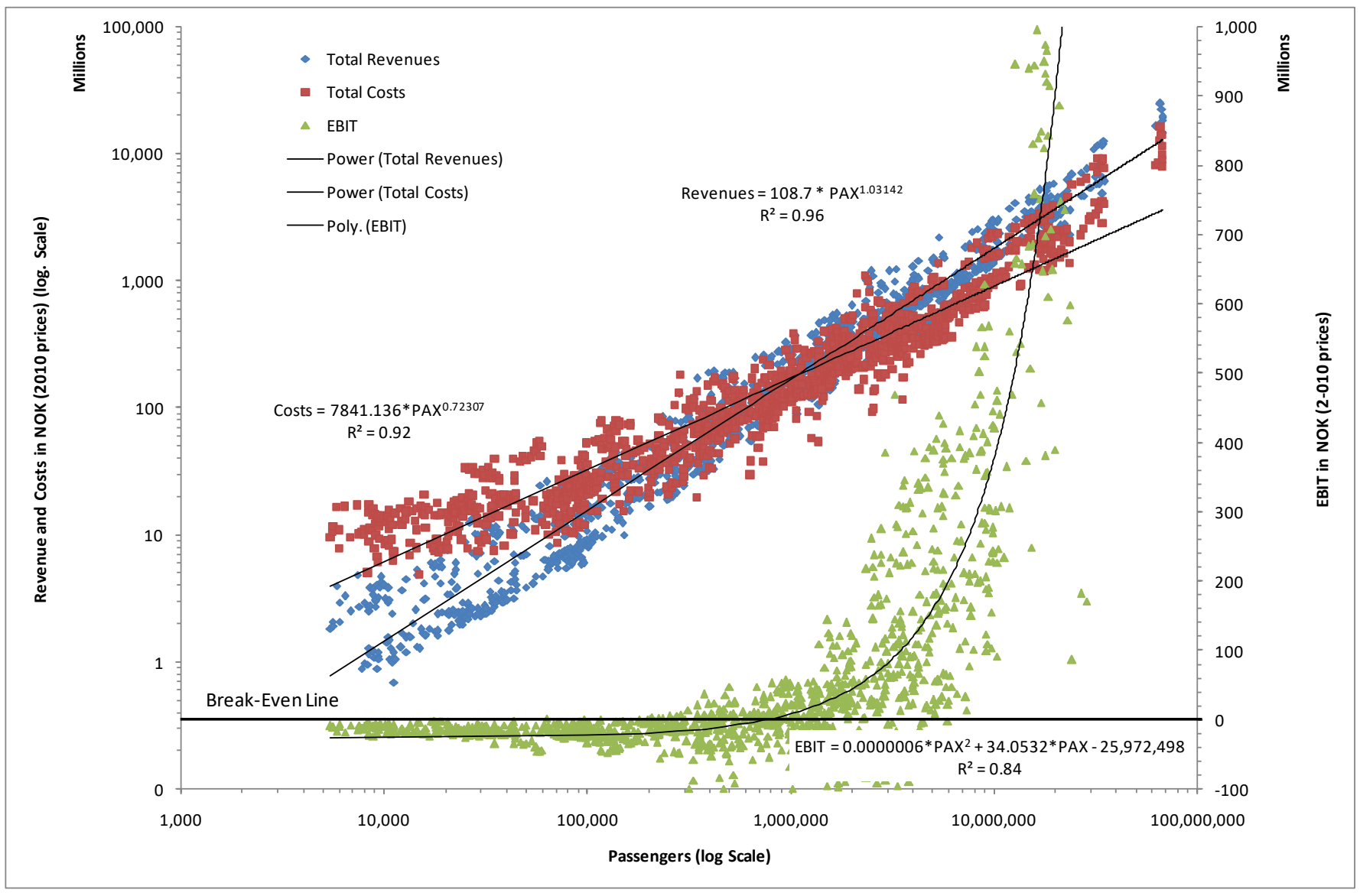

**Fig. 36.** Revenues, costs and EBIT for sample airports in the years 2002 to 2010 in Norwegian Kroners (PPP-adjusted base Norway in 2010 prices) (Source: Own survey data)

When plotting the adjusted and cleaned dataset of the 139 European airports (see the table in the Appendix), it is observed that especially small and regional airports lie below the break-even line of (unit) profits and usually do not report any profits in subsequent years (**Fig. 36**). There are very few airports which are *volatile* with regard to their profitability, showing deficits as well as profits in subsequent years. Another main observation is decreasing returns to scale on a unit profit basis (i.e. EBIT per PAX), but constant (or even increasing) returns to scale on absolute profits (i.e. EBIT). With increasing output these returns to scale pick up below the break-even line at volumes of 300.000 to 400.000 PAX per year, which can be observed in **Fig. 36** and subsequent **Fig. 39** and **Fig. 41** in Section 5.3. With the approximation functions in **Fig. 36** the *average* industry break-even point is calculated at 1.06 million PAX when total operating costs equal total operating revenues, and is about 750.000 PAX, when EBIT is equal to zero.

In the literature one can find different assumptions regarding the airport industry break-even point. For example, in the 1970's Doganis and Thompson (1975) have calculated the break-even point for British airports to be around 3 million passengers. This might be a result of a much smaller sample and different distribution of airport sizes. They also used a different definition of profits, hence breakeven, which includes interest expenses. Heymann and Vollenkemper (2005) (citing an unavailable study by the European Commission) estimates the break-even point for European airports to be in the broad range of 500.000 to 2 million passengers. Our figures support this estimation, since the calculated break-even point falls into that range. Certainly, the break-even point estimates vary with the definition of *profits* and which costs or revenues are included in it. Our data is cleaned from "outside" revenue sources such as state grants or "government transfers."

Koopmans (2008) has empirically tested the correlations between the closely related profitability measure "earnings before interests, taxes, depreciation and amortization" (EBITDA) per workload unit (WLU = 1 PAX or 1/10 tons of cargo) and found it a good descriptor of "operational airport performance." Hence, this strongly implies that *EBIT per PAX* could be a sufficiently good descriptor of airport managerial efficiency. This is a similar result as shown by Vogel and Graham (2010) who found that "profit per WLU" correlates significantly and highly significantly to ten of fifteen of the studied airport performance measures, such as return on investments (ROI), EBITDA margin and asset turnover. Doganis and Thompson (1975) have used absolute "surplus and deficits" (including interests) relative to the output level in WLU as a descriptor of "managerial effectiveness" and for determining the break-even point. Their work focused mainly on the structure and balance of revenues and costs. Gritta, Adams and Adrangi (2006) have chosen EBIT and its variation over time as a measure to quantify the "business risk" of a firm and to construct financial performance measures in their study "of the effects of operating and financial leverage on […] Major U.S. Air carriers' rate of return." In conclusion sufficient evidence was collected to support the claim that annual EBIT per PAX serves as an adequate approximation of airport productivity and relative efficiency.

However, Gillen and Lall (1997) dispute the usefulness of profitability measures and state that these are "totally misleading, […] given the unique position of airports." In light of our graphical and numerical analysis of data this judgment may not be accurate, since we observe strong "industry-wide" (Doganis and Thompson 1975), trends in profit generating ability with regard to the output level. Hence, we consider for the "unique position" related to airport size. It should be noted that our results only relate to European airports. The situation may look differently for North-American airports or airports on other continents, which may exhibit contrasting cost and revenue structures.

Firstly, we argue that using EBIT as a single aggregate output measure has the advantage that this figure includes all required operating costs, generated income and necessary investments in infrastructure and provision of capacity. In a simplified sense annual (positive or negative) EBIT is defined as revenues minus costs minus depreciation of assets. In our case, certain types of non-operating revenues or costs have been deducted from the EBIT figures, e.g. earnings from trading or government sources or costs for air traffic control services. For many airports EBIT (net result, EBITDA or some other definition of profits) is generally published in profit-and-loss statements as part of annual reports or financial statements, because it represents a common figure in finance. For reasons of comparability, the collected nominal figures are adjusted for currency, purchasing power parity (PPP) and inflation. As base currency Norwegian Kroners (NOK) was chosen. The financial data has been PPP-adjusted relative to Norway to account for different price and wage levels between countries, as well as inflation adjusted to 2010 prices.

Second, we limit the operational variables only to the input number of passengers, as this figure represents the original (origin, transfer and destination) demand, and is usually easily available. Divisions by passenger characteristics, such as international, domestic, business or leisure, have not been made, since these ultimately reflect groups of people with different spending behavior, service requirements and associated costs, which need to be targeted commercially by the airport business and charges model. In Doganis and Thompson (1973) and in Doganis and Thompson (1975) it was emphasized that airport management has only limited control over “externally defined” factors, such as demand, i.e. level of passengers departing from or arriving at a particular location, and indivisible costs. Therefore, the airport managements’ prime function is to balance (expected) revenues and cost, and to maximize output (EBIT) given the level of input (PAX). To generate demand for certain destinations is understood as a major function of airline marketing and route development but should be closely coordinated with the airport management in a collaborative decision-making (CDM) setting.

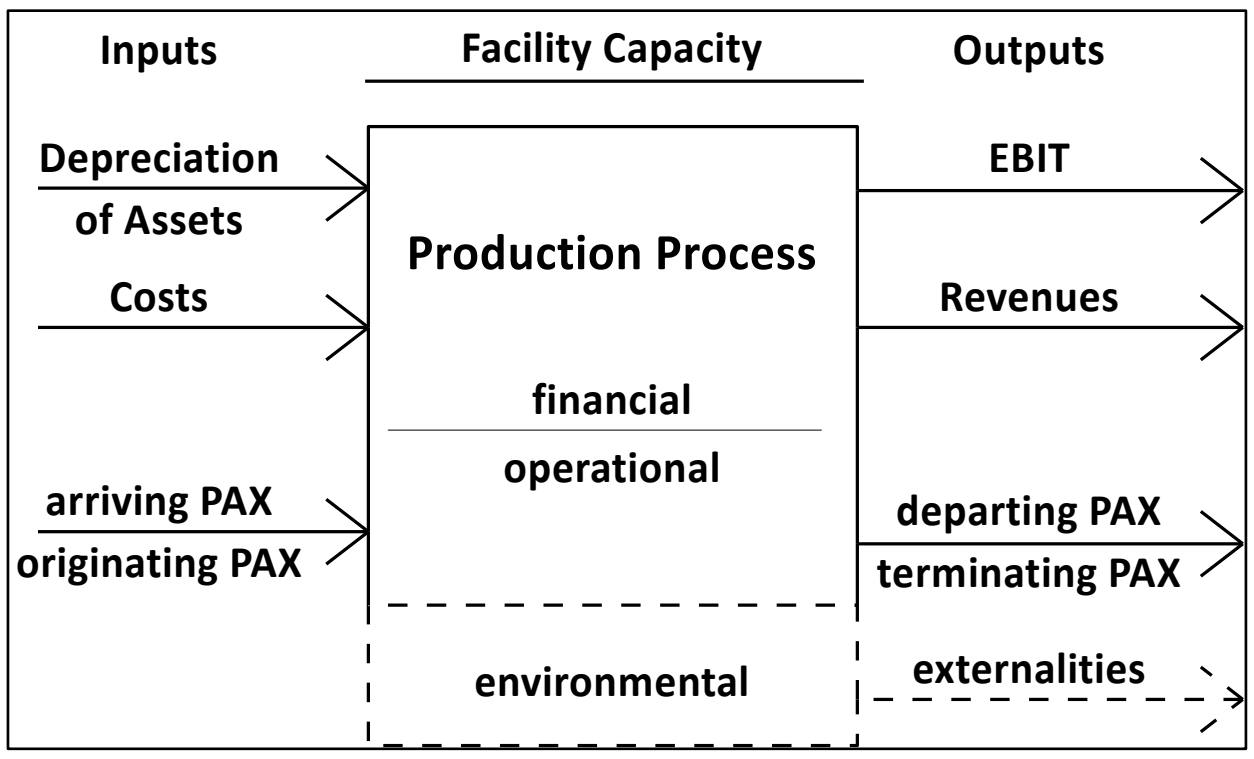

**Fig. 37.** Core inputs and outputs of the airport production process (Source: Own illustration)

**Fig. 37** exemplarily shows the main inputs and outputs of the airport production process, yet it is not so clear which side needs to be maximized or minimized in an optimization program. Typically, the inputs are minimized and/or the outputs are maximized, for example in an output-oriented DEA. In **Fig. 37** one may define all costs, including depreciation, as a financial input, and all revenues as a financial output. On the operational side of the production process arriving or originating passengers, who request to be served by the airport, may be viewed as an operational input, and all departing and terminating passengers, who have been served by the airports may be viewed as an operational output. Therefore, total PAX show input and output characteristics, in our case these are defined as inputs. Externalities, such as noise and delay, may be viewed as "undesirable output" and its *reciprocal* (i.e. inverse) may be maximized (Adler, Liebert and Yazhemsky 2012). Additionally all passenger related facilities place a limit on the technical capacity in relation to a minimum level of service (Bubalo and Daduna 2012).

The data points can have five characteristic locations in relation to the profitability envelope and the break-even line. **Fig. 37** shows theoretically the different stages of development of an airport with regard to increase in size and profitability. For example Reinhardt (1973) has used a similar function for the break-even analysis of the "TriStar" aircraft development program of Lockheed in 1970's, where the amount of future sales of aircraft determine the financial feasibility (i.e. positive net present value) of the project.

Group 1 would represent loss-making airports, with data points below the reference profitability benchmark. The second group would include loss-making airports which define the profitability envelope and represent a benchmark. The third group consists of one (or more) airports, which break even exactly. More realistically it is this airport in the data set, which is first making profits in EBIT per PAX, with minimum demand in terms of number of passengers.

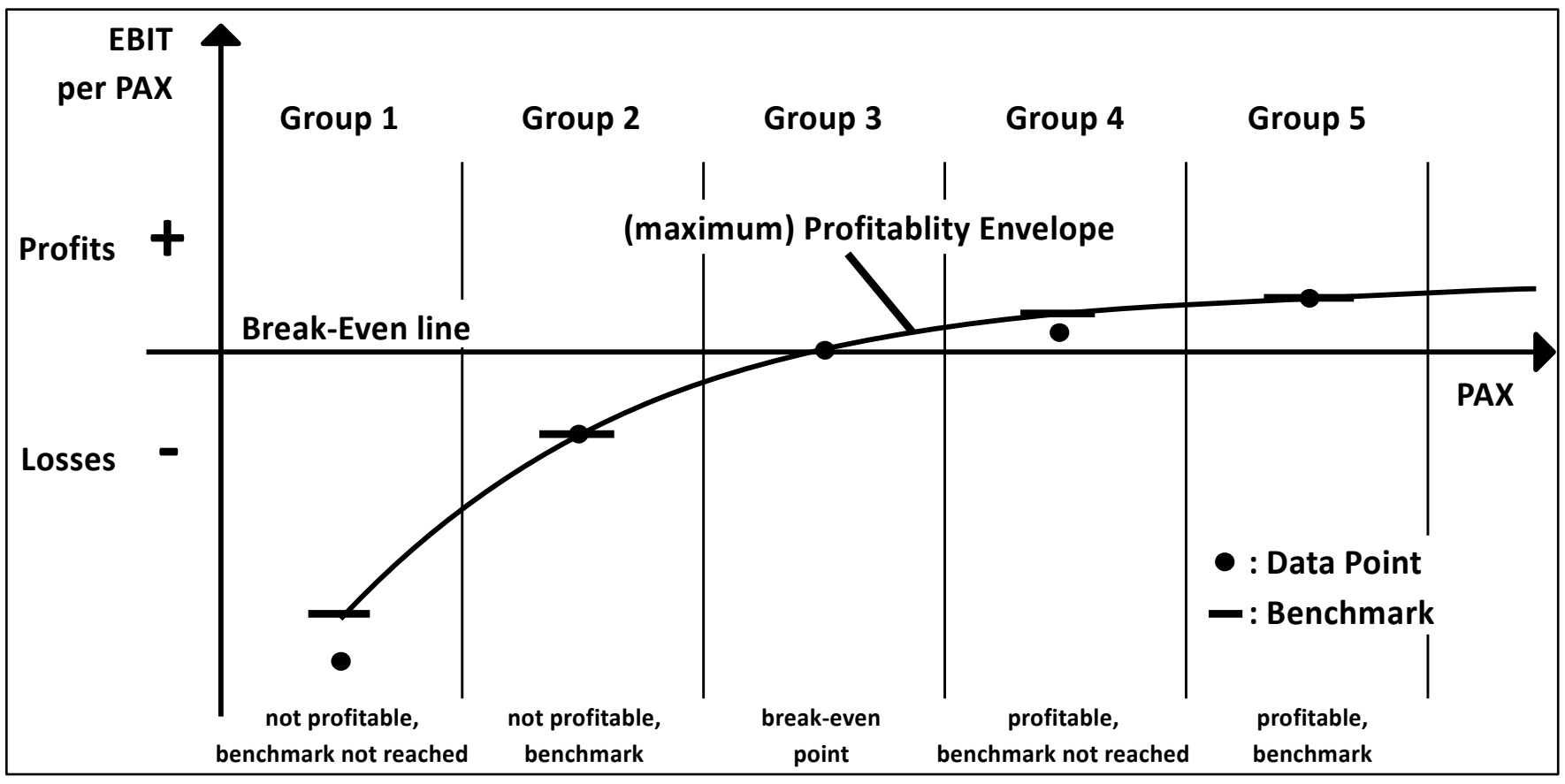

**Fig. 38.** Data points in relation to break-even line, (industry) profitability envelope and break-even point (Source: Own illustration)

The fourth group consists of profitable airports, which are relatively inefficient regarding their position below the (maximum) profitability envelope. Similar to the first group, it is possible to calculate productivity gains, if the benchmarks would be reached (see Section 5.4). The fifth group represents superior airports, which are profitable and define the profitability envelope. These airports show maximum profitability relative to their output level and comparable peers; therefore, these truly represent best-practices. It is this group of airports, where one could expect the largest interest from private investors, as these promise the highest dividends and ROI. Under the assumption of future air traffic growth at different locations one can assume that airports will *mature* over time and will place themselves in different groups. Through streamlined management activities they may switch from group 1, the underperformers, to the peer group of best practices in group 2. At this stage it is not expected for them to operate profitably. As growth continues, they may pass the break-even point and become member of group 4. Eventually these airports may set the benchmark regarding their passenger level and profits, hence become members of superior group 5. In summary, the members of groups 2, 3 and 5 represent the *cornerstones* of the discrete profitability envelope (**Fig. 38**).

The strategies behind the required number of total flights measured in annual air transport movements (ATM) have not been analyzed here. It is assumed that frequencies of flights and available seats per route are under the control of carriers which try to match capacity to demand on each flight on a particular route by maximizing load factors. Subsidized public service obligation (PSO) route networks are certainly an exception, where whichever operating costs of the commissioned airline are covered by the state to serve a particular network and frequency (Bubalo 2012). Certainly, the aircraft sizes in each carrier's fleet put a significant limitation to matching air transport demand exactly, since the ability of a flight breaking even, and economies of scale must be considered by the airline management. Therefore, we find that the applicability of ATM as a holistic operational variable for airports is not given, since this measure is related only to *airside* Inputs and Outputs. The airport management has only limited control over the actual number of movements.

**Table 18.** Airports with relevant average share of cargo >20% of total Workload Units (WLU) across the years 2002 to 2010 (Own survey data)

| Airport | IATA Code | Share of Cargo of total WLU |
|---|---|---|
| Leipzig | LEJ | > 60% |
| Cologne | CGN | 41% |
| East-Midlands | EMA | 38% |
| Bergamo | BGY | 30% |
| Benbecula | BEB | 28% |
| Brussels | BRU | 27% |
| Rennes | RNS | 23% |
| Billund | BLL | 21% |

Note: One WLU equals one passenger or 100 kg of cargo. In the literature the usage of this combined measure and the equivalence of passengers to cargo is disputed, especially in terms of different handling costs, revenues and infrastructure requirements.

To account for airports, which make most of their turnover with the handling of cargo, the use of the aggregate measure WLU may be considered as an alternative operational measure, i.e. EBIT per WLU. Sensitivity analyses across the sample only showed minor differences between the trends of average EBIT per PAX and average EBIT per WLU. For the full data sample, the average amount of cargo accounts for about 5% of the total WLU, with eight airports having shares of more than 20% of the total WLU (**Table 18**). The Norwegian airports Hasvik (HAA), Røst (RET) and Svalbard (LYR) are the only ones having significant shares of cargo of up to 10% cargo of total WLU. For reasons of clarity we do not discuss the alternative measure WLU any further in this article.

### 5.3 Application of the profitability envelope on the dataset

In our heterogeneous sample of airports across different countries, time, and sizes we relate average units of output to average units of input. This gives us the profitability ratio between the surpluses or losses in EBIT divided by the number of passengers, namely EBIT per PAX. It is state-of-the-art in current research to divide the airport sample into sub-sets, classes or “clusters” (Doganis and Thompson 1973, Doganis and Thompson 1975) by certain characteristics, such as *hubs* (above 2 million) or regional airports (below 2 million passengers). Since this kind of division is always arbitrary, we chose to present the ratio on a continuous scale. The distribution of passenger demand covers a large range of airports different in size, therefore, a logarithmic (base 10) scale has been chosen for displaying the data points.

We observe in **Fig. 39** that growth in number of passengers is a strong driver of more than proportional increases in EBIT per PAX. However, the chosen profitability ratio stagnates at a certain level of saturation, after which it appears the average EBIT per PAX can only marginally be increased any further. The maximum profitability benchmark has been observed at London-City (LCY) airport with 144 NOK per passenger at a level of 3.3 million passengers in 2008. LCY defined the maximum benchmark for the years 2007 and 2009 as well, with passenger levels of 2.9 or 2.8 million passengers and a related EBIT per PAX of 133 and 99 NOK, respectively. LCY airport is operating a short take-off and landing (STOL) runway and has limited terminal and aircraft parking stand capacity, which makes this achievement of high average profitability even more special. This example shows that the airport management at LCY airport has reached a high value-added productivity by fully exploiting its available resources.

It can be concluded that airports with a critical level around the break-even point of more than, say 300.000 to 1 million passengers, are becoming increasingly interesting for investors, but also request less regulation for achieving maximum profitability. A marginal increase in annual number of passengers directly leads to increasing absolute profits and returns to scale at this high level of output (see **Fig. 36**). However, from a welfare state point of view, the application of price-cap regulation to profitable airports (group four and five) may be an advantageous instrument. This regulation limits monopoly power and decelerates growth. Therefore, it could balance airport revenues from charges and from non-aeronautical sources against true (societal) costs.

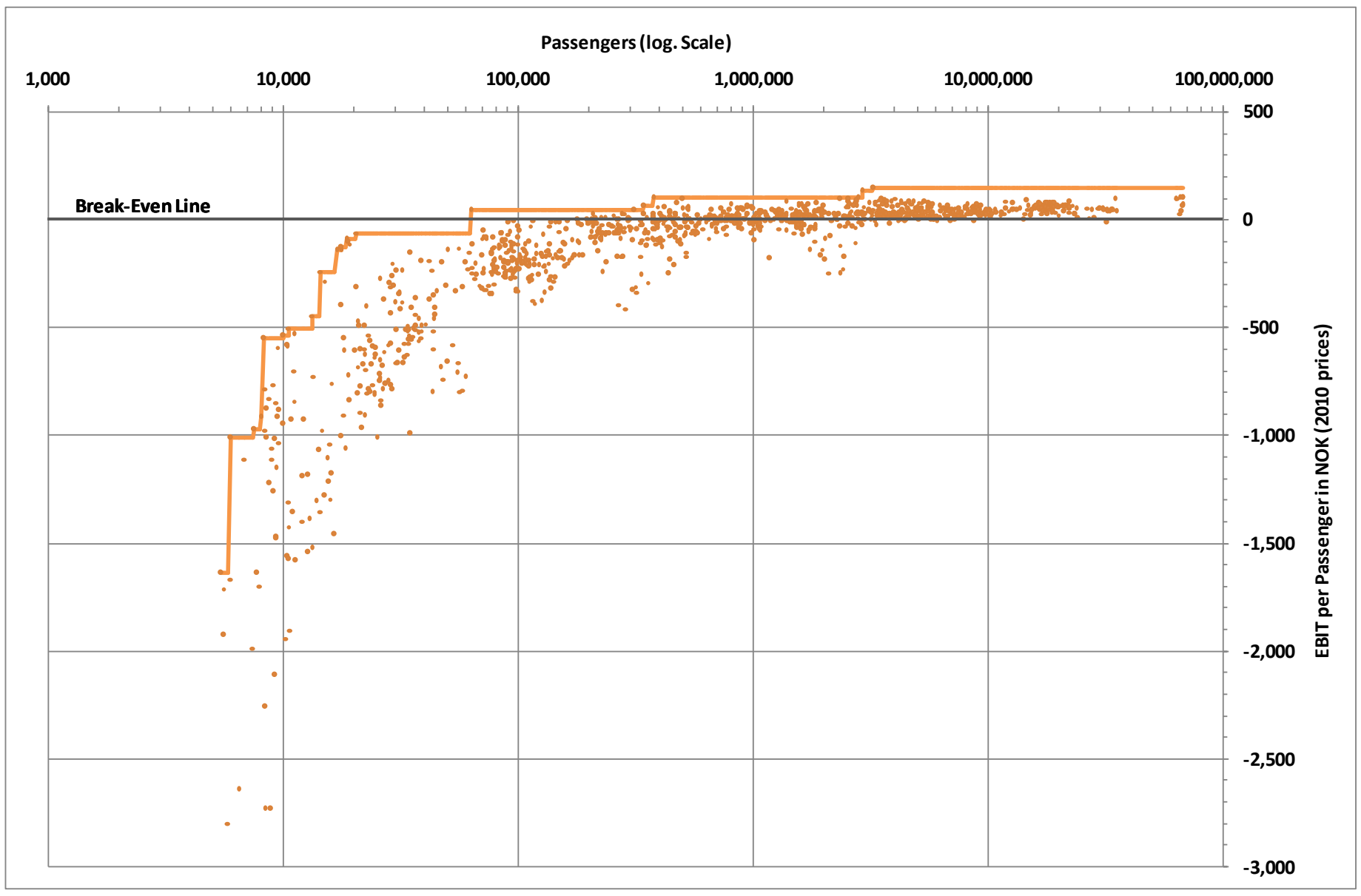

**Fig. 39.** Profitability and profitability envelope by airport size for the years 2002 to 2010 (all airports in the Appendix, PPP-adjusted base Norway in 2010 prices; 2010 without Italy and France) (Source: Own survey data)

For loss-making airports (in groups one to three) a *laissez-faire* approach may be appropriate, if social obligations for transportation services are not of public interest. Otherwise, there is a lack of incentives and degrees of freedom for airport management to change the revenue and cost structure to become profitable. Like private entrepreneurs, the state may provide start-up funds or low-interest grants in initial stages of airport development. Even airport closures should remain as a last option if an airport has no financially sustainable strategy.

To derive the *profitability envelope* exemplarily shown in **Fig. 38**, we made use of a simple algorithm, which plots the envelope over the maximum data points. The result of the algorithm provides profitability benchmarks for each level of passengers, thus giving a feasible *strategic target* for the airport management at underperforming airports. Trying to reach the benchmark should be the main motivation of all management efforts.

```
Program Profitability_Benchmarks

DIM n, PAX(n) as NATURAL
DIM EBIT_per_PAX(n), BENCHMARK(n) as REAL

n = Number of airports in the sample
i = 1 to n

# Sort table by PAX column in ascending order
PAX_i < PAX_i+1 < ... < PAX_n

# The first entry in the 'EBIT_per_PAX' column is equal to
the initial starting point for the Envelope.
i = 1
BENCHMARK_i = EBIT_per_PAX_i

# From the second entry onwards, each new entry is
compared, if it is larger than the last stored Benchmark
(in third column). If yes, it is set as new Benchmark. If
not, the last Benchmark remains.
FOR i = 2 TO n STEP 1
        IF     EBIT_per_PAX_i > BENCHMARK_i-1
        THEN   BENCHMARK_i = EBIT_per_PAX_i
        ELSE   BENCHMARK_i = BENCHMARK_i-1
NEXT
END.
```

**Fig. 40.** The algorithm in Pseudo-code outlines the required steps to create the *profitability envelope*. Three table columns are required for PAX, EBIT per PAX and the running benchmark

## 5.4 Break-even analysis

We want to get clearer answers on where the best-practice break-even point lies in each year. More particularly, we ask at which level of demand the observations surpass the break-even line. Thus, at which critical level of demand full coverage of operating cost could possibly be managed. We have not attempted to calculate the break-even points exactly by interpolation between the last observation below breakeven and the first observation above breakeven, because we recognize when working with real data, the quality is commonly mixed and certain limitations of accuracy exist. Therefore, all presented figures and thresholds are *real* and accurate, given the resources. For a more mathematical and theoretical treatment in the future, such as the derivation of the profitability envelope or an approximation of an exact function, a smoother construction of the curve by interpolation seems appropriate.

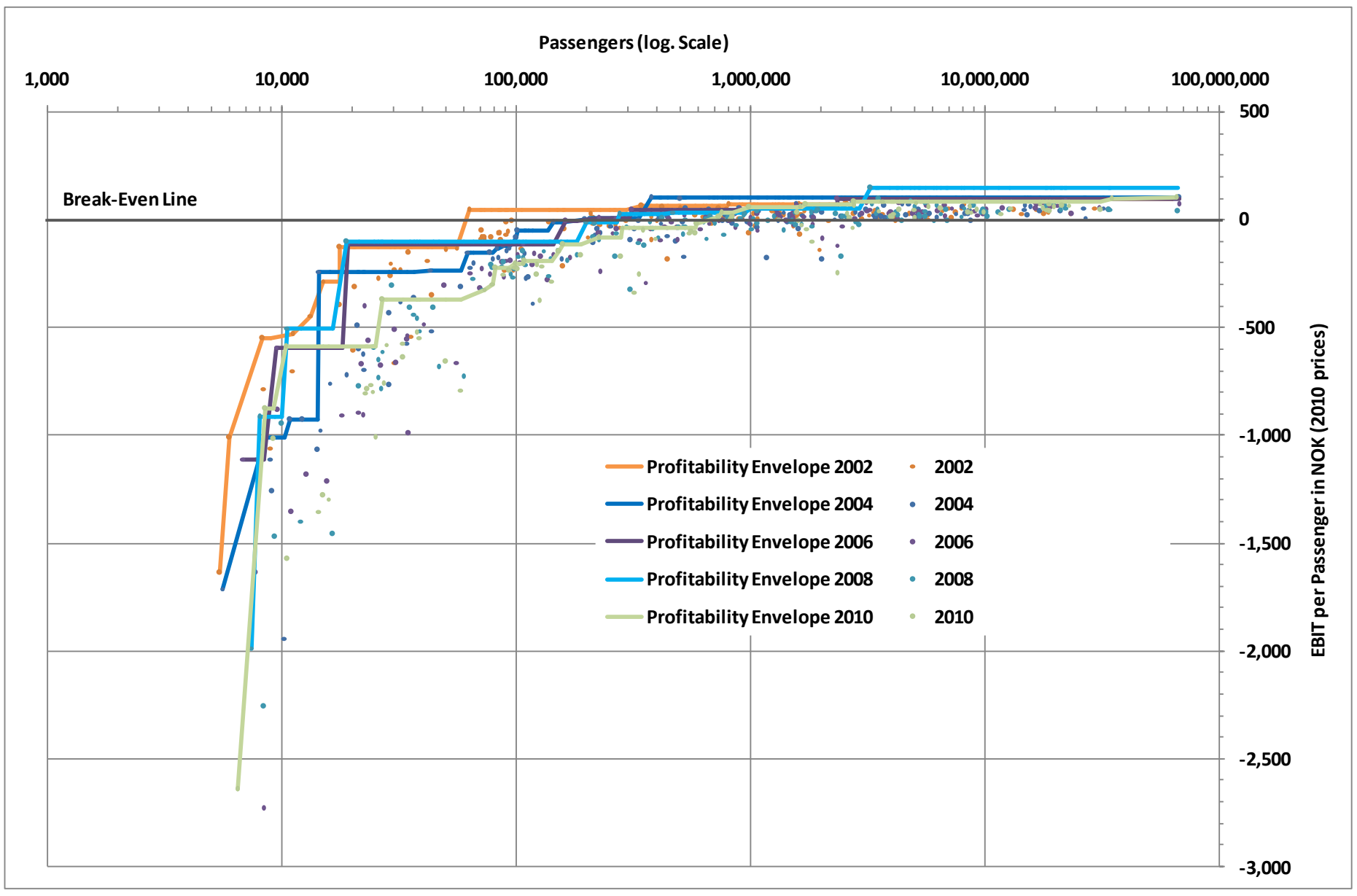

**Fig. 41.** Sample airports profitability trend by selected years (PPP-adjusted base Norway in 2010 prices; 2010 without Italy and France) (Source: Own survey data)

By dividing the sample under study into years we can derive more precise benchmarks for particular years which can be related to the observations. **Fig. 41** gives an overview of the profitability envelope change for even years between 2002 and 2010. As it can be seen the part of the envelope below the break-even line shifts downwards, which could be interpreted, that, in general, the profitability worsened for the low-demand airports. The break-even point moves right suggesting that a higher level of passengers is required to increase the chances of breaking even on the EBIT level, while the top part of the envelope stagnates at average EBIT levels of 60 to 100 NOK per PAX, with London-City (LCY) airport with 144 NOK per passenger being the exception.

In 2002 airports on the lower tail were more profitable than in the following years. The break-even point is located between such low passenger volumes as 17,680 and 63,000 defined by French airports Aurillac (AUR) and Bergerac-Roumaniere (EGC). Already in 2003 the break-even point significantly shifts to the right towards larger passenger volumes and is then situated somewhere between the observations Pescara (PSR) airport with about 300,000 passengers and an average EBIT of -3 NOK per passenger and Forli (FRL) airport (both in Italy) with about 350,000 passengers and an average EBIT of 4 NOK per passenger.

In the years 2004 to 2009 the break-even point lies between top performing French airports in the range of between 180,000 and 290,000 passengers. In 2010, where Italian and French airport data is missing, the break-even point is shifted even more significantly to the right and lies approximately between German airport Friedrichshafen (FDH) with 590,000 passengers and an average EBIT of -23 NOK per PAX and British airport Exeter (EXT) with 737,000 passengers and an average EBIT of 12 NOK per PAX. This leads to the conclusion that France is managing its low-demand airports very effectively, while achieving break even with around 200,000 passengers (**Table 19**).

**Table 19.** Airports frequently defining the profitability envelope (2010 without Italy and France)

| **Benchmark Airports** | **IATA Code** | **Country** | **2002** | **2003** | **2004** | **2005** | **2006** | **2007** | **2008** | **2009** | **2010** | **Number of Times defining the Envelope** |
|---|---|---|---|---|---|---|---|---|---|---|---|---|
| Tiree | TRE | UK | 1 | 1 | 1 | 1 | 1 | 1 | 1 | 1 | 1 | 9 |
| Aurillac | AUR | France | 1 | 1 | 1 | 1 | 1 | 1 | 1 | 1 | | 8 |
| Barra | BRR | UK | | 1 | 1 | 1 | 1 | 1 | 1 | 1 | 1 | 8 |
| Dinard-Pleurtuit-Saint-Malo | DNR | France | | 1 | 1 | 1 | 1 | 1 | 1 | 1 | | 7 |
| London City | LCY | UK | 1 | | | | 1 | 1 | 1 | 1 | 1 | 6 |
| Bournemouth | BOH | UK | | 1 | | 1 | 1 | | | 1 | 1 | 5 |
| Bergerac-Roumaniere | EGC | France | 1 | 1 | 1 | 1 | | 1 | | | | 5 |
| Southampton | SOU | UK | 1 | 1 | | 1 | 1 | | | | 1 | 5 |
| Graz | GRZ | Austria | | | | 1 | | 1 | 1 | 1 | 1 | 5 |
| London Heathrow | LHR | UK | | 1 | 1 | 1 | | | | | 1 | 4 |
| Stavanger | SVG | Norway | 1 | 1 | | 1 | | | | | 1 | 4 |
| Fagernes | VDB | Norway | 1 | | | | | | 1 | 1 | 1 | 4 |
| Caen-Carpiquet | CFR | France | | 1 | 1 | 1 | | | | | | 3 |
| Berlevåg | BVG | Norway | 1 | 1 | 1 | | | | | | | 3 |
| Svolvær | SVJ | Norway | | 1 | 1 | | | | | | 1 | 3 |
| Rennes | RNS | France | | 1 | 1 | 1 | | | | | | 3 |
| Exeter | EXT | UK | 1 | 1 | | | | | | | 1 | 3 |
| Hasvik | HAA | Norway | 1 | 1 | 1 | | | | | | | 3 |
| Bristol | BRS | UK | 1 | 1 | | | | | | | | 2 |
| Lorient-Lann-Bihoue | LRT | France | | | 1 | | 1 | | | | | 2 |
| Stokmarknes | SKN | Norway | | | 1 | | | | | | 1 | 2 |
| Røst | RET | Norway | 1 | 1 | | | | | | | | 2 |
| Ørsta-Volda | HOV | Norway | | | 1 | | | | | | 1 | 2 |
| Calvi-Sainte-Catherine | CLY | France | | | | | | | 1 | 1 | | 2 |
| Vadsø | VDS | Norway | | 1 | 1 | | | | | | | 2 |
| La-Rochelle-Ile De Re | LRH | France | | | 1 | 1 | | | | | | 2 |

The EBIT per PAX benchmarks back the argument, that volatile airports, which achieve profitability in certain years, but make losses in other years, should receive special attention from the airport operator, mainly towards increasing profitability or demand (for example by pushing route development) in order for the airport to reach breakeven or remain profitable in the long-term (**Table 19**). The observed trend of the profitability envelope is clearly downwards for low demand airports below one million passengers and upwards for the profitable airports above the break-even point. Obviously, profitability on the EBIT level could mainly be achieved either by reducing operating costs, such as staff costs, or by increasing charges by *Ramsey-Pricing* in cases of high elasticity of demand (please see lecture material on Marginal Cost Pricing from Prof. Dr. Hans-Martin Niemeier). Furthermore, profitability may be increased by a higher degree of commercialization, such as revenues from non-aeronautical sources, or by reducing depreciation, through postponing investments or extending economical lives. The ideal balance of these factors should ideally be under the responsibility of the airport management, in order to set incentives. In such cases where the airports lie below the profitability benchmarks (groups two and four), we do not expect the airports to have positive EBIT *per se*, but rather to achieve the respective benchmark by analyzing the business model of its best-practice peers.

Furthermore, we calculated the potential relative "efficiency gains" (Adler, Liebert and Yazhemski 2012) for each airport in a hypothetical scenario, if the according profitability benchmarks of other European airports would be reached. Reaching these gains means that increasingly fewer transactions would be made in form of cross-subsidies to loss-making airports in the system and increasingly more profits could be paid in dividends (or would be subject to taxes, hence would increase social welfare). We highly recommend a long-term sustainable efficiency policy, 1.) by continuously benchmarking with comparable peer airports, 2.) by analyzing business models of national and international best-practices and 3.) by trying to adjust profits (hence, finding the ideal balance between revenues and costs) to the best-practices. Consequently, cross-subsidizations from profitable to loss-making airports will be reduced, particularly in a national system or in a multi-airport organization. This effect is exemplarily shown by a cumulative curve for the sample of 139 European airports, showing cumulative EBIT (ordered from smallest to largest).

In **Fig. 42** it can be observed that 80% to 90% of the European airports are not profitable. In 2009 only the top 13% of the airports made a profit after depreciation. Overall, the 139 airports made a substantial system profit of 12 to 22 billion Kroners (1.5 to 2.75 billion Euros[1]). This picture changes significantly, if we account for the potential efficiency gains, which amount to about 28 to 38 billion Kroners (3.5 to 4.75 billion Euros) in absolute terms or increases of between 233 to 317%. The order of magnitude of these calculations shows that efforts to increase efficiency are a worthwhile endeavor, especially for larger systems of airports.

When including the feasible profitability gains in the cumulative distribution curves in **Fig. 42**, the immediate relief on the loss-making airports underneath the break-even line can be seen. With the gains included in the 2009 figures, significantly less overall cross-subsidies would be required for the loss-making airports. In this case the proportions are much different and about 50% of the 139 airports report profits. In more detail the results for the 139 European airports are as follows: In 2005, 54% of the airports report losses, which are compensated by the profits of the next 36% higher ranked airports. The break-even point lies at the rank of the top 18% of the airports. With efficiency gains in place the proportions would change significantly and 33% of the airports would report losses. These losses are compensated by the next 22% higher ranked airports; therefore, the break-even point lies at 55%. The top 45% airports in this case contribute to the system profits of 41.3 billion Kroners (~5.15 billion Euros).

[1] Average nominal exchange rate in 2010: 1 Norwegian Kroner = 0.124921205 Euro (Source: OECD)

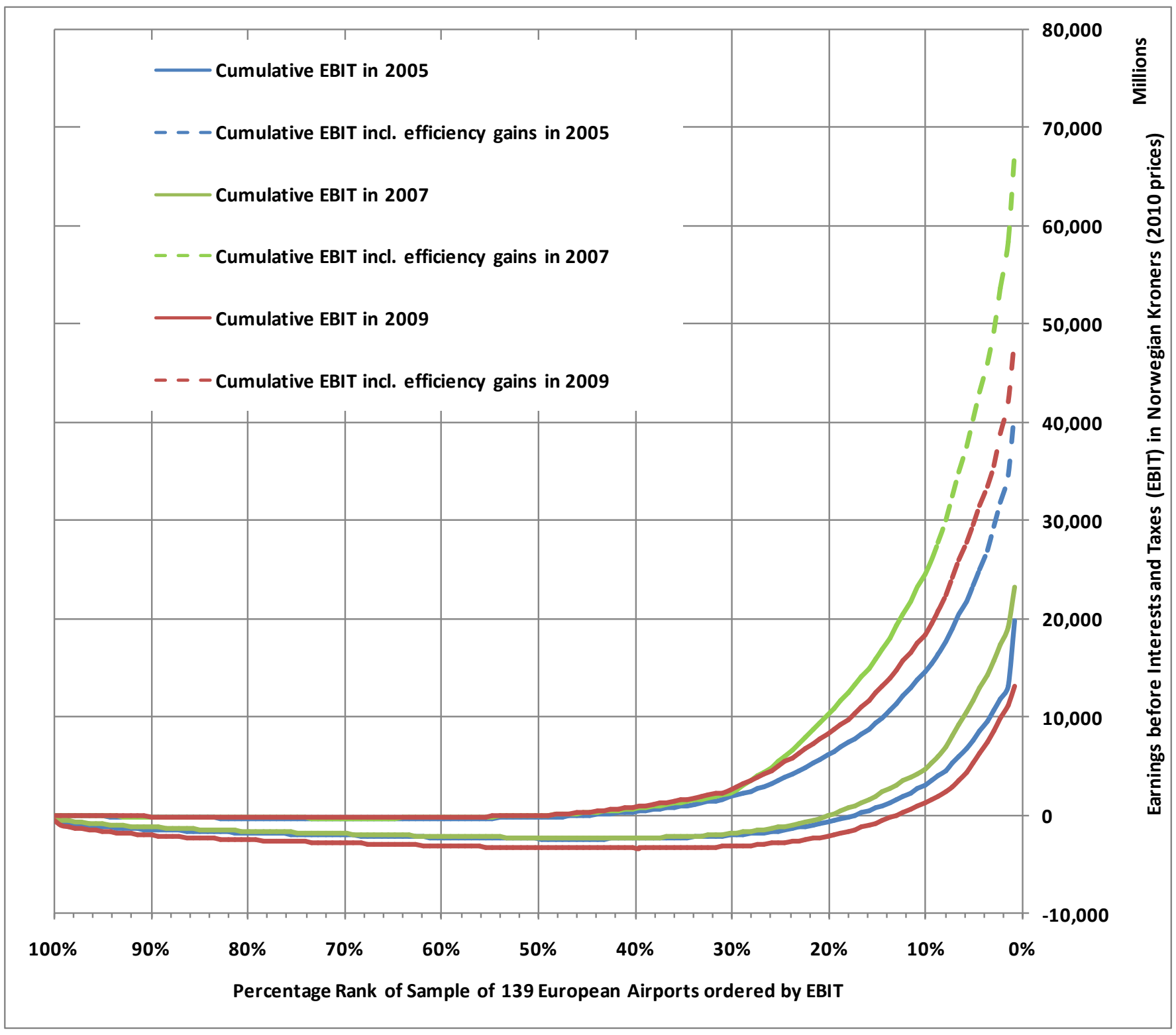

**Fig. 42.** Cumulative EBIT (in NOK, 2010 prices) with and without potential efficiency gains for 139 European Airports 2005, 2007, 2009 (Source: Own survey data)

In the strong financial year of 2007, the efficiency gains have also the most significant impact on the feasible cumulative profits. In this year, 55% of the airports report losses, which are offset by 25% higher ranked airports which are reporting profits. We find the break-even point at around the top 20% of the sample, which generate the system profits of around 23.2 billion Kroners (~2.9 billion Euros). If the efficiency gains are included in the 2007 data, we could expect the largest increases in system profitability. Again, 33% (100% – 67%) of the lowest ranked airports would report just losses, which are compensated by profits of the next 13% of airports to reach the system equilibrium. In 2007, 50% of all airports, if managed efficiently, could contribute to the total system profit of 67.4 billion Kroners (~8.42 Billion Euros). The system profits would almost *triple* by these profitability improvements.

In this sense, large airport operators could *ex-ante* postulate clear strategic management targets, whereas its achievements may *ex-post* be evaluated, for example, in cost-benefit analyses regarding policy changes such as changes in the structure of airport charges schemes. These efforts should be continuously conducted and closely monitored.

### 5.5 Conclusions and outlook

It is the aim of this article to present a method of comparison and to spot best-practice airports regarding profitability. We wanted to give some numerical evidence of “managerial effectiveness” (Doganis and Thompson 1975) and airport performance. It was recognized that a benchmarking approach needs the right choice of peers to give meaningful results, therefore a continuous scale rather than an arbitrary classification has been chosen, where we find the corresponding peers along the scale. Thereby particular justice is done towards small airports, especially in the region below 100.000 passengers, which have quite different characteristics compared to airports with, say, 100.000 to 1 million passengers and above. As it was shown with empirical data, airports with more than 1 million passengers can be expected to operate above breakeven, thus making profits. These profit-making airports show sharply increasing returns to scale on the absolute EBIT level, although the marginal returns to scale are decreasing. This means that the effort to exploit each additional passenger is much higher at large airports, than it is at small airports, where it may be much easier to increase the revenues per passenger. However, any small increase in revenues per passenger leads directly to increasing profits at large airports, because fixed costs and depreciation are already covered. With the construction of a maximum possibility frontier, which here is named ‘profitability envelope’, it is relatively straightforward to generate appropriate benchmarks and to recognize trends and shifts of the curve over time. In this way, airport management can formulate realistic strategic targets towards profitability. It can for example be calculated which level of profits per passenger may be feasible given a passenger demand forecast.

To be able to replicate this analysis, large amounts of data are needed, which mostly do not lie in the public domain. This is a major drawback of the presented method, because in practice the viewpoint may be one-dimensional. However, from the airport management perspective just peers with comparable passenger figures are required, which lowers the data collecting effort drastically. Therefore, our method is more suitable for large airport system operators, regulators, or other large-scale decision-makers.

If the fundamental effects, which are presented in this paper, are understood, it may seem reasonable for future research to aggregate the numerical evidence further (as was exemplarily shown with the cumulative curves in **Fig. 42** towards even fewer control variables, such as the *Gini*-coefficient, to measure the changes over time in the distribution of airport profits in a multi-airport environment.

## Appendix

**Table 20.** Airport Sample

| Country | Airport Name | IATA | Country | Airport Name | IATA |
|---|---|---|---|---|---|
| Austria | Graz | GRZ | Norway | Evenes (Harstad-Narvik) | EVE |
| Austria | Salzburg | SZG | Norway | Fagernes | VDB |
| Austria | Vienna | VIE | Norway | Florø | FRO |
| Belgium | Brüssel | BRU | Norway | Førde | FDE |
| Denmark | Billund | BLL | Norway | Hammerfest | HFT |
| Denmark | Copenhagen | CPH | Norway | Hasvik | HAA |
| France | Ajaccio | AJA | Norway | Haugesund | HAU |
| France | Aurillac | AUR | Norway | Honningsvåg | HVG |
| France | Bastia | BIA | Norway | Kirkenes | KKN |
| France | Bergerac-Roumaniere | EGC | Norway | Kristiansand | KRS |
| France | Biarritz | BIQ | Norway | Kristiansund | KSU |
| France | Brest | BES | Norway | Leknes | LKN |
| France | Caen-Carpiquet | CFR | Norway | Mehamn | MEH |
| France | Calvi-Sainte-Catherine | CLY | Norway | Mo I Rana | MQN |
| France | Dinard-Pleurtuit-Saint-Malo | DNR | Norway | Molde | MOL |
| France | Figari,Sud-Corse | FSC | Norway | Mosjøen | MJF |
| France | Grenoble-Isère Airport | GNB | Norway | Namsos | OSY |
| France | La-Rochelle-Ile De Re | LRH | Norway | Narvik | NVK |
| France | Lille | LIL | Norway | Ørsta-Volda | HOV |
| France | Limoges-Bellegarde | LIG | Norway | Oslo Gardemoen | OSL |
| France | Lorient-Lann-Bihoue | LRT | Norway | Røros | RRS |
| France | Lyon | LYS | Norway | Rørvik | RVK |
| France | Marseille | MRS | Norway | Røst | RET |
| France | Montpellier | MPL | Norway | Sandane | SDN |
| France | Nimes-Garons | FNI | Norway | Sandnessjøen | SSJ |
| France | Pau-Pyrénées | PUF | Norway | Sogndal | SOG |
| France | Perpignan-Rivesaltes | PGF | Norway | Sørkjosen | SOJ |
| France | Rennes | RNS | Norway | Stavanger | SVG |
| France | Tarbes-Lourdes-Pyrénés | LDE | Norway | Stokmarknes | SKN |
| France | Toulon-Hyères | TLN | Norway | Svalbard | LYR |
| Germany | Bremen | BRE | Norway | Svolvær | SVJ |
| Germany | Dortmund | DTM | Norway | Torp | TRF |
| Germany | Dresden | DRS | Norway | Tromsø | TOS |
| Germany | Düsseldorf | DUS | Norway | Trondheim | TRD |
| Germany | Erfurt | ERF | Norway | Vadsø | VDS |
| Germany | Friedrichshafen | FDH | Norway | Vardø | VAW |
| Germany | Hamburg | HAM | Switzerland | Geneva | GVA |
| Germany | Hannover | HAJ | Switzerland | Zürich | ZRH |
| Germany | Köln-Bonn | CGN | United Kingdom | Aberdeen | ABZ |
| Germany | Leipzig | LEJ | United Kingdom | Belfast International | BFS |
| Germany | Muenster | FMO | United Kingdom | Birmingham | BHX |
| Germany | München | MUC | United Kingdom | Bournemouth | BOH |
| Germany | Nürnberg | NUE | United Kingdom | Bristol | BRS |
| Germany | Stuttgart | STR | United Kingdom | Durham Tees Valley | MME |
| Italy | Alghero | AHO | United Kingdom | East Midlands | EMA |
| Italy | Bergamo | BGY | United Kingdom | Edinburgh | EDI |
| Italy | Bologna | BLQ | United Kingdom | Exeter | EXT |
| Italy | Cagliari | CAG | United Kingdom | Glasgow | GLA |
| Italy | Catania | CTA | United Kingdom | Humberside | HUY |

| Country | Airport Name | IATA | Country | Airport Name | IATA |
|---|---|---|---|---|---|
| Italy | Florence | FLR | United Kingdom | Leeds/Bradford | LBA |
| Italy | Forli | FRL | United Kingdom | Liverpool | LPL |
| Italy | Genoa | GOA | United Kingdom | London City | LCY |
| Italy | Lamezia Terme | SUF | United Kingdom | London Gatwick | LGW |
| Italy | Naples | NAP | United Kingdom | London Heathrow | LHR |
| Italy | Palermo | PMO | United Kingdom | London Luton | LTN |
| Italy | Pescara | PSR | United Kingdom | London Stansted | STN |
| Italy | Pisa | PSA | United Kingdom | Manchester | MAN |
| Italy | Trapani | TPS | United Kingdom | Newcastle | NCL |
| Italy | Turin | TRN | United Kingdom | Southampton | SOU |
| Italy | Venice | VCE | United Kingdom | Barra | BRR |
| Norway | Ålesund | AES | United Kingdom | Benbecula | BEB |
| Norway | Alta | ALF | United Kingdom | Campbeltown | CAL |
| Norway | Andøya | ANX | United Kingdom | Inverness | INV |
| Norway | Banak (Lakselv) | LKL | United Kingdom | Islay | ILY |
| Norway | Bardufoss | BDU | United Kingdom | Kirkwall | KOI |
| Norway | Båtsfjord | BJF | United Kingdom | Stornoway | SYY |
| Norway | Bergen | BGO | United Kingdom | Sumburgh | LSI |
| Norway | Berlevåg | BVG | United Kingdom | Tiree | TRE |
| Norway | Bodø | BOO | United Kingdom | Wick | WIC |
| Norway | Brønnøysund | BNN | | | |

# 6 Economic outlook for an air cargo market in the Baltic Sea Region[1]

Branko Bubalo

**Abstract.** The following final report was commissioned by University of Applied Sciences Wismar as part of the Baltic.AirCargo.Net research project. The project ran three years and was part of the European Union's Baltic Sea Region Programme 2007-2013. The project was funded by the EU Commission's European Regional Development Fund (ERDF). The author has written the final report for the project and presented important takeaways at the final conference of the Baltic.AirCargo.Project at University of Applied Sciences in Wismar in summer 2013. We highlight potential commodities that could be traded by air cargo and which could boost the local economies in the larger Baltic Sea Region (BSR). We gave insights into the volume and flow of traded products within Europe. It was evident that only products with a large value per ton seem adequate to justify transport by air. The advantage of air transport is clearly the quick distribution of products regionally and globally, whereas traditional modes of transport, e.g. by trucking, require lengthy processes especially at the EU/Non-EU borders. We give recommendations on which products are predestined for transport by air and which air routes may need to be established and expanded. We made use of Origin-Destination matrices, similar to the methods presented by Wassily Leontief in his landmark publication Input-Output Economics.

## 6.1 Introduction

The Baltic Sea Region (BSR) is a peripheral region in the North-Eastern hemisphere of Europe and is defined for the purposes of this study by the following eleven countries (and parts thereof; in alphabetical order): Belarus, Denmark, Estonia, Finland, Germany, Latvia, Lithuania, Norway, Poland, Russia (incl. exclave Kaliningrad) and Sweden (**Fig. 43**). Norway qualifies as a group member due to its role in Scandinavia and because of its connection to the Baltic Sea through the Skagerrak in the North of Denmark. However, we have omitted some Norwegian airports in the West and far North of the country in our study.

The Baltic States with Belarus, Estonia, Latvia, and Lithuania have their historic roots as former states of the Soviet Union. When the Soviet Union collapsed in the early nineteen-nineties, several former states sought for independence.

[1] The author is indebted to Prof. Gunnar Prause, Anatoli Beifert and Laima Gerlitz for funding and assistance.

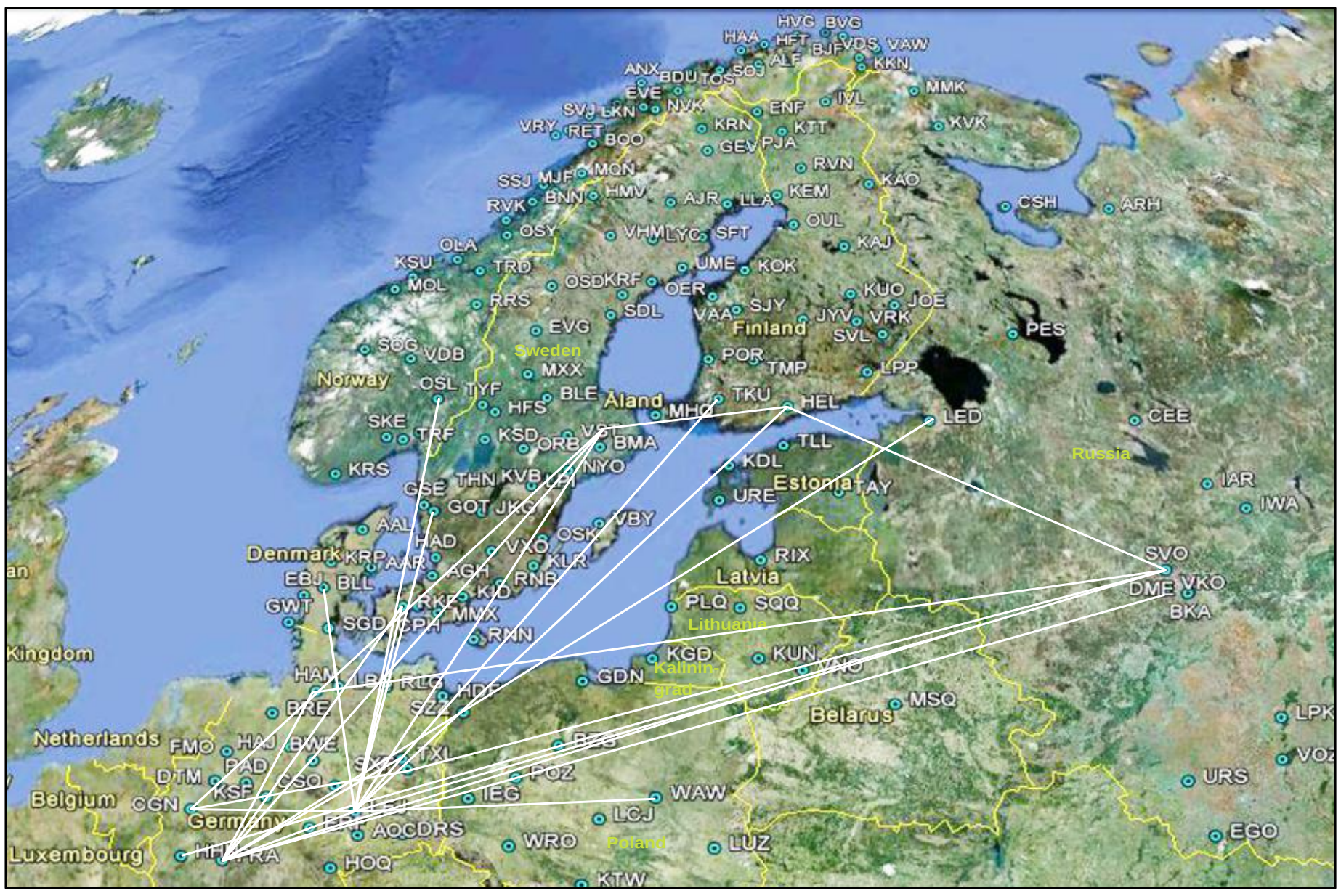

**Fig. 43.** Map of the Baltic Sea Region, sample airports and intra-regional main cargo routes today (Map from Google Earth; Route data from Eurostat, OAG and FlightStats)

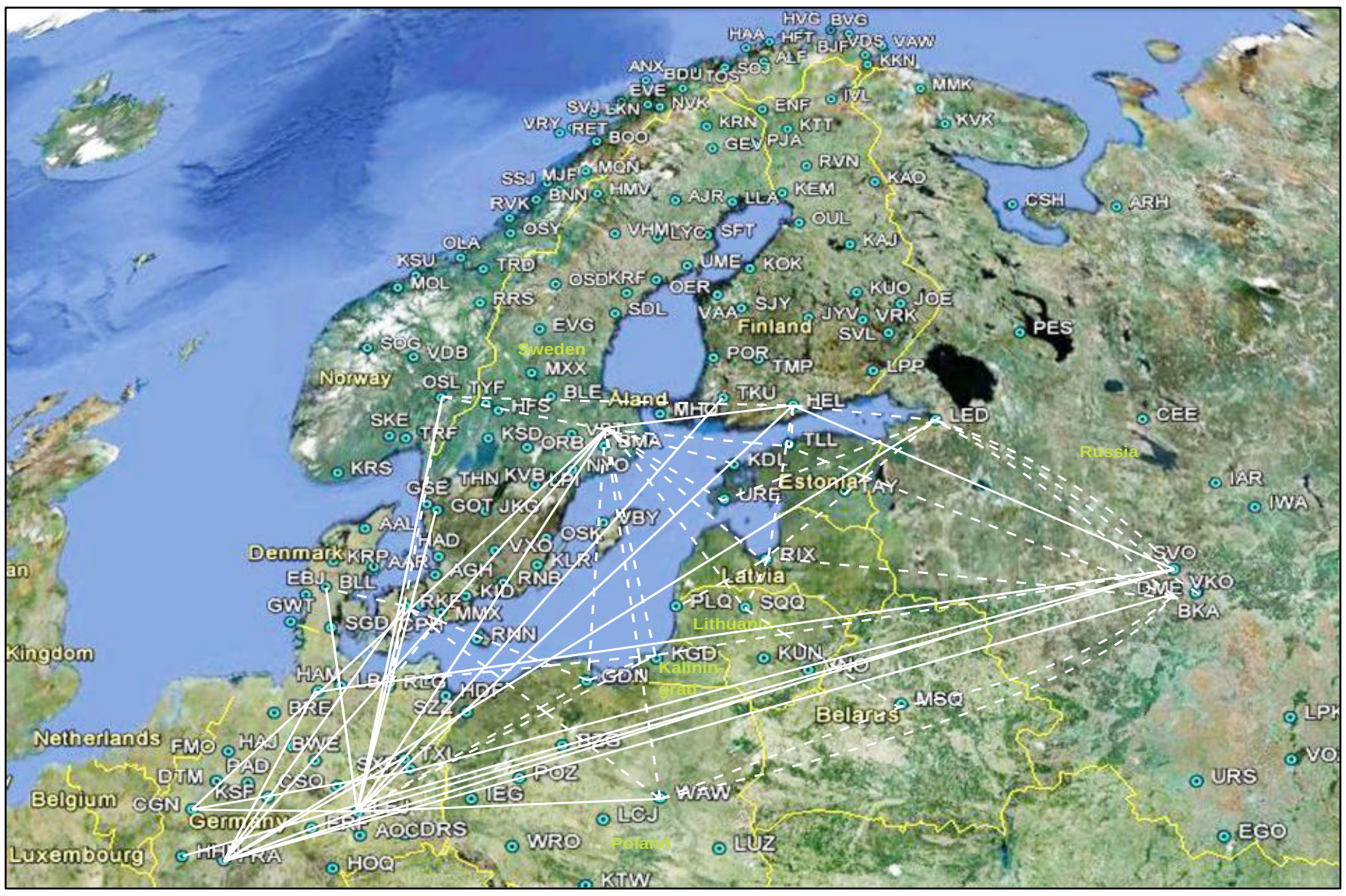

**Fig. 44.** Map of the Baltic Sea Region, sample airports and intra-regional main cargo routes in a hypothetical 2030 scenario (Map from Google Earth; Route data from Eurostat, OAG and FlightStats)

Today, all the Baltic States, except for Belarus, are either EU-members or member candidates. Belarus remains tied closely to Russia mainly because both countries have a long partnership in the production, control, and distribution of petroleum products over a network of pipelines. These pipelines bring the products, such as mineral fuels from the distilleries in Russia or Belarus to western-Europe or the harbors of the Baltic Sea for further shipping. Some Baltic States serve as transit states in this logistic chain.

The aim of this study is to aggregate numerical evidence of the financial and physical interaction, i.e. the trade and transportation of goods and passengers, in the BSR countries and among its relevant airports. We will analyze within and outside regional trade and traveling of people, and we will give ideas regarding development potentials for a Baltic air cargo market.

Available literature and data on the BSR are scarce, especially regarding detailed trade with Belarus or Russia; forecasts or regional development strategies are inconsistent, missing or not well defined. Difficulties arise since we want to observe countries across political structures. This study will fill this gap in the recent literature and will contribute to the on-going discussions and to the existing political and economic strategies for the BSR (Sprūds 2012), which are being initiated on the European level. The European Union (EU) strongly supports the development of the markets in the BSR. EU development funds, as for example through the Baltic Sea Region development program 2007-2013 (http://eu.baltic.net/), promote the economic growth in the region. With regard to the specific air cargo market in the BSR our literature review is to a large degree based on previous reports written within the Baltic.AirCargo.Net (http://www.balticaircargo.net/) research project. We appreciate the previous work by Mączka (2013) and Ivanov and Stigaard (2013) for initial thoughts and direction. In order to systematically define the region under study we chose not to include all regions and its airports within the national boundaries, we rather arbitrary set the dimensions of the BSR by the coordinates of the German airports Cologne (CGN) for the western and Hahn (HHN) for the Southern boundary. Mehamn airport (MEH) in Norway defines the Northern and Russian airport Ivanovo (IWA) defines the Eastern boundary of the region (**Table 21**). In total the BSR consists of 290 civil airports (i.e. having a 3-letter IATA code) of which 176 airports (61%) have regular scheduled traffic (**Table 22**).[1] The full list of airports can be found in the Appendix (**Table 34**). When looking at the distribution of airports across the BSR (**Fig. 45**), we find that, as expected, Germany has a very high concentration of airports per area of land in the lower left grids. Sweden has many secondary airports concentrated in the south-west, where also its main hub Stockholm-Arlanda airport (ARN) is located.

[1] We define regular scheduled traffic by at least 1 flight observation per week in the reporting period March 16th to 22nd 2009 (OAG), April 26th to May 6th and May 22nd to 31st 2013 (FlightStats).

**Table 21.** Region defining airports (FlightStats 2013)

| City | Airport Name | IATA Code | ICAO Code | Longitude | Latitude | Country |
|---|---|---|---|---|---|---|
| Mehamn | Mehamn | MEH | ENMR | 71.03333 | 27.833332 | Norway |
| Hahn | Frankfurt – Hahn | HHN | EDFH | 49.948334 | 7.264167 | Germany |
| Cologne | Cologne Bonn | CGN | EDDK | 50.878365 | 7.122224 | Germany |
| Ivanovo | Ivanovo | IWA | | 56.942955 | 40.944546 | Russia |

Finland's airports to the contrary are evenly dispersed among its landmass. Norwegian airports are thinly dispersed in the north-western outskirts of the BSR. Surprisingly, Russia only has a few active airports with reported scheduled traffic (18 of 33 region total) located in the bordering region to the neighboring European BSR partner countries, however, two airports, Kaliningrad (KGD) and Saint Petersburg (LED), are on the Baltic Sea.

**Table 22.** Airports under study in the Baltic Sea Region (OAG, FlightStats)

| Countries | Number of Airports in the Region[1] | Number of Airports with scheduled Traffic[2] |
|---|---|---|
| **Belarus** | 7 | 1 |
| **Denmark** | 18 | 9 |
| **Estonia** | 6 | 4 |
| **Finland** | 31 | 22 |
| **Germany** | 66 | 23 |
| **Latvia** | 3 | 1 |
| **Lithuania** | 6 | 4 |
| **Norway** | 46 | 41 |
| **Poland** | 18 | 12 |
| **Russia** | 33 | 18 |
| **Sweden** | 56 | 41 |
| **Sum** | **290** | **176** |

Notes: (1) Only airports with IATA Code and located within the region defined by the following coordinates have been considered (Northern boundary: 71.03333° N; Southern boundary: 49.948334° N; Western boundary: 7.122224° E; Eastern boundary: 40.944546° E).
(2) Recorded Traffic in the reporting period.

Most of the traffic from or to Russia flows through the Moscow airports Sheremetyevo (SVO) and Domodedovo (DME) in the far east of the defined region; therefore, these provide connectivity to the remote parts of Russia and the Far East. Poland has 18 airports of which twelve have regular scheduled traffic.

The "core" Baltic states Belarus, Estonia, Latvia, and Lithuania have 22 airports in total of which ten (one, four, one and four, respectively) have regular scheduled traffic. Most of those airports are located near the Baltic Sea (with obvious exception of Belarus). Denmark's nine active airports are surrounded by close competitors from Germany, Sweden, and Norway. Despite the competition its main hub Copenhagen airport (CPH) has a large share of cargo traffic and ranks seventh in terms of number of cargo flights, behind German airports Frankfurt international (FRA; 1.), Frankfurt-Hahn (HHN; 3.), Halle/Leipzig (LEJ; 4.) and Köln/Bonn (CGN; 6.) and the Russian airports SVO (2.) and DME (5.). CPH in Denmark holds a long record of being voted 'best international airport' by (passenger) surveys, for example in the popular Airport Council International (ACI) Airport Service Quality (ASQ) ranking (http://www.airportservicequality.aero/).

Based on our sample of 176 airports some original data was collected from several sources. It must be clear that only some specific areas of interest can be tackled under this descriptive approach and therefore the expressed views may be limited. However, we tried to gather as much evidence as is available to enrich the current discussions. The author of one of the pre-studies concludes that a holistic analysis is currently impossible due to poor data availability (Mączka 2013). Yet, it was possible to analyze data on a variety of topics relevant to air traffic in general, and air cargo traffic in particular. For example, in Chapter 6.2 we shed light on the industries engaged in international trade in the BSR and on the type of traded products transported by different modes of transport. One of the goals is to find adequate traded products, which would justify and offset the high costs of transportation by air. Based on flight schedule data and aggregated annual data we could furthermore give evidence to the interconnection between BSR countries in air travel (and air cargo) by their Origin-Destination matrix and main routes (**Fig. 43** und **Fig. 44**). Chapter 6.4 will give details on these interconnections.

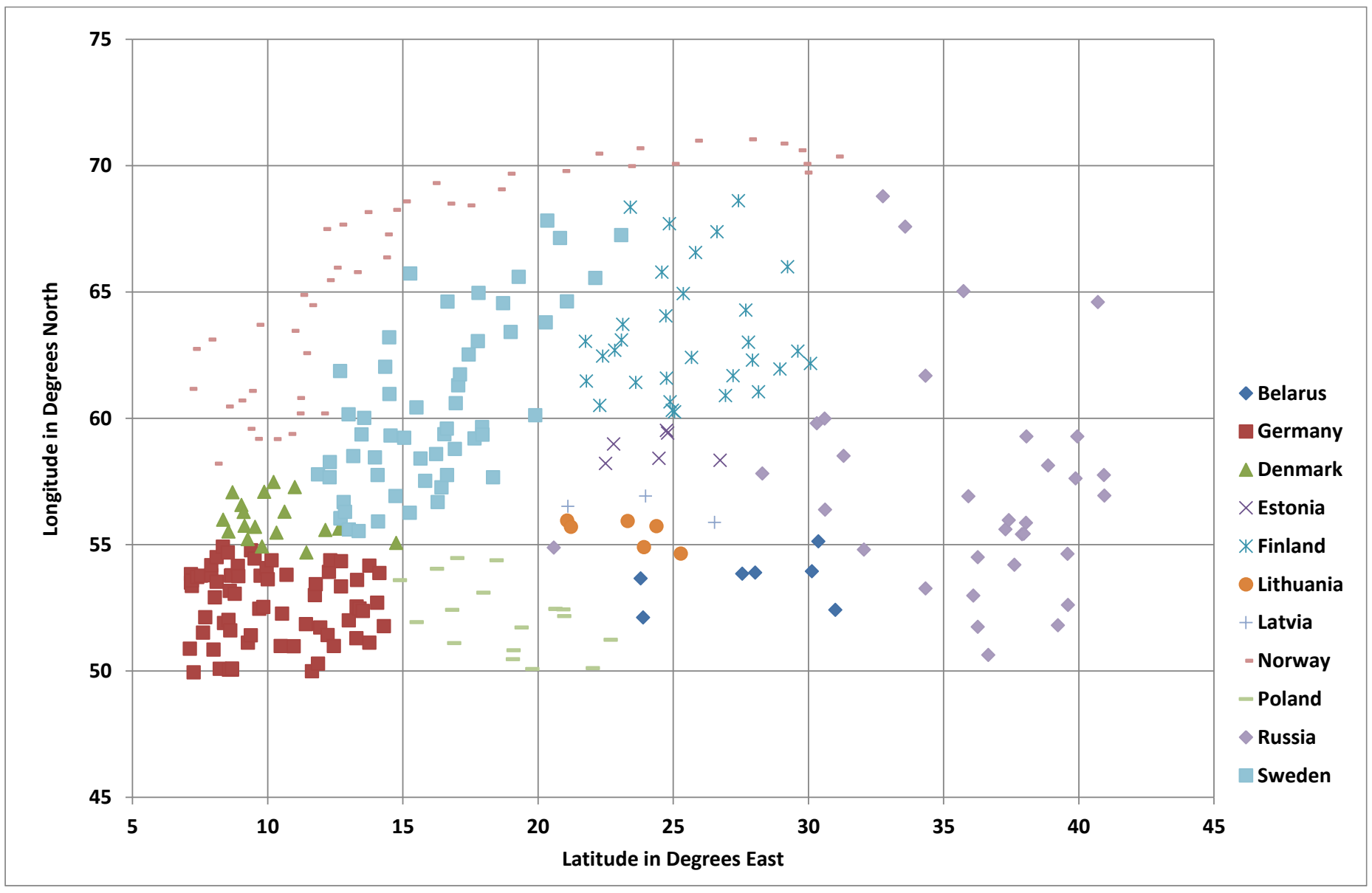

**Fig. 45.** Distribution of Airports in the Baltic Sea Region (Own Illustration, scale and projection).

We collected and compared information on the level of landing charges for cargo aircraft at the main hubs in the region so we can make qualified judgements regarding price competition. In a developing market such as the BSR airlines will try to find the most economical connections for transporting goods (and passengers) from point A to point B. The savings in transportation costs will be passed on to the customers thus will directly influence their decision making. Chapter 6.3 will highlight the total turnaround costs for cargo aircraft operating at airports in the BSR. The structure and content of the charges schemes is tremendously different across airports in the region. We found very simple pricing schemes regarding airport services for cargo aircraft, of which some only contained landing or take-off charges. In contrast some airports collect charges for many additional services, such as noise abatement, security, parking, emissions and so on. These differences make it very difficult to compare charges across airports. For example most airports in Poland publish charges schemes in which interestingly growth- or incentive schemes are in place to promote growth and to specifically attract Low-Cost Carriers (LCCs).[1] LCCs have the potential to generate large amounts of traffic (movements and passengers) which makes them attractive to regional airport. In these cases, usually the income for the airport stems to a lesser degree from aeronautical charges (landing or passenger charges), but to a larger degree from non-aeronautical income (Bubalo and Gaggero 2015).[2] However, air cargo plays an insignificant role in LCCs operations, which relies on tight flight schedules and quick aircraft turnarounds (frequently below 30 minutes). Cargo airlines usually have turnaround times of a few hours (but depending on aircraft size on average less than three hours) between arrival and subsequent departure.

We also investigated in Chapter 6.3 the economics of airports in the BSR and particularly at which level of demand, expressed in Workload Units (WLU = 1 PAX or 100kg of Cargo[3]), a financial break-even could be reached (on the example of airports in Sweden, Poland and Norway). We assume the payment of subsidies (to the disadvantage of the public and surrounding communities) in such cases, where airports operate below the break-even point and make operating losses. If the "critical mass" or demand level is surpassed an airport operates above the break-even point and can generate operating profits. In Sweden only four (out of 41) airports serve a relevant amount of cargo (>5% of Total WLU) and show best-in-class profitability. Three of which make profits (Stockholm-Arlanda [ARN], Goteborg [GOT] and Malmö [MMX]), but Jönköping (JKG) makes a loss. We will go into more detail in Chapter 6.3.

[1] Low-cost carriers (LCCs) represent airlines which offer the most basic services to their customers ("No-Frills"), however, at a very low ticket fare. This business model originates in the U.S. with Southwest airlines and emerged in Europe during the late nineties with Ryanair and Easy Jet. Low fares are realized through low operating costs, e.g. low salaries and a single aircraft type fleet, and a fast turn-around policy.

[2] Non-aeronautical income is typically generated by additional services for the passenger, such as (concessions from) food and beverages, shopping, or car parking etc.

[3] This equivalence can be disputed from an operational perspective but is a lack of alternatives and evidence to the contrary.

Our data originates to a large degree from publicly available sources. We have obtained annual traffic and international trade data from Eurostat, the statistical bureau of the EU. For an oversight on recent airport activity, flight schedules including aircraft type and carrier were downloaded from Official Airline Guide (OAG) and from FlightStats.com, both provide flight schedule information for research or business purposes. EUROCONTROL, the European Organization for the Safety of Air Navigation, hosts airport and airspace related regulations, charts and manuals through the European AIS Database (EAD; http://www.eurocontrol.int/ead). Other data on airport finances is from public reports (Näringsdepartementet 2007, GAP 2012, Samferdselsdepartementet 2013), annual reports, financial statements and questionnaires collected within the German Airport Performance (GAP) research project.

### 6.2 Baltic Sea Region and its economy

The European Union led by the European Commission (EC) is pushing to promote growth and economic prosperity in the BSR, for example through programs such as the European Regional Development Fund (ERDF). The ERDF website states: This fund "aims to strengthen economic and social cohesion in the European Union by correcting imbalances between its regions. In short, the ERDF finances:

- Direct aid to investments in companies (SMEs) to create sustainable jobs;
- Infrastructures linked notably to research and innovation, telecommunications, environment, energy and transport;
- Financial instruments (capital risk funds, local development funds, etc.) to support regional and local development and to foster cooperation between towns and regions [and]
- Technical assistance measures."[1]

The ERDF also contributes most of the funds for the Baltic Sea Region development programme 2007-2013.

To underline the importance for the EU to maintain social stability in the former Russian Baltic states further evidence is given. Just recently during the global banking crisis in 2008 and 2009, the Baltic state Latvia was hard hit by the events taking place in the U.S. and in Europe. Despite the years of high growth rates, which remained between 5% and 10% growth of Gross Domestic Product (GDP) over a period of at least ten years, the economy in Latvia suddenly declined (Figure 4). The extreme downturn of the economy of -17% expressed in change of GDP between 2008 and 2009, led to various national problems, such as an increasing unemployment rate and decreasing incomes (Staehr 2012).[2] However, Latvia restored its economy through some drastic cost saving programs in government spending. Loans from the European

[1] http://ec.europa.eu/regional_policy/thefunds/regional/index_en.cfm

[2] http://ec.europa.eu/economy_finance/publications/occasional_paper/2012/op120_en.htm

Union or the International Monetary Fund (IMF) were not requested, but both institutions would have supported the Latvian economy if it were needed.

Figure 5 makes it clear how tightly connected the economies of the Baltic States are. The patterns of the average growth rates are remarkably similar, especially Estonia's compared to Lithuania's. The Latvian trend line is slightly above the others during boom years but is hitting rock bottom during the bust years.

The trends for Czech Republic and Poland are more "moderate", with growth rates between -1% and 5% over the years, whereby the recession in 2009 seemed to have hit Czech Republic only mildly and shortly, and Poland was hit none at all (similar to Germany).

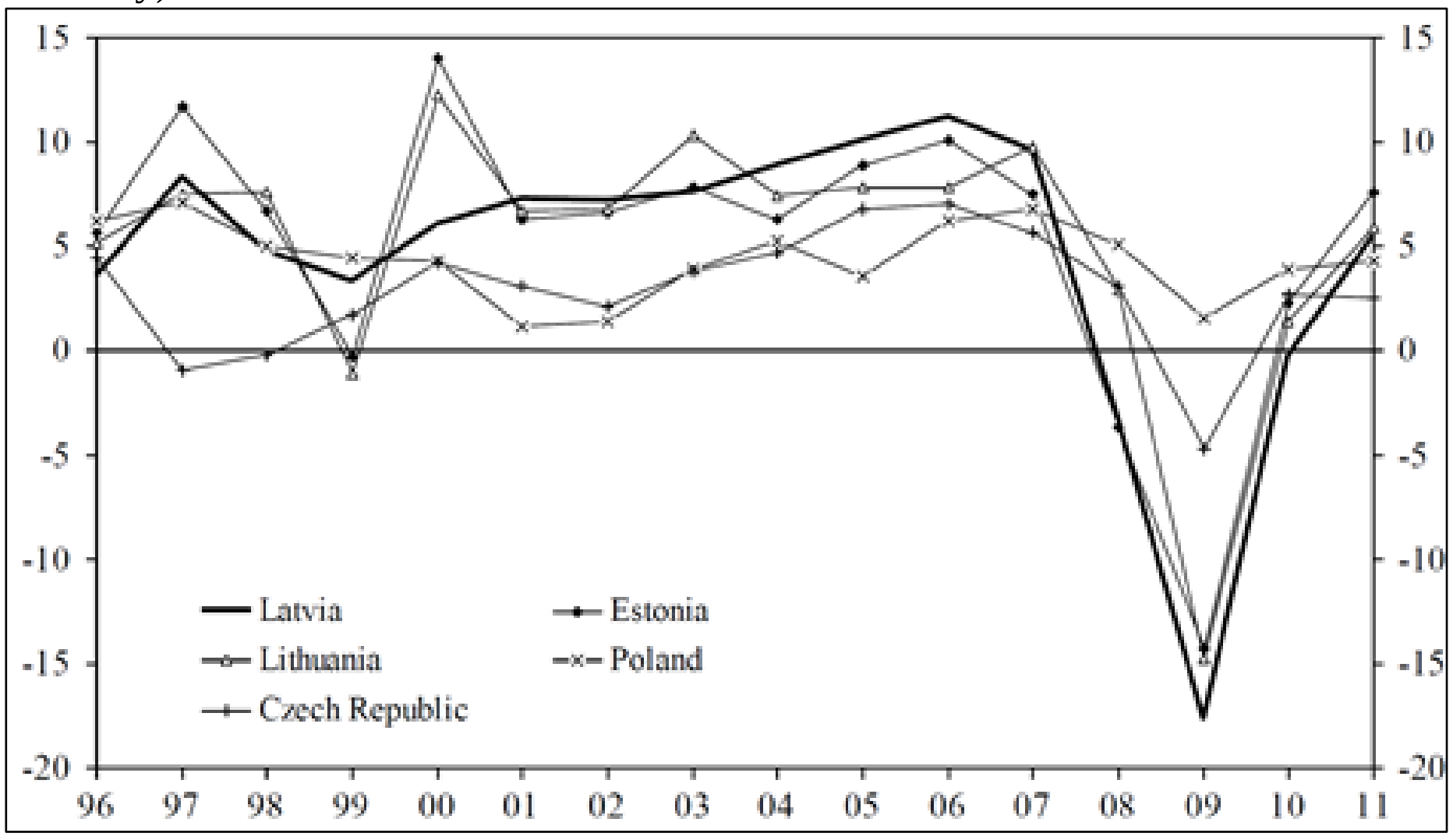

**Fig. 46.** Annual GDP growth in Latvia and four other CEE countries, percentage (Source: EU Commission 2012)

We want to put an emphasis on GDP and GDP growth, since it is widely recognized that changes in GDP have a direct consequence on air travel demand (**Fig. 46**).[1] This means that an increase of GDP also leads to an increase of air traffic demand, and a fall in GDP leads to a decrease in air traffic demand (Airbus 2012). However, the trend of both indicators over time show that the air traffic demand reacts sensitive to a change in GDP, and the up and down swings are more extreme in the case of air traffic. We assume that air cargo traffic derives from general air traffic demand, therefore the correlation between air cargo traffic and GDP is believed to be valid as well. With reference to **Fig. 47** the consequences are that only a stable and growing economy leads to an increase in air traffic in general, or air cargo traffic. The recession in 2008 and 2009 clearly dampened the developments in the Baltic air cargo sector. In 2011 world traffic growth stabilized at around 5% per annum (**Fig. 47**).

[1] http://www.airbus.com/company/market/forecast/

For the transformation of the BSR into a self-sustained and prosperous region it is important to have a broad range of traded, manufactured, and consumed products. We analyzed the data with the aim to finding markets and types of products in the BSR, which are suited for air transport and which are priced high enough to compensate the transportation costs.

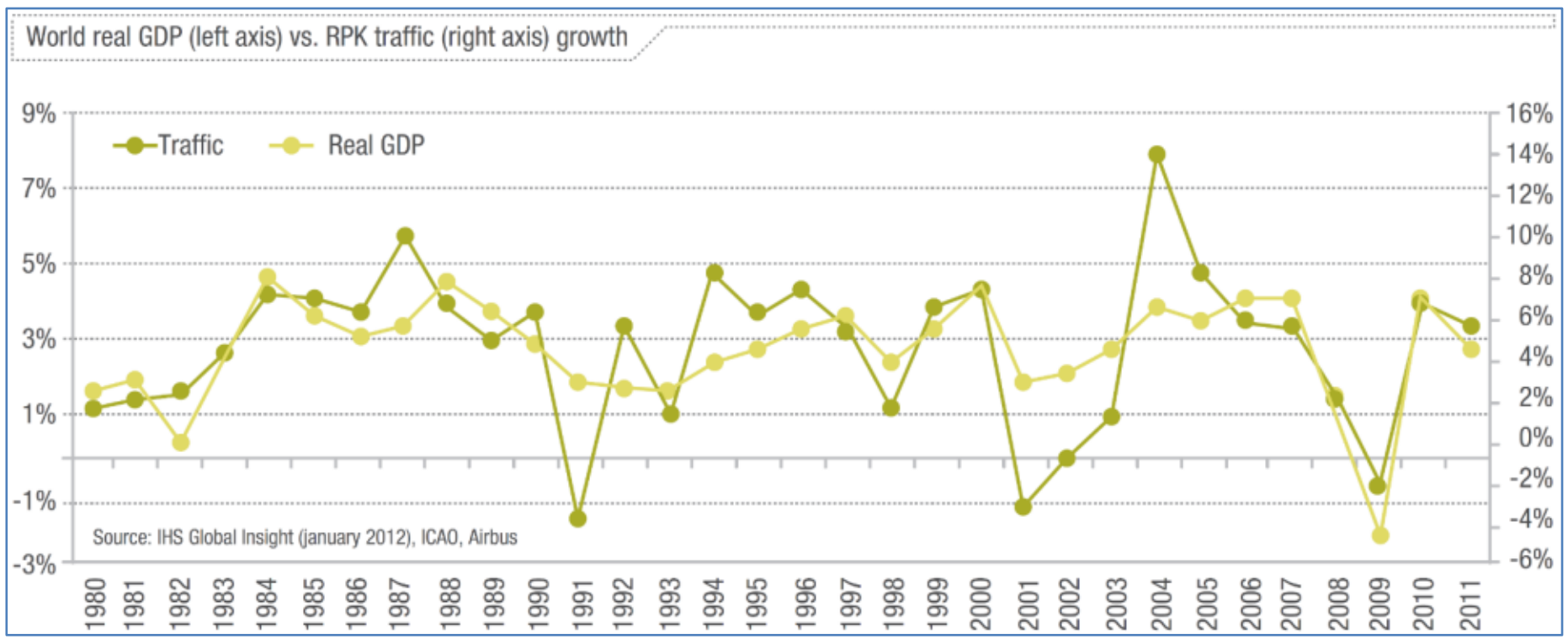

**Fig. 47.** How air traffic is correlated to development of the economy (Airbus 2012).

Ivanov and Stigaard (2013) point to "highly perishable goods", such as fresh foods, flowers and other products which are time critical. For them, one of the main advantages of air cargo transportation is the high "speed and frequency".

For isolating products with a large "value per unit of weight" we have analyzed data from Eurostat's international trading database Comext.[1] In particular we have focused on products on the main category of "machinery, transport equipment, manufactured and miscellaneous articles." Additionally, to the findings of Ivanov and Stigaard (2013) we have found electronic products and valuable industrial spare parts suitable for air transportation. This does not lay in the fact that these products are "time critical" in their distribution, but rather so valuable that a safe and secure transport across national borders needs to be guaranteed. The next two sections will give some results on this analysis.

## International trade

In order to understand the structure of the economies in the BSR and the Baltic States we analyzed the international trade flows with regard to quantities imported and exported, and its unit value, measured in Euro per Ton. Unfortunately, our data is limited to the countries of the European Union (EU), Denmark, Estonia, Finland, Latvia, Lithuania, Poland, Sweden, and Germany. Data on Norway is missing in this part of the study, however, international trade data on Russia and Belarus is published in statistical dossiers. The available data will give insights which countries in the region import or export products with a high value compared to its quantity. Especially in air cargo transportation the goal is to find products, which have a high unit value, since weight is an important factor. Air cargo transportation is only economically feasible if the high value per unit of weight justifies transportation through this mode.

---

[1] http://epp.eurostat.ec.europa.eu/NavTree_prod/everybody/BulkDownloadListing?sort=1&dir=comext

It is evident from **Table 23** that imports to the BSR states from within the EU are quite harmonized regarding its unit values. We also observe that the total imported quantities differ dramatically across the countries under study, mainly due to its different size in population and economic strength. The quantities, e.g. in 2012, range from less than 10 million tons (in Latvia) to over 360 million tons (in Germany). However, in the same year the unit values for the traded goods within the EU only fluctuate between 1,387 Euro per Ton in Latvia and 1,905 Euro per Ton in Denmark (**Table 23**). The proportional upward trend in general price and value levels can be observed at all countries imports and exports.

**Table 23.** Trade partners and flows for countries in the BSR in 2012 (Eurostat)

| | | **Flow** | | | |
|---|---|---|---|---|---|
| | | **Export** | | **Import** | |
| **Country** | **Trade Partner** | **Quantity in Tons (in millions)** | **Unit Value (in Euros per Ton)** | **Quantity in Tons (in millions)** | **Unit Value (in Euros per Ton)** |
| **Denmark** | **Extra EU 27** | 7.74 | 3,892 | 19.82 | 1,057 |
| | **Intra EU 27** | 30.57 | 1,703 | 26.58 | 1,905 |
| **Estonia** | **Extra EU 27** | 3.78 | 1,131 | 3.08 | 893 |
| | **Intra EU 27** | 9.12 | 907 | 6.26 | 1,759 |
| **Finland** | **Extra EU 27** | 16.48 | 1,601 | 33.28 | 665 |
| | **Intra EU 27** | 27.26 | 1,117 | 23.46 | 1,586 |
| **Germany** | **Extra EU 27** | 87.86 | 5,344 | 231.60 | 1,432 |
| | **Intra EU 27** | 286.42 | 2,184 | 372.14 | 1,550 |
| **Latvia** | **Extra EU 27** | 4.17 | 962 | 4.90 | 598 |
| | **Intra EU 27** | 14.93 | 466 | 7.54 | 1,387 |
| **Lithuania** | **Extra EU 27** | 7.09 | 1,284 | 17.67 | 613 |
| | **Intra EU 27** | 18.00 | 776 | 8.77 | 1,624 |
| **Poland** | **Extra EU 27** | 17.76 | 1,951 | 63.29 | 791 |
| | **Intra EU 27** | 67.97 | 1,591 | 59.64 | 1,719 |
| **Sweden** | **Extra EU 27** | 32.95 | 1,757 | 34.11 | 1,219 |
| | **Intra EU 27** | 59.10 | 1,291 | 47.11 | 1,802 |

A strong increase in price and value level for imports can be observed in Estonia. In this country the unit values have increased 75% for imports from inside the EU (**Fig. 48**) and 100% for import from outside the EU (**Fig. 50**) between 2009 and 2012. The import quantities from the EU are similar across the three Baltic States, Lithuania, Latvia and Estonia. Lithuania's exports are very strong among the Baltic States (**Fig. 49** and **Fig. 51**). As we will see below Lithuania's imports and exports from outside of the EU are probably due to petroleum products which transit the country originating from Belarus and Russia.

Regarding the exports to the 27 EU member states (**Fig. 50**) we can observe that the unit values already differ considerably between the BSR states. Although the total quantities exported to the EU are similarly distributed as the imports (with notable exceptions particularly on the part of Latvia and Lithuania, which have a very advantageous trade balance in the EU of 7.54 million tons in imports and 14.93 million tons in exports, and of 8.77 million tons in imports and 18 million tons in exports, respectively), the unit values for exported products range from 598 Euro per Ton in Latvia to 2,184 Euro per Ton in Germany.

When we look at the imports from (**Fig. 50**) and exports to (**Fig. 51**) outside the EU, another picture emerges. Particularly in the case of Germany there is a huge imbalance towards exports to outside of the EU in terms of quantity. In 2012 Germany imported 231.6 million tons of products from outside the EU, but it only exported 87.9 million tons of products to countries outside the EU. However, this trade imbalance is compensated by the remarkably high values per ton of exports. Exports are roughly four times as valuable as the imports per ton of products -5,344 Euro per ton of exports compared to 1,432 Euro per ton of imports (to/from outside the EU).

This is to prove why we put emphasis towards "high value per weight" products, which may be suitable for air cargo transport. Denmark shows a similar phenomenon with a value of 1,057 Euro per ton for its Extra-EU imports and a value of 3,892 Euro per ton for its exports.

Sweden to the contrary has a more "harmonized" balance between the trade partners from the EU and from outside. The traded quantities in Sweden are about similar in import and exports with differences only in value. Trade within the EU is about 50 million tons and with outside of the EU at about 30 million tons (2. highest exports to outside the EU, after Germany). Exports within the EU are lower in value than the imports; in trade with outside the EU unit values are higher for exports than the imports (**Fig. 48** to **Fig. 51**).

Poland shows strong trade activities and ranks second (after Germany) in total imports and in EU exports. Poland is able to take advantage of low value imports (791 Euros per ton) in contrast to its outside EU exports (1,951 Euros per ton) in 2012 (**Fig. 50** and **Fig. 51**).

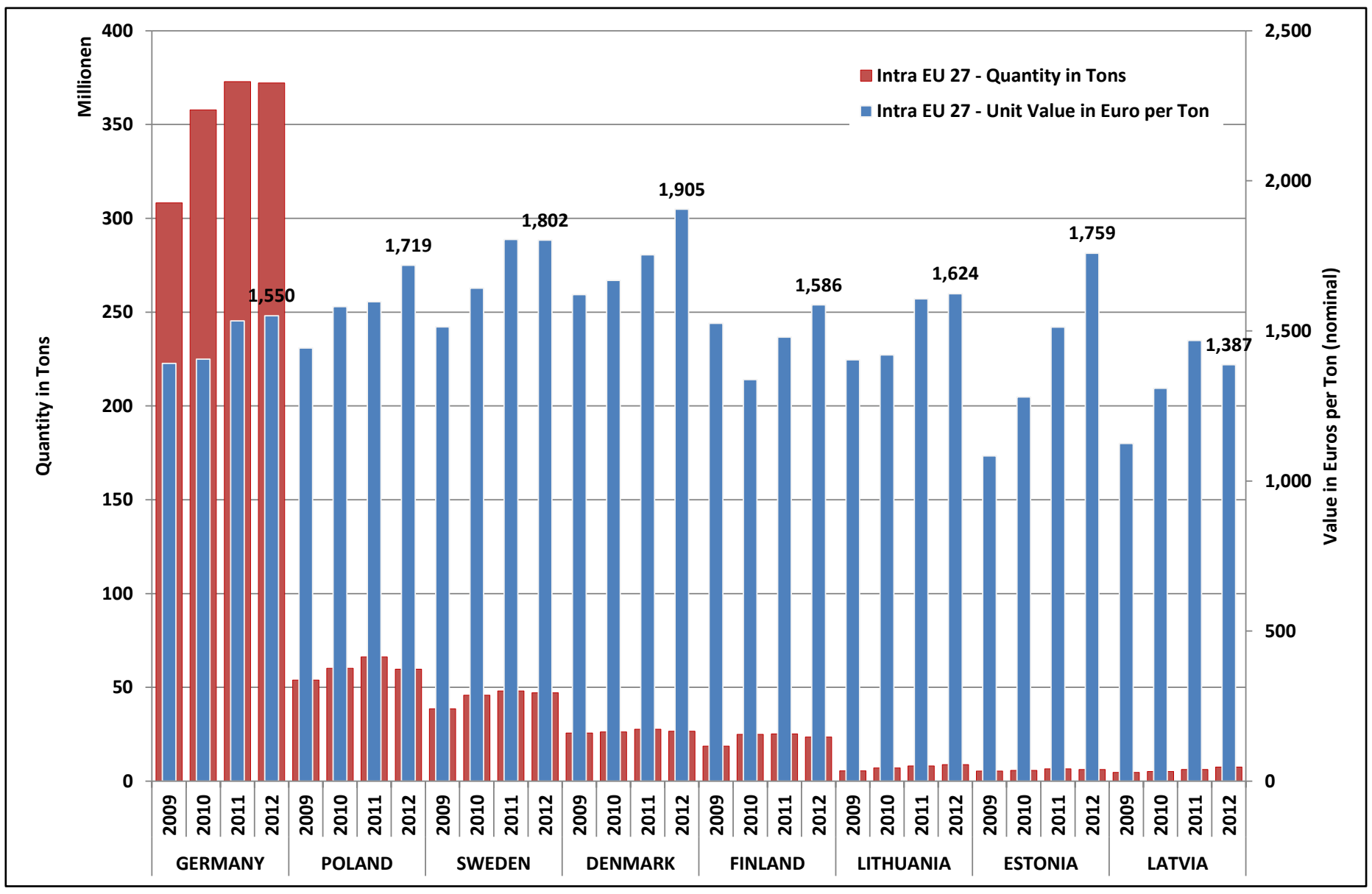

**Fig. 48.** Imports from the EU in 2009 to 2012 (Own illustration with data from Eurostat)

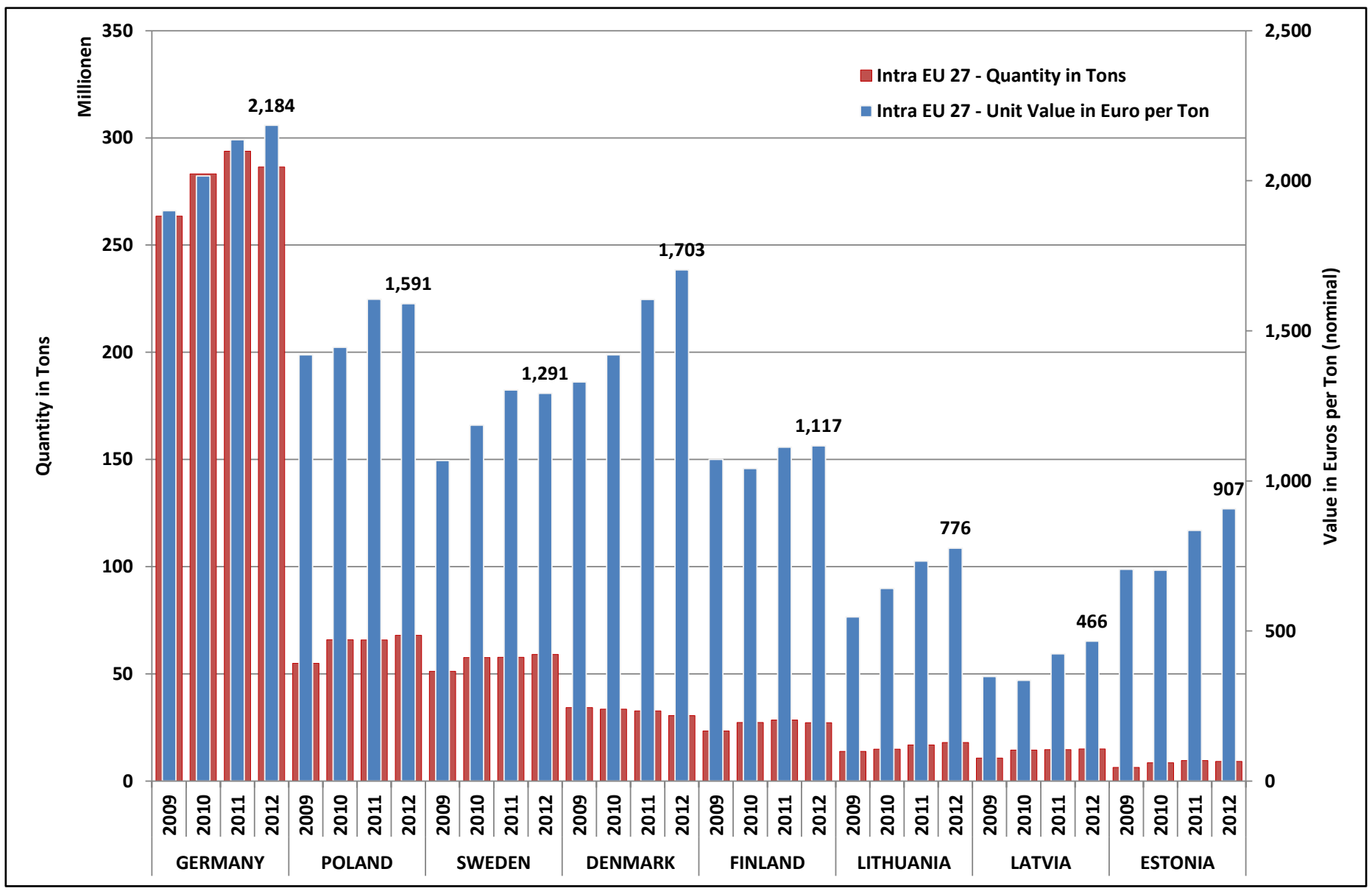

**Fig. 49.** Exports to the EU in 2009 to 2012 (Own illustration with data from Eurostat)

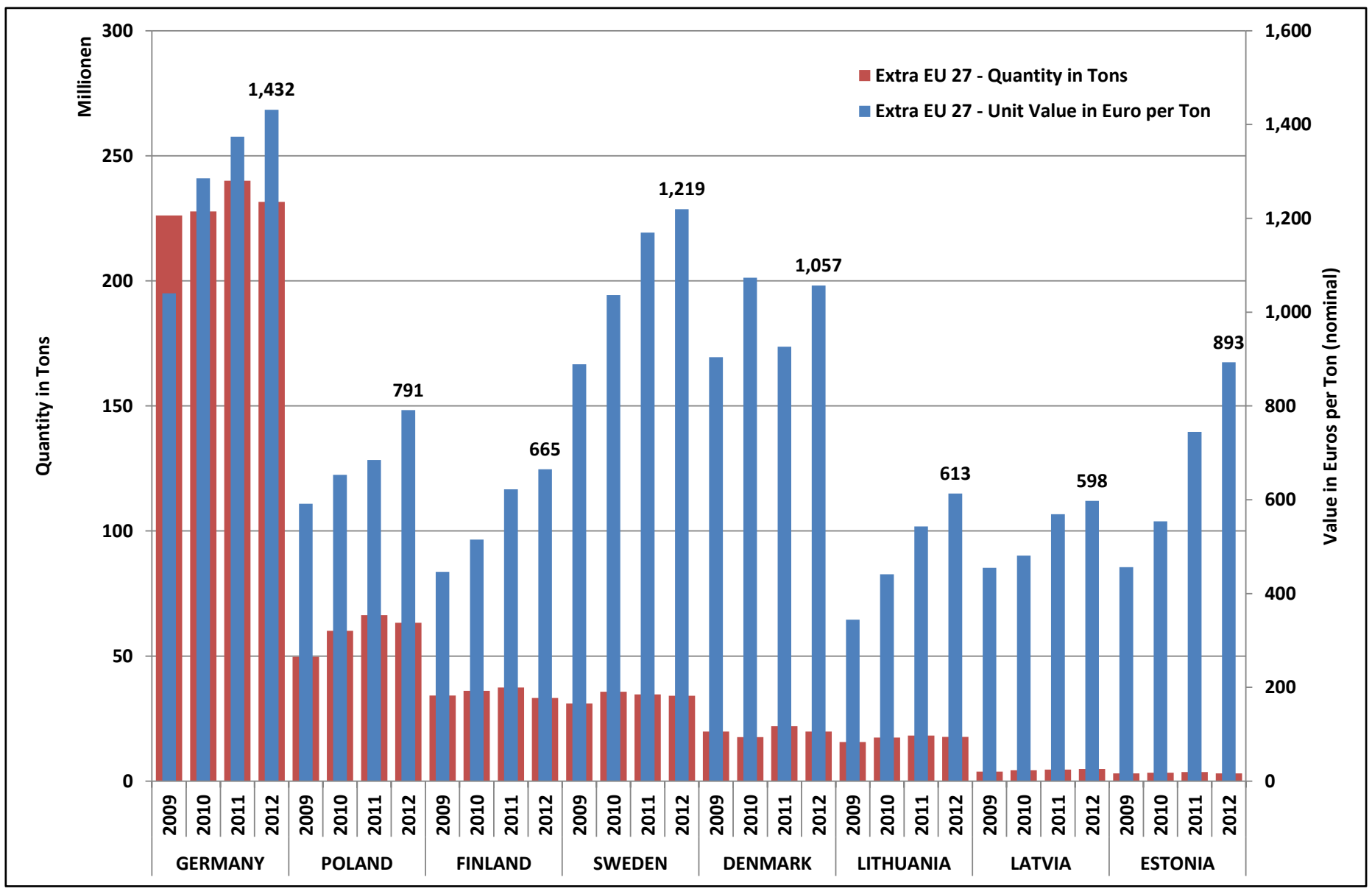

**Fig. 50.** Imports from outside the EU in 2009 to 2012 (Own Illustration with data from Eurostat)

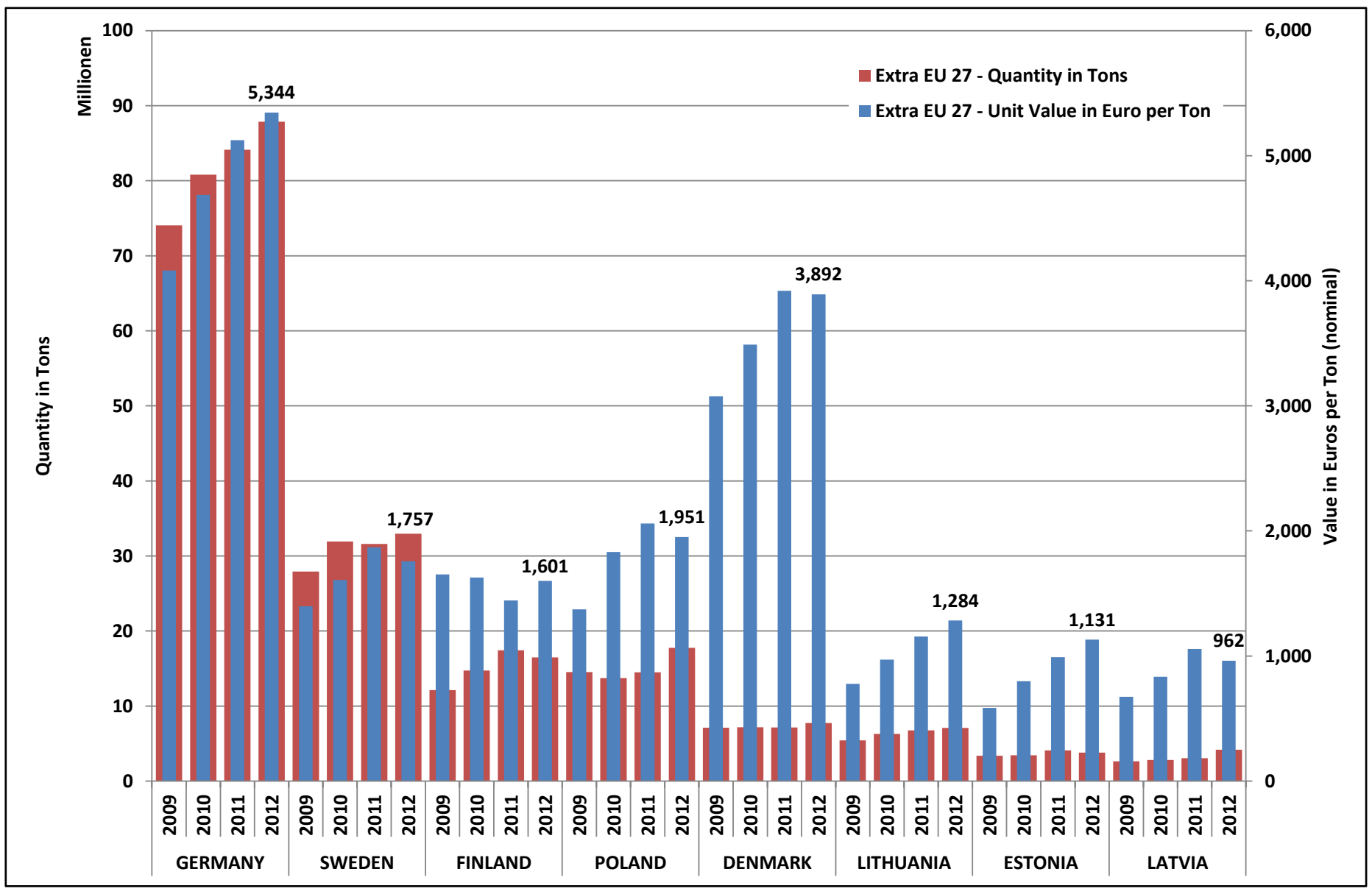

**Fig. 51.** Exports to outside the EU in 2009 to 2012 (Own Illustration with data from Eurostat)

### Industries and trade products

At an aggregate level, we will find trade statistics broken down by mode of transport. Originally, we wanted to have this information in more detail, so we would be able to find only such products traded by air, which are most popular across the BSR countries[1] and which have the largest values per unit of weight. In **Fig. 52** and **Fig. 53** we see the structure of the trade flows on average by mode of transport and product category. For example, we see that petroleum products are mostly transported via rail and sea. Metal products are typically transported by rail, but to a lesser extent by road. “Foodstuffs and animal fodder” are transported mostly by road.

Air transportation is used throughout the countries for the broad class of “machinery, transport equipment and manufactured articles.” Most end-consumer goods and high-tech products which are imported to the EU fall into this category of which more than 90% are transported by air (**Fig. 52**). A report by Lufthansa Consulting (2009) has revealed that in Canada imports and exports transported by air have similar percentages. In Canada, the product groups “Other manufactured […] goods”, “Machinery and electrical equipment” and “Aircraft and other transport equipment” represent roughly 90% to 95 % of the imports and exports, respectively (Lufthansa Consulting 2009, p. 60).

Other notable shares in total exports from the EU by air transportation come from “chemicals” and “agricultural products […]”. In the case of Sweden, we see in Figure 20 that 30% of all exports traded by air are “chemicals” products.

Petroleum products (as part of the “Mineral Products” category) flowing into the EU originate to a large extent from Russia and from Belarus. In 2012, the EU imports “Mineral Products” with a value of 2.4 billion Euros, which is 54.2% of the total imports from Belarus. Other major categories include chemical (12.2%) and base metal (10.8%) products. EU imports from Russia show a similar structure, where “Mineral Products” with a value of 163.7 billion Euros represent 77% of the total imports from Russia. The range of “Mineral […]” and “Base metals […]” products that flow from Russia to the EU represent 28.4% and 10.4% of total EU imports. We could therefore conclude that the EU is largely dependent on Russia’s gas and oil for energy production.

On the other hand, 50% of Belarus’ imports from the EU are “Machinery and mechanical appliances, electrical equipment […]” (34.6%) and “Vehicles, aircraft, vessels […]” (14.8%). The case for Russia is similar, because in 2012 also 50% of all imports from the EU to Russia are “Machinery […]” (31.3%) and “Vehicles […]” (18.4%). As it seems, both countries have few high-tech products with a large unit value to export to the EU.[2,3]

---

[1] We originally aimed for an analysis on the NUTS (Nomenclature of Territorial Units for Statistics) regions level, to be able to know relevant product categories per each airport catchment area, but due to time and data restrictions we had to limit this analysis to country averages. Further publications may investigate this topic in more detail.

[2] http://trade.ec.europa.eu/doclib/docs/2006/september/tradoc_113440.pdf

[3] http://trade.ec.europa.eu/doclib/docs/2006/september/tradoc_113351.pdf

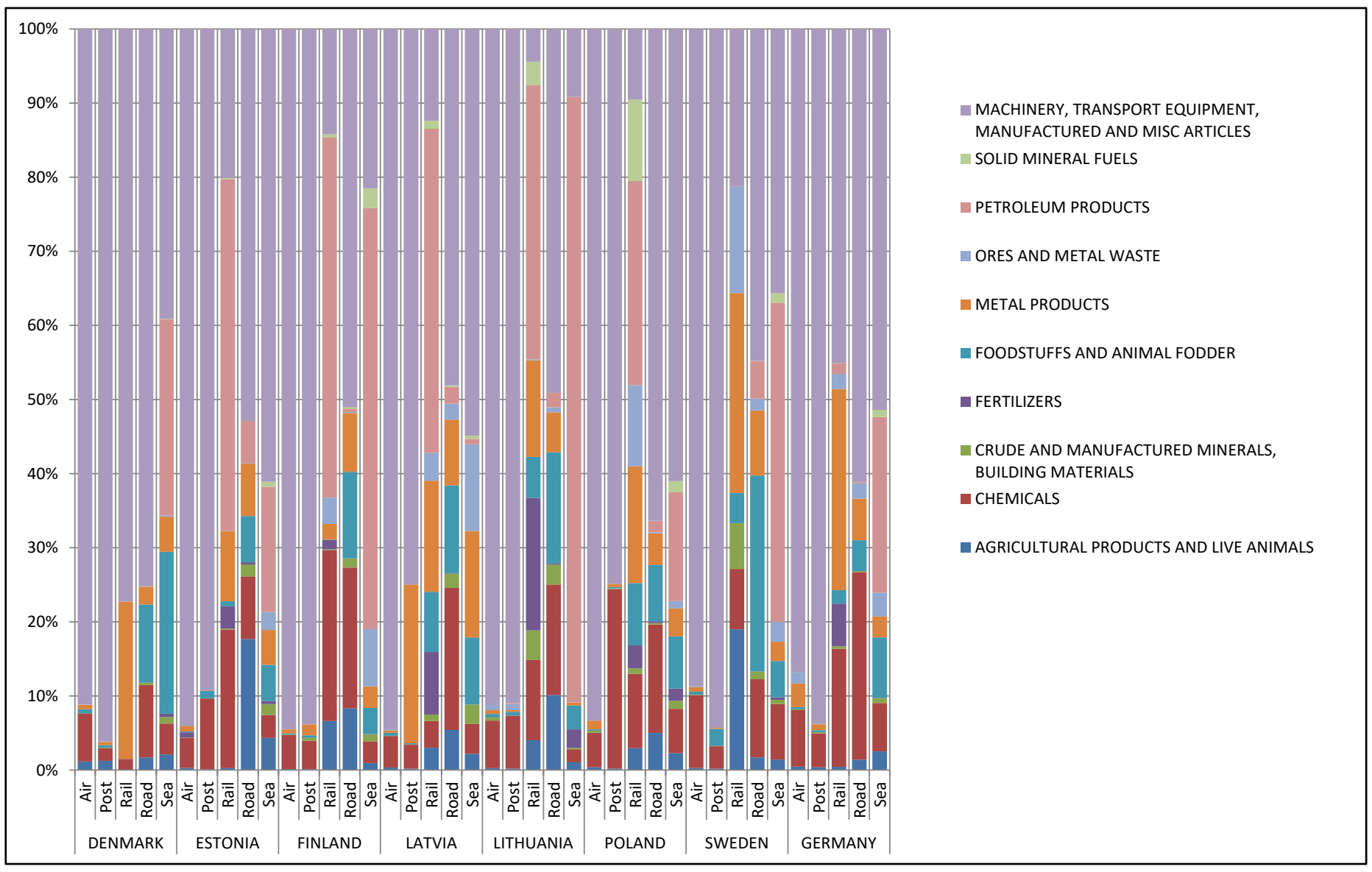

**Fig. 52.** Import Structure, Modal Split and Trade Partners (Eurostat Comext)

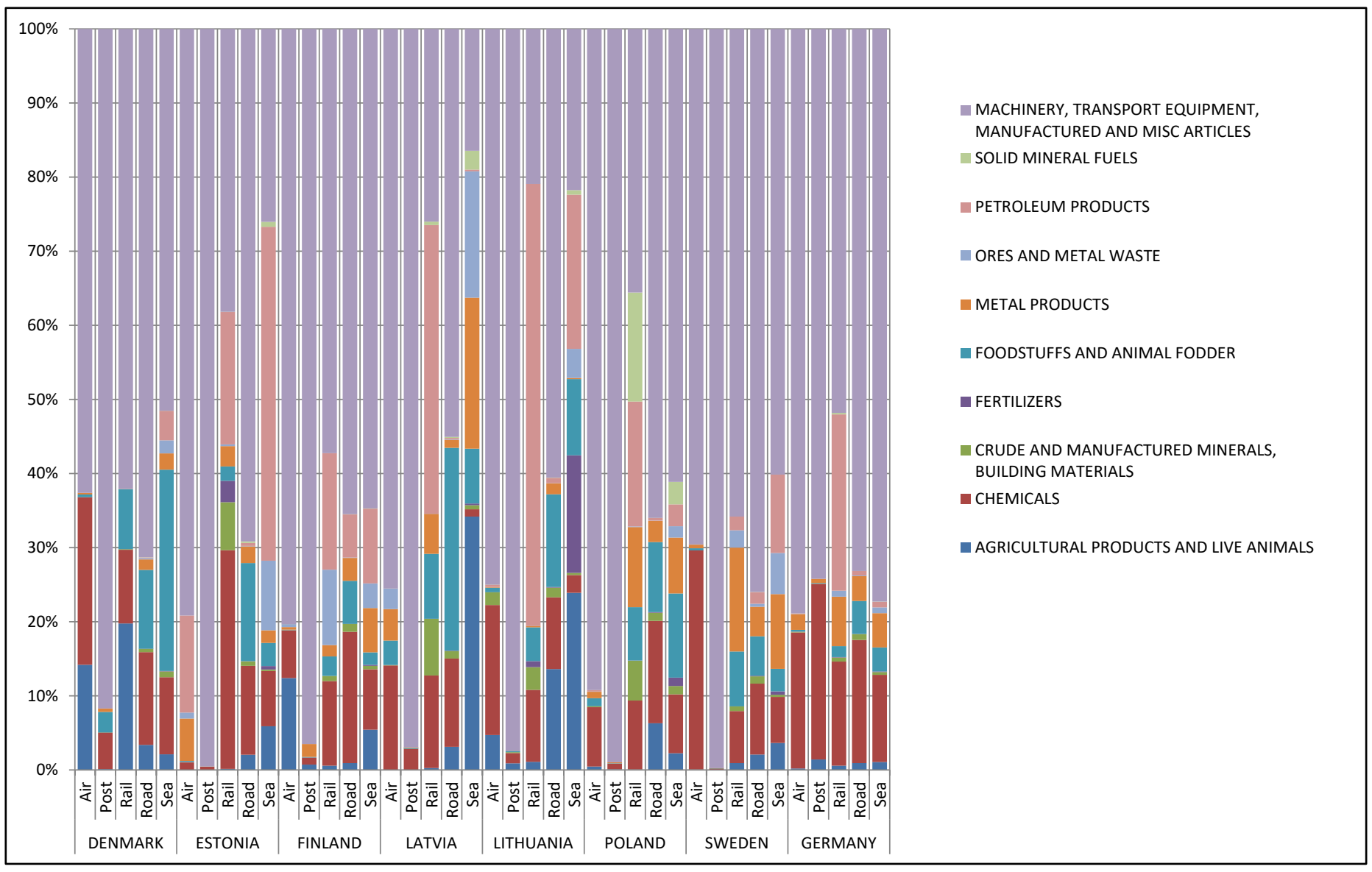

**Fig. 53.** Export Structure, Modal Split and Trade Partners (Eurostat Comext)

From the *Comext* database[1] from Eurostat we obtained figures by product category, sorted by the Combined Nomenclature (CN) codes which are used in international trade with the EU. We limit the analysis to the products we define for the broad category

[1] http://epp.eurostat.ec.europa.eu/newxtweb/

"machinery and manufactured articles." The database consists of traded quantities and values of over 20,000 different product categories, listed by an 8-digit CN code. We filtered all products based on assumptions regarding its high unit value with product numbers starting with 09 (coffee, tea etc.), 24 (tobacco and substitutes), 85 (electrical machinery and parts), 88 (aircraft, spacecraft and parts), 93 (arms and ammunition) and 95 (toys, games and sports requisites).

We purposely excluded agricultural, animal, chemical, industrial, textile, metal, furniture products, since we believe these kinds of products are mostly traded in large quantities under economies of scale through terrestrial modes of transport (e.g. rail, road or ships). The outcome of this filtering is **Table 24**, where we created an overview of the top 25 products in the BSR, which have the largest exported and imported values in Euros in descending order from top to bottom and at the same time a minimum unit value of above 175,000 Euros per ton. As we can see it is mostly consumer electronics and other electronic devices that dominate the list. Another range of products are "aeroplanes and helicopters" or spare parts for such. It may be that we will find suitable products for air cargo transportation from **Table 24** and that forwarders or other small and medium enterprises (SME) engaged in logistics may strategically focus on this range of products. Airports may want to specialize and build their storage and distributional facilities accordingly to the demand of traded high unit value products in their country.

Of course, we do not want to promote a "Baltic Silicon Valley", especially not dispersed information technology (IT) clusters in every BSR country, but it should not be forgotten in national policy and strategy, that research and development of highly competitive products, such as electronics, computer or aeronautical products, require a strong foundation in education, personal and political liberty, creativity, financing and entrepreneurship.

In the EU, the actual mass production of products and the trade thereof will be less important in the future than the development of new products. However, for international trade and transit routes through the EU, enough capacities for the supply chains and the distribution networks will be essential.

Nowadays it is not surprising that we find "mobile telephones" as the number one traded mass product category in terms of quantity, value and unit value. We find other popular consumer products, such as digital cameras (10.), telephone sets (16.) and media devices, such as CD's and DVD's (18.) in the ranking (**Table 24**).

Especially regarding the reproduction of digital media, we could imagine the BSR to become specialist and leader in the EU, due to (still) low labour costs, but unlimited access to automated, computer and other production equipment from the EU for building factories. Despite the on-going trend towards online web-based products, these kinds of reproduction technologies for digital media are nonetheless important for the timely and synchronized distribution of movies, software or audio. We believe that the shipping of electronic, media and other digital products is especially suited for air cargo transportation, either by the "flying truck" concept or not, not least for security and marketing reasons (e.g. the leaking of products before official product launch date). The development of digital products also requires state-of-the-art IT and communications technology, such as data centers. Constant innovation, exchange of people and knowledge and short product cycles are essential for developing “Web 2.0” services and products. We believe that IT could bring many advantages to the BSR countries, e.g. with respect to growth of demand and GDP.

**Table 24.** Top traded products (imported and exported) in the BSR by unit value and quantity (Eurostat Comext database)

| Rank | Products |
|---|---|
| 1 | Mobile Telephones |
| 2 | Aeroplanes And Other Powered Aircraft >15.000 Kg |
| 3 | Relays For A Voltage > 60 V But <= 1.000 V |
| 4 | Microphones And Stands Therefor |
| 5 | Electronic Integrated Circuits As Processors And Controllers |
| 6 | Radio Navigational Aid Apparatus |
| 7 | Apparatus For The Transmission Or Reception Of Voice, Images Or Other Data |
| 8 | Parts Of Aeroplanes Or Helicopters |
| 9 | Machines For The Reception, Conversion And Transmission Or Regeneration Of Voice, Images Or Other Data |
| 10 | Digital Cameras |
| 11 | Parts Of Telephone Sets, Telephones For Cellular Networks Or For Other Wireless Networks |
| 12 | Parts For Aircraft |
| 13 | Electrical Signaling, Safety Or Traffic Control Equipment For Railways Or Tramways |
| 14 | Powered Aircraft > 2.000 Kg But <= 15.000 Kg |
| 15 | Fixed Electrical Capacitors, Ceramic Dielectric, Multilayer |
| 16 | Telephone Sets |
| 17 | Parts Of Electrical Signaling, Safety Or Traffic Control Equipment |
| 18 | Optical Media, Recorded, For Reproducing Representations Of Instructions, Data, Sound, And Image |
| 19 | Cards And Tags Incorporating Only One Electronic Integrated Circuit "Smart Cards" |
| 20 | Filament Lamps For Motorcycles Or Other Motor Vehicles |
| 21 | Transistors With A Dissipation Rate >= 1 W |
| 22 | Sealed Beam Lamp Units |
| 23 | Solid-State, Non-Volatile Data Storage Devices For Recording Data From An External Source [Flash Memory Cards Or Flash Electronic Storage Cards] |
| 24 | Inductors Of A Kind Used With Telecommunication Apparatus And For Power Supplies For Automatic Data-Processing Machines |
| 25 | Helicopters Of an Unladen Weight <= 2.000 Kg |

## 6.3 Baltic air cargo feeder and hub airports

In the given timeframe it was impossible to gather all information on the main airports in the region. The airport industry is not as transparent as one could wish, especially regarding airport economics. However, we build a small database based on newly and previously collected airport related information in the German Airport Performance (GAP) research project.[1]

**Break-even analysis – case study: Sweden**

In the framework of this study we try to give credible evidence of the profitability of small and large airports and of the location of the break-even point, at which an airport has operating costs equal to its revenues. The finding of a benchmark for a break-even point in the BSR leads to conclusions about the needed 'critical mass' of demand, which is needed for operating an airport profitably. As others did in previous studies, e.g. Doganis and Thompson (1975), we calculated the break-even point with regard to the demand expressed in Workload Units (WLU). Although this measure is not free from criticism, especially the equivalence of one passenger to 100 kg (or one tenth of a ton) of cargo, we have chosen this measure for this analysis as it accounts for cargo quantities.

Based on historical data on Sweden (Näringsdepartementet 2007, Swedavia annual reports) we can draw slightly different conclusions as GAP (2012) and Bubalo (2012a), as a break-even is plausible in the range of Visby (VBY), Angelholm (AGH) and Umeå (UME) airports with around 320, 400 and 820 thousand annual WLUs (e.g. in 2007; see **Table 25**). Interpolated from these benchmarks this means we have found profitable airports far below 1 million passengers per year, at around 500 thousand PAX for the case of Sweden (**Fig. 54**). In the cases where *profitability benchmarks* are defined, however, the role of cargo is insignificant at the airports with levels lower than 0.4% share in total WLUs.

The distribution of airports in Sweden over size (in WLUs) and *unit profits* (expressed in *earnings before taxes* [EBT] per WLU) is comparable with distributions of previous studies (e.g. GAP 2012, or Bubalo 2012a), where instead *earnings before interests and taxes* [EBIT] was chosen). However, one important distinction has to be made in comparison with other countries, e.g. Norway (GAP 2012), that the trend curve enveloping the profitability benchmarks remains remarkably *static* over time (**Fig. 54**). Over a similar period, airports in the Norwegian *Avinor* system are subject to much higher increases in operating costs over the years, thus resulting falling profitability, especially in the low range of below 100,000 PAX.

[1] We want to thank our former colleagues from the GAP project for collecting and systematizing the data, especially Tolga Ülkü, Mikhail Zolotko, Roman Pashkin, Savia Hasanova, Keith Lukwago, Sascha Michalski and Ivana Strycekova.

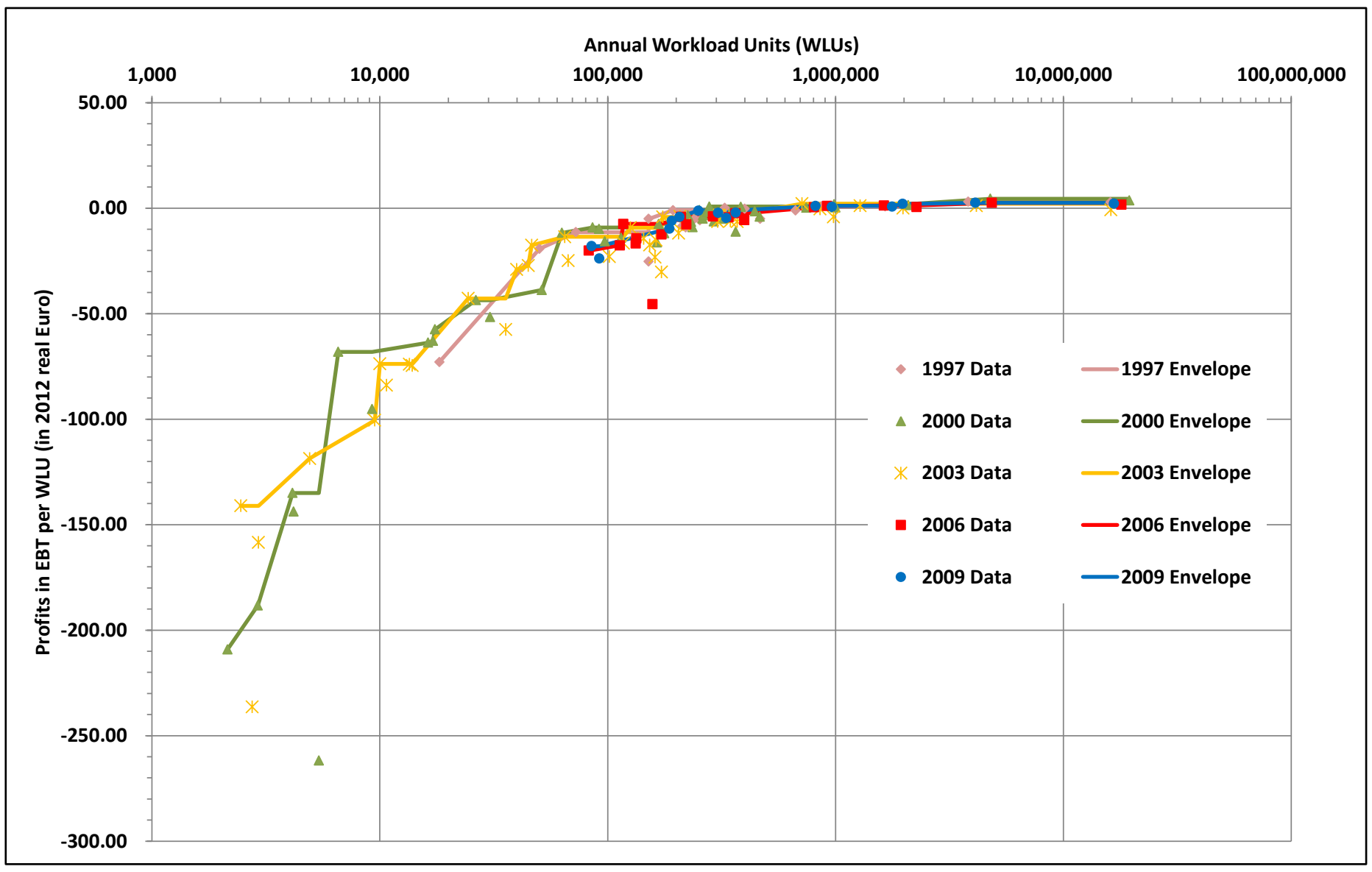

**Fig. 54.** Profitability of airports in Sweden in 1997 to 2009 (in 2012 real Euro; $\log_{10}$ scale on the x-axis; Own illustration based on Näringsdepartementet [2007] and Swedavia / LFV annual reports)

After the construction of a *profitability envelope*[1] (see Bubalo 2012a) for different years we can now identify the curve defining airports, which represent the best-in-class benchmarks (**Table 25**). We may then be able to spot the root causes why these airports are more profitable (or make less losses) than their similar sized peers. We believe that "benchmarking" regarding finding the *best-practice* among similar airports provides a good starting point for managerial progress.

[1] We construct the envelope by our iterative algorithm which is similar to the discrete Hill-climbing method:

1. Build the dataset of n observations with three columns: WLUs (column 1), EBIT per WLUs (column 2), and calculated Profitability benchmark (column 3).
2. Sort the data by WLUs (in column 1) in ascending order: WLU (1) < WLU (2) < …< WLU (*n*).
3. Initialize Profitability Benchmark (1) (in column 3) = EBIT per WLU (1) (in column 2).
4. For *i* = 2 to *n*
5. Compare in column 3, if EBIT per PAX (*i*) is > previous Profitability Benchmark (*i*-1).
6. If yes, then Profitability Benchmark (*i*) = EBIT per PAX (*i*);
7. If not, then Profitability Benchmark (*i*) = Profitability Benchmark (*i*-1)
8. Next (go to step 4, *i* increases by 1).

**Table 25.** Swedish benchmark airports with regard to profitability and airport size (Näringsdepartementet 2007; Swedavia Annual Reports).

| Year | Airport Name | IATA | ICAO | WLUs | Benchmark in EBT per WLU in EUR | Percentage Domestic Passengers | Cargo as a percentage of WLU |
|---|---|---|---|---|---|---|---|
| 2006 | Norrköping | NRK | ESSP | 82,640 | -20.12 | 2.50% | 0.10% |
| | Karlstad | KSD | ESOK | 113,061 | -17.62 | 63.60% | 0.00% |
| | Halmstad | HAD | ESMT | 117,185 | -7.45 | 96.00% | 0.00% |
| | Visby | VBY | ESSV | 290,035 | -3.98 | 97.20% | 0.00% |
| | Angelholm Helsingborg | AGH | ESTA | 364,037 | -2.41 | 99.90% | 0.10% |
| | Umeå | UME | ESNU | 805,155 | 0.42 | 95.40% | 0.40% |
| | Luleå | LLA | ESPA | 918,445 | 1.02 | 94.30% | 0.50% |
| | Stockholm-Bromma | BMA | ESSB | 1,632,808 | 1.25 | 92.30% | 0.00% |
| | Goteborg Landvetter | GOT | ESGG | 4,855,217 | 2.53 | 31.50% | 11.90% |
| 2007 | Karlstad | KSD | ESOK | 119,722 | -19.46 | 51.50% | 0.20% |
| | Jönköping | JKG | ESGJ | 126,079 | -12.86 | 63.50% | 15.10% |
| | Kiruna | KRN | ESNQ | 192,807 | -9.46 | 98.50% | 0.50% |
| | Ronneby | RNB | ESDF | 220,102 | -2.76 | 99.20% | 0.10% |
| | Visby | VBY | ESSV | 318,148 | -2.23 | 95.80% | 0.20% |
| | Angelholm Helsingborg | AGH | ESTA | 395,310 | 0.26 | 97.70% | 0.10% |
| | Umeå | UME | ESNU | 814,533 | 1.49 | 94.80% | 0.40% |
| | Stockholm-Bromma | BMA | ESSB | 1,804,994 | 1.85 | 92.30% | 0.10% |
| | Malmö | MMX | ESMS | 2,300,857 | 1.94 | 60.50% | 18.80% |
| | Goteborg Landvetter | GOT | ESGG | 4,953,864 | 3.89 | 28.60% | 12.10% |
| | Stockholm-Arlanda | ARN | ESSA | 18,887,094 | 4.06 | 28.20% | 5.30% |
| 2008 | Jönköping | JKG | ESGJ | 89,901 | -56.52 | 70.10% | 14.80% |
| | Karlstad | KSD | ESOK | 119,012 | -44.34 | 49.80% | 0.20% |
| | Örnsköldsvik | OER | ESNO | 145,465 | -38.29 | 97.90% | 0.30% |
| | Ronneby | RNB | ESDF | 207,212 | -3.77 | 99.00% | 0.10% |
| | Visby | VBY | ESSV | 325,447 | -2.7 | 96.10% | 0.30% |
| | Umeå | UME | ESNU | 825,079 | 1.18 | 95.00% | 0.20% |
| | Stockholm-Bromma | BMA | ESSB | 1,855,805 | 1.63 | 92.20% | 0.20% |

| Year | Airport Name | IATA | ICAO | WLUs | Benchmark in EBT per WLU in EUR | Percentage Domestic Passengers | Cargo as a percentage of WLU |
|---|---|---|---|---|---|---|---|
| | Goteborg Landvetter | GOT | ESGG | 4,792,913 | 3.94 | 27.60% | 10.30% |
| | Karlstad | KSD | ESOK | 85,040 | -17.97 | 51.00% | 0.20% |
| 2009 | Kiruna | KRN | ESNQ | 186,700 | -9.72 | 98.20% | 0.30% |
| | Ronneby | RNB | ESDF | 191,338 | -5.99 | 99.70% | 0.10% |
| | Skellefteå | SFT | ESNS | 206,681 | -4.16 | 95.10% | 0.50% |
| | Sundsvall Harnosand | SDL | ESNN | 250,112 | -1.15 | 92.30% | 0.50% |
| | Umeå | UME | ESNU | 816,722 | 1.05 | 95.80% | 0.20% |
| | Stockholm-Bromma | BMA | ESSB | 1,970,329 | 2.04 | 92.10% | 0.10% |
| | Goteborg Landvetter | GOT | ESGG | 4,112,519 | 2.53 | 27.60% | 10.40% |

When analyzing the results of the benchmarks in Sweden over the years 1997 to 2009 we realize that there are several airports frequently defining the envelope. The top ten airports in this respect are Gothenburg-Landvetter (GOT), Angelholm (AGH), Kallinge (RNB), Umeå (UME), Trollhättan (THN), Kramfors (KRF), Halmstad (HAD), Luleå (LLA), Sveg (EVG) and Torsby (TYF) (**Table 25**).

Airports in the BSR may compare their revenues and cost structures against these Swedish benchmark airports, given the similar range of demand expressed in annual WLU's.

### Comparison of charges for cargo traffic in the BSR

In estimating the potential for air cargo market in the BSR we want to take another aspect into consideration, which is *airport competition*. As we have seen by the distribution in **Fig. 45**, airports in the BSR, especially in the "core" Baltics States (Belarus, Estonia, Latvia and Lithuania), are located close to each other to have overlapping catchment areas and similar distances to the nearest connecting hubs. Therefore, airports need to seek a *comparative advantage* over other airports and countries with which they compete. Pricing is only one such tactic, whereby an airport could attract potential customers.

**Fig. 55** shows the distribution of cargo flights across the BSR of the observations in the reported period. About half of all cargo flights (47%) pass through Frankfurt am Main International airport (FRA). About 22% of the flights travel through Moscow-Sheremetyevo (SVO) airport. The main market participants in the BSR are Germany with a total market share of 59% (with airports Frankfurt [FRA], Frankfurt-Hahn [HHN], Halle/Leipzig [LEJ] and Köln/Bonn [CGN]), followed by Russia with a share of 26.4 % (with airports Moscow-Sheremetyevo [SVO], Moscow-Domodedovo [DME], Saint Petersburg-Pulkovo [LED] and Vnukovo [VKO] airports), Denmark with a share of 3.4% (with Copenhagen [CPH] airport), Finland with a market share of 3.3% (with Helsinki [HEL] airport), Norway with a share of 1.6% (with Oslo [OSL] airport) and Sweden with a market share of 1.5% (with Stockholm-Arlanda [ARN] and Gothenburg-Landvetter [GOT] airport).

Broken down to the average daily movement numbers we observe only 127 average daily cargo flights in the reporting period at the sample BSR airports. This is indeed evidence for a very small existing market, but with a strong growth potential. If an average growth rate of 5% per annum can be maintained in the long-term in the BSR, then we may expect a doubling of the daily movements after 15 years and an increase by 164 daily movements (129%) to a level of 291 daily movements by 2030.[1]

If the distribution among the airports remains proportional over the years (which is quite unlikely, since some airports will grow stronger than others) this means, for example, about 76 additional daily flights in FRA by 2030.

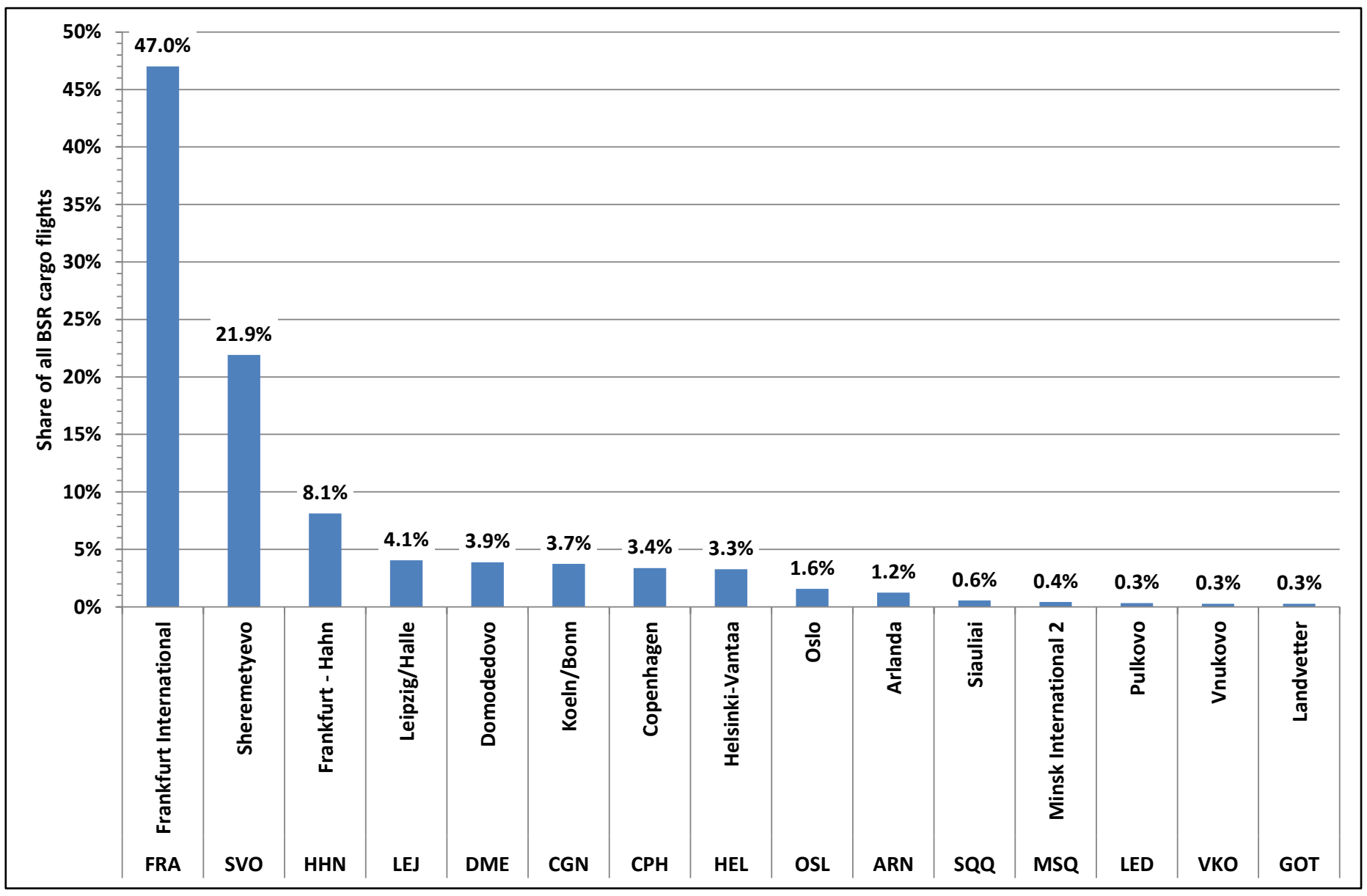

**Fig. 55.** Distribution of Cargo-flights in the BSR (Own illustration based on 232,814 flight observations in the reporting period from FlightStats.com)

[1] For calculating the multiplier $k$ for the forecast, we used the "*compound interest*" formula $k = (1+i)^n$, where $i$ is the average growth rate and $n$ are the number of years. From the base year 2013 until 2030 it is $n$=17 years: $k$=$1.05^{17}$=2.29 or 129%.

The total increase will then be: 127 flights (in 2013) * 2.29 = 290 flights (in 2030).

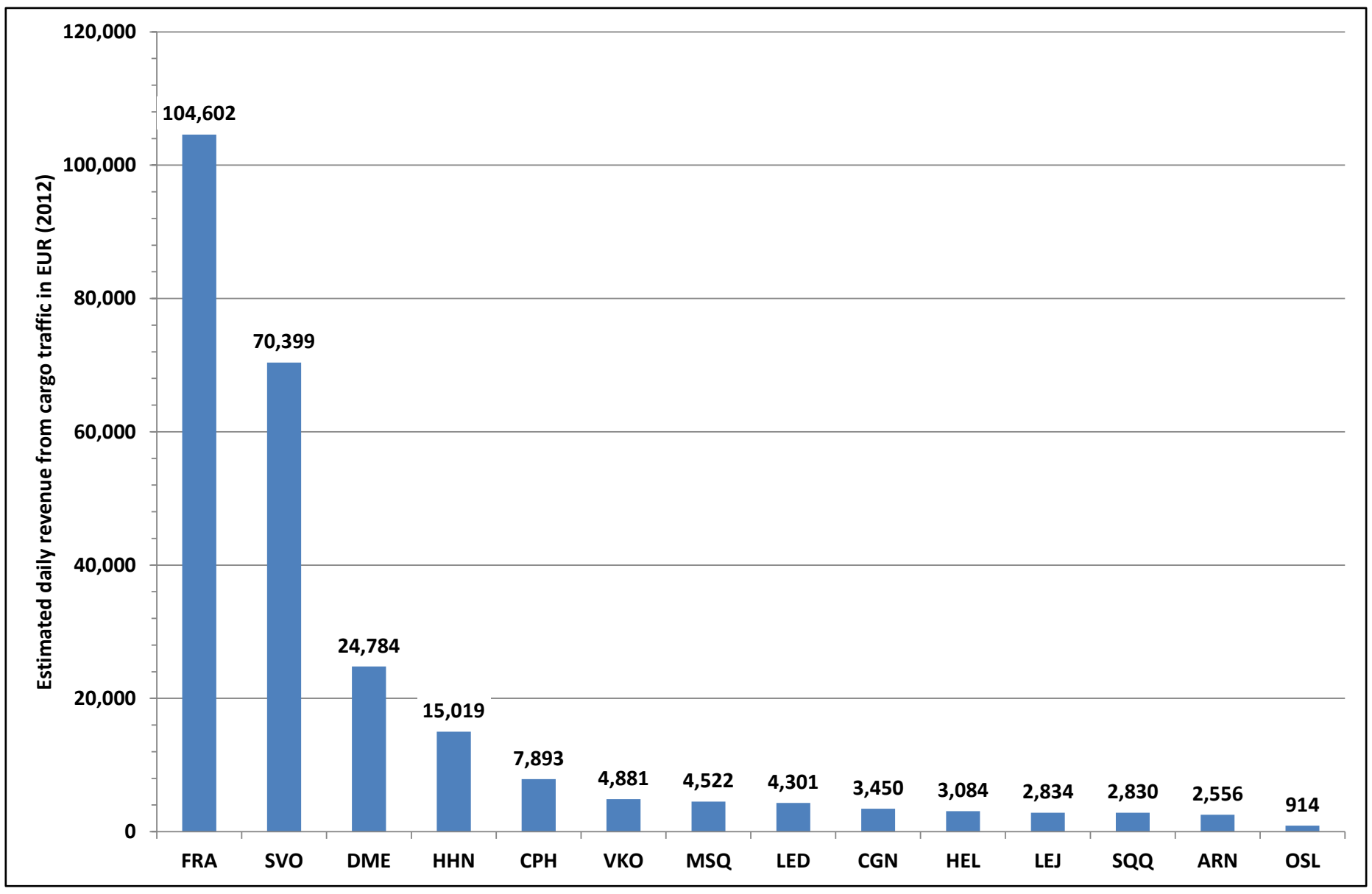

**Fig. 56.** Average Daily Revenues from Cargo Traffic based on Airport Traffic Mix (Own calculations based on data from FlightStats, OAG, and country specific AIP GEN 4.1 from EUROCONTROL)

Translated into revenues from charges of cargo traffic, based on the calculation described in **Table 26** and **Table 27** below, an increase by 129% of today's levels will *ceteris paribus*, for example at FRA, lead to an increase from 104,602 Euros *per day* in 2013 (**Fig. 56**) to 239,539 Euros *per day* by 2030. The daily revenues presented in **Fig. 56** show a similar distribution as the number of flights in the BSR. However, the ranking has slightly changed due to the price differences in the turnaround charge. FRA and SVO are first and second in terms of ranking by number of flights and by average revenues. Due to the higher charges per turnaround, DME is fifth in terms of flights, but third in terms of charges. It would be challenging for the airport and interesting for us, when e.g. HHN decides to change the charges scheme and how this would affect the income. HHN ranks third by number of cargo movements in the BSR but ranks fourth in terms of daily charges revenue.

We have compared the charges schemes for a range of main cargo airports in the BSR, and we have calculated turnaround charges for the types of cargo aircraft, which are used in the region (**Table 26**).

From EUROCONTROL's "European AIS Database" (EAD) we were able to obtain the charges schemes for the main airports in the BSR. These are reported under the file code AIP GEN 4.1 for each country. We recognized that these schemes are very inhomogeneous in their structure and in their user-friendly applicability. On the one hand, we find manuals with a very simple structure such as at OSL (which has in fact only a landing charge based on the maximum take-off weight (MTOW) for cargo aircraft [however, optional additional de-icing charges during winter operations]), on the other hand, we find very complicated and complex charges regulations such as for FRA (which has many different charges, separately payable for landing and take-off, and based on the based on membership in certain lists or aircraft and engine certificates).

Parking of less than three hours is usually free for cargo aircraft at most sample airports. Some airports even exempt parking charges of up to 5 hours (CGN, Germany) or 6 hours (MSQ, Belarus) for heavy cargo aircraft. By studying the airport flight schedules regarding turnaround times of cargo aircraft we find that these rarely take longer than three hours. It is obvious that cargo airlines adjust their turnaround times, such that these are less than the free amount, to save extra charges.

Some airports have charges schemes which are supportive of the operations and the business model of the main airline. HHN, for example charges "0,00" Euros per ton MTOW for each take-off and landing, if a turnaround time of less than 30 minutes can be achieved. This clearly favors the business model of based Low-Cost Carrier (LCC) *Ryanair*.

As we can see it is possible to create and to incentivize the growth in demand for cargo and passengers by intelligently balancing the structure of the charges schemes. The airport customers will appreciate simple charges schemes, which are transparent and cost based. It should be left to regulators or politicians to implement control mechanisms to prevent cases of "price dumping", where airports purposely charge order of magnitudes less than would be required to economically break even. In the case of cargo charges at HHN airport one could argue in that way, since the charges for *typical* turnarounds are fourth lowest in the BSR (behind OSL, FRA and ARN) (**Table 26**). However, HHN has implemented charges which are insufficient to cover all costs. In 2011 according to the German "Bundesanzeiger"[1] (where annual reports of limited companies in Germany are being published) "Flughafen Frankfurt-Hahn GmbH" has posed a pre-tax loss (EBT[2]) of -10.6 million Euros at a total turnover of 53 million Euros and costs of 63.4 million Euros (20 million Euros of raw materials and supply, 18.2 million Euros of *labour costs*, 13.3 million Euros of *Depreciation and Amortization*, 6.7 million Euros in *other costs* and 5 million in *finance costs*, mainly from *interests*). In order to break even, HHN would have to increase their charges by around 20%. Lowering their costs, e.g. for supplies, would allow a lower increase in charges.

In our opinion airport managers who are responsible for developing airport charges schemes, should be more creative and innovative. Poland represents a *good practice* example, where we find creatively designed and transparent charges manuals, which additionally include growth or bonus schemes. In these cases, the airport user only pays for the services it uses and is encouraged to expand operations. In this report we lay out the basis for further investigation into *airport charges* within the BSR.

We tried to give an account of actual charges paid at selected airports in the region. Our approach takes statistical information from representative traffic samples. This enables us to have order-of-magnitude figures on frequency, average turnaround times, aircraft mix (see **Table 27**) and approximate payload. Previous investigations usually compare certain (arbitrary) examples of aircraft types across airports, but we have isolated the actual equipment in use by the main carriers at the main cargo routes within the region. Please see **Table 36** in the Appendix on the full account of cargo carriers, aircraft types and flown routes in the BSR.

[1] http://www.bundesanzeiger.de/

[2] Earnings before taxes

For our calculations which lead to the numbers in **Table 26**, we have multiplied the variables from our statistical analysis with the rates published in the charges manuals. The "GEN 4.1" publications as part of the Aeronautical Information Publications (AIPs) consist of charge information and can be found on the web for most airports. For the case of cargo aircraft, the airport charges consist usually of a landing or take-off charge, a parking charge and a security charges. Sometimes additional charges are collected for extra services or emissions, therefore in the cases of FRA, CGN and others we have put aircraft types into noise ("Bonus lists") and wingspan classes. Passenger aircraft usually pay a passenger charge in addition.

For the calculations of the landing (or take-off) charge the aircraft MTOW is multiplied by the relevant charge per MTOW. Then the parking charge based on the average turnaround time is added. In our sample the average turnaround time was in most cases below the limit for free parking (e.g. less than 3 hours), thus no parking charge was added. The time limit for free parking is usually less restrictive for cargo aircraft than for passenger aircraft.

Russia's and Belarus's charges schemes are among the simplest structured, however for each airport individually adjusted. Some, e.g. German, airports could learn from these countries in terms of clarity and simplicity.

To make further judgements we want to directly compare charges for a typical turnaround at the "cargo-only" airports in the BSR, with the actual equipment currently in use is applied to the calculations.

We gathered some aircraft specific data from Official Airline Guide (OAG) databases, so we have variables for MTOW and cargo capacity in tons for specific aircraft types. Aircraft characteristics are averaged over all airlines having similar equipment in operation under the identical aircraft code. With aircraft information from the flight schedules we can base our calculations on realistic assumptions. **Table 26** shows the cargo aircraft types operating in the BSR and the calculated turnaround charges per airport. As our calculation have shown OSL airport in Norway has by far the lowest turnaround charges for cargo aircraft in the Region, at least for heavy equipment operated over 100 tons MTOW. Cargo aircraft with a MTOW of less than 100 tons turn around cheaper at e.g. FRA, ARN or HHN (without considering emission charges).

We want to sketch our approach of assuming turnaround charges and daily revenues on the example of Siaulial (SQQ) airport, given the traffic mix (**Table 27**) and the charges scheme.[1] SQQ is the only "cargo-only" airport in the sample and is therefore a suitable benchmark in place for airports with a similar business model, e.g. Schwerin-Parchim (SZW/EDOP) airport. With one take-off and one landing of a cargo aircraft on average per day, SQQ airport earns approximately 2,830 Euro per day (**Fig. 56**). With its share of 0.6% of all cargo flights in the BSR it ranks 11th (**Fig. 55**) with its amount of daily revenues the airport ranks 12th (out of 15). Over the whole reporting period we have measure a traffic mix of 83.3% of Airbus A300-600 (Freighter) aircraft (ABY) with 171 tons MTOW and 16.7% of Boeing 747-400F aircraft (74Y) with 397 tons (**Table 26**). The average turnaround charge is then composed of 3,451 Euros landing charge and 345 Euros security charge for a 74Y and 1,883 Euros landing charge and 188 Euros security charge for an ABY.[1] Since only landings are charged the average daily revenues from cargo traffic of 2,830 Euros are very low.

[1] http://www.siauliai-airport.com/airport/main-airport-fees

If we are to recommend a strategy about “ideal” charges schemes in the BSR, or at least among the “core” Baltic States, we would encourage implementing similar structured schemes, including *incentive schemes*. We feel that it is unnecessary to have so diversely structured price lists in place. When the BSR countries discuss how such a harmonized scheme would have to look like it may be reasonable to consider FRA airport’s approach with the unbundling of services under the *polluter pays principle*. Instead of a MTOW-based landing charge each take-off and landing at FRA is subject to a noise charge which is specified based on aircraft characteristics and engine type. In our opinion FRA’s charges scheme is supposed to be fair among the different airport users, so each customer is charged accordingly to the amount of noise or emissions emitted, payload, aircraft MTOW and wingspan.

[1] The “Follow me” charge of 17.38 Euros has been omitted.

**Table 26.** Total turnaround costs in EUR (2013 prices) for cargo aircraft at BSR airports (Own calculations based on data from FlightStats, OAG, AIP GEN 4.1 from EUROCONTROL)[1]

| Aircraft Type (Freighter Variants) | A/C Code | Freight Capacity in tons | MTOW | OSL | FRA* | ARN** | HHN | CGN | LEJ | HEL | SVO | CPH | SQQ | LED | DME | VKO | MSQ |
|---|---|---|---|---|---|---|---|---|---|---|---|---|---|---|---|---|---|
| Boeing 747-8*** | 74N | 130 | 442 | 820 | 2076 | 2090 | 2281 | 2788 | 2811 | 3133 | 3588 | 3789 | 4226 | 5160 | 5434 | 5434 | 6093 |
| Boeing 747 (All Types) | 74F | 109 | 397 | 737 | 2802 | 1881 | 2049 | 2647 | 2525 | 2827 | 3223 | 3404 | 3796 | 4635 | 4881 | 4881 | 5473 |
| Boeing 747-400F | 74Y | 130 | 397 | 737 | 2836 | 1881 | 2049 | 2791 | 2525 | 2827 | 3223 | 3404 | 3796 | 4635 | 4881 | 4881 | 5473 |
| Boeing 777 (All Types)*** | 77F | 104 | 348 | 646 | 1516 | 1653 | 1796 | 2565 | 2213 | 2493 | 2825 | 2984 | 3328 | 4063 | 4278 | 4278 | 4798 |
| Boeing 777-200F | 77X | 104 | 347 | 644 | 1525 | 1648 | 1791 | 2563 | 2207 | 2487 | 2817 | 2975 | 3318 | 4051 | 4266 | 4266 | 4784 |
| Boeing (Douglas) MD-11 | M1F | 93 | 286 | 531 | 1772 | 1364 | 1476 | 2949 | 2277 | 2072 | 2322 | 2452 | 2735 | 3339 | 3516 | 3516 | 3943 |
| Airbus A330 (All Types) | 33F | 20 | 230 | 427 | 1191 | 1103 | 1187 | 1658 | 1463 | 1691 | 1867 | 1972 | 2199 | 2685 | 2828 | 2828 | 3171 |
| Airbus A330-200 | 33X | 20 | 230 | 427 | 1191 | 1103 | 1187 | 1658 | 1463 | 1691 | 1867 | 1972 | 2199 | 2685 | 2828 | 2828 | 3171 |
| Airbus Industrie A300 (All Types) | ABF | 51 | 171 | 317 | 1141 | 828 | 882 | 1591 | 1088 | 1290 | 1388 | 1466 | 2071 | 1996 | 2102 | 2102 | 2357 |
| Airbus Industrie A300-600 | ABY | 55 | 171 | 317 | 1115 | 828 | 882 | 1619 | 1088 | 1290 | 1388 | 1466 | 2071 | 1996 | 2102 | 2102 | 2357 |
| Airbus Industrie A300B4/-C4/-F4 | ABX | 51 | 165 | 306 | 1098 | 800 | 851 | 1558 | 1049 | 1249 | 1340 | 1415 | 1998 | 1926 | 2028 | 2028 | 2275 |
| Airbus Industrie A310-300 | 31Y | 38 | 164 | 761 | 1067 | 796 | 846 | 1464 | 1043 | 1242 | 1331 | 1406 | 1986 | 1915 | 2016 | 2016 | 2261 |
| Boeing 757-200PF | 75F | 93 | 116 | 538 | 702 | 572 | 599 | 1550 | 738 | 916 | 942 | 995 | 1405 | 1354 | 1426 | 1426 | 1599 |
| Boeing 737-300 | 73Y | 15 | 63 | 585 | 385 | 325 | 325 | 635 | 401 | 555 | 511 | 540 | 843 | 736 | 775 | 775 | 869 |
| Antonov An-12 | ANF | 20 | 61 | 566 | 1652 | 316 | 971 | 828 | 486 | 542 | 495 | 523 | 817 | 712 | 750 | 750 | 841 |
| Boeing 737 | 73F | 18 | 58 | 538 | 413 | 302 | 299 | 616 | 369 | 521 | 471 | 497 | 887 | 677 | 713 | 713 | 800 |
| Boeing 737-400 | 73P | 18 | 58 | 538 | 652 | 302 | 299 | 616 | 369 | 521 | 471 | 497 | 887 | 677 | 713 | 713 | 800 |

Notes:
*Turnaround at 70% Cargo Load Factor
**Turnaround (w/o Noise charges)
***Assumptions
[1] NOx charges not included in the calculations.

**Table 27.** Aircraft mix at BSR airports with the main share of cargo flights (Own calculations based on data from FlightStats, OAG)

| **Aircraft Type** | **MTOW** | **FRA** | **SVO** | **DME** | **HHN** | **CPH** | **HEL** | **OSL** | **ARN** | **SQQ** | **LED** | **MSQ** | **VKO** | **CGN** | **LEJ** |
|---|---|---|---|---|---|---|---|---|---|---|---|---|---|---|---|
| 74N | 442 | 4.5% | 15.8% | 19.0% | - | - | - | - | - | - | - | - | - | 12.3% | - |
| 74F | 397 | 12.3% | - | - | - | - | - | - | - | - | - | - | - | 7.4% | - |
| 74Y | 397 | 16.6% | 37.3% | 76.2% | 29.0% | 50.7% | - | 58.8% | 44.4% | 16.7% | 42.9% | - | 100% | - | 4.5% |
| 77F | 348 | 5.4% | - | - | - | 16.4% | - | - | - | - | - | - | - | - | 12.5% |
| 77X | 347 | 6.9% | - | 4.8% | 0.6% | 24.7% | - | 41.2% | - | - | 57.1% | - | - | - | - |
| M1F | 286 | 38.8% | 42.7% | - | 65.3% | - | 56.3% | - | - | - | - | - | - | - | - |
| 33F | 230 | 1.2% | - | - | - | - | - | - | - | - | - | - | - | - | - |
| 33X | 230 | 1.6% | - | - | 1.7% | - | - | - | - | - | - | - | - | - | - |
| ABF | 171 | 4.0% | - | - | - | 8.2% | - | - | - | - | - | - | - | - | 39.8% |
| ABY | 171 | - | - | - | - | - | 8.5% | - | - | 83.3% | - | - | - | 12.3% | - |
| ABX | 165 | - | - | - | - | - | - | - | - | - | - | - | - | 12.3% | - |
| 31Y | 164 | 4.3% | 4.2% | - | - | - | 18.3% | - | 55.6% | - | - | 100% | - | 7.4% | - |
| 75F | 116 | 4.3% | - | - | - | - | - | - | - | - | - | - | - | - | 26.1% |
| 73Y | 63 | 0.2% | - | - | - | - | - | - | - | - | - | - | - | - | - |
| ANF | 61 | - | - | - | 3.4% | - | - | - | - | - | - | - | - | - | - |
| 73F | 58 | - | - | - | - | - | - | - | - | - | - | - | - | 23.5% | - |
| 73P | 58 | - | - | - | - | - | 16.9% | - | - | - | - | - | - | 24.7% | 17.0% |
| Sum | | 100% | 100% | 100% | 100% | 100% | 100% | 100% | 100% | 100% | 100% | 100% | 100% | 100% | 100% |

The comparative advantage at equally sized airports would then solely be based on other factors, such as *level of service*, *infrastructure and capacity*, *location*, and *connectivity*. It is unreasonable to have an airline, which is flying an aircraft between a city pair, pay a different price for a similar infrastructure, e.g. a runway and a parking position. The airline should value connectivity and level-of-service through the payment of an extra price, e.g. a congestion fee during peak hours and arrival waves, where connectivity is highest because of dense hub operations.

### Public Service Obligation routes in the BSR

Consequently, from lack of connectivity and a poor self-sustained market for air transport services, we find subsidized traffic in some of the peripheral regions in Europe (and internationally). Such services are financed by public service obligations (PSOs) whereby the government offers funds for the provision of air transport operations to those airlines, which pose the best bid and apply with the lowest price on selected routes. A call for tender is typically published through relevant European channels. In Europe several countries and its airports rely to a different degree on PSO traffic (Bubalo 2012b) (Please see **Table 37** in the Appendix for details).

France, Italy, Norway and Greece have published 48, 41, 40 and 31 calls for tender, respectively (**Fig. 57**). In the BSR it is surprising that Finland, which has a similar geography as the Scandinavian partner countries Norway and Sweden, only offers 3 calls for tender on domestic routes, compared to 40 and 12 call for tender, respectively. Germany has published three call for tender on the routes Erfurt (ERF) - Munich (MUC), Hof/Plauen (HOQ) - Frankfurt (am Main) (FRA) and Rostock-Laage (RLG) - Munich (MUC), but only the first route was served by the now bankrupt carrier Cirrus Airlines.[1]

PSOs provide a credible instrument whereby unprofitable basic social services can be economically supported. Under "market forces" these organizations, such as regional airports and its serving airlines, would disappear, due to the lack of demand and the high operational costs (e.g. fuel and labour costs).

The PSOs are offered by the government (e.g. the ministry of transport) to airlines for serving routes. Through the subsidies, which may be the amount of economic losses incurred by air carrier operations, the scheduled air traffic is kept vital between city pairs. The only flow of income for some airports which solely serve PSO traffic is through airport charges paid by (subsidized) airlines.

[1] Cirrus Airlines quit its operations on the PSO route ERF-MUC on 23rd of December 2011. The regional carrier had to file for bankruptcy in January 2012 (http://de.wikipedia.org/wiki/Cirrus_Airlines).

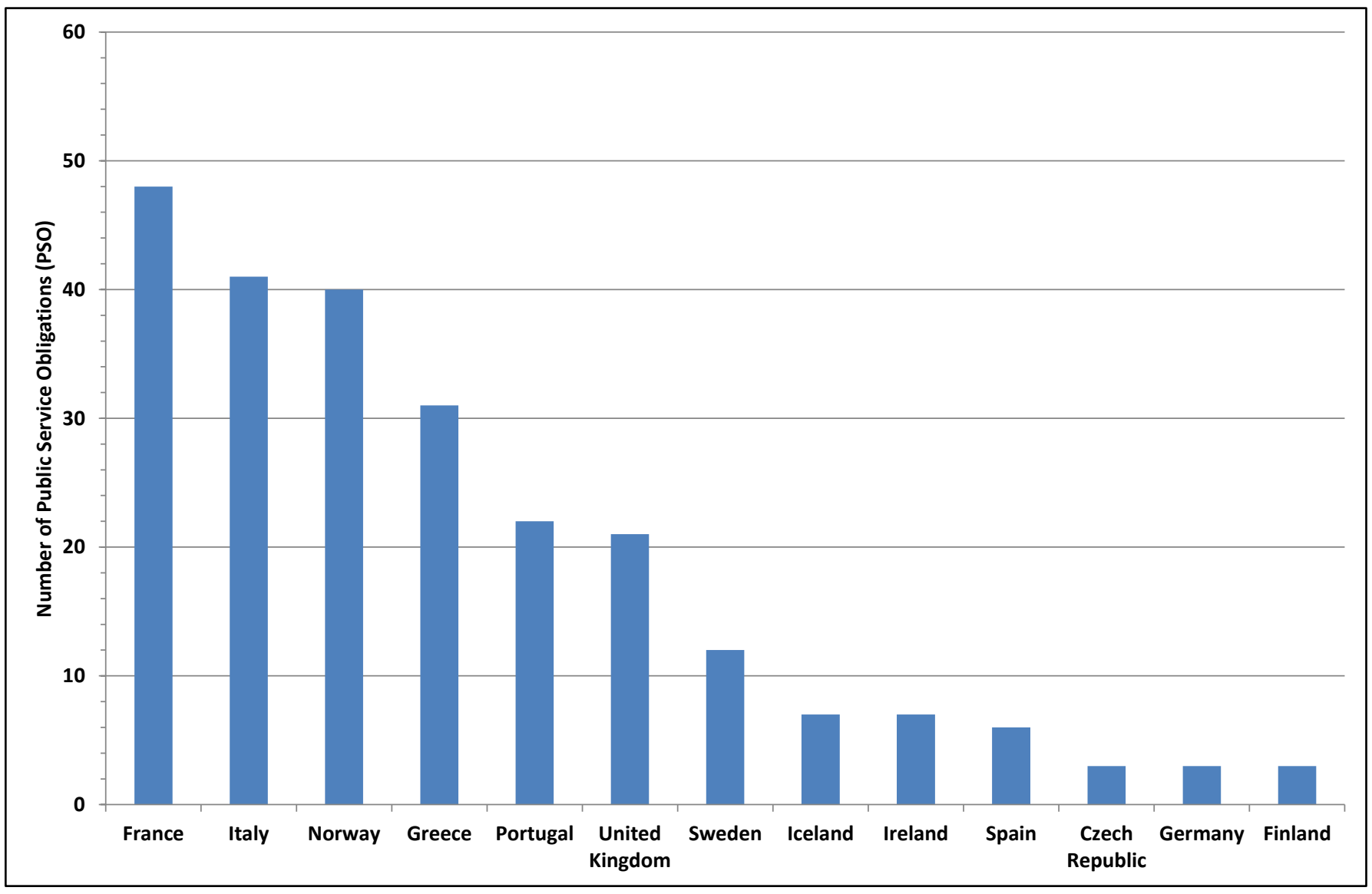

**Fig. 57.** Call for tender for PSO routes by European country (European Commission DG Mobility and Transport 2013)

## Small and large airport economics and benchmarks

In this section we want to summarize how costs and revenues of 63 airports in the BSR region[1] are structured. We divide airports by specific classes for which we give averages. The main observation is that airports operating below a "critical mass" of demand, measured by number of passengers (PAX) or Workload Units (WLUs) rarely break-even, i.e. total costs (including depreciation) are larger than total revenues. As we have seen for the case study of Swedish airports in section **6.3**, we expect the "break-even point" to lay at around 500,000 to 1 million passengers per year for best-in-class airports. On average, we see in **Table 28** that, in the class of airports with below 1 million (but above 100,000) passengers per year, the earnings before interest and taxes (EBIT) are still negative (a loss of 9 Euros per PAX) at an average annual throughput of 339.000 passengers. In the class of small airports below 100,000 PAX (but above 10,000 PAX) the losses are even greater with (-) 42 Euros per PAX at an average annual throughput of 45,000 PAX.

From **Table 28** we see that only the class of below 10 million PAX (but above 1 million) a break-even could be expected and airports will on average be able to generate profits. In this class the average annual throughput is 3.4 million PAX per year and the average EBIT per PAX is zero, thus airports with a larger throughput may be able to generate profits. In the class above 10 million PAX per year profitability should be certain. On average these airports operate 237 thousand flights per year and have an annual throughput of 21.45 million passengers, who generate 276.8 million Euros in aeronautical revenues[1], 222.7 million Euros in non-aeronautical revenues[2] and 137.2 million Euros in Profits (EBIT). Additionally, these airports handle 162 thousand tons of cargo.

[1] We drew information from the financial database of the German Airport Performance (GAP) research project. The sample consists of airports in Denmark, Estonia, Germany Norway, Poland and Sweden. See Bubalo (2012a), GAP (2012), Samferdselsdepartementet (2013) and **Table 28** to **Table 30** for details. Statistics regarding some airports cannot be disclosed due to confidentiality reasons.

Of a sub-sample of 63 airports on average these have an annual throughput of 2.7 million PAX, 35 thousand tons of cargo and 37,000 flights (in 2010). Given the throughput it can be expected that 431 Euro per flight in aeronautical revenues and 5 Euro per PAX in non-aeronautical revenues can be generated. Remarkably in 2010, considering all the efforts involved in delivering air transport services to tens of millions of consumers (167 million PAX in 2010 in this small sub-sample), the 63 airports produce on average a loss of -18 Euros per passenger.

**Table 28.** Descriptive statistics of airports in the BSR (2010)[3]

| | | **Airport Class** | | | | |
|---|---|---|---|---|---|---|
| | **Class Averages** | **<100** | **<1,000** | **<10,000** | **≥ 10,000** | **Totals/ Average** |
| | | **in thousand passengers** | | | | |
| | Observations | 24 | 18 | 16 | 5 | 63 |
| in thousands | Average Passengers | 45 | 339 | 3,438 | 21,453 | 2,690 |
| | Staff Costs | 801 | 2,138 | 21,385 | 131,060 | 16,749 |
| | Aeronautical Revenues | 342 | 2,895 | 49,015 | 276,786 | 35,373 |
| | Non-Aeronautical Revenues | 137 | 1,744 | 28,983 | 222,734 | 25,588 |
| | Earnings before Interests and Taxes | -1,232 | -1,793 | 6,570 | 137,261 | 11,581 |
| | Total Flights | 4 | 10 | 55 | 237 | 37 |
| | Total Cargo in tons | 0 | 1 | 88 | 162 | 35 |
| | Aircraft Parking Positions | 2 | 6 | 31 | 77 | 16 |
| | Aeronautical Revenues per Flight | 89 | 290 | 883 | 1,131 | 431 |
| | Non-Aeronautical Revenue per PAX | 3 | 5 | 8 | 9 | 5 |
| | Staff Costs per PAX | 25 | 7 | 6 | 6 | 14 |
| | Total Revenues per PAX | 11 | 13 | 22 | 22 | 15 |
| | Total Costs per PAX | 41 | 15 | 17 | 13 | 25 |
| | EBIT per PAX | -42 | -9 | 0 | 6 | -18 |

These losses are seldom covered by the profits produced at the few main hubs through cross-subsidization within large airport operators' portfolios, but rather through public subsidies from neighboring municipalities. The municipalities are frequently owners of parts or all of the, usually small, airport organization. However, transactions to cover deficits are often criticized by the public in the communities surrounding these small airports but justified by the authorities by arguing with 'multiplier effects', 'tax revenue generation' and 'job motors'. It is plausible to find many examples of inefficient public spending on small airports in Europe and in the BSR, for example, because of: (1.) lack of rigid cost control, (2.) lack of revenue generation innovations, (3.) lack of incentivized charges schemes for airlines, (4.) unnecessary airport competition over overlapping catchment areas, (5.) passive route development, (6.) mismanagement or (7.) under-investments.

In **Table 29** we present some class averages for Norwegian airports. In this case it is also apparent that on average regional airports with passenger numbers below 1 million are unable to make a profit. However, based on 2011 data from

[1] Revenues from air traffic related charges, e.g. for aircraft landing and parking or passenger safety.

[2] Earned revenues from passengers' consumption and demanded services, e.g. car parking, food and beverages or shopping.

[3] Source: compiled by the author.

Samferdselsdepartementet (2013; pages 97-108) we have isolated two examples in the Norwegian airport system operated by AVINOR, which serve below 1 million PAX and surpassed the break-even point. The airport Ålesund (ALF) and Kristiansand (KRS) make a unit profit (EBIT) of 1.82 Euros per PAX and 3.25 Euros per PAX at passenger levels of 910 thousand and 952 thousand PAX, respectively. In contrast, another airport, Tromsø (TOS), served 1.7 million PAX in 2011, but did not manage to operate profitably (-1.3 Euros per PAX). Bodø airport, with 1.56 million PAX in 2011, broke-even exactly. Differences between the latter two examples compared to the former two "benchmarks" may result from higher depreciation, due to recent investments in new or upgraded infrastructure. Costs, especially labor costs per employee, should be in a similar range within the same organization.

**Table 29.** Descriptive statistics of airports in Norway in 2010 (GAP 2012)

| | **Country** | **Norway** | | | | |
|---|---|---|---|---|---|---|
| | **Year** | **2010** | | | | |
| | | **Airport Class** | | | | |
| | **Class Averages** | **<100** | **<1,000** | **<10,000** | **≥ 10,000** | **Country Average** |
| | | | **in thousand passengers** | | | |
| | Observations | 23 | 16 | 6 | * | * |
| in thousands | Average Passengers | 46 | 324 | 2,854 | 19,091 | 923 |
| | Staff Costs | 802 | 1,719 | 5,698 | 34,134 | 2,484 |
| | Aeronautical Revenues | 318 | 2,471 | 22,571 | 146,053 | 7,138 |
| | Non-Aeronautical Revenues | 132 | 1,590 | 16,553 | 198,009 | 7,083 |
| | Earnings before Interests and Taxes | -1,259 | -1,299 | 15,544 | 125,042 | 3,665 |
| | Total Flights | 3 | 9 | 59 | 219 | 17 |
| | Total Cargo in tons | 0 | 0 | 4 | 86 | 3 |
| | Aircraft Parking Positions | 1 | 6 | 19 | 64 | 6 |
| | Aeronautical Revenues per Flight | 88 | 250 | 358 | 666 | 192 |
| | Non-Aeronautical Revenue per PAX | 3 | 4 | 6 | 10 | 4 |
| | Staff Costs per PAX | 25 | 6 | 2 | 2 | 15 |
| | Total Revenues per PAX | 10 | 12 | 13 | 18 | 11 |
| | Total Costs per PAX | 41 | 13 | 6 | 7 | 26 |
| | EBIT per PAX | -43 | -8 | 4 | 7 | -24 |

Notes: Financial variables in 2012 real Euros.

*Confidential

**Table 29** shows that Norwegian airports with below 10 million PAX (but above or equal to 1 million PAX) had a profitability of 4 Euros per PAX at an average level of 2.85 million PAX per year in 2010.

The four airports in Poland, of which we have data in the GAP financial database, belong to the class of airports above or equal to 1 million PAX, but below 10 million (**Table 30**). On average the Polish airports have an annual throughput of 1.7 million PAX (or 1.73 million WLUs), but these produce a unit loss of -1 Euro per PAX (or -1.15 Euros per WLU). However, this amount of loss cannot be considered exceptionally large. It is reasonable to assume that the Polish airports in the class of between 1 and 10 million will operate profitably in the future. This depends also on how the global, European, and national economies and demand patterns develop.

**Table 30.** Descriptive Statistics of Airports in Poland in 2009 (GAP 2012)

| | **Country** | **Poland** |
|---|---|---|
| | **Year** | **2009** |
| | **Class Average** | **Airport Class <10,000,000** |
| | Observations | 4 |
| in thousands | Total Passengers | 1,704 |
| in thousands | Staff Costs | 11,031 |
| in thousands | Aeronautical Revenues | 25,124 |
| in thousands | Non-Aeronautical Revenues | 5,750 |
| in thousands | Earnings before Interests and Taxes | -1,998 |
| in thousands | Total Flights | 24 |
| in thousands | Total Cargo in tons | 3 |
| | Aircraft Parking Positions | 16 |
| | Aeronautical Revenues per Flight | 1,049 |
| | Non-Aeronautical Revenue per PAX | 4 |
| | Staff Costs per PAX | 7 |
| | Total Revenues per PAX | 19 |
| | Total Costs per PAX | 87 |
| | EBIT per PAX | -1 |

### 6.4 Input-Output analysis based on air traffic flows

We tried to focus on the most relevant airports (and cargo routes) as found in the ranking in **Table 34** in the Appendix. We use micro information from flight schedules to conduct a "bottom up" analysis. The micro information is aggregated to the needed macro level. With detailed flight schedule information at hand, we were able to study the linkages and destinations of the air traffic network in the BSR and outside. We were interested how the air cargo traffic is distributed among the BSR countries and airports. Primarily we investigated the patterns of cargo flights originating in the BSR and its main destinations. **Table 31** shows the distribution by country in an Origin-Destination Matrix. In our representation the originating countries are listed in the top horizontal row and the destination countries are listed vertically on the left.

As we can see Germany[1] accounts for 59% of the total cargo flights in the BSR of which 44% have their destination airport outside of the BSR and 15% have their destination airport in the BSR. Domestic traffic from German airports in the BSR to other German airports accounts for 3.2% of total traffic in the BSR. 12% of all flights in the BSR region originate in Germany and fly to its main destination Russia. In the other direction 8.7% of all cargo flights fly from Russia to Germany (**Table 31**).

Russia accounts for 31% of total cargo traffic in the BSR of which 19% flies to countries outside the BSR and 12% fly to destinations inside the BSR. Please see **Table 36** in the Appendix for cargo airlines operating at airports and routes in the Baltic Sea Region. **Table 32** finally displays the traffic distribution of all cargo flights flying between airports in the BSR during the reporting period. Cargo flights to airports outside the BSR are aggregated under "Extra BSR". The most frequented

[1] On the country level no divisions between German destinations have been made (**Table 31**). On the airport level only the German airports which are framed by the airports CGN in the West and HHN in the South have been considered as part of the "BSR", all others as "Non-BSR". On the country level no divisions between German destinations have been made (**Table 32**).

route is between Sheremetyevo (SVO) to Frankfurt-Hahn (HHN) with 3.3 % of all cargo flights.

**Table 31.** Origin/destination map of cargo traffic in the BSR by country (FlightStats)

| **Share of Cargo Traffic** | **Origin Country** | | | | | | | | |
|---|---|---|---|---|---|---|---|---|---|
| **Destination Country** | **Germany** | **Russia** | **Denmark** | **Finland** | **Norway** | **Sweden** | **Lithuania** | **Belarus** | **Sum** |
| **BSR Countries** | **15.5%** | **12.5%** | **0.1%** | **0.9%** | - | **0.4%** | - | - | **29.4%** |
| **Russia** | **12.2%** | **3.4%** | - | **0.5%** | - | - | - | - | **16.0%** |
| **Germany** | **3.2%** | **8.7%** | **0.1%** | - | - | - | - | - | **12.0%** |
| **Finland** | - | **0.4%** | - | - | - | **0.4%** | - | - | **0.8%** |
| **Sweden** | - | - | - | **0.4%** | - | - | - | - | **0.4%** |
| **Denmark** | **0.1%** | - | - | - | - | - | - | - | **0.1%** |
| **No-BSR countries** | **43.9%** | **18.7%** | **2.9%** | **2.0%** | **1.4%** | **0.9%** | **0.5%** | **0.4%** | **70.6%** |
| China | 4.2% | 10.3% | 1.0% | 0.4% | - | - | - | - | 15.9% |
| United States of America | 4.6% | 0.4% | 0.2% | 0.2% | - | - | - | - | 5.5% |
| Turkey | 3.4% | 0.2% | - | 0.7% | - | 0.2% | 0.4% | 0.1% | 5.1% |
| United Arab Emirates | 3.7% | - | 0.7% | - | - | 0.1% | - | - | 4.6% |
| South Korea | 1.4% | 2.0% | 0.2% | - | 0.6% | 0.2% | - | - | 4.5% |
| United Kingdom | 4.0% | 0.2% | 0.3% | - | - | - | - | - | 4.5% |
| Netherlands | 0.9% | 2.1% | 0.4% | 0.1% | - | 0.4% | - | - | 3.9% |
| India | 3.2% | - | - | 0.2% | - | - | - | - | 3.4% |
| Senegal | 2.3% | - | - | - | - | - | - | - | 2.3% |
| Spain | 1.5% | 0.4% | - | - | - | - | - | - | 1.9% |
| Saudi Arabia | 1.6% | - | - | - | - | - | - | - | 1.6% |
| Austria | 0.4% | 0.4% | - | - | 0.5% | - | - | 0.2% | 1.6% |
| Azerbaijan | 1.5% | - | - | - | - | - | - | - | 1.5% |
| Kazakhstan | 1.3% | - | - | - | - | - | - | - | 1.3% |
| Kenya | 1.2% | - | - | - | - | - | - | - | 1.2% |
| Italy | 0.4% | 0.6% | - | - | 0.1% | - | - | - | 1.1% |
| Qatar | 1.1% | - | - | - | 0.1% | - | - | - | 1.1% |
| France | 0.6% | 0.4% | - | - | - | - | - | - | 1.0% |
| Japan | - | 0.9% | - | - | - | - | - | - | 1.0% |
| **Sum** | **59.4%** | **31.2%** | **3.0%** | **2.9%** | **1.4%** | **1.3%** | **0.5%** | **0.4%** | **100.0%** |

**Table 32.** Detailed origin/destination matrix of cargo traffic in the BSR by airport (Source: FlightStats.com)

| Share of Total Cargo Flights | | Origin | | | | | | | | | | | | | | | | | |
|---|---|---|---|---|---|---|---|---|---|---|---|---|---|---|---|---|---|---|---|
| | | BSR | | | | | | | | | | | | | | | BSR Sum | Extra BSR | Sum |
| | | Germany | | | | Russia | | | | Finland | Denmark | Norway | Sweden | | Belarus | Lithuania | | | |
| | | Frankfurt International | Frankfurt - Hahn | Koeln/Bonn | Leipzig/Halle | Sheremetyevo | Domodedovo | Pulkovo | Vnukovo | Helsinki-Vantaa | Copenhagen | Oslo | Arlanda | Landvetter | Minsk International 2 | Siauliai | | | |
| **Destination** | **IATA** | **FRA** | **HHN** | **CGN** | **LEJ** | **SVO** | **DME** | **LED** | **VKO** | **HEL** | **CPH** | **OSL** | **ARN** | **GOT** | **MSQ** | **SQQ** | | | |
| **BSR** | | **2.7%** | **3.6%** | | **2.9%** | **6.6%** | **1.2%** | **0.2%** | | **1.3%** | | | | | | | **18.5%** | **4.5%** | **22.9%** |
| **Germany** | | | | | **2.9%** | **6.1%** | **1.2%** | **0.2%** | | | | | | | | | **10.4%** | **2.6%** | **13.0%** |
| **Frankfurt International** | FRA | | | | 2.1% | 2.8% | 1.2% | 0.2% | | | | | | | | | 6.3% | 2.0% | 8.3% |
| **Frankfurt - Hahn** | HHN | | | | | 3.3% | | | | | | | | | | | 3.3% | 0.6% | 3.9% |
| **Koeln/Bonn** | CGN | | | | 0.8% | | | | | | | | | | | | 0.8% | | 0.8% |
| **Russia** | | **2.5%** | **3.6%** | | | | | | | **0.5%** | | | | | | | **6.6%** | **1.9%** | **8.5%** |
| **Sheremetyevo** | SVO | 2.1% | 3.6% | | | | | | | 0.5% | | | | | | | 6.2% | 1.9% | 8.1% |
| **Domodedovo** | DME | 0.4% | | | | | | | | | | | | | | | 0.4% | | 0.4% |
| **Sweden** | | | | | | | | | | **0.8%** | | | | | | | **0.8%** | | **0.8%** |
| **Arlanda** | ARN | | | | | | | | | 0.8% | | | | | | | 0.8% | | 0.8% |
| **Finland** | | | | | | **0.5%** | | | | | | | | | | | **0.5%** | | **0.5%** |
| **Helsinki-Vantaa** | HEL | | | | | 0.5% | | | | | | | | | | | 0.5% | | 0.5% |
| **Denmark** | | **0.2%** | | | | | | | | | | | | | | | **0.2%** | | **0.2%** |
| **Copenhagen** | CPH | 0.2% | | | | | | | | | | | | | | | 0.2% | | 0.2% |
| **Extra BSR** | | **39.3%** | **4.0%** | **4.1%** | **1.2%** | **13.5%** | **2.3%** | **0.2%** | **0.2%** | **1.9%** | **3.0%** | **1.3%** | **0.9%** | **0.2%** | **0.5%** | **0.5%** | **73.1%** | **4.0%** | **77.1%** |
| **Sum** | | **42.0%** | **7.6%** | **4.1%** | **4.1%** | **20.1%** | **3.5%** | **0.4%** | **0.2%** | **3.1%** | **3.0%** | **1.3%** | **0.9%** | **0.2%** | **0.5%** | **0.5%** | **91.5%** | **8.5%** | **100.0%** |

## 6.5 Baltic air cargo market forecasts and trade potential

Similar to **Fig. 58** by the Baltic Transport Outlook (BTO) 2030 we tried in **Fig. 44** in the introduction to capture how the economic outlook for the BSR into the future could be presented by a map. Both maps show the greater Baltic Sea region, with the exception that we took more countries and airports into consideration for our study. The full list of countries airports can be found in **Table 34** in the Appendix.

The map of the "BTO 2030 strategic network" includes existing and planned "terminals, ports and airports", thus the analysis is very broad and includes many modes of transport, such as rail and sea. In comparison with our work BTO 2030 gives only little attention to air cargo traffic. In **Fig. 44** we have included proposed routes, which partly emerged from the total traffic analysis, passenger and cargo (see **Table 33**). In connection with passenger traffic we see the greatest potential for the development of air cargo transport.

It is much easier for the airport management if passengers already provide a solid economical foundation. The demands and potential incomes are small in comparison to passenger traffic. Taking the times into consideration a cargo airplane usually blocks a parking position (about 2 to 3 hours on average), this space could be used for at least twice as many passenger flights in the same time period. LCCs and some smaller regional carriers, such as Scandinavian Wideroe (with turnaround times less than 30 minutes) could even manage to operate up to 6 flights in three hours on one parking position. This blocking of space should proportionally to time be penalized. However, in practice parking less than three hours is usually exempted from aircraft parking charges.

With Origin and Destination matrices calculated on a regular basis, such as **Table 33**, forecast can be conducted very easily and reliably. Changes can be included based on changes in the flight distribution. There is great certainty that in the short- and mid-term the structure of the distribution does not change much, so it is possible to proportionally increase the traffic presented in **Table 33** for a rough forecast into, say, the next one or two years. A more sophisticated analysis would require extrapolating from data and applying to the distribution pattern the growth rates for each location separately.

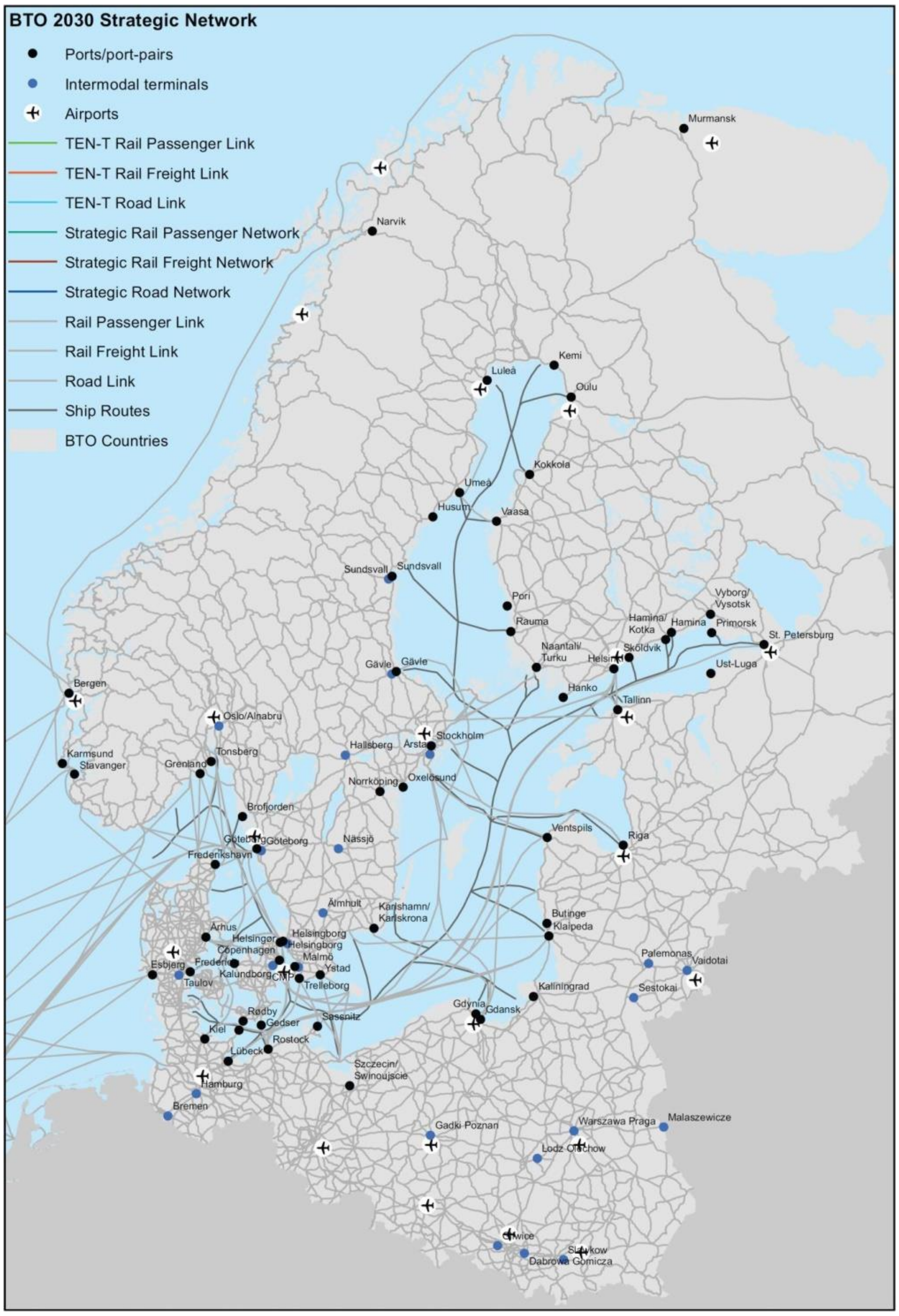

**Fig. 58.** Baltic Transport Outlook 2030 strategic map terminals, ports and airports (BTO 2011)

**Table 33.** Distribution of observed flights between origins and destinations of the „core" Baltic states Belarus, Estonia, Latvia and Lithuania (Source: FlightStats.com).

| Distribution of Flights | Origin | | | | | | | | | | | |
|---|---|---|---|---|---|---|---|---|---|---|---|---|
| | Country | LV | EE | | | EE SUM | LT | | | LT SUM | BY | Total Sum |
| Destination | IATA | RIX | TLL | URE | KDL | | VNO | PLQ | KUN | | MSQ | |
| FI | HEL | 45 | 50 | | | 50 | 25 | | | 25 | | 120 |
| FI | TMP | 21 | | | | | | | | | | 21 |
| FI | OUL | 4 | | | | | | | | | | 4 |
| FI | KUO | 3 | | | | | | | | | | 3 |
| EE | TLL | 37 | | 19 | 12 | 31 | 17 | | | 17 | 3 | 88 |
| EE | URE | | 19 | | | 19 | | | | | | 19 |
| EE | KDL | | 12 | | | 12 | | | | | | 12 |
| RU | SVO | 14 | 3 | | | 3 | 5 | | | 5 | 21 | 43 |
| RU | KGD | 14 | | | | | | | | | 7 | 21 |
| RU | LED | 12 | | | | | | | | | 4 | 16 |
| RU | DME | | | | | | | | | | 14 | 14 |
| RU | VKO | | | | | | 7 | | | 7 | | 7 |
| SE | ARN | 28 | 31 | | | 31 | 12 | | | 12 | | 71 |
| SE | NYO | 14 | | | | | | | | | | 14 |
| SE | GOT | 4 | 4 | | | 4 | | | | | | 8 |
| SE | MMX | | | | | | | | 5 | 5 | | 5 |
| SE | VXO | | | | | | 2 | | | 2 | | 2 |
| DE | FRA | 7 | 10 | | | 10 | 14 | | | 14 | 9 | 40 |
| DE | TXL | 19 | | | | | 5 | | | 5 | | 24 |
| DE | HAM | 7 | | | | | | | | | | 7 |
| DE | HHN | 3 | | | | | | | 3 | 3 | | 6 |
| DE | HAJ | 5 | | | | | | | | | | 5 |
| DE | BRE | 3 | | | | | | | | | | 3 |
| DE | SXF | | | | | | | | | | 3 | 3 |
| DK | CPH | 21 | 19 | | | 19 | 21 | 14 | | 35 | | 75 |
| DK | BLL | 3 | | | | | | | | | | 3 |
| LV | RIX | | 37 | | | 37 | 33 | | | 33 | 7 | 77 |
| LT | VNO | 38 | 17 | | | 17 | | | | | | 55 |
| NO | OSL | 17 | 11 | | | 11 | 3 | | | 3 | | 31 |
| NO | TRD | 2 | | | | | | | | | | 2 |
| NO | RYG | | | | | | | 2 | | 2 | | 2 |
| PL | WAW | 11 | | | | | 11 | | | 11 | 7 | 29 |
| BY | MSQ | 7 | 3 | | | 3 | | | | | | 10 |
| Total Sum | | 339 | 216 | 19 | 12 | 247 | 155 | 16 | 8 | 179 | 75 | 840 |

## 6.6 Recommendations and conclusions

Unless airports in the BSR reach break-even demand levels, these will rely to a large degree on different forms of subsidies. Subsidies may be justifiable to the public as start-up funds but should in scope be limited to the short and mid-term. Similar to an inefficient PSO framework, the instrument of (cross-) subsidies might lead to a "vicious cycle" of ever-increasing costs, if there is no incentive for airport management to reduce or stabilize costs. This means costs will most likely increase over the years, for example because of an increase in average salaries, but demand may stagnate at some maximum level, because the origin/destination demand is limited by low population in the catchment area of regional airports. Small airports and its management should therefore reduce the manpower, which is necessary for serving aircraft, to an absolute minimum and reduce operating hours as much as possible around the schedule. We are aware of the 'basic airport' business model for small airports in Sweden, but unfortunately, we were unable to find some specific details on this strategy. However, we know that *multitasking* plays a role in this model, where employees could have different skills and conduct different jobs at the airport, e.g. baggage handling and plumbing. Outside of operating hours Prior Permission Required (PPR) operations might be possible, where no manned tower is needed.

It is critical for airport management to attract as many customers (passengers or cargo forwarders) as possible. Therefore, airport management should have some decisive power regarding airport charges, so it is able to adjust its pricing schemes to the local demand characteristics. This should also include incentives for increasing growth as it is state of affairs at many airports in Poland.

This report has shown that, in general, air cargo plays only an insignificant role at most airports in the BSR[1] and that additional passenger traffic is needed to reach profitability. It will be very difficult to reach break-even demand levels by cargo alone, especially for new entrants into the market which are yet unknown to forwarders regarding existing business networks and experience.

Cargo traffic in the BSR has only a realistic future when there is a concentrated, coordinated and collaborative effort in managing the commodity and passenger flows through the main cargo hubs. We have not found any examples of profitable "cargo-only" airports in any dataset. We are aware of cargo-only airports, such as Liege (LGG) in Belgium or Siaulial (SQQ) airport in Lithuania. It remains unknown if these airports make profits from their operations. We would recommend developing belly cargo in addition to regular passenger traffic as feeder traffic into the main hubs, which could then be subsidized through a PSO framework in initial stages of development. However, as Ivanov and Stigaard (2013) have pointed out the transporting of belly cargo is inefficient for Low-Cost carriers (LCCs) operating at secondary and regional airports due to their short turn-around policy (see also Bubalo and Gaggero 2015). The handling of cargo would need to speed up enormously if it were anticipated to be transported by LCCs to meet their requirements.

In addition, we would like to see incentive programs regarding airport landing charges as it is the case in Poland, which has growth and bonus schemes as part of the airport charges manuals. While focusing solely on cargo traffic, we believe there is a great risk of business failure (see e.g. Schwerin-Parchim airport case), if an investor does not provide timely action in form of necessary investments (e.g. in cargo storage and handling facilities or other essential infrastructure).

[1] We have only isolated Siauliai (SQQ) airport in Lithuania as a 'Cargo-only' airport in the BSR.

## Appendix

**Table 34.** List of airports in the BSR

| City and Country Code | Airport Name | IATA | ICAO | Latitude | Longitude | Country |
|---|---|---|---|---|---|---|
| Minsk BY | Minsk International 2 | MSQ | UMMS | 53.889725 | 28.032442 | Belarus |
| Altenburg DE | Altenburg Nobitz | AOC | EDAC | 50.983334 | 12.45 | Germany |
| Berlin DE | Tegel | TXL | EDDT | 52.553944 | 13.291722 | Germany |
| Berlin DE | Schoenefeld | SXF | EDDB | 52.370277 | 13.521388 | Germany |
| Braunschweig DE | Braunschweig | BWE | EDVE | 52.266666 | 10.533333 | Germany |
| Bremen DE | Bremen | BRE | EDDW | 53.05297 | 8.785352 | Germany |
| Cochstedt DE | Cochstedt | CSO | | 51.85 | 11.416667 | Germany |
| Cologne DE | Cologne Bonn | CGN | EDDK | 50.878365 | 7.122224 | Germany |
| Dortmund DE | Dortmund | DTM | EDLW | 51.514828 | 7.613139 | Germany |
| Dresden DE | Dresden | DRS | EDDC | 51.124332 | 13.766082 | Germany |
| Erfurt DE | Erfurt | ERF | EDDE | 50.974915 | 10.961163 | Germany |
| Frankfurt DE | Frankfurt am Main | FRA | EDDF | 50.050735 | 8.570773 | Germany |
| Hahn DE | Frankfurt – Hahn | HHN | EDFH | 49.948334 | 7.264167 | Germany |
| Hamburg DE | Hamburg | HAM | EDDH | 53.63128 | 10.006414 | Germany |
| Hamburg DE | Blankensee | LBC | EDHL | 53.80527 | 10.701162 | Germany |
| Hanover DE | Hanover | HAJ | EDDV | 52.459255 | 9.694766 | Germany |
| Heringsdorf DE | Heringsdorf | HDF | EDAH | 53.87825 | 14.138242 | Germany |
| Hof DE | Hof | HOQ | EDQM | 50.289165 | 11.862222 | Germany |
| Kassel DE | Kassel-Calden | KSF | EDVK | 51.415855 | 9.380858 | Germany |
| Leipzig/Halle DE | Leipzig/Halle | LEJ | EDDP | 51.42026 | 12.221202 | Germany |
| Muenster DE | Muenster | FMO | EDDG | 52.130054 | 7.694928 | Germany |
| Paderborn DE | Paderborn | PAD | EDLP | 51.610332 | 8.619832 | Germany |
| Rostock-Laage DE | Laage | RLG | ETNL | 53.92 | 12.266667 | Germany |
| Westerland DE | Westerland - Sylt | GWT | EDXW | 54.91528 | 8.343056 | Germany |
| Aalborg DK | Aalborg | AAL | EKYT | 57.08655 | 9.872241 | Denmark |
| Aarhus DK | Tirstrup | AAR | EKAH | 56.30823 | 10.626351 | Denmark |
| Billund DK | Billund | BLL | EKBI | 55.747383 | 9.147874 | Denmark |
| Bornholm DK | Bornholm | RNN | EKRN | 55.065556 | 14.757778 | Denmark |
| Copenhagen DK | Roskilde | RKE | EKRK | 55.583332 | 12.133333 | Denmark |
| Copenhagen DK | Copenhagen | CPH | EKCH | 55.62905 | 12.647601 | Denmark |
| Esbjerg DK | Esbjerg | EBJ | EKEB | 55.52143 | 8.549062 | Denmark |
| Karup DK | Karup | KRP | EKKA | 56.3 | 9.116667 | Denmark |
| Sonderborg DK | Sonderborg | SGD | EKSB | 54.93028 | 9.794722 | Denmark |
| Kardla EE | Kardla | KDL | EEKA | 58.983334 | 22.8 | Estonia |
| Kuressaare EE | Kuressaare | URE | EEKE | 58.216667 | 22.5 | Estonia |
| Tallinn EE | Tallinn | TLL | EETN | 59.416622 | 24.798702 | Estonia |
| Tartu EE | Tartu | TAY | EETU | 58.333332 | 26.733334 | Estonia |
| Enontekio FI | Enontekio | ENF | EFET | 68.35 | 23.416668 | Finland |
| Helsinki FI | Helsinki-Vantaa | HEL | EFHK | 60.31795 | 24.96645 | Finland |
| Ivalo FI | Ivalo | IVL | EFIV | 68.611115 | 27.415556 | Finland |
| Joensuu FI | Joensuu | JOE | EFJO | 62.656788 | 29.61354 | Finland |
| Jyvaskyla FI | Jyvaskyla | JYV | EFJY | 62.40362 | 25.68143 | Finland |
| Kajaani FI | Kajaani | KAJ | EFKI | 64.27778 | 27.688889 | Finland |
| Kemi/Tornio FI | Kemi/Tornio | KEM | EFKE | 65.78006 | 24.57677 | Finland |
| Kittila FI | Kittila | KTT | EFKT | 67.695946 | 24.859028 | Finland |

| City and Country Code | Airport Name | IATA | ICAO | Latitude | Longitude | Country |
|---|---|---|---|---|---|---|
| Kokkola/ Pietarsaari FI | Kruunupyy | KOK | EFKK | 63.718838 | 23.133068 | Finland |
| Kuopio FI | Kuopio | KUO | EFKU | 63.008907 | 27.788696 | Finland |
| Kuusamo FI | Kuusamo | KAO | EFKS | 65.99028 | 29.233889 | Finland |
| Lappeenranta FI | Lappeenranta | LPP | EFLP | 61.046112 | 28.156668 | Finland |
| Mariehamn AX | Mariehamn | MHQ | EFMA | 60.123333 | 19.896667 | Finland |
| Oulu FI | Oulu | OUL | EFOU | 64.93012 | 25.375425 | Finland |
| Pori FI | Pori | POR | EFPO | 61.46866 | 21.791382 | Finland |
| Rovaniemi FI | Rovaniemi | RVN | EFRO | 66.559044 | 25.829609 | Finland |
| Savonlinna FI | Savonlinna | SVL | EFSA | 61.95 | 28.95 | Finland |
| Seinajoki FI | Ilmajoki | SJY | EFIL | 62.6932 | 22.835285 | Finland |
| Tampere FI | Tampere-Pirkkala | TMP | EFTP | 61.42045 | 23.617563 | Finland |
| Turku FI | Turku | TKU | EFTU | 60.512794 | 22.28098 | Finland |
| Vaasa FI | Vaasa | VAA | EFVA | 63.04355 | 21.760122 | Finland |
| Varkaus FI | Varkaus | VRK | EFVR | 62.3 | 27.933332 | Finland |
| Kaunas LT | Kaunas | KUN | EYKA | 54.9 | 23.916668 | Lithuania |
| Palanga LT | Palanga | PLQ | EYPA | 55.95 | 21.083332 | Lithuania |
| Siauliai LT | Siauliai | SQQ | EYSA | 55.933334 | 23.316668 | Lithuania |
| Vilnius LT | Vilnius | VNO | EYVI | 54.643078 | 25.279606 | Lithuania |
| Riga LV | Riga | RIX | EVRA | 56.92208 | 23.979807 | Latvia |
| Alta NO | Alta | ALF | ENAT | 69.977165 | 23.355808 | Norway |
| Andenes NO | Andenes | ANX | ENAN | 69.30492 | 16.133326 | Norway |
| Bardufoss NO | Bardufoss | BDU | ENDU | 69.056114 | 18.54 | Norway |
| Batsfjord NO | Batsfjord | BJF | ENBS | 70.6 | 29.666668 | Norway |
| Berlevag NO | Berlevag | BVG | ENBV | 70.86667 | 29 | Norway |
| Bodo NO | Bodo | BOO | ENBO | 67.27262 | 14.367839 | Norway |
| Bronnoysund NO | Bronnoy | BNN | ENBN | 65.4617 | 12.215772 | Norway |
| Fagernes NO | Valdres | VDB | ENFG | 61.083332 | 9.333333 | Norway |
| Hammerfest NO | Hammerfest | HFT | ENHF | 70.67999 | 23.675867 | Norway |
| Harstad-Narvik NO | Harstad/Narvik, Evenes | EVE | ENEV | 68.496666 | 16.679722 | Norway |
| Hasvik NO | Hasvik | HAA | ENHK | 70.46667 | 22.15 | Norway |
| Honningsvag NO | Valan | HVG | ENHV | 70.98333 | 25.833332 | Norway |
| Kirkenes NO | Hoeybuktmoen | KKN | ENKR | 69.7235 | 29.891184 | Norway |
| Kristiansand NO | Kjevik | KRS | ENCN | 58.20255 | 8.073732 | Norway |
| Kristiansund NO | Kvernberget | KSU | ENKB | 63.114723 | 7.844444 | Norway |
| Lakselv NO | Banak | LKL | ENNA | 70.06778 | 24.973612 | Norway |
| Leknes NO | Leknes | LKN | KLKN | 68.15421 | 13.614864 | Norway |
| Mehamn NO | Mehamn | MEH | ENMR | 71.03333 | 27.833332 | Norway |
| Mo I Rana NO | Mo I Rana | MQN | ENRA | 66.36465 | 14.302748 | Norway |
| Molde NO | Aro | MOL | ENML | 62.747303 | 7.262118 | Norway |
| Mosjoen NO | Kjaerstad | MJF | ENMS | 65.78439 | 13.218328 | Norway |
| Namsos NO | Namsos | OSY | ENNM | 64.47273 | 11.570002 | Norway |
| Narvik NO | Framnes | NVK | ENNK | 68.425 | 17.425 | Norway |
| Orland NO | Orland | OLA | ENOL | 63.7 | 9.616667 | Norway |
| Oslo NO | Sandefjord | TRF | ENTO | 59.178085 | 10.251807 | Norway |
| Oslo NO | Oslo Gardermoen | OSL | ENGM | 60.19419 | 11.100411 | Norway |
| Roervik NO | Ryumsjoen | RVK | ENRM | 64.88333 | 11.233333 | Norway |
| Roros NO | Roros | RRS | ENRO | 62.579166 | 11.345556 | Norway |
| Rost NO | Stolport | RET | ENRS | 67.48333 | 12.083333 | Norway |
| Rygge NO | Moss-Rygge | RYG | ENRY | 59.37903 | 10.800161 | Norway |
| Sandnessjoen NO | Stokka | SSJ | ENST | 65.959946 | 12.476518 | Norway |
| Skien NO | Skien | SKE | ENSN | 59.182777 | 9.5625 | Norway |
| Sogndal NO | Haukasen | SOG | ENSG | 61.158127 | 7.136988 | Norway |

| City and Country Code | Airport Name | IATA | ICAO | Latitude | Longitude | Country |
|---|---|---|---|---|---|---|
| Sorkjosen NO | Sorkjosen | SOJ | ENSR | 69.78333 | 20.933332 | Norway |
| Stokmarknes NO | Skagen | SKN | ENSK | 68.57915 | 15.032921 | Norway |
| Svolvaer NO | Helle | SVJ | ENSH | 68.24498 | 14.667774 | Norway |
| Tromso NO | Tromso/Langnes | TOS | ENTC | 69.67983 | 18.907343 | Norway |
| Trondheim NO | Vaernes | TRD | ENVA | 63.454285 | 10.917863 | Norway |
| Vadso NO | Vadso | VDS | ENVD | 70.065 | 29.845278 | Norway |
| Vaeroy NO | Stolport | VRY | | 67.666664 | 12.683333 | Norway |
| Vardoe NO | Vardoe | VAW | ENSS | 70.35472 | 31.045555 | Norway |
| Bydgoszcz PL | Bydgoszcz | BZG | EPBY | 53.09667 | 17.978611 | Poland |
| Gdansk PL | Rebiechowo | GDN | EPGD | 54.380978 | 18.468655 | Poland |
| Katowice PL | Katowice | KTW | EPKT | 50.470833 | 19.07403 | Poland |
| Krakow PL | J. Paul II International Krakow-Bali | KRK | EPKK | 50.075493 | 19.793743 | Poland |
| Lodz PL | Lodz Lublinek | LCJ | EPLL | 51.721943 | 19.398333 | Poland |
| Poznan PL | Lawica | POZ | EPPO | 52.414326 | 16.828844 | Poland |
| Rzeszow PL | Jasionka | RZE | EPRZ | 50.11525 | 22.03133 | Poland |
| Swidnik PL | Lublin | LUZ | | 51.231945 | 22.690277 | Poland |
| Szczecin PL | Goleniow | SZZ | EPSC | 53.593525 | 14.894611 | Poland |
| Warsaw PL | Frederic Chopin | WAW | EPWA | 52.170906 | 20.97329 | Poland |
| Wroclaw PL | Strachowice | WRO | EPWR | 51.10482 | 16.899403 | Poland |
| Zielona Gora PL | Babimost | IEG | EPZG | 51.933334 | 15.516667 | Poland |
| Arkhangelsk RU | Arkhangelsk | ARH | ULAA | 64.594795 | 40.711903 | Russia |
| Belgorod RU | Belgorod | EGO | UUOB | 50.63333 | 36.65 | Russia |
| Cherepovets RU | Cherepovets | CEE | | 59.283333 | 38.066666 | Russia |
| Ivanovo RU | Ivanovo | IWA | | 56.942955 | 40.944546 | Russia |
| Kaliningrad RU | Kaliningrad | KGD | UMKK | 54.882656 | 20.586645 | Russia |
| Kirovsk RU | Kirovsk | KVK | | 67.583336 | 33.583332 | Russia |
| Kursk RU | Kursk | URS | UUOK | 51.75 | 36.266666 | Russia |
| Lipetsk RU | Lipetsk | LPK | | 52.61667 | 39.6 | Russia |
| Moscow RU | Vnukovo | VKO | UUWW | 55.60315 | 37.2921 | Russia |
| Moscow RU | Sheremetyevo International | SVO | UUEE | 55.966324 | 37.416573 | Russia |
| Moscow RU | Moscow Domodedovo | DME | UUDD | 55.414566 | 37.899494 | Russia |
| Moscow RU | Bykovo | BKA | UUBB | 55.433334 | 37.966667 | Russia |
| Murmansk RU | Murmansk | MMK | ULMM | 68.785095 | 32.759155 | Russia |
| Petrozavodsk RU | Petrozavodsk | PES | ULPB | 61.683334 | 34.333332 | Russia |
| Saint Petersburg RU | Pulkovo | LED | ULLI | 59.806084 | 30.3083 | Russia |
| Solovetsky RU | Solovetsky | CSH | ULAS | 65.02944 | 35.733334 | Russia |
| Voronezh RU | Voronezh | VOZ | UUOO | 51.812355 | 39.226997 | Russia |
| Yaroslavl RU | Yaroslavl | IAR | | 57.61667 | 39.88333 | Russia |
| Angelholm/ Helsingborg SE | Angelholm | AGH | ESDB | 56.293056 | 12.8625 | Sweden |
| Arvidsjaur SE | Arvidsjaur | AJR | ESNX | 65.59139 | 19.285557 | Sweden |
| Borlange/Falun SE | Dala | BLE | ESSD | 60.42973 | 15.50826 | Sweden |
| Gallivare SE | Gallivare | GEV | ESNG | 67.134445 | 20.816668 | Sweden |
| Gothenburg SE | Saeve | GSE | ESGP | 57.77775 | 11.864513 | Sweden |
| Gothenburg SE | Landvetter | GOT | ESGG | 57.66664 | 12.294878 | Sweden |
| Hagfors SE | Hagfors | HFS | ESOH | 60.0175 | 13.569167 | Sweden |
| Halmstad SE | Halmstad | HAD | ESMT | 56.680935 | 12.815005 | Sweden |
| Hemavan SE | Hemavan | HMV | ESUT | 65.727776 | 15.273056 | Sweden |
| Jonkoping SE | Axamo | JKG | ESGJ | 57.7501 | 14.070497 | Sweden |

| City and Country Code | Airport Name | IATA | ICAO | Latitude | Longitude | Country |
|---|---|---|---|---|---|---|
| Kalmar SE | Kalmar Oland | KLR | ESMQ | 56.685 | 16.287222 | Sweden |
| Karlstad SE | Karlstad | KSD | ESOK | 59.360283 | 13.472059 | Sweden |
| Kiruna SE | Kiruna | KRN | ESNQ | 67.82222 | 20.345833 | Sweden |
| Kramfors SE | Kramfors | KRF | ESNK | 63.049442 | 17.772778 | Sweden |
| Kristianstad SE | Kristianstad | KID | ESMK | 55.919445 | 14.088889 | Sweden |
| Linkoping SE | Linkoping City | LPI | ESSL | 58.406944 | 15.656944 | Sweden |
| Lulea SE | Kallax | LLA | ESPA | 65.54939 | 22.123587 | Sweden |
| Lycksele SE | Lycksele | LYC | ESNL | 64.55071 | 18.70967 | Sweden |
| Malmo SE | Sturup | MMX | ESMS | 55.538757 | 13.363727 | Sweden |
| Mora SE | Mora | MXX | ESKM | 60.95812 | 14.504529 | Sweden |
| Norrkoping SE | Kungsangen | NRK | ESSP | 58.583298 | 16.232393 | Sweden |
| Orebro SE | Orebro-Bofors | ORB | ESOE | 59.225758 | 15.047543 | Sweden |
| Ornskoldsvik SE | Ornskoldsvik | OER | ESNO | 63.412582 | 18.992073 | Sweden |
| Oskarshamn SE | Oskarshamn | OSK | ESMO | 57.266666 | 16.433332 | Sweden |
| Ostersund SE | Froesoe | OSD | ESPC | 63.198612 | 14.494444 | Sweden |
| Pajala SE | Pajala | PJA | ESUP | 67.246666 | 23.075 | Sweden |
| Ronneby SE | Kallinge | RNB | ESDF | 56.25833 | 15.261111 | Sweden |
| Skelleftea SE | Skelleftea | SFT | ESNS | 64.62251 | 21.068548 | Sweden |
| Skovde SE | Skovde | KVB | ESGR | 58.45 | 13.966667 | Sweden |
| Stockholm SE | Vasteras/Hasslo | VST | ESOW | 59.58917 | 16.630556 | Sweden |
| Stockholm SE | Skavsta | NYO | ESKN | 58.78425 | 16.922867 | Sweden |
| Stockholm SE | Stockholm Arlanda | ARN | ESSA | 59.64982 | 17.930365 | Sweden |
| Stockholm SE | Bromma | BMA | ESSB | 59.35566 | 17.94608 | Sweden |
| Sundsvall SE | Sundsvall/ Harnosand | SDL | ESNN | 62.52165 | 17.438147 | Sweden |
| Sveg SE | Sveg | EVG | ESND | 62.033333 | 14.35 | Sweden |
| Torsby SE | Torsby | TYF | ESST | 60.154484 | 12.99661 | Sweden |
| Trollhattan SE | Trollhattan | THN | ESGT | 58.266666 | 12.3 | Sweden |
| Umea SE | Umea | UME | ESNU | 63.79333 | 20.28954 | Sweden |
| Vaxjo SE | Vaxjo | VXO | ESMX | 56.925095 | 14.732046 | Sweden |
| Vilhelmina SE | Vilhelmina | VHM | ESNV | 64.61667 | 16.65 | Sweden |
| Visby SE | Visby | VBY | ESSV | 57.660446 | 18.338154 | Sweden |

**Table 35.** Ranking of most frequented airports in the BSR

| Rank | Airport Name | Country Code | IATA Code | Reported Days | Landed MTOW in tons per Day | Landings per Day | Departing Seats per Day |
|---|---|---|---|---|---|---|---|
| 1 | Frankfurt International | DE | FRA | 20 | 62,171 | 585 | 100,472 |
| 2 | Sheremetyevo | RU | SVO | 20 | 29,844 | 278 | 46,942 |
| 3 | Domodedovo | RU | DME | 20 | 24,274 | 278 | 43,165 |
| 4 | Copenhagen | DK | CPH | 20 | 20,738 | 312 | 38,639 |
| 5 | Oslo | NO | OSL | 20 | 19,273 | 300 | 35,524 |
| 6 | Arlanda | SE | ARN | 20 | 17,408 | 275 | 31,415 |
| 7 | Tegel | DE | TXL | 20 | 16,280 | 223 | 32,105 |
| 8 | Helsinki-Vantaa | FI | HEL | 20 | 12,780 | 201 | 24,064 |
| 9 | Fuhlsbuettel | DE | HAM | 20 | 11,795 | 169 | 22,757 |
| 10 | Pulkovo | RU | LED | 20 | 10,040 | 143 | 19,619 |
| 11 | Vnukovo | RU | VKO | 20 | 9,048 | 140 | 18,193 |
| 12 | Frederic Chopin | PL | WAW | 20 | 8,076 | 156 | 17,184 |
| 13 | Koeln/Bonn | DE | CGN | 20 | 7,414 | 102 | 14,033 |
| 14 | Schoenefeld | DE | SXF | 20 | 4,922 | 65 | 9,267 |
| 15 | Hanover | DE | HAJ | 20 | 4,084 | 68 | 8,318 |

| Rank | Airport Name | Country Code | IATA Code | Reported Days | Landed MTOW in tons per Day | Landings per Day | Departing Seats per Day |
|---|---|---|---|---|---|---|---|
| 16 | Riga | LV | RIX | 20 | 3,954 | 84 | 8,759 |
| 17 | Landvetter | SE | GOT | 9 | 3,869 | 74 | 7,519 |
| 18 | Vaernes | NO | TRD | 20 | 3,446 | 71 | 6,679 |
| 19 | Frankfurt - Hahn | DE | HHN | 20 | 3,337 | 33 | 4,514 |
| 20 | J. Paul II Balice | PL | KRK | 20 | 2,570 | 47 | 5,527 |
| 21 | Rebiechowo | PL | GDN | 20 | 1,983 | 38 | 4,282 |
| 22 | Bremen | DE | BRE | 20 | 1,953 | 34 | 4,073 |
| 23 | Leipzig/Halle | DE | LEJ | 20 | 1,888 | 30 | 3,208 |
| 24 | Bromma | SE | BMA | 17 | 1,753 | 59 | 4,148 |
| 25 | Vilnius | LT | VNO | 20 | 1,749 | 32 | 3,830 |
| 26 | Skavsta | SE | NYO | 20 | 1,688 | 24 | 3,755 |
| 27 | Sturup | SE | MMX | 20 | 1,614 | 27 | 3,127 |
| 28 | Tromso/Langnes | NO | TOS | 20 | 1,605 | 40 | 3,336 |
| 29 | Sandefjord | NO | TRF | 20 | 1,584 | 32 | 3,333 |
| 30 | Rygge Ab | NO | RYG | 20 | 1,584 | 24 | 3,577 |
| 31 | Bodo | NO | BOO | 20 | 1,533 | 44 | 3,088 |
| 32 | Dortmund | DE | DTM | 20 | 1,529 | 20 | 3,086 |
| 33 | Minsk International 2 | BY | MSQ | 20 | 1,523 | 32 | 3,105 |
| 34 | Billund | DK | BLL | 20 | 1,513 | 38 | 3,463 |
| 35 | Ulemiste | EE | TLL | 20 | 1,479 | 38 | 3,331 |
| 36 | Pyrzowice | PL | KTW | 20 | 1,323 | 21 | 2,844 |
| 37 | Dresden | DE | DRS | 20 | 1,295 | 24 | 2,721 |
| 38 | Strachowice | PL | WRO | 20 | 1,254 | 25 | 2,839 |
| 39 | Kaliningrad | RU | KGD | 20 | 1,120 | 18 | 2,274 |
| 40 | Aalborg | DK | AAL | 20 | 1,088 | 20 | 2,239 |
| 41 | Kjevik | NO | KRS | 20 | 927 | 22 | 2,049 |
| 42 | Kallax | SE | LLA | 20 | 840 | 17 | 1,476 |
| 43 | Umea | SE | UME | 20 | 805 | 17 | 1,533 |
| 44 | Lawica | PL | POZ | 20 | 736 | 15 | 1,647 |
| 45 | Oulu | FI | OUL | 20 | 716 | 13 | 1,403 |
| 46 | Saeve | SE | GSE | 9 | 641 | 9 | 1,433 |
| 47 | Paderborn | DE | PAD | 20 | 547 | 9 | 1,172 |
| 48 | Arkhangelsk | RU | ARH | 20 | 523 | 13 | 1,096 |
| 49 | Evenes | NO | EVE | 20 | 495 | 10 | 952 |
| 50 | Kaunas | LT | KUN | 20 | 494 | 7 | 1,126 |
| 51 | Muenster | DE | FMO | 20 | 475 | 10 | 1,244 |
| 52 | Murmansk | RU | MMK | 20 | 444 | 9 | 987 |
| 53 | Jasionka | PL | RZE | 20 | 433 | 9 | 1,003 |
| 54 | Aro | NO | MOL | 20 | 420 | 9 | 823 |
| 55 | Sundsvall/Harnosand | SE | SDL | 17 | 397 | 11 | 694 |
| 56 | Tirstrup | DK | AAR | 20 | 365 | 11 | 885 |
| 57 | Tampere-Pirkkala | FI | TMP | 20 | 358 | 9 | 938 |
| 58 | Angelholm | SE | AGH | 20 | 330 | 9 | 729 |
| 59 | Kvernberget | NO | KSU | 20 | 323 | 8 | 666 |
| 60 | Froesoe | SE | OSD | 20 | 307 | 6 | 543 |

**Table 36.** Cargo airlines operating at airports and routes in the BSR

| Country | Airline | Aircraft Type | MTOW | Airport Name | O/D Country Name | O/D Airport Name | O/D IATA |
|---|---|---|---|---|---|---|---|
| Belarus | Turkish Airlines | 31Y | 164 | Minsk International 2 | Austria | Vienna International | VIE |
| | | | | | Turkey | Ataturk | IST |
| Denmark | Air China Limited | 74Y | 397 | Copenhagen | China | Capital | PEK |
| | | | | | | Pu Dong | PVG |
| | British Airways | ABF | 171 | Copenhagen | United Kingdom | East Midlands | EMA |
| | | | | | | Heathrow | LHR |
| | China Cargo | 77F | 348 | Copenhagen | China | Pu Dong | PVG |
| | Emirates | 77X | 347 | Copenhagen | United Arab Emirates | Dubai | DXB |
| | | | | | United States of America | O'Hare International | ORD |
| | Korean Air | 74Y | 397 | Copenhagen | Netherlands | Amsterdam-Schiphol | AMS |
| | | | | | South Korea | Incheon International | ICN |
| | | 77X | 347 | Copenhagen | Germany | Frankfurt International | FRA |
| | | | | | South Korea | Incheon International | ICN |
| | Singapore Airlines | 74Y | 397 | Copenhagen | Netherlands | Amsterdam-Schiphol | AMS |
| | | | | | United Arab Emirates | Sharjah | SHJ |
| | | | | | United Kingdom | Heathrow | LHR |
| Finland | Aeroflot | M1F | 286 | Helsinki-Vantaa | Russia | Sheremetyevo | SVO |
| | Finnair | M1F | 286 | Helsinki-Vantaa | Belgium | Brussels | BRU |
| | | | | | China | Hong Kong International | HKG |
| | | | | | India | Chhatrapati Shivaji International | BOM |
| | | | | | United States of America | John F. Kennedy International | JFK |
| | | | | | | O'Hare International | ORD |

| Country | Airline | Aircraft Type | MTOW | Airport Name | O/D Country Name | O/D Airport Name | O/D IATA |
|---|---|---|---|---|---|---|---|
| | MNG Airlines Cargo | 73P | 58 | Helsinki-Vantaa | Netherlands | Amsterdam-Schiphol | AMS |
| | | | | | Turkey | Ataturk | IST |
| | | ABY | 171 | Helsinki-Vantaa | Turkey | Ataturk | IST |
| | Turkish Airlines | 31Y | 164 | Helsinki-Vantaa | Sweden | Arlanda | ARN |
| | | | | | Turkey | Ataturk | IST |
| Germany | Aeroflot | M1F | 286 | Frankfurt - Hahn | Russia | Sheremetyevo | SVO |
| | | | | | | Tolmachevo | OVB |
| | Air Armenia | ANF | 61 | Frankfurt - Hahn | Armenia | Yerevan | EVN |
| | Air China Limited | 74Y | 397 | Frankfurt International | China | Capital | PEK |
| | | | | | | Chongqing | CKG |
| | Air France | ABX | 165 | Koeln/Bonn | France | Charles De Gaulle | CDG |
| | | | | | Turkey | Ataturk | IST |
| | Air Malta | 75F | 116 | Frankfurt International | Malta | Luqa | MLA |
| | AirBridge Cargo | 74N | 442 | Frankfurt International | Russia | Domodedovo | DME |
| | | | | | | Sheremetyevo | SVO |
| | | 74Y | 397 | Frankfurt International | Russia | Domodedovo | DME |
| | | | | | | Ekaterinburg | SVX |
| | | | | | | Sheremetyevo | SVO |
| | Asiana Airlines | 74Y | 397 | Frankfurt International | Austria | Vienna International | VIE |
| | | | | | South Korea | Incheon International | ICN |
| | | | | | United Kingdom | Stansted | STN |
| | Bluebird Cargo | 73F | 58 | Koeln/Bonn | Iceland | Keflavik International | KEF |
| | British Airways | 74F | 397 | Frankfurt International | India | Indira Gandhi International | DEL |
| | | | | | United Kingdom | Stansted | STN |
| | | | | Koeln/Bonn | Georgia | Novo Alexeyevka | TBS |
| | | | | | United Kingdom | Stansted | STN |
| | | 74N | 442 | Frankfurt International | China | Pu Dong | PVG |
| | | | | | United Kingdom | Stansted | STN |

| Country | Airline | Aircraft Type | MTOW | Airport Name | O/D Country Name | O/D Airport Name | O/D IATA |
|---|---|---|---|---|---|---|---|
| | | | | | United States of America | Hartsfield-Jackson Atlanta International | ATL |
| | | | | | | O'Hare International | ORD |
| | | 75F | 116 | Frankfurt International | Germany | Leipzig/Halle | LEJ |
| | | | | Leipzig/Halle | Bulgaria | Sofia | SOF |
| | | | | | Germany | Frankfurt International | FRA |
| | | | | | Hungary | Ferihegy | BUD |
| | | ABF | 171 | Frankfurt International | Germany | Leipzig/Halle | LEJ |
| | | | | | United Kingdom | East Midlands | EMA |
| | | | | | | Heathrow | LHR |
| | | | | | | Luton | LTN |
| | | | | Leipzig/Halle | Germany | Frankfurt International | FRA |
| | | | | | Spain | Barajas | MAD |
| | | | | | Ukraine | Kiev-Zhulhany | IEV |
| | | | | | United Kingdom | Heathrow | LHR |
| | C.A.L. Cargo Airlines | 74F | 397 | Frankfurt International | Belgium | Bierset | LGG |
| | | | | | Israel | Ben Gurion International | TLV |
| | Cathay Pacific Airways | 74F | 397 | Frankfurt International | France | Charles De Gaulle | CDG |
| | | | | | India | Chennai | MAA |
| | | | | | | Chhatrapati Shivaji International | BOM |
| | | | | | Italy | Malpensa | MXP |
| | | | | | Netherlands | Amsterdam-Schiphol | AMS |
| | | | | | United Arab Emirates | Dubai | DXB |
| | | | | | United Kingdom | Manchester International | MAN |
| | China Airlines | 74Y | 397 | Frankfurt International | Taiwan | Taiwan Taoyuan International | TPE |
| | | | | | United Arab Emirates | Abu Dhabi International | AUH |
| | China Southern | 77F | 348 | Frankfurt International | China | Pu Dong | PVG |

| Country | Airline | Aircraft Type | MTOW | Airport Name | O/D Country Name | O/D Airport Name | O/D IATA |
|---|---|---|---|---|---|---|---|
| | China Southern Airlines | 77F | 348 | Frankfurt International | China | Pu Dong | PVG |
| | Emirates | 77X | 347 | Frankfurt International | Ghana | Kotoka | ACC |
| | | | | | Libya | Tripoli International | TIP |
| | | | | | Russia | Domodedovo | DME |
| | | | | | Senegal | Yoff | DKR |
| | | | | | United Arab Emirates | Dubai | DXB |
| | Etihad Airways | 33F | 230 | Frankfurt International | United Arab Emirates | Abu Dhabi International | AUH |
| | | 74F | 397 | Frankfurt International | United Arab Emirates | Abu Dhabi International | AUH |
| | EVA Air | 74Y | 397 | Frankfurt International | India | Indira Gandhi International | DEL |
| | Gemini Air Cargo | 31Y | 164 | Frankfurt International | Russia | Sheremetyevo | SVO |
| | Global Supply Systems | 74N | 442 | Frankfurt International | China | Hong Kong International | HKG |
| | | | | | | Pu Dong | PVG |
| | | | | | Gabon | Hartsfield-Jackson Atlanta International | ATL |
| | | | | | India | Indira Gandhi International | DEL |
| | | | | | United Kingdom | Stansted | STN |
| | | | | | United States of America | Hartsfield-Jackson Atlanta International | ATL |
| | | | | | | O'Hare International | ORD |
| | | | | Koeln/Bonn | Georgia | Novo Alexeyevka | TBS |
| | | | | | Spain | Barajas | MAD |
| | | | | | United Kingdom | Stansted | STN |
| | Iberia | 75F | 116 | Frankfurt International | Spain | Barajas | MAD |
| | Korean Air | 74Y | 397 | Frankfurt International | Russia | Sheremetyevo | SVO |
| | | | | | South Korea | Incheon International | ICN |
| | | 77X | 347 | Frankfurt International | Denmark | Copenhagen | CPH |
| | | | | | Russia | Pulkovo | LED |
| | | | | | South Korea | Incheon International | ICN |

| Country | Airline | Aircraft Type | MTOW | Airport Name | O/D Country Name | O/D Airport Name | O/D IATA |
|---|---|---|---|---|---|---|---|
| | Lan Chile Cargo | 77F | 348 | Frankfurt International | Brazil | Viracopos | VCP |
| | Lufthansa | 73Y | 63 | Frankfurt International | Italy | Montichiari | VBS |
| | | M1F | 286 | Frankfurt International | Canada | Los Angeles International | LAX |
| | | | | | | Pearson International | YYZ |
| | | | | | Egypt | Cairo International | CAI |
| | | | | | Gabon | Hartsfield-Jackson Atlanta International | ATL |
| | | | | | India | Bengaluru International | BLR |
| | | | | | | Chennai | MAA |
| | | | | | | Chhatrapati Shivaji International | BOM |
| | | | | | | Indira Gandhi International | DEL |
| | | | | | Ireland | Shannon | SNN |
| | | | | | Israel | Ben Gurion International | TLV |
| | | | | | Kazakhstan | Almaty | ALA |
| | | | | | Kenya | Jomo Kenyatta International | NBO |
| | | | | | Russia | Krasnojarsk | KJA |
| | | | | | | Sheremetyevo | SVO |
| | | | | | Saudi Arabia | King Abdulaziz International | JED |
| | | | | | | King Fahad International | DMM |
| | | | | | | King Khaled International | RUH |
| | | | | | Senegal | Yoff | DKR |
| | | | | | Turkey | Ataturk | IST |
| | | | | | United Arab Emirates | Sharjah | SHJ |
| | | | | | United Kingdom | Manchester International | MAN |
| | | | | | United States of America | Dallas/Ft. Worth International | DFW |
| | | | | | | Detroit Metropolitan Wayne County | DTW |
| | | | | | | George Bush Intercontinental | IAH |

| Country | Airline | Aircraft Type | MTOW | Airport Name | O/D Country Name | O/D Airport Name | O/D IATA |
|---|---|---|---|---|---|---|---|
| | | | | | | Hartsfield-Jackson International Atlanta | ATL |
| | | | | | | John F. Kennedy International | JFK |
| | | | | | | O'Hare International | ORD |
| | | | | | Unites States of America/Puerto Rico | Borinquen | BQN |
| | Malaysia Airline | 74Y | 397 | Frankfurt International | Azerbaijan | Heydar Aliyev International | GYD |
| | MNG Airlines Cargo | 73P | 58 | Koeln/Bonn | Germany | Leipzig/Halle | LEJ |
| | | | | | Turkey | Ataturk | IST |
| | | | | Leipzig/Halle | Germany | Koeln/Bonn | CGN |
| | | | | | Turkey | Ataturk | IST |
| | | ABY | 171 | Koeln/Bonn | France | Charles De Gaulle | CDG |
| | | | | | Turkey | Ataturk | IST |
| | Nippon Cargo | 74Y | 397 | Frankfurt - Hahn | Italy | Malpensa | MXP |
| | | | | | Japan | Narita | NRT |
| | | | | | Netherlands | Amsterdam-Schiphol | AMS |
| | Polar Air Cargo | 74Y | 397 | Leipzig/Halle | Bahrain | Bahrain International | BAH |
| | | | | | United States of America | Cincinnati/Northern Kentucky | CVG |
| | Qatar Airways | 33X | 230 | Frankfurt - Hahn | Qatar | Doha | DOH |
| | | | | Frankfurt International | Qatar | Doha | DOH |
| | | 77X | 347 | Frankfurt - Hahn | Qatar | Doha | DOH |
| | | | | Frankfurt International | Qatar | Doha | DOH |
| | Royal Jordanian | 31Y | 164 | Frankfurt International | Jordan | Queen Alia International | AMM |
| | Saudia | 74F | 397 | Frankfurt International | Saudi Arabia | King Abdulaziz International | JED |
| | | | | | | King Fahad International | DMM |
| | | | | | | King Khaled International | RUH |
| | Silk Way Airlines | 74Y | 397 | Frankfurt - Hahn | Azerbaijan | Heydar Aliyev International | GYD |

| Country | Airline | Aircraft Type | MTOW | Airport Name | O/D Country Name | O/D Airport Name | O/D IATA |
|---|---|---|---|---|---|---|---|
| | Southern Air, Inc. | 77F | 348 | Frankfurt International | China | Hong Kong International | HKG |
| | | | | | India | Chennai | MAA |
| | | | | | Thailand | Suvarnabhumi | BKK |
| | | | | | United Arab Emirates | Dubai | DXB |
| | | | | | | Sharjah | SHJ |
| | | | | Leipzig/Halle | China | Hong Kong International | HKG |
| | | | | | United States of America | Los Angeles International | LAX |
| | Thai Airways International | 74F | 397 | Frankfurt International | India | Indira Gandhi International | DEL |
| | | | | | Thailand | Suvarnabhumi | BKK |
| | TNT Airways | 74Y | 397 | Dusseldorf | Germany | Frankfurt International | FRA |
| | | | | | United Arab Emirates | Dubai | DXB |
| | | | | Frankfurt International | Germany | Dusseldorf | DUS |
| | | | | | United Arab Emirates | Dubai | DXB |
| | Turkish Airlines | 31Y | 164 | Frankfurt International | Turkey | Ataturk | IST |
| | | | | Koeln/Bonn | Turkey | Ataturk | IST |
| | | 33X | 230 | Frankfurt International | Turkey | Ataturk | IST |
| | Yangtze River Express | 74Y | 397 | Frankfurt - Hahn | China | Capital | PEK |
| | | | | | | Pu Dong | PVG |
| | | | | | Luxembourg | Luxembourg | LUX |
| Lithuania | El Al | 74Y | 397 | Siauliai | Israel | Ben Gurion International | TLV |
| | MNG Airlines Cargo | ABY | 171 | Siauliai | Turkey | Ataturk | IST |
| Norway | Asiana Airlines | 74Y | 397 | Oslo | Austria | Vienna International | VIE |
| | | | | | Italy | Malpensa | MXP |
| | | | | | South Korea | Incheon International | ICN |
| | Korean Air | 74Y | 397 | Oslo | Austria | Vienna International | VIE |
| | | | | | South Korea | Incheon International | ICN |

| Country | Airline | Aircraft Type | MTOW | Airport Name | O/D Country Name | O/D Airport Name | O/D IATA |
|---|---|---|---|---|---|---|---|
| | | 77X | 347 | Oslo | Austria | Vienna International | VIE |
| | | | | | South Korea | Incheon International | ICN |
| | Qatar Airways | 77X | 347 | Oslo | Belgium | Bierset | LGG |
| | | | | | Qatar | Doha | DOH |
| Russia | Aeroflot | M1F | 286 | Sheremetyevo | China | Capital | PEK |
| | | | | | | Hong Kong International | HKG |
| | | | | | | Pu Dong | PVG |
| | | | | | Finland | Helsinki-Vantaa | HEL |
| | | | | | Germany | Frankfurt - Hahn | HHN |
| | | | | | Russia | Tolmachevo | OVB |
| | | | | Tolmachevo | China | Hong Kong International | HKG |
| | | | | | | Pu Dong | PVG |
| | | | | | Germany | Frankfurt - Hahn | HHN |
| | | | | | Japan | Narita | NRT |
| | | | | | Russia | Sheremetyevo | SVO |
| | | | | | South Korea | Incheon International | ICN |
| | Air China Limited | 74Y | 397 | Tolmachevo | China | Chengdu | CTU |
| | | | | | | Chongqing | CKG |
| | | | | | Netherlands | Amsterdam-Schiphol | AMS |
| | AirBridge Cargo | 74N | 442 | Domodedovo | China | Hong Kong International | HKG |
| | | | | | | Pu Dong | PVG |
| | | | | | France | Charles De Gaulle | CDG |
| | | | | | Germany | Frankfurt International | FRA |
| | | | | | Netherlands | Amsterdam-Schiphol | AMS |
| | | | | Sheremetyevo | China | Hong Kong International | HKG |
| | | | | | | Pu Dong | PVG |
| | | | | | France | Charles De Gaulle | CDG |
| | | | | | Germany | Frankfurt International | FRA |

| Country | Airline | Aircraft Type | MTOW | Airport Name | O/D Country Name | O/D Airport Name | O/D IATA |
|---|---|---|---|---|---|---|---|
| | | | | | Netherlands | Amsterdam-Schiphol | AMS |
| | | 74Y | 397 | Domodedovo | China | Capital | PEK |
| | | | | | | Hong Kong International | HKG |
| | | | | | | Pu Dong | PVG |
| | | | | | | Zhengzhou | CGO |
| | | | | | France | Charles De Gaulle | CDG |
| | | | | | Germany | Frankfurt International | FRA |
| | | | | | Japan | Narita | NRT |
| | | | | | Netherlands | Amsterdam-Schiphol | AMS |
| | | | | | United States of America | O'Hare International | ORD |
| | | | | Ekaterinburg | China | Chengdu | CTU |
| | | | | | | Pu Dong | PVG |
| | | | | | | Zhengzhou | CGO |
| | | | | | Germany | Frankfurt International | FRA |
| | | | | Novyy | China | Zhengzhou | CGO |
| | | | | Sheremetyevo | China | Capital | PEK |
| | | | | | | Chengdu | CTU |
| | | | | | | Hong Kong International | HKG |
| | | | | | | Pu Dong | PVG |
| | | | | | | Zhengzhou | CGO |
| | | | | | France | Charles De Gaulle | CDG |
| | | | | | Germany | Frankfurt International | FRA |
| | | | | | Italy | Malpensa | MXP |
| | | | | | Japan | Narita | NRT |
| | | | | | Netherlands | Amsterdam-Schiphol | AMS |
| | | | | | South Korea | Incheon International | ICN |
| | | | | | Spain | Zaragoza | ZAZ |
| | | | | | United States of America | O'Hare International | ORD |

| Country | Airline | Aircraft Type | MTOW | Airport Name | O/D Country Name | O/D Airport Name | O/D IATA |
|---|---|---|---|---|---|---|---|
| | | | | Tolmachevo | China | Chengdu | CTU |
| | | | | Vnukovo | China | Hong Kong International | HKG |
| | | | | | Netherlands | Amsterdam-Schiphol | AMS |
| | Asiana Airlines | 74Y | 397 | Domodedovo | Austria | Vienna International | VIE |
| | | | | | South Korea | Incheon International | ICN |
| | | | | | United Kingdom | Stansted | STN |
| | | | | Pulkovo | Austria | Vienna International | VIE |
| | | | | | South Korea | Incheon International | ICN |
| | Emirates | 77X | 347 | Domodedovo | Germany | Frankfurt International | FRA |
| | | | | | United Arab Emirates | Dubai | DXB |
| | Gemini Air Cargo | 31Y | 164 | Sheremetyevo | Germany | Frankfurt International | FRA |
| | | | | | Mongolia | Chinggis Khaan International | ULN |
| | Korean Air | 74Y | 397 | Sheremetyevo | Germany | Frankfurt International | FRA |
| | | | | | South Korea | Incheon International | ICN |
| | | 77X | 347 | Pulkovo | Germany | Frankfurt International | FRA |
| | | | | | South Korea | Incheon International | ICN |
| | Lufthansa | M1F | 286 | Krasnojarsk | China | Capital | PEK |
| | | | | | | Pu Dong | PVG |
| | | | | | | Shenyang | SHE |
| | | | | | Germany | Frankfurt International | FRA |
| | | | | | Japan | Kansai International | KIX |
| | | | | | | Narita | NRT |
| | | | | | South Korea | Incheon International | ICN |
| | | | | Sheremetyevo | Germany | Frankfurt International | FRA |
| | | | | | Japan | Narita | NRT |
| | MNG Airlines Cargo | 73P | 58 | Ekaterinburg | Turkey | Ataturk | IST |
| | RusLine | CR2 | 23 | Ivanovo | Russia | Domodedovo | DME |

| Country | Airline | Aircraft Type | MTOW | Airport Name | O/D Country Name | O/D Airport Name | O/D IATA |
|---|---|---|---|---|---|---|---|
| | Yangtze River Express | 74Y | 397 | Tolmachevo | China | Pu Dong | PVG |
| | | | | | Luxembourg | Luxembourg | LUX |
| Sweden | Korean Air | 74Y | 397 | Arlanda | Netherlands | Amsterdam-Schiphol | AMS |
| | | | | | South Korea | Incheon International | ICN |
| | TNT Airways | 74Y | 397 | Landvetter | Netherlands | Amsterdam-Schiphol | AMS |
| | | | | | United Arab Emirates | Dubai | DXB |
| | Turkish Airlines | 31Y | 164 | Arlanda | Finland | Helsinki-Vantaa | HEL |
| | | | | | Turkey | Ataturk | IST |

**Table 37.** Public Service Obligation routes in the BSR (EU Commission 2013)

| Country | Origin | Destination | Single operator | Market access |
|---|---|---|---|---|
| Finland | Helsinki | Savonlinna | Finncomm | RESTRICTED |
| | | Varkaus | Finncomm | RESTRICTED |
| | Mariehamn/Åland | Stockholm | Air Åland; new: NextJet | RESTRICTED |
| Germany | Erfurt | München | Cirrus Airlines | RESTRICTED |
| | Hof/Plauen | Frankfurt (am Main) | ongoing call for tender | RESTRICTED |
| | Rostock-Laage | München | repealed | RESTRICTED |
| Norway | Bergen | Fagernes | | OPEN |
| | | Florø | Danish Air Transport | RESTRICTED |
| | | Førde | Wideroe | RESTRICTED |
| | | Ørsta-Volda | Wideroe | RESTRICTED |
| | | Sandane | Wideroe | RESTRICTED |
| | | Sogndal | Wideroe | RESTRICTED |
| | Bodø | Andenes | Wideroe | RESTRICTED |
| | | Brønnøysund | Wideroe | RESTRICTED |
| | | Leknes | Wideroe | RESTRICTED |
| | | Mo i Rana | Wideroe | RESTRICTED |
| | | Mosjøen | Wideroe | RESTRICTED |
| | | Narvik | Wideroe | RESTRICTED |
| | | Røst | Wideroe | RESTRICTED |
| | | Sandnessjøen | Wideroe | RESTRICTED |
| | | Stokmarknes | | OPEN |
| | | Svolvær | Wideroe | RESTRICTED |
| | | Værøy (heliport) | Lufttransport | RESTRICTED |
| | Hasvik | Hammerfest | Wideroe | RESTRICTED |
| | | Tromsø | Wideroe | RESTRICTED |
| | Ørsta-Volda | Ålesund | | OPEN |
| | Oslo | Brønnøysund | | OPEN |
| | | Fagernes | DOT LT | RESTRICTED |
| | | Florø | Danish Air Transport | RESTRICTED |
| | | Førde | Wideroe | RESTRICTED |
| | | Mo i Rana | | OPEN |
| | | Mosjøen | | OPEN |
| | | Ørsta-Volda | Wideroe | RESTRICTED |
| | | Røros | DOT LT | RESTRICTED |
| | | Sandane | Wideroe | RESTRICTED |
| | | Sandnessjøen | | OPEN |
| | | Sogndal | Wideroe | RESTRICTED |
| | Tromsø | Andenes | Wideroe | RESTRICTED |
| | | Lakselv | Wideroe | RESTRICTED |
| | | Sørkjosen | Wideroe | RESTRICTED |
| | Trondheim | Brønnøysund | Wideroe | RESTRICTED |
| | | Mo i Rana | Wideroe | RESTRICTED |
| | | Mosjøen | Wideroe | RESTRICTED |
| | | Namsos | Wideroe | RESTRICTED |
| | | Rorvik | Wideroe | RESTRICTED |

| Country | Origin | Destination | Single operator | Market access |
|---|---|---|---|---|
| | | Sandnessjøen | Wideroe | RESTRICTED |
| Sweden | Östersund | Umea | Avies Sverige AB - Svenske Direktflyg AB | RESTRICTED |
| | Pajala | Lulea | Avies Sverige AB - Svenske Direktflyg AB | RESTRICTED |
| | Stockholm (Arlanda) | Arvidsjaur | NEX Time Jet AB | RESTRICTED |
| | | Gällivare | NEX Time Jet AB | RESTRICTED |
| | | Hagfors | | RESTRICTED |
| | | Hemavan | NEX Time Jet AB | RESTRICTED |
| | | Lycksele | NEX Time Jet AB | RESTRICTED |
| | | Sveg | Avies Sverige AB - Svenske Direktflyg AB | RESTRICTED |
| | | Torsby | Avies Sverige AB - Svenske Direktflyg AB | RESTRICTED |
| | | Vilhelmina | NEX Time Jet AB | RESTRICTED |
| | Storuman | Stockholm (Arlanda) | | RESTRICTED |

# 7 Social costs of Public Service Obligation (PSO) routes - calculating subsidies of regional flights in Norway[1,2]

Branko Bubalo

***Abstract*****.** In this note we describe an iterative procedure of how to estimate unit costs per leg of a public service obligation (PSO) route network if certain data is publicly available. The aim of this approach is to make judgments in benchmarking and in regulation if revenues, costs, and profits (or losses) per flight and its distribution among these route networks are typical compared to carriers serving networks under a competitive regime. Particularly this note aims to set a reference for the question if market or bargaining powers are abused and to which extent. This work is thus especially important for PSO cases where a particular network cannot be operated in a profitable manner; therefore, its routes are offered to monopoly providers in a bidding competition and (in most cases) the service needs to be publicly subsidized. We shall apply the procedure on origin-destination matrices from tender documents published by the Norwegian Ministry of Transport and Communications. The Ministry covers incurred losses produced by the bidding and winning carrier. As a first result we can observe that the PSO allocations show indications of an inefficient allocation process reflected in more than three-fold quoted costs on PSO routes above estimated market levels.

## 7.1 Introduction

In Europe *public service obligations* (PSO) grant exclusive access to air transport markets in low demand regions (European Commission 2010; The European

[1] The author is indebted to Mr. Thomas Tørmo and Mr. Per Kolstad from the Ministry of Transport and Communications in Norway, who financed and initiated parts of this research and who presented the original problem of regional air transport access and funding.

[2] This note covers the technical part of a lecture held in Berlin on June 20th 2012 for the German Airport Performance (GAP) Project with the title: "How to break the vicious circle?—Monopoly bidding for public service obligation route networks in Norway". The author also participated in the ICCL.

Parliament and the Council 2008). However, it is yet unclear if society in general benefits from connectivity through subsidized air traffic compared to its 'costs' (OECD 2003). In the past this question has already been discussed by others elsewhere with regard to Europe (Williams and Pagliari 2004) or with regard to Norway (Bråthen et al. 2001; Lian 2002).We highlight a case where the Norwegian government experiences ever increasing subsidy trends over subsequent tender periods, because there is low or no competition among scheduled air transport service providers in the bidding process (Bråthen 2011; Lian et al. 2010; Williamson 1976). The question arises, how to change the status quo of the tender rules, which fail to include mechanisms against bids from a monopoly supplier? In theory such dominant positions can lead to abuse of market power. Given the required specialization to serve airports and routes in remote regions, which in public utilities theory leads to 'market entry barriers' (Demsetz 1968) and 'natural monopoly supply' (Williamson 1976), we do not expect exact market compliant distance costs, but 'moderate' costs and profits per (revenue) passenger-kilometer (RPK). As revealed by industry benchmarks we argue that it is feasible to operate the PSO network less costly as at present (**Fig. 59**).

Although comparisons among different airlines with contrasting business models, such as low-fare (Ryanair, Southwest Airlines or Norwegian) or cargo (Fed-Ex), and operating under different economies of scale must be made with caution, we want to emphasize that the unit costs of carrier Widerøe of 4 Norwegian Kroners (NOK; 1 NOK corresponds to about 12.8 Eurocent in 2011) per passenger-kilometer (CRPK) (IATA 2012; Widerøe's Flyveselskap 2010) across all activities are indeed in the upper range. FlyBe, a carrier serving regional routes in northern United Kingdom, manages to operate its network with average unit costs of 1.6 NOK per RPK. Larger relevant airlines serving Scandinavia, such as SAS, Blue 1 and Norwegian have unit costs as low as 1.4 NOK, 1.1 NOK and 0.6 NOK per RPK, respectively (**Fig. 59**). As our calculations reveal the current operating costs, including a profit margin, of 13.6 NOK per RPK (claimed by operator Widerøe) for serving the network of Finnmark and North-Troms are three-fold above the uppermost benchmark. If these costs are justified given the short distances and, therefore, higher associated airport charges and take-off and landing fuel consumption cannot be answered with certainty. However, to make valid judgments we approach the problem two-fold:

a.) Initially, we assume the same average unit costs (CRPK) across the whole Finnmark and North-Troms network.

b.) Secondly, we distribute unit costs non-linearly across the network, based on the distribution of unit revenues over flown distance.

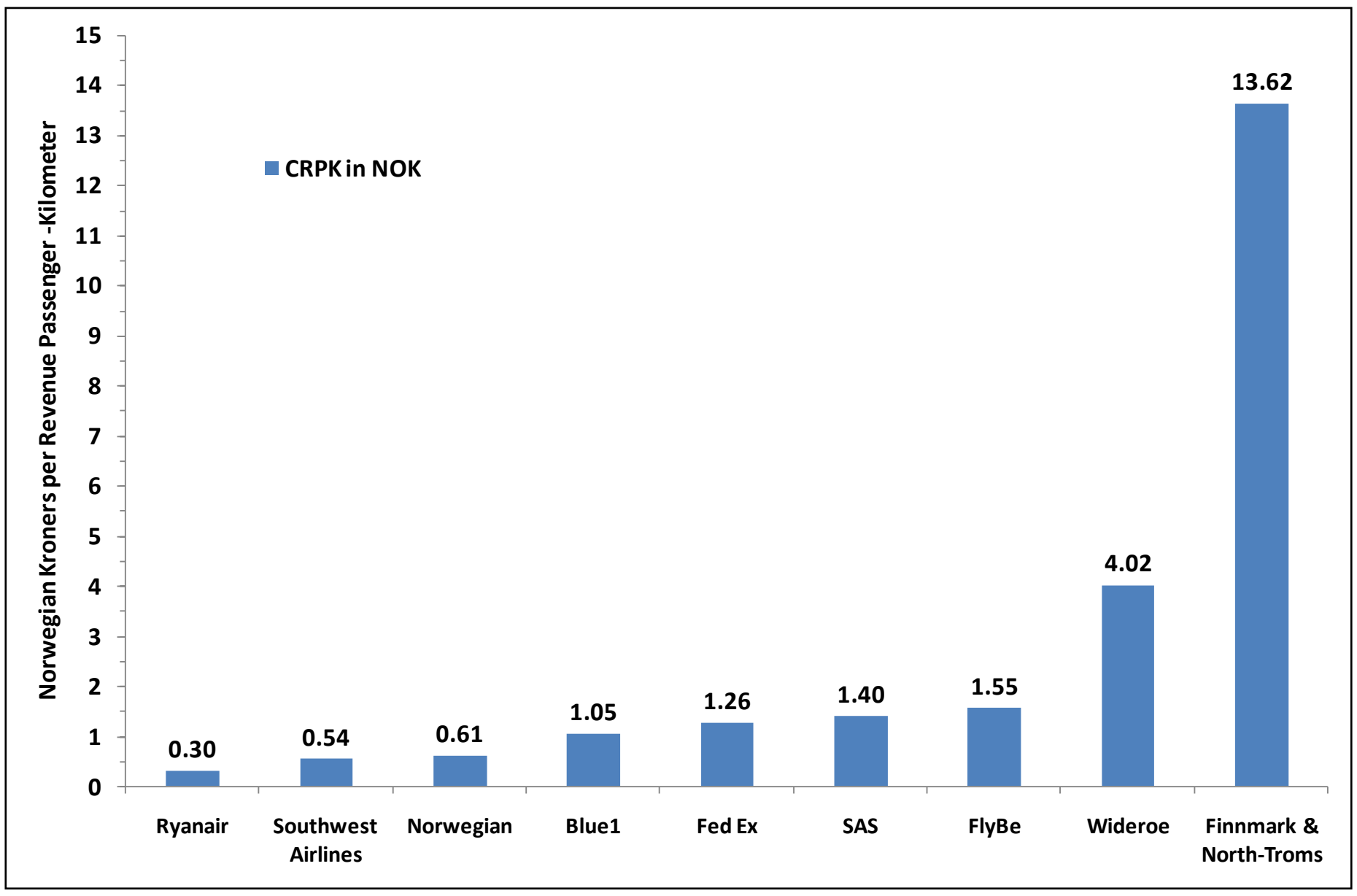

**Fig. 59.** Unit costs of international airlines in NOK per RPK (2011 prices; own illustration from company annual reports (IATA 2012; Norwegian Air Shuttle 2010; Southwest Airlines 2010; Widerøe's Flyveselskap 2010).

We demonstrate in Section 7.3 how to calculate the deviation from the break-even point given constant revenues and iterating, i.e. increasing, costs. Hence subsidies by route or in total over a point-to-point network can be calculated. Demand and revenue data is derived from publicly available origin-destination tables (The Ministry of Transport and Communications 2006, 2008, 2009, 2010, 2012a, 2012b), which are linked to working matrices in this bottom-up approach. Such small scale 'What if?' simulations on matrices now and in the past aid investigation and decision making in policy and regulation (Dantzig 1951). Similar to Cost-Volume-Profit (CVP) analyses in managerial accounting in our case we seek answers regarding level of subsidies under changing cost scenarios.

## 7.2 Problem description and the Norwegian setting

The current Norwegian airport network is kept vital to a large degree by cross subsidization within the *Avinor* airport system consisting of 46 airports (GAP 2013). These airports function as the nodes for origin, destination and transfer traffic on peripheral routes. The connected low demand airports are linked through a PSO framework. This form of interrelated governmental support of traffic strongly affects the cost side of the Norwegian air transport system, which is leaving its balance towards increasing PSO compensations paid by the Ministry of Transport and Communication. On the one hand the Ministry receives profits (if there are any after cross-subsidizing loss-making airports [GAP 2013]) from its airport system in form of dividends, whereas on the other hand it subsidizes the regional air traffic through PSO routes (if operating costs plus carrier's profit margin are larger than operating revenues). In this note we explore the role of the demand side and we provide some critical determinants of evaluating costs and profitability of the Norwegian PSO network.

**PSO routes and carriers**

Currently about 1.1 million passengers or 10 % of domestic air passengers travel on PSO routes in Norway. PSO routes have been established by the Norwegian government in form of public tenders from 1997 onwards to guarantee air service to the population residing in peripheral regions. Within a competitive market these demands would unlikely to be served. According to EU-Regulation 1008/2008 on PSO routes (The European Parliament and the Council 2008), these can be tendered out to one carrier, restricting market access by competitors, but may in fact be tendered under various rules (Bråthen 2011; Santana 2009). The PSO routes in Norway are served primarily by the regional network carrier Widerøe (a SAS subsidiary, currently on sale to investors), but also by Danish Air Transport and its Lithuania based subsidiary DOT LT (European Commission 2010), and the helicopter shuttle service Lufttransport. The respective fleets consist of mainly small aircraft, such as the Bombardier DHC-8–103 with 39 seats and a maximum take-off weight (MTOW) of 16 t, the DHC-8–311 with 50 seats and 19.5 t MTOW or the DHC-8-402 with 78 seats and 29.5 t MTOW. The Widerøe fleet consists of 19, seven and seven aircraft from each type and these are between two and 22 years old. Danish Air Transport mainly flies five about 25-years old ATR-42 or two ATR-72 aircraft with 46 or 68 seats and about 17 or 22 t MTOW, respectively. DOT LT flies only on two routes between Oslo and Fagernes, and Oslo and Røros, with its two about 25-years old Saab 340A aircraft with 34 seats and 12.7 t MTOW. The shuttle service between Bodø and Værоy is flown by two Eurocopter with 4.3 t MTOW and about six seats (JP Airline-Fleets International 2009).

**STOL runways**

TØI report 1116/2010 (Lian et al. 2010) appropriately points to the fact that the numerous *short take-off and landing* (STOL) runways (<1200 m) in the PSO network pose a significant limit to carrier competition for the offered tenders, since only very few European carriers are able to serve these runways with aircraft within their fleet given the tender requirements (among those that would qualify operating at STOL airports we find only the following airlines with their number of DHC-8-100- or -300-type aircraft in brackets: Air Nostrum (11) and SATA Air Açores (2) (IATA 2012; JP Airline-Fleets International 2009). Therefore, Widerøe being the sole provider of adequate aircraft types and numbers in Europe may exploit its quasi-monopolistic bargaining power when posing bids for such routes, by declaring operating costs and eventually required route compensations higher than current market levels. One reason for such actions could be cross-subsidization of other routes in the Widerøe network where one may find more competition on flown routes, thus lower revenues and profits.

Widerøe may, therefore, use the PSO contracts as a 'secure' income stream to maintain long-term profitability. According to Lian (2002) about 50 % of Widerøe's operations are based on PSO tenders.

## Degree of dependency of airports on PSO traffic

In Norway 19 airports (out of around 50) rely almost fully on PSO traffic (>90 % of total passengers) (GAP 2013). The dependency on PSO traffic is strengthened by a low number of available destinations and a mandatory amount of daily performed flights (European Commission 2010; GAP 2013; Hansson 2007; Lian 2002, 2010; Lian et al. 2010; Lian and Rønnevik 2011; The European Parliament and the Council 2008; The Ministry of Transport and Communications 2006, 2008, 2009, 2010, 2012a, 2012b), which limits the consumer choices of travelling. Since airport managers at regional airports 'get used' to the scheduled PSO flights, there are less incentives to actively push route developments besides PSO traffic. In order to decrease the degree of dependency on PSO flights, the EU commission is debating certain funding especially for developing regional air traffic (European Commission 2005). Other researchers propose independent 'route development funds' (Hanssen 2007) for supporting regional airports or airlines. According to Hansson (2007) this proved to be successful for developing air transport in Scotland. However, in Norway the main social target is the provision of connectivity to the main and regional hubs and capitals, for reaching offices or hospitals.

When analyzing the PSO network economics, structure and regulations, we observe a strong (financial) interdependency between the different stakeholders, namely the passengers, carriers, airports and its operator Avinor, and the Ministry of Transport and Communications. With the right incentives, and under a competitive regime, carriers would try to lower their operating costs to reach profitability or to minimize the required PSO compensations (OECD 2003). Avinor and its airports should have the equal objective to minimize operating costs or to maximize revenues while serving aircraft and providing a minimum level of service, such as clear runways and possible operations under most weather conditions (GAP 2013). Clearly, the current airport system allows such inefficiencies, mainly due to lack of competition and cost sensitivity. Air navigation services also in its reduced form of Aerodrome Flight Information Service (AFIS) is generally provided by the airport tower and air traffic control centers and must be internally purchased by the airports within Avinor. These mandatory services can be quite costly for small airports.

## Subsidies for PSO traffic

The Ministry of Transport and Communication has the function of subsidizing PSO route losses, but it may in return get dividends from the profits generated by the Avinor airport system. **Table 38** shows the amounts of subsidies required to run the PSO network between 2007 and 2011, which increased by 46 % from 474.0 to 692.6 million

NOK per operating year and translate into an average increase of around 10% per year (GAP 2013).

Norwegian airports rely to different degrees on public subsidies through PSO routes. In some cases, such as at Florø, Sognal or Røros airport, PSO routes determine exclusively the amount of traffic which is generated at these airports (GAP 2013). However, not all aeronautical revenues received from PSO traffic are *per se* subsidies. It is theoretically possible for airlines to bid for “zero tenders” (The Ministry of Transport and Communication 2012), which governs the assumption that a particular PSO route could be served profitably solely by the income from passenger revenues on these routes without requiring compensation of operating losses by the Ministry of Transport and Communications. As we can see from **Table 38** more than half a billion Norwegian Kroners are spent on PSO routes by the Ministry of Transport and Communications each year. Primarily, the larger portion of these subsidies flows to the airline operating these routes in order to cover its operating and overhead costs. Secondarily, some of the subsidies then flow to the airports in form of aeronautical charges (GAP 2013).

**Table 38.** Required PSO subsidies

| Operating Year | Subsidy for PSO service in million Norwegian Kroners | Year-on-Year change |
|---|---|---|
| **2007** | 474.0 | - |
| **2008** | 509.8 | +7.6% |
| **2009** | 589.6 | +15.7% |
| **2010** | 656.6 | +11.4% |
| **2011** | 692.6 | +5.5% |
| **Increase 2007 to 2011** | **+218.6** | **+46.1 %** |

**Note: Kroners in nominal Prices**

### Looking ahead

The recent National Transport Plan 2014–2023 assessed what kind of infrastructure is necessary for the public to be able to reach regional hubs and nearby airports. Some of the proposed measures are relevant for increasing competition between carriers serving these airports or between alternative modes of transport. Among the improvements are runway extensions at certain STOL airports, bridges, tunnels, road access, ferry or train services and the construction of new alternative airports (Avinor 2012). Runway or apron related improvements can have a direct impact on airport capacity and aircraft landing and take-off suitability.

## 7.3 The impact of airline costs on PSO compensations

In the long run subsidies should be reduced and ultimately abolished and market forces alone should be able to deliver the needed services. There is a risk that earlier subsidized activities will eventually stop because these remain unprofitable. Yet, if the market is distorted it is impossible to predict if this risk is going to become serious. In this section we put attention to airline operating costs on PSO routes. A change of these costs has a direct impact on the level of PSO subsidies.

### The case of the Finnmark and North-Troms Network and the helicopter route Væroy-Bodø

In **Fig. 60** we can see that a network of regional airports serving PSO flights (in our case the Finnmark and North-Troms network in Norway) does not imply low connectivity or a distinctive hub-and-spoke network. To the contrary, we observe strong interconnections between the airports in a point-to-point network.

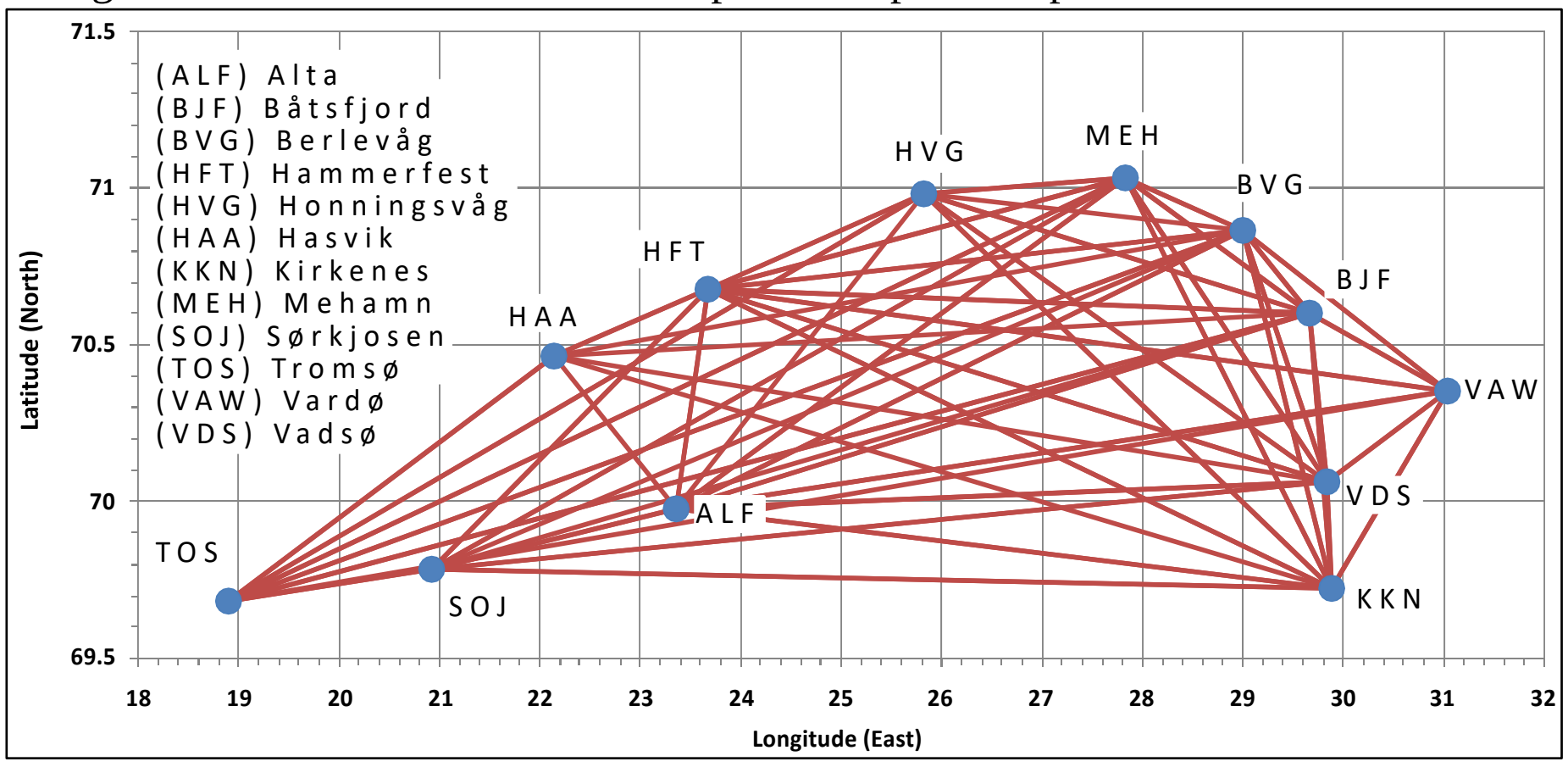

**Fig. 60.** Point-to-point demands between airports in Finnmark and North-Troms (Own illustration; Not to scale)

In order to understand the economics of regional PSO routes in Norway, we will explore a few critical determinants of such networks on the example of the Finnmark and North-Troms network, conducted by Widerøe, and the Helicopter shuttle route between Væroy and Bodø, conducted by Lufttransport. Particularly the northern regional routes could be regarded as a 'social luxury' with regard to their compensation costs in relation to transported annual passengers. Since a service is desired from the public, but large growth rates of traffic numbers in these sparsely populated areas seem unrealistic, it is necessary to stabilize the costs given the level of average fares and demand (GAP 2013).

The analyzed routes require more than one-third of the total PSO compensation only with an equivalent share of 13 % of the total domestic demand. About 240 million Norwegian Kroners (200 million NOK for Finnmark and North-Troms routes and 40 million for the service Væroy-Bodø) are spent annually in compensation on about 145,000 passengers flying these routes (**Table 39**) (GAP 2013; Lian et al. 2010; The Ministry of Transport and Communications 2006, 2008, 2009, 2010, 2012a, 2012b).

### Set up of the matrices or tables

It is helpful to construct origin-destination matrices or tables for the calculations. Choosing square matrices (**Table 41**) has the advantage that the information is presented in a compact form. Already in this small-scale example with 12 airports we deal with 98 routes. A 12-by-12 matrix can be easily presented, calculated and analyzed, but a list with 98 rows or columns can only be read with difficulty. However, a table has certain advantages regarding the calculation steps. In our case we need to manipulate the given two origin-destination matrices in 8 steps, which means we need 10 separate matrices (plus the orthodromic "great circle" distance matrix), compared to 11 rows or columns in a table setting.

**Table 39.** Demand, Revenues and Subsidies on Finnmark and North-Troms and Væroy-Bodø PSO routes; Own Illustration from (Lian et al. 2010; The Ministry of Transport and Communications 2012a)

| Finnmark and North-Troms | Passengers | Average Fares in NOK per PAX | Subsidy in NOK per PAX (2011/2012) | Compensation in million NOK (2011/2012) |
|---|---|---|---|---|
| 2007/2008 | 114,357 | 477 | 1,553 | 177.6 |
| 2008/2009 | 133,598 | 485 | 1,493 | 199.5 |
| Væroy-Bodø (85 km) | | | | |
| 2008/2009 | 9,063 | 570 | 4,341 | 39.3 |

In short we arrange the given data from the tender documents (see The Ministry of Transport and Communications 2006, 2009, 2012a) in the following fashion or as exposed in **Table 40**:

- Given:

  PAX-O-D-Matrix
  Revenue-O-D-Matrix

- Key:

  Km-Distances-O-D-Matrix

- Formulas:

  Revenue Passenger-Kilometer (RPK) = PAX × Km-Distance
  Revenue per RPK (RRPK; Yield) = Revenue / RPK

- Unknowns:

Profit (+) or Subsidy (-) per RPK = Cost per RPK (CRPK) – RRPK (see **Fig. 61** and **Fig. 62**)

- Example Steps in Scenario I:

a) Set average unit costs to 0. Increase CRPK step wise on all routes simultaneously, until sum of route subsidies (or profits) equals total subsidies.

b) Set distance-based scaling parameter to 0. Increase parameter step wise, until sum of all route subsidies (or profits) equals total subsidies.

**Calculations**

Certainly, the escalation of 10 % annual increase in subsidies is worrying from a social point of view (**Table 38**). Therefore, we analyzed unit revenues and costs on a 'per route' basis in the given documents to understand on which routes most subsidy per passenger is spent. This is a slightly different approach than the one taken in Lian (2010) regarding the changes in routing and pricing in the Norwegian PSO system, and Santana (2009) or Lian and Rønnevik (2011), focusing on the impact on operational costs of carriers and different regulations in PSO networks in comparison with the Essential Air Services (EAS) program in the U.S., where tenders can be cancelled or be given to competitors operating with lower cost during mid period (Bråthen 2011), however, a simpler approach than the one taken in chapter four and five in TØI report 1116/2010 (Lian et al. 2010).

In the airline industry it is common to base the costs and revenues on either the capacity, measured in available seat-kilometers (ASK), or the demand, measured in RPK. Due to the lack of reliable capacity data, we have based our financial figures on the RPK's per route in the operating year 2008 and 2009. Therefore, we calculated costs per RPK (CRPK), revenues per RPK (RRPK) and subsidies or profits per RPK (**Table 40**).

With the given data we calculated the RRPKs for each individual route *i* (**Fig. 61** and **Fig. 62**). From the RRPKs ($rrpk_i$) we subtracted a) average CRPK ($\bar{c}$) or b) non-linearly distributed $c_i^{\lambda}$, where $c_i^{\lambda}$ has the same monotonic decreasing slope over great-circle distance as the RRPKs. We use the distance-based approximation formula $c_i^{\lambda} = \frac{\lambda}{{d_i}^{0.36}}$, where λ is our scaling parameter and $d_i$ is the route great-circle distance in kilometers (**Table 40**). The RRPK's are distributed over distance by $\lambda = 21.03$.

**Table 40.** Calculation table – values, symbols and formulas

| Column | Variable | Data | Unit | Index | | | | Total or Average | Values |
|---|---|---|---|---|---|---|---|---|---|
| I | Route ID | | $i$ | *(1)* | *(2)* | | *(n)* | $n = N$ | $N = 98$ |
| II | Route Code (Origin & Destination) | GIVEN | $od_i$ | $od_1$ | $od_2$ | | $od_n$ | *(String)* | - |
| III | Great Circle Distance (in Kilometers) | GIVEN | $d_i$ | $d_1$ | $d_2$ | *(…)* | $d_n$ | $\frac{1}{n}\sum_{i=1}^{n} d_i = \bar{d}$ | $\bar{d} = 185\ km$ |
| IV | Route Revenues | GIVEN | $r_i$ | $r_1$ | $r_2$ | | $r_n$ | $\sum_{i=1}^{n} r_i = R$ | $R = 64{,}846{,}000\ NOK$ |
| V | Route Passengers | GIVEN | $p_i$ | $p_1$ | $p_2$ | | $p_n$ | $\sum_{i=1}^{n} p_i = P$ | $P = 133{,}598\ PAX$ |
| VI | Total Route Network Subsidy | GIVEN | $S$ | | | | | $S$ | $S = -199{,}491{,}000$ |
| VII | Route Revenue Passenger-Kilometer (RPK) | V× III | $rpk_i = p_i \times d_i$ | $rpk_1$ | $rpk_2$ | | $rpk_n$ | $\sum_{i=1}^{n} rpk_i = RPK$ | $RPK = 19{,}418{,}968\ RPK$ |
| VIII | Average Ticket Fare | IV ÷ V | $f_i = r_i \div p_i$ | $f_1$ | $f_2$ | | $f_n$ | $\frac{1}{n}\sum_{i=1}^{n} f_i = \bar{f}$ | $\bar{f} = 485\ NOK$ |
| IX | Route Revenues per RPK (RRPK) | IV ÷VII | $rrpk_i = r_i \div rpk_i$ | $rrpk_1$ | $rrpk_2$ | | $rrpk_n$ | $\frac{1}{n}\sum_{i=1}^{n} rrpk_i = \overline{rrpk}$ | $\overline{rrpk} = 3.74\ NOK$ per RPK |

**Table 40.** Calculation table – values, symbols and formulas (continued)

| Column | Variable | Data | Unit | Index | | | | Total or Average | Values | | |
|---|---|---|---|---|---|---|---|---|---|---|---|
| | | | $i$ | *(1)* | *(2)* | *(…)* | *(n)* | $n = N$ | Scenario | | |
| | | | | | | | | | I | II | III |
| X a | Average Unit Costs per RPK (CRPK) a) | UNKNOWN | $\bar{c}$ | $\bar{c}$ | $\bar{c}$ | | $\bar{c}$ | $\bar{c}_a$ | $\bar{c}_a = 13.62$ | $\bar{c}_a = 4.02$ | $\bar{c}_a = 3.34$ |
| X b | Distance-based Unit Costs per RPK (CRPK) b) (Scaling Parameter $\lambda \geq 0$) | UNKNOWN | $c_i^{\lambda} = \frac{\lambda}{{d_i}^{0.36}}$ | $c_1^{\lambda}$ | $c_2^{\lambda}$ | | $c_n^{\lambda}$ | $\sum_{i=1}^{n}(\frac{rpk_i}{RPK}c_i^{\lambda}) = \bar{c}_b$ | $\bar{c}_b = 13.62$ $\lambda^{I} = 85.75$ | $\bar{c}_b = 4.02$ $\lambda^{II} = 23.62$ | $\bar{c}_b = 3.34$ $\lambda^{III} = 21.03$ |
| XI a | Unit Subsidies or Profits at CRPK a) | IX – X a | $s_i^a = rrpk_i - \bar{c}$ | $s_1^a$ | $s_2^a$ | | $s_n^a$ | $s^a$ | | | |
| XI b | Unit Subsidies or Profits at *distance-based* CRPK b) | IX – X b | $s_i^b = rrpk_i - c_i$ | $s_1^b$ | $s_2^b$ | *(…)* | $s_n^b$ | $s^b$ | | | |
| XII a | Route Subsidies or Profits a) | XI a × IX | $S_i^a = s_i^a \times rpk_i$ | $S_1^a$ | $S_2^a$ | | $S_n^a$ | $\sum_{i=1}^{n} S_i^a = S^a = S^b \leq S$ | $S^a = S^b =$ *-199.5 million NOK* | $S^a =$ *-13.2 million NOK* | $S^a = 0$ |
| XII b | Route Subsidies or Profits b) | XI b × IX | $S_i^b = s_i^b \times rpk_i$ | $S_1^b$ | $S_2^b$ | | $S_n^b$ | $\sum_{i=1}^{n} S_i^b = S^b = S^a \leq S$ | | $S^b =$ *-7.98 million NOK* | $S^b = 0$ |
| XIII a | Route Subsidies per PAX a) | XII a ÷ V | $Sp_i^a = S_i^a \div p_i$ | $Sp_1^a$ | $Sp_2^a$ | | $Sp_n^a$ | $SP^a$ | | | |
| XIII b | Route Subsidies per PAX b) | XII b ÷ V | $Sp_i^b = S_i^b \div p_i$ | $Sp_1^b$ | $Sp_2^b$ | | $Sp_n^b$ | $SP^b$ | | | |

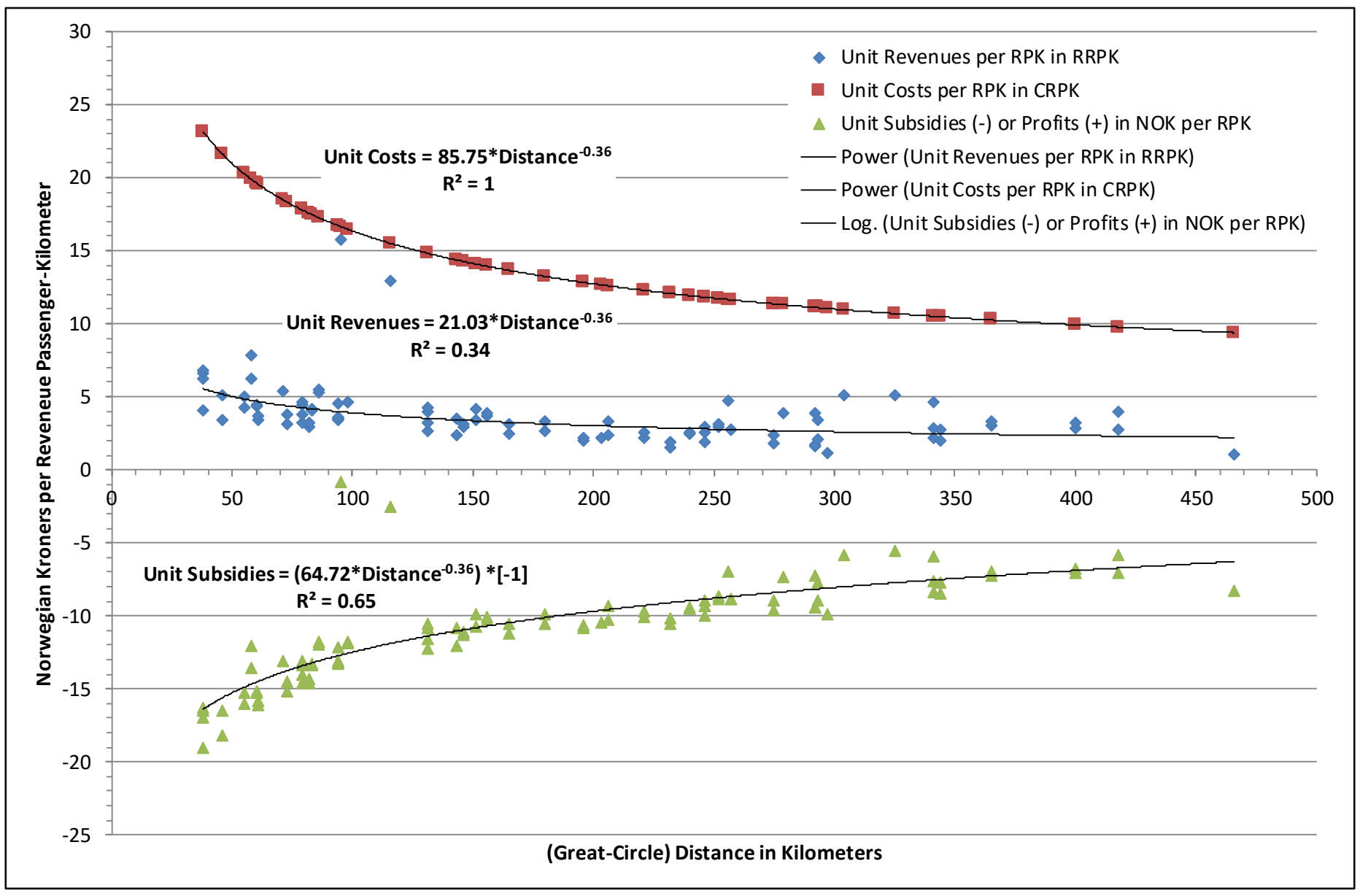

**Fig. 61.** Revenues and Subsidies in Finnmark and North-Troms Network at calculated current costs (Scenario I b: Average CRPK = 14.6 NOK at $\lambda = 85.75$).

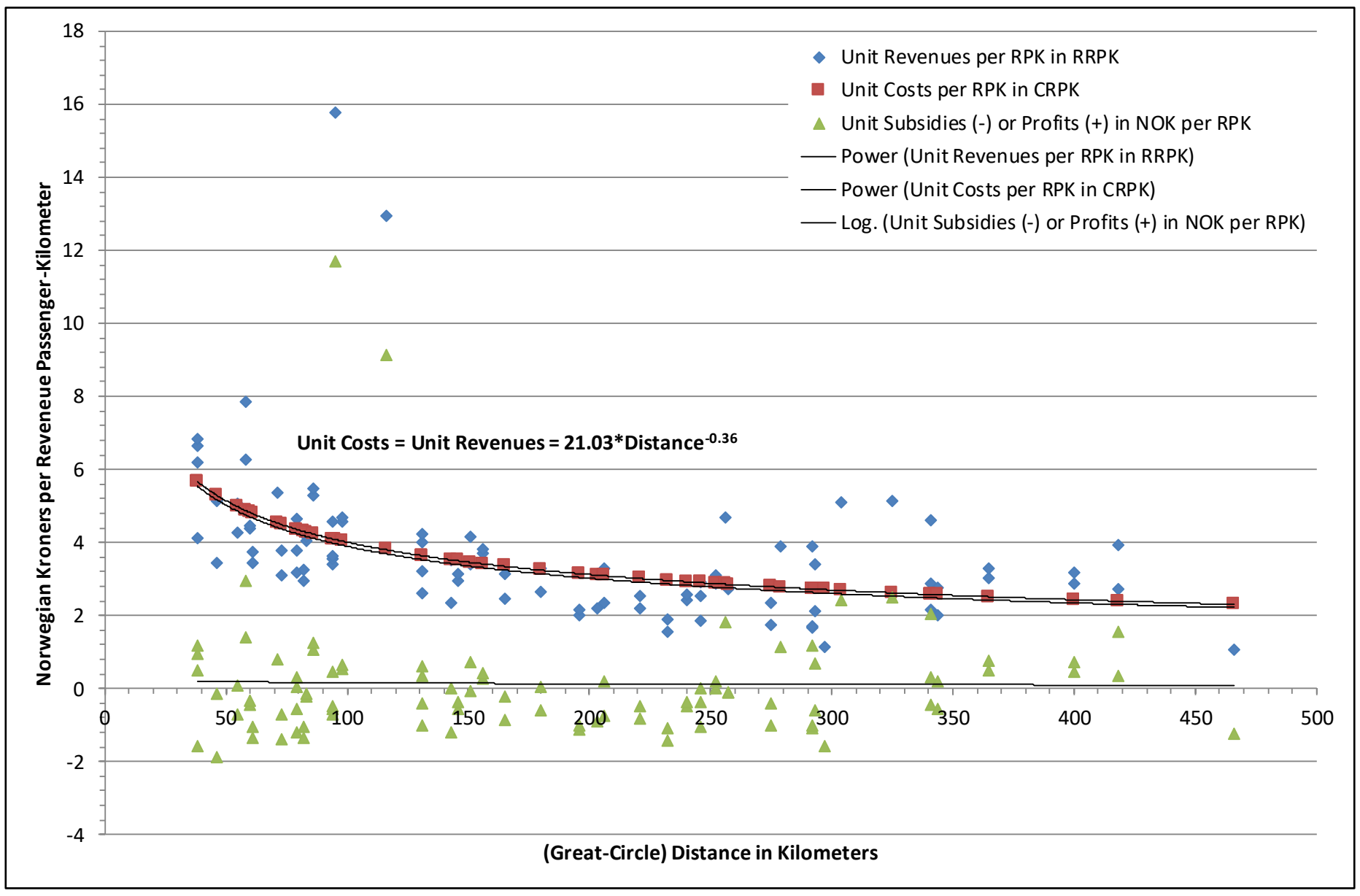

**Fig. 62.** Revenues and Subsidies in Finnmark and North-Troms Network at assumed break-even costs (Scenario III b: Average CRPK = 3.34 NOK at $\lambda = 21.03$).

In **Fig. 61** we show the case (Scenario I) where we have given RRPKs and calculated CRPKs, such that the sum of difference multiplied by the individual RPKs ($rpk_i$) gives us the total subsidies spent on the network of $S = (-)199{,}5$ million NOK (GAP 2013; Lian et al. 2010). Weighted average unit costs in this case are 13.62 NOK per RPK, compared to an average RRPK of 3.74 NOK per RPK. This difference leads to a deficit of about 10 NOK per RPK which needs to be subsidized by the Ministry of Transport and Communications.

Scenario II reveals how much lower the subsidies would be, if Widerøe would claim its own network costs of 4.02 NOK per RPK (CRPK). Here we find that a) the linearly distributed CRPK would result in a deficit of $S^a = 13.2$ million NOK and b) for non-linearly distributed CRPK would result in $S^b = 7.98$ million NOK. Therefore, we estimate a feasible level of just 4 % of the current subsidies.

In another Scenario III (**Fig. 62**) we calculated the *break-even costs*, which is the point where CRPK equals RRPK and we require *zero subsidies*. This was the case with weighted average unit costs of 3.34 NOK per RPK. Ideally, this cost level could be used by the Ministry of Transport and Communications to negotiate the tenders and required compensations with potential providers. **Fig. 63** shows the full relationship between the level of total subsidies and unit operating costs.

**Table 41.** Subsidy per passenger in NOK on routes in Finnmark and North-Troms network (April 2008 – March 2009); Own illustration from The Ministry of Transport and Communications (2012a)

| | | | | TO | | | | | | | | | | | | |
|---|---|---|---|---|---|---|---|---|---|---|---|---|---|---|---|---|
| | **Latitude** | **Longitude** | **FROM** | **ALF** | **BJF** | **BVG** | **HFT** | **HVG** | **HAA** | **KKN** | **MEH** | **SOJ** | **TOS** | **VAW** | **VDS** | **Average** |
| **Alta** | 69.977 | 23.356 | **ALF** | - | 2,450 | 2,364 | 1,105 | | | 2,175 | | 1,181 | | 2,251 | 2,284 | 2,251 |
| **Berlevåg** | 70.6 | 29.667 | **BJF** | | - | 644 | 2,153 | 1,622 | | 1,155 | 1,196 | | 2,434 | 698 | 910 | 1,484 |
| **Båtsfjord** | 70.867 | 29 | **BVG** | 2,443 | 723 | - | 2,117 | | | 1,418 | 836 | | 2,698 | 1,237 | 1,230 | 1,568 |
| **Hammerfest** | 70.68 | 23.676 | **HFT** | 1,154 | 2,231 | 2,091 | - | 1,012 | 980 | 2,279 | 1,574 | 1,721 | | 2,472 | 2,248 | 1,581 |
| **Honningsvåg** | 70.983 | 25.833 | **HVG** | | 1,653 | 297 | 1,028 | - | | 2,112 | 1,108 | | 2,612 | | 1,904 | 1,131 |
| **Hasvik** | 70.467 | 22.15 | **HAA** | 932 | 2,060 | 1,782 | 963 | | - | 1,780 | | | 1,498 | | 2,107 | 1,265 |
| **Kirkenes** | 69.724 | 29.891 | **KKN** | 2,223 | 1,163 | 1,386 | 2,286 | 1,920 | | - | 1,734 | 2,650 | | 1,106 | 620 | 1,141 |
| **Mehamn** | 71.033 | 27.833 | **MEH** | 2,123 | 1,172 | 757 | 1,595 | 1,060 | | 1,843 | - | 2,744 | 2,637 | | 1,521 | 1,627 |
| **Sørkjosen** | 69.783 | 20.933 | **SOJ** | 81 | 2,011 | 1,807 | 1,550 | | | 2,913 | 2,761 | - | 1,036 | | 2,600 | 1,124 |
| **Tromsø** | 69.68 | 18.907 | **TOS** | | 2,940 | 2,817 | | 2,946 | 1,615 | | 2,545 | 1,057 | - | 3,875 | | 1,620 |
| **Vardø** | 70.355 | 31.046 | **VAW** | 2,627 | 789 | 1,142 | 2,638 | | | 1,114 | | | | - | 837 | 1,534 |
| **Vadsø** | 70.065 | 29.845 | **VDS** | 2,195 | 915 | 1,252 | 2,275 | 1,785 | | 627 | 1,599 | 2,849 | | 880 | - | 1,484 |
| | | | **Average** | 2,001 | 1,653 | 1,953 | 1,726 | 1,131 | 1,174 | 1,181 | 1,735 | 1,069 | 1,253 | 1,062 | 1,496 | 1,493 |
| **Number of Passengers** | | | | | | | | | | | | | | | | **133,598** |
| **Revenue Passenger-Kilometers (RPK)** | | | | | | | | | | | | | | | | **19,412,681** |
| **Average Weighted Costs per RPK (CRPK)(in NOK)** | | | | | | | | | | | | | | | | **13.62** |
| **Scaling Parameter λ** | | | | | | | | | | | | | | | | **85.75** |
| **Operating Costs (in NOK)** | | | | | | | | | | | | | | | | **264,329,000** |
| **(-) Passenger Revenues (in NOK)** | | | | | | | | | | | | | | | | **64,838,000** |
| **(=) Total Route Network Subsidy per Year (in NOK)** | | | | | | | | | | | | | | | | **199,491,000** |

### 7.4 Results and conclusions

Considering the historic revenues and passenger figures per individual route from the public tender documents and the subsidies per route area from a presentation published by the Ministry of Transport and Communications and further evidence from (Lian et al. 2010), we were able to calculate the relevant revenues per RPK (RRPK). In an iterative fashion we increased the unit costs per RPK until the level of total route area subsidy of ca. 200 million NOK for the Finnmark and North-Troms network was reached. **Fig. 63** shows that the carrier Widerøe declared unit costs (and a profit margin) of around 13.6 NOK per RPK (CRPK), given the 2008/2009 levels of revenue and demand. We also see the break-even point at a cost level of 3.34 NOK per RPK.

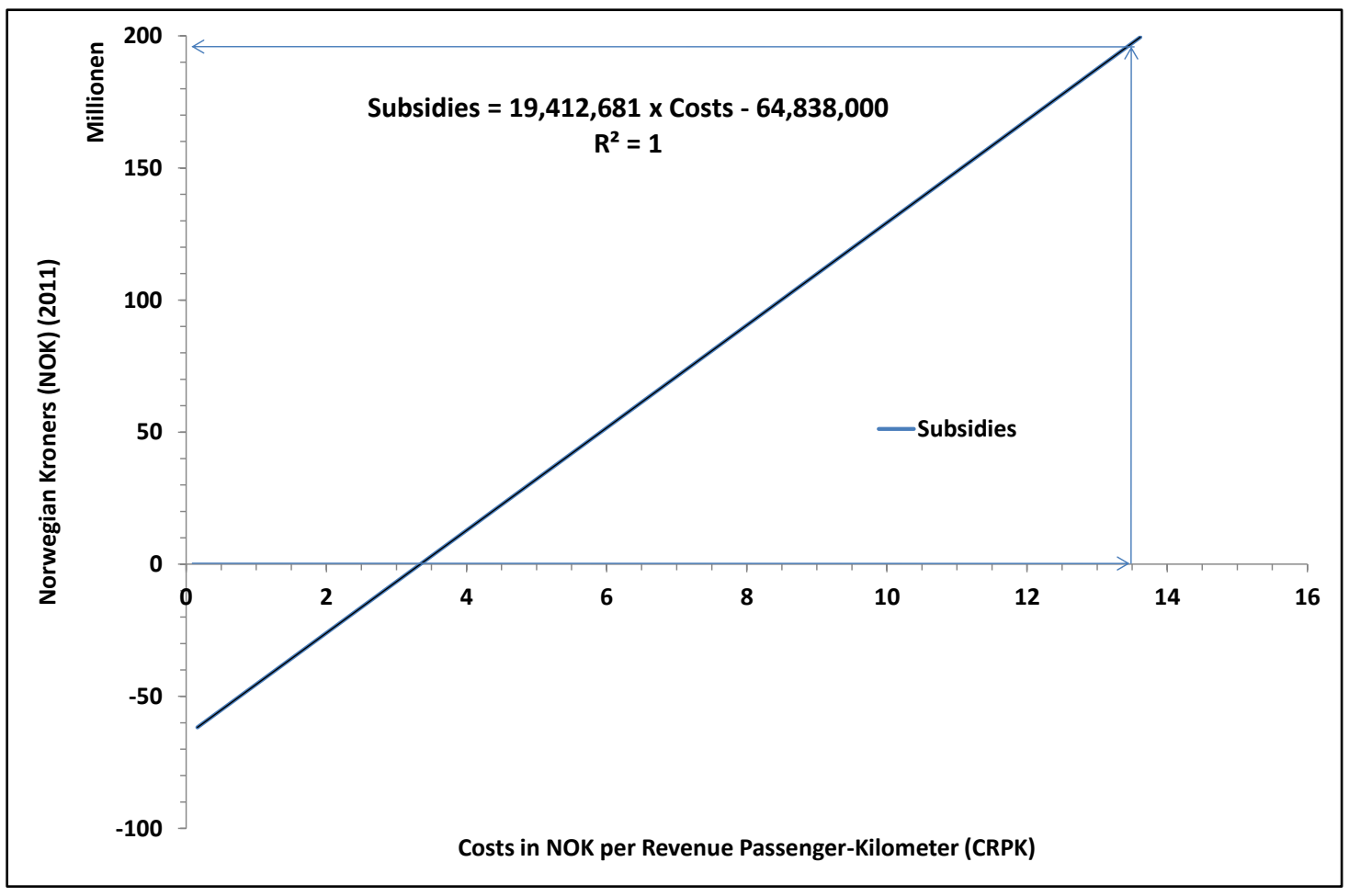

**Fig. 63.** Dependency of route network subsidy on carrier unit costs (incl. profit margin) for the case of Widerøe operating the Finnmark and North-Troms network (Own illustration; Note: Here the signs are reversed such that total subsidies are positive and profits are negative)

In **Table 41** we can observe strongly varying subsidy levels among the 98 routes of between 81 NOK (SOJ→ALF) and 3,875 NOK (TOS→VAW) per route and per passenger, at an average of 1,493 NOK per Passenger for the whole Finnmark and North-Troms network. None of the connections in this network seems to be profitable.

On the routeVæroy↔Bodø we can expect revenues of 6.7 NOK per RPK, which stand against approximated cost of 58 NOK per RPK, leaving a subsidy of 51.3 NOK per RPK. For the 85 km distance between the airports and the given demand of 9,063 passengers this amounts to the total route subsidy of around 39.5 million NOK. On this particular route the average fare of 570 NOK (which is 69 % of the maximum fare requested by the Ministry of Transport and Communications [The Ministry of Transport and Communications 2008, 2010]) and only amounts to 13 % of the required subsidy of 4,341 NOK per passenger for this island shuttle service.

Of course, there are good reasons to stimulate regional development by supporting air traffic services, but as the regulations are currently set a lot of money seems to be wasted. The reasons for this partly lie in the fact that there is not enough competition for serving subsidies PSO routes, which gives the few bidders a too dominant position in making their cost claims. The authorities have the only alternative to withdraw the PSO tenders and to risk a stop of services or to compensate whichever claimed costs. With the EAS framework in the US we find an alternative program where it is possible to cancel tenders mid-period, when another carrier has a more attractive cost base (at the same requested level-of-service) (Bråthen 2011). Another measure could be the breaking up of networks by separating individual routes from them, when smaller players offer operations on these routes with lower or no compensations. This would circumvent the need for a large fleet in strongly interconnected networks as the one described in this article. Strong point-to-point connections may be excluded from the tender if these prove to be operated profitably by the current provider in subsequent years. This would require even more transparency of the authorities regarding the publication of detailed operating costs and revenues on the route level so that airlines in the market gain better operational insights before posing bids for tenders on offer.

# Part III: Airport Capacity and Sustainable Operations

In Part III of this dissertation new approaches to managing airport capacity and sustainability are presented. To solve real world operational problems modelling techniques, such as simulation and optimization, are necessary that support decision making in airport management and in air transport planning. In the previous parts fundamentals of airport benchmarking and air transport economics were highlighted. These analyses were commonly based on historic data. In this part we use modelling techniques to analyze present and future scenarios. Some forecasting techniques also rely on past data, but the scenarios in this part make use of present baseline scenarios, that display the current operations, versus which possible future scenarios are compared. The purpose of this part is to show that it is feasible to use a sophisticated modelling technique such as simulation for benchmarking. Oftentimes it is impossible to get a good estimate for capacity in benchmarking studies, as published figures could be far from the truth. When an airport under study or a future state of the airport is benchmarked against other airports simulation provides sufficiently good capacity measures. A positive side effect of such simulation studies is, that airport operations could be streamlined and hotspots or bottlenecks on the current or planned airfield or in the airspace can be identified. By formulating traffic rules or by resequencing flights it is possible to reduce queues at the runway(s) or in the airspace. Simple simulation models can be built in a short time, which is necessary in the benchmarking context, where a lot of small model details might not alter the main results and recommendations but could be frustratingly complicated to include. In the final chapter the discussion is opened towards considering delay propagation in air transport networks, that could build up during daily operations and that could only partially be mitigated by airlines. Different carriers operate different networks which leads to the question which network structure is superior to the other regarding the propagation of delay. For a benchmarking study done in the most meaningful way, many interactions outside every airport's own operations should be considered. An airport cannot be analyzed as an isolated entity, but must be seen in context with its environment, such as its users, competitors, and the immediate neighborhood.

# 8 Airport capacity and demand calculations by simulation - the case of Berlin-Brandenburg International airport[1,2]

Branko Bubalo and Joachim R. Daduna

**Abstract.** Airports are vital parts of traffic infrastructure and networks in global and dynamic economies securing inter- and transcontinental mobility of goods and people in spatially dislocated market structures. The operational capacity of an airport must be dimensioned under a long-term strategic view as its productivity is determined by available infrastructure. Often expansion projects for, e.g., an additional runway require a timeframe of up to twenty years for negotiation, planning and construction. The correction of existing or future bottlenecks will be increasingly difficult, partly due to public and political opposition and environmental awareness. From this point of view we critically examine the published planning figures and forecasts of demand and capacity of the currently constructed Berlin-Brandenburg International airport over a 20-year timeframe. Our methodology is based on a computer simulation of an independent parallel runway in segregated mode. Increasing traffic and changing traffic mix are simulated with the airport and airspace modeling tool SIMMOD, which provides the output data to calculate the capacity utilization and chosen level of service indicator minutes of average delay per flight. The simulation has shown that the practical capacity of 76 flights per hour is the maximum demand to be served under the defined assumptions. We discuss our findings and compare our results with other airports operating a similar runway layout.

## 8.1 Problem description

Mobility of passengers and goods is an essential basis for the development of economic market structures resulting from the spatial dislocation of production facilities and labor in industries as well as in services and retail trade. Adequate systems and capacities are mandatory to serve the demand for transport. Each situation requires its specific technical and functional framework for mode and type of transport. In this respect, the characteristics of quality rating (Daduna 2009) of transport modes play an essential role because those can give an answer concerning case-specific ability and level of service (LOS). In this context the traffic infrastructure and the technical development of equipment is of great importance for transport mode performance and quality.

[1] B. Bubalo (corresponding author)
Berlin School of Economics and Law / German Airport Performance,
Badensche Str. 52, 10825 Berlin, Germany
E-mail: branko.bubalo@googlemail.com
[2] J. R. Daduna
Berlin School of Economics and Law, Badensche Str. 52, 10825 Berlin, Germany
E-mail: daduna@hwr-berlin.de

Especially when dealing with inter- and transcontinental transport of passengers and (small sized and high value, as well as special) cargo, air transport is the dominant mode. Competitive advantages mainly result from high speed and therefore shorter travel times between origins and destinations, which are associated with convenience and cost reductions (such as tied capital costs and insurance costs). Disadvantages of air transport services are clearly the limited aircraft capacity and the comparably costly tonnage. Airlines must overcome the trade-off between raising air transport capacities to attain economies of scale and reducing fares due to a strong increase of low-fare competition.

In the foreground are market-entry barriers resulting from the need for large capital investment and skilled work force for new market entrants as well as from current and forecasted slot constraints at some of the largest hub-airports. These slot-coordinated (IATA 2011, pp. 11) airports, e.g. London Heathrow Airport (LHR), Frankfurt Rhine-Main Airport (FRA) and Munich Franz Josef Strauss-Airport (MUC), are operating at declared capacity during peak hours and are considered congested (EUROCONTROL 2009b, p. 64). However, for certain products and market segments there is no substitute for air transport, which leads de facto to monopoly situation for certain routes and origin / destination markets. Therefore, a decrease in LOS and an increase of queuing and turnaround times will be tolerated by the airlines at congested major hubs.

When designing national and international Air Transport Systems (ATS), the airport productivity (Bubalo 2010) and particularly the runway capacities play a key role. On the one hand airports are core elements of infrastructure representing the nodes of air transport networks, and on the other hand they serve as gateways in connection with the land transport modes in pre- and on-carriage services. The costly closure of many airports and also the airspace over northern Europe in April 2010 resulting from the ash cloud of a volcano eruption in Iceland the importance of ATS in connecting international markets and retail trade.

The airside capacity is determined by a variety of factors such as local topography and weather conditions, as well as configuration of the runways, runway exits, taxiway system and aircraft parking stands. Airside capacity could also be limited by Air Traffic Control (ATC) if for example the availability of specially trained air traffic controllers or specific radar equipment is restricted. On the landside airport capacity is typically constrained by airport facilities, ground handling performance as well as road and rail accessibility.

The enhancement of airside capacity, especially the realization of additional runway capacity, requires lengthy planning, negotiation and approval processes prior to construction. In addition large private and public investments are needed, which require detailed forecasts of future demand and Cost-Benefit Analyses (CBA) with the objective to outweigh the disadvantages versus the benefits of a particular airport expansion project.

The focus of our study is to analyze the impact of different scenarios and increasing levels of demand on the runway throughput. We will derive estimates of capacity utilization, congestion delays and LOS, as well as delay costs (IATA 2004, pp. 157; Bäuerle et al. 2007; EUROCONTROL 2009a, pp. 9; NEXTOR 2010, pp. 14). Based on a given runway layout and international safety regulations a simulation approach is applied in this study (ATAC / HNTB 2006, pp. 5). Only by employing simulations it is possible to analyze and calculate airside capacity and queuing phenomena in such a complex and dynamic environment as that of an airport (Bazargan, Fleming and Subramanian 2002, pp. 1235).

"What if?" questions can be answered, such as 'What are the consequences if the planned maximum capacity is not adequate for serving forecasted demand?' We will take the case of the Berlin-Brandenburg International airport (BER), which is currently under construction (current construction plans were drawn up in the late 1990's) and which is expected to go into operation October 31st, 2020. It is located on the south–eastern periphery of Berlin partly on the territory of the existing airport Berlin-Schönefeld (SXF).

The newly constructed BER with a maximum of 83 take-off and landing slots per hour (MSWV 2004, p. 222) is going to replace the current airports Berlin-Tegel (TXL) and SXF. The former is currently operating two dependent closed spaced parallel runways with a capacity of 52 hourly slots and the latter is operating a single runway with a capacity of 26 hourly slots. It is questioned if the installed parallel runway configuration is sufficient for unrestricted future traffic growth. In our view it is more likely that we will experience strongly increasing congestion delays at levels as low as 60% above current traffic levels. This will lead to slot constraints in the near future due to limited runway availability at BER, e.g. ten years after opening assuming 5% annual growth of traffic.

The planned airside configuration will be critically examined with regard to the future development of demand, assuming a range of growth patterns. Finally, BER will be benchmarked against data from MUC and LHR airports, both operating two independent parallel runways.

## 8.2 Methodology

In Berlin, the capital of the Federal Republic of Germany, there is a political debate regarding not only the capacity of the new BER airport but also the future relevance of air traffic. There are reasonable doubts that a parallel runway system with two runways replacing the currently operating airports TXL and SXF with a combined total of three runways will be adequate. The available capacity in the Berlin area has already been reduced, because of the closure of the Berlin-Tempelhof (THF) airport, which was mainly serving small and business jets on its two (short) runways until 2008.

The key question is which ultimate level of demand can be served at the BER airport given the proposed parallel runway layout (**Fig. 64**) and when this maximum level is expected to be reached. The runways will have a length of 3,600 and 4,000 meters, and a separation of 1,900 meters; also they will be staggered by 1,250 meters (MSWV 2004, p. 222). The aerial charts of the proposed airport layout come from the public planning documents. The associated coordinates of the planned infrastructure and the nodes for our model are extracted from the Google Earth software environment and could be regarded sufficiently accurate.

Requested and served demand will be examined by hour of day with regard to LOS, measured in minutes of average delay per flight. Different scenarios with various aircraft mixes are derived from the baseline flight schedule. The required calculations are made with the Visual SIMMOD software package (AirportTools 2003) in connection with established SIMMOD tools and recommendations from the Federal Aviation Administration (FAA) (FAA 1995).

The airside airport layout structure will be simulated as a staggered independent far parallel runway system (MSWV 2004, pp. 409), where arrival and departures are served separately on the two runways in segregated mode. This is the main mode at Europe's busiest airport, LHR. Official planning documents for the BER airport assume an ultimate capacity of 83 flights in the peak hour (MSWV 2004, p. 222) (under mixed-mode operations (ARC 2010)). Other assessments in the documents reveal a capacity of 90 movements in the peak hour (MSWV 2004, p. 334), as it is the case at MUC with a similar runway layout.

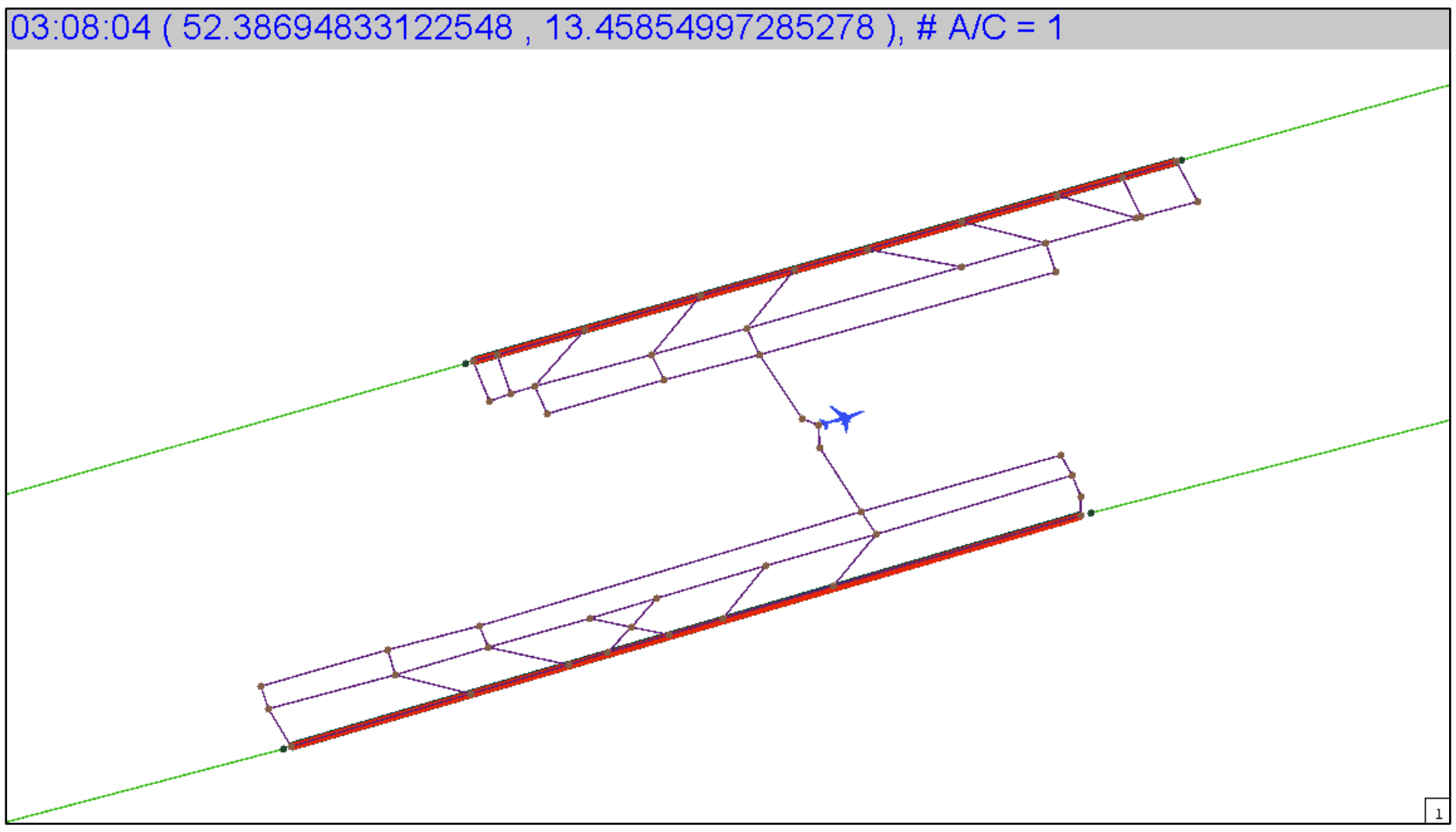

**Fig. 64.** Basic layout BER airport

MUC airport is currently able to operate at this capacity only over short periods of time during the day during waves of arriving or departing flights (DLR 2006, p. 18 and pp. 27). Segregated and mixed-mode operations at LHR and MUC will be further discussed in Section 0 (Atkin, Burke and Greenwood 2010).

Besides mode of operation other main limiting factors regarding airport airside operations capacity are the number of available independent (greater than 1,380 m of lateral separation) parallel runways, nighttime curfews (between 23:00 and 7:00) or other operating or environmental restrictions, separation minima between succeeding aircraft, runway occupancy times, and the setting of LOS. For the majority of flights at European airports ATC apply instrument flight rules (IFR) with separation minima of at least 2.5 nautical miles (NM) on final approach (Hansen 2002, p. 77).

### Applied scenarios

The basic input for the SIMMOD simulation is the combined flight schedule with 635 flights which were operated at the Berlin airports TXL and SXF on the design day (Thursday, June 26th 2008). Diverted and cancelled flights, representing 10% of the daily movement, have been excluded from the original peak schedule. Overcoming the decrease in aircraft movements in 2009, the traffic volume in 2010 attained previous levels of 2008, so the magnitude of the input data remains valid for the further computational tests (**Table 42**). The representative flight schedule consists of the following basic data: Origin, destination, scheduled departure or arrival time, aircraft type and flight number. These flights are separated by three wake turbulence categories (WTC) based on Maximum Take-off Weight (MTOW) of each aircraft type. The categories HEAVY (H), with an MTOW greater than 136 tons, MEDIUM (M), with a MTOW between 7 and 136 tons, and LIGHT (L), with a MTOW below 7 tons define the types to calculate the aircraft mix at a given airport over certain periods of time (in our case on the design day) (Horonjeff, McKelvey, Sproule and Young 2010, p. 90).

**Table 42.** Daily and hourly flights by wake turbulence category

| Hour of day<br>WTC | 00 | 01 | 02 | 03 | 04 | 05 | 06 | 07 | 08 | 09 | 10 | 11 | 12 | 13 | 14 | 15 | 16 | 17 | 18 | 19 | 20 | 21 | 22 | 23 | Sum |
|---|---|---|---|---|---|---|---|---|---|---|---|---|---|---|---|---|---|---|---|---|---|---|---|---|---|
| Heavy | | | | | | | | 1 | 1 | 2 | | 1 | 1 | 1 | 1 | 1 | | | 1 | 1 | | | 1 | | 12 |
| Medium | 2 | 1 | | 1 | 1 | 2 | 24 | 26 | 46 | 43 | 34 | 39 | 25 | 31 | 32 | 32 | 37 | 41 | 42 | 44 | 46 | 35 | 21 | 4 | 609 |
| Light | | | | | | | | 2 | 1 | | 2 | 1 | | 1 | | 2 | | 1 | 2 | 1 | 1 | | | | 14 |
| Sum | 2 | 1 | 0 | 1 | 1 | 2 | 24 | 29 | 48 | 45 | 36 | 41 | 26 | 33 | 33 | 35 | 37 | 42 | 45 | 46 | 47 | 35 | 22 | 4 | 635 |

The sequencing of aircraft in the airspace needs to be seen as an important influence on the airside capacity and performance of an airport system (de Neufville and Odoni 2003, p. 395; Hansen 2002, p. 78). The WTC defines minimal separation requirements between two subsequent departures in seconds and between two subsequent arrivals in NM (IATA 2004, p. 167). In the case of simulating mixed-mode operations on the same runway (which has not been modeled in this study), separation matrices for subsequent departures after arrivals and vice versa must be applied (de Neufville and Odoni 2003, 377 pp.; Horonjeff et al. 2010, p. 113; NASA 1998, p. 5). It should be noted that different aircraft speeds may require even further separation minima during approach than the minima listed in **Table 43** (Hansen 2002, p. 78; NASA 1998, pp. B-3). However, in this study these listed separation minima have been used in our *first-come–first-served* (FCFS) model (Bubalo 2011).

Since the separation minima for IFR flights among the three categories are not symmetric, the sequential order by type and number of aircraft during an arrival bank results in different hypothetical sequence lengths, which in turn influences the capacity of an airport (IATA 2004, p. 168; Klempert and Wikenhauser 2009; Romano, Santillo and Zoppoli 2008). This can be observed in an example of combinations of H and L category aircraft. For the series of six arrivals on the same runway H → L → H → L → H → L the calculated sequence length is (6 + 3 + 6 + 3 + 6 NM =) 24 NM (Average = 4.8 NM), whereas for the rearranged series L → L → L → H → H → H the resulting sequence length is reduced by 7 NM to (3 + 3 + 3 + 4 + 4 NM =) 17 NM (Average = 3.4 NM), therefore two more aircraft could be fitted into the original sequence length (Newell 1982, pp. 78; Atkin et al. 2010). This intuitive example shows the increase in capacity due to aircraft re-sequencing and the benefits of bundling succeeding flights of the same category, which could have short-run benefits, e.g. by reducing congestion in times when several aircraft arrive simultaneously (Klempert and Wikenhauser 2009).

Therefore, different aircraft or traffic mixes are examined in six scenarios testing possible future shifts of preference for certain aircraft types (IATA 2004). We have modeled many possibilities of succeeding approaching and departing aircraft of different WTC in the baseline Scenario 0 and Scenario I to V.

Sequencing is useful at the tactical level in keeping the average separation minimum close to the absolute minimum of currently three NM (with some exceptional occurrences of separation minima of 2.5 NM at airports having sophisticated landing systems) (see **Table 43**). As it can be seen in **Fig. 65** the arrival (or departure) capacity not only depends on the average separation minima applied by ATC, but also depends on the average landing (or departure) speeds. While the capacity could vary by six to seven flights per hour for any given mix of aircraft in the Scenarios 0 to V, it could furthermore vary by eight or nine flights per hour with approach speeds between 250 and 300 kilometers per hour.

**Table 43.** Wake turbulence related aircraft separation minima under IFR

| **Sequence** | | **Arrival - Arrival (in NM)** | | | **Departure – Departure (in seconds)** | | |
|---|---|---|---|---|---|---|---|
| Trailing | Leading Maximum take-off weight (MTOW) (in tonnes) | Heavy | Medium | Light | Heavy | Medium | Light |
| Heavy | > 136 | 4 | 5 | 6 | 120 | 120 | 120 |
| Medium | 7 - 136 | 3 | 3 | 5 | 60 | 60 | 120 |
| Light | < 7 | 3 | 3 | 3 | 50 | 50 | 50 |

By applying the capacity formula, *runway capacity* (C) (in flights per hour) = *average speed* (V) (in kilometers per hour) / *separation* (D) (in kilometers) (Gosling, Kanafani and Hockaday 1981, p. 51), it turns out that the overall range of landing capacity could span from 38 to 53 flights per hour in the short run, given the range of speeds. These capacities could be achieved by the controller by sequencing exactly the given schedule and by applying the respective separation minima.

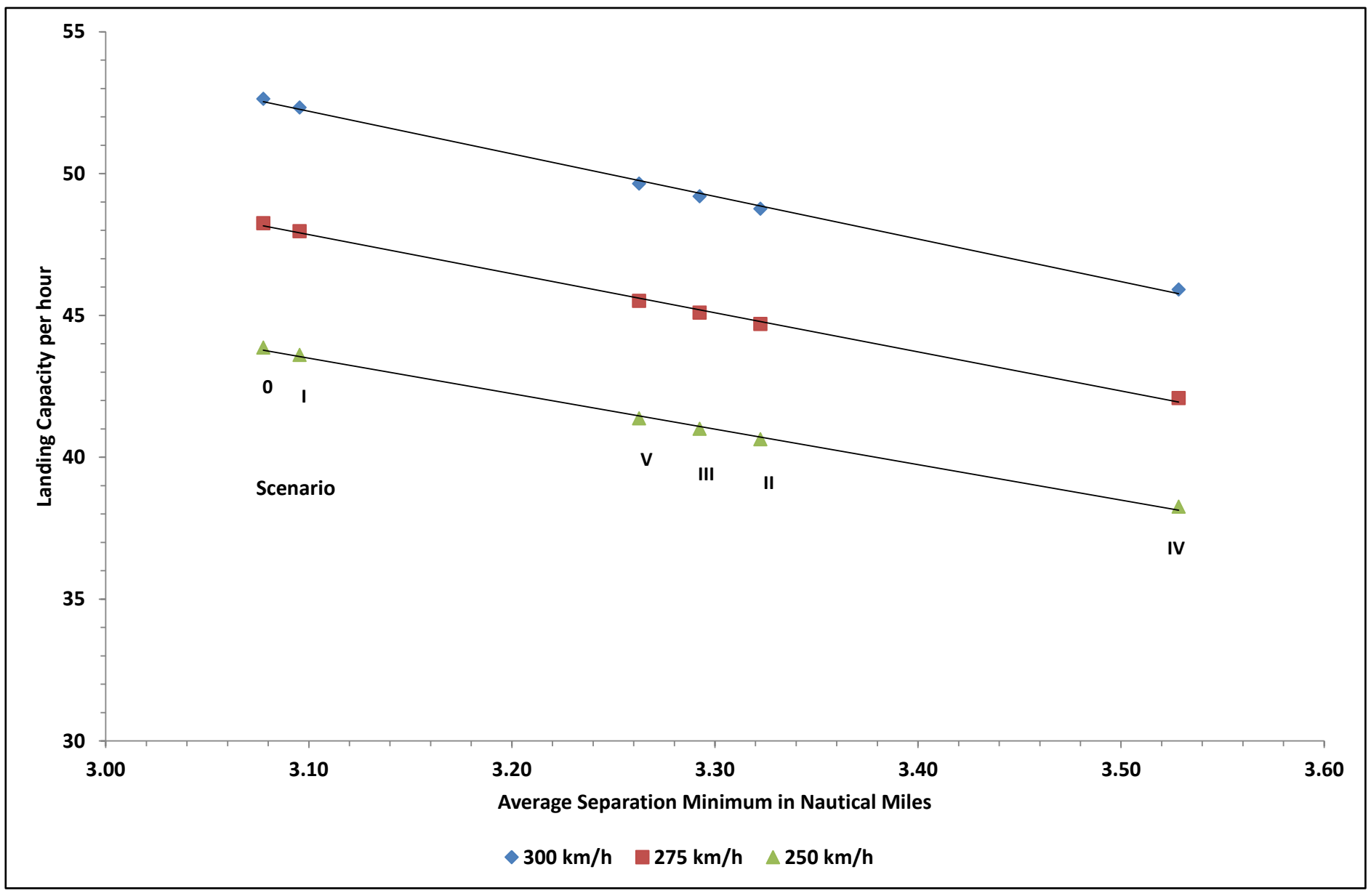

**Fig. 65.** Variation of theoretical landing capacity of Scenario 0 to V

In **Fig. 65** we show that the current aircraft mix (Scenario 0) is quite beneficial with respect to improving capacity. The aircraft mix given by the schedules of Scenario 0 and I provide sequences with theoretically the highest capacities, followed by Scenarios V, III and II with intermediate capacities in the range. Lastly we can observe that the aircraft mix used in Scenario IV represents the least efficient (assuming the same average approach speeds).

**Table 44.** Calculated average separation minima and landing capacities

| | **Average separation minima** | | **Capacity at approach speed in flights per hour** | | |
|---|---|---|---|---|---|
| **Scenario** | **in NM** | **in km** | **300 km/h** | **275 km/h** | **250 km/h** |
| Scenario 0 | 3.08 | 5.70 | 53 | 48 | 44 |
| Scenario I | 3.10 | 5.73 | 52 | 48 | 44 |
| Scenario II | 3.32 | 6.15 | 49 | 45 | 41 |
| Scenario III | 3.29 | 6.10 | 49 | 45 | 41 |
| Scenario IV | 3.53 | 6.53 | 46 | 42 | 38 |
| Scenario V | 3.26 | 6.04 | 50 | 46 | 41 |

The average separation minimum of the input schedule for Scenario 0 of 3.08 NM (or the consequent maximum capacity C of 53 landings or departures or the minimum interval of 1/C = 67.9 seconds between flights at 300 kilometers per hour approach or departure speed) could be regarded (with the present aircraft mix) as a benchmark for the sequencing by ATC. **Table 44** shows the calculated landing capacity by taking the average theoretical separation minimum of the sequence of all daily arriving flights.

In addition it can be seen in the computed figures in **Table 44** that the theoretical average separation minimum of the arrival sequence in the baseline Scenario 0 is 3.08 NM, which corresponds to a landing capacity in a perfectly scheduled sequence of 53 landings per hour at 300 km/h approach speed, 48 landings at 275 km/h approach speed and 44 landings at 250 km/h approach speed. The furthest average separation is calculated for Scenario IV with 3.53 NM and a corresponding capacity of 46, 42 and 38 arrivals per hour at 300, 275 and 250 km/h average approach speed, respectively.

Initially a baseline scenario is created directly from the design day schedule which consists of the combined schedules of TXL and SXF airports (Scenario 0). Before modeling five additional scenarios with varying WTC shares (Scenario I to V) (see **Table 45**) we randomly substituted aircraft types in the baseline design flight schedule with other aircraft types of higher or lower WTC to reflect the changes in the WTC distribution. The changes in the aircraft mix which may arise in the future in comparison to the baseline Scenario 0 can be seen in **Table 45**.

A single mathematical expression describing this distribution of aircraft types at a given airport over a certain period of time is the Mix Index (MI). The MI can be used for long-range capacity planning (FAA 1995, p. 3). To calculate the MI for the share of M category aircraft, the three-fold weighted share of H category aircraft is added to account for its impact on capacity utilization (Klempert and Wikenhauser 2009) and is used in this context for comparisons based on a single indicator (see de Neufville and Odoni 2003, pp. 391; Horonjeff et al. 2010, pp. 515).

**Table 45.** Scenario data

| **Scenario** | **0** | | **I** | | **II** | | **III** | | **IV** | | **V** | |
|---|---|---|---|---|---|---|---|---|---|---|---|---|
| %-Share / # | % | # | % | # | % | # | % | # | % | # | % | # |
| Heavy (H) | 2 | 12 | 5 | 32 | 15 | 95 | 5 | 32 | 20 | 127 | 2 | 13 |
| Medium (M) | 96 | 609 | 95 | 603 | 80 | 508 | 84 | 533 | 65 | 413 | 84 | 533 |
| Light (L) | 2 | 14 | 0 | 0 | 5 | 32 | 11 | 70 | 17 | 95 | 14 | 89 |
| Sum | 100 | 635 | 100 | 635 | 100 | 635 | 100 | 635 | 100 | 635 | 100 | 635 |
| Mix Index (MI)* | 102 | | 110 | | 125 | | 99 | | 125 | | 90 | |

* MI = (3 * %-share of H) + %-share of M

The baseline flight schedule and the flight schedules of the scenarios serve as the main input for the SIMMOD simulations. In the six scenarios, varying in WTC shares and MI, the 635 baseline movements are subject to a cloning probability of producing a replica of themselves during the simulation runs, reflecting growth in demand. Zero growth is set for Scenario 0 as the original baseline scenario. Further tests include a 20% decline in traffic, and incremental increases of 20% up to 200% of baseline demand. Altogether 660 simulation runs were conducted and analyzed for the six scenarios, eleven growth rates and the ten iterations of each simulation. With a probability distribution to the injection time of each flight into the simulation and the cloning probability to simulate traffic growth we modeled random FCFS sequences which vary from the original sequence order of the scenario schedules.

The objective of simulating the scenarios with varying aircraft mixes and levels of demand is the determination of a maximum throughput of the planned airport airside configuration. More specifically the objective is to estimate at which rate of demand and under which scenario the service rate is exceeded, flights are delayed and the average delays increase above predefined LOS.

This threshold, where a stable flow of traffic can be maintained over extended periods of time, is defined as the practical or sustainable capacity of the airport (Bazargan et al. 2002, p. 1236; de Neufville and Odoni 2003, p. 448; Mensen 2007, pp. 379).

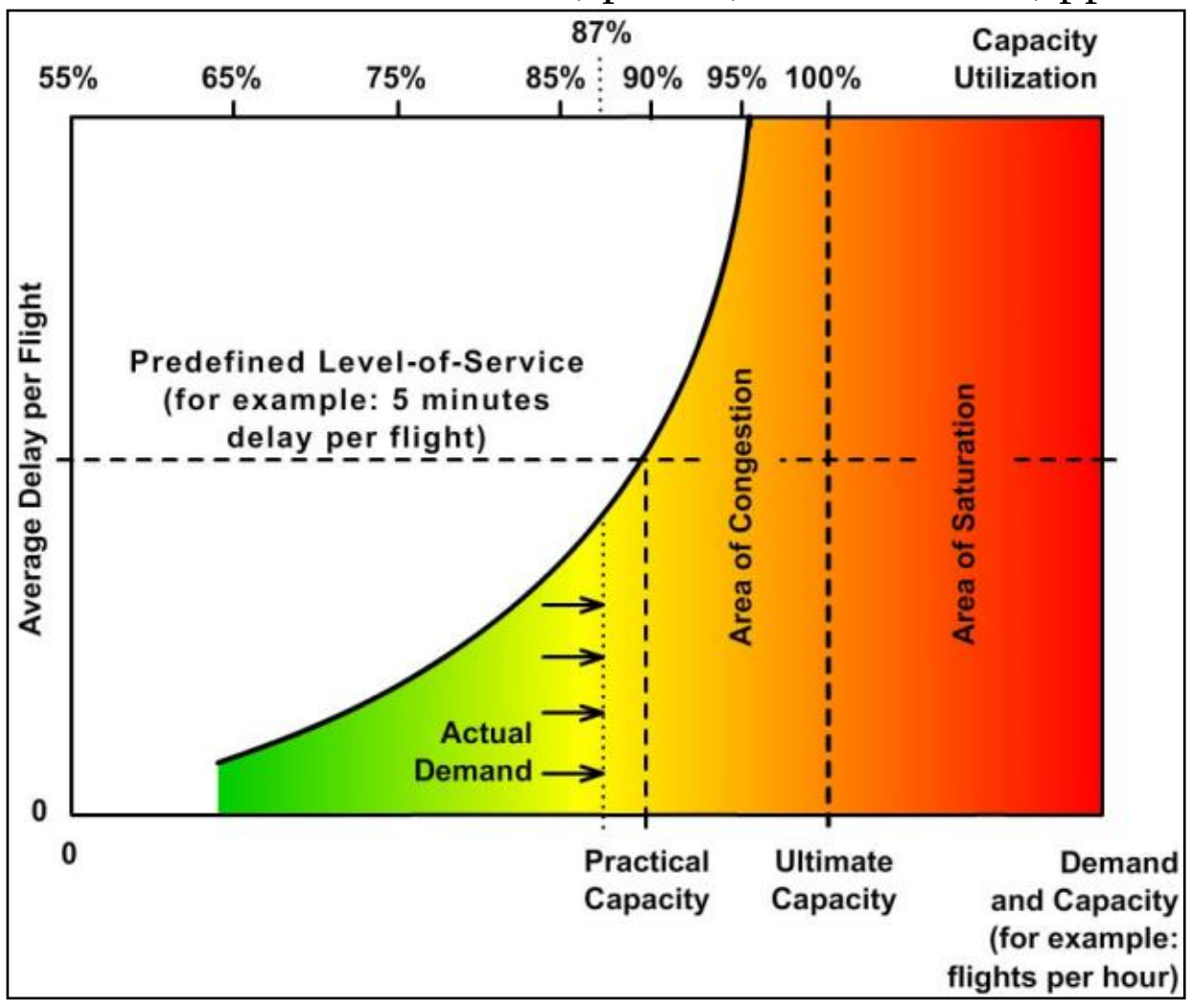

**Fig. 66.** Relationship between demand, capacity and delay

The (theoretical) relationship between the development of demand, capacity and its utilization, and delay is shown in **Fig. 66**. To describe the level of congestion at BER airport, a LOS of below six minutes average delay per flight on design day has been defined. Suggested parameters in the literature vary from four minutes in de Neufville and Odoni (2003, p. 448) to eight minutes in Mensen (2007, p. 388).

## Computational results

In the context of this contribution, only the main findings from simulation Scenario 0 can be presented in detail. However, these results may safely be regarded as representative of all the other computations. In addition a comparison of different scenario-based capacity developments is shown.

From simulating the baseline Scenario 0 and subsequent runs with decreasing and incrementally increasing traffic, we get the daily and peak hour demand, throughput capacity, capacity utilization, and occurring waiting times (delay), which are the basis for subsequent calculations. When aggregating the data and analyzing the flow of aircraft in the animation displaying the results of the SIMMOD simulation, we observe the creation of capacity bottlenecks. When the baseline demand increases above certain capacity levels, considerable delays in queuing aircraft emerge (NASA 1998, pp. 2-2; FAA 1995, p. 4).

At 60% growth and around 1,050 flights on design day, it is evident that throughput is exceeded by demand, and consequently the average delay increases sharply above a LOS of six minutes. This however would imply that, related to a planning horizon of one year, about 383,000 flights have to be operated, which would already exceed the calculation basis of 360,000 flights (MSWV 2004, p. 222).

Estimating the performance indicator average delay per flight at a money value of € 42 per minute (EUROCONTROL 2009a, pp. 9), exponentially rising delay costs are expected with increasing capacity utilization and delays. Ultimately, bottlenecks and delays lead to flight cancellations, due to unavailable replacement flight crews or aircraft, which represent serious disturbances in the daily airline fleet turnarounds. Cancellations are costly for airlines and inconvenient for the passenger. Their frequent occurrence will certainly affect airlines' profitability.

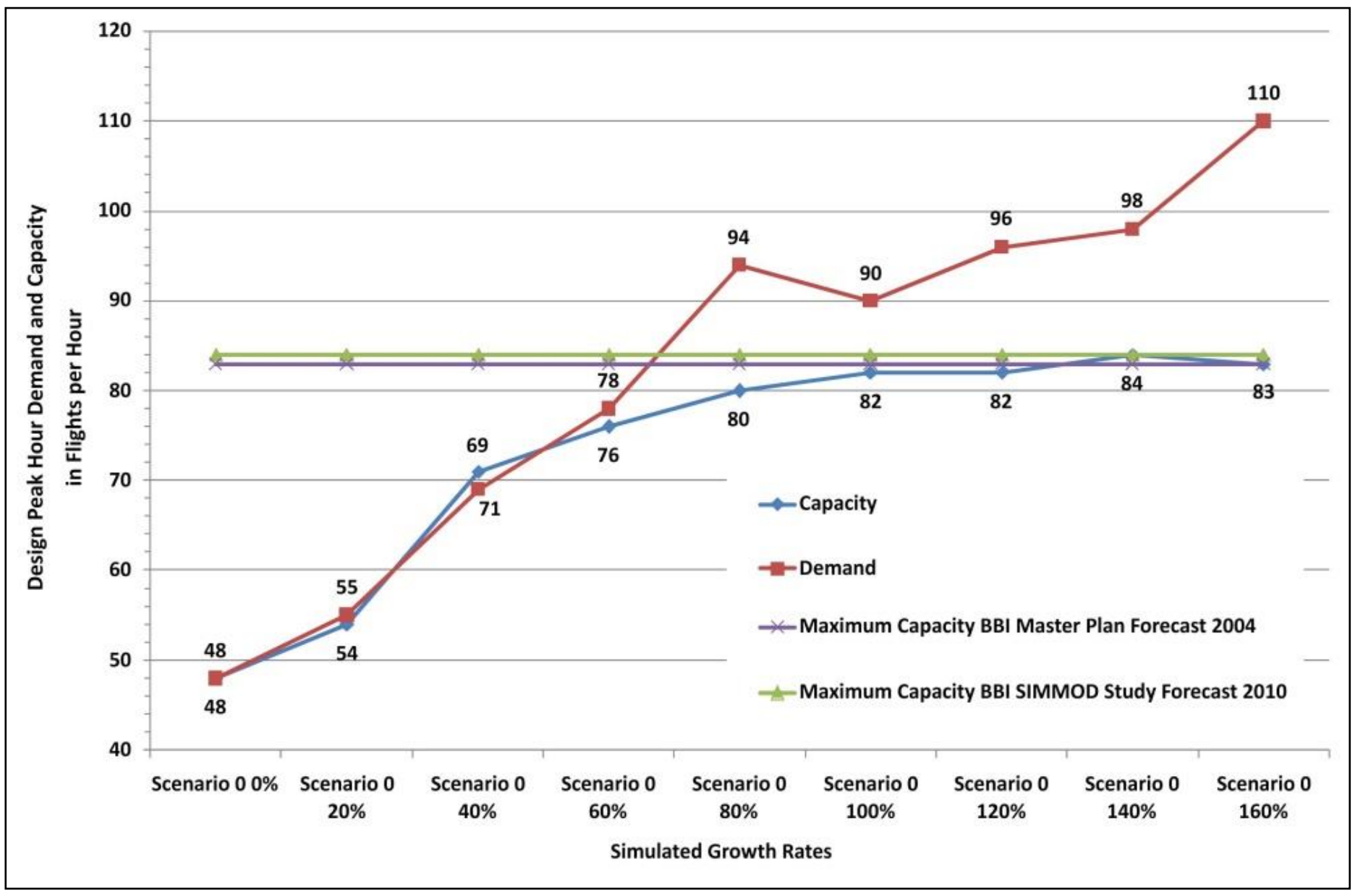

**Fig. 67.** Development of peak hour capacity and demand (Scenario 0)

**Table 46.** Computational results from Scenario 0

| Growth rate (%) | Flights per day | Design peak hour demand | Design peak hour capacity | Capacity utilization (demand / maximum capacity) | Average delay per flight (minutes) | Daily delay (minutes) | Daily delay costs (thousand €)[a] | Cancellation of flights[b] |
|---|---|---|---|---|---|---|---|---|
| -20 | 511 | 40 | 40 | 48% | 1.1 | 543 | 22.8 | 0 |
| 0 | 635 | 48 | 48 | 57% | 1.4 | 887 | 37.3 | 0 |
| 20 | 758 | 55 | 54 | 65% | 2.3 | 1,760 | 73.9 | 0 |
| 40 | 886 | 69 | 71 | 82% | 3.7 | 3,287 | 138.1 | 0 |
| 60 | 1,012 | 78 | 76 | 93% | 5.9 | 5,955 | 250.1 | 0 |
| 80 | 1,145 | 94 | 80 | 112% | 11.5 | 13,223 | 555.4 | 0 |
| 100 | 1,270 | 90 | 82 | 107% | 21.2 | 26,968 | 1,132.7 | 1 |
| 120 | 1,400 | 96 | 82 | 114% | 26.8 | 37,501 | 1,575.0 | 134 |
| 140 | 1,517 | 98 | 84 | 117% | 27.4 | 41,538 | 1,744.6 | 440 |
| 160 | 1,639 | 110 | 83 | 131% | 58.2 | 95,364 | 4,005.3 | 807 |

[a]Costs per minute of delay: € 42 approximated from EUROCONTROL (Standard Inputs for CBA Analysis)

[b]Cancellations resulting from longer than threshold waiting times and therefore conflicting with flight injections.

Depending on aircraft size the costs may vary between €3,400 and €75,000 per cancellation (EUROCONTROL 2009a, pp. 9). The direct relationship between increasing peak levels of demand and resulting delays and costs from congestion is presented in **Table 46**. In **Fig. 67** the development of capacity and demand over the incremental growth rates can be clearly followed.

While performing various simulation runs with incrementally increasing growth rates of daily traffic a capacity bottleneck could be observed in the SIMMOD results. As demand increases beyond a threshold of around 76 flights in the peak hour and at a growth rate of 60% above the base year traffic level (in Scenario 0). As the number of flights waiting to land or to takeoff continually increases further than this threshold, peak capacity is clearly exceeded (**Fig. 67**). In reality (and in the simulation processes) this divergence of demand and capacity is solved because of the fact that arrivals wait in the holding airspace and in the departure queue near the runway entries (or remain at the aircraft parking positions when the slots could not be secured) (de Neufville and Odoni 2003, p. 17), the waiting time and length of queues increase.

A runway can only be occupied by one aircraft at a time, which ultimately sets the capacity to the service time, consisting of runway occupancy time (ROT) and a safety time buffer, depending on aircraft weight and type of operation (see Section 8.1). For instance, the maximum peak hour capacity of 76 flights per hour corresponds to a service time of 1/76 flights per hour, which equals about 47 seconds per flight, whereas the 78 flights during the peak hour request service every 46 seconds. Another measure of congestion is the ratio of demand and capacity, the capacity utilization. In the presented example the demand of 78 flights per hour is divided by the practical capacity of 76 flights per hour, resulting in a capacity utilization of 103%. Alternatively, the ultimate capacity could be taken as the denominator, so in this case the capacity utilization equals 93% (78 divided by 84 flights per hour) (Bubalo 2011). Therefore, peak hour demand cannot be served with current peak hour capacity beyond a 60% growth rate) (**Table 46** and **Fig. 67**).

By defining a LOS (e.g. below six minutes of average delay per flight) and by balancing the capacities of the various airport processes (e.g. aircraft and ground handling, ATC capabilities and flight restrictions) each airport has a specific operational limit (per hour)—this is the practical capacity (IATA 2011, p. 17]).

This practical capacity can be used as declared capacity and then translated into available landing rights for a determined number of scheduled flights per hour at an airport over the course of the day. These are defined as airport slots (de Neufville and Odoni, p. 373; EU Commission 2007, p. 7; IATA 2011, pp. 11; Theiss 2008). Airport slots are coordinated by appointed national slot coordinators in bi-annual Schedules Conferences (SC) organized by the International Air Transport Association (IATA) (IATA 2011).

When demand is higher than available hourly slots, airlines typically try to adjust their schedules, routes and frequencies to find desired slots at some earlier or later time of the day. So eventually at congested and slot-coordinated airports demand is capped during peak hours and additional demand could fill the idle capacity periods in off-peak hours. If such shifts are not possible, these flights-in-demand will have to be rejected.

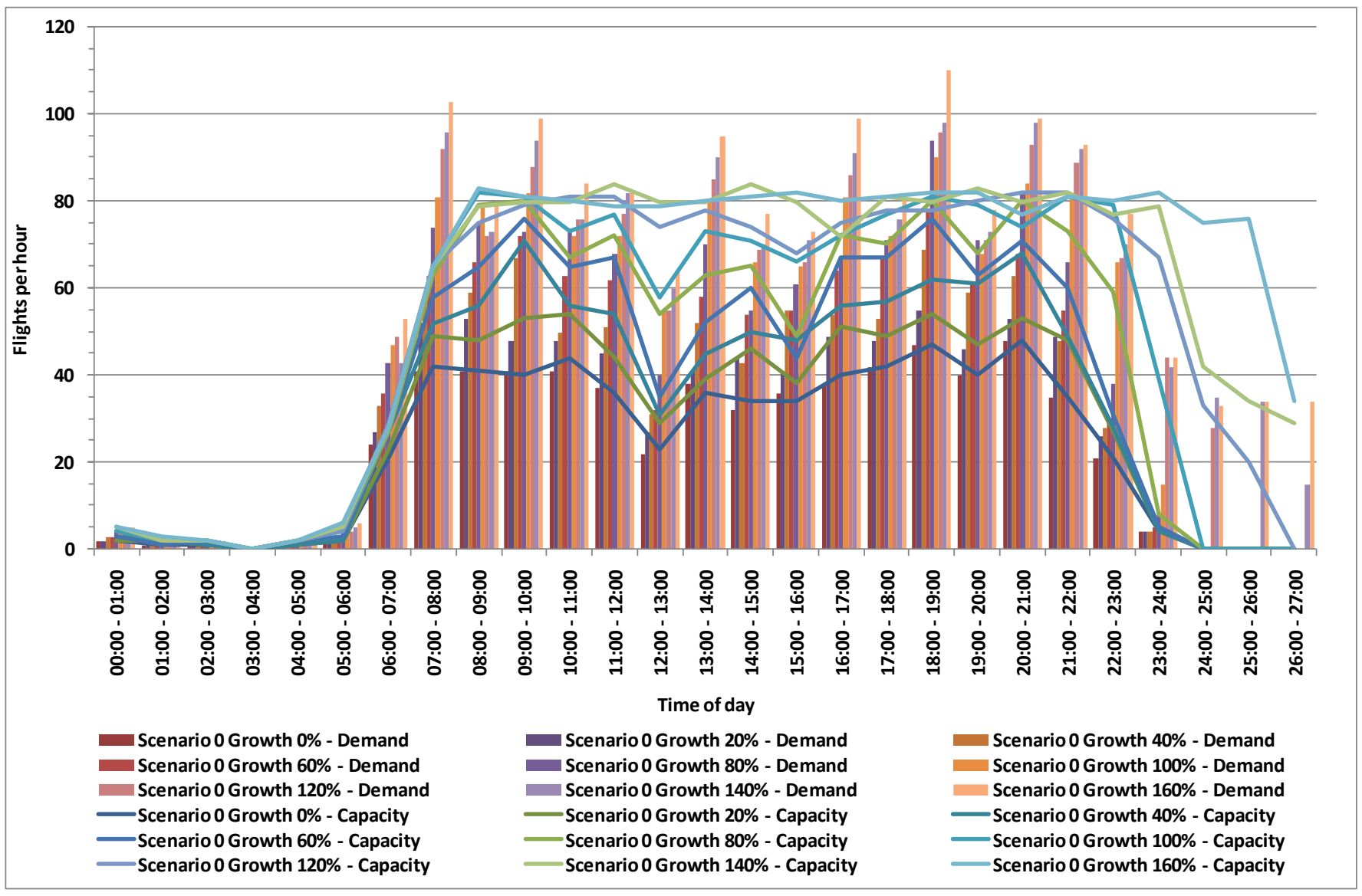

**Fig. 68.** Daily pattern of capacity and demand

In **Fig. 68** this effect is demonstrated clearly by showing the unconstrained (theoretical) demand at BER airport expressed in hourly requested movements and the constrained supply expressed in maximum hourly throughput. During the peak hour from 18:00 to 19:00 in Scenario 0 the demand grows continuously from 48 to 110 requested flights per hour compared to the peak capacity of 48 to 84 flights per hour. At growth rates beyond 80% delayed demand is shifted into the night hours after 23:00.

A comparison of the results of all analyzed scenarios show trends strongly growing exponential functions for the daily number of flights and the computed delays, which demonstrate a significant correlation (de Neufville and Odoni 2003, p. 449; Bäuerle, Engelhardt-Funke and Kolonko 2007; Horonjeff et al. 2010, p. 488). For all scenarios, the clearly recognizable sharp increase of average delay starts beyond 1,000 daily operations (**Fig. 69**). With the help of the simulation runs, it is shown that delays mainly occur at three bottlenecks. These are on the one hand (for arrivals) the entry into the airspace holding stack, and on the other hand (for departures) the departures queue of the runway or else directly at the gate (Atkin et al. 2010). The very close correlation of the functions below 1,000 daily flights proves that the aircraft mixes within the given range used in the scenarios do not have a significant impact on congestion (under current conditions and rules). In Scenario II and IV with an increasing number of category L and H flights (both scenarios have a MI of 125%), in the range of 1,000 to 1,300 daily flights at around 80% growth, higher average delays per flight can be seen more clearly than in other scenarios (**Fig. 69**).

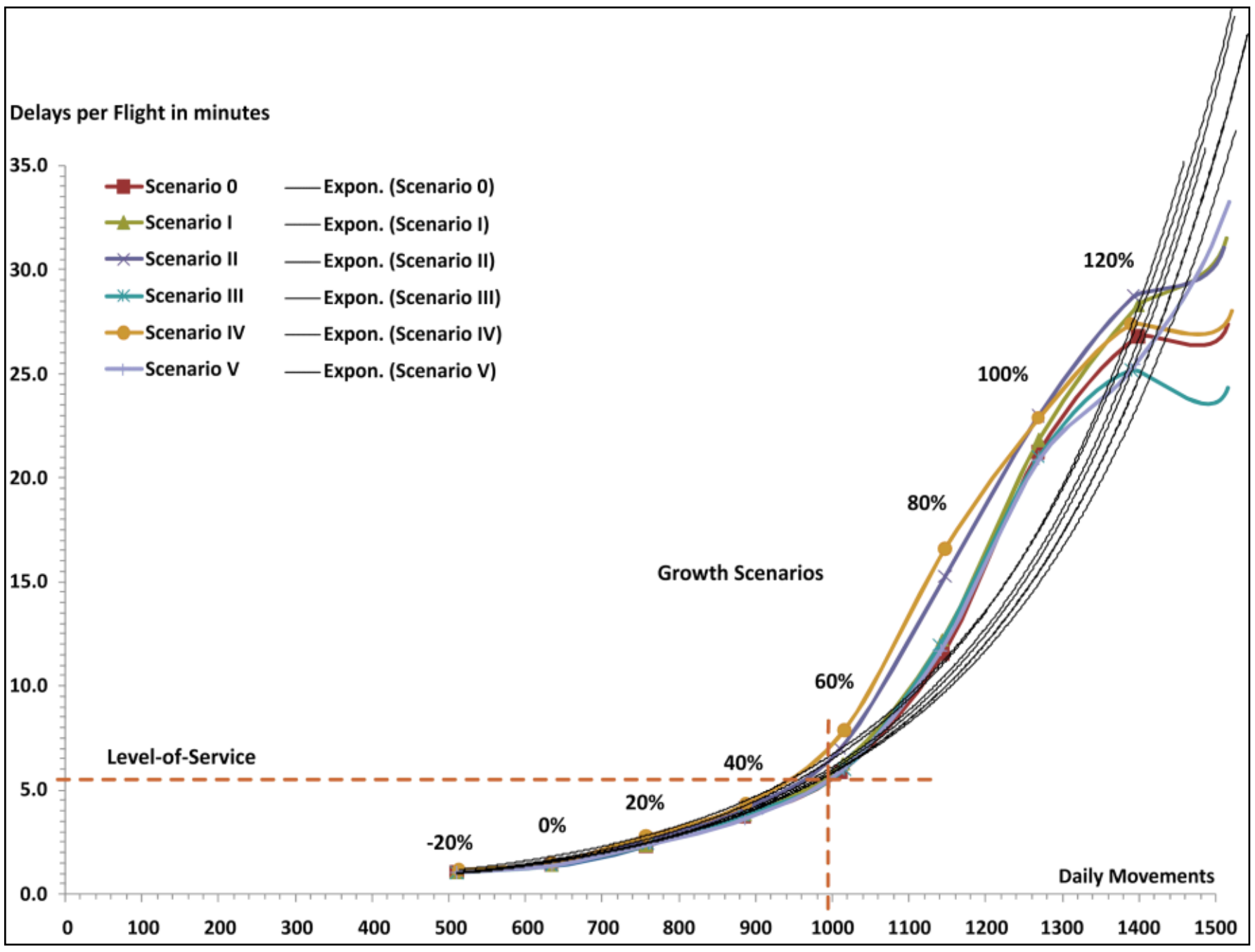

**Fig. 69.** Comparison of delay extends and number of flights

Indeed, the predominant M aircraft class is the reason for a reduction in variation of wake-vortex related separation minima. Consequently, the variation from the mean separation increases with the variation of the percentage of the M aircraft category in the traffic mix (**Fig. 65** and **Table 45**). Therefore, the current aircraft mix at both Berlin airports in Scenario 0, with a remarkably high share of category M aircraft and carrying approximately 100 passengers per flight, is beneficial regarding airport capacity and LOS.

## 8.3 Analysis and evaluation

The results for Scenario 0 (starting with the demand development of 2010) have shown that a cumulative growth above 60% from the baseline demand results in mid- and long-term capacity shortages (**Fig. 70**). The time when these bottlenecks occur depends on (assumed) annual growth rates. In our case study we assume a range of average annual growth rates between 3% and 6%. Therefore, we expect the practical capacity to be reached by as early as 2018, but in any case, by 2026 at the latest. A growth rate of 6% can be seen as realistic, since the Berlin airports (TXL, SXF, and THF) experienced a traffic growth of 36% between 2003 and 2008. This corresponds to an annual increase in traffic of 5.3%, but without accounting for the additional growth effects which will emerge from the future role of BER airport as an international hub.

With regard to the future development and the market position of the BER airport, this study reveals serious planning failures by the authorities involved. The possible risk of the occurrence of capacity bottlenecks six to eight years following the opening of the airport in 2012 raises some questions. An essential aspect of this is the underestimation of the future demand and economical attractiveness of the Berlin-Brandenburg region and its main airport (Heymann 2008), as it is stated in the official forecasts and master planning documents. The traffic increase at BER airport can be even more critical, if the airport will be used as a hub for flights to and from Scandinavia and Eastern Europe and /or if airlines (e.g. Emirates airline) or an airline alliance (e.g. oneworld) decide to use BER as their main German base (MSWV 2004, pp. 343). If this will happen the number of transfer passengers at BER airport is going to increase strongly.

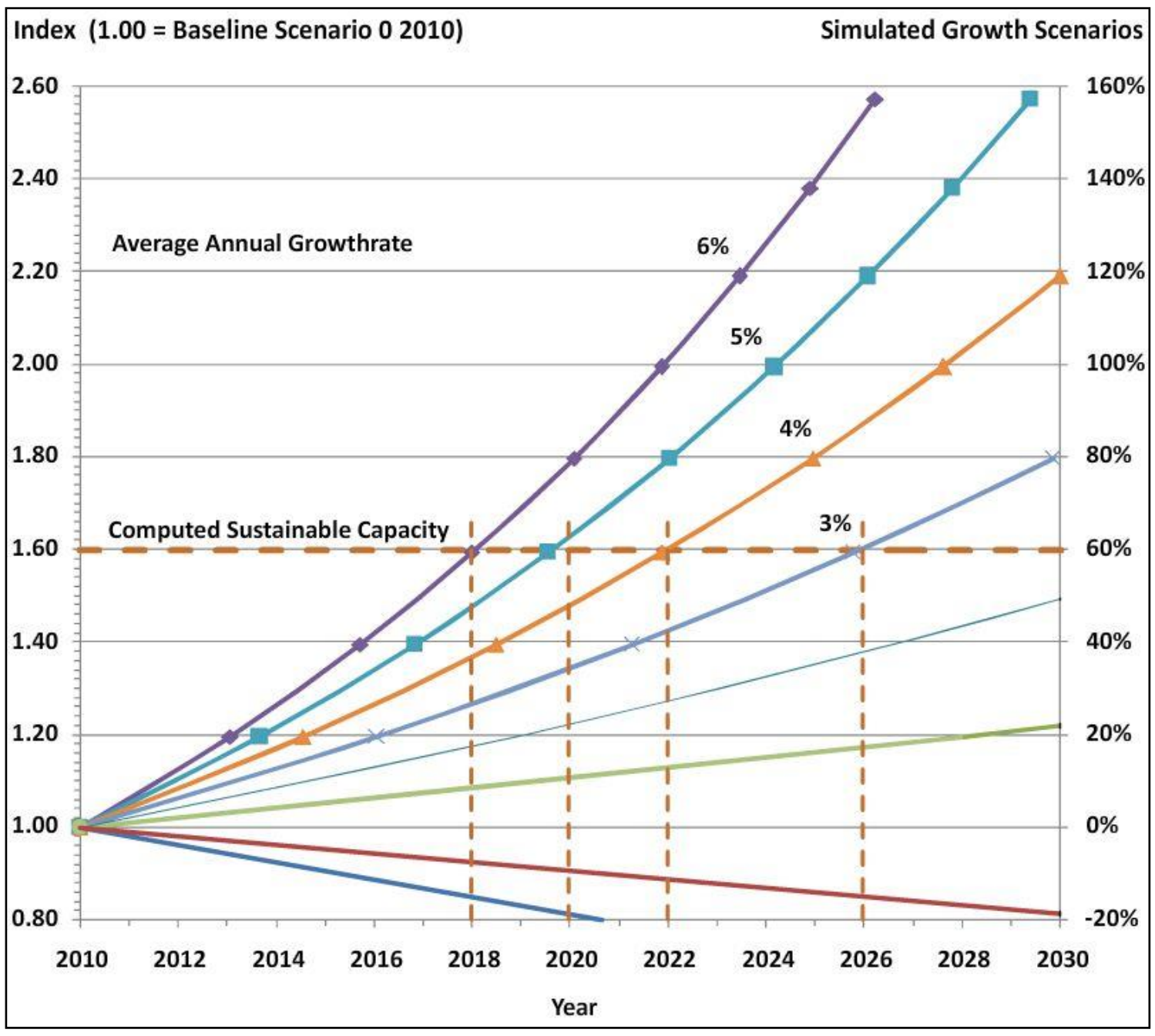

**Fig. 70.** Influences of different (annual) growth rates on forecasts

Certain similarities can be observed when comparing the developments presented for BER airport with other already highly utilized European airports also operating a far parallel runway configuration. Particularly the airports LHR and MUC will be used as benchmarks, which will be briefly explained below.

- LHR: Operating an independent far parallel runway system in segregated mode (due to federal regulations, which provide respite time from aircraft noise and pollution during daytime for the airport community). LHR is able to use both runways in mixed mode only in rare exceptions to ease morning congestion under peak demand (Atkin et al. 2010; BAA 2009, p. 16). In principal at this mode of operation the same ultimate capacity can be identified as calculated for the BER airport. The actual peak throughput of up to 100 flights per hour and 1,550 daily flights can only be achieved at the expense of LOS and congestion delays (Bubalo 2011). Over the year LHR airport operates on average 90 flights (45 departures and 45 arrivals) per hour and approximately 1,300 daily flights, allowing a minimum LOS of 10 minutes delay per flight (NATS 2007a, pp. 5). Operating the airport at such a low LOS, the system is extremely sensitive to operational disturbances (de Neufville and Odoni, p. 448; NATS 2007b, pp. 7).

- MUC: Operating as a staggered independent far parallel runway system in mixed mode (during peak banks of connecting arrivals and departures of airline hub operations). On average MUC has an hourly throughput of 84 flights (42 arrivals and 42 departures) and approximately 1,100 daily flights. During peaks, MUC operates under mixed mode, where the airport can achieve and schedule 90 flights per hour (e.g. 60 arrivals and 30 departures per hour), under the premise that arrivals and departures are not equally distributed during these hours (see **Fig. 71**) (Gilbo 1993).

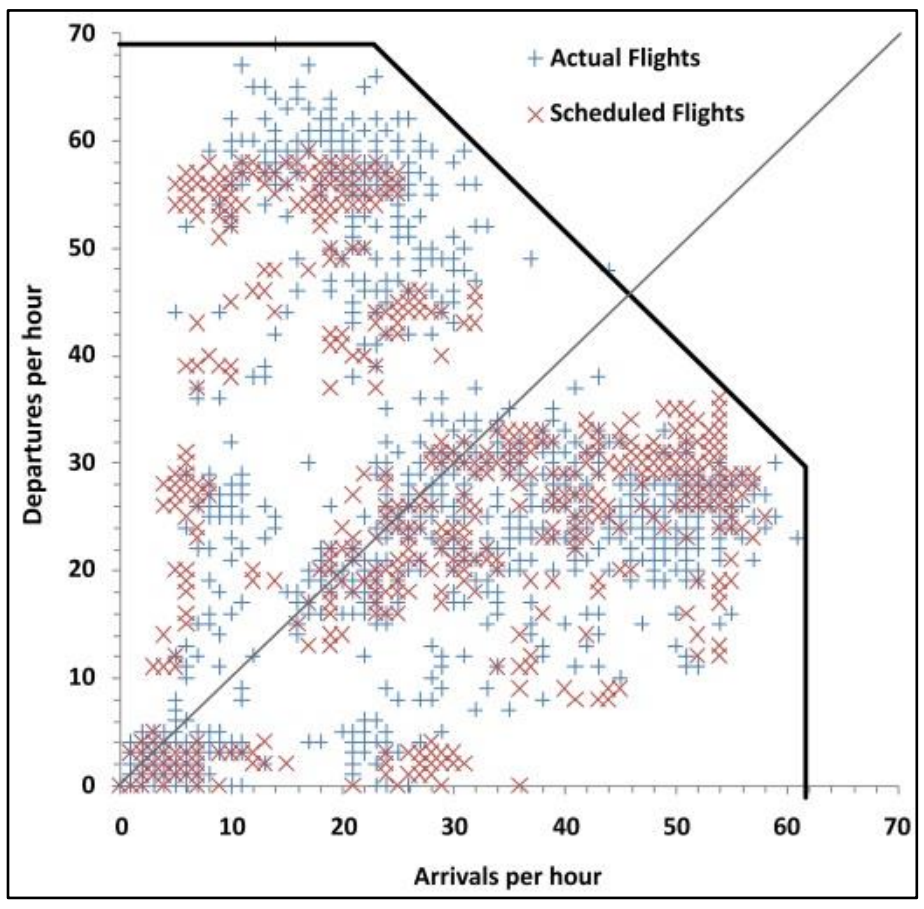

**Fig. 71.** Capacity envelope for traffic in April, May and June 2011 at MUC

During higher traffic density it is questionable if a mixed-mode operation of the runways would be sustainable over longer periods of time (de Neufville and Odoni 2003, pp. 394). This mode of operation has a direct impact on both: the ATC workload and the required surveillance equipment and overall safety, since departures have to be fitted into incoming arrival streams on the same runway. This reduces safety buffers between parallel and succeeding flights in the event of missed approaches, runway incursions or other unintended events (NATS 2007b). The airport system is becoming vulnerable with regard to small changes in capacity or demand, e.g. during changes in weather or by unscheduled (e.g. charter or general aviation) flights.

Furthermore, the surveillance of air and ground movements by ATC becomes much more complicated. On the ground, on the apron and the taxiways aircraft movements in opposite directions have to be coordinated by ground control, which leads to highly increased coordination complexity and staff requirements. But it should be noted that additional terminal facilities, necessary to accommodate the (long-term) desired volume of 30 million passengers at BER by 2023 (MSWV 2004, p. 384), will reduce precious apron area and will lead to substantial ‘taxiing inefficiencies’ (Caves and Gosling 1999, p. 96).

Compared to segregated mode, mixed-mode operations of the runways, spreads aircraft noise over a larger part of the countryside, due to the resulting simultaneous take-offs and landings from parallel runways (Janić 2007, pp. 14). This is one of the main reasons for not allowing mixed-mode operations at LHR, where the government ruled that the community has a right to respite from aircraft noise (Atkin, Burke, Greenwood and Reeson 2009). At LHR take-off and landing runways are consequently alternated daily at around 15:00 by an openly published scheme.

Segregated mode parallel runway operations at an airport can be seen as second-best choice with regard to capacity, since only one incoming and one outgoing flow is used, compared to mixed-mode operations, where each runway has one incoming and outgoing flow, thus four flows in total. Regarding improving safety, sustainability, and respite from noise the segregated mode, as it is modeled in this study, is arguably the most suitable mode of operation.

The comparison between LHR, MUC and BER shows that with a similar runway configuration, but under different modes of operation and with different demand patterns the same ultimate capacity can be expected. Although mixed mode operation is viewed as mandatory by planners of BER for successfully managing the airport (ARC 2010), we observe that LHR achieves a higher throughput in segregated mode than MUC in mixed mode. This leads to the conclusion that mixed mode is not a necessity for operating an airport. Note that during peak periods in the morning (between 6:00 and 7:00) mixed-mode operations (parallel approaches with Tactically Enhanced Arrival Measures [TEAM] in place) are allowed at LHR to ease congestion, thus the second runway is used for “overflow” capacity from the primary runway (de Neufville and Odoni 2003, p. 395). Permanent mixed mode operations at LHR would provide around 15% increase of airside capacity (NATS 2007b, p. 4).

Since the benchmark airports MUC and LHR are already operating at full capacity, expansion plans for both airports have begun. In 2007 MUC started a feasibility study regarding alternative locations for a third runway, which is expected to go into operation in 2020. LHR had similar expansion plans, but these resulted in the plan for a full length third runway north of the existing runway system being scrapped. This discussion implies that generally in the future it will certainly be difficult to expand runway capacity at congested airports in line with growing demand. Therefore, delays can be expected to increase in the future, since, while technical developments and further improvements in operational processes can temporarily prevent capacity bottlenecks for a few years, they cannot in principle by themselves solve the capacity problem.

## 8.4 Conclusions and outlook

To avoid the risk of capacity bottlenecks and congestion at BER (especially at the demand peaks), one must ask what threshold values are suggested. The simulation has shown that a LOS below six minutes of average delay, resulting in approximately 1,000 daily movements and 76 flights in the peak hour, must be seen as a practical limitation. At this stage, daily delays already cumulate to about 6,000 minutes and 250,000C delay costs. Lowering the LOS to ten minutes, as is the case at LHR, depending on the aircraft mix between 50 and 150 additional daily flights could be operated, but with the consequence of more than doubling delay times and delay costs.

From an engineering (and as a matter of fact from a purely economic) viewpoint, our results have shown, that the benefits of re-sequencing are minimal with regard to its effort and costs, compared to the economic advantages from expanding capacity by providing new infrastructure, particularly when it is already highly utilized. In the long run there is only so much the air traffic controllers and the airport management can do to minimize variation from the scheduled times and spacing between aircraft to increase capacity under growing demand. A sequencing of flights with similar speeds might prove to be efficient, especially in cases where these aircraft are waiting for take-off (or landing), which is of course more likely at times of high congestion and a diverse mix of aircraft types being present at the airport.

Re-sequencing leads to incurred delays on flights and those delay costs may outweigh the benefits of increasing short-run capacity. A sequencing by 'hierarchical order' from H to M to L with highest to least priority may be regarded as 'ideal', since the L aircraft category needs furthest separation, e.g. following an H aircraft, but also hinders following faster aircraft with its usually lower arrival and departure speed. This hierarchical order from H to L should be considered already during the scheduling of flights and slots, where it is advisable to schedule similar sized aircraft in a homogeneous sequence, i.e. H → H → M → M → L → L. In the future it is plausible that the separation minima could be reduced through the emergence of new aircraft surveillance technologies and the detection of wake-turbulences and their propagation. Therefore, a further increase in runway capacity is possible.

It must be noted that airport expansion projects not only have a technical and operational dimension, but even more so have a political and economic aspect concerning the proposed expansion location. Therefore, this predictable lack of capacity must be included in German national traffic planning and has to be put on the political agenda with high priority within a clearly short-term horizon. Otherwise opportunity costs will rise during periods in which runway capacity is unable to satisfy all demand (Röhl 2007). These costs refer not only to occurring bottlenecks in air traffic, but in this case also to the considerable negative impact on economic and political development. More specifically long-term unconstrained growth of BER as an international air traffic hub is important for a prosperous Berlin-Brandenburg region with the capital of the Federal Republic of Germany at its center.

When looking at the time-consuming and long-term procedures involved in planning and realizing large transportation infrastructure projects (not only in the Federal Republic of Germany), the discussion for a demand-oriented and timely expansion of BER beyond the current runway layout should already have been started. It is possible that we will witness this discussion soon after the opening of BER in June 2012 when the single airport has to serve the aggregated demand of currently two airports.

# 9 Simulating airside capacity enhancements at Oslo-Gardermoen airport post-2017[1]

*Measuring the effect on level of service of adding a pair of high-speed runway exits to the existing runways.*

Branko Bubalo

This report was commissioned by Avinor, based in Oslo, Norway in early 2013. Between October 2013 and December 2016, we conducted a simulation study for the 2050 Masterplan of Oslo-Gardermoen airport. The author would like to express appreciation to Mr. Kjell-Arne Sakshaug, who was project manager, to Mr. Svein Bratlie and Mr. Bjarne Aurstad from the airport planning department, and to Mr. Erik Gløersen, who provided airport maps and geographic coordinates. I also would like to thank Mr. Hans Ole Sagstuen for support and administrative support. Presented maps have been provided by Avinor.

## 9.1 Executive summary

This report was commissioned by Norwegian airport and air traffic control operator *Avinor* to solve the question, if adding one *Rapid Exit Taxiway* (RET) on each of the parallel runways and simultaneously reducing the final approach separation creates any measurable benefit to customers at Oslo-Gardermoen airport (OSL) with regard to overall *level of service* (LoS).

We used a computer simulation approach to model a full operational day under Cat. I weather conditions as a reference day or baseline scenario against which different scenarios are benchmarked. This „Without RETs“ scenario model reflects the situation at OSL in 2017, when ongoing construction projects, such as the first phase of terminal extension Pier-North, the extension of taxiway Whiskey and a set of remote parking stand at taxiway Uniform, will be completed. We test how the installed infrastructure will perform under increases in traffic.

We simulated *Mixed Parallel Operations (MPO)*, where landings and take-offs take place on both runways at the same time, and traffic flow in runway direction 01 (from South-West to North-East). In daily traffic management the generally used minimum separation between succeeding landings is 6NM. This distance is sufficient to safely fit departures into the arrival interval.

The main basis for our simulation model is the design peak day traffic schedule from June 14, 2013 with 776 flights. Furthermore, we include the current and planned airport and airspace layout and corresponding traffic rules. In this way we build a realistic model which, we think, behaves similar to the actual airport system.

---

[1] Final revised and authorized version, Berlin, December 7th 2016.

When the „Without RETs“ model is set up, we increase traffic to a level which we anticipate in 2017 and measure and record the average operational delay, which all aircraft in the model accumulate over the design peak day. After an initial simulation run of this baseline scenario, we observed a bottleneck at the taxiway junction taxiway Mike and November (see Figure 5), which led to opposite traffic flow on parts of taxiway Victor and Papa. Therefore, to reduce the chance of unexpected head-on conflicts on taxiways, it is recommended to investigate the proposed design of the access from the RET 01L to taxiway M and P south of A6.

In a second step we create a scenario called "With RETs" scenario which we compare against the initial „Without RETs" scenario. We activate and deactivate those links in our model, which reflect changes in the infrastructure in connection with implementing the RETs (i.e. opening the RETs and simultaneous closing of Exit A5). Speed limit on the RETs is 50 knots (92.6 km/h).

Between the scenarios the ground traffic rules are the same. However, when RETs are in place in the "With RETs" scenario, the minimum lateral staggering distance is reduced from 3NM to 2NM and the minimum in-trail separation between succeeding landings is reduced from 6NM to 5NM. A minimum in-trail separation of 5NM is an estimate based on the assumption that not all aircraft will be able to use the RETs. At the same time this separation distance gives Air Traffic Control (ATC) enough predictability to fit take-offs between succeeding landings.

In the simulation our quantitative measure of LoS is *aircraft delay*. For OSL airport and the purpose of this study we set a maximum acceptable delay (or minimum LoS) of 5 minutes per flight[1]. When average delay rises above this value, the airport is likely saturated during the peak hours. We want to reveal how much growth is possible at OSL until the predefined LoS cannot be sustained over a full operating day.

In both model scenarios we are interested at which traffic level and in which hour during the main operating period (between 06:00 and 24:00)[2] "average delay per flight" reaches or surpasses 5 minutes, and passenger and airline service quality is deteriorating. This information may be used in the scheduling process and in dialogs with airlines.

First, we ran the two models and recorded the output statistics given the 2017 traffic levels. Then we started increasing traffic in the model in increments of 10% above the initial level of 950 flights per operational day and repeated the measurement of delay. In this way we can make predictions, at which traffic level the airport becomes congested and when our predefined average LoS of 5 minutes delay per flight is reached.

[1] We measure LoS as the averaged delay per flight over the period from 00:00 to 24:00.

[2] During nighttime (between 00:00 and 06:00) traffic and delay is usually negligible. However, some delay is generated during nighttime, when several mail and/or cargo flights arrive at OSL within a short period of time and have simultaneous unloading and loading operations.

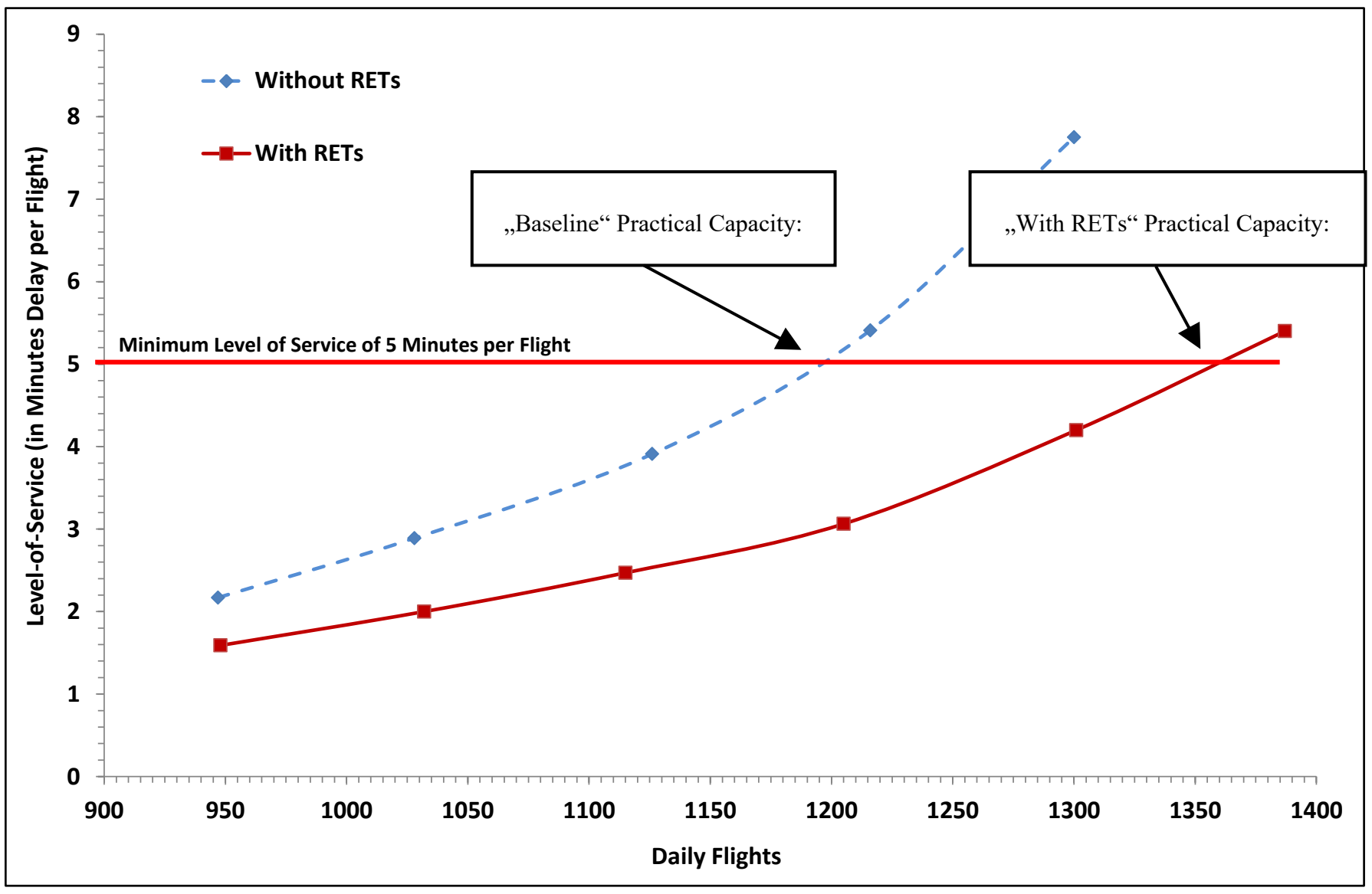

**Fig. 72.** Daily Total Flights and Level-of-Service Relationship

By reviewing and interpreting the simulation reports we recommend constructing the proposed RETs on each of the runways, 01L and 01R. We found a clear positive effect from activating the RETs. The figure above from chapter 9.4 shows exactly the expected pattern of the daily demand and LoS relationship.

While in the “Without RETs“ scenario 1,202 flights per day can be operated within the given limit, we calculated a practical capacity of 1,351 flights (assuming capacity constraints regarding parking stands and taxiways are solved) in the “With RETs” scenario. This means that about 150 additional daily flights can be operated.

We found that the airport benefits considerably from the addition of the RETs, because it allows a faster runway turnoff speed and a reduced approach separation. We measure a capacity benefit in the peak hour by up to 10% after the addition of the RETs. We measure this effect especially at a growth rate beyond 20% above the Baseline, when a demand of 1,000 flights per day is surpassed.

To make sure all bottlenecks and hotspots in the airside are known to the airport management we recommend conducting a simulation study for the opposite runway 19. During operations in runway direction 19, which usually occurs for a few hours daily, the aircraft flow in the airspace or while taxiing would look differently than in the presented model analyzing operations in 01 direction.

We further recommend carrying out a study on the effects of adding another set of RETs to the runways which are suitable for heavy or small aircraft in the fleet mix. We assume that the aircraft mix is likely to change at OSL in the future due to an increasing percentage of long-haul flights.

In conclusion, during the main operating hours between 6:00 and 24:00 a current daily practical capacity of 1,202 flights per day and 93 flights in the peak hour can be assumed for the „Without RETs” scenario.

After the construction of the RETs on runways 01L and 01R and after reducing the final approach separation in the "With RETs" scenario we can expect a practical capacity of 1,351 flights per day and 102 flights per hour, which means an additional capacity benefit of 12.4% per day and 9.7% in the peak hour when the RETs are activated compared to the 2017 baseline and base year.

At the maximum level of traffic additional airside infrastructure such as parking stand capacity will be required, and runway capacity becomes a limiting factor.

## 9.2 Introduction

This report covers a simulation study of the current and proposed post-2017 airfield and airspace design at Oslo-Gardermoen airport (OSL). In this report we outline the experimental design and explain the model inputs in Chapter 9.3, we deliver the results in Chapter 9.4, draw conclusions and deliver recommendations in Chapter 9.5.

In particular we measure the effect of installing high speed exits (here: rapid exit taxiways (RETs)) at both parallel runways and reducing approach separation on the same approach path with regard to average delay (Gosling, Kanafani and Hockaday 1981).

We use the airport and airspace simulation model SIMMOD as the main research tool for this study. The SIMMOD software environment was initially commissioned by the Federal Aviation Administration (FAA) in the U.S. to model the national airports and airspace and is now continuously maintained by the FAA and aviation analysis experts ATAC Corporation with the SimmodPLUS! software package.

In the last twenty years the SIMMOD framework has proven to deliver results in countless preceding studies (Bubalo and Daduna 2012)[1,2]. We believe that conducting an airport and airspace simulation study is essential whenever there is a need in answers regarding the capacity and demand relationship (NASA 1998; IATA 2004).

SIMMOD is essentially a queuing model based on a network of links and nodes, which represent the airport and its surrounding airspace. We observe the accumulation of delay caused by travelling agents in the system. Over the course of the modelled design peak day the agents, i.e. aircraft, travel through the network of interlinked nodes from points of origin to points of destination. We expect most delay to accumulate at the corresponding queues at nodes, where aircraft wait to be served. These nodes or "servers" are the runway departure queues, the gates, taxiway intersections or airspace routes.

---

[1] Los Angeles Airport Study from July 2012, LAX SPAS Report App F-2 Alt SIMMOD Simulation Final: http://www.lawa.org/

[2] Oakland Airport Masterplan Preparation Study from January 2006: http://www.oaklandairport.com/masterplan_oak/pdf/masterplan/march2006/chapters/appendix_I.pdf

We build sets of links and nodes, for all significant airfield areas at OSL airport. Each set requires certain priority rules to automatically resolve traffic conflicts while the simulation is running. The simplest priority rule is first-come-first-served (FCFS) for simple intersections, but larger areas, such as the taxiway Kilo and Lima cul-de-sacs, require complex rules to let certain aircraft pass an area first, while others have to wait.

The challenge in building such a sophisticated model as the one presented here is to mitigate all potential head-on conflicts with aircraft. Head-on conflicts that cannot be solved by the simulation automatically lead to “gridlock” situations, where aircraft cannot pass each other or an intersection thus they wait an infinite amount of time at a position and block subsequent traffic. Especially when traffic is increasing those conflict situations increase as well.

### Airport capacity and level of service

The main purpose of this simulation study is to define the (practical) airport capacity, i.e. the number of aircraft movements over time when the amount of average delay per flight reaches 5 minutes (see FAA 1983; de Neufville and Odoni 2003; Horonjeff, McKelvey, Sproule and Young 2010). In the case of hourly capacity, we average the delay in a particular hour over all flights in that hour, whereas in the case of daily capacity, we average the accumulated delay in a 24 hour period over all daily flights.

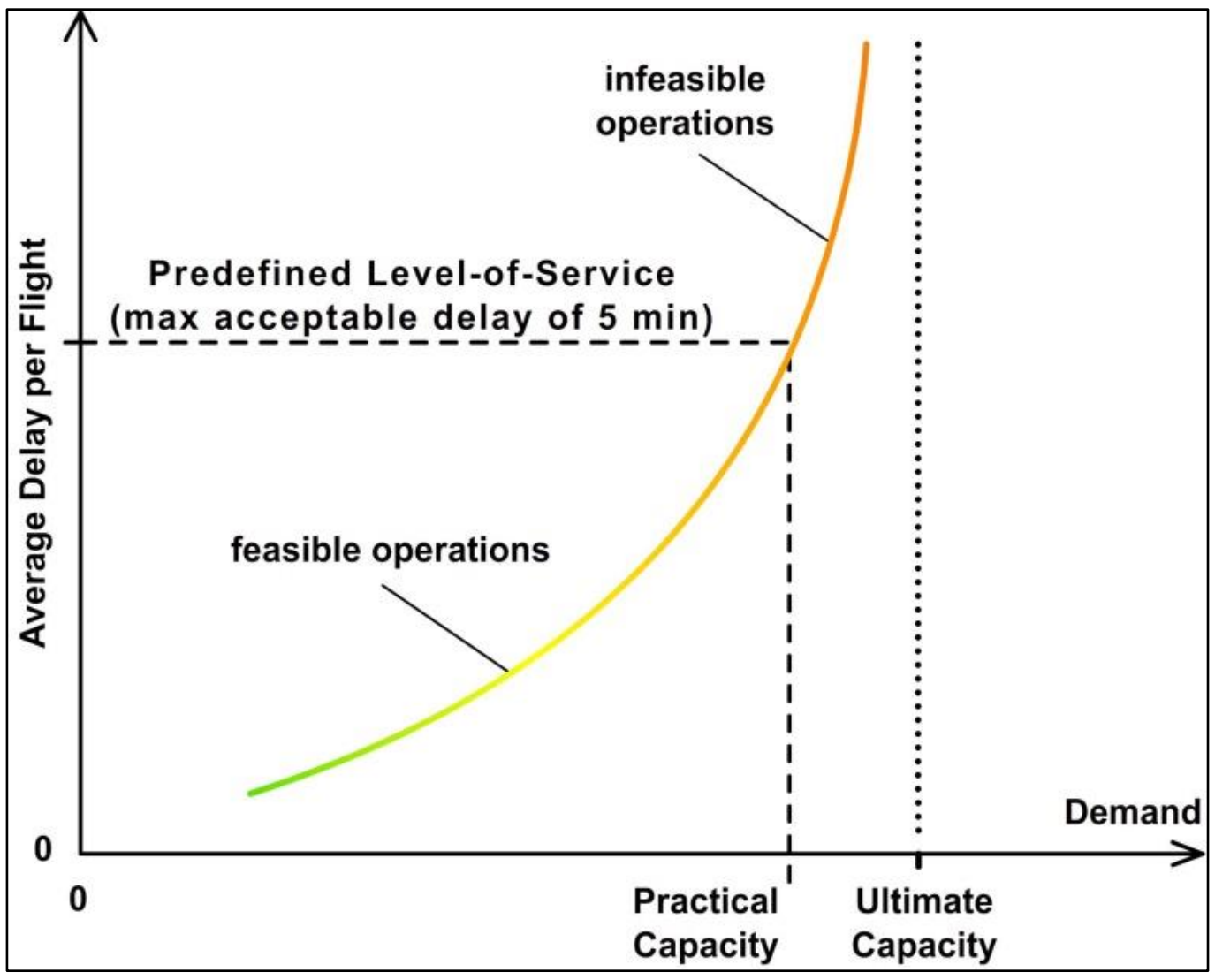

**Fig. 73.** Theoretical Relationship between Demand, Capacity and Delay (modified from Horonjeff et. al. 2010, p. 488)

The amount of delay becomes an expression for the LoS of the airport based on premises such as installed aircraft stands, taxiway layout and number of runways (NASA 1998).

**Fig. 73** shows the theoretical relationship between demand and average delay. As traffic increases and capacity is approaching average delay increases *non-linearly*. This means, for example, while demand increases 10% delay increases by 30%. We define the practical capacity of OSL airport at a traffic level, where LoS is on average 5 minutes delay per flight. We will look at the relation between number of aircraft movements and average delay in Chapter 9.4.

LoS is inversely related to that of average delay. This means, when a customer experiences increasing delay (because of queuing of an aircraft or other waiting times), LoS is decreasing, and *vice versa*. In this reasoning, we find the best LoS, when average delay is close to 0 and flights are operated exactly as scheduled, and the worst LoS, when average delay is beyond our 5 minutes threshold and is at some point approaching infinity (e.g. a flight has a very long delay and must eventually be cancelled).

According to **Fig. 73**, airport management is responsible that operations take place within the feasible limits of practical capacity when average LoS at OSL is 5 minutes delay per flight (and passenger) or better.

It should be pointed out that the measured delay is purely operational delay because of limited physical resources and traffic rules. In practice more sources of delay add to this level, such as weather related and propagating delay, human reaction times, technical failures and passenger behavior.

**Prerequisites**

Before an airport model can be designed, the modeler needs to have reliable documents available, which deliver good statistics or measurements for inclusion as a probability distribution into the model.

The presented simulation model is an enhancement to a pre-existing model which was designed by York Aviation from Manchester, U.K. We have crosschecked pre-existing data, for example with available data on runway threshold or parking position coordinates.

Airspace fixes have been checked by the author, modified or added, if needed. Countless traffic rules on taxiway junctions, at gates or regarding push-back procedures were added to the model. The existing model has been enhanced considerably, especially regarding a pre-2050 three parallel runways layout including several terminal gate capacity additions. The presented model is "ready" for simulating hypothetical future scenarios, from, e.g., the 2012 to 2050 OSL airport masterplan.[1]

[1] Oslo-Gardermoen 2012-2050 Masterplan report: https://avinor.no/globalassets/_oslo-lufthavn/om-oslo-lufthavn/om-oss/rapporter/masterplan_2012-2015.pdf

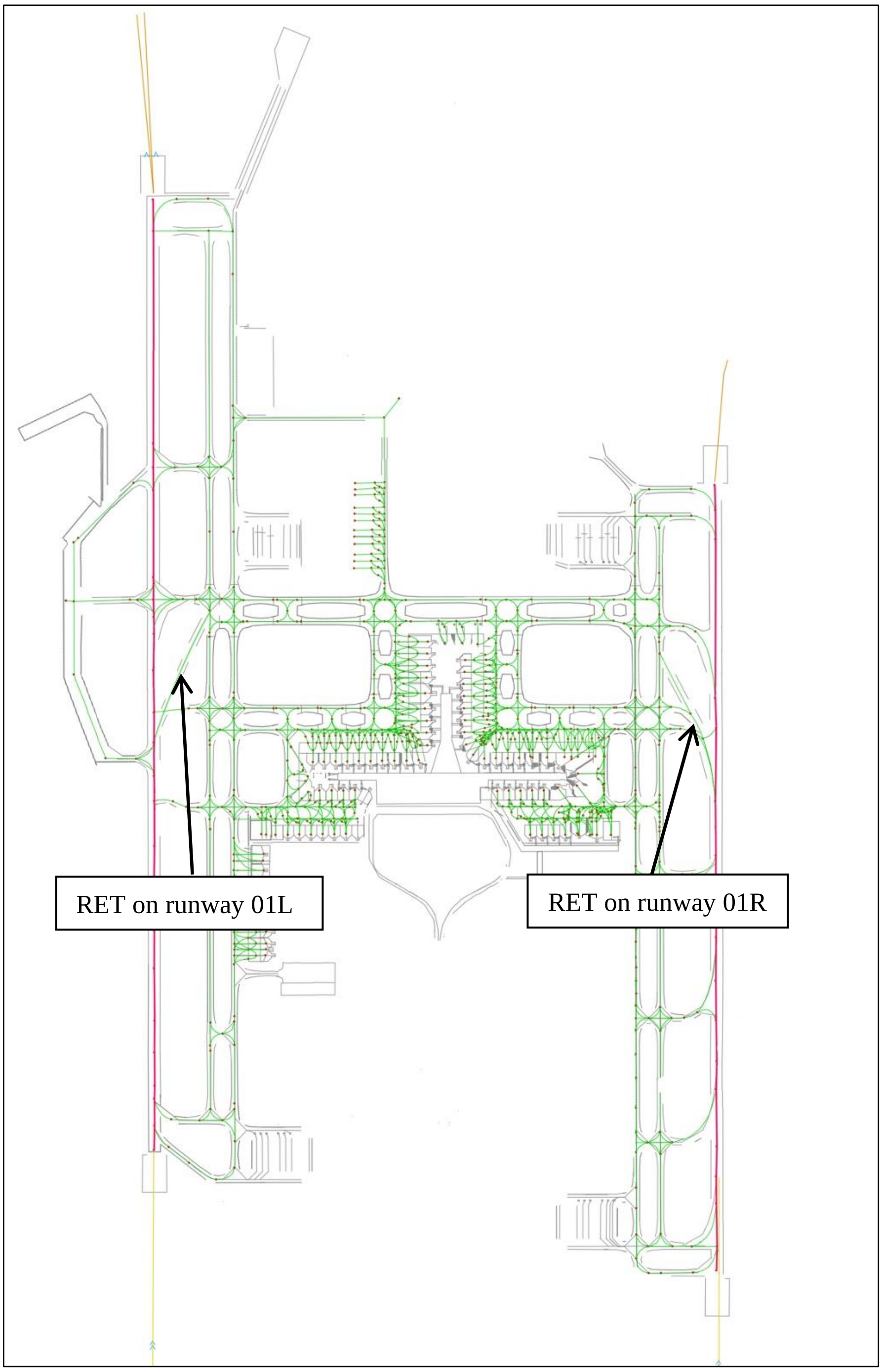

**Fig. 74.** SIMMOD model layout for OSL airport

For this report we have limited the simulation to the current parallel runway configuration of OSL (**Fig. 74**). However, we included additional taxiway and gate stand capacity into the model, representing several anticipated airport construction projects, which are scheduled to be finished by our base year 2017.

The main research question to be answered in this report is: "What is the capacity and level-of-service benefit for the airport customers, before and after a Rapid Exit Taxiway (RET) is added on each of the two parallel runways in the proposed locations?"

### 9.3 Model description

This study compares two model scenarios:

1. In our baseline scenario, named "Without RETs" scenario, we assume the available infrastructure, traffic levels and operational characteristics in 2017[1] and a minimum aircraft separation of 6 NM on final approach.

2. In the "With RETs" scenario we assume the same initial conditions compared to the baseline scenario, except we activate one Rapid Exit Taxiway (RET) for each runway in 01 direction (runway 01L and 01R), and we reduce the minimum final approach separation to 5 NM.

#### The "Without RETs" scenario

Compared to the airport today (as of 2016) we assume a slightly different airport layout in 2017. Some ongoing airport construction projects and airfield modifications will be finished by 2017. This change in infrastructure has been included into our model as well as proposed changes post-2017, such as the RETs. However, we activate only those airfield modifications which will be in place in our base year, and we leave later modifications deactivated. In this way we get a clear picture of which LoS we can expect without the addition of RETs (or other infrastructure).

Our simulation model is a predictive model which relies heavily on the quality of the input documentation. We use maps that show the traffic flow during 01 runway direction operations as an input into the SIMMOD model (**Fig. 75**). Both runways were used for both landings and take-offs (Mixed Parallel Operations - MPO). Traffic rules where input in a way as the author believed a human ground controller as part of ATC would have solved an aircraft movement problem.

[1] We created the traffic demand in 2017 based on a design peak day schedule from June 14th, 2013. We used changing growth rates over the day to "artificially" increase demand in the schedule. Flights are added stochastically during the simulation in SIMMOD with the "cloning function", this means a particular flight has a certain probability that a duplicate is injected into the simulation.

There are a few areas on the taxiway system of OSL, where one can observe "hotspots" and bottlenecks. At these areas and junctions there is a higher risk of traffic flow problems, incursions or accidents. When taxiing speed is low, such problems could create gridlocks and aircraft block each other on intersections and larger taxiway areas. We observed gridlocks frequently in the model that could likely occur under real peak traffic conditions.[1]

The taxiway L1 as part of the Lima taxiway on the east of the Terminal is an area were gridlocks occurred very often during the modelling of the growth scenarios. Aircraft that originate from or head towards a specific gate on the L1 or L2 taxiways need to pass a one lane taxiway segment between taxiway Tango and Gate 49. If additional push backs are conducted from Gate 188 and 187 into taxiway Foxtrot this could lead to major congestion or gridlocks in the "Lima area" (**Fig. 90**).

Potential conflict situations occur more often during the scenarios with a high level of traffic, such as the 40% growth above baseline scenarios.

Taxiing delay could occur more frequently at the following bottlenecks when demand rises:

a. Operations involving gates 185 to 188, because these gates serve mainly Heavy aircraft which require long push back paths into taxiway Foxtrot or Lima.
b. Operations involving gates 171, 171L and 172, because aircraft might interact with taxiing aircraft on a one lane segment on taxiway Kilo.
c. Operations at gate 50, 51 and 53, because aircraft are frequently pushed back into taxiway Foxtrot and blocking other traffic heading towards taxiway Lima.
d. Operations involving mail flights on the stands on taxiway Charlie, because departing aircraft need to cross runway 01L. Most of these flights are scheduled during the night, therefore pose delay just among the mail and cargo flights during this time of day.
e. Operations involving the proposed parking stands gate 707 to 712 on taxiway Uniform. The four stands planned for phase 1 are accessed from taxiway Victor on a one lane segment. Aircraft need to wait on the main taxiways Victor or Papa and may block other aircraft when aircraft are departing from taxiway Uniform. For phase 2 it is recommended to consider constructing a double lane taxiway perhaps extending taxiway Delta.

A human air traffic controller solves tactic and operational problems far more elegantly than any intelligence that could be included in our model. The challenge to define traffic rules in the model lies in the generalization of the rules so these fit to most possible conflict situations. This must be impossible without a highly developed *artificial intelligence* (AI).

[1] Towing operations, which would be necessary if gridlock situation would happen in reality, were neglected in our model. We tried to implement traffic prioritization rules to avoid gridlock situations as often as possible under different traffic conditions.

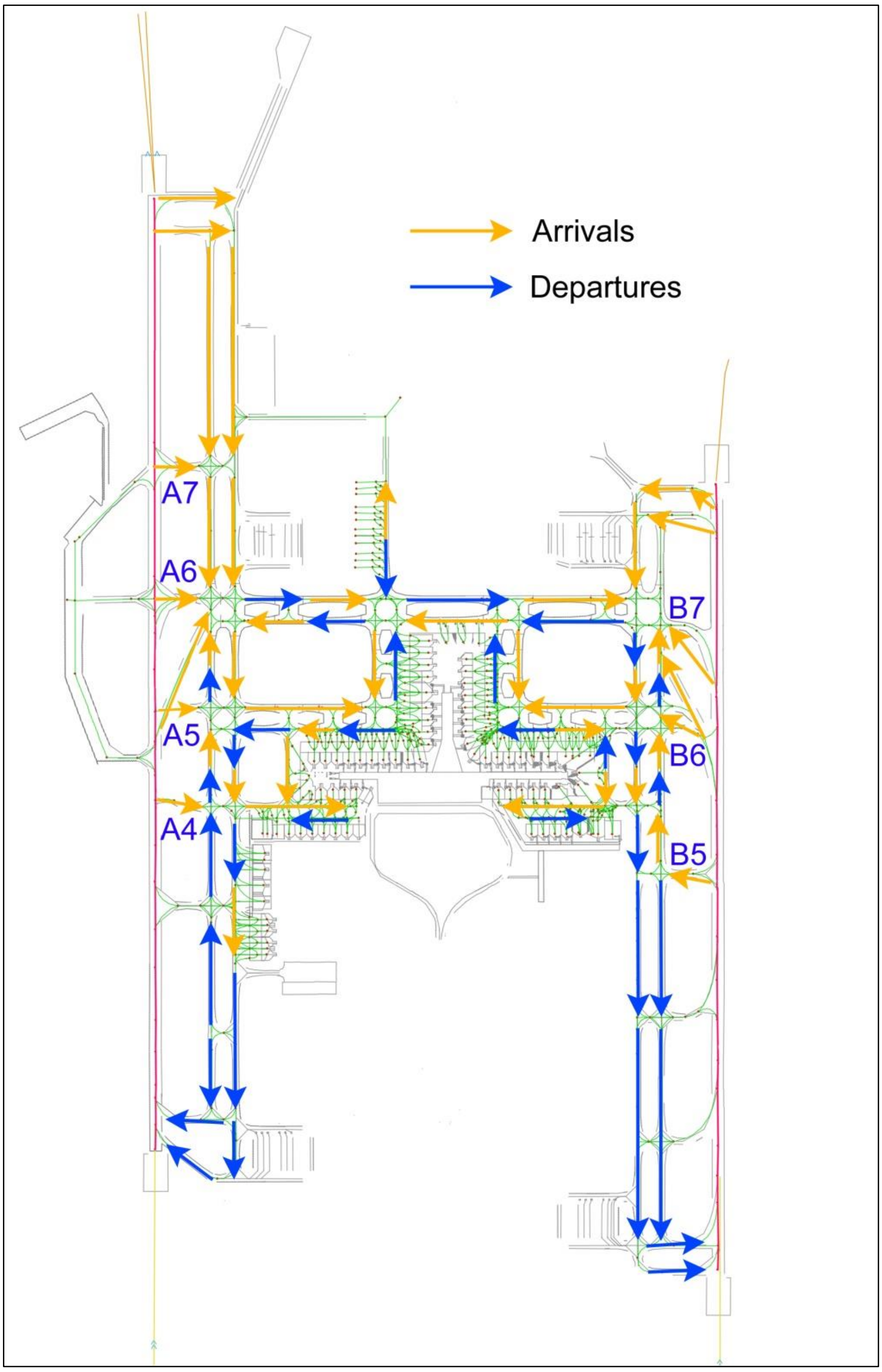

**Fig. 75.** Taxiing Traffic Flow in 01 Runway Direction

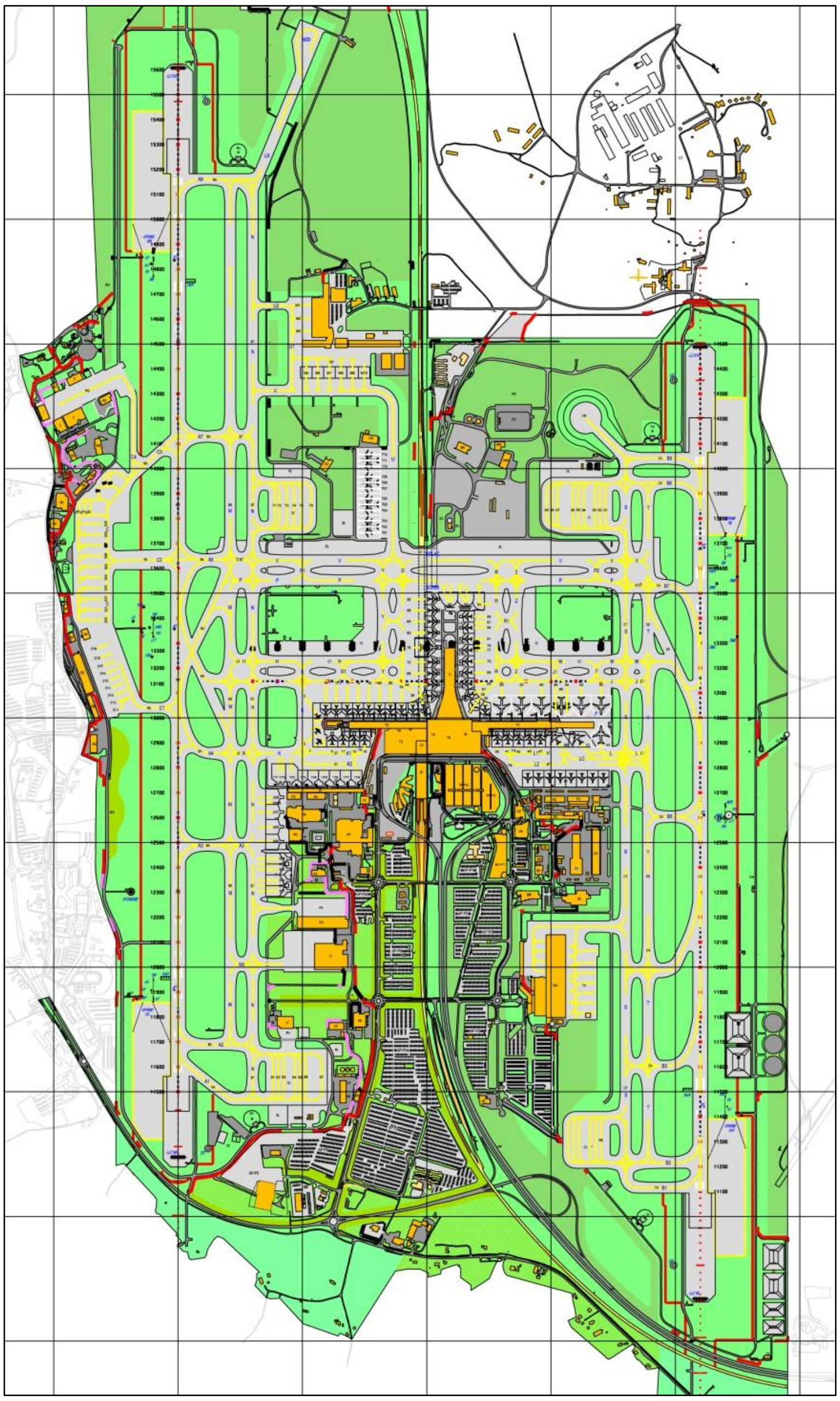

**Fig. 76.** Taxiway and Gate Map for OSL (Avinor)

Another main input for the model is the aircraft pushback manual[1]. In it we find sketches of different push back operations from the gates and parking positions. Most push back operations are exactly modeled as described in the push back manual. At some gates different push back procedures for different aircraft classes are implemented.

The map shown in **Fig. 76** shows the layout of the airport with exact terminal and gate locations. Each gate is limited by a maximum aircraft size class (e.g. class C, D or E) that is permitted to park at a specified gate or parking position. This information is included in the gate choice logic in the simulation model. The positions of the stop bars on the apron have been taken into consideration and build into the model. Every node in the simulation model could be a node in the system where aircraft wait for service, because other aircraft use a particular junction, are pushed back from a gate or taxi along a single lane taxiway.

### The "With RETs" scenario

**Fig. 77** and **Fig. 78** show the design of the proposed RETs. This design is carefully replicated in the "With RETs" scenario. The effect of the construction of the RETs is measured by activating the links and nodes representing the RETs in the model and by comparing the simulation outputs before and after activation.

**Fig. 77** shows the RET on runway 01L. This high speed runway exit allows a landing aircraft to exit the runway earlier and with a higher velocity than the current "Without RETs" situation. We see that the old exit A5 will be closed and replaced by the RET, which enters the main taxiway system at taxiway Victor. A preliminary simulation showed, that the design of this RET was leading to a congestion of opposite traffic in the junction with taxiway Papa. The early access to taxiway Mike was then closed in the simulation, and the flow of traffic was lead on to taxiway Victor. This appeared to be a better solution.

**Fig. 78** shows the proposed RET on runway 01R. In this case the old exit B6 will be closed and replaced by a high-speed runway exit. In **Fig. 78** it is highlighted that the extension of taxiway Whiskey to taxiway Tango will be necessary for aircraft to exit the runway over the proposed RET. The aircraft can then enter the taxiway system towards taxiways Whiskey, Sierra, and Tango.

On both proposed RETs we allow an exit speed of up to 50 knots, whereas on the main taxiways we only allow a maximum taxiing speed of between 15 to 20 knots. Near the gates we allow a taxiing speed of 5 knots or less.

[1] Internal document: "Push back manual for Oslo Lufthavn", original title: "Prosedyre for push-back ved Oslo lufthavn"

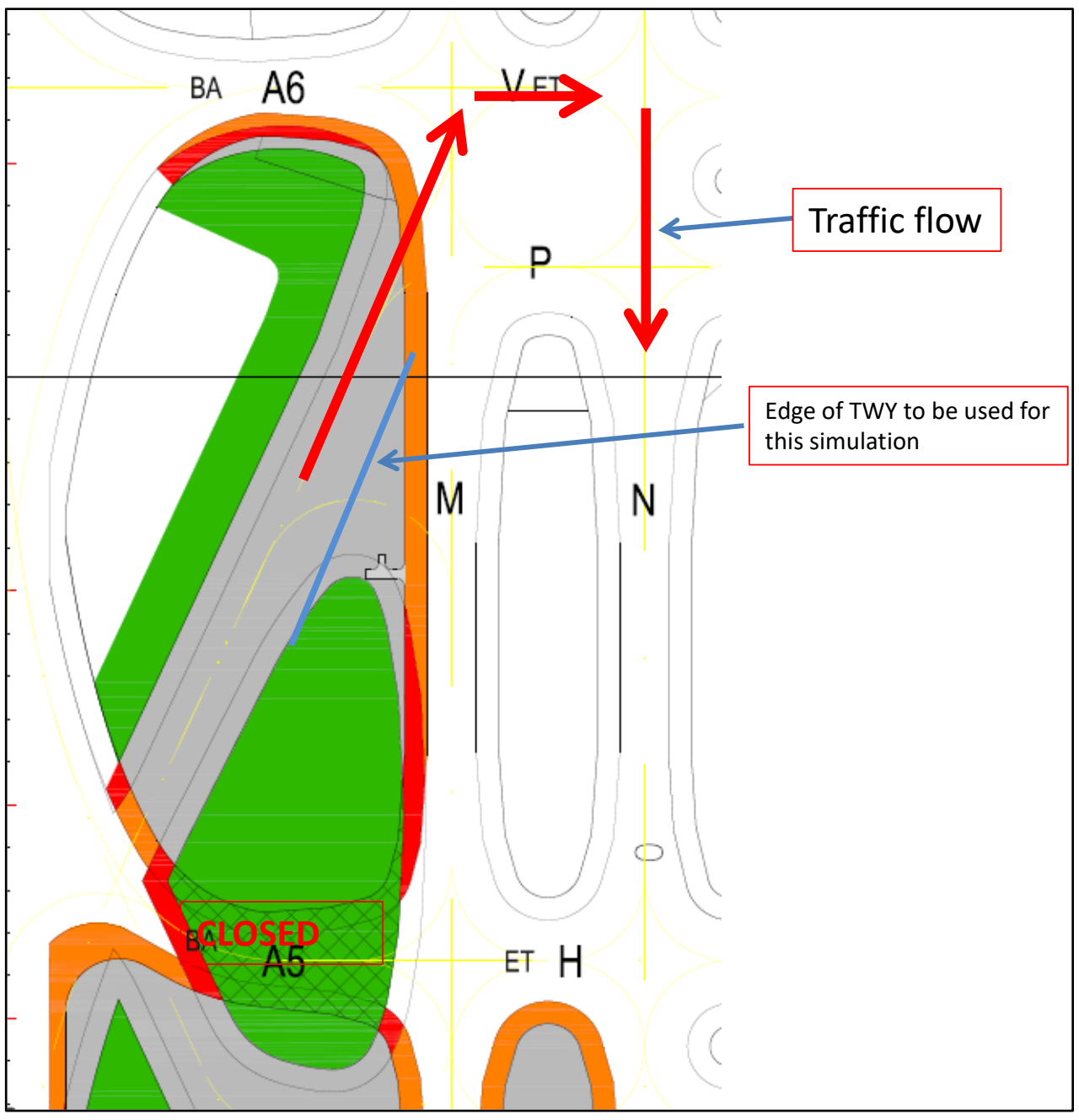

**Fig. 77.** RET on Runway 01L (Avinor)

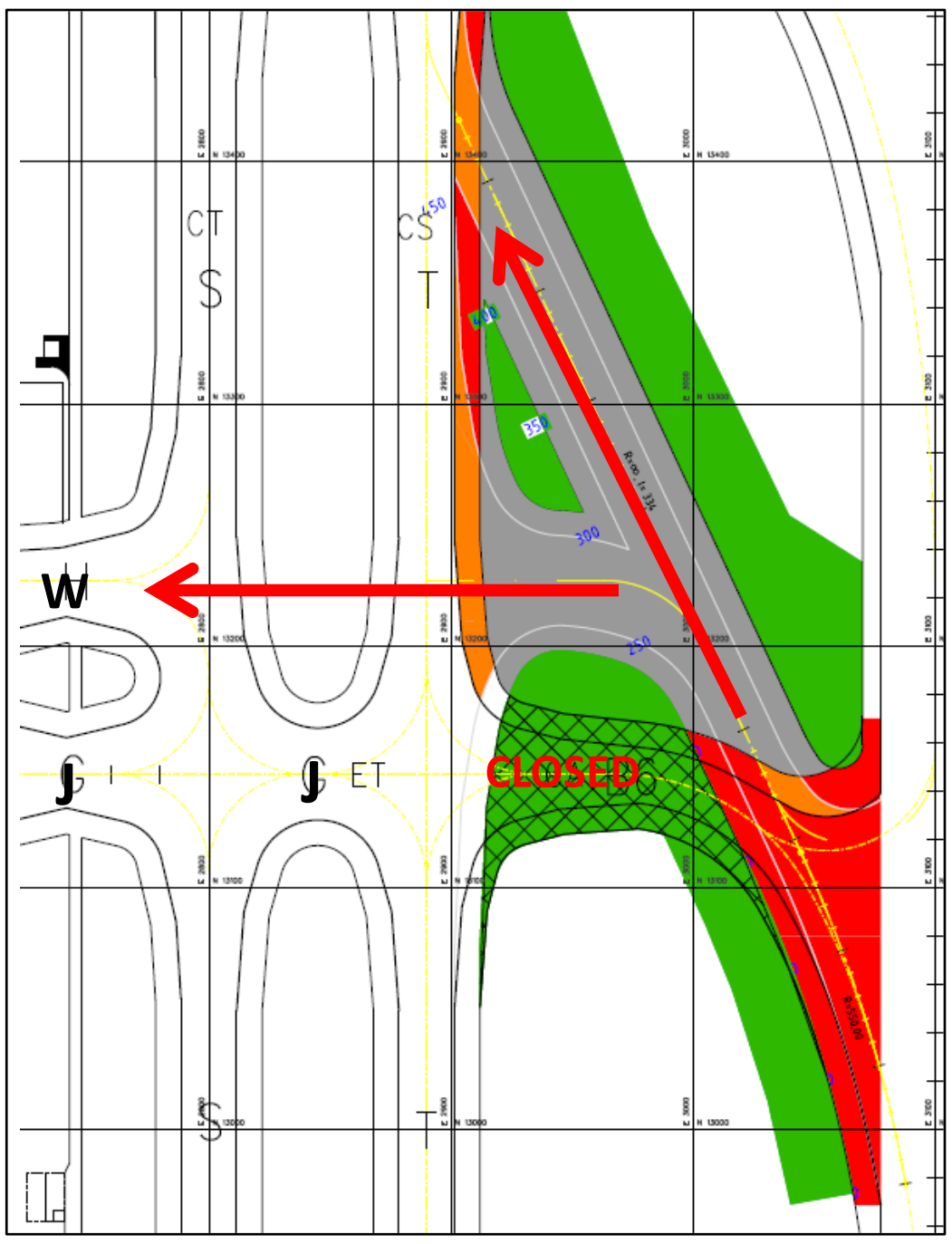

**Fig. 78.** RET on Runway 01R (Avinor)

**Probability distributions**

Another set of prerequisites are airport statistics, which we include into the simulation as probability distributions. These can be elementary in modelling an existing airport system. It is essential to have some probability distributions from observations or from the literature in the model so that it behaves realistically. The robustness of a simulation model depends mostly on the behavior of the system under random influences and on the ability of the model to solve conflict situations automatically. Each iteration starts with a set of "seed numbers". These seeds influence the generation of (pseudo-)random number in the model and values picked from the probability distributions. By using the same seeds throughout our experiments, we can make sure that the model and all iterations are replicable. In a purely random model this would not be possible.

In the "Without RETs“ model we have, for example, forced the landing aircraft to use the current exit distribution on the runways 01L and 01R. In the data from March 2014 we observe that 27% of flights landing on runway 01L use Exit A4, 55% use Exit A5 and 12% use Exit A6 (**Fig. 79**). On runway 01R we observe that 13% of flights landing on this runway use Exit B5, 68% use Exit B6 and 19% use Exit B7 (**Fig. 80**).

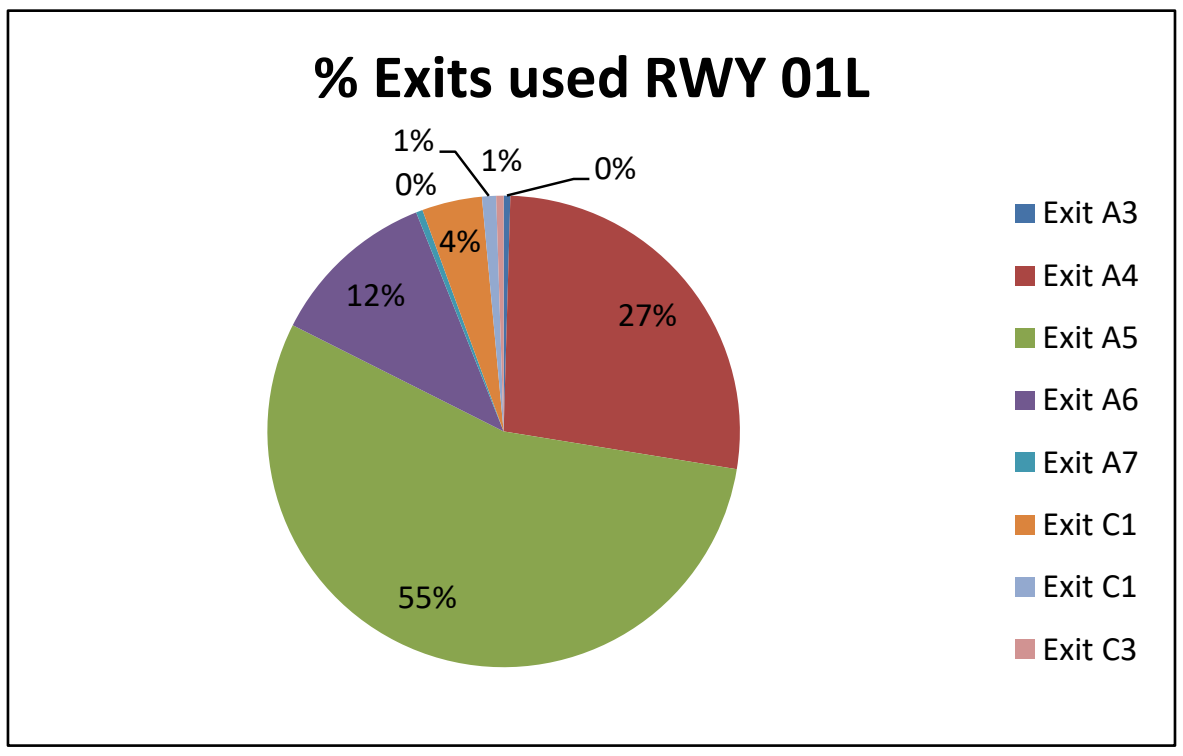

**Fig. 79.** Baseline exit usage on runway 01L (Avinor)

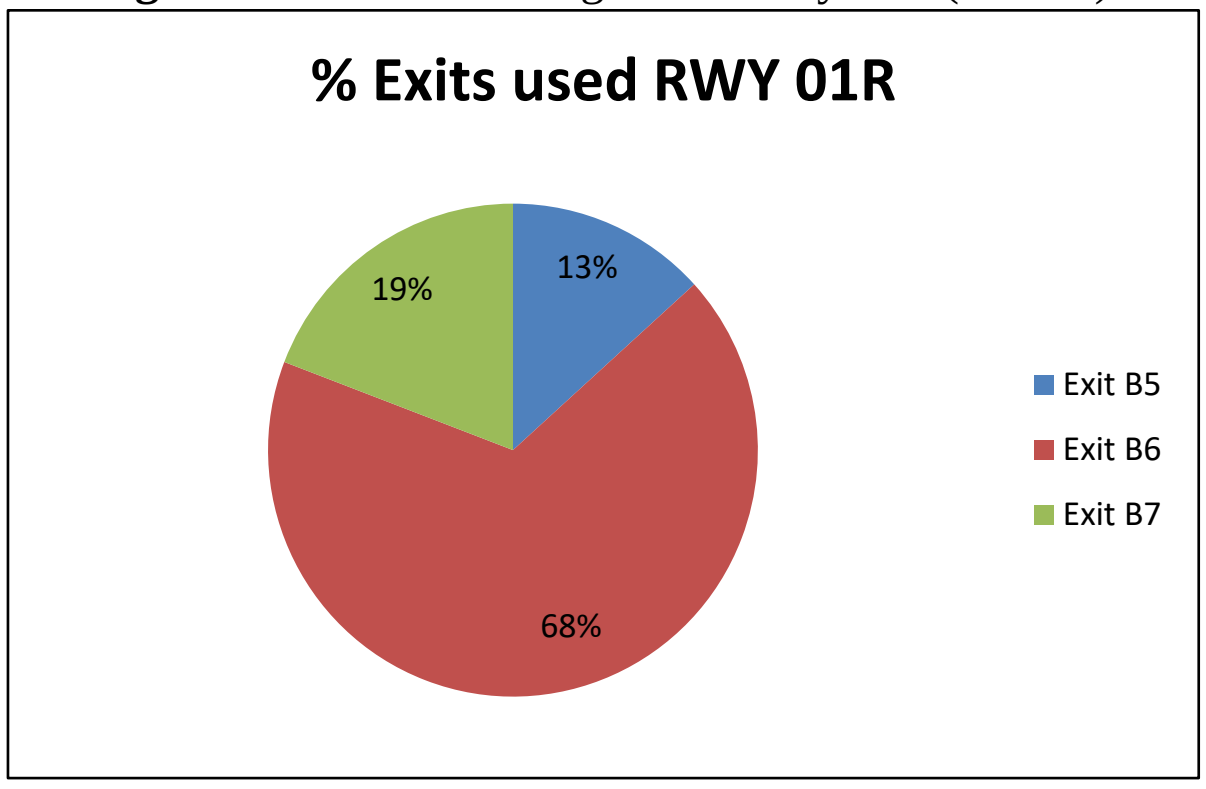

**Fig. 80.** Baseline exit usage on runway 01R (Avinor)

The constraint to let arrivals choose an exit based on the exit usage distribution has been removed in the "With RETs" scenario. In this scenario aircraft choose runway exits based on their landing roll distance distribution (which itself is a distribution calculated from observations made at OSL).

More than 80% of the landing flights use the proposed RETs to exit the runways (e.g. instead of using the closed exits A5 and B6, or a later exit such as A6 or B7) because of the higher permitted exit speed of up to 50 knots on the 30° angled RET (compared to 15 knots on regular 90° angled runway exits).

It would be interesting to study the effect of a different future aircraft mix and the need of another pair of RETs, which, for example, can be used by either more Small (i.e. because of more growth in regional traffic) or Heavy aircraft (i.e. because of more growth in long-haul, cargo or military traffic).

Our main inputs for building the airspace and airfield link network in the model are maps and charts from the airport. Additionally, satellite or aerial imagery could be used. Some information is publicly available such is the case with Aeronautical Information Publications (AIP). These publications reflect the layout of the airport and airspace, including Standard Instrument Departure (SID) and Standard Terminal Arrival (STAR) routes[1]. The main purpose of these documents is to inform pilots about a particular airport and the route to an airport, or about the ground or airspace procedures and structure. We have used this information to make our SIMMOD model more accurate.

**Fig. 74** shows that the presented simulation model consists at the core of a set of nodes which are connected by a network of links. In this way we can build a virtual model which represents OSL airport and its immediate airspace, including flight paths, single flight vectors and directionality, taxiways, gate positions and runways and its thresholds.

Furthermore, it is required to define traffic rules within the model for the agents travelling through this network. Each agent, i.e. aircraft, travels from an injection node to a destination node and may have to follow certain procedures to solve traffic conflicts along the way.

In the case of arrivals, flights are injected at the outermost airspace fixes of the terminal control area (TMA), move along the standard instrument arrival routes (STARs), then land on the runways and taxi to their gate positions. At the gates aircraft may park for a certain period until they are loaded for the next flight.

In our model most arrivals are linked to a corresponding departure to model full aircraft turnaround operations. We used an algorithm which compared aircraft characteristics such as airline and aircraft type. For each arrival the algorithm searched in the schedule for a matching departure in succeeding flights, meeting these characteristics and by allowing a minimum turnaround time window of 15 minutes. Before the algorithm can be executed the design day schedule must be sorted by actual gate on/off times, which served as our reference time.

[1] Airport and airspace charts and publications can be downloaded from the European AIS Database (EAD) developed by Eurocontrol: https://www.ead.eurocontrol.int/

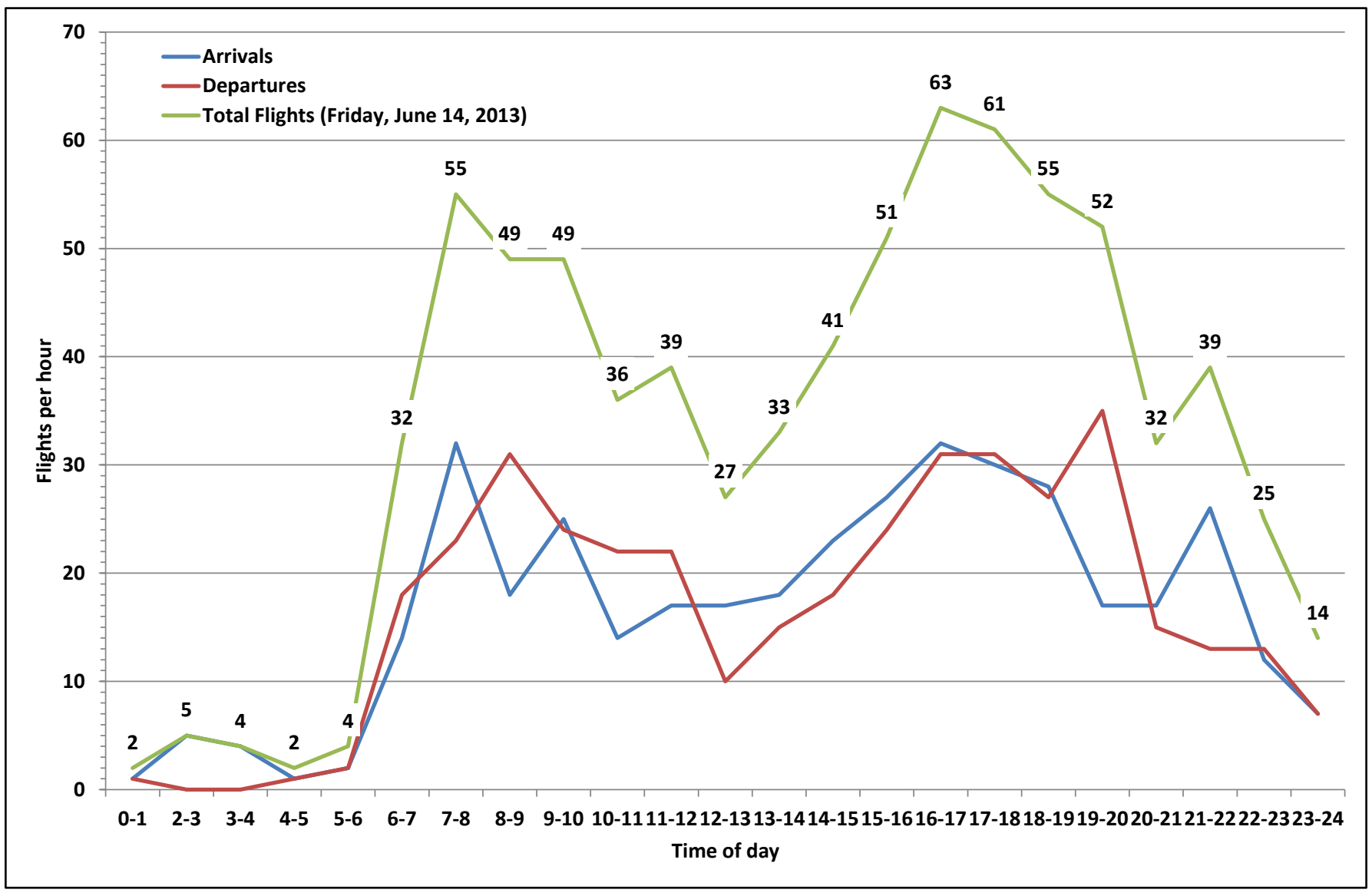

**Fig. 81.** Input Schedule from Design Peak Day June 14th 2013

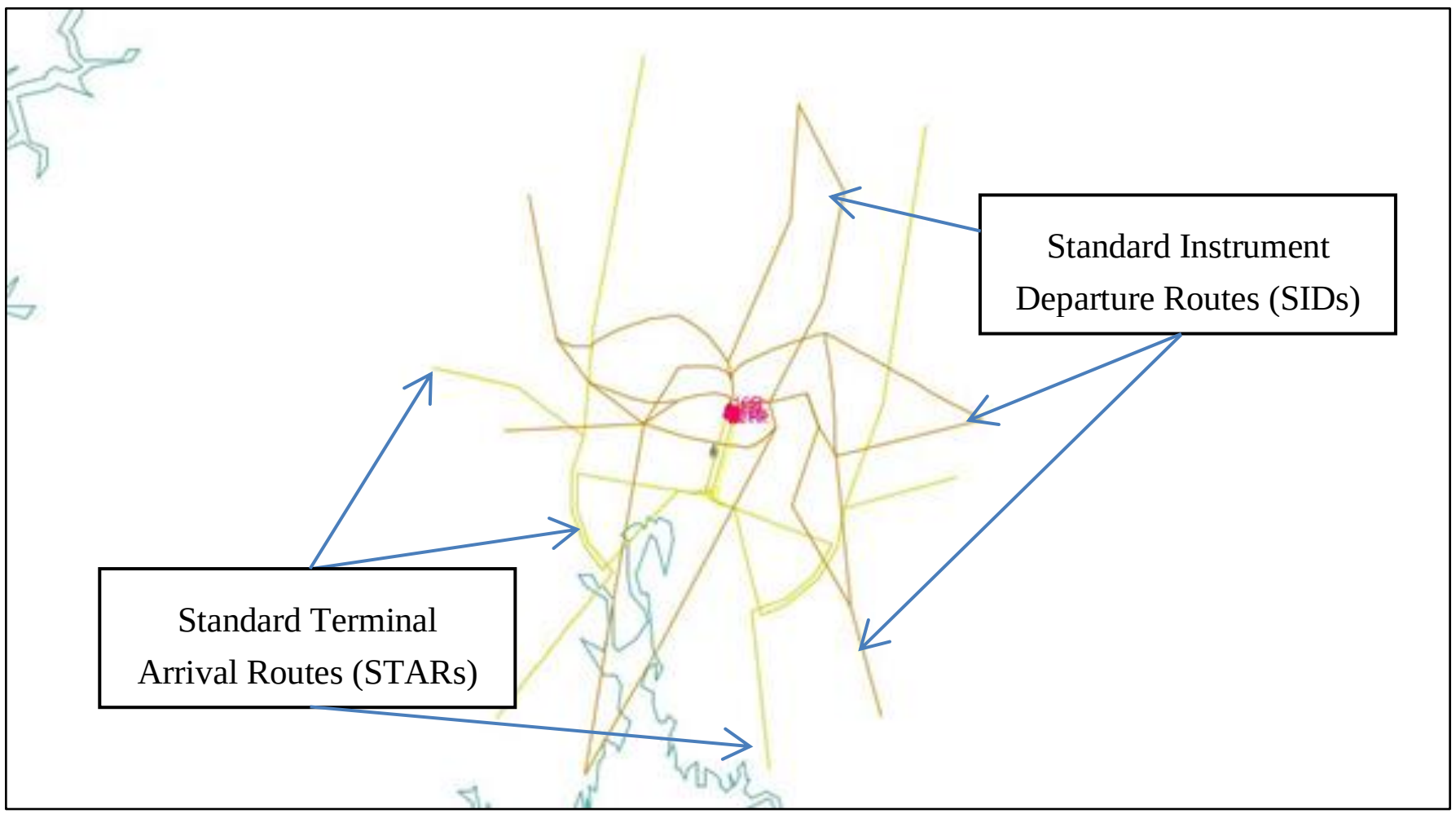

**Fig. 82.** OSL SIMMOD Model Airspace Layout

The linking of flights has been conducted on the available design peak day schedule of June 14$^{th}$ 2013 (**Fig. 81**). This schedule consisted of 776 flights on that day. The SID and STAR routes being used by aircraft flying certain city-pairs to or from OSL has been extracted from a different schedule.

In the case of departures these are injected at the gates, are (usually) being pushed back from the parking positions, then travel to the departure queues, take-off from the runway and move along the Standard Instrument Departure routes (SIDs) to the fixes at the outer boundary of the TMA (**Fig. 82**).
We limit the model in such a way that we assume Category (Cat.) I weather conditions. We simulate only operational direction 01 on the parallel runways 01L and 01R in MPO.

We assume a wind direction for which the 01 runway direction is representative. In a later study operational direction 19 should be modelled. Most likely a set of RETs will be constructed for that runway direction. It will be interesting to compare the effects for both directions 01 and 19.

### Simulation plan

Our reference is the "Without RETs" scenario (a) which reflects the state of OSL airport in 2017. In this scenario Pier North is completed and the respective aircraft parking positions are available. Additionally, the first phase of remote parking positions along taxiway Uniform is completed, making additionally four Large (class C and D) or two Heavy aircraft positions (class E) available.

The first step of construction of the Terminal 2 (T2) project is also in place. Furthermore, there are several taxiway extensions in place, such as the extension of taxiway Mike between exits A7 and A9, the extension of taxiway Golf towards taxiway Mike and the extension of taxiway Whisky towards taxiway Tango. We outline the scenarios in more detail below.

The Baseline scenario "Without RETs" (a) consists of the following characteristics:

a)

- Airport layout in 2017, with remote parking stands along taxiway U (phase 1) and current runway exit design.
- The runway is operated in MPO, i.e. mixed mode, which means departures are sequenced between succeeding arrivals on the runways.
- A separation of 6.0 NM between successive arrivals on the same runway.
- The minimum permitted lateral stagger is 3.0 NM.

Scenario (b) is named "With RETs" scenario and has these characteristics:

b)

- Similar airfield and airspace layout as in a) with proposed RETs activated on runways 01L and 01R.
- Exits A5 and B6 are closed.
- An in-trail separation between successive arrivals on the same runway of 5.0 NM.
- The minimum permissible lateral stagger could be as little as 2.0 NM.

The baseline scenario "Without RETs" (a) is then compared against the scenario (b) "With RETs".

## 9.4 Results

For simulating traffic growth (which reflects the growth of traffic at OSL in the future) we used the *cloning function*[1] within SIMMOD. We assumed different growth rates or cloning probabilities during different periods of the day (**Table 50**). Every six minutes a new cloning probability is activated. This leads to better stochastic spreading of cloned flights over time and to less bunching of flights in certain periods, as we aim to preserve the characteristic of the original schedule[2].

The daily growth rate summed to increments of 10%, 20%, 30% and 40% above the baseline level (which itself was increased by a certain percentage above the 2013 design peak day schedule). The two main objectives in simulating growth were to retain the daily demand pattern and to give emphasis on traffic growth during certain periods of the day in future schedules.

We ran the simulation through several iterations until we have collected at least ten runs without gridlocks (a traffic blocking situation involving two or more aircraft that cannot be solved automatically by the simulation and its internally programmed traffic rules).

We observe an increase of gridlock situations due to the traffic complexities involved in the higher growth scenarios. In the worst case we could only use 10 out of 30 total iterations when simulating the 40% growth scenarios. Ten successful complete runs of the simulation over the modeled design day were a critical requirement, not least for the full collection of the necessary figures from the output reports.

### Daily scheduled flights

The Baseline scenario includes on average 472 arrivals, 474 departures and 947 total flights (**Table 47**) or, say, about 950 daily flights. These numbers of flights were increased in 10% increments until either the simulation completely gridlocked or the level of service of 5 minutes was reached. The maximum number of flights we simulated in the model is about 1,300 flights per day in the 40% growth scenarios (**Table 47**). A usable iteration could be processes for a 50% growth scenario in the case where the RETs were added. The constraints are at one point defined by the given underlying traffic rules or by the activated infrastructure. We identify a lack of aircraft parking stands and lack of runway capacity as the main limiting factors for practical capacity.

---

[1] This function is modified by the SETCLONE table within SIMMODPlus.

[2] We want to thank Mr. Christoph Schneider from Munich airport for this suggestion.

**Table 47.** Daily Scheduled Flights in all Scenarios

| Scenario | Baseline - Without RETs | | | With RETs | | |
|---|---|---|---|---|---|---|
| | Arrivals | Departures | Total | Arrivals | Departures | Total |
| Baseyear | 472 | 474 | 947 | 474 | 474 | 948 |
| 10% | 512 | 516 | 1028 | 514 | 518 | 1032 |
| 20% | 559 | 566 | 1126 | 554 | 561 | 1115 |
| 30% | 603 | 614 | 1216 | 596 | 609 | 1205 |
| 40% | 639 | 660 | 1300 | 642 | 659 | 1301 |

**Level of service for arrivals and departures in different scenarios**

As we increase the traffic level in the baseline scenario model “Without RETs“, we observe an increase in the level of delay, i.e. a decrease of LoS (**Table 48**).

At a growth level of 20% (or 559 arrivals per day [**Table 47**] or 44 arrivals in the peak hour [**Table 49**]) we find a LoS of 6 minutes delay per arrival, which means that our predefined average LoS of 5 minutes per flight cannot be met at this traffic level (**Table 48**). The corresponding delay curve for arrivals is shown in **Fig. 83.** Daily Arriving Flights and Delay Relationship shows that the LoS is reached at 528 arrivals per day in the “Without RETs“ scenario.

In comparison, **Fig. 83** reveals that in the “With RETs” scenario LoS is not reached before 612 arrivals per day or a growth rate of about 30% above the base year traffic level of 474 arrivals per peak day.

**Table 48** shows that departing flights seem to meet the set LoS in both scenarios and under all growth increments (see **Fig. 84**). At 40% growth the level of delay is 2.1 minutes in the “With RETs” scenario compared to about 3 minutes per departure in the “Without RETs” scenario.

**Table 48.** LoS in Minutes Delay per Flight in the Scenarios

| Scenario | Baseline - Without RETs | | | With RETs | | |
|---|---|---|---|---|---|---|
| | Arrivals | Departures | Total | Arrivals | Departures | Total |
| Baseyear | 3.21 | 1.15 | 2.17 | 2.10 | 1.10 | 1.59 |
| 10% | 4.34 | 1.48 | 2.89 | 2.76 | 1.27 | 2.00 |
| 20% | 6.00 | 1.87 | 3.91 | 3.51 | 1.45 | 2.47 |
| 30% | 8.40 | 2.45 | 5.41 | 4.45 | 1.71 | 3.06 |
| 40% | 12.60 | 3.03 | 7.75 | 6.38 | 2.12 | 4.20 |

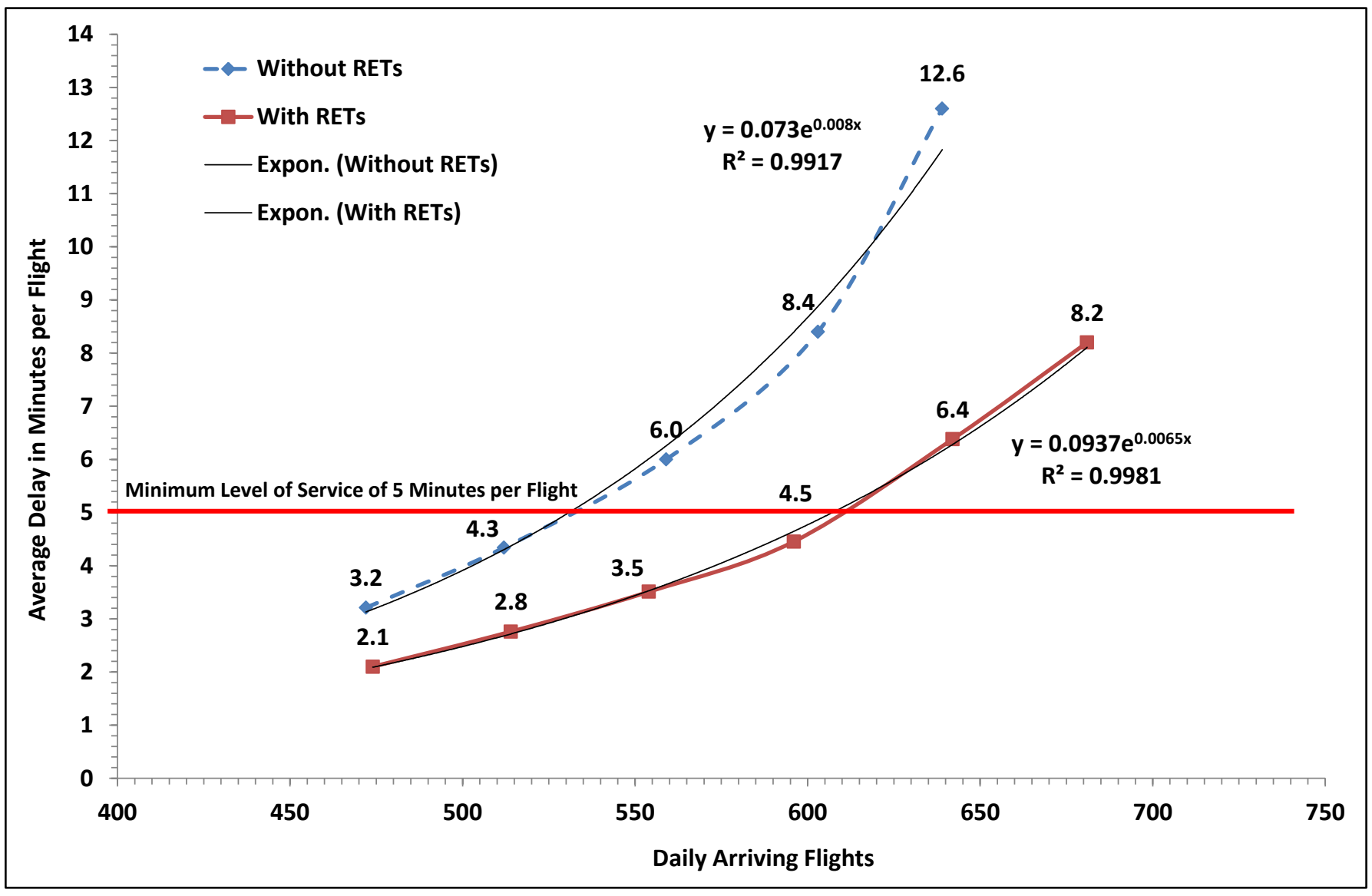

**Fig. 83.** Daily Arriving Flights and Delay Relationship

This means, with the RETs in place OSL can operate about 150 additional flights per day (or up to 84 additional arrivals or 81 departures per day, assuming a similar demand pattern and LoS). After the activation of the RETs in the "With RETs" scenario we can meet the set LoS of 5 minutes average delay per flight under all simulated traffic levels.

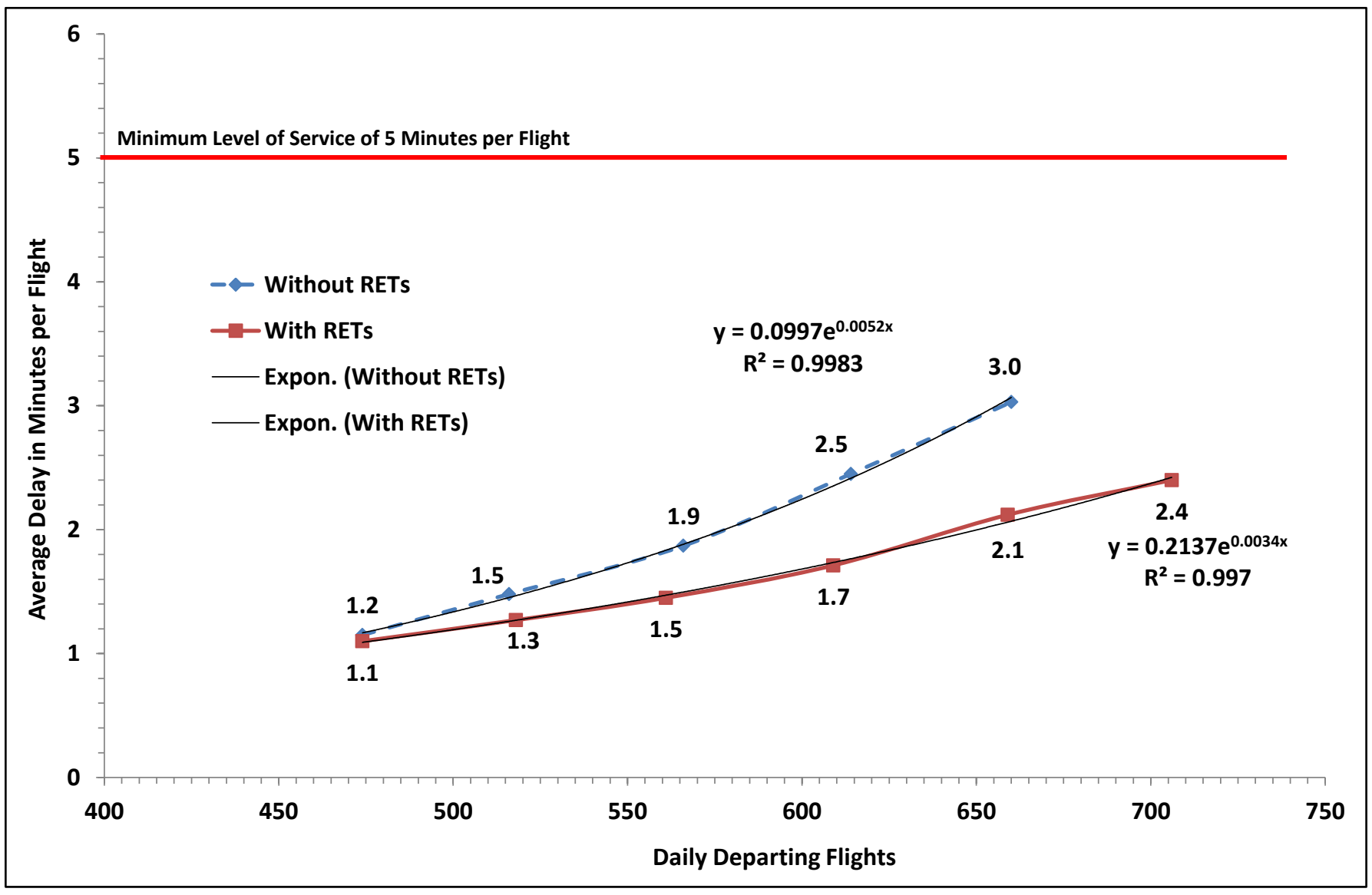

**Fig. 84.** Daily Departing Flights and Delay Relationship

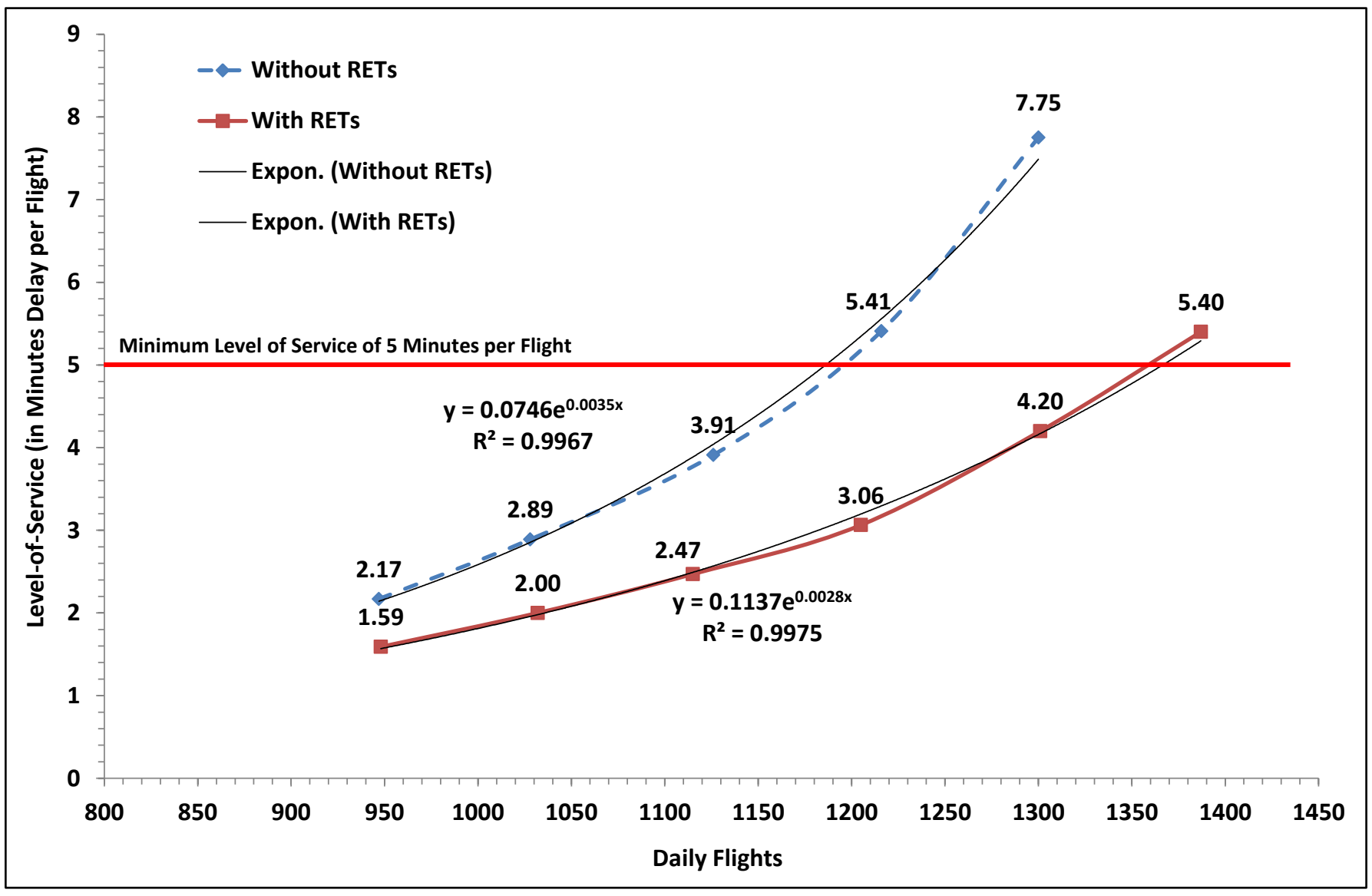

**Fig. 85.** Daily Total Flights and Delay Relationship

**Fig. 85** shows the exponential delay curves for total flights in both scenarios under all growth increments. We added a trend line for our data points and a best fit formula.

In the "Without RETs" scenario we observe that the LoS of 5 minutes delay per flight is passed at the practical capacity of about 1,202 daily flights

$$Average\ delay\ per\ flight\ =\ 0.0746 * e^{(0.0035*daily\ flights)}. \quad (1)$$

We solve the estimated formula for the "With RETs" scenario

$$Average\ delay\ per\ flight\ =\ 0.1137e^{(0.0028*daily\ flights)} \quad (2)$$

to calculate a practical capacity of 1,351 flights per day, when the permitted LoS of 5 minutes per flight is reached.

We calculate a maximum benefit in practical capacity of about 150 additional flights per day or +12.4% at the maximum tolerable LoS, after the addition of the RETs. Delay can be reduced by almost 50% in the "With RETs" scenario compared to the "Without RETs" scenario, from 7.75 to 4.2 minutes per flights, at a 40% growth rate (**Fig. 85**).

This means practical capacity will be reached at +27% above the traffic level of 2017 under the "Without RETs" scenario, whereas in the "With RETs" scenario practical capacity will be reached only at levels of +43% above the level of 2017.

## Causes of delay

In **Fig. 86** we can see that the causes of delay are similar under all scenarios and growth increments. However, the causes of delay differ between arrivals and departures.

In the "Without RETs" scenario about 55% of total departure delay is caused by departure queue delay. About 40% of departure delay is caused by ground delay (e.g. aircraft waiting on intersections or for an available gate or a push-back procedure of another aircraft). A small share of less than 5% of total delay is caused by airspace delay.

Most critical cause of delay for arriving flights is airspace delay which has a share of more than 95% of total delay. Less than 5% of total delay by arrivals is ground delay.

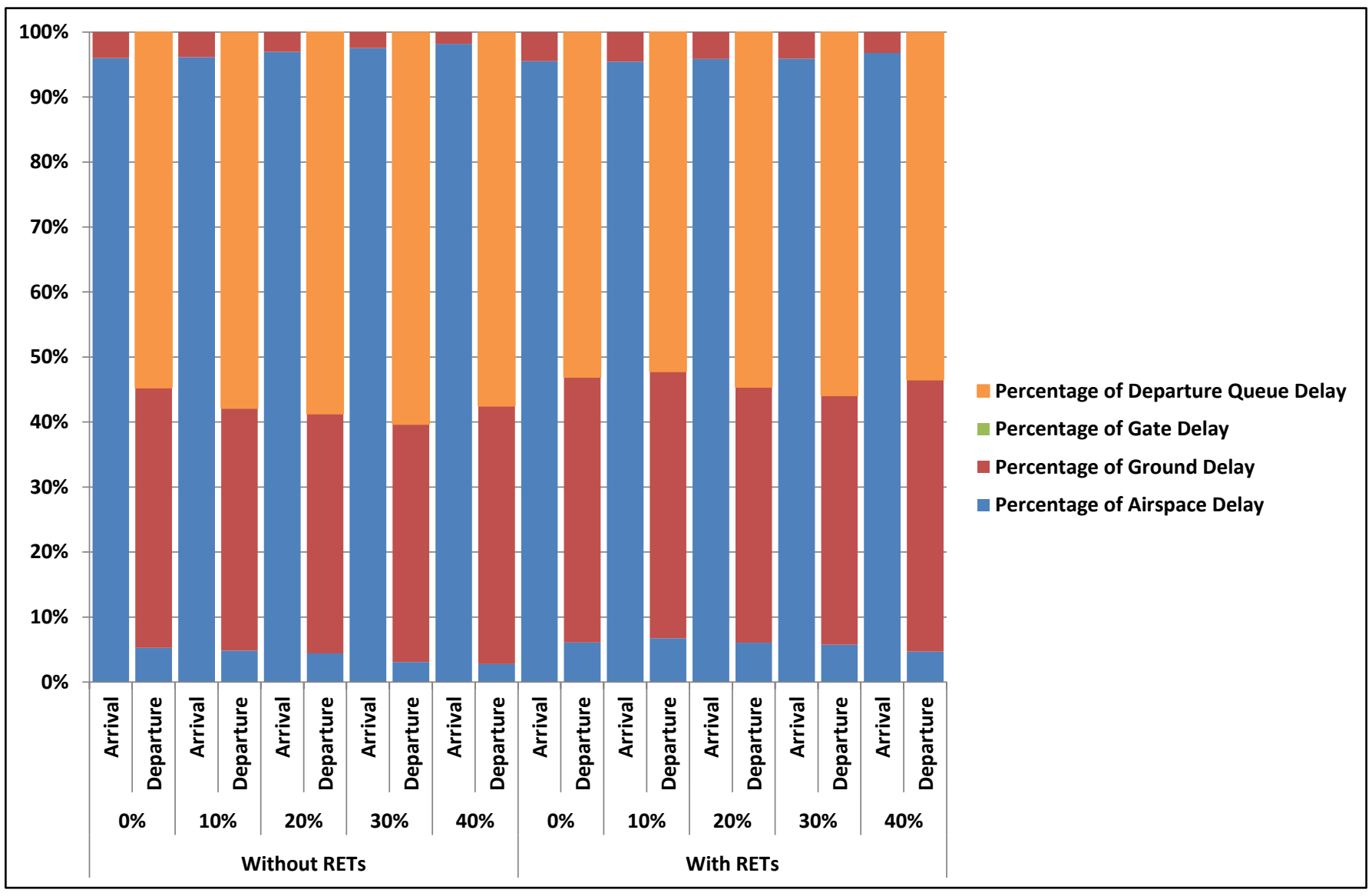

**Fig. 86.** Percentages of delay reasons for each scenario

The main difference between the scenarios is the slightly reduced percentage of departure queue and airspace delay in the "With RETs" scenario. This means due to the activation of the RETs, departing aircraft are released earlier from the departure queue and start their take-off roll, or aircraft arrive in shorter intervals and reduce waiting time in the airspace.

## Hourly pattern of capacity and level of service

**Fig. 87** and **Fig. 88** show the pattern of hourly processed flights and the amount of average delay per hour. **Fig. 87** shows the pattern for the "Without RETs" scenario, and the 10%, 20%, 30% and 40% traffic growth increments. We observe that between 10:00 and 11:00 our predefined threshold value for LoS of 5 minutes delay per flight is surpassed when 10% traffic growth is simulated.

**Fig. 87** shows some flattening of capacity, i.e. throughput, when growth is larger than 30% above the baseline. The demand pattern approaches the maximum limit of 93 flights per hour in the "Without RETs" scenario. LoS is above average in the morning hours between 8:00 and 11:00 when 20% traffic growth is reached. The more detailed numbers in **Fig. 87** show that 93 flights are operated in the peak hour between 9:00 and 10:00. However, this capacity comes at an unacceptable average cost of 18 minutes of delay per flight.

Between 7:00 and 12:00 the LoS decreases considerably, while delays increase. Already at 20% growth in traffic above the 2017 level the LoS cannot be met during the morning hours between 8:00 and 11.00, when we observe delays between 6 and 8 minutes per flight.

It should be mentioned that we let traffic grow stronger during the peak hours (see **Table 50**), because we believe that demand for morning and evening flights will grow comparably faster than the demand during off-peak the rest of the day.

In comparison the "With RETs" scenario in **Fig. 88** shows that the pattern of hourly flow can grow unconstrained throughout different stages of growth – Baseline to 40 % above the baseline. We were able to simulate growth increments until 50% above the 2017 baseline level.[1] LoS in the "With RETs" scenario case is only surpassed from 8:00 to 9:00 and from 10:00 to 11:00 at growth rates larger than 30%. At 40% growth, we see some more congestion in the morning periods between 7:00 and 12:00. The most severe delay can be observed between 10:00 and 11:00 when LoS reaches 8.8 minutes per flight.

At the maximum we experience an increase in occurrences of gridlock situations, e.g. due to the lack of parking stands or due to the increase of the number of complex pushback procedures. The maximum throughput capacity in the "With RETs" scenario is 102 flights per hour between 8:00 and 9:00 at a cost of a LoS of 7.7 minutes of delay per flight.

[1] We observed only one successful simulation run out of 50 iterations at 50% growth, therefore we dropped this growth level from **Fig. 88**.

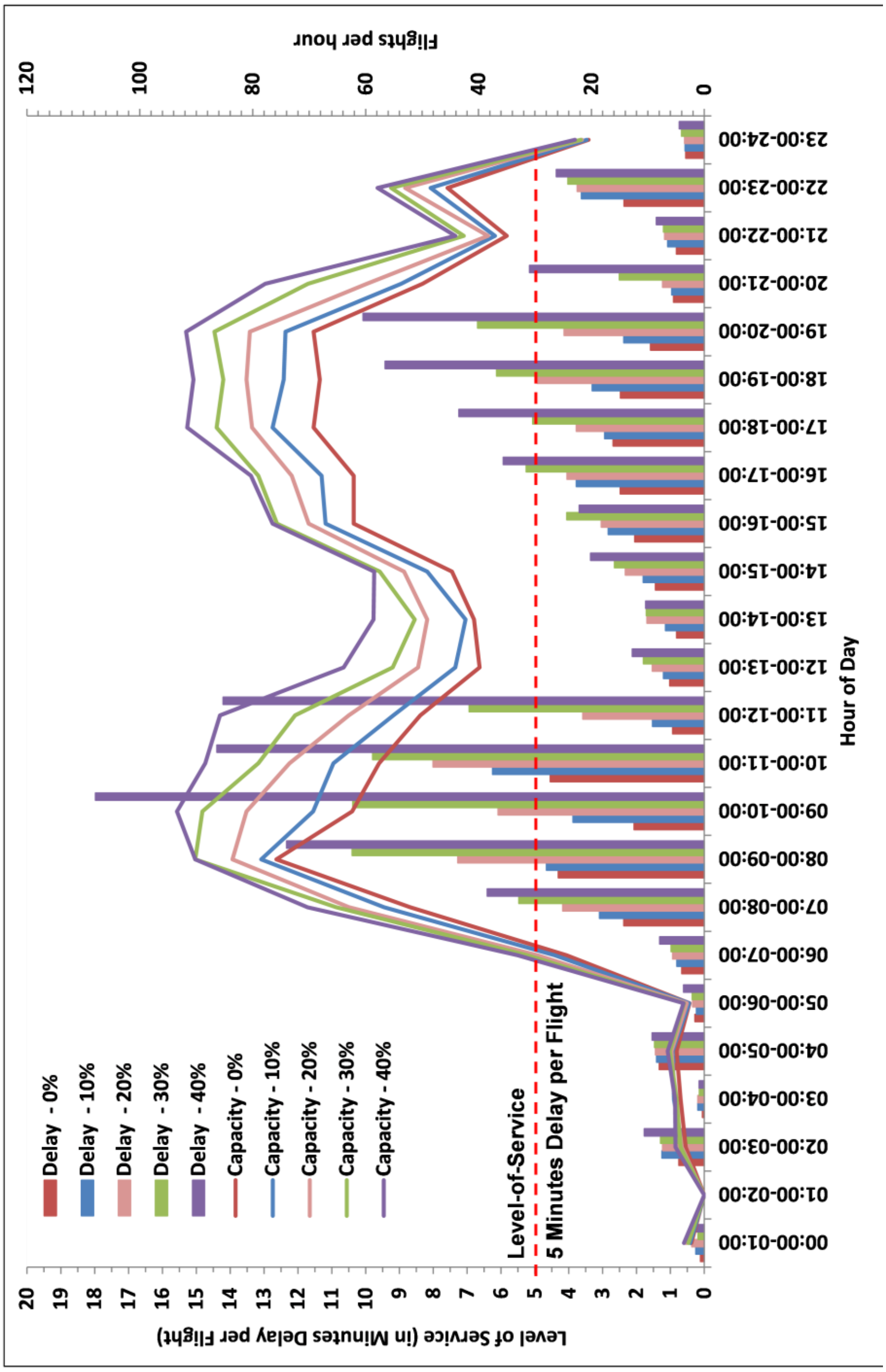

**Fig. 87.** Hourly Flights and LoS in the "Without RETs" scenario.

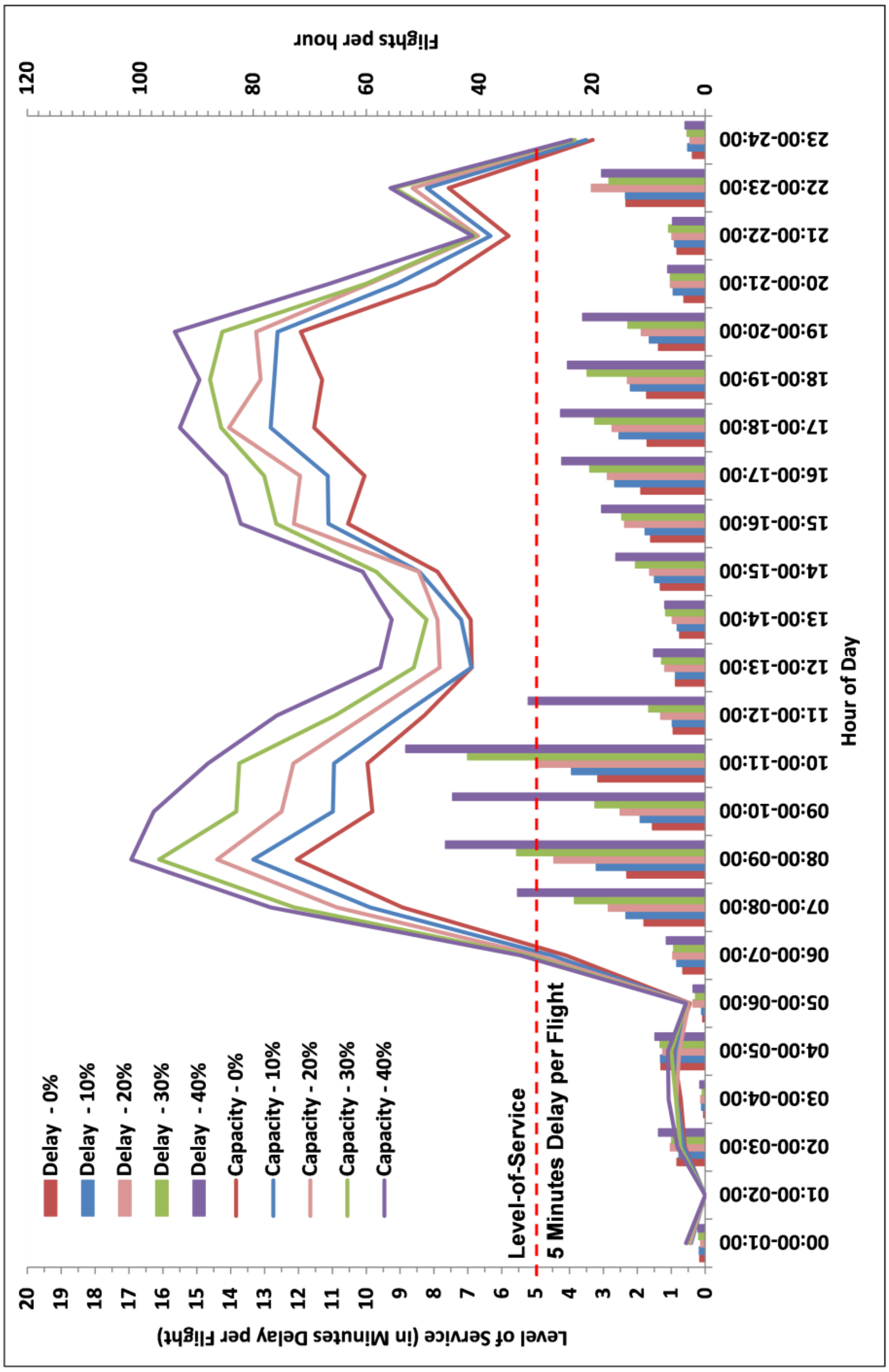

**Fig. 88.** Hourly Flights and LoS in the “ With RETs” scenario.

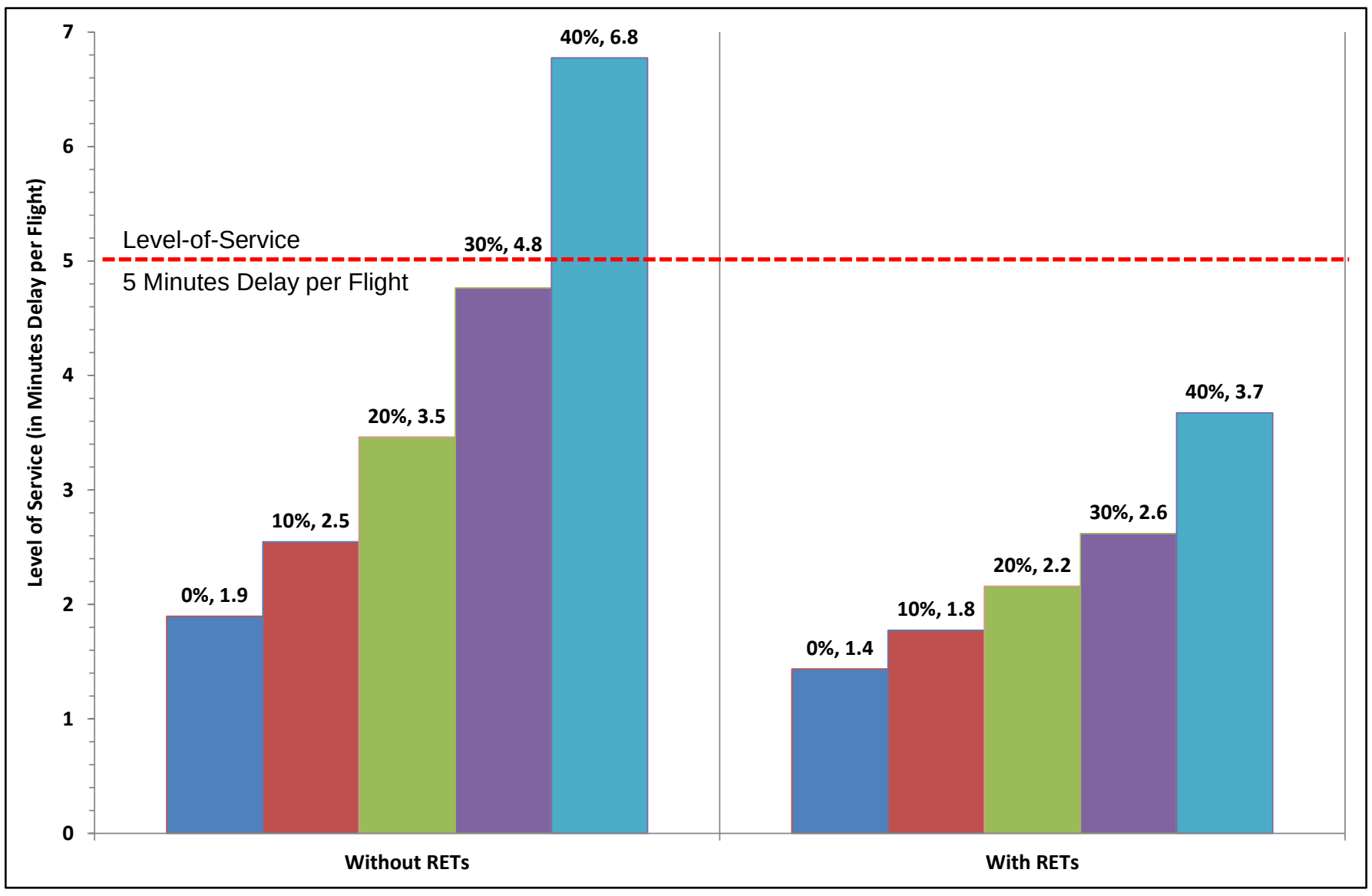

**Fig. 89.** Traffic Growth above baseline and LoS

We observe the average delay during the "core" operating hours from 6:00 to 24:00 and see that delay is much lower at different growth increments in the "With RETs" scenario and almost cut half in the growth scenarios beyond 20% growth above the base year 2017 level of about 1,120 daily flights (**Fig. 89**). For example, LoS has improved from 3.5 minutes of average delay per flight in the "Without RETs" scenario compared to 2.2 minutes of average delay in the "With RETs" scenario.

As **Table 49** shows we indeed observe an increase in hourly capacity after the addition of the RETs on both runways by 9.7%. However, this maximum benefit can only be observed at growth rates of 40% above the base year level. At the baseline level of about 950 daily flights, we observe a slight decrease of -5.3% in peak hour practical capacity after the activation of the RETs, however starting at 10% growth above the baseline the decline in LoS turns into a benefit of about +1.3% increase in peak hour capacity. This benefit is expanded at each level of growth but can clearly be measured at a growth rate of 30% above the 2017 base year levels. The airport can serve 7 additional movements, bringing the total to 97 flights during the peak hour (**Table 49**).

**Table 49.** Peak Hour Capacity Differences between the Scenarios

| Growth Rate | Without RETs Arrivals | With RETs Arrivals | % Difference | Without RETs Departures | With RETs Departures | % Difference | Without RETs Total Flights | With RETs Total Flights | % Difference |
|---|---|---|---|---|---|---|---|---|---|
| - | 41 | 39 | -4.9% | 42 | 44 | 4.8% | 76 | 72 | -5.3% |
| 10% | 41 | 42 | 2.4% | 43 | 45 | 4.7% | 79 | 80 | 1.3% |
| 20% | 44 | 46 | 4.5% | 48 | 48 | 0.0% | 84 | 86 | 2.4% |
| 30% | 45 | 50 | 11.1% | 49 | 52 | 6.1% | 90 | 97 | 7.8% |
| 40% | 47 | 53 | 12.7% | 50 | 57 | 14.0% | 93 | 102 | 9.7% |

## 9.5 Conclusions and recommendations

The numerous results from the simulation output prove our recommendation to construct the proposed two RETs. We base our opinion on the capacity benefit of 149 additional flights per day (+12.4%) and 9 additional flights in the peak hour (+9.7%) that OSL can operate at the same LoS when a set of RETs is in place compared to when the set of RETs is not in place.

To reduce the chance of unexpected head-on conflicts on taxiways, it is recommended to look into the proposed design of the access from the RET 01L to taxiway M and P south of A6.

From the output reports of the simulation software we can draw some valuable conclusions. The model reveals that in the morning hours between 8:00 and 11:00 and particularly between 8:00 and 9:00 peak demand causes most delay. These delays magnify when traffic is increasing (**Table 51**).

The "Without RETs" scenario shows already some congestion at a growth level of 10% above the 2017 base year level at around 1,030 flights per day. We observe congestions during two peak periods, one in the morning and one in the afternoon. These periods extend over two, three or more hours until traffic is decreasing midday or late at night, respectively.

The "With RETs" scenario in comparison does not show any violation of the 5-minute average delay before a 20% increase of traffic (ca. 1,120 flights per day) when we find congestion in the peak hours between 8:00 and 9:00 or 10:00 and 11:00. The overall LoS in this scenario is almost twice as good as in comparison to the "Without RETs" scenario (see **Fig. 89**).

LoS increases considerably when the RETs are in place. Arriving aircraft can land more frequently since most aircraft can depart the runway at a turnoff point which is located nearer to the runway threshold and at a faster speed. During MPO the reduction in runway occupancy time (ROT) due to the RETs leads to a reduction of airspace and departure queue delay. Average delay is almost half the level in the "With RETs" scenario compared to the "Without RETs" scenario at the higher growth scenarios.

In this study we mainly put an emphasis on differences in customer satisfaction measured in LoS. For this study we use the model to compare two modelled scenarios and look at the differences. It is beyond the scope of this study to evaluate weather or other impacts on aircraft landing and take-off roll distributions. It may be studied how these distributions change under inclement weather. We also neglect the direct delay costs to passenger and airlines related to the measured differences in LoS.

It should be clear from this study that any reduction of ROT for aircraft departing the runway on RETs can only alleviate LoS and increase practical capacity for a limited time, in a situation where traffic and passenger demand grows continuously. Given our assumptions, we observe that in order to serve future demand, additional runway and parking stand capacity needs to be installed at OSL beyond a level of 1,300 flights per design peak day.

We recommend carrying out a study on different locations of RETs and its linkage to the taxiway system, this could include the effects of adding a second set of RETs to the runways for more large or heavy aircraft, whose share in the aircraft mix could increase in the future, e.g. due to a larger percentage of long-haul flights.

Two RETs per runway would give ATC more choices for approaching aircraft and would lead to the use of even shorter distances between succeeding landings in the long term, because of increased controller experience and confidence. The use of only one RET could lead to the doubt of its worth, but this depends on experiences with the first set of RETs and its utilization.

In order to make sure all bottlenecks and “hotspots” in the airfield and airspace layout are known we also recommend conducting a simulation study for the opposite 19 runway direction, where the flow of taxiing aircraft would look differently than in the presented model.

## Appendix

**Table 50.** Clone factors used in the simulation to model future traffic growth

| Time of Day Buckets | Baseline | 10% | 20% | 30% | 40% | 50% | 60% |
|---|---|---|---|---|---|---|---|
| **0** | 0.20 | 0.30 | 0.40 | 0.50 | 0.60 | 0.70 | 0.80 |
| **1** | 0.20 | 0.30 | 0.40 | 0.50 | 0.60 | 0.70 | 0.80 |
| **2** | 0.20 | 0.30 | 0.40 | 0.50 | 0.60 | 0.70 | 0.80 |
| **3** | 0.20 | 0.30 | 0.40 | 0.50 | 0.60 | 0.70 | 0.80 |
| **4** | 0.20 | 0.30 | 0.40 | 0.50 | 0.60 | 0.70 | 0.80 |
| **5** | 0.40 | 0.60 | 0.80 | 1.00 | 1.20 | 1.40 | 1.60 |
| **6** | 0.40 | 0.60 | 0.80 | 1.00 | 1.20 | 1.40 | 1.60 |
| **7** | 0.30 | 0.45 | 0.60 | 0.75 | 0.90 | 1.05 | 1.20 |
| **8** | 0.30 | 0.45 | 0.60 | 0.75 | 0.90 | 1.05 | 1.20 |
| **9** | 0.40 | 0.60 | 0.80 | 1.00 | 1.20 | 1.40 | 1.60 |
| **10** | 0.40 | 0.60 | 0.80 | 1.00 | 1.20 | 1.40 | 1.60 |
| **11** | 0.20 | 0.30 | 0.40 | 0.50 | 0.60 | 0.70 | 0.80 |
| **12** | 0.20 | 0.30 | 0.40 | 0.50 | 0.60 | 0.70 | 0.80 |
| **13** | 0.20 | 0.30 | 0.40 | 0.50 | 0.60 | 0.70 | 0.80 |
| **14** | 0.20 | 0.30 | 0.40 | 0.50 | 0.60 | 0.70 | 0.80 |
| **15** | 0.20 | 0.30 | 0.40 | 0.50 | 0.60 | 0.70 | 0.80 |
| **16** | 0.20 | 0.30 | 0.40 | 0.50 | 0.60 | 0.70 | 0.80 |
| **17** | 0.20 | 0.30 | 0.40 | 0.50 | 0.60 | 0.70 | 0.80 |
| **18** | 0.20 | 0.30 | 0.40 | 0.50 | 0.60 | 0.70 | 0.80 |
| **19** | 0.20 | 0.30 | 0.40 | 0.50 | 0.60 | 0.70 | 0.80 |
| **20** | 0.10 | 0.15 | 0.20 | 0.25 | 0.30 | 0.35 | 0.40 |
| **21** | 0.10 | 0.15 | 0.20 | 0.25 | 0.30 | 0.35 | 0.40 |
| **22** | 0.10 | 0.15 | 0.20 | 0.25 | 0.30 | 0.35 | 0.40 |
| **23** | 0.10 | 0.15 | 0.20 | 0.25 | 0.30 | 0.35 | 0.40 |
| **24** | 0.10 | 0.15 | 0.20 | 0.25 | 0.30 | 0.35 | 0.40 |

**Table 51.** Hourly Capacity and LoS in all Scenarios

| | **Baseline – Without RETs** | | | | | | | | | | **With RETs** | | | | | | | | | |
|---|---|---|---|---|---|---|---|---|---|---|---|---|---|---|---|---|---|---|---|---|
| | **Hourly Capacity** | | | | | **Average Delay** | | | | | **Hourly Capacity** | | | | | **Average Delay** | | | | |
| **Time of Day** | **0%** | **10%** | **20%** | **30%** | **40%** | **0%** | **10%** | **20%** | **30%** | **40%** | **0%** | **10%** | **20%** | **30%** | **40%** | **0%** | **10%** | **20%** | **30%** | **40%** |
| 00:00-01:00 | 2 | 3 | 3 | 3 | 4 | 0.1 | 0.3 | 0.4 | 0.2 | 0.4 | 3 | 3 | 3 | 3 | 3 | 0.2 | 0.2 | 0.1 | 0.2 | 0.2 |
| 01:00-02:00 | 0 | 0 | 0 | 0 | 0 | 0.0 | 0.0 | 0.0 | 0.0 | 0.0 | 0 | 0 | 0 | 0 | 0 | 0.0 | 0.0 | 0.0 | 0.0 | 0.0 |
| 02:00-03:00 | 3 | 4 | 4 | 4 | 5 | 0.8 | 1.3 | 1.2 | 1.3 | 1.8 | 4 | 4 | 4 | 4 | 5 | 0.8 | 0.8 | 1.0 | 1.0 | 1.4 |
| 03:00-04:00 | 4 | 5 | 5 | 5 | 5 | 0.1 | 0.2 | 0.2 | 0.2 | 0.2 | 4 | 5 | 5 | 5 | 7 | 0.0 | 0.1 | 0.1 | 0.1 | 0.2 |
| 04:00-05:00 | 5 | 6 | 6 | 6 | 7 | 1.3 | 1.4 | 1.4 | 1.5 | 1.5 | 6 | 5 | 5 | 6 | 7 | 1.3 | 1.3 | 1.3 | 1.3 | 1.5 |
| 05:00-06:00 | 3 | 3 | 3 | 4 | 4 | 0.3 | 0.2 | 0.3 | 0.4 | 0.6 | 3 | 3 | 3 | 3 | 3 | 0.1 | 0.1 | 0.4 | 0.3 | 0.4 |
| 06:00-07:00 | 24 | 27 | 30 | 31 | 33 | 0.7 | 0.8 | 0.9 | 1.0 | 1.3 | 25 | 27 | 30 | 32 | 33 | 0.7 | 0.8 | 1.0 | 0.9 | 1.2 |
| 07:00-08:00 | 52 | 57 | 63 | 65 | 70 | 2.4 | 3.1 | 4.2 | 5.5 | 6.4 | 54 | 59 | 65 | 73 | 77 | 1.8 | 2.3 | 2.9 | 3.9 | 5.5 |
| 08:00-09:00 | 76 | 79 | 84 | 90 | 90 | 4.3 | 4.7 | 7.3 | 10.4 | 12.3 | 72 | 80 | 86 | 97 | 102 | 2.3 | 3.2 | 4.5 | 5.6 | 7.7 |
| 09:00-10:00 | 62 | 69 | 81 | 89 | 93 | 2.1 | 3.9 | 6.1 | 10.4 | 18.0 | 59 | 66 | 75 | 83 | 98 | 1.6 | 1.9 | 2.5 | 3.3 | 7.5 |
| 10:00-11:00 | 58 | 66 | 74 | 79 | 88 | 4.6 | 6.3 | 8.0 | 9.8 | 14.4 | 60 | 66 | 73 | 83 | 88 | 3.2 | 3.9 | 5.0 | 7.0 | 8.8 |
| 11:00-12:00 | 50 | 55 | 63 | 73 | 86 | 0.9 | 1.5 | 3.6 | 6.9 | 14.2 | 50 | 54 | 60 | 66 | 76 | 0.9 | 1.0 | 1.3 | 1.7 | 5.2 |
| 12:00-13:00 | 40 | 44 | 51 | 55 | 64 | 1.0 | 1.2 | 1.5 | 1.8 | 2.1 | 41 | 41 | 47 | 52 | 58 | 0.9 | 0.9 | 1.2 | 1.3 | 1.5 |
| 13:00-14:00 | 41 | 42 | 49 | 51 | 59 | 0.8 | 1.2 | 1.7 | 1.7 | 1.7 | 42 | 43 | 47 | 49 | 56 | 0.8 | 0.8 | 1.0 | 1.2 | 1.2 |
| 14:00-15:00 | 45 | 49 | 53 | 58 | 59 | 1.4 | 1.8 | 2.3 | 2.7 | 3.4 | 47 | 51 | 51 | 58 | 61 | 1.3 | 1.5 | 1.6 | 2.1 | 2.6 |
| 15:00-16:00 | 62 | 67 | 70 | 76 | 77 | 2.1 | 2.8 | 3.0 | 4.1 | 3.7 | 63 | 67 | 73 | 76 | 82 | 1.6 | 1.8 | 2.4 | 2.5 | 3.1 |
| 16:00-17:00 | 62 | 68 | 73 | 79 | 80 | 2.5 | 3.8 | 4.1 | 5.3 | 5.9 | 60 | 67 | 72 | 78 | 85 | 1.9 | 2.7 | 2.9 | 3.4 | 4.2 |
| 17:00-18:00 | 69 | 77 | 80 | 86 | 92 | 2.7 | 2.9 | 3.8 | 5.1 | 7.2 | 69 | 77 | 84 | 86 | 93 | 1.7 | 2.5 | 2.7 | 3.3 | 4.3 |
| 18:00-19:00 | 68 | 75 | 81 | 85 | 91 | 2.5 | 3.3 | 4.9 | 6.1 | 9.4 | 68 | 76 | 79 | 88 | 90 | 1.7 | 2.2 | 2.3 | 3.5 | 4.1 |
| 19:00-20:00 | 69 | 74 | 81 | 87 | 92 | 1.6 | 2.4 | 4.1 | 6.7 | 10.1 | 72 | 76 | 80 | 86 | 94 | 1.4 | 1.7 | 1.9 | 2.3 | 3.6 |
| 20:00-21:00 | 50 | 54 | 60 | 70 | 78 | 0.9 | 1.0 | 1.2 | 2.5 | 5.2 | 48 | 55 | 60 | 60 | 67 | 0.6 | 0.9 | 1.0 | 1.0 | 1.1 |
| 21:00-22:00 | 35 | 37 | 38 | 43 | 44 | 0.8 | 1.1 | 1.2 | 1.2 | 1.4 | 35 | 38 | 40 | 41 | 41 | 0.8 | 0.9 | 1.0 | 1.1 | 1.0 |
| 22:00-23:00 | 46 | 49 | 53 | 56 | 58 | 2.4 | 3.6 | 3.8 | 4.0 | 4.4 | 45 | 49 | 52 | 55 | 56 | 2.3 | 2.4 | 3.4 | 2.8 | 3.1 |
| 23:00-24:00 | 21 | 21 | 22 | 22 | 23 | 0.5 | 0.6 | 0.6 | 0.7 | 0.7 | 20 | 21 | 23 | 23 | 24 | 0.4 | 0.5 | 0.5 | 0.5 | 0.6 |
| **Sum** | 947 | 1028 | 1126 | 1216 | 1300 | | | | | | 948 | 1032 | 1116 | 1205 | 1302 | | | | | |
| **Maximum** | 76 | 79 | 84 | 90 | 93 | | | | | | 72 | 80 | 86 | 97 | 102 | | | | | |

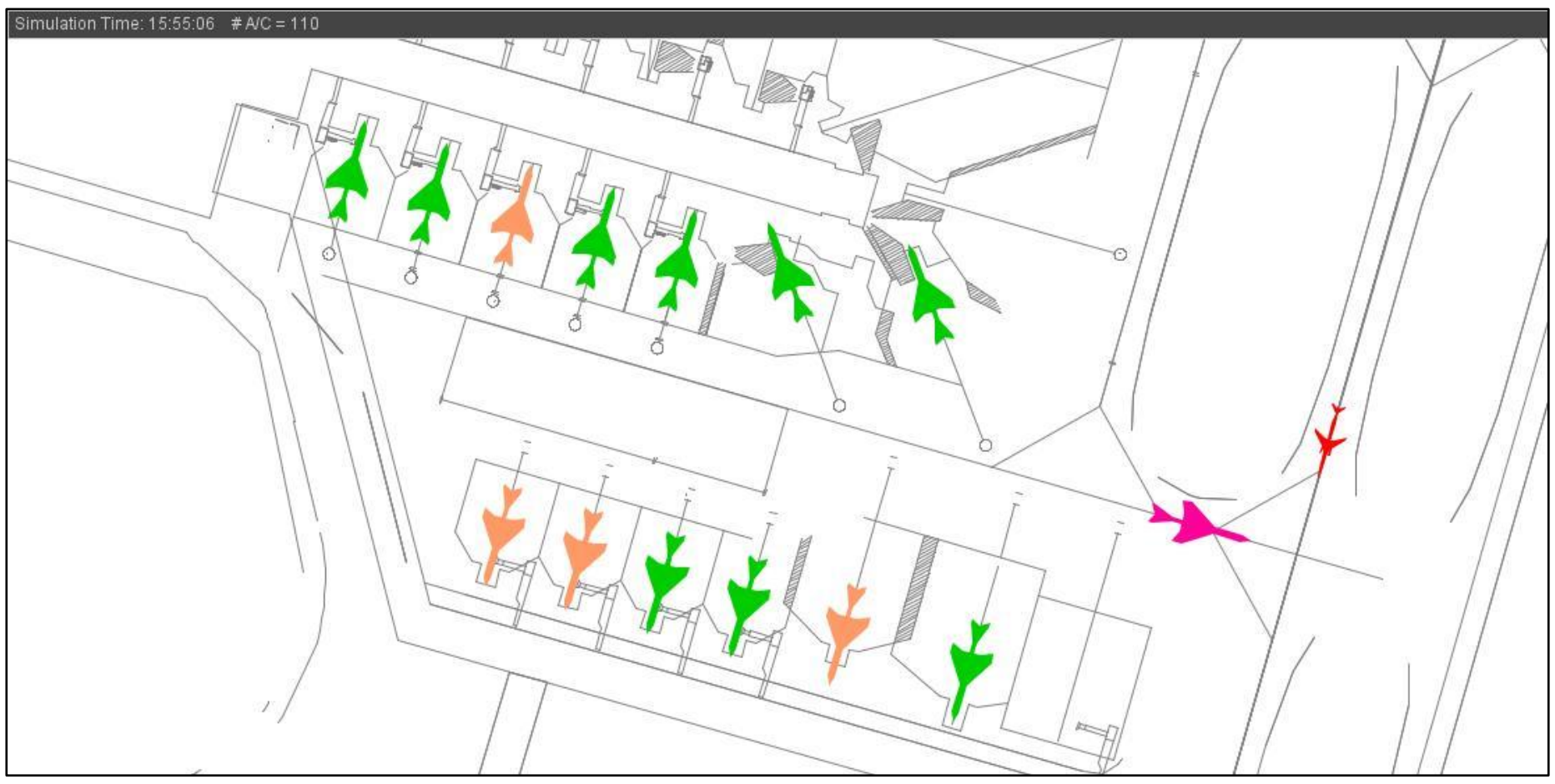

**Fig. 90.** Exemplary gridlock situation on taxiway Lima at 40% growth

An aircraft on taxiway Sierra wants to enter taxiway Lima, but encounters head on traffic on L1. The waiting aircraft will gridlock the simulation because the gates 39 to 51 where it wants to park are occupied. This flight is not allowed to park at gate 188, which became available by a departing Heavy aircraft. However, a human air traffic controller would decide otherwise, or certain constraints in the traffic rules within the model need to be applied or removed.

# 10 Reducing airport emissions with coordinated pushback processes: A case study

Branko Bubalo, Frederik Schulte, and Stefan Voß

Institute of Information Systems, University of Hamburg, Germany
{branko.bubalo,frederik.schulte,stefan.voss@uni-hamburg.de}

**Abstract.** Empirical research has shown that airside ground operations imply a significant percentage of overall airport-related emissions. Among those operations, taxiing is one of the most emission-intensive processes, directly related to the initial pushback process that has a significant impact on the taxiing duration of departing flights. Possible approaches for an effective management of pushbacks at an airport are simulation and optimization models. Airside operations at major airports involve a complex interplay of many operations and parties and therefore need to be planned in a coordinated fashion. Yet, existing approaches have not been applied in a comprehensive planning environment for airside operations. In this work, we develop an algorithm-based relocation approach for pushback vehicles that enables an effective minimization of delays and emissions during the taxiing process. As a result alternative sequences of departing flights are evaluated against each other to find the ones with least total emissions and delay. These algorithms are applied in a simulation environment and evaluated against real-world cases. Preliminary results demonstrate that we are able to solve the underlying pushback routing problem in appropriate computational times for dynamic decision support needed at airports.

## 10.1 Introduction

Airport emissions have recently received plenty of attention by regulators, airport operators, and researchers aiming to foster environmental sustainability that is threatened by emissions and delays caused in the aviation industry. Apart from the reduction of noise and gaseous emissions like carbon dioxide ($CO_2$), other emissions like, e.g., oxides of nitrogen ($NO_x$) need to be considered as well. Currently, the (German) automobile industry is troubled by intense public debates. This industry failed to reduce $NO_x$ emitted by diesel motors according to the permitted threshold values.

Regarding airports, empirical research has shown that airside ground operations form the biggest share of overall airport-related emissions (British Airports Authority 2011; Ekici, Yalin, Altuntas and Karakoc 2013; Winther et al. 2015). Among those ground operations, aircraft that are taxiing, i.e., traveling to the runways from the aircraft parking position or vice versa, are the largest contributors to pollution. At some airports, over 40% of ground-based aircraft emissions are related to taxiing (BAA 2011). Therefore, various approaches to reduce taxiing times and emissions have recently been introduced (Simaiakas and Balakrishnan 2009, and Wollenheit and Mühlhausen 2013).

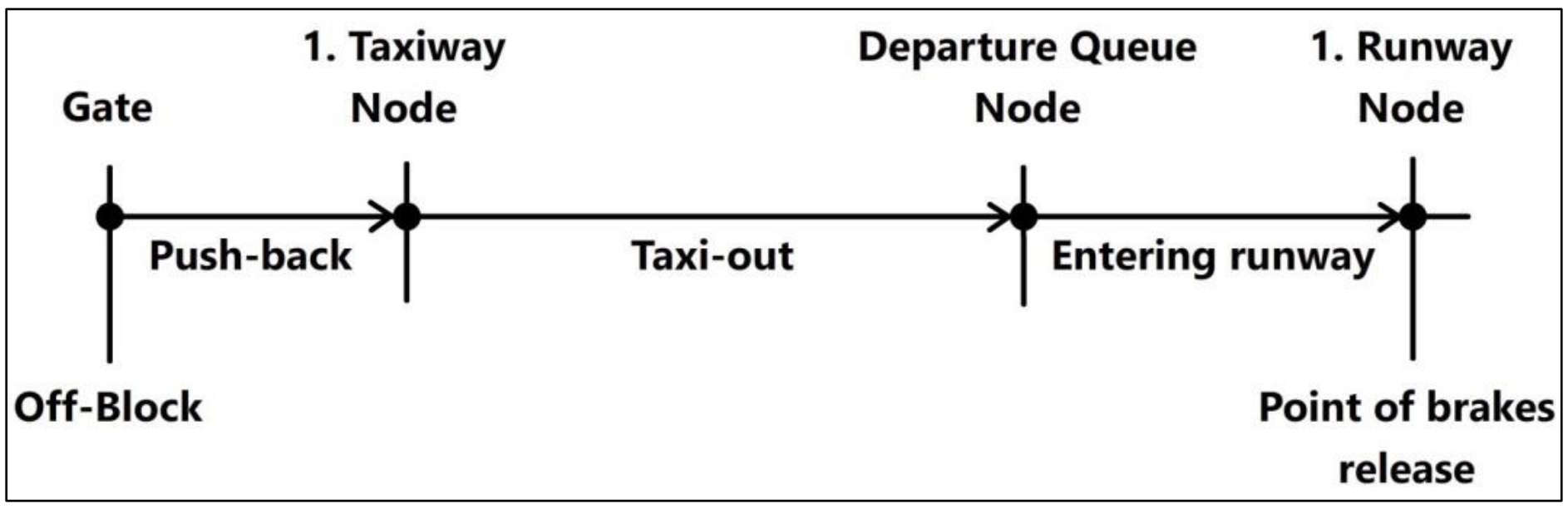

**Fig. 91.** Relation of pushback and taxiing process during Taxi-Out

These studies have focused on the development of alternative engine power settings or pushback frequencies to control the taxiing duration, but did not consider approaches to grant the availability of pushback vehicles in a holistic planning model for airside operations. The taxiing time of departing flights is directly related to the initiating pushback processes (Simaiakis and Balakrishnan 2010). Possible approaches to tackle this problem are simulation and optimization models for an effective management of pushbacks as a resource at the airport. Promising relocation models and adoptions of the vehicle routing problem (VRP) (Schwarze and Voß 2015) have been proposed for related problems in different domains (Schulte 2017). Airside operations at major airports involve a complex interplay of many operations and parties (Smith, Ehmke, Mattfeld, Waning and Hellmann 2014) and thus need to be planned and optimized in a coordinated fashion, especially as air traffic growth will lead to increasing demand, quicker aircraft turnarounds and requests for available pushback vehicles in higher frequencies. Yet, existing approaches have not been applied in a comprehensive planning environment for airside operations. They rather take pushback availability for granted in taxiing planning (Simaiakis and Balakrishnan 2009). However, we regard this view as too narrow.

With the constant increase in air traffic, airports are facing capacity problems. This can be due to bottlenecks on each and every level of airport operations. Optimization methods for specific airport processes are increasingly utilized by many large airports. However, many processes occur in parallel and make more complex optimization models necessary, which can consider multiple parallel processes simultaneously (Stergianos et al. 2015). This general observation also holds in the regarded case of the pushback control problem considering taxiing emissions as **Fig. 91** illustrates. A pushback vehicle pushes an aircraft into a taxiway and initiates the taxiing process. Hence, the pushback defines when an aircraft starts the taxi-out phase and enters the departure queue. However, with an increasing amount of aircraft in queue the taxiing time and delay of an aircraft grows non-linearly—an effect that is explained in the general theory of load-dependent lead times (Pahl, Voß and Woodruff 2007) and has been confirmed for the taxiing process in aviation (Simaiakis and Balakrishnan 2010). In line with the increased delay, evitable costs for fuel burn and emissions result from a too large departure queue of aircraft. To address the issue of bad timing in pushback control, we propose to plan taxiing in order to minimize delay and emissions and to schedule pushbacks on that basis. For this approach pushback vehicles need to be relocated according to the position of aircraft scheduled for taxiing. In this work, we develop an algorithm-based relocation approach for pushback vehicles that enables an effective minimization of delays, due to waiting for an available truck, and emissions during the taxiing process. These algorithms are applied in a realistic simulation environment for airside operations and evaluated for the real-world case of Oslo Airport, Norway, which is operated by public airport operator AVINOR. The subsequent Section 10.2 reviews related work and highlights current challenges in the domain. In Section 10.3 we introduce the simulation used for the emission evaluation. Section 10.4 proposes several vehicle routing models for the pushback relocation, a metaheuristic for quick solutions within the dynamic airport environment, and quantitative results for instances available in literature (Schwarze and Voß 2015). Section 10.5 elaborates on the promising conceptual integration of the simulation and the routing model, before Section 10.6 draws conclusions and discusses future work.

## 10.2 Related work

Research related to this study falls into three basic categories: (I.) Estimation of airport ground emissions, especially taxiing emissions, (II.) general work on airside operations including taxiing and pushback processes as well as (III.) routing of airport vehicles. In describing the *System for assessing Aviations Global Emissions* (SAGE), Kim et al. (2007) lay the fundamentals for modeling emission inventories related to aircraft fuel burn. Setting the focus on take-off activities, Zhu, Fanning, Yu, Zhang and Froines (2011) demonstrate the local air quality impacts at the Los Angeles International Airport. Following these studies, Koudis, Hu, Majumdar, Jones and Stettler (2017) show that reduced thrust takeoff operations can reduce fuel consumption and pollutant emissions in the studied case of London Heathrow airport. Moreover, several authors examine taxiing emissions in detail. Nikoleris, Gupta and Kistler (2011) as well as Khadilkar and Balakrishnan (2012) apply different approaches to estimate taxiing emissions. Yang, Dong, Lin and Hu (2016) extend these findings by predicting the market potential and environmental benefits of deploying electric taxis in Nanjing, China. With a focus on airside operations, Atkin, Maere, Burke and Greenwood (2012) address the pushback time allocation problem at London Heathrow airport. Stergianos, Atkin, Schittekat, Nordlander, Gerada and Morvan (2015) follow up in this direction by analyzing the effect of pushback delays on the routing and scheduling problem of aircraft, while Mori (2014) applies a reinforcement learning approach to obtain optimal pushback times facing uncertainties at busy airports. Similarly, Balakrishnan, Ganesan and Sherry (2010) use reinforcement learning algorithms for predicting aircraft taxi-out times. Related to routing problems at an airport, Guépet, Briant, Gayon and Acuna-Agost (2016) provide a mixed-integer programming formulation for the aircraft ground routing problem. Summarizing, the literature review reveals promising approaches in all three categories mentioned initially. Nevertheless, the integration of various interconnected problems at an airport appears to remain an important challenge for future research, and that applies also for the considered problems of pushback routing and taxiing.

### 10.3 Simulation model

In order to analyze the dependencies between pushback control and aircraft taxiing, we have built a simulation model for the practical case of Oslo airport. The simulation model captures the queuing behavior of aircraft during the taxiing process and thus allows us to analyze different schedules of pushback control with respect to taxiing delay, i.e. lead times, fuel burn (costs), and emissions. We have constructed the simulation model based on real-world input data. The link-and-node network model is a representation of Oslo airport and its Terminal Maneuvering Area (TMA) on a design peak day in the year 2017. This means, all arrival and departure routes within the terminal area, and all runways, taxiways, and parking stands which are expected to be installed by 2017 are included in the model, even infrastructure that was still under construction during the preparation of this study. For our taxiing times and delay calculations we made use of the Airport and Airspace Simulation Model (SIMMOD) version 4.7.9 that is developed by the U.S. Federal Aviation Administration. We have run the simulation engine on a 64-bit Windows 10 machine with 8 GB RAM and an Intel Core i5-3337U 1.8 GHz processor. A run of 100 iterations took less than seven minutes, of which pure calculation time (without additional time required for user input in cases when gridlocks occur in the simulation) is estimated at less than three minutes.

Based on documents from the airport, we were able to build a model which behaves sufficiently realistic and which shows similar characteristics as the real airport. We included the flight schedule, the aircraft type mix, the air and ground movements into our model which was continuously approved by feedback talks and discussion with local experts. Arrivals in the schedule are linked to the respective departure in the schedule to model the full aircraft turnaround. The taxiing of aircraft on the apron is modeled by the SIMMOD logic, which directs aircraft to take the quickest path from the runway exits to the gates after landing or from the gates to the runway departure queues prior departure. An important prerequisite in the simulation is the programming of traffic rules at taxiway crossings and larger tarmac areas to prevent collision conflicts. When aircraft share certain single lane taxiways in both directions this could lead to head-on situations, where the simulation needs rules of how to prioritize aircraft to pass such bottlenecks. Otherwise it could happen that aircraft around bottlenecks and hotspots gridlock, i.e., they indefinitely block each other. The simulation is unable to solve gridlock situations automatically, it is only able to predict potential conflict situations and let aircraft wait at predefined nodes until they can pass one or many taxiway segments unhindered. We solved many gridlock problems by including additional traffic rules at intersections and at certain taxiways, but we stopped, when more than 50% of all iterations were gridlock free.

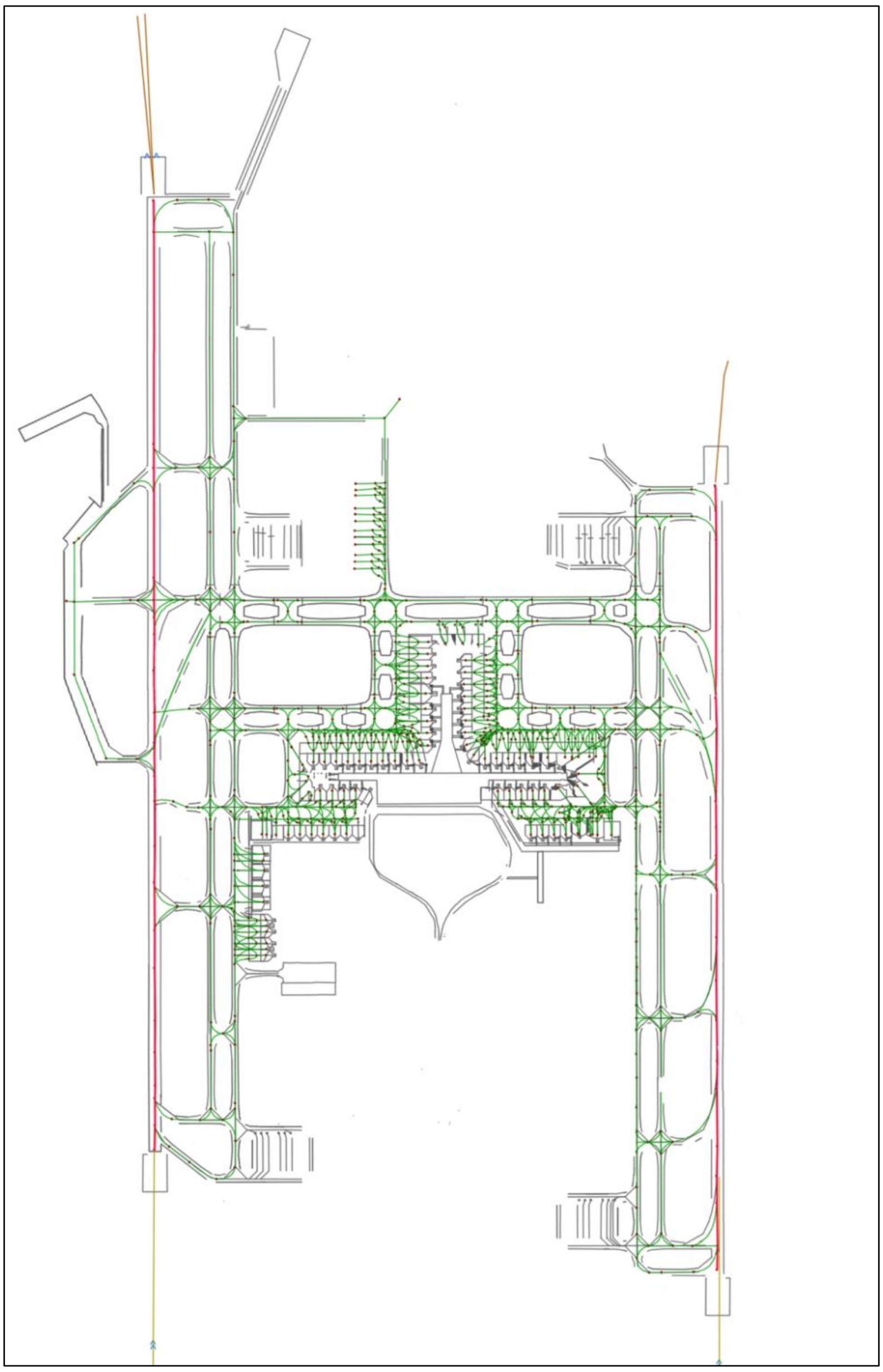

**Fig. 92.** A simulation model for Oslo airport (Source: Avinor and VisualSIMMOD)

Oslo airport operates two independent parallel runways. The runways are operated under mixed mode, which means departures and arrivals are served in sequence on both runways during the same hours of day. This type of operation offers more capacity for the airport, as long as air traffic control (ATC) permits departures to be put between two consecutive arrivals or vice versa. If the gaps between two flights are becoming narrow, long queues can form at the departure queues, where aircraft wait for approval to enter the runway to start the take-off roll.

In our model we defined six different concourses, i.e., groups of gates which are located near to each other at different terminals. We have defined concourses for gates and parking stands around taxiway Kilo and taxiway Lima, one East and one West of the new North pier, one remote concourse for mail flights and one concourse with a few stands for heavy cargo aircraft as shown in **Fig. 92**. The main purpose why we run simulations is its output in form of taxiing durations and delays for each flight when traveling from the gates to the runways, while crossing the airfield and paths of other aircraft. From the SIMMOD output we can extract an extensive amount of useful data for further computational experiments. The taxiing phase from and to the runways can be broken down in segments, by pushback, taxi-out, taxi-in, departure queue arrival, departure queue departure, runway exit times, etc.

In order to calculate emissions, such as hydrocarbon (*HC*), carbon monoxide (*CO*), $NO_x$ and $CO_2$ for each aircraft for a particular airport, we need the amount of burned fuel as a basis for further emission estimations (Winther et al. 2015). The process to calculate emissions for this study was conducted as follows:

1. 1. Step: We take the time records (e.g., pushback times and departure ground delay) from SIMMOD.
2. 2. Step: Calculate fuel burn during the taxiing (in kilograms per second) from off-block time at the gate to the time (point) of brakes release at the runway departure queue, before the aircraft turns into the runway and starts its take-off run.
3. 3. Step: Calculate *HC*, *CO* and $NO_x$ from fuel flow quantity (in grams per kilogram fuel).
4. 4. Step: Multiply emission by the number of engines per aircraft.
5. 5. Step: Calculate $CO_2$ emissions from fuel flow by applying factor 3.157 (kilogram $CO_2$ per kilogram fuel) (Jardine 2009).

We iterate the above sequence of calculations. We first ran 20, later up to 100 iterations. Due to the complex interactions of aircraft on the modeled airfield of Oslo airport we experienced several gridlock problems, where aircraft block each other on the apron and the situation could not be solved by the SIMMOD logic. In such cases queuing aircraft accumulate quickly and block other aircraft on taxiways in the area. When threshold values for very long waiting times of blocked aircraft are reached, the simulation will stop. We analyzed only simulation runs that were gridlock-free, which means that all scheduled flights during the day were simulated. The design peak day schedule from June 14, 2013 had 776 flights. To make the model and its results more realistic and robust, we included random factors in form of probability distributions into the model, mainly for actual departures occurring within a 15-minute time window after the scheduled departure time, for landing and take-off roll distance and for arriving aircraft choosing a particular gate at a defined concourse. In this study, growth above the amount of traffic in the design peak day schedule is neglected.

We intentionally include aircraft delay experienced along the taxiway path from the beginning of the pushback process at the gate until the exiting of the departure queue at the runway. Thus, we want to increase traffic and minimize total taxiing time (by changing the pushback sequence). In this study, we only examine the effect of randomly re-sequencing flights within a 15-minute time window. This gives us an indication of the minimal and maximal externalities (delay and emissions) and their respective flight and pushback sequence. The pushback begins at the scheduled departure time plus a random time between 0 and 15 minutes, assuming a uniform distribution. In this study the pushback duration has been set at 180 seconds for all gates, except for self-maneuvering aircraft stands, where a pushback truck is not needed. In subsequent studies we want to model the pushback procedures more gate specific.

We take the taxiing times from the simulation and apply the factors shown in **Table 52**. At first, we calculate the amount of burned fuel during the taxiing out phase for different aircraft types. We observe that more than 60% of the aircraft flying into OSL are Boeing 737 type aircraft. **Table 52** gives us the factors for fuel flow in kilograms per second at idle engine conditions, which means the engine thrust is set between 0% and 7% (in case of turbo prop aircraft, e.g. DHC8, DHC6 or SF340) of the maximum. This number must then be multiplied by the number of engines. The amounts of gaseous emissions (in grams) correspond to the amount of fuel burned (in kilograms), thus we can now apply the factors for *HC*, *CO*, or $NO_x$ from **Table 52**. The last emission type we want to quantify is $CO_2$. As presented in Step 5, we derive the amount of $CO_2$ (in kilograms) by multiplying the total fuel burned during taxi-out (in kilograms) times 3.157 (Jardine 2009).

**Table 53** gives us an overview of the calculated quantities for nine different iterations. The taxiing out duration varies between 2,937 and 3,048 minutes (or around 50 hours) per day, of which about 7% are waiting times (delay). This results directly in between 40.12 and 42.01 tons of burned fuel for all daily departures. Not surprisingly $CO_2$ emissions have the highest amounts of all emissions, with between 126.7 and 132.6 tons for all 382 daily departing flights. Amounts of *HC* are calculated between 252.2 and 327.1 kilograms, amounts of *CO* between 1032.3 and 1119.6 kilograms and amounts of $NO_x$ between 134.3 and 140.99 kilograms. We have ranked the iterations by total externalities, delay and emissions (**Table 53**). Our results seem to be of the same order of magnitude compared to similar studies, such as Winther et al. (2015). From all run iterations, iteration 13 performs best with regard to total externalities. We ranked the iterations by each category, and then by overall performance. Iteration 13 reveals 382 daily departures, which need 2938 minutes of taxi-out time. On average, aircraft taxi out time to one of the two parallel runways is approximately 8 minutes, of which around 3 minutes are pushback time.

**Table 52.** OSL aircraft mix and engine emission factors. Source: Avinor, ICAO[1] and FOI[2]

| Aircraft Type | Percentage of Total Flights | Engine Identification | No of Engines | Hydrocarbon (*HC*) emission index at idle condition | Carbon Monoxide (*CO*) emission index at idle condition | Oxides of nitrogen ($NO_x$) emission index at idle condition | Fuel flow at idle condition |
|---|---|---|---|---|---|---|---|
| | % | | | g/kg | g/kg | g/kg | kg/sec |
| 737 | 38.4% | JT8D-9 series | 2 | 3.12 | 14.1 | 2.9 | 0.132 |
| 737300 | 25.8% | CFM56-3-B1 | 2 | 2.28 | 34.4 | 3.9 | 0.114 |
| A320 | 9.6% | CFM56-5-A1 | 2 | 1.40 | 17.6 | 4.0 | 0.101 |
| DHC8 | 8.0% | PW123 | 2 | 1.40 | 17.6 | 3.1 | 0.043 |
| 737500 | 6.8% | CFM56-3-B1 | 2 | 2.28 | 34.4 | 3.9 | 0.114 |
| SF340 | 4.0% | CT7-9B | 2 | 3.52 | 27.7 | 1.7 | 0.019 |
| F10062 | 2.4% | TAY Mk620-15 | 2 | 3.40 | 24.1 | 2.5 | 0.110 |
| 720 | 1.4% | JT3D-3B | 4 | 112.00 | 98.0 | 2.5 | 0.135 |
| 757PW | 1.0% | PW2037 | 2 | 1.92 | 22.4 | 4.1 | 0.152 |
| DHC6 | 0.9% | PT6A-67 | 2 | 3.10 | 48.3 | 1.8 | 0.019 |
| 747400 | 0.4% | PW4056 | 4 | 1.92 | 21.9 | 4.8 | 0.208 |
| 747SP | 0.4% | JT9D-7 | 4 | 36.50 | 84.1 | 3.1 | 0.210 |
| 767300 | 0.3% | PW4060 | 2 | 1.66 | 20.3 | 4.9 | 0.213 |
| MD82 | 0.3% | JT8D-217 series | 2 | 3.33 | 12.3 | 3.7 | 0.137 |
| A300 | 0.1% | CF6-50C2 | 2 | 2.72 | 24.0 | 3.4 | 0.163 |

[1] https://www.easa.europa.eu/document-library/icao-aircraft-engine-emissions-databank [last accessed on 20/05/2017]

[2] https://www.foi.se/en/our-knowledge/aeronautics-and-air-combat-simulation/fois-confidential-database-for-turboprop-engine-emissions.html [last accessed on 20/05/2017]

**Table 53.** Calculated emission for different iterations (ranked by lowest to highest total emissions).

| Rank | Iteration | Taxi-out duration | Taxi-out fuel burn | *HC* | *CO* | *$NO_x$* | *$CO_2$* | Ground delay | Departures | Average delay per departure |
|---|---|---|---|---|---|---|---|---|---|---|
| | | min. | tons | kg | kg | kg | tons | min. | | |
| 1 | 13 | 2937.8 | 40.12 | 252.17 | 1032.31 | 134.38 | 126.66 | 209 | 382 | 0.55 |
| 2 | 1 | 2937.4 | 40.29 | 284.28 | 1079.82 | 135.19 | 127.21 | 280 | 381 | 0.73 |
| 3 | 17 | 2964.3 | 40.40 | 294.09 | 1082.50 | 134.32 | 127.54 | 277 | 381 | 0.73 |
| 4 | 14 | 2943.7 | 40.90 | 287.85 | 1088.50 | 137.62 | 129.12 | 311 | 382 | 0.81 |
| 5 | 18 | 2981.4 | 40.94 | 323.48 | 1100.78 | 135.75 | 129.26 | 357 | 382 | 0.93 |
| 6 | 5 | 3033.1 | 41.04 | 284.07 | 1062.36 | 136.69 | 129.56 | 304 | 382 | 0.80 |
| 7 | 2 | 2999.0 | 41.04 | 301.71 | 1090.96 | 136.47 | 129.57 | 337 | 382 | 0.88 |
| 8 | 12 | 2978.3 | 41.38 | 327.12 | 1119.62 | 138.48 | 130.63 | 245 | 382 | 0.64 |
| 9 | 9 | 3048.3 | 42.01 | 279.48 | 1108.66 | 140.99 | 132.63 | 293 | 381 | 0.77 |

It should be noted that the simulation represents a "perfect" environment, i.e., weather effects were not taken into account and we assume that aircraft always travel the maximum allowed speed limit, which may be 5, 10 or 15 knots etc., depending on the type and location of the taxiway.

Aircraft are usually placed at a gate close to the landing runway or directed to a runway close to the parking stand in case of arrivals. This procedure that is administered by ATC shall minimize the number of aircraft traveling around the whole airfield. We followed the same approach when the simulation was programmed, particularly in such cases, when the information was not given in the original schedule from Oslo airport.

## 10.4 Dynamic vehicle routing model

The pushback relocation problem described in the introduction has been modeled as skill vehicle routing problem (skill VRP) to reflect different qualifications of pushback vehicles to serve aircraft types. Next, we introduce two ways to model a skill pushback routing problem (Section A Skill VRP formulation for pushback vehicles) and present a *large neighborhood search* (LNS) metaheuristic for the dynamic application (Section A large neighborhood search algorithm) as well as results for benchmark instances from literature (Section Computational results).

### A Skill VRP formulation for pushback vehicles

This skill VRP is a variant of the site dependent VRP; see, e.g., Cordeau and Laporte (2001). Following Schwarze and Voß (2015), we adopt the skill VRP formulation to the pushback routing and control introduced in the beginning. Assuming $G = (N, A)$ as a directed graph, with $|N| = n$ and $|A| = m$. Let each node $j$, $j \neq 1$, represent a service requirement by an aircraft, and $s_j$ denote the skill level required by node $j$ for the associated pushback service by a specific aircraft. On the other hand, node 1 denotes the depot. We assume a set $P$ of available pushback vehicles, each one operating at a certain skill level, were $s_p$ denotes the skill level of pushback vehicle $p \in P$. Furthermore, we suppose that the service requirement at node $j$ can be operated by any pushback vehicle having a skill level of at least $s_j$, for $j \in N \setminus \{1\}$, and $S$ denotes the set of skills given by the union of the skill requirements sets at the nodes and the skill sets associated with the pushback vehicles, i.e., $S = s_i : i \in N \cup s_p : p \in P$. Given non-negative skill-dependent traveling costs $c_{ij}^{p}$ for each $(i,j) \in A$ and pushback vehicle $p \in P$, we study the problem of defining the tour vehicles, each one starting and ending at node 1, in such a way that each service requirement of the considered aircraft is fulfilled by exactly one pushback vehicle, and the skill level constraints are satisfied. Moreover, we define $c_{ij}^{p}$ as the time needed by $p$ to traverse edge $(i, j)$ and $a_i, b_i \geq 0$ as the lower and upper bounds of the time window for node $i$. Finally, let the operation time $o_i \geq 0$ be the time needed to carry out the service at $i$ and $M$ be a large number. For each $(i, j) \in A$ and $p \in P$ with $s_p \geq \max\{s_i, s_j\}$ denote:

$$x_{ij}^p = \begin{cases} 1 \text{ if } (i,j) \in A \text{ is in the tour of vehicle } p, \\ 0 \text{ otherwise} \end{cases}$$

and

$$z_k^p = \begin{cases} 1 \text{ if } k \in V \setminus \{1\} \text{ is served by vehicle } p, \\ 0 \text{ otherwise.} \end{cases}$$

Moreover, let $y_{ij}$ be a non-negative flow variable for each $(i, j) \in A$ and $w_i^p \geq 0$ the time when the vehicle starts a service at node $i \in V \setminus \{1\}$.

The problem is then given as:

$$\min \sum_{(i,j)\in A} \sum_{p:s_p \geq \max\{s_i,s_j\}} c_{ij}^p x_{ij}^p \tag{1}$$

$$\sum_{i\in N} \sum_{p:s_p \geq max\{s_i,s_j\}} x_{ij}^p = 1 \quad \forall j \neq 1 \tag{2}$$

$$\sum_{i\in N:s_p \geq s_i} x_{ij}^p = \sum_{i\in N:s_p \geq s_i} x_{ji}^p \quad \forall j \neq 1, p: s_p \geq s_j \tag{3}$$

$$\sum_{i\in N} y_{1j} = n - 1 \tag{4}$$

$$\sum_{i\in N} y_{ij} - \sum_{i\in N} y_{ji} = 1 \quad \forall j \neq 1 \tag{5}$$

$$y_{ij} \leq (n-1) \sum_{p:s_p \geq \max\{s_i,s_j\}} x_{ij}^p \quad (i,j) \in A \tag{6}$$

$$w_i^p + o_i + c_{ij} - w_j^p \leq M\left(1 - x_{ij}^p\right) \ \forall (i,j) \in A: \neq 1; p: s_p \geq \max\{s_i, s_j\} \tag{7}$$

$$a_i z_k^p \leq w_k^p \leq b_k z_k^p \quad \forall k \neq 1, p: s_p \geq s_j \tag{8}$$

$$w_k^p \geq 0 \quad \forall k \neq 1, p: s_p \geq s_j \tag{9}$$

$$x_{ij}^p \in \{0,1\} \ (i,j) \in A, p: s_p \geq \max\{s_i, s_j\} \tag{10}$$

$$y_{ij} \geq 0 \quad (i,j) \in A \tag{11}$$

$$z_k^p \in \{0, 1\} \ \forall k \neq 1, p: s_p \geq s_j \tag{12}$$

The objective function (1) minimizes the total routing costs for the pushback vehicles. Constraints (2) and (3) guarantee that each node is served by an appropriate vehicle and that tours for the vehicles establish are established. The flow constraints (4) and (5) in combination with constraints (6), linking the decision variables, break potential sub tours. Constraints (7) and (8) enforce time windows, while constraints (9) - (12) are standard restrictions on the decision variables.

### A large neighborhood search algorithm

An initial CPLEX implementation of the introduced mathematical program for the problem leads to impractical computational times, even for small instances. For planning in the dynamic environment of an airport computationally efficient algorithms for the defined pushback routing problem are desirable. We have therefore developed a LNS algorithm for the problem. LNS algorithms have been successfully applied to VRPs with time windows (Pisinger and Ropke 2010). The core idea of an LNS heuristic is a large neighborhood that enables the algorithm to explore the solution space easily, even if the instance is tightly constrained. This is usually much harder with small neighborhoods (Pisinger and Ropke 2010).

Let $U$ be the set of feasible solutions for the pushback routing problem, then $u \in U$ is a single solution and $N(u)$ be the neighborhood of solution $u$, defined as the set of solutions that can be reached by applying the *destroy and repair* methods typical for LNS. The function $d(\cdot)$ is the destroy method, while $r(\cdot)$ is the repair method. That means, $d(u)$ returns a modified version of $u$ that is partly destroyed. In the case skill VRP for pushback vehicles, the destroy method removes a percentage of random gates, i.e. nodes of the VRP. The method $r(\cdot)$ repairs partly destroyed solutions, i.e. it returns a feasible solution constructed of the destroyed one. In our case, the repair method applies a greedy heuristic that gradually selects gates with the lowest cost and adds them to the solution.

The LNS procedure is illustrated in Algorithm 1. The algorithm uses three variables. $u^b$ is the best solution, $u$ is the current solution, and $u^t$ is a temporary solution that can either become the current solution or get discarded. The algorithm starts by initializing the global best solution $u^b$ using a feasible solution $u$.

This solution enters a loop in which the destroy method ($d(\cdot)$) and then the repair method ($r(\cdot)$) are repeatedly applied to obtain new solutions $u^t$. Then this solution is evaluated based on some criterion; we have used cost-improving solutions as a default criterion. If it is accepted, the current solution is updated. The accept function can be implemented in different ways. Next $x^t$ is evaluated, comparing its costs $c(u^t)$ to the costs of the best solution $c(u^b)$. The value $c(u)$ obviously corresponds to the objective function value of the model. If costs can be reduced, $u^b$ is updated. After that, the termination condition - a time limit in our case - is verified. Finally, the best solution found is returned.

**Algorithm 1** An LNS algorithm for the pushback routing problem

```
1:  u ← feasible solution
2:  u^b ← initialize(u)
3:  repeat
4:      u^t ← r(d(u))
5:      if accept(u^t , u) then
6:          u ← u^t
7:      end if
8:      if c(u^t) < c(u^b) then
9:          u^b ← u^t
10:     end if
11: until stop criterion
12: return u^b
```

**Computational results**

The numerical experiments have been executed on a computer equipped with an AMD Opteron Processor 6272 2.1 GHz and 128 GB of RAM under Windows Server 2012. We use the instances introduced by Schwarze and Voß (2015) that assume an airport with 17 gates and 6 pushback vehicles. The skills of the vehicles are $s_1 = s_2 = s_3 = s_4 = 2$ and $s_5 = s_6 = 3$. At each node (gate) there is a single aircraft that requires a pushback service with a certain skill level. We have three aircraft that require skill 1, seven requiring skill 2 and another seven requiring skill 3. The time windows start between time units 0 and 100 and have a constant length of 25 time units.

We have conducted the scenarios assuming *skill levels* and *skill sets* for the pushback vehicles. That means, vehicles with skill levels have downwards compatible skills and can serve all aircraft with skill levels less than or equal to their own skill, and vehicles with skill sets must exactly match the skill requirements. **Table 54** and **Table 55** display the results for both scenarios. The preliminary results demonstrate that we can solve all considered instances in less than three seconds with the implemented LNS.

Though the instances are relatively small, this confirms, that the LNS provides solutions in adequate computational time for the dynamic application. Instances of the described problem reflect a decentral organization at airports where different companies operate different fleets of pushback vehicles. Moreover, the implementation of skill sets or skill levels has no significant impact on the computational times, and computational times are consistently low for different initial solutions.

**Table 54.** Results for skill levels

| # | Routes | Jobs | Time | Costs |
|---|---|---|---|---|
| 1 | 9 | 17 | 2,65 | 328,44 |
| 2 | 9 | 17 | 2,61 | 512,64 |
| 3 | 8 | 17 | 2,58 | 873,69 |
| 4 | 9 | 17 | 2,59 | 382,47 |
| 5 | 8 | 17 | 2,49 | 719,19 |

**Table 55.** Results for skill sets

| # | Routes | Jobs | Time | Costs |
|---|---|---|---|---|
| 1 | 9 | 17 | 2,69 | 480,71 |
| 2 | 9 | 17 | 2,76 | 567,23 |
| 3 | 8 | 17 | 2,33 | 1297,27 |
| 4 | 10 | 17 | 2,48 | 590,23 |
| 5 | 8 | 17 | 2,34 | 982,21 |

## 10.5 Conceptual integration

To leverage the fuel, cost, and emission saving potential in taxiing, the pushback routing and control model must be linked to the queuing model for the taxiing process. **Fig. 93** sketches the integration of the respective models presented in Section 10.3 and Section 10.4. First, the simulation model built in SIMMOD is used to determine load-dependent emission curves. These curves basically model aircraft taxiing times based on the number of aircraft in the taxiing process - a concept known as load-dependent lead times in production planning (Pahl, Voß and Woodruff 2007). With the information from the simulation model, this model is extended to capture emissions as well. Next, the flight schedule is iteratively improved to minimize costs and emissions based on the queuing situation modeled using the load-dependent performance curves. This schedule is then given to the pushback routing and control model (skill VRP) to implement the flight schedule for the pushback processes, i.e., define the time windows for the routing problem. In the case of dynamic changes, this process is repeated. Finally, the simulation model evaluates how the plan is conducted, again considering costs, emissions, and delays. The implicit management strategies used here are changes in the flight sequence and thresholds for the taxiing queue. The final evaluation step considers costs and emissions, while emissions are weighted using various emission price levels discussed in literature (Edenhofer et al. 2015). This means that delays or waiting times are implicitly considered in terms costs caused by them.

## 10.6 Conclusion

Current research has emphasized the importance of the taxiing process for the overall aircraft emissions. Several studies appeared to capture and mitigate taxiing emissions. However, most approaches have not taken the interconnection into account of airport processes such as the pushback process. In this work, we have developed a simulation model for taxiing lead times and related costs as well as emissions. Furthermore, we have presented a pushback control and routing model based on skill VRP formulation, and implemented an LNS algorithm for fast solutions in the dynamic airport environment. We have also shown how the simulation model and the routing model can be integrated to fully leverage the cost and emission saving potential. Nevertheless, the presented results only refer to small benchmark results available in literature. Future work will therefore cover extensive numerical experiments with real-world data and comprehensive quantitative analysis of the integration of both the simulation and the routing model.

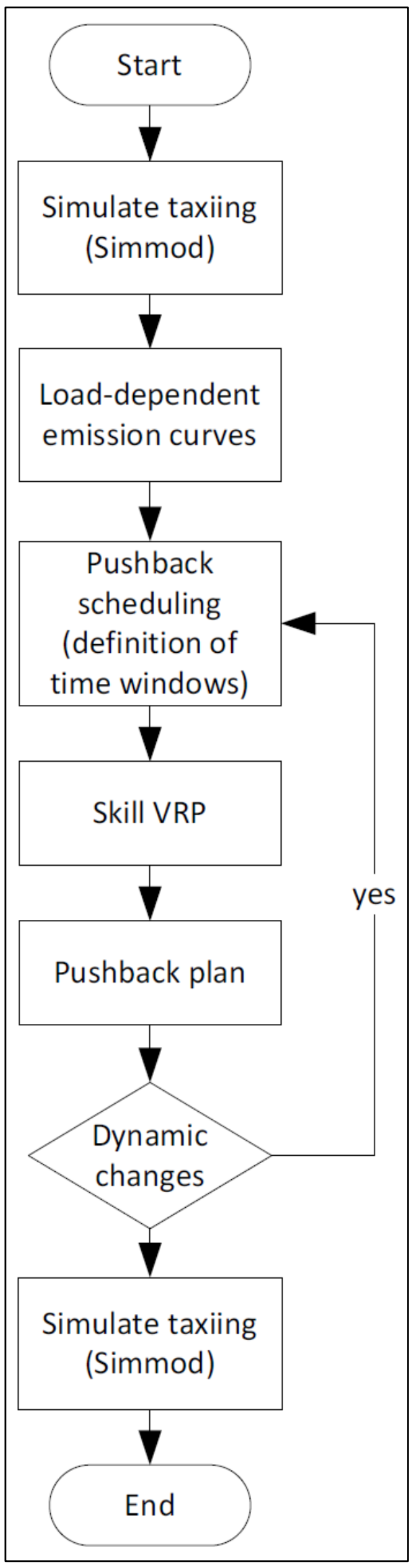

**Fig. 93.** Conceptual integration flow chart

# 11 Service quality and degree of interconnectedness in air transport networks*

Branko Bubalo†
University of Hamburg

Alberto A. Gaggero‡
University of Pavia

**Abstract.** This paper presents an application of an index of interconnectedness in air transport networks and studies empirically how the service quality of carriers, measured by average minutes of arrival delay, differs with a different network structure. We run panel data fixed effects on a sample of 2,526,180 flights operated within Europe during the period April 2015 to March 2016. We observe that carriers with contrasting business models, no frills versus full service, prefer different network types. Our main finding is that airlines operating a network with hub-and-spoke characteristics are more successful to manage delay, especially if the flight originates at one of their hubs, compared to airlines that operate in a network with point-to-point characteristics. Our findings support the hypothesis that carriers at hubs, where the carriers have a dominant position, internalize congestion into their own operations at these airports.

## 11.1 Introduction

The airline industry is characterized by two different business models: the low-cost carrier (LCC) model and the full-service carrier (FSC) model. The former typically features no frills service, single passenger class, extra fees for checked bags and low baggage weight allowance in exchange for a lower fare than FSCs. The latter historically represents the traditional model to run airline business and aims to provide a higher service quality than LCCs by offering multiple fare classes for the same flight (typically

* We thank two anonymous reviewers whose helpful comments contribute to improve the paper. We are grateful to Consuelo Nava and to the 2018 ATRS conference participants in Seoul. This paper was written partially while Gaggero was visiting the Department of Economics at the University of California Irvine in Fall 2018, Gaggero gratefully acknowledges the kind hospitality provided by Jan K. Brueckner and his Department. All errors are ours.

† University of Hamburg, Institute for Information Systems, Von-Melle-Park 5, 20146 Hamburg, Germany; telephone +49 40 42838 3064; fax +49 40 42838 5535; email: branko.bubalo@googlemail.com. *Corresponding author*.

‡ Department of Economics and Management, University of Pavia, Via S. Felice 5, 27100 Pavia, Italy; telephone +39 0382 986226; fax +39 0382 986228; email: alberto.gaggero@unipv.it.

first class, business and economy), exclusive airport lounges to first class and business passengers, better cabin services (complimentary, and often diverse, meal service, in-flight entertainment, etc.).

Recent developments in the air transport industry, however, show that these two business models have begun to converge, especially in the short-haul market (Klophaus et al., 2012; Daft and Albers, 2015). For instance, as of December 7$^{th}$, 2017 Air France and KLM introduce the "Light" fare, which does not include checked baggage in the ticket price. Under such a fare the checked baggage becomes a paid option in a similar fashion to what LCCs normally do. This practice is also adopted by American Airlines for their U.S. domestic flights.

On the LCCs side, besides the "standard" fare, easyJet sells seats also offering the "Flexi" fare, which comprises add-ons that are payable under the standard fare option (e.g. checked baggage, seat selection, priority boarding, etc.). In the U.S. Southwest Airlines, together with its cheapest "Wanna Get Away" class, make available more expensive fare classes, "Anytime" and "Business Select", which differ in terms of the type of benefits they include (i.e. weight allowance of checked bags, refundability, free beverages on board, etc.). On this line easyJet and Southwest Airlines mimic the practice of FSCs to offer more than one fare class at any given point in time.[1]

Despite these examples of business-model convergence, LCCs and FSCs still operate under a different network structure. Current paradigm is that FSCs prefer a hub-and-spoke (H&S) system, where the non-hub endpoints of the spoke are seamlessly linked through the stopover at the hub(s). Air France, KLM, British Airways, Lufthansa are examples of H&S airlines. LCCs, on the contrary, tend to adopt a point-to-point (P2P) system in which each destination is linked to some other destinations of the carrier's network by a non-stop flight connection; easyJet and Ryanair are examples of European carriers operating under such a system. Connecting flights beyond one destination are seldom offered to LCC passengers, whereas FSCs allow the booking of complex flight itineraries.[2]

One interesting research question is whether the differences in the network structure can affect the quality of service, measured in average delay, even though the business models are now more similar than in the past.

The answer to this question is not trivial because, as our analysis shows, the short-haul network structure of several airlines cannot be merely reduced to a pure H&S or a pure P2P system. This is because even if an airline operates under a P2P system, the endpoints of each route could be either the endpoints of several or of few routes; this feature would make the carrier's network more or less interconnected, thus reflecting to a smaller or to a larger extent the structure of a H&S system despite the fact that the network structure is declared P2P. In other words, it is important to assess, if an airline operates a network showing a more H&S characteristic or a more P2P topology. In our view the measure connectivity is not a binary variable, rather a continuous variable that differs among carriers.

---

1 It is worth mentioning that LCCs are also starting to explore the long-haul market, which is traditionally the domain of FSCs, thus pushing the convergence of business model even further. For example, Norwegian Air operates several U.S.-Europe routes; Ryanair, in partnership with Air Europa, has begun selling flights to the U.S. and Latin America from various European cities.

2 Technically all destinations of a FSC are reachable with one single airline ticket.

For this reason, in our analysis we construct an index of network interconnectedness to better measure the transition between H&S and P2P. Our index is then employed to assess which network system is more successful in reducing the arrival delay at the destination.

Using a panel of 2,526,180 flights operated on the European short-haul market during the period April 2015 to March 2016, we find that the closer the airline operates to a H&S system, the more effective is in reducing the propagating delay (i.e. the delay due to a late-arriving aircraft), especially if the flight originates from the airline's own hub. This finding relates to the fact that at its hub the airline is in a better position to absorb the propagating delay. The carrier tries to internalize any congestion to keep the schedule on-time (Brueckner, 2005). Indeed, at the hub the airline can more easily find crew substitutes in case of crew shortage or replace an aircraft if the arriving aircraft is too late for a subsequent scheduled connecting flight. Moreover, technical staff are normally based at the hub, hence they can quickly fix any technical problem. Also, at the hub the airline normally has an inventory of the replacement parts or a related service provider, since aircraft maintenance is normally carried out at the hub.

Our previous studies (Bubalo and Gaggero, 2015) have shown, that any change in average delay can have a significant impact. On the one hand a reduction in delay, even in marginal quantities, reduces fuel burn, fuel cost and emissions when multiplied across thousands of flights. On the other hand, an increase in delay has a nonlinear effect on the network in which such a flight is operating. We observe that delays occurring in a congested network, such we have in Europe, leads to an increase in propagating delay, that is cascading through other airports and affecting other flights in the network. These flights may or may not be operated by the same airline. At the end of an operating day, flights could build up delay that was inherited from other flights operating at different locations and at earlier times in the network. We, therefore, put attention to any (small) efficiency benefit, such as a decrease in average delay, that could be achieved through managerial change.

The contribution of our work to the existing literature is twofold. First, it presents a simple index of interconnectedness which can easily be replicated to better assess to what extent an airline network operates closer to a H&S or to a P2P system. Second, the richness of our data allows us to track the movement of each aircraft, uniquely identified by its registration number, through the network and, hence, we can study how delay propagates in different network systems. Furthermore, our analysis contributes to the discussion, if airlines internalize its externalities, in our case traffic congestion. Congestion may either be self-imposed or be caused by other carriers at neuralgic points in the network (Brueckner, 2002b, 2005; Santos and Robin, 2010).

The rest of the paper is structure as follows: the next section reviews the main contributions of the literature. The data collection process is described in Section 3, followed by the empirical analysis in Section 4, which first describes our index of interconnectedness together with some descriptive statistics and then continues with the econometric investigation and discussion of the results. Finally, Section 5 summarizes and concludes the paper.

## 11.2 Literature review

In general, a very broad literature is available on networks, especially with applications to transport or information systems. We drew from interesting papers outside our main specialization to gain new insights. Particularly in Miranda et al. (2018) we found a fitting network indicator from the field of digital signal processing that serves our purpose. We used graph (edge) density, i.e. the average number of neighbors to a node in a carrier network, as our main measure to define connectivity. In Guimera et al. (2005) this measure is called the average 'degree'. We scaled our connectivity measure to a distribution between 0 and 100, where 0 represents a hypothetical network with no connections from any airport at all, a number close to 0 defines a network that has H&S characteristics, a number approaching 100 defines a P2P network and 100 represents a fully connected network, i.e. complete graph.[1]

Lawson et al. (1997) look at queuing delay in transportation network using input/output diagrams. Regarding the airline industry, Kim and Hansen (2013) discuss more generally demand, throughput and delay in capacity constrained server systems that develop a backlog of queuing flights with an application to U.S. airports. Cook and Goodwin (2008) compare the operating costs of H&S networks versus P2P networks. They argue that the benefits of a H&S structure are ultimately limited by high operating costs. A more recent paper by Trapote-Barreira et al. (2016) collects several indicators to define network structure. These indicators are applied to networks in the airline industry to enable a better management of complex systems. Zanin and Lillo (2013) discuss the characteristics of H&S versus P2P networks and they analyze the topology of international air transport networks as either unweighted or weighted graphs. Earlier work from Burghouwt and Hakfoort (2001) applies a cluster analysis to the European air transport network. In their work airports are grouped by primary hubs, secondary hubs, medium-sized airports, small airports and very small airports. In our view, an arbitrary grouping of airports does not adequately reflect the structure and interconnectedness of a network.

[1] Our chosen indicator has some limitations when the number of airports in a carrier network is very small or for certain special topologies. The discrepancy between a P2P and H&S network structure, expressed in our connectivity measure, increases with the number of airports in the network. Therefore, this measure is best suited when the structure of two carriers serving a similar sufficiently large number of airports in their respective network are compared. In our study the top half of the carriers have a sufficiently large number of airports in the network. We made trial calculations of the indicator 'hierarchical degree' level 2 (Miranda et al., 2018), which means we count the neighbors of the direct neighbors of a node. We found that calculating this alternative measure is not practical, since a double counting of nodes must be avoided. Another weakness is that our indicator does not account for how the number of neighbors of an airport is distributed among all airports in the network, which means theoretically a circular path with an average number of two neighbors which are uniformly distributed among all airports could have a similar connectivity measure than a network with several spokes connected to a hub. In reality most networks are 'scale-free', which means some nodes have a much larger than average number of neighbors. The distribution of number of neighbors in scale-free networks usually follows a power-law distribution, which can be observed for most airline networks in our sample. Please see Guimera et al. (2005) and Latg´e-Roucolle et al. (2018) for a more complete discussion on advantages and limitations of different descriptors of network structure, such as the concept of "betweenness centrality".

With respect to delay propagation in air transport, we were inspired by groundwork of e.g. AhmadBeygi et al. (2010), Forbes and Lederman (2010), Wang (2015), Fleurquin et al. (2013) and Fleurquin et al. (2014). Especially Wang (2015) stresses the important connection between business model, network structure and level of service. He puts an emphasis on how delay can be reduced by carriers at the arrival airport, with an analysis of turnaround times and scheduled buffer times in the flight schedules. He finds that sufficient turnaround plus buffer times are particularly important at hub airports where feeder flights, which are typically connected to intercontinental flights, arrive. AhmadBeygi et al. (2010) take a similar path, however, they use operating costs and a change in scheduling to derive a model which increases operational productivity in air transport networks. Forbes and Lederman (2010) look at the extend of vertical integration of major airlines in the U.S. on on-time performance. They differentiate between integrated and independent regional carriers, which are carriers either owned or commissioned through contracts by a major airline, respectively. They find that owned regional carriers have a better on-time performance, measured in average departure delay. Their data set consists of flight schedules and on-time performance data. Network structure is not discussed in Forbes and Lederman (2010), but 'hub' airports are represented by a dummy variable that was defined by the authors. Brueckner (2002a, 2002b, 2005) shows that U.S. airlines internalize the congestion costs of an additional flight in their own network. He compares H&S and 'fully-connected' (read P2P) networks. He finds that the network structure chosen by carriers has a large impact on costs and scheduling efforts. These works only cover the U.S. airline market.

Ren and Li (2018) use aircraft tracking data to develop characteristics of airline networks. They compare the network characteristics between the U.S. and the Chinese market and try to optimize the number of routes in each network by clustering methods. They find that compared to the great-circle route between two airports, aircraft follow the routes more directly in China, due to airspace restrictions, than in the U.S., where flights have more diverse trajectories on each domestic route.

Regarding Europe the literature on air transport network operations and performance is scarce. Klophaus et al. (2012) compare services between LCCs and FSCs in Europe and conclude that the observed airlines converge with their service offerings into a 'hybrid' airline business model.

Fundamental work regarding delay and networks in Europe comes from Burghouwt and de Wit (2005), Santos and Robin (2010) and Bubalo and Gaggero (2015). Santos and Robin (2010) provide an excellent study in which aggregated delay data on European airports from 2000 to 2004 was analyzed. They found that the distribution of delay was U-shaped across hub size, which means delay was lowest at medium-sized European hubs. Regarding European airports they found similar evidence that carriers internalize congestion compared to what Brueckner (2005) found for the case of U.S.

airports. Another interesting finding is that on-time performance at the point of origin increases with the slot coordination level.[1] The study carried out by Santos and Robin (2010) has two main limitations, namely the arbitrary choice of hub size, based on the number of destinations, and the lack of using flight-by-flight data.

Our study could be viewed as an extension to our previous study in (Bubalo and Gaggero, 2015), where we investigated service quality related to competition, particularly from LCCs, and market power at European airports. Brueckner (2002a) includes similar fundamental work related to the U.S. market.

Sternberg et al. (2016) offer interesting insights into airline and airport networks and performance in the Brazilian air transport market. Besides covering an important geographical area in South America, they find that delay from incoming flights into Brazilian airports can hardly be absorbed on the ground during the turnaround. This situation worsens under inclement weather and traffic congestion, which lead to increasing delay propagation in the Brazilian air transport network. Latgé-Roucolle et al. (2018) follow a similar approach to what we present in this paper, however their analysis lacks performance or service quality indicators such as flight delay.

Therefore, to the best of our knowledge, our analysis is unique in combining several aspects of the air transport network operations in a coherent analysis with respect to Europe. Another novelty in our work is the precise identification of aircraft by its registration number, by which we can calculate the amount of propagation delay carried by each flight in the network. The relevance of connectivity in airline networks in Europe is underlined by the frequently published 'Airport Industry Connectivity Report' from (Airport Council International Europe, 2018), which is the main airport association in Europe and represents more than 500 airports.

## 11.3 Data collection

For this study we have programmed several web crawlers, which collected data and information from the Internet. The very first crawler collected flight positioning data from a data provider that offers real-time and historic tracking of global flights. We carefully build the crawler to be non-intrusive, as it was limited in the number queries per minute. We imitated user behavior to retrieve all information. The main difference to a human user is, that we can run the crawler continuously over prolonged periods of time.

With the first crawler we collected from flightradar24.com the ADS-B positioning signals of most global flights. Together with the position, we also retrieved the aircraft registration (or tail) number, which is essential to our analysis.

[1] The slot coordination level expresses the severity of congestion at an airport. An airport with coordination level 3, for example, has insufficient capacity during peak hours most of the day, which means its hourly landing and take-off permission, i.e. slots, are limited and allocated to airlines within the slot coordination process. A level 2 coordinated airport is reaching capacity only at certain periods of the day or during specific seasons and has the freedom to choose, if it wants to participate in the slot allocation. A level 1 coordinated airport is in general able to meet demand but needs to monitor its operations.

The coverage of received positioning signals over land is very complete and accurate, except for very remote regions. Over sea and far away from the shores we have hardly any coverage. Over the ocean we observe large gaps in the positioning signal database. To solve this issue and identify as many active aircraft as possible, we repeated the position sampling over two weeks every minute during the day, so that unobserved aircraft at some point may be observed in the airspace. Over this period, we identified 20,455 different aircraft, that operated globally during our initial observation period.

With the list of aircraft registration numbers, the second web crawler queried flightradar24.com approximately every third day to follow the flight rotations through airports and time. It took about three days, until the list of aircraft registration numbers was completely queried. The schedule data of each aircraft was available for about five days. As mentioned above, we wanted to collect the data in a non-intrusive way. We were able to collect the flight schedule, including actual and scheduled times, of most flights operated by international passenger airlines. This means we are able to observe exactly in which sequence a given aircraft travels from one airport of origin to a destination, how long this aircraft remains at the airport for the turnaround, and how long it is delayed during flight and at the ground.

Our collected database consists of more than 17.2 million global flight observations that were gathered over a period from April 2015 to March 2016. For our list of aircraft, we were able to record continuous flight schedule data over the collecting period. With the flight schedule information we can calculate the delay comparing the scheduled to the actual times of each flight. By knowing the registration number of each flight, we can observe the buildup or decrease of delay of each aircraft over the day of operation. By accounting for the delay of the arriving aircraft at the airport, we know how delay propagates through the network, from one airport to another, because delayed incoming flight cannot only impact its own itinerary but could also delay other connecting flights operated by the same carrier.

Our sample focuses on the European market, which we believe is a good system to test our model. The market is well developed, but still comprises a large number of both FSCs, since each European country has a flag carrier, and LCCs, which have spread out following the enlargement of the European Union.

Finally, as in Bubalo and Gaggero (2015) or Alderighi and Gaggero (2018), we include weather variables to account for their external effects on flight operations. In order to collect weather data, so called “METAR” (Meteorological Terminal Aviation Routine Weather Report), we have built another web crawler which queried weatherunderground.com for each airport observed during our collection period. This resulted in the collection of 9.2 million weather data records. Weather data is typically published in 30 minutes intervals, which means for each airport in our sample, we collected weather information twice each hour during the whole collection period.

## 11.4 Empirical analysis

In this section we first present the connectivity index to describe differently structured networks and secondly, we conduct some related descriptive and graphical analysis based on this index. Then, we detail our econometric model followed by the estimation and discussion of the results.

**Connectivity index**

One of the main goals prior to our econometric analysis is the construction of a measure which can adequately describe the extent of interconnectedness in different airline networks, differentiating between H&S and P2P networks.

With our data set we have an accurate overview of the airports served by each carrier in its own network. By analyzing all flown routes between airports, we know which airports are connected and to which extent, i.e. on which airport and city pairs flights are offered.

We understand H&S networks as graphs that have one or a few nodes connected to most of the other airports in the carrier's network, whereas we understand that P2P airports are more uniformly connected among each other. As examples, consider **Fig. 94** to **Fig. 97**, which present the different network structures for the European destinations served by British Airways (**Fig. 94**), Lufthansa (**Fig. 95**), easyJet (**Fig. 96**) and Ryanair (**Fig. 97**) during our sample period.

British Airways show a well-defined H&S structure since it is practically impossible to travel mainland Europe without a layover in London. Lufthansa is another example of H&S network, but with a dual-city hub structure: Frankfurt and Munich. Ryanair and easyJet, on the contrary, are P2P carriers since they show a very spread network, despite some focal points that can be identified.[1]

[1] They represent the airline's "operational bases" where aircraft are parked overnight or from where backup aircraft, maintenance equipment and technicians can be requested, but these are no "hubs" where flights are concentrated and connected to other hubs. Examples of such operational bases are London Stansted or Bergamo for Ryanair; London Gatwick or Milan Malpensa for easyJet.

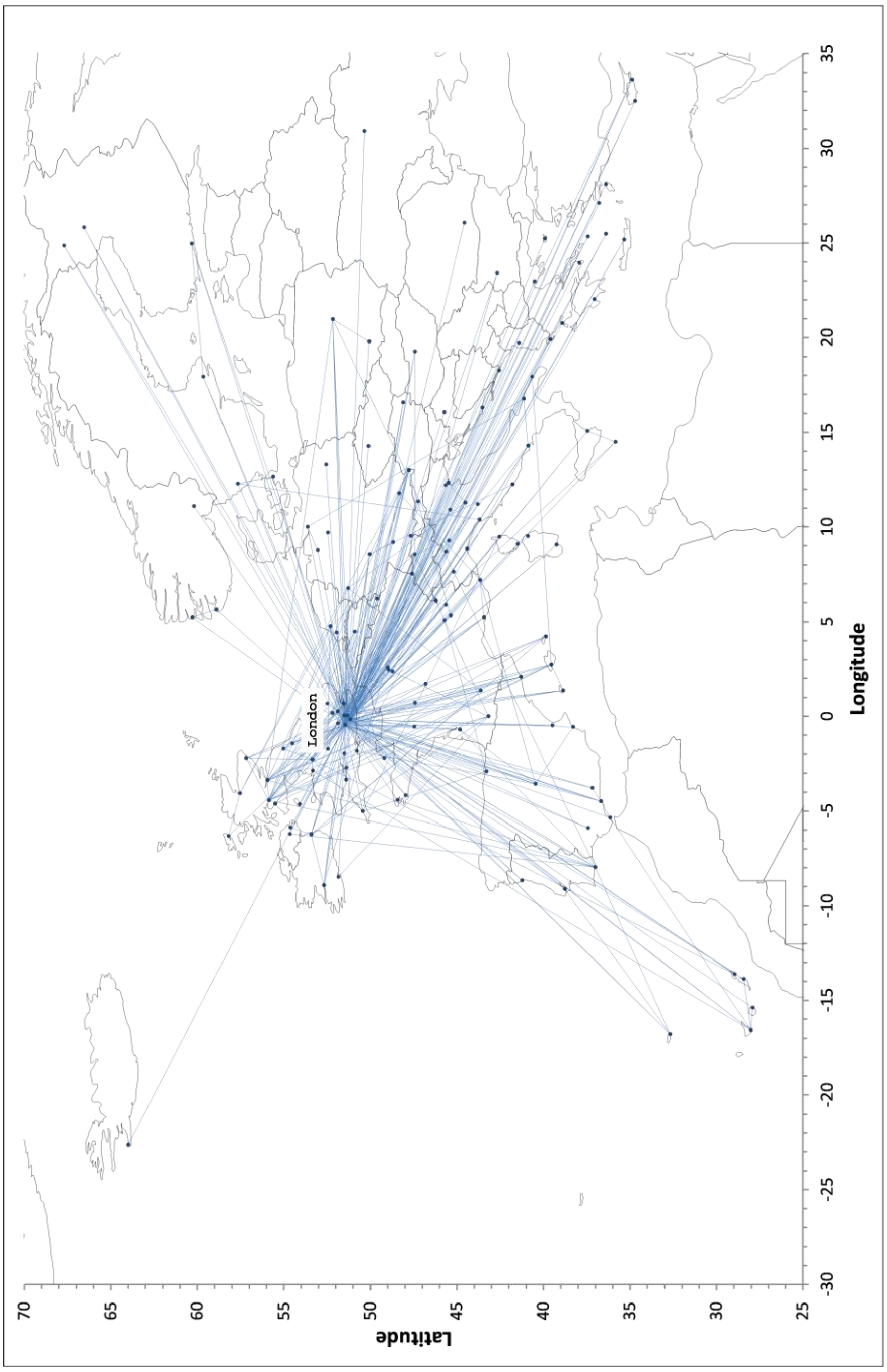

**Fig. 94.** British Airways` European Network (H&S)

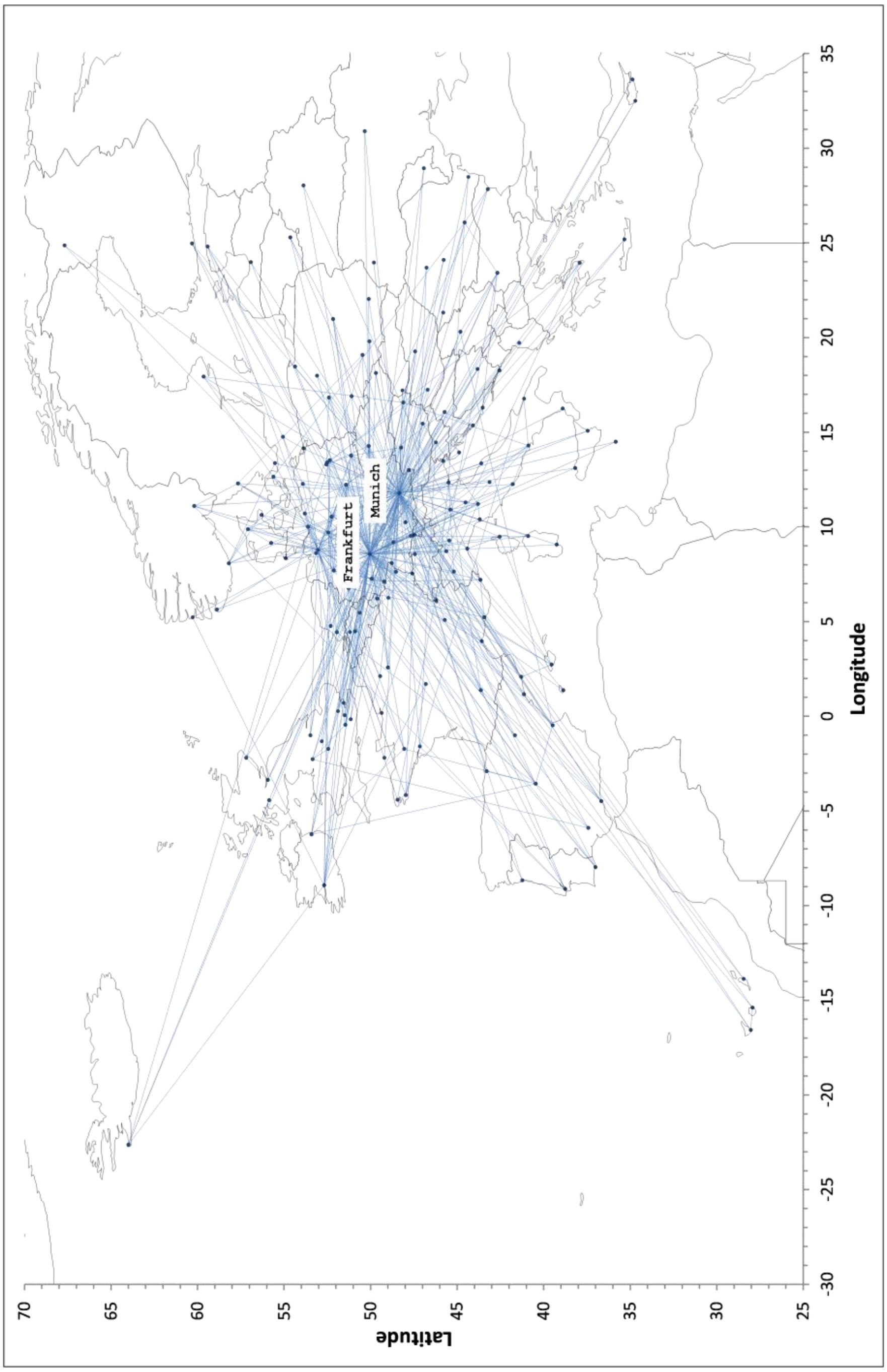

**Fig. 95.** Lufthansa`s European Network (H&S)

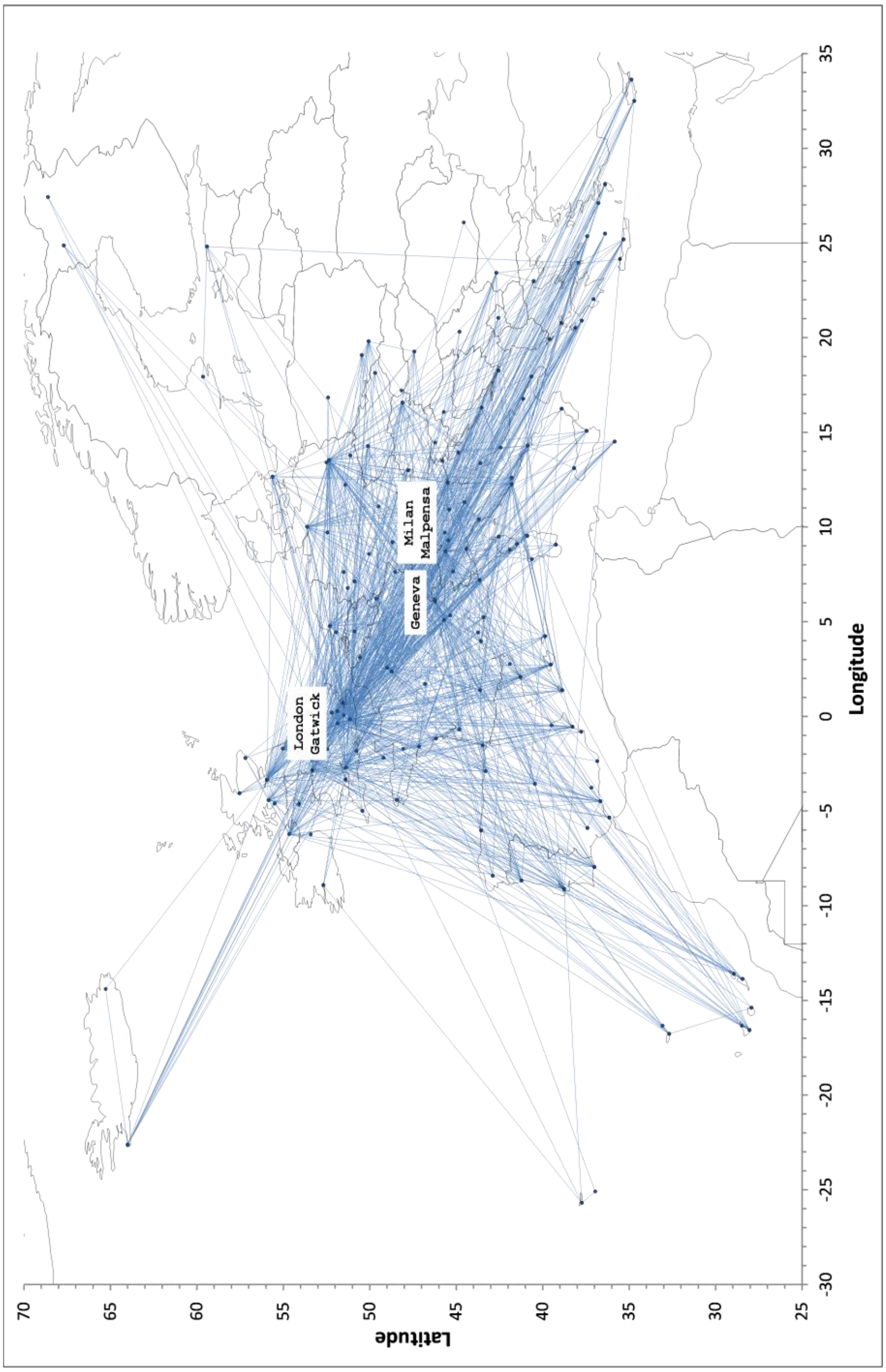

**Fig. 96.** easyJet`s European Network (P2P)

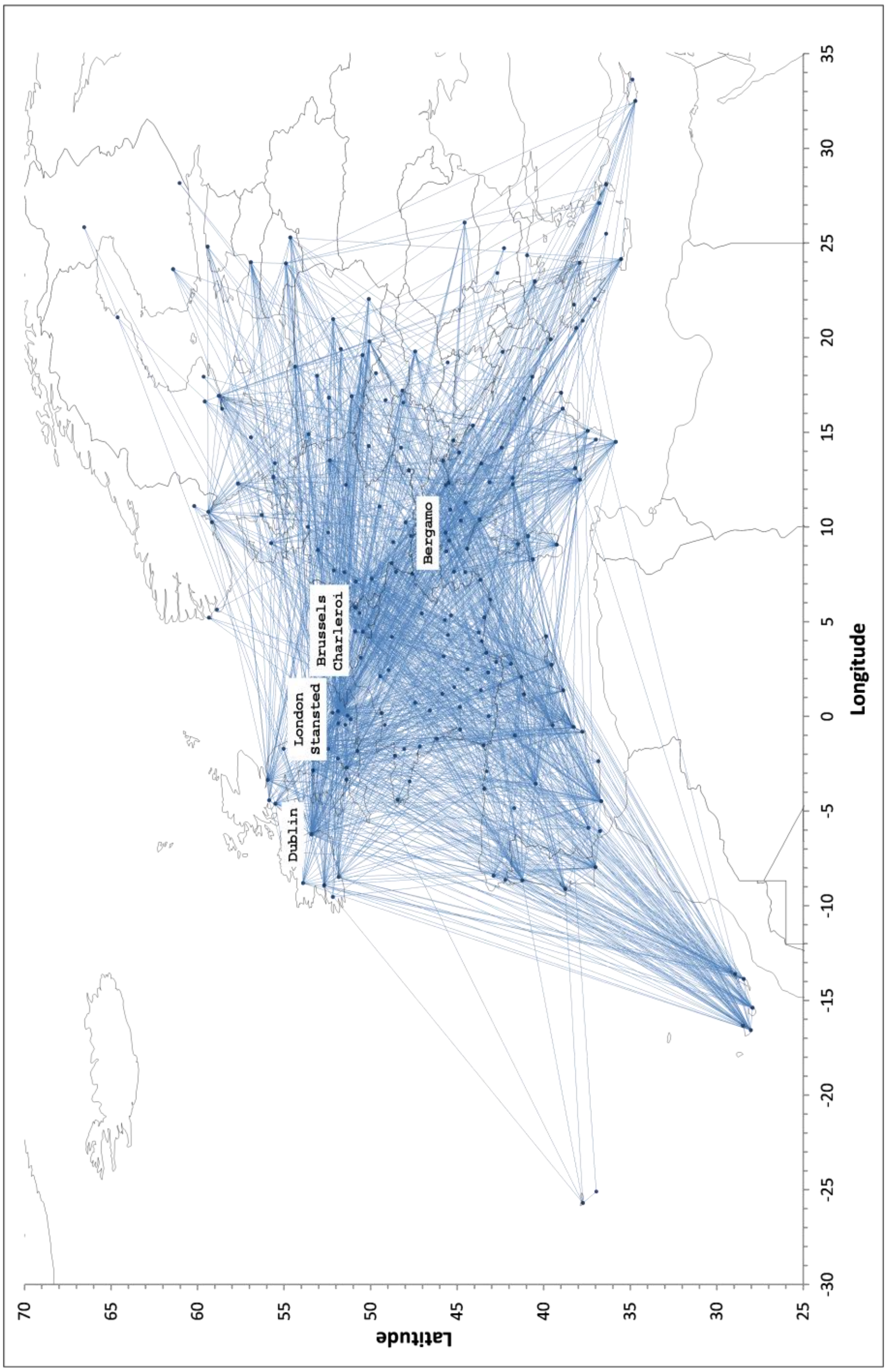

**Fig. 97.** Ryanair`s European Network (P2P)

Our index of interconnectedness, or in short connectivity index *Cindex*, for carrier *c* is summarized by the following formula:

$$\text{Cindex}_C = \frac{\sum_{p=1}^{N} k_p^c}{N^c(N^c-1)} * 100\% = \frac{\overline{k_p^c}}{N^c-1} * 100\% \qquad (1)$$

where $k_p^c$ is the number of carrier *c*'s destinations from airport *p*; $N^c$ is the total number of airports in the network of carrier *c* and $\overline{k_p^c}$.

The index technically can range between 0 and 100. The closer to zero, the more an airline operates under H&S system, the closer to 100 the more an airline operates under P2P system. The index is calculated for each carrier in the sample monthly to consider any seasonality effect in the choice of the number of destinations served by each carrier.

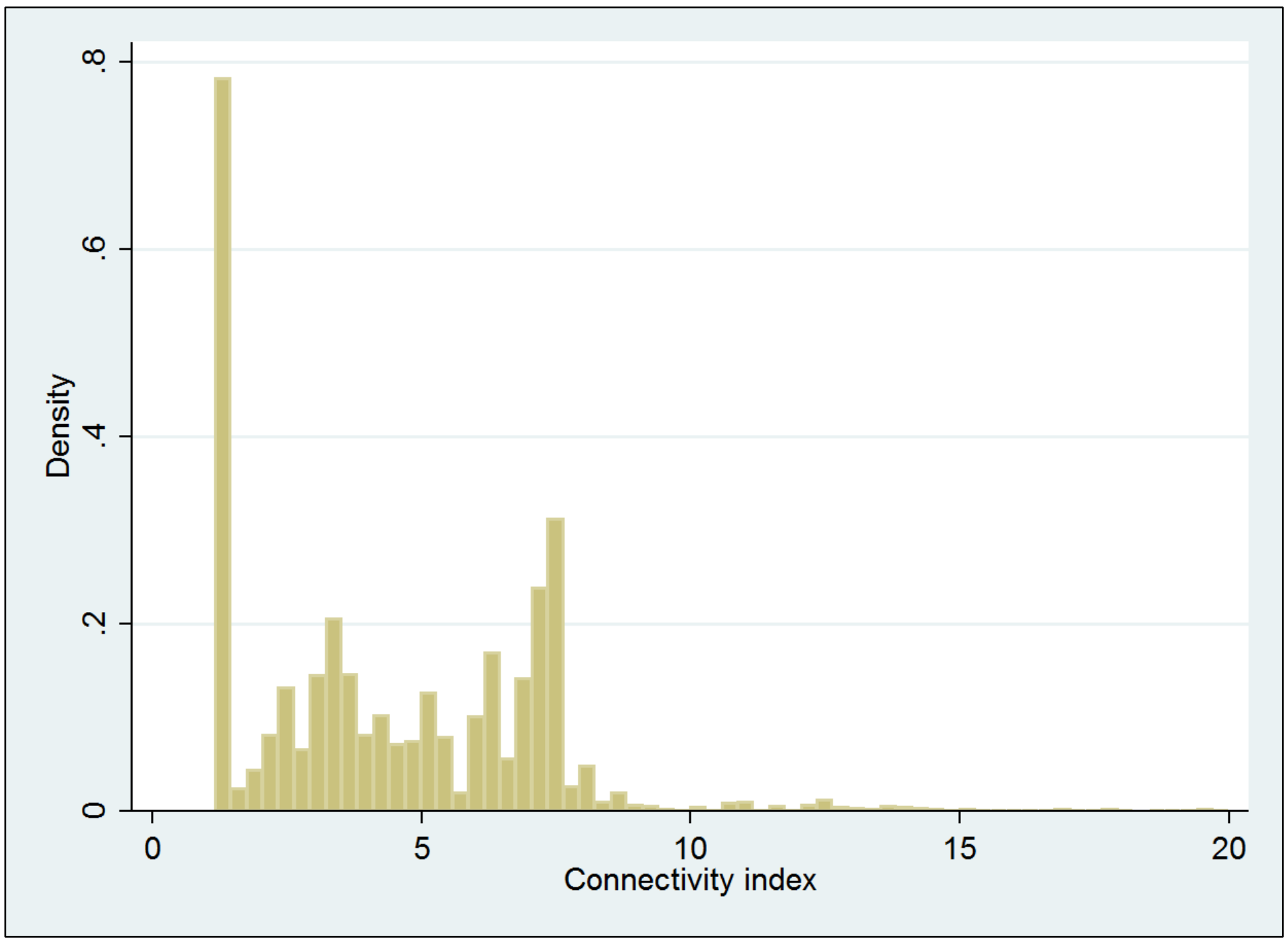

**Fig. 98.** Distribution of *Cindex*

The **Fig. 98** displays practically the entire distribution of *Cindex*, which ranges from a minimum of 1.16 to a maximum of 66.67.[1] **Table 56** reports the average value and standard error of *Cindex* during the sample period for each airline included in our sample. A higher standard error denotes a larger extent of variation of *Cindex* within the airline and therefore implicitly suggests that the carrier's route map is characterized by higher seasonality. Indeed, major flag carriers like Air France, British Airways and Lufthansa show the lowest standard errors, because they are weakly affected by seasonality since they generally serve their destinations all year around.

[1] To make the diagram more visible, we exclude from the figure the values of *Cindex* above 20, which corresponds to the 99.7th percentile of *Cindex*.

**Table 56** separates the airlines between FSCs and LCCs.[1] The weighted average connectivity index is 3.04 for FSCs and 6.63 for LCCs.[2] These two numbers well summarize **Table 56**, which indicates that in general FSCs tend to operate through a H&S system, while LCCs are more inclined towards a P2P system. This difference, however, is not always clear, if we compare airlines individually. For example, Vueling, a Spanish LCC, has an average *Cindex* four time smaller than the one of Croatian Airlines, the Croatian flag carrier; HOP!, a regional carrier subsidiary of Air France, and Air Dolomiti, a feeder airline of Lufthansa, are both FSCs, but present a stronger P2P network structure (have a higher average *Cindex*) than easyJet and Ryanair, the two major LCCs in Europe.

Thus, it is not possible to argue that all FSCs are pure H&S airlines and that all LCCs are pure P2P airlines. For this reason, it is necessary to consider *Cindex* when studying the relationship between on-time performance and type of business model of a carrier.

### Econometric model

Our econometric model aims to explain the determinants of flight delay through panel data fixed-effects. The panel individual is defined by the combination of aircraft and flight day, while the time dimension of the panel is the sequence of flights that a given aircraft operates during the day.[3] For this reason, the panel fixed-effect is both aircraft and airline specific.

The dependent variable of our econometric model is represented by the minutes of arrival delay of the aircraft at the destination and is explained by a set of independent variables that can be broadly classified as flight, route, airport and weather specific. More specifically, the list of included regressors is as follows.

---

1 LCCs are identified based on the list provided by the International Civil Aviation Organization (available at https://www.icao.int/sustainability/documents/lcc-list.pdf), while FSCs are the counterfactual, typically formed by flag carriers or their regional affiliates.

2 The weights in the computation of the mean are given by the number of flights operated by each airline.

3 As an example consider the journeys made by Air France aircraft with tail registration FGRHH on June 24th, 2014: the first flight of the day is Paris-Vienna, followed by Vienna-Paris, Paris-Munich, Munich-Paris, Paris-Geneva and Geneva-Paris, which is the last flight operated by the FGRHH aircraft on June 24th, 2014. FGRHH and June 24th, 2014 identify the panel individual, while the panel time is given by the sequence *1* for the flight Paris-Vienna, *2* for the flight Vienna-Paris and so on up to *6* for the flight Geneva-Paris.

**Table 56.** Average *Cindex* by airlines with standard error in parentheses

PANEL A -Full-Service Carriers

| | | | | | | | | |
|---|---|---|---|---|---|---|---|---|
| Adria Airways | 6.02 | (1.51) | Austrian Airlines | 2.39 | (0.18) | Icelandair | 4.71 | (0.92) |
| Aegean | 4.57 | (0.63) | BMI Regional | 6.84 | (0.96) | KLM | 1.37 | (0.03) |
| Aer Lingus | 4.10 | (0.63) | Belavia | 3.31 | (1.84) | LOT | 3.63 | (0.61) |
| Aigle Azur | 14.23 | (2.48) | Binter Canarias | 21.09 | (3.41) | Lufthansa | 1.38 | (0.04) |
| Air Baltic | 5.38 | (1.79) | British Airways | 1.22 | (0.07) | Luxair | 4.19 | (1.46) |
| Air Berlin | 5.16 | (0.44) | Brussels Airlines | 2.78 | (0.36) | Montenegro Airlines | 14.49 | (4.23) |
| Air Corsica | 24.93 | (5.83) | Bulgaria Air | 6.31 | (0.71) | Primera Air | 12.82 | (5.29) |
| Air Dolomiti | 19.63 | (6.52) | Croatia Airlines | 12.00 | (3.69) | SAS | 3.51 | (0.22) |
| Air Europa | 5.12 | (1.19) | Czech Airlines | 5.33 | (0.96) | SATA Air Açores | 18.71 | (2.76) |
| Air France | 1.38 | (0.05) | Edelweiss Air | 6.45 | (1.47) | Swiss | 2.19 | (0.38) |
| Air Malta | 5.52 | (1.38) | Estonian Air | 8.06 | (2.88) | TAP Air Portugal | 3.00 | (0.12) |
| Air Moldova | 8.68 | (1.96) | Eurowings | 5.97 | (1.90) | TAROM | 5.13 | (0.68) |
| Air Serbia | 5.11 | (1.06) | Finnair | 2.40 | (0.27) | TUI fly | 4.69 | (0.28) |
| Alitalia | 3.29 | (0.36) | HOP! | 9.13 | (2.02) | Thomas Cook | 5.58 | (0.74) |
| Atlantic Airways | 20.87 | (7.21) | Iberia | 2.24 | (0.14) | Ukraine Int. Airlines | 4.27 | (0.40) |

PANEL B -Low-Cost Carriers

| | | | | | | | | |
|---|---|---|---|---|---|---|---|---|
| Blue Air | 7.78 | (1.84) | Meridiana | 7.49 | (0.97) | Travel Service | 3.11 | (0.54) |
| Blue Panorama | 8.70 | (1.69) | Monarch Airlines | 13.31 | (1.36) | Volotea | 8.26 | (2.26) |
| Condor | 6.48 | (0.70) | Niki | 4.62 | (0.89) | Vueling | 3.91 | (0.82) |
| Flybe | 9.64 | (1.37) | Norwegian Air | 5.29 | (0.55) | WOW Air | 13.39 | (2.82) |
| Germania Express | 5.18 | (0.85) | Ryanair | 6.95 | (0.58) | Wizz Air | 6.92 | (0.40) |
| Germanwings | 5.36 | (1.33) | SunExpress | 17.00 | (1.92) | easyJet | 7.27 | (0.48) |
| Jet2.com | 13.25 | (1.15) | Transavia | 3.46 | (0.68) | | | |

*Propagation* is the minutes of arrival delay of the incoming aircraft. This may generate a knock-on effect in the flights downstream operated by a carrier: the late arrival of the incoming flight leaves out less time for ground operations which may cause the late departure and eventually the late arrival of the next flight operated by the same aircraft; the second flight will be the incoming flight of the third flight, thus the late arrival of the second flight may affect the punctuality of the third flight, and so on. At a hub, a late arrival could not only impact the punctuality of operations of an aircraft, it may also impact connecting flights of the own carrier or even other airlines. This externality may impact other competitors at an airport. Since especially LCCs have tight turnaround times to offload the arriving aircraft and prepare it for the next departure, any schedule disruption may affect later flights during the day. *Propagation*, therefore, aims to capture the effect of propagating delay, which, as we argue in this paper, can be different depending on the different network structure.

The time of departure is coded into four dummy variables equal to 1 if the departure time is respectively in the *Early morning* (0.00−6.59 am), *Morning* (7.00−11.59 am), *Afternoon* (12.00 am−4.59 pm), and *Evening* (5.00−11.59 pm). *Early morning* is the omitted category in the regression.

*Distance* is the route length, measured in 100 km; while *Route market shr* is the carrier's route market share, measured using the number of flights operated on the route during the day of the scheduled departure of the flight.

*Airport mkt shr - orig.* and *Airport mkt shr - dest.* are respectively the carrier's airport market shares at origin and at destination, also computed based on the number of scheduled flights at the airports.[1]

*Airport congestion* is the number of flights landing at the airport within the one-hour window when the observed aircraft is expected to land, to control for the negative externality of congested times on arrival delay (Mazzeo, 2003; Ater, 2012).

*Airport volume* is the total number of flights at the airport during the entire day, to consider the capacity of the airport (Mayer and Sinai, 2003; Rupp, 2009).

A set of weather variables both at the origin and at the destination to control for the meteorological effect on flight operations. Similarly to Bubalo and Gaggero (2015), we classify weather into the following categories: *Cloud*, *Fog*, *Haze*, *Rain*, *Snow*, *Sun* and *Thunder*, each identified by a dummy variable; *Sun* is set as the omitted category in the regression. It is worth mentioning that the weather conditions at the origin and at the destination are sampled within a half-hour window prior to departure and arrival, respectively. This accuracy is relevant to measure the weather at the time when it is likely to affect the flight.[2]

*Deicing operations* is a dummy variable equal to one if, still within the aforementioned half-hour window from departure, the temperature is below 0 degrees

---

[1] See Brueckner (2002a) for a detailed study on airport market power and airport congestion.

[2] With very few exceptions, namely Bubalo and Gaggero (2015) and Alderighi and Gaggero (2018), most of the existing literature uses an average daily value of the meteorological condition and therefore may imprecisely measure the effect of weather on airline delay. Consider for example a foggy morning which clears in the day: the average meteorological condition would most likely be clear sky, however the on-time performance of morning flights would probably be severely affected by the fog, which an average daily measure cannot capture.

Celsius. This variable aims to account for possible deicing operations at the airport of origin, irrespective of the actual meteorological condition.

The descriptive statistics of the variables included in our econometric analysis are reported in **Table 57**.

Finally, the standard errors are clustered by origin-date to allow the residuals of different flights, even of different airlines, originating from the same airport in a given day to be correlated. This clustering strategy takes into consideration those unobserved external shocks that may be common amongst the flights of each cluster: for example, a national strike or a technical disruption at a given airport is likely to affect any flight taking off from that airport on the day such shock occurs.

**Table 57.** Descriptive statistics (number of observations: 2,526,180)

| Variable | Mean | Std. Err. | Min | Max |
|---|---|---|---|---|
| Delay | 5.39 | 14.02 | 0.00 | 146.00 |
| Propagation | 3.92 | 11.92 | 0.00 | 146.00 |
| Morning | 0.30 | 0.46 | 0.00 | 1.00 |
| Afternoon | 0.30 | 0.46 | 0.00 | 1.00 |
| Evening | 0.22 | 0.42 | 0.00 | 1.00 |
| Distance | 9.91 | 6.30 | 1.01 | 51.34 |
| Route market share | 0.74 | 0.28 | 0.03 | 1.00 |
| Airport congestion | 13.43 | 33543 | 1.00 | 67.00 |
| Airport volume | 212.87 | 184.30 | 1.00 | 689.00 |
| Airport mkt shr - orig. | 0.30 | 0.25 | 0.00 | 1.00 |
| Airport mkt shr - dest. | 0.30 | 0.25 | 0.00 | 1.00 |
| Cloud - origin | 0.53 | 0.50 | 0.00 | 1.00 |
| Cloud - destination | 0.53 | 0.50 | 0.00 | 1.00 |
| Fog - origin | 0.03 | 0.17 | 0.00 | 1.00 |
| Fog - destination | 0.02 | 0.15 | 0.00 | 1.00 |
| Haze - origin | 0.03 | 0.18 | 0.00 | 1.00 |
| Haze - destination | 0.03 | 0.18 | 0.00 | 1.00 |
| Thunder - origin | 0.00 | 0.05 | 0.00 | 1.00 |
| Thunder - destination | 0.00 | 0.05 | 0.00 | 1.00 |
| Rain - origin | 0.13 | 0.33 | 0.00 | 1.00 |
| Rain - destination | 0.13 | 0.33 | 0.00 | 1.00 |
| Snow - origin | 0.01 | 0.11 | 0.00 | 1.00 |
| Snow - destination | 0.01 | 0.11 | 0.00 | 1.00 |
| De-icing operations | 0.06 | 0.24 | 0.00 | 1.00 |
| Cindex | 4.57 | 2.93 | 1.16 | 66.67 |

## 11.5 Results

This paper aims to investigate whether a different extent of network structure, either closer to H&S or closer to P2P, can have a different impact on how the delay propagates through the network. To test such a hypothesis, we evenly split our sample into three sub-samples, A, B and C, according to the different values of *Cindex*. Sub-sample A collects those airlines whose *Cindex* is smaller or equal to its thirty-third percentile; sub-sample B is for observations strictly above the thirty-third percentile and below or equal to the sixty-sixth percentile of *Cindex*; sub-sample C comprises the observations strictly above the sixty-sixth percentile.[1]

**Table 58.** The determinants of flight delay

| Carriers | (1) All | (2) H&S | (3) Hybrid | (4) P2P |
|---|---|---|---|---|
| Propagation | 0.687*** | 0.200*** | 0.251*** | 0.319*** |
| Morning | 0.449*** | 1.643*** | 1.645*** | 1.670*** |
| Afternoon | -0.308*** | 2.186*** | 2.059*** | 1.500*** |
| Evening | -0.199*** | 1.700*** | 2.552*** | 2.829*** |
| Distance | 0.075*** | 0.055*** | 0.069*** | 0.046*** |
| Route market share | 0.022 | 0.551*** | 0.545*** | 0.185*** |
| Airport mkt shr - orig. | -0.555*** | -0.300*** | -0.040 | -0.112 |
| Airport mkt shr - dest. | -0.755*** | -0.135 | 0.118 | 0.257*** |
| Airport congestion | 0.033*** | 0.027*** | 0.052*** | 0.037*** |
| Airport volume | 0.002*** | 0.002*** | 0.001*** | 0.003*** |
| Cloud - origin | -0.293*** | -0.011 | -0.072* | 0.121*** |
| Cloud - destination | 0.198*** | 0.365*** | 0.401*** | 0.381*** |
| Fog - origin | 1.395*** | 1.836*** | 1.880*** | 1.905*** |
| Fog - destination | 3.166*** | 2.967*** | 3.046*** | 2.807*** |
| Haze - origin | -0.447*** | 0.114 | 0.173* | 0.195** |
| Haze - destination | 0.003 | 0.801*** | 0.475*** | 0.493*** |
| Rain - origin | 0.012 | 0.306*** | 0.487*** | 0.357*** |
| Rain - destination | 0.624*** | 0.833*** | 0.818*** | 0.616*** |
| Snow - origin | 5.709*** | 5.593*** | 5.600*** | 5.477*** |
| Snow - destination | 1.108*** | 1.207*** | 1.874*** | 1.000*** |
| Thunder - origin | 5.069*** | 5.114*** | 4.430*** | 4.481*** |
| Thunder - destination | 5.304*** | 5.362*** | 3.650*** | 3.157*** |
| De-icing operations | 0.826*** | 1.940*** | 1.345*** | 1.330*** |
| Cindex | 0.102*** | | | |
| Constant | 0.598*** | 0.096 | 0.418*** | 0.500*** |
| Aircraft fixed-effects | | ✓ | ✓ | ✓ |
| Carrier fixed-effects | | ✓ | ✓ | ✓ |
| $R^2$ | 0.352 | 0.240 | 0.312 | 0.391 |
| Observations | 2526180 | 837,690 | 832,538 | 855,952 |

(a) Dependent variable: minutes of arrival delay of the flight.
(b) The standard errors, not reported to save space, are clustered by origin-date.
(c) ***, ** and * denote statistical significance at 1%, at 5% and at 10% level.

[1] The thirty-third and sixty-sixth percentile values of *Cindex* are respectively 29.21 and 59.95.

Because of the way in which *Cindex* is constructed, low values our connectivity index identify a network structure close to a H&S system, whilst high values of *Cindex* point towards a P2P system. Therefore, sub-sample A gathers airlines which operates under H&S system, sub-sample C is for the case of P2P, and sub-sample B represents the hybrid case.

The estimates of these sub-samples are presented in the last three columns of Table 3: (2) is for H&S, (3) for hybrid, and (4) for P2P. Column (1) reports also the full sample estimate including *Cindex* amongst the regressors. It is worth pointing out that *Cindex* is time-invariant within the panel and therefore its inclusion in the regression is possible after removing the panel fixed-effects. For this reason, column (1) should be considered with caution.

*Propagation* is statistically significant across all specifications. The estimated coefficient is positive and less than one, implying that only part of the incoming-aircraft delay is transferred to the observed flight delay.[1] This is because airlines are able to catch up with some of the accrued delay and thus tackle the delay propagation through their network. Interestingly, the monotonic increasing magnitude of *Propagation* as we move from column (2) to column (4), i.e. as we move from H&S systems towards P2P systems, indicates that H&S systems are more capable in reducing the delay propagation than hybrid and P2P systems. A delay of half hour of the incoming aircraft translates into six minutes of delay of the observed flight under a H&S system. This number increases to about eight minutes in the hybrid case and to about ten minutes in the P2P case.

Although the estimated impact of six to ten minutes may appear quantitative small in absolute terms, it is not in relative terms since the average delay in our sample is about five minutes (see **Table 57**), or more generally since almost for three quarters its density *Delay* is zero (*Delay* turns positive, to 1 minute, at the 71$^{st}$ percentile). Moreover, other empirical studies have documented smaller marginal effect on delay-related variables. For example, Forbes and Lederman (2010) find that the departure delay of regional carriers integrated with major airlines is on average 2.3 minutes shorter than that of major airlines using only independent regional carriers.

The reason why we find that the delay propagates more in a P2P network system may hinge on the fact H&S allows to better tackle the delay than P2P, both in the case the flight originates at the spoke and in the case the flight originates at the hub.

First, consider the spoke perspective. If a technical problem on a flight occurs on the spoke, the airline can send a replacement part or an engineer to fix it directly from the hub which, in a H&S network, is likely to be linked to the spoke by more than one flight per day. In a P2P system, instead, this is less likely to be possible. As an example, consider Genoa Airport in Italy, which is served, amongst others, by Air France and Ryanair. Air France has two flight connections from Genoa to Charles de Gaulle airport, the airline's hub. Ryanair instead flies to Genoa from Bari, London Stansted (one of the bases of Ryanair) and Trapani, and, more importantly, not always on a daily basis. If a technical problem occurs it becomes more difficult for Ryanair to solve it, because its technicians would have to fly in from London or, alternatively, they can arrive from

[1] This result is smaller than what found by Wang (2015) who conducted a similar study to ours using U.S. domestic market data.

Bergamo, the nearest operational base of Ryanair to Genoa. In any case, it will take longer time to solve the problem.

Second, consider the hub perspective. At the hub there are technicians and mechanical replacements, the airline can switch crew in case of shortage, or can find an aircraft replacement, e.g. if the arriving aircraft is very late or breaks down. For these reasons, if the flight originates at the hub, the airline is in a position to better reduce the delay propagation. We will refer to this effect as “hub externality”.

Finally, although column (1) ignores any panel specific heterogeneity, i.e. it does not control for the panel fixed effects, it shows that higher values of *Cindex*, i.e. network system closer to P2P, are associated to higher values of arrival delay, in line with joint findings of the other columns in the table.

As far as the effect of weather on delay is concerned, the expectations are met overall, since the weather variables are in practically all the cases positive and statistically significant. Recalling that clear sky is the reference category, the positive sign on each of the other weather variable indicates that worse meteorological conditions imply more delay. This effect is of stronger magnitude when the weather is expected to severely affect the journey, i.e. under fog (about five minutes of overall delay), snow (about 7 minutes overall) and thunder (about eight to ten minutes overall).[1] Moreover, in line with the expectations, de-icing operations are to increase the delay further, on average by less than two minutes.[2]

The positive sign on the departure time dummies indicates that early morning flights (the reference category) are the most punctual flights. Interestingly, and in line with the argument that airlines better manage the delay at their hubs, the pattern on these variables in the H&S column is different the columns of hybrid and P2P carriers, which instead behave similarly. If the flight operates under a P2P system or under a hybrid system, it scheduled to go sequentially from one place to another, without any possibility to break such sequence. The delay is more likely to propagate during the day as showed by the fact that the coefficient on *Morning* is smaller than the coefficient on *Afternoon,* which is smaller than the coefficient on *Evening*. Under a H&S system, instead, the airline may be able to reduce some of the delay propagation, as explained previously. Indeed, we observe that in the evening, when more short haul flights are likely to leave the hub to reach the spokes, the delay reduces relative to the afternoon flights, although it still remains higher than the delay on the early morning flights.

---

1 These number are obtained by adding the two estimated coefficients at origin and destination. For instance, 5 minutes of overall delay in case of fog come from 2 minutes at the departure and 3 minutes at the arrival.

2 With the exceptions of Bubalo and Gaggero (2015) and Alderighi and Gaggero (2018), the existing empirical literature on flight delay uses average weather conditions during the day, thus it is unable to quantify precisely the effect of weather disruption on airline operations. Think for example to a very foggy morning that clears out during the day: the on-time performance of morning flights will probably be very poor, despite the day will be on average reported as clear. Since we measure the weather condition within a +/-30-minute interval from the takeoff and landing scheduled time, we can identify with great accuracy, and therefore to quantity, the effect of weather on flight delay.

*Distance* is positive, showing that long-distance flights are on average less punctual than short distance flights. It is worth mentioning, however that this effect is practically negligible since a 1000 km increase in distance would lead to about half minute of delay.

Congested and larger airports, *Airport congestion* and *Airport volume* variables, are associated to more minutes of delay, in line with the expectations and with the findings of the existing empirical literature (Mazzeo, 2003; Bubalo and Gaggero, 2015).

*Route market share* is positive and statistically significant suggesting that the lower the flight substitutes available on the route, the worse is the quality of the service (Mazzeo, 2003). We acknowledge that this estimate could suffer from an endogeneity bias, if a better on-time performance increases the airline's market share. That is, if on one hand a monopolistic or highly concentrated routes may experience lower service quality, on the other hand longer delays deteriorate the reputation of the airline which may eventually lose market share. For this reason, the endogeneity that may arise is described as reverse causation or simultaneity.[1] The effect of service quality on market share, however, is not fully simultaneous, because it requires some time to take place. In other words, poor on-time performance at time *t* does not affect the market share at time *t* but can affect the market share at time *t + 1*. This reason, combined with the fact that our sampling window is relatively short for such a reverse causation to occur, implies that in our sample the endogeneity concern may not be too severe.

As far as the competition at the airport is concerned, in line with the findings on route market share, one would expect that a higher market share at the airport leads to more market power of the firms and therefore to less quality of the service, i.e. longer delay. This expectation is partially met at the destination; when we focus on the airport of origin, however, we observe a negative coefficient, which is statistically significant only in the case of H&S, column (2). This finding indicates that the higher the airport share at the origin, the lower the delay. Because H&S carriers have a very large share at their hub airport, *Airport mkt shr - orig.* may capture the aforementioned hub externality, if the flight departs from the hub the airline has more chances to absorb the delay.

In order to test the robustness of this the hub externality, we replace *Airport mkt shr - orig.* with a dummy equal to one if the airport of origin is the hub of the observed airline, *Hub - origin*, and re-estimate our model.[2] If our argument is correct we should observe a negative sign on *Hub - origin* for H&S sub-sample and a positive sign on P2P sub-sample, whilst we have no a priori expectations on the hybrid sub-sample, since the hub effect could be mitigated by the hybrid system.

As **Table 59** shows, *Hub - origin* is negative and statistically significant on the H&S sub-sample with a p-value equal to 5.2 percent, whilst it turns positive and statistically significant for the P2P sub-sample and also on hybrid sub-sample. This finding confirms our idea that H&S airlines are more effective in tackling the delay at their hub. For completeness the table reports also the pooled OLS estimate in column (5), where *Hub - origin* is found negative and statistically significant.

---

[1] A detailed discussion of this problem in regressions of fares on delays has been conducted by Forbes (2008), who studies the impact of an exogenous shock to product quality on the prices of 18 routes from LaGuardia Airport sampled in the years 1999 and 2000.

[2] The hub is identified by inspecting official documents of the airline or the company overview page provided by seatguru.com (see http://www.seatguru.com/browseairlines/browseairlines.php). An airline can be characterized by a single-city hub, like Air France with CGD, or by a dual-city hub, like Lufthansa with FRA and MUC.

**Table 59.** The effect of hubs on flight delay

| Carriers | (5) All | (6) H&S | (7) Hybrid | (8) P2P |
|---|---|---|---|---|
| Propagation | 0.687*** | 0.200*** | 0.251*** | 0.319*** |
| Morning | 0.456*** | 1.644*** | 1.669*** | 1.679*** |
| Afternoon | -0.304*** | 2.186*** | 2.054*** | 1.505*** |
| Evening | -0.194*** | 1.700*** | 2.559*** | 2.839*** |
| Distance | 0.077*** | 0.057*** | 0.076*** | 0.047*** |
| Route mkt share | -0.063 | 0.526*** | 0.395*** | 0.151** |
| Hub - origin | -0.309*** | -0.115* | 0.912*** | 0.474*** |
| Airport mkt shr - dest. | -0.808*** | -0.031 | 1.083*** | 0.354*** |
| Airport congestion | 0.033*** | 0.027*** | 0.049*** | 0.036*** |
| Airport volume | 0.003*** | 0.002*** | 0.001*** | 0.003*** |
| Cloudy - origin | -0.297*** | -0.011 | -0.053 | 0.121*** |
| Cloudy - destination | 0.205*** | 0.367*** | 0.392*** | 0.382*** |
| Foggy - origin | 1.397*** | 1.838*** | 1.955*** | 1.903*** |
| Foggy - destination | 3.181*** | 2.968*** | 2.999*** | 2.812*** |
| Hazy - origin | -0.435*** | 0.120 | 0.198** | 0.194** |
| Hazy - destination | 0.012 | 0.799*** | 0.467*** | 0.495*** |
| Rainy - origin | 0.002 | 0.306*** | 0.503*** | 0.356*** |
| Rainy - destination | 0.632*** | 0.835*** | 0.816*** | 0.615*** |
| Snowy - origin | 5.704*** | 5.588*** | 5.595*** | 5.477*** |
| Snowy - destination | 1.109*** | 1.204*** | 1.839*** | 1.007*** |
| Thundery - origin | 5.064*** | 5.114*** | 4.465*** | 4.483*** |
| Thundery - destination | 5.311*** | 5.365*** | 3.645*** | 3.157*** |
| Deicing operations | 0.809*** | 1.937*** | 1.361*** | 1.311*** |
| Cindex | 0.095*** | | | |
| Constant | 0.528*** | -0.007 | 0.016 | 0.427*** |
| Aircraft fixed-effects | | ✓ | ✓ | ✓ |
| Carrier fixed-effects | | ✓ | ✓ | ✓ |
| $R^2$ | 0.352 | 0.240 | 0.309 | 0.391 |
| Observations | 2,526,180 | 837,690 | 832,538 | 855,952 |

(a) Dependent variable: minutes of arrival delay of the flight.
(b) The standard errors, not reported to save space, are clustered by origin-date.
(c) ***, ** and * denote statistical significance at 1%, at 5% and at 10% level.

It is also worth noticing that the inclusion of *Hub - origin* does not alter the results on the other regressors: that it, no qualitative discrepancy is observed with respect to **Table**

**58**, except for a better level of statistical significance on the airport market share at the destination in column (7).

Finally, we test whether the estimated coefficients of the sub-samples are statistically different, both for **Table 58** and for **Table 59**. The results, reported in the Appendix, show that the difference on *Propagation*, the variable of interest, remains always statistically significant from zero below the 1% threshold.

## 11.6 Conclusion

In this paper we have studied how a different network structure can impact the quality of service, measured in minutes of delay. We have shown that the distinction between a pure hub-and-spoke (H&S) structure and a pure point-to-point (P2P) structure is an oversimplification. Network structures compared between the two business models FSC and LCC reveal certain similarities, but also differences.

For this reason, we have first constructed a connectivity index based on the entire airline network (short-haul and long-haul destinations) to measure the extent to which the network structure of each airline can be classified closer to H&S or closer to P2P. Then, we have focused our analysis on the short-haul market and considered only the routes within Europe, to test whether network structures closer to H&S are more efficient in managing delay, than network structures closer to P2P. We recognized some limitations regarding our descriptor for network structure. In future research we plan to investigate, if alternative measures are better suited to describe interconnectedness in networks.

The novelty of our data is that we were able to trace the aircraft through the carrier's network and thus we were able to give an estimate for the delay propagation. Using European data, we find that airlines succeed in reducing the delay propagation on their network, however, and more importantly, H&S and P2P structures do not have the same impact on delay. Our econometric analysis indicates that the closer a carrier operates to a H&S structure, the better it internalizes congestion, measured in delay. We explain this result mainly with the fact that if the flight originates at the airline's hub, the carrier can better reduce or absorb the delay. This because at its hub the airline can switch crew in case of shortage; it can more easily find a plane to substitute the arriving aircraft if the latter is very late or breaks down, so the scheduled flight cannot be operated by that aircraft. Moreover, the airline's technicians are based at the hub and therefore are quickly ready to work on any technical problem that may arise. Critical replacement parts may be stored at such a hub if necessary.

Future work could replicate the present analysis for other continents, for instance considering the U.S. domestic market, for which the U.S. Department of Transportation provides several statistics on the on-time performance of the carriers. A further research could also be based on the long-haul market, although the empirical analysis may be more difficult, because it should somehow consider that airlines can also re-route passengers to handle the delay or a missed connection. This concern is less severe in the short-haul market, as we did in our analysis.

Finally, and possibly related to the long-haul market, it will be interesting to include in the model the effect of global airline alliances, as recent literature shows that global airline alliances may affect the quality of service (Alderighi and Gaggero, 2018; Brueckner and Flores-Fillol, 2018).

## Appendix

**Table 60** reports the *t*-statistics of the test of the difference of the estimated coefficients across columns (2) to (4) of **Table 58**. Because the statistical difference can be tested between the estimates of the same coefficient through two different models, we have three possible combinations (H&S versus Hybrid, H&S versus P2P, and Hybrid versus P2P) and, therefore, three columns in the table. The null hypothesis to be tested is that the difference between the estimated coefficients is zero, i.e. that the two estimated coefficients between a pair of columns are statistically the same.

**Table 60.** Statistical difference of the estimated coefficients in **Table 58**

| | H&S-Hybrid | H&S-P2P | Hybrid-P2P |
|---|---|---|---|
| Propagation | -6.587*** | -15.633*** | -10.168*** |
| Morning | -0.038 | -0.413 | -0.381 |
| Afternoon | 1578 | 9.156*** | 8.061*** |
| Evening | -10.165*** | -13.76*** | -3.459*** |
| Distance | -2.527** | 1.711* | 4.592*** |
| Route market share | 0.057 | 3.683*** | 3.666*** |
| Airport mkt shr - orig. | -1.858* | -1587 | 0.588 |
| Airport mkt shr - dest. | -1.731* | -3.19*** | -1.13 |
| Airport congestion | -5.664*** | -2.075** | 3.167*** |
| Airport volume | 3.906*** | -1.929* | -5.416*** |
| Cloud - origin | 1.03 | -2.371** | -3.497*** |
| Cloud - destination | -0.746 | -0.348 | 0.428 |
| Fog - origin | -0.184 | -0.316 | -0.123 |
| Fog - destination | -0.426 | 0.924 | 1.31 |
| Haze - origin | -0.434 | -0.63 | -0.166 |
| Haze - destination | 2.839*** | 2.831*** | -0.164 |
| Rain - origin | -1.838* | -0.531 | 1.544 |
| Rain - destination | 0.198 | 3.083*** | 2.934*** |
| Snow - origin | -0.014 | 0.214 | 0.268 |
| Snow - destination | -2.729*** | 0.788 | 3.388*** |
| Thunder - origin | 0.613 | 0.627 | -0.067 |
| Thunder - destination | 3.306*** | 4.498*** | 1.094 |
| Deicing operations | 4.271*** | 4.098*** | 0.11 |
| Constant | -2.117** | -2.772*** | -0.62 |

(a) Null hypothesis: $\beta_i - \beta_j = 0$, where $\beta$ is the estimated coefficient of **Table 58**; *i* and *j* correspond to the columns of **Table 58**, identified respectively by the prefix and suffix of the column heading in the present table.
(b) The table reports the *t*-statistics, while ***, ** and * respectively denotes the statistical significance at 1%, at 5% and at 10% level.

As **Table 61** shows, the effect of the variable of interest, Propagation, is found to be statistically different (below a 1% significance level) across the three models. As far as the other regressors are concerned, their behavior is in line with the expectations. First, the difference between the estimated coefficients on the weather variables is often statistically not significant. This result makes sense, because weather affects the on-time performance of any carrier, irrespective of its H&S, Hybrid or P2P classification.

**Table 61.** Statistical difference of the estimated coefficients in **Table 59**

| | H&S-Hybrid | H&S-P2P | Hybrid-P2P |
|---|---|---|---|
| Propagation | -6.671*** | -15.641*** | -10.086*** |
| Morning | -0.365 | -0.541 | -0.154 |
| Afternoon | 1.654* | 9.103*** | 7.931*** |
| Evening | -10.249*** | -13.89*** | -3.498*** |
| Distance | -3.686*** | 1.992** | 6.158*** |
| Route market share | 1349 | 3.888*** | 2.573** |
| Hub - orig. | -13.574*** | -5.651*** | 4.475*** |
| Airport mkt shr - dest. | -8.35*** | -3.369*** | 7.112*** |
| Airport congestion | -4.883*** | -1.891* | 2.622*** |
| Airport volume | 4.349*** | -1.935* | -5.877*** |
| Cloud - origin | 0.708 | -2.36** | -3.151*** |
| Cloud - destination | -0.528 | -0.311 | 0.237 |
| Fog - origin | -0.495 | -0.296 | 0.251 |
| Fog - destination | -0.166 | 0.901 | 1.026 |
| Haze - origin | -0.571 | -0.571 | 0.038 |
| Haze - destination | 2.899*** | 2.799*** | -0.261 |
| Rain - origin | -2.012** | -0.531 | 1.748* |
| Rain - destination | 0.267 | 3.126*** | 2.905*** |
| Snow - origin | -0.014 | 0.205 | 0.257 |
| Snow - destination | -2.601*** | 0.748 | 3.225*** |
| Thunder - origin | 0.582 | 0.626 | -0.023 |
| Thunder - destination | 3.326*** | 4.505*** | 1085 |
| Deicing operations | 4.143*** | 4.207*** | 0.374 |
| Constant | -0.165 | -3.171*** | -3.361*** |

(a) Null hypothesis: $\beta i - \beta j = 0$, where $\beta$ is the estimated coefficient of **Table 59**; *i* and *j* correspond to the columns of **Table 59**, identified respectively by the prefix and suffix of the column heading in the present table.
(b) The table reports the *t*-statistics, while ***, ** and * respectively denotes the statistical significance at 1%, at 5% and at 10% level.

Second, the effect of the time of departure is statistically different across the models, but only later in the day. No statistical difference is found for morning and partially afternoon flights. This finding is in line with the presumption that different network structures handle delay differently and the effect of propagating delay strengths on late-in-the-day flights.

Third, the stars denoting statistical significance difference across the models are frequently observed on airport/route variables, reflecting the different impact that airport/route variables may have on delay depending on the carrier's classification.

Finally, the results do not change when the statistical difference of the estimated coefficients is checked for columns (6) to (8) of **Table 59** (see **Table 61**).

# 12 Conclusions

In this dissertation, several innovative modelling techniques and methodologies were presented, such as airport simulation and benchmarking based on performance measures, to tackle management problems at airports. There is still some more work to do to fully understand the complexity of an airport. In modelling for performance evaluation purposes, the adequate system boundaries must be defined to make the solving of certain problems in air transportation manageable. Similar considerations must be made when choosing the right time frame regarding a benchmarking analysis. Questions to be answered are: Which timeline of historic financial and operations data is available and could be acquired in which time? How much historic data is necessary to answer the posed questions? Is it necessary to start a three-year research project, if a question could be solved on the "back of an envelope" with sufficient certainty? What granularity, level of detail and level of aggregation is necessary for our analysis?

When the airport system boundaries for answering a certain performance-related question are not clearly defined, the amount of information to build a model and for analysis could be overwhelming. However, certain long-range *strategic* problems, such as found in airport master planning, require longer planning durations and lots of interrelated information. It is, therefore, recommended to use more sophisticated methods such as simulation, optimization, or heuristic methods to solve certain problems.

Through simulation models extensive knowledge about process details in aircraft operations can be gathered. Like passenger flow simulations (that have not been covered in this dissertation) the dynamics of the system, involving either passengers or aircraft or both, can be closely observed. From these model observations it is evident where in the system queues build up when demand is larger than service capacity at a certain time. It is the responsibility of the airport managers to provide sufficient capacity for its current and future users. By reducing waiting time, the service quality increases.

Passengers are more satisfied with the provided service when they experience less waiting through all required processes in the terminal or on board and aircraft burn less fuel and emit less emissions when queuing and waiting on the airside is reduced. At the same time airport managers must find ways to make a profit from its operations by exploring innovative paths in finding alternative sources of income besides airport charges. Some airports are successful with attracting additional general aviation traffic, other airports provide facilities for air cargo logistics and many airports serve as shopping or entertainment centers.

Two case studies involve approaches to solve capacity related problems, the capacity estimation in the master planning for the new Berlin-Brandenburg airport prior its opening in October 2020 (Chapter 8) and the step wise capacity expansion of Oslo-Gardermoen airport until 2050 (Chapter 9). Simulation tools[1] are applied to these problems to support and validate the master planning process. It is possible to measure waiting time spans and total travel time on segments that guide the flow of aircraft.

[1] We used the simulation and modeling tool SIMMOD, that is developed by the FAA, and its derivatives VisualSIMMOD and SIMMODPlus. There exist, however, several different

*Delay* is the main output for airport capacity analyses because queues of aircraft waiting for service could impact the airside flow of traffic and accumulate significant delay costs, from additional fuel burn and other operating or passenger opportunity costs. It is evident from observations in reality and in models that when capacity is reached delays increase exponentially.

In Chapter 9 the Oslo-Gardermoen simulation model is used to simulate the effect on total aircraft delay by resequencing flights waiting for push back service. The simulation outputs sufficient information to calculate aircraft ground emissions while taxiing or while waiting in idle position for a push back truck. By connecting data sheets on aircraft engine fuel burn flow and emissions composition for each stage of flight in the post-processing total aircraft ground emissions of several different emitted gases are estimated for an airport.

In Chapter 4 and in Chapter 11 statistical methods are applied to analyze the data and to derive our results and recommendations on competition and service quality. In Chapter 5 we present a new method in airport benchmarking, where we show how to construct and interpret the *profitability envelope* (or *profitability frontier*), which is based on prior work of some of my colleagues regarding the Data Envelopment Analysis.

In Chapter 6 we apply our own version of a single industry multi country and multi airport *Input-Output Analysis*, a method developed by Prof. Wassily Leontief to model industry interconnections. In this chapter we analyzed the interconnections of airports in the Baltic Sea Region served by air cargo carriers. We calculated the flow of cargo aircraft between airports and the degree of dominance of airports serving air cargo traffic in the region.

We examine larger air transport systems, such as the whole of Europe (Chapter 4, Chapter 5 and Chapter 11) the Baltic Sea Region (Chapter 6) or very remote European regions, such as the North of Norway (Chapter 7). For most studies we collected our own data from public sources on the Internet. We use computer algorithms to collect recent, sometimes real-time, data that we could not obtain from other sources.

The tools that are presented can be used to make operations more efficient. Extensive data, for example traffic volume and delay observations, are used to explain inefficiencies that were identified during our research at the airports under study.

This dissertation highlights a broad variety of different topics in global and European air transport. Certainly, the saying “if you can’t measure it, you can’t manage it” by Dr. W. Edwards Deming is true in many ways, especially in our approach to *airport benchmarking*. Although we present new ways to collect data, such as by automatic query methods of information from the Web, we must apply some sound human judgement to derive the answers to our problems at hand. Measures calculated from the data might be an indication that some processes operate inefficiently, but we need to look closely into each case to find the root causes why performance measures of some airports are worse than compared to the best-in-class of their peers. In practice we found simulation tool, such as CAST (Comprehensive Airport Simulation Technology), TAAM (Total Airspace and Airport Modeller) and AirTOp (Air Traffic Optimization).

cases where poor measurements and inadequate analyses were officially published. To solve data problems, it is advised to communicate directly with airport management. We also recommend seeking for several sources of similar data to be able to make some cross and plausibility checks. In treating financial data *checksums* have been used to calculate and compare these data to the published totals. Sometimes our calculated ratios revealed outliers compared to our expectations. Airport management has a much better insight into their own processes and data collection, it could therefore assist in finding the causes for bad behaving ratios or outliers in the data.

Another problem is the publication of data at different levels of aggregation. Especially when dealing with financial data we ran into difficulties, when we tried to disaggregate data for specific analyses. When data is highly aggregated it could be difficult to find the main performance drivers. Most public sources publish, for example, just total annual costs and revenues of an airport, but analyses often require financial data broken down into smaller segments, for example, aeronautical and non-aeronautical revenues, to get better insight into airport financial performance. When airports across different nations in Europe or internationally are compared, we need to make sure that the data is in the best way comparable. We developed our own methods to adjust financial data. Financial data is adjusted relative to a base year, base country, and base currency. We commonly use tables on inflation, exchange rate and purchasing power parity (PPP) by the *World Bank*, *International Monetary Fund* (IMF) or by *Eurostat*.

In many cases it is possible to collect a range of data and indicators on financial and operational performance of airports. Sometimes the amount of data is overwhelming, sometimes available data is very scarce. By building automatically collecting algorithms, i.e. *web crawlers*, valuable data from the Internet could be obtained[1]. However, the dynamics of economics and air transport traffic are quite different from each other. In the case of income and cost data of airports, we commonly had only annual data available, so that data was published and collected for several subsequent years, whereas in the case of operational data, public data with different temporal granularity aggregated by hour, day or month was available. We chose the best suited data depending on the purpose of the analysis.

For most recent studies flight position data was used to get a better understanding of the flow of aircraft in the vicinity of airports. Aircraft navigation data that is broadcasted from the ADS-B transmitter on board of most civil aircraft includes aircraft type, coordinates, height, speed, altitude and heading. We plan to work more extensively with such data in the future for even better studies on financial and operational performance of airports, considering the airport and airspace capacity, aircraft delay, and environmental impact.

---

[1] In most cases the collected data was "raw" in a way that we had to transform it into a usable format by text, symbol, and number manipulation. To achieve that we used our own text cleaning algorithms and string manipulation formulas. Our data sets are as complete as they are presented to the public. Limitations are therefore mostly technical, when data was not fully reported or transmitted from the original source.

# 13 Addendum

In this addendum the single most important data collection method for this dissertation is presented. We made use of this or similar *web crawlers* as tools to assemble a large data collection that we used for our analyses. This code was first written and used end of 2008 for my diploma thesis. The web crawlers were used to collect airport schedule data, weather data and aircraft specific data.

Furthermore, airport coordinates and flight positions were assembled, but the program could be used for various purposes. The most extensive data collection was conducted for the paper "Low-cost carrier competition and airline service quality in Europe" which we presented in Chapter 4. For this study more than 12 million flights were observed during a period of 20 months. For a more recent study we collected more than 17 million flight observations over a period of ten months. For recent studies we automatized the data collection as far as possible, so we entered the realms of *Big Data*. Usually we just had to resume the program, when the Internet connection was lost, or the software had crashed. We wrote the code in Visual Basic for Applications (VBA) within Microsoft Excel. It was therefore possible to download data directly into spreadsheets. In later steps we added algorithms either for data cleaning and formatting or for further text and number manipulation.

The name of the program which was mostly used for crawling data from the Internet is "Airport Flight Plan Query". Two main inputs are needed: 1.) A list of airports identified by their three-letter IATA Code and 2.) the dates that are required for the analysis (**Table 62**). The website that we downloaded the data from will not be disclosed here, but we recognize that many researchers nowadays exploit sites and adequate security measures were put in place to protect against data scraping. It is recommended not to stress the webservers too much. One query per one to ten seconds to the servers might be an adequate limit. When the web crawler would collect the data too fast, a random time variable of at least 1 second was added to add some respite time for the webservers. We tried a "gentle" approach in the data collection.

As can be seen in **Table 62** the downloaded data was formatted into tabular format with the following headers for our output variables, e.g. origin, destination, scheduled departure or arrival times, actual departure or arrival times, flight number, aircraft type, delay and flight status. Four different worksheets had been used to "translate" the raw data into the final table by using Microsoft Excels internal formulas for time, date, and text manipulation. This temporary data is cleared before a new query takes place. The flow chart in **Fig. 99** displays the internal working of the algorithm.

Regarding the METAR (Meteorological Aerodrome Report) weather data, we needed a list of dates and airports as well, with the slight difference that we needed the airport codes as four-letter ICAO codes. It was possible to collect airport specific data on meteorological conditions, such as temperature, humidity, air pressure, visibility, wind direction, wind speed or general description of weather conditions.

The weather data was matched to the flight schedule data by (universal) time, date, and airport code. This data was combined into one dataset and then analyzed.

**Table 62.** Workbooks and sheets of the "Flightplan" and weather data web crawlers

| Workbook | Worksheet | Input Variables | Output Variables |
|---|---|---|---|
| AirportFlightPlanDataQuery.xlsm | Sample | IATA airport code<br>Date | |
| | Sheet 1 - 4 | | (Temporary Data) |
| | Date | | IATA Code_Date<br>Origin<br>Destination<br>Airline<br>Flight Number<br>Scheduled Time<br>Actual Time<br>Gate<br>Aircraft Type<br>Status Text<br>Arrival/Departure Flag<br>Aircraft Weight Class<br>Declared Delays >15 Min<br>Seats per Aircraft<br>Airline Name<br>Destination City Name |
| | Date_weather | | ICAO Code_Date<br>Time (CET)<br>Temperature<br>Windchill<br>Dew Point<br>Humidity<br>Pressure<br>Visibility<br>Wind Direction<br>Wind Speed<br>Gust Speed<br>Precipitation<br>Events<br>Conditions |

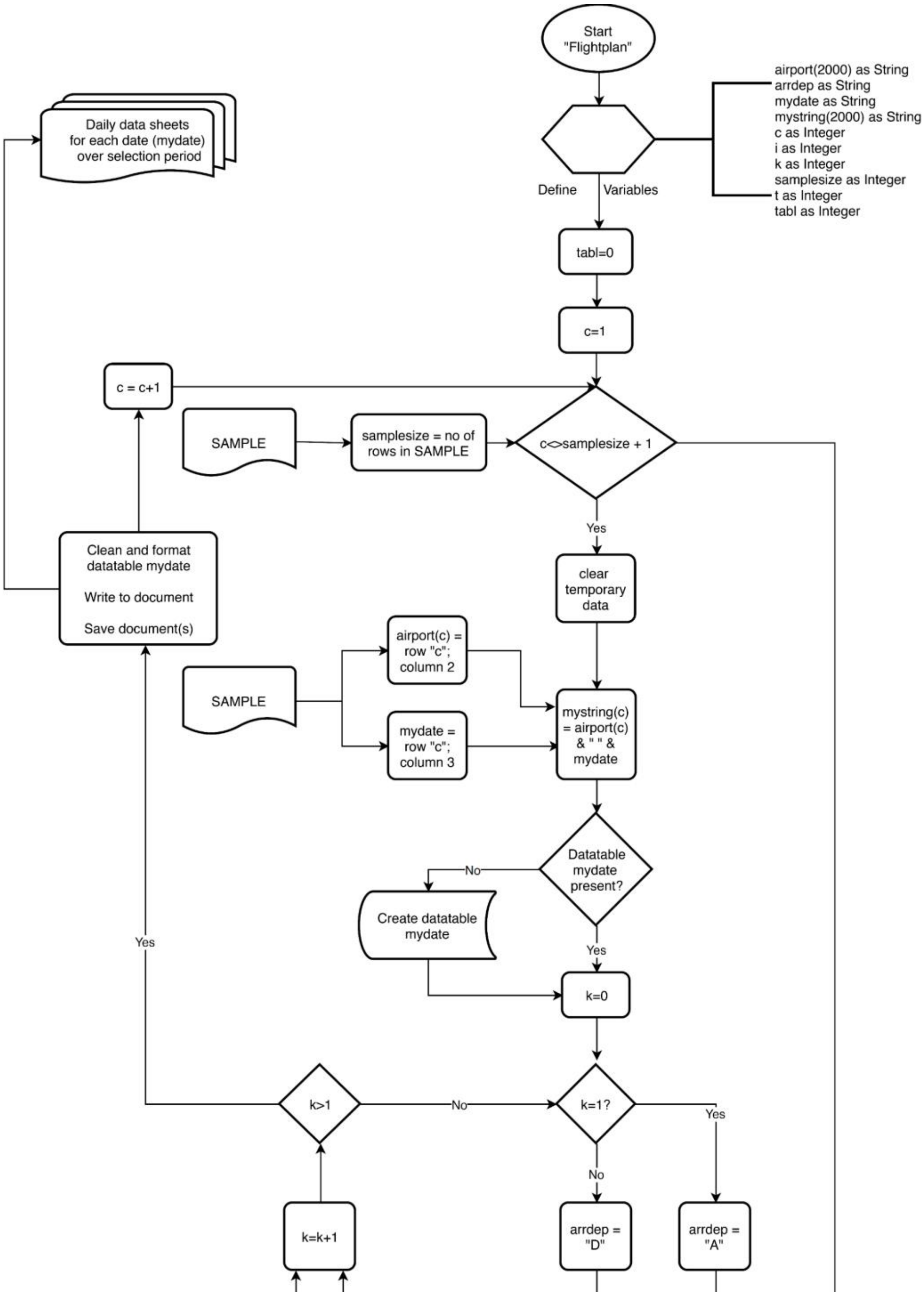
Start
"Flightplan"
airport(2000) as String
arrdep as String
mydate as String
mystring(2000) as String
c as Integer
i as Integer
k as Integer
samplesize as Integer
t as Integer
tabl as Integer
Define
Variables
Daily data sheets
for each date (mydate)
over selection period
tabl=0
c=1
c = c+1
SAMPLE
samplesize = no of rows in SAMPLE
c<>samplesize + 1
Yes
Clean and format datatable mydate
Write to document
Save document(s)
clear temporary data
airport(c) = row "c"; column 2
SAMPLE
mydate = row "c"; column 3
mystring(c) = airport(c) & " " & mydate
Datatable mydate present?
No
Create datatable mydate
Yes
Yes
k=0
k>1
No
k=1?
Yes
No
k=k+1
arrdep = "D"
arrdep = "A"

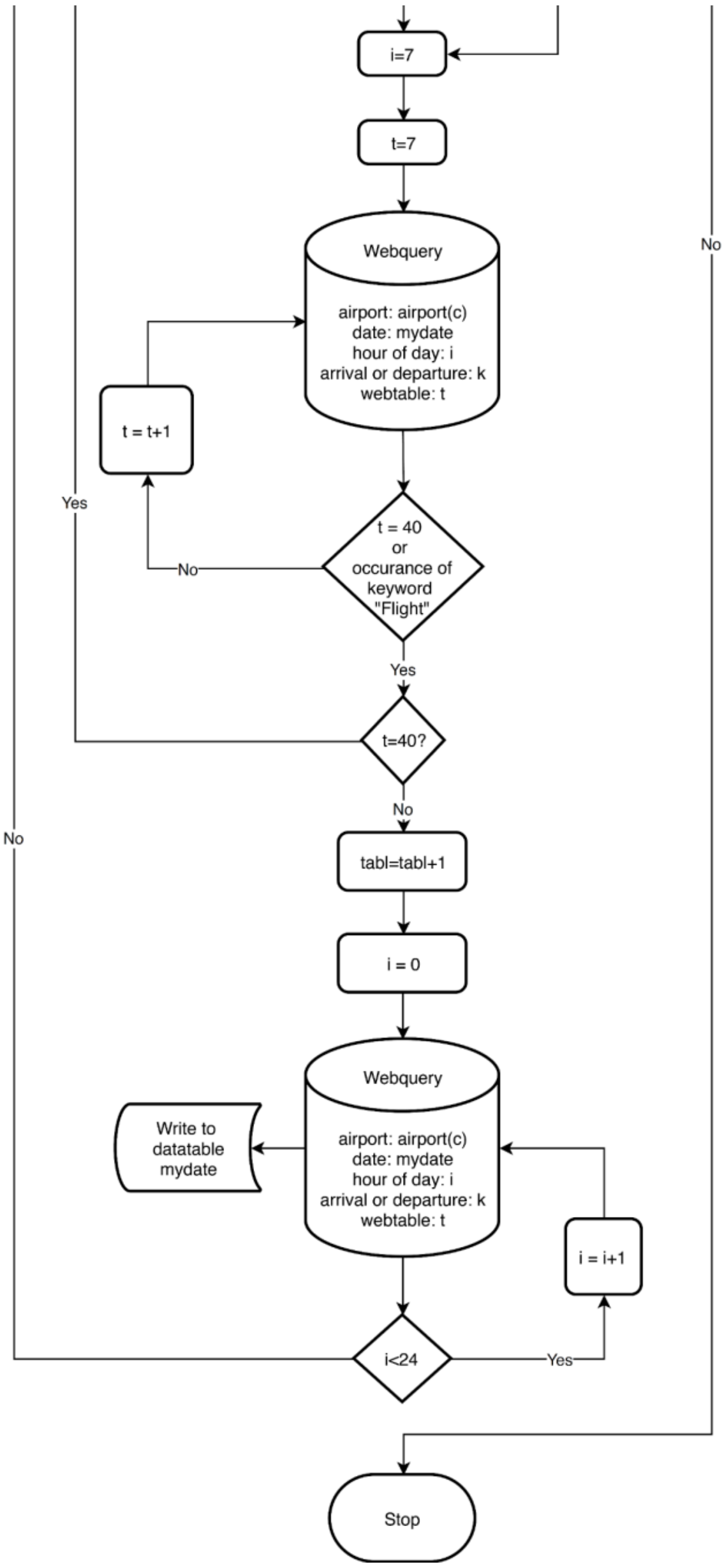

**Fig. 99.** The “Flightplan” webcrawler as Flow Chart.

## Overview of published chapters and share of research work

The following **Table 63** provides the work share done by the author in collaboration with other researchers as part of this dissertation. We list the title of each publication, the publication status (if, where and when the manuscript had been published) and the work share done by each of the authors.

**Table 63.** Published chapters and research share

| **Title of the chapter and author(s)** | **Publication status** | **Work share** |
|---|---|---|
| "Airport evolution and capacity forecasting" by Branko Bubalo | Submitted to *Research in Transportation Business & Management* Volume 1, Issue 1, in 2011, but the manuscript was rejected. http://www.gapprojekt.de/downloads/gap_papers/airportevolution.pdf | Bubalo 100% |
| "Airport punctuality, congestion and delay– The scope for benchmarking" by Branko Bubalo | Published in the *Aerlines* magazine, volume 50 in June 2011. https://www.scribd.com/document/116741160/50-Bubalo-Airport-Punctuality-Congestion-and-Delay | Bubalo 100% |
| "Airside productivity of selected European airports" by Branko Bubalo | Published in the *conference proceedings of the Air Transport Research Society* (ATRS) World Conference 2010. http://www.gapprojekt.de/downloads/gap_papers/airsideprod.pdf | Bubalo 100% |
| "Low-cost carrier competition and airline service quality in Europe" by Alberto Gaggero and Branko Bubalo | Published in *Transport Policy*, Volume 43 in October 2015. https://doi.org/10.1016/j.tranpol.2015.05.015 | Gaggero 50%, Bubalo 50% |
| "Benchmarking European airports based on a profitability envelope" by Branko Bubalo | Published in *Lecture Notes in Computer Science,* Volume 7555, Computational Logistics in 2012. https://doi.org/10.1007/978-3-642-33587-7_13 | Bubalo 100% |
| “Economic Outlook for the Air Cargo Market in the Baltic Sea Region” by Branko Bubalo | Published as a book chapter in *Air Cargo Role for Regional Development and Accessibility in the Baltic Sea Region – Handbook of the EU projects Baltic Bird and Baltic.AirCargo.Net in the framework of the Baltic Sea Region Programme 2007–2013* by Berliner Wissenschaftsverlag (BWV) in 2014. https://www.bwv-verlag.de/detailview?no=2032 [last accessed 03.04.2020] | Bubalo 100% |
| “Social costs of public service obligation routes - calculating subsidies of regional flights in Norway” by Branko Bubalo | Published in *NETNOMICS*, Volume 13, Issue 2 by Springer in March 2013. https://doi.org/10.1007/s11066-013-9074-8 | Bubalo 100% |
| “Airport capacity and demand calculations by simulation—the case of Berlin-Brandenburg International Airport” by Prof. Dr. Joachim Daduna and Branko Bubalo | Published in *NETNOMICS*, Vol. 12, Issue 3 by Springer in January 2012. https://doi.org/10.1007/s11066-011-9065-6 | Daduna 50%, Bubalo 50% |
| “Simulating airside capacity enhancements at Oslo-Gardermoen airport post-2017 - Measuring the effect on level of service by adding a pair of high-speed runway exits” by Branko Bubalo | Internal scientific report published by AVINOR in December 2016. | Bubalo 100% |
| "Reducing Airport Emissions with Coordinated Pushback Processes: A Case Study" by Frederik Schulte, Prof. Dr. Stefan Voß and Branko Bubalo | Published in *Lecture Notes in Computer Science* (LNCS), vol. 10572 by Springer in October 2017. https://doi.org/10.1007/978-3-319-68496-3_38 | Schulte 33,3%, Voß 33,3%, Bubalo 33,3% |
| "Service quality and degree of interconnectedness in air transport networks" by Alberto Gaggero and Branko Bubalo | Submitted to *Transportation Research Part A: Policy and Practice* in December 2019 and is currently under revision. | Gaggero 50%, Bubalo 50% |

## Affidavit

I, Branko Bubalo, hereby declare, in lieu of an oath, that I have written the dissertation titled

„*Airport Capacity and Performance in Europe - A study of transport economics, service quality and sustainability*"

autonomously - and if in cooperation with other scientists as described in **Table 63** according to § 6 Abs. 4 of the doctoral regulations of the Faculty of Business Administration dated July 9, 2014 and that I did not use any other aids than those I indicated herein.

The parts taken literally or by sense from other works than mine are marked as such. I assure that I did not take advantage of any commercial doctoral consultation nor was my work accepted or judged insufficient in an earlier doctoral procedure at home or abroad.

Berlin, August 13th, 2020

______________________ ____________________________________

Place, Date Signature